\newcommand{\ifdraft}{nodraft}

\newcommand{\ifpsdraft}{nodraft}

\newcommand{\ifpslabel}{nolabel}

\newcommand{\ifchsign}{nochsign}
 
\def\twopagedg{1}

\def\draftlabel{0} 

\newcommand{\draftnr}{$3.1$}
\newcommand{\vdate}{December 1996}
\newcommand{\vthdate}{December 1996}
\newcommand{\cernnr}{96--52}
\newcommand{\beq}{\begin{equation}}
\newcommand{\eeq}{\end{equation}}
\newcommand{\beqn}{\begin{eqnarray}}
\newcommand{\eeqn}{\end{eqnarray}}

\newcommand{\epem}{$\mbox{\it e}^+ \mbox{\it e}^-${}}

\newcommand{\nonu}{\nonumber\\}
\newcommand{\porder}[1]{\mbox{${\cal O}(#1)$}}

\newcommand{\GeV}{\mbox{GeV}}

\newcommand{\dd}{\mbox{d}}
\newcommand{\dps}{\mbox{dPS}}

\newcommand{\qb}{\overline{q}}
\newcommand{\Qb}{\overline{Q}}
\newcommand{\msbar}{$\overline{\mbox{MS}}$}
\newcommand{\xiM}{\xi_{\scriptscriptstyle M}}
\newcommand{\XB}{{x_B}}
\newcommand{\SH}{S_H}
\newcommand{\PP}{P}
\newcommand{\hh}{h}
\newcommand{\EE}{P_0}
\newcommand{\EH}{h_0}
\newcommand{\Pii}{P}
\newcommand{\IHQ}{{\mbox{\scriptsize (IHQ)}}}
\newcommand{\sIHQ}{{\mbox{\tiny (IHQ)}}}
\newcommand{\lettlab}{\rm}
\newcommand{\zshortparallel}{
{\scriptscriptstyle\parallel}
}

 \newread\iteminfile
 \openin\iteminfile\jobname.ite  
 \ifeof\iteminfile
 \else
    \closein\iteminfile
    \input\jobname.ite
 \fi
              
\newcounter{itemcounter}
\newwrite\itemfile 
\immediate\openout\itemfile=\jobname.ite
              
\newcommand{\footn}[1]{
\immediate\write\itemfile{\string\def
\string\itemlabel#1{\arabic{footnote}}}}
\def\ifundefined#1{\expandafter\ifx\csname#1\endcsname\relax}
                           
\newcommand{\itemr}[1]{\ifundefined{itemlabel#1}\message
{ <:\string#1: undefined> }???\else\csname itemlabel#1\endcsname\fi}

\def\dgcleardoublepage{
\clearpage
\message{cleardoublepage (\arabic{page})}
\ifodd\value{page}
\message{******************* isodd *************************}
\else
\message{******************* iseven ************************}
\markh{}
\thispagestyle{empty}
\cleardoublepage
\fi
}

\newcommand{\fullline}{
\unitlength0.4mm
\begin{picture}(13,4)
\linethickness{0.3mm}
\put(-1,2.0){\line(1,0){15}}
\thinlines
\end{picture}
}

\newcommand{\dashline}{
\unitlength0.4mm
\begin{picture}(20,4)
\linethickness{0.3mm}
\put(-1,2.0){\line(1,0){4}}
\put(8,2.0){\line(1,0){4}}
\put(17,2.0){\line(1,0){4}}
\thinlines
\end{picture}
}

\newcommand{\dotline}{
\unitlength0.4mm
\begin{picture}(9,4)
\linethickness{0.3mm}
\put(-1,2.0){\line(1,0){1}}
\put(4,2.0){\line(1,0){1}}
\put(9,2.0){\line(1,0){1}}
\thinlines
\end{picture}
}

\newcommand{\longdashline}{
\unitlength0.4mm
\begin{picture}(22,4)
\linethickness{0.3mm}
\put(-1,2.0){\line(1,0){10}}
\put(13,2.0){\line(1,0){10}}
\thinlines
\end{picture}
}

\newcommand{\dashdotline}{
\unitlength0.4mm
\begin{picture}(17,4)
\linethickness{0.3mm}
\put(-1,2.0){\line(1,0){5}}
\put(8,2.0){\line(1,0){1}}
\put(13,2.0){\line(1,0){5}}
\thinlines
\end{picture}
}

\newcommand{\dotdotline}{
\unitlength0.4mm
\begin{picture}(12,4)
\linethickness{0.3mm}
\put(-1,2.0){\line(1,0){1}}
\put(2,2.0){\line(1,0){1}}
\put(9,2.0){\line(1,0){1}}
\put(12,2.0){\line(1,0){1}}
\thinlines
\end{picture}
}

\newcommand{\dashdotdotline}{
\unitlength0.4mm
\begin{picture}(11,4)
\linethickness{0.3mm}
\put(-1,2.0){\line(1,0){5}}
\put(8,2.0){\line(1,0){1}}
\put(11,2.0){\line(1,0){1}}
\thinlines
\end{picture}
}

\newcommand{\dashdashdotdotline}{
\unitlength0.4mm
\begin{picture}(22,4)
\linethickness{0.3mm}
\put(-1,2.0){\line(1,0){5}}
\put(9,2.0){\line(1,0){5}}
\put(19,2.0){\line(1,0){1}}
\put(22,2.0){\line(1,0){1}}
\thinlines
\end{picture}
}

\newcommand{\dashdashline}{
\unitlength0.4mm
\begin{picture}(21,4)
\linethickness{0.3mm}
\put(-1,2.0){\line(1,0){3}}
\put(4,2.0){\line(1,0){3}}
\put(14,2.0){\line(1,0){3}}
\put(19,2.0){\line(1,0){3}}
\thinlines
\end{picture}
}

\newcommand{\fullcircle}{
$\bullet$
}

\newcommand{\opencircle}{
$\circ$
}

\newcommand{\fulltriangle}{
$\blacktriangle$
}

\newcommand{\opentriangle}{
$\vartriangle$
}

\newcommand{\fullsquare}{
${\scriptscriptstyle\blacksquare}$
}

\newcommand{\opensquare}{
${\scriptscriptstyle\square}$
}

\newcommand{\dgpicture}[2]{
\begin{picture}(#1,#2)
\thicklines
{\ifthenelse{\equal{\ifdraft}{draft}}%
{
\put(0.,0.){\dashbox(#1,#2){}}
}%
{}%
}
}

\newcommand{\dgbullet}{{\vspace{0.2ex}$\bullet${}}}
\newcommand{\dgbulleta}{---}

\newcommand{\chsign}[1]{%
{\ifthenelse{\equal{\ifchsign}{chsign}}%
{
{\sf {\Large$\bullet$$\bullet$$\bullet$} #1 
                         {\Large$\bullet$$\bullet$$\bullet$} }%
}%
{}%
}}

\newcommand{\smallmark}[1]{
\marginpar{\fbox{\vspace{0.0cm}{\scriptsize #1}}}}

\newcommand{\labelm}[1]{%
\label{#1}%
\ifthenelse{\equal{\ifdraft}{draft}}%
{\smallmark{#1}}%
{}%
}

\newcommand{\labelmm}[1]{%
\label{#1}%
\ifthenelse{\equal{\ifdraft}{draft}}%
{\protect\fbox{\sf #1}}%
{}%
}

\newcommand{\labelmmm}[1]{%
\label{#1}%
}

\newcommand{\beqm}[1]{%
\ifthenelse{\equal{\ifdraft}{draft}}%
{\smallmark{#1}}%
{}%
\beq \label{#1}}

\newcommand{\beqnm}[1]{%
\ifthenelse{\equal{\ifdraft}{draft}}%
{\smallmark{#1}}%
{}%
\beqn \label{#1}}

\newcommand{\sN}{{\Bbb N}}

\newcommand{\ds}{\displaystyle}

\newcommand{\subabc}[4]{{#1_{+[{#2},#3]#4}}}
\newcommand{\suba}[3]{#1_{+[\underline{#2},#3]}}
\newcommand{\subal}[4]{#1_{+#2[\underline{#3},#4]}}
\newcommand{\subb}[3]{#1_{+[#2,\underline{#3}]}}
\newcommand{\subbl}[4]{#1_{+#2[#3,\underline{#4}]}}

\newcommand{\mathl}{\!\!\!\!\!\!\!\!\!\!\!\!\!\!\!\!}
\newcommand{\mathr}{\,\,\,\,\,\,\,\,\,\,\,\,\,\,\,\,}

\newcommand{\shiftcaption}{\vspace{-0.3cm}}

\documentstyle[11pt,twoside,epsfig,ifthen,citesort]{article}

\ifthenelse{\equal{\ifpsdraft}{draft}}{\psdraft}{}

\ifthenelse{\equal{\ifpslabel}{nolabel}}
{
   \newcommand{\epsfigdg}[2]{\epsfig{figure=#1,#2}}
}{
   \newcommand{\epsfigdg}[2]{
   \unitlength1mm
   \begin{picture}(10,10)
   \put(0,0){\epsfig{figure=#1,#2}}
   \put(1,1){\fbox{\Large \tt #1}}
   \end{picture}
   }
}

\catcode`\@=11

 \font\tenmsx=msam10 scaled \magstep1
 \font\sevenmsx=msam8
 \font\fivemsx=msam6
 \font\tenmsy=msbm10 scaled \magstep1
 \font\sevenmsy=msbm8
 \font\fivemsy=msbm6

\newfam\msxfam
\newfam\msyfam
\textfont\msxfam=\tenmsx  \scriptfont\msxfam=\sevenmsx
  \scriptscriptfont\msxfam=\fivemsx
\textfont\msyfam=\tenmsy  \scriptfont\msyfam=\sevenmsy
  \scriptscriptfont\msyfam=\fivemsy

\def\hexnumber@#1{\ifnum#1<10 \number#1\else
 \ifnum#1=10 A\else\ifnum#1=11 B\else\ifnum#1=12 C\else
 \ifnum#1=13 D\else\ifnum#1=14 E\else\ifnum#1=15 F\fi\fi\fi\fi\fi\fi\fi}

\def\msx@{\hexnumber@\msxfam}
\def\msy@{\hexnumber@\msyfam}
\mathchardef\boxdot="2\msx@00
\mathchardef\boxplus="2\msx@01
\mathchardef\boxtimes="2\msx@02
\mathchardef\square="0\msx@03
\mathchardef\blacksquare="0\msx@04
\mathchardef\centerdot="2\msx@05
\mathchardef\lozenge="0\msx@06
\mathchardef\blacklozenge="0\msx@07
\mathchardef\circlearrowright="3\msx@08
\mathchardef\circlearrowleft="3\msx@09
\mathchardef\rightleftharpoons="3\msx@0A
\mathchardef\leftrightharpoons="3\msx@0B
\mathchardef\boxminus="2\msx@0C
\mathchardef\Vdash="3\msx@0D
\mathchardef\Vvdash="3\msx@0E
\mathchardef\vDash="3\msx@0F
\mathchardef\twoheadrightarrow="3\msx@10
\mathchardef\twoheadleftarrow="3\msx@11
\mathchardef\leftleftarrows="3\msx@12
\mathchardef\rightrightarrows="3\msx@13
\mathchardef\upuparrows="3\msx@14
\mathchardef\downdownarrows="3\msx@15
\mathchardef\upharpoonright="3\msx@16

\mathchardef\downharpoonright="3\msx@17
\mathchardef\upharpoonleft="3\msx@18
\mathchardef\downharpoonleft="3\msx@19
\mathchardef\rightarrowtail="3\msx@1A
\mathchardef\leftarrowtail="3\msx@1B
\mathchardef\leftrightarrows="3\msx@1C
\mathchardef\rightleftarrows="3\msx@1D
\mathchardef\Lsh="3\msx@1E
\mathchardef\Rsh="3\msx@1F
\mathchardef\rightsquigarrow="3\msx@20
\mathchardef\leftrightsquigarrow="3\msx@21
\mathchardef\looparrowleft="3\msx@22
\mathchardef\looparrowright="3\msx@23
\mathchardef\circeq="3\msx@24
\mathchardef\succsim="3\msx@25
\mathchardef\gtrsim="3\msx@26
\mathchardef\gtrapprox="3\msx@27
\mathchardef\multimap="3\msx@28
\mathchardef\therefore="3\msx@29
\mathchardef\because="3\msx@2A
\mathchardef\doteqdot="3\msx@2B

\mathchardef\triangleq="3\msx@2C
\mathchardef\precsim="3\msx@2D
\mathchardef\lesssim="3\msx@2E
\mathchardef\lessapprox="3\msx@2F
\mathchardef\eqslantless="3\msx@30
\mathchardef\eqslantgtr="3\msx@31
\mathchardef\curlyeqprec="3\msx@32
\mathchardef\curlyeqsucc="3\msx@33
\mathchardef\preccurlyeq="3\msx@34
\mathchardef\leqq="3\msx@35
\mathchardef\leqslant="3\msx@36
\mathchardef\lessgtr="3\msx@37
\mathchardef\backprime="0\msx@38
\mathchardef\risingdotseq="3\msx@3A
\mathchardef\fallingdotseq="3\msx@3B
\mathchardef\succcurlyeq="3\msx@3C
\mathchardef\geqq="3\msx@3D
\mathchardef\geqslant="3\msx@3E
\mathchardef\gtrless="3\msx@3F
\mathchardef\sqsubset="3\msx@40
\mathchardef\sqsupset="3\msx@41
\mathchardef\vartriangleright="3\msx@42
\mathchardef\vartriangleleft="3\msx@43
\mathchardef\trianglerighteq="3\msx@44
\mathchardef\trianglelefteq="3\msx@45
\mathchardef\bigstar="0\msx@46
\mathchardef\between="3\msx@47
\mathchardef\blacktriangledown="0\msx@48
\mathchardef\blacktriangleright="3\msx@49
\mathchardef\blacktriangleleft="3\msx@4A
\mathchardef\vartriangle="3\msx@4D
\mathchardef\blacktriangle="0\msx@4E
\mathchardef\triangledown="0\msx@4F
\mathchardef\eqcirc="3\msx@50
\mathchardef\lesseqgtr="3\msx@51
\mathchardef\gtreqless="3\msx@52
\mathchardef\lesseqqgtr="3\msx@53
\mathchardef\gtreqqless="3\msx@54
\mathchardef\Rrightarrow="3\msx@56
\mathchardef\Lleftarrow="3\msx@57
\mathchardef\veebar="2\msx@59
\mathchardef\barwedge="2\msx@5A
\mathchardef\doublebarwedge="2\msx@5B
\mathchardef\angle="0\msx@5C
\mathchardef\measuredangle="0\msx@5D
\mathchardef\sphericalangle="0\msx@5E
\mathchardef\varpropto="3\msx@5F
\mathchardef\smallsmile="3\msx@60
\mathchardef\smallfrown="3\msx@61
\mathchardef\Subset="3\msx@62
\mathchardef\Supset="3\msx@63
\mathchardef\Cup="2\msx@64

\mathchardef\Cap="2\msx@65

\mathchardef\curlywedge="2\msx@66
\mathchardef\curlyvee="2\msx@67
\mathchardef\leftthreetimes="2\msx@68
\mathchardef\rightthreetimes="2\msx@69
\mathchardef\subseteqq="3\msx@6A
\mathchardef\supseteqq="3\msx@6B
\mathchardef\bumpeq="3\msx@6C
\mathchardef\Bumpeq="3\msx@6D
\mathchardef\lll="3\msx@6E

\mathchardef\ggg="3\msx@6F

\mathchardef\circledS="0\msx@73
\mathchardef\pitchfork="3\msx@74
\mathchardef\dotplus="2\msx@75
\mathchardef\backsim="3\msx@76
\mathchardef\backsimeq="3\msx@77
\mathchardef\complement="0\msx@7B
\mathchardef\intercal="2\msx@7C
\mathchardef\circledcirc="2\msx@7D
\mathchardef\circledast="2\msx@7E
\mathchardef\circleddash="2\msx@7F
\def\ulcorner{\delimiter"4\msx@70\msx@70 }
\def\urcorner{\delimiter"5\msx@71\msx@71 }
\def\llcorner{\delimiter"4\msx@78\msx@78 }
\def\lrcorner{\delimiter"5\msx@79\msx@79 }
\def\yen{\mathhexbox\msx@55 }
\def\checkmark{\mathhexbox\msx@58 }
\def\circledR{\mathhexbox\msx@72 }
\def\maltese{\mathhexbox\msx@7A }
\mathchardef\lvertneqq="3\msy@00
\mathchardef\gvertneqq="3\msy@01
\mathchardef\nleq="3\msy@02
\mathchardef\ngeq="3\msy@03
\mathchardef\nless="3\msy@04
\mathchardef\ngtr="3\msy@05
\mathchardef\nprec="3\msy@06
\mathchardef\nsucc="3\msy@07
\mathchardef\lneqq="3\msy@08
\mathchardef\gneqq="3\msy@09
\mathchardef\nleqslant="3\msy@0A
\mathchardef\ngeqslant="3\msy@0B
\mathchardef\lneq="3\msy@0C
\mathchardef\gneq="3\msy@0D
\mathchardef\npreceq="3\msy@0E
\mathchardef\nsucceq="3\msy@0F
\mathchardef\precnsim="3\msy@10
\mathchardef\succnsim="3\msy@11
\mathchardef\lnsim="3\msy@12
\mathchardef\gnsim="3\msy@13
\mathchardef\nleqq="3\msy@14
\mathchardef\ngeqq="3\msy@15
\mathchardef\precneqq="3\msy@16
\mathchardef\succneqq="3\msy@17
\mathchardef\precnapprox="3\msy@18
\mathchardef\succnapprox="3\msy@19
\mathchardef\lnapprox="3\msy@1A
\mathchardef\gnapprox="3\msy@1B
\mathchardef\nsim="3\msy@1C
\mathchardef\napprox="3\msy@1D
\mathchardef\varsubsetneq="3\msy@20
\mathchardef\varsupsetneq="3\msy@21
\mathchardef\nsubseteqq="3\msy@22
\mathchardef\nsupseteqq="3\msy@23
\mathchardef\subsetneqq="3\msy@24
\mathchardef\supsetneqq="3\msy@25
\mathchardef\varsubsetneqq="3\msy@26
\mathchardef\varsupsetneqq="3\msy@27
\mathchardef\subsetneq="3\msy@28
\mathchardef\supsetneq="3\msy@29
\mathchardef\nsubseteq="3\msy@2A
\mathchardef\nsupseteq="3\msy@2B
\mathchardef\nparallel="3\msy@2C
\mathchardef\nmid="3\msy@2D
\mathchardef\nshortmid="3\msy@2E
\mathchardef\nshortparallel="3\msy@2F
\mathchardef\nvdash="3\msy@30
\mathchardef\nVdash="3\msy@31
\mathchardef\nvDash="3\msy@32
\mathchardef\nVDash="3\msy@33
\mathchardef\ntrianglerighteq="3\msy@34
\mathchardef\ntrianglelefteq="3\msy@35
\mathchardef\ntriangleleft="3\msy@36
\mathchardef\ntriangleright="3\msy@37
\mathchardef\nleftarrow="3\msy@38
\mathchardef\nrightarrow="3\msy@39
\mathchardef\nLeftarrow="3\msy@3A
\mathchardef\nRightarrow="3\msy@3B
\mathchardef\nLeftrightarrow="3\msy@3C
\mathchardef\nleftrightarrow="3\msy@3D
\mathchardef\divideontimes="2\msy@3E
\mathchardef\varnothing="0\msy@3F
\mathchardef\nexists="0\msy@40
\mathchardef\mho="0\msy@66
\mathchardef\thorn="0\msy@67
\mathchardef\beth="0\msy@69
\mathchardef\gimel="0\msy@6A
\mathchardef\daleth="0\msy@6B
\mathchardef\lessdot="3\msy@6C
\mathchardef\gtrdot="3\msy@6D
\mathchardef\ltimes="2\msy@6E
\mathchardef\rtimes="2\msy@6F
\mathchardef\shortmid="3\msy@70
\mathchardef\shortparallel="3\msy@71
\mathchardef\smallsetminus="2\msy@72
\mathchardef\thicksim="3\msy@73
\mathchardef\thickapprox="3\msy@74
\mathchardef\approxeq="3\msy@75
\mathchardef\succapprox="3\msy@76
\mathchardef\precapprox="3\msy@77
\mathchardef\curvearrowleft="3\msy@78
\mathchardef\curvearrowright="3\msy@79
\mathchardef\digamma="0\msy@7A
\mathchardef\varkappa="0\msy@7B
\mathchardef\hslash="0\msy@7D
\mathchardef\hbar="0\msy@7E
\mathchardef\backepsilon="3\msy@7F
\def\Bbb{\ifmmode\let\next\Bbb@\else
 \def\next{\errmessage{Use \string\Bbb\space only in math mode}}\fi\next}
\def\Bbb@#1{{\Bbb@@{#1}}}
\def\Bbb@@#1{\fam\msyfam#1}

\catcode`\@=12
\font\teneusmf=eufm10 scaled 1200
\font\seveneusmf=eufm8
\font\fiveeusmf=eufm6
\newfam\eusmffam
\textfont\eusmffam=\teneusmf
\scriptfont\eusmffam=\seveneusmf
\scriptscriptfont\eusmffam=\fiveeusmf

\font\teneusm=eusm10 scaled 1200
\font\seveneusm=eusm8
\font\fiveeusm=eusm6
\newfam\eusmfam
\textfont\eusmfam=\teneusm
\scriptfont\eusmfam=\seveneusm
\scriptscriptfont\eusmfam=\fiveeusm

\font\teneusmc=cmsy10 scaled 1200
\font\seveneusmc=cmsy8
\font\fiveeusmc=cmsy6
\newfam\eusmcfam
\textfont\eusmcfam=\teneusmc
\scriptfont\eusmcfam=\seveneusmc
\scriptscriptfont\eusmcfam=\fiveeusmc

\newcommand{\markitr}[1]{\unitlength1mm
\protect\begin{picture}(1,1)
\protect\thinlines
\protect\put(0,-1){\protect\line(1,0){160}}
\protect\put(0,0){{\sc #1}}
\protect\end{picture}
}

\newcommand{\markitl}[1]{\unitlength1mm
\protect\begin{picture}(0,1)
\protect\thinlines
\protect\put(1,-1){\protect\line(-1,0){160}}
\protect\put(1.5,0){\protect\makebox[0mm][r]{{\sc}}}
\protect\end{picture}
}

\newcommand{\markh}[1]{
\markboth{\markitl{#1}}{\markitr{#1}}
\thispagestyle{myheadings}}

\newlength{\sectionnumbersize}
\newlength{\sectionsize}
\begin{document}


\textfloatsep 20pt plus 10pt minus 4pt
\intextsep 14pt plus 10pt minus 4pt

\newcommand{\dgsa}[1]{\section{{#1}}\vspace{-0.2cm}}
\newcommand{\dgsb}[1]{\subsection{{#1}}\vspace{-0.2cm}}
\newcommand{\dgsbhide}[2]
{\subsection{{#1}}#2\vspace{-0.2cm}}
\newcommand{\dgsm}[1]{\noindent{\Large\bf{#1}}}


\newcommand{\dgtitle}{
                      Heavy-Quark Production \\
                      in the Target Fragmentation Region}

\newcommand{\dgabstract}{
Fixed-target experiments permit the study of hadron production in the target 
fragmentation region. It is expected that the tagging of specific particles in 
the target fragments can be employed to introduce a bias in the hard scattering 
process towards a specific flavour content. The case of hadrons containing a 
heavy quark is particularly attractive because of the clear experimental 
signatures and the applicability of perturbative QCD. The standard approach to 
one-particle inclusive processes based on fragmentation functions is valid in 
the current fragmentation region and for large transverse momenta~$p_T$
in the target fragmentation region, but it fails for particle production at 
small~$p_T$ in the target fragmentation region. A collinear singularity, which 
cannot be absorbed in the standard way into the phenomenological distribution 
functions, prohibits the application of this procedure. This situation is 
remedied by the introduction of a new set of distribution functions, the target 
fragmentation functions. They describe particle production in the target 
fragmentation region, and can be viewed as correlated distribution functions 
in the momentum fractions of the observed particle and of the parton initiating 
the hard scattering process. It is shown in a next-to-leading-order calculation
for the case of deeply inelastic lepton--nucleon scattering that the additional 
singularity can be consistently absorbed into the renormalized target 
fragmentation functions on the one-loop level. The formalism is derived in 
detail and is applied to the production of heavy quarks. The renormalization 
group equation of the target fragmentation functions for the perturbative 
contribution is solved numerically, and the results of a case study for deeply 
inelastic lepton--nucleon scattering at DESY (H1 and ZEUS at HERA), at CERN
(NA47) and at Fermilab (E665) are discussed. We also comment briefly on the 
case of an intrinsic heavy-quark content of the proton.
}

\newcommand{\newaddr}{{\em Address after February 1, 1997:
Paul Scherrer Institute, Department of Nuclear and Particle Physics,
Theory Group, CH-5232 Villigen PSI, Switzerland.}}

\newcommand{\dgemail}{{\em Electronic
mail address: Dirk.Graudenz\char64{}cern.ch}}

\newcommand{\dgwww}{{\em WWW URL:
http://wwwcn.cern.ch/$\sim$graudenz/index.html}}


\thispagestyle{empty}

\renewcommand{\thefootnote}{\fnsymbol{footnote}}
\setcounter{footnote}{0}

\hfill
\hspace*{\fill}
\begin{minipage}[t]{5cm}
\hfill CERN--TH/\cernnr
\end{minipage}
\vspace{0.5cm}

\ifnum\draftlabel=0
   \vspace{0.5cm}
\fi

\begin{center}

{\LARGE\bf \dgtitle{}$\,$\footnote[3]{{\em Habilitationsschrift, 
           Universit\"{a}t Hamburg, 1996 (submitted in April 1996).}}\\
}

\vspace{1.0cm}

{\bf Dirk Graudenz}$\;$\footnote[1]{\dgemail}$\;$\footnote[5]{
\dgwww}$\;$\footnote[4]{\newaddr}\\
\vspace{0.1cm}
{\it Theoretical Physics Division, CERN\\
CH--1211 Geneva 23\\
}

\end{center}

\vspace{0.5cm}

\begin{center}
\begin{minipage}{14cm}

\begin{center} {\bf Abstract}\end{center}

{
\small
\dgabstract
}

\renewcommand{\baselinestretch}{1.0}

\end{minipage}

\vspace{0.5cm}
\ifnum\draftlabel=1
\fbox{\fbox{Draft \draftnr, printed on \today; not for circulation.}}
\fi
\vspace{0.5cm}

\end{center}

\vfill
\noindent
\begin{minipage}[t]{5cm}
CERN--TH/\cernnr\\
\vdate
\end{minipage}
\vspace{1cm}

\clearpage

\thispagestyle{empty}
{\ }
\newpage



\thispagestyle{empty}

\renewcommand{\thefootnote}{\fnsymbol{footnote}}
\setcounter{footnote}{0}

{\ }

\vspace{2cm}

\begin{center}

{\LARGE\bf \dgtitle{}\\}

\vspace{1.5cm}

{\bf Dirk Graudenz}\\
\vspace{0.1cm}
{\it Theoretical Physics Division, CERN\\
Geneva, Switzerland\\
}

\vspace{1.3cm}

\vthdate

\end{center}

{\ }
\clearpage
\thispagestyle{empty}
{\ }

{\ }
\clearpage
\thispagestyle{empty}
{\ }

\vspace{2cm}
\footnotetext[0]{{\em Mailing address: Theoretical Physics Division, CERN, 
CH-1211 Geneva 23, Switzerland.}}
\footnotetext[0]{\newaddr}
\footnotetext[0]{\dgemail}
\footnotetext[0]{\dgwww}

\vspace*{\fill}

\begin{center}
\begin{minipage}{16cm}

\begin{center} {\bf Abstract}\end{center}

{
\dgabstract
}
\renewcommand{\baselinestretch}{1.0}

\end{minipage}
\end{center}

\vspace{0.0cm}

{\ }
\thispagestyle{empty}
\clearpage

\clearpage

\ifnum\twopagedg=1
{\ }
\thispagestyle{empty}
\clearpage
\else
\fi


\ifthenelse{\equal{\ifdraft}{draft}}
{
\newpage
\setcounter{page}{1}
\thispagestyle{empty}
\markh{Changes...}
{\bf Changes still to be done:}
\begin{itemize}
\item Citations with numbers, not in the form ABC99.
\end{itemize}
 
\noindent
The Editor.

\dgcleardoublepage
}

\pagenumbering{roman}
\setcounter{page}{1}
 
\renewcommand{\thefootnote}{\arabic{footnote}}
\setcounter{footnote}{0}

\ifnum\twopagedg=1
\else
\dgcleardoublepage{
\fi

\pagestyle{empty}
{
\tableofcontents
}
\cleardoublepage



\pagestyle{myheadings}


\pagenumbering{arabic}
\setcounter{page}{1}
\markh{Introduction}
\dgsa{Introduction}
\labelm{intro}
{
Fixed-target experiments permit the study of hadron production
in the target fragmentation region\footnote{
The terms ``current fragmentation region'' and ``target
fragmentation region'' will be defined in Section~\ref{FDIS}.}.
Indeed, to mention only a few, 
results have been obtained in deeply inelastic 
lepton--nucleon scattering
for the production of $K$~mesons and $\Lambda$~hyperons
\cite{1,2,3,4}, and in hadron--hadron
scattering\footnote{For a review, see also Ref.\ \cite{5}.} 
for the production of $D$~mesons
\cite{6,7,8,9,10}.
A future experiment \cite{11,12} 
with excellent particle identification
capabilities will make it possible to investigate the case of
polarized deeply inelastic muon--nucleon scattering.
One might be interested in a study of particle production
in the target fragmentation region for at least three reasons. 
First of all, the fragmentation of a nucleon after being hit by a 
highly energetic probe is interesting {\it per se}.
Apart from phenomenological models 
\cite{13,14,15,16,17,18}, 
little is known of the dynamics involved in the fragmentation
of a composite object, except for special cases,
for instance 
diffractive scattering,
where the target system is either untouched or merely excited.
In non-diffractive scattering,
the colour flux between the current system and the remnant of the
target nucleon makes a theoretical description a complicated problem.
Experimental results and a description in the framework of
Quantum Chromodynamics (QCD)
could shed light on the mechanisms involved.
Secondly, one might hope to learn something about the nucleon state itself
by tagging specific hadrons in the remnant. The analysis of, 
for example, $\Lambda$ 
production in the target fragmentation region in polarized lepton--nucleon 
scattering is expected to reveal information about the spin carried by
the strange-quark sea \cite{19}.
Thirdly, the tagging of particles in the target fragments provides 
additional information on the hard scattering process. 
An event sample with for instance a proton in the proton remnant
in electron--proton scattering 
is expected to be enhanced with gluon-initiated processes.
This kind of mechanism therefore permits a bias in the
hard scattering process towards a specific flavour content.
These points illustrate that it 
is worth while to study particle production in the target 
fragmentation region in some detail.


In this paper we consider the production of heavy 
quarks\footnote{For our purposes, 
heavy quarks are the bottom quark and the charm quark,
collectively denoted by~$Q$.
We will also denote the photon virtuality in deeply inelastic
scattering by~$Q$; the meaning will be clear from the context.}
in deeply inelastic lepton--nucleon scattering with an emphasis on the
target fragmentation region.
Heavy quarks are well suited for two reasons.
{}From an experimental point of view, mesons containing a heavy quark,
e.g.\ $D$~and $D^*$~mesons, are easily tagged. 
On the theoretical side, the large masses of heavy quarks allow for the 
application of perturbative QCD, and thus predictions of
experimental quantities that do not rely completely on measured 
fragmentation functions
are possible,
contrary to the case, for example, of mesons 
built of 
only light valence quarks.

The production of heavy quarks
in the current fragmentation region
or at finite transverse momentum~$p_T$ in the target fragmentation region
is accessible by means of perturbation theory.
In contrast, heavy-quark production at small~$p_T$ in the target 
fragmentation region is commonly described by non-perturbative mechanisms
or by phenomenological models.
It is the purpose of this paper to develop a framework that allows
a unified and coherent 
description of heavy-quark production, 
in the current and target fragmentation 
regions, in terms of convolutions of  
perturbatively calculable
hard scattering cross sections with phenomenological
distribution functions. To this end, a new set of distribution functions, 
the target fragmentation functions, are employed.
In this introduction, we first
give a short review
of two theoretical approaches to heavy-quark production.
We then comment briefly on the problem of factorization
in the case of semi-inclusive processes 
in the framework of QCD.
Throughout the paper, the basic process under consideration will be 
deeply inelastic lepton--nucleon scattering. The developed formalism is, 
however, more general, and applies to
photoproduction in lepton--nucleon scattering 
and to hadron--hadron collisions as well.
}


\dgsb{Heavy-Quark Production}
\labelm{HQP}
Heavy-quark production in lepton--nucleon
scattering
has recently been studied in great detail
for the cases of both photoproduction
\cite{20,21,22,23,24,25}
and deeply inelastic scattering
\cite{26,27,28,29,30,31,32,33}.
The systematics of heavy-quark production
in lepton--nucleon scattering, including the target fragmentation region,
has been developed 
in Ref.\ \cite{34}, with some emphasis
on the Fock state picture of the nucleon wave function.
In principle, there are two different approaches to determine
heavy-quark
production cross sections: (a) by directly calculating Feynman diagrams with
heavy quarks as final-state particles, as has been done
for instance in
Refs.~\cite{26,27,32}, and (b) by applying the approach
of perturbative heavy-quark
fragmentation functions\footnote{We use the terminology 
``heavy-quark fragmentation function'' for fragmentation functions
of partons into heavy quarks, contrary to for example Ref.\ \cite{35},
where this term refers to the fragmentation function of heavy quarks
into hadrons.
}, where the amplitudes that are calculated 
have only massless partons as final-state particles, but are convoluted 
with heavy-quark fragmentation functions, as used,
for instance, 
in Ref.~\cite{23}. 
Both approaches have their merits and drawbacks.
A direct calculation to fixed order (a) allows for the determination 
of expectation values of arbitrary infrared-safe 
observables involving all momenta of outgoing partons. 
It requires, however, a very careful
treatment of scales and the matching of various kinematical 
regimes \cite{28,29,31} in order to obtain reliable results for
differential cross sections
close to, as well as far above, threshold.
In particular, if the heavy quark is assumed to contribute via 
a non-vanishing heavy-quark density of the incident nucleon
\cite{36},
and if 
the parton densities are evolved under the assumption of massless
quarks, a subtraction procedure has to be applied to 
match the parton densities with 
hard scattering cross sections based on massive quarks.
The 
fragmentation function approach (b) allows the resummation of
terms of the form $\left[\alpha_s \ln(\mu^2/m^2)\right]^n$ by means of
the renormalization group equation for fragmentation functions
and is thus expected to be valid for a large range of scales~$\mu$. 
Here~$\mu$ is the factorization scale,
typically set to~$p_T$ in photoproduction, and~$m$ is the heavy-quark mass.
However, the fragmentation function picture is
restricted to more inclusive observables.
The heavy-quark fragmentation functions $D_{Q/i}(x,\mu^2)$ themselves 
are calculated in perturbation theory for $\mu\approx m$,
and then evolved by means of the renormalization group equation 
to arbitrary scales~$\mu$ \cite{37}.

The approaches mentioned so far are applicable only in the current
fragmentation region or at finite~$p_T$ in the target fragmentation
region; the produced heavy quarks are always part of the hard scattering
process, either directly as external lines of Feynman diagrams, or
indirectly in the case of heavy-quark fragmentation functions.
The hard scattering process being calculated in perturbative QCD,
it is clear that possible non-perturbative mechanisms of heavy-quark
production in the target fragmentation region 
cannot be taken into account in this way.

The goal of the present paper is to develop a unified picture for
heavy-quark production in both the current and target fragmentation 
regions, based on the fragmentation function approach,
for scattering processes with a hard subprocess accessible 
in perturbative QCD. The inclusion 
of the target fragmentation region in the analysis makes it necessary
to introduce a new type of distribution function, the heavy-quark 
target fragmentation function. This function is a means to parametrize
non-perturbative effects of heavy-quark production
in the target fragmentation region 
in a way similar to that in which parton densities parametrize
the non-perturbative bound-state dynamics of the proton.
Therefore, the new approach relies on a possible {\it measurement} of these
distribution functions. 
The advantage is that the developed formalism is a natural
extension of the standard one in QCD involving fragmentation functions
and parton densities.
Because of the lack of an experimental parametrization of heavy-quark
target fragmentation functions, we restrict our numerical
study to the perturbative contributions
that arise from the 
perturbative radiation of heavy quarks in the backward 
direction\footnote{The backward direction ($x_F<0$) is the
direction of the target fragments.
Please note that the conventions at HERA are such that the 
forward direction of the HERA machine 
corresponds to the backward direction in our notation.
In the case of colliders, the target fragmentation region
has a certain overlap with the beam pipe, and therefore measurements 
are difficult, so that special-purpose detectors
would certainly be necessary to tag heavy flavours in the 
remnant jet. Fixed-target experiments are different: the strong boost
in the direction of the incoming projectile makes it possible for
detectors to cover 
effectively $4\pi$ of solid angle.
},
and to parametrizations based on the hypothesis of intrinsic heavy quarks
in the proton \cite{38,39}.

In the approach based on heavy quarks as external lines in the matrix
elements, soft and collinear singularities are regulated by the
non-vanishing heavy-quark mass~$m$ in the heavy-quark propagators. 
The inclusive 
cross sections are thus finite even at small transverse momenta, 
but they contain terms of the form
$\ln(m^2/Q^2)$, thus making the perturbative result unreliable
for masses~$m$ very different from~$Q$. As already mentioned, 
this problem is taken care of in the fragmentation function approach, 
where these possibly large logarithmic terms are resummed.
There, the parton level hard scattering
matrix element itself is calculated for massless quarks. The outgoing partons, 
quarks and gluons, are then assumed to fragment into the observed
heavy quark. Because of the vanishing parton masses, the matrix elements
suffer from severe soft and collinear singularities. 
Soft and final-state collinear singularities 
related to parton lines not attached to distribution functions
of the 
real corrections and 
corresponding singularities
of the virtual corrections cancel \cite{40,41,42,43}, whereas the
remaining collinear singularities have to be removed 
in a different way.
In the
current fragmentation region, they are taken care
of by means of a suitable redefinition of the renormalized 
parton densities and fragmentation functions in terms of 
the bare ones \cite{44,45,46}. It will turn out that 
a similar mechanism is at work in the target fragmentation region,
where in this case the renormalized quantity is the target
fragmentation function.

\dgsb{Factorization in Deeply Inelastic Scattering}
\labelm{FDIS}
Because of the property of
asymptotic freedom
\cite{47,48,49,50}, 
hard scattering cross sections on the parton level 
for large scales
can be reliably calculated in perturbative QCD.
However, typical hadronic mass scales are too small for perturbation 
theory to be valid. The confinement of partons within hadrons
and
the process of hadronization in scattering processes are long distance 
phenomena and cannot yet be calculated within the theory from first 
principles.
Non-perturbative phenomena, 
such as the dynamics of bound states of hadrons in the
initial state and the 
subsequent
formation of final-state hadrons 
in the fragmentation process, 
therefore have to be incorporated 
into the theory in a way that makes them compatible 
with perturbatively calculable short-distance processes.

In the case of inclusive 
deeply inelastic lepton--nucleon scattering, the technical tool 
is the operator product expansion 
\cite{51,52,53,54,55,56,50}, which permits to separate
the non-perturbative regime, expressed in terms of 
expectation values 
of local operators, 
{}from the perturbative regime, represented by 
calculable coefficient functions, by the observation that the processes
are light-cone dominated
in the leading twist approximation.
In the parton picture\footnote{See for example 
the comprehensive reviews by Buras, Reya and Altarelli 
\cite{57,58,59}.}, 
the expectation values
are certain Mellin moments of parton densities.
Essential in this approach is that a sum over all possible hadronic final 
states is performed. The local operators then arise in the operator product
expansion of a commutator of two local currents.
The operator product expansion is not applicable in cases where the
final state is not totally inclusive, as for instance 
in the case of one-particle
inclusive processes, where more general operators than a 
commutator of two local currents have to be considered.

\begin{figure}[htb] \unitlength 1mm
\begin{center}
\dgpicture{159}{50}

\put(43,0){\epsfigdg{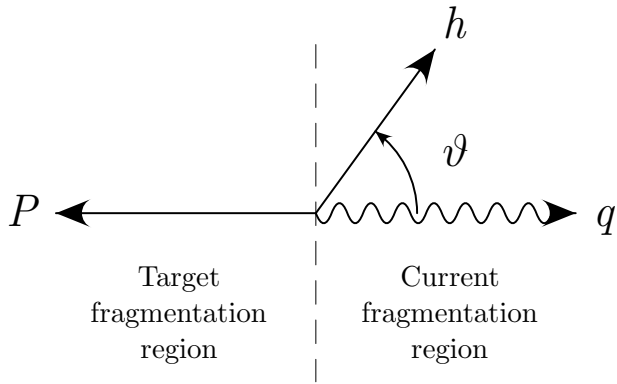}{width=70mm}}
\put(45,13){\begin{minipage}[t]{3cm}
                  {\begin{center}Target\\ fragmentation\\ region
                   \end{center}}
            \end{minipage}
           }
\put(81,13){\begin{minipage}[t]{3cm}
                  {\begin{center}Current\\ fragmentation\\ region
                   \end{center}}
            \end{minipage}
           }
\put(37,21){\LARGE\it P}
\put(115,21){\LARGE\it q}
\put(95,46){\LARGE\it h}
\put(95,29){\LARGE $\vartheta$}

\end{picture}
\end{center}
\shiftcaption
\caption[Current and Target Fragmentation Regions]
{\labelmm{ctfrpic} {\it The current and target fragmentation regions
in deeply inelastic scattering.
}}   
\end{figure}

There are two kinematical 
regions where a hadron~$h$ may be produced \cite{60}: the 
current fragmentation region and the target fragmentation region, 
see Fig.~\ref{ctfrpic}.
Usually, the phase space 
of the hadronic final state is 
divided up into three regions: the current region, 
the target region and the plateau region.
Particle production in the parton model picture 
is then assumed to be due
to the fragmentation of the current jet, of 
the target jet, and of wee partons
emitted from the current quark, respectively. 
In the QCD-improved parton model,
the current quark may emit a hard parton, e.g.\ a gluon,
and thus the clear assignment 
of the quark to the current jet is no longer possible.
We therefore use the convenient 
terminology that the current and target fragmentation 
regions simply correspond to the two hemispheres containing the
virtual photon and the incident nucleon, respectively,
in the hadronic centre-of-mass
frame.
Thus the production cross section can be written as
\beqm{curtag}
\sigma\,=\,\sigma_{\mbox{\scriptsize current}}\,+
\sigma_{\mbox{\scriptsize target}}.
\eeq
If the hadron~$h$ is produced in the current fragmentation 
region or at finite~$p_T$ in the target fragmentation region, then the 
factorization theorems of perturbative QCD
apply
and allow the statement that collinear singularities
are always of a form 
that makes it possible to define universal, process-independent parton 
densities and fragmentation functions
\cite{61,62,63,64,65,66,67,68,69}.
As a consequence,
the one-particle inclusive 
cross section in the case of
deeply inelastic scattering\footnote{
For two incident hadrons and the observation of more than one particle,
similar formulae hold.}
can eventually be cast into the
form 
\beqnm{cpp}
\sigma=\int \frac{\dd\xi}{\xi}\int\frac{\dd\eta}{\eta}
f^r(\xi,\mu_f^2)\, D^r(\eta,\mu_D^2) 
\,\sigma_{\mbox{\scriptsize hard}}(\xi,\eta,\mu_f^2,\mu_D^2),
\eeqn
where~$f^r$ and~$D^r$ are the renormalized, physical parton density
and fragmentation function, respectively,~$\mu_f$ and~$\mu_D$
are factorization scales and $\sigma_{\mbox{\scriptsize hard}}$ is the 
mass-factorized parton-level scattering cross section.


We write 
$\sigma_{\mbox{\scriptsize current}}$ 
formally as
\beqm{scur}
\sigma_{\mbox{\scriptsize current}}=\sigma_{fD}, 
\eeq
where
the phase space in $\sigma_{fD}$ is suitably restricted
to the current fragmentation region, with
\beqm{cur}
\sigma_{fD}=
\sigma^{\mbox{\scriptsize hard}}_{fD} \otimes f \otimes D,
\eeq
see Fig.~\ref{twomech}a.
The next-to-leading-order analysis has been performed in 
Ref.\ \cite{70}\footnote{
See also Ref.\ \cite{71}, and for the 
polarized case Ref.\ \cite{72}.}, 
and from an experimental
point of view, this kind of process has been studied
in great detail. 

\begin{figure}[htb] \unitlength 1mm
\begin{center}
\dgpicture{159}{65}

\put(18,14){\epsfigdg{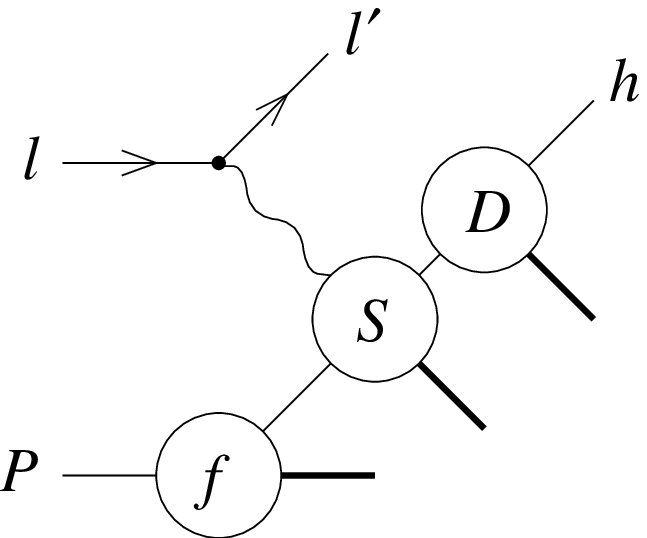}{width=60mm}}
\put(28,0){\lettlab (a)}

\put(93,8){\epsfigdg{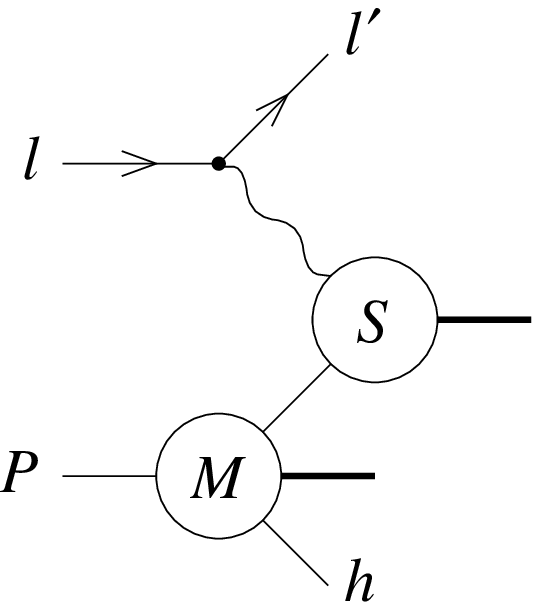}{width=50mm}}
\put(103,0){\lettlab (b)}

\end{picture}
\end{center}
\shiftcaption
\caption[Two Mechanisms for Particle Production]
{\labelmm{twomech} {\it Diagrams corresponding to the production of
particles in the current fragmentation region (a) and in the
target fragmentation region (b) in deeply inelastic lepton--nucleon 
scattering. Here $l$ and $l^\prime$ are the incoming and outgoing lepton,
respectively, $P$ is the incoming nucleon, 
$h$ stands for the observed hadron
and the solid lines for the other particles in the hadronic final state;
$S$ is the parton-level 
scattering cross section, and~$f$, $D$ and~$M$ are 
a parton density, a fragmentation function
and a target fragmentation function, respectively.
}}   
\end{figure}

The case of particle production in the target fragmentation region has
received some interest from the theoretical side
\cite{73,74,75}.
In Refs.\ \cite{73,74}, the problem is considered under the explicit
assumption, going back to Ref.\ \cite{76}, of light-cone dominance of
the commutator of time-ordered and anti-time-ordered products
of the electromagnetic current and interpolating fields
for the produced hadron.
As a consequence, a scaling relation in~$x_B$ and in the
longitudinal momentum fraction of the observed hadron can be
derived.
It would be interesting to formulate
this kind of process in the framework of perturbative QCD in the spirit
of the factorization theorems, which unfortunately do not extend to this
situation.
Recently, a new class of distribution functions, called 
``fracture functions'', 
and denoted by $M_{i,h/P}(\xi,\zeta,\mu^2)$,
has been introduced
\cite{77}. 
Fracture functions are target fragmentation functions since
they describe the production of hadrons in the target fragmentation 
region of, for example, a nucleon~$P$ 
in terms of the correlated probability density for finding,
at a specified scale~$\mu$, a parton~$i$
scattered in the hard process and a hadron~$h$ 
in the nucleon remnant with given momentum fractions
$\xi$ and~$\zeta$, respectively,
relative to the momentum of the incoming 
nucleon\footnote{
The word 
``fracture'', as introduced in Ref.~\cite{77}, is a combination of the words
``fragmentation'' and ``structure'', 
describing the hybrid status of target fragmentation functions as
being related to fragmentation functions and structure functions.
Because the term ``structure function'' refers to an object different
from the one defined by ``parton density'', and because target
fragmentation functions are best described as a hybrid of
fragmentation functions and parton densities, we refrain from the
use of the term ``fracture function''.}.
They fulfil a renormalization group equation which, in contrast
to the one for parton densities and fragmentation functions, 
contains also an inhomogeneous term. 
The additional source term is given by a convolution
of a parton density and a fragmentation function with an 
unsubtracted splitting function. The contribution
stemming from this term will be denoted ``perturbative''.

Within the QCD-improved parton model, 
the basic process has the structure
of the diagram depicted in Fig.~\ref{twomech}b,
corresponding to a cross section
\beqm{tar}
\sigma_{M}=
\sigma^{\mbox{\scriptsize hard}}_M \otimes M.
\eeq
The total cross section in the target fragmentation region is thus
\beqm{star}
\sigma_{\mbox{\scriptsize target}}=\sigma_{M}+\sigma_{fD},
\eeq 
where the phase space in $\sigma_{fD}$ is restricted to the
target fragmentation region.
QCD corrections 
to the leading-order processes in the target fragmentation region
have been calculated, and it has been 
shown that in next-to-leading order,
by a suitable definition of renormalized target fragmentation
functions in terms of the bare ones, all collinear singularities can be 
consistently absorbed into renormalized distribution functions
$f^r$, $D^r$ and $M^r$ \cite{78}. 
Recently, the case of polarized deeply inelastic lepton--nucleon
scattering
has been treated as well \cite{79}, with similar conclusions
regarding the consistent redefinition of distribution functions.

The splitting of the cross section~$\sigma$ into 
$\sigma_{fD}$ and $\sigma_{M}$
in Eq.~(\ref{star}) 
for particles produced in the target fragmentation region
depends on the chosen 
factorization scheme,
because the contribution from certain finite terms
may be hidden in any of the renormalized distribution functions.
In the processes under consideration here, beyond the leading order,
$\sigma_{fD}$ 
contributes to particle production in the target fragmentation region, 
in kinematical regimes where a parton attached to a fragmentation 
function is collinear to the incident parton of the hard scattering
process.
Only the sum of the two contributions from Eqs.~(\ref{cur}) 
and~(\ref{tar}) is physical and
gives, in a fixed order of QCD perturbation theory, 
a consistent description.
Related to this is the fact that the evolution equation for the target
fragmentation functions $M^r$ is inhomogeneous and receives contributions
in the form of a source term from $f^r$ and $D^r$.
As a consequence, a finite renormalization of $f^r$ and $D^r$
modifies the scale evolution of $M^r$, and thus gives rise to a change
in the contribution $\sigma_{\mbox{\scriptsize $M$}}$.
This effect is, however, present only beyond the
next-to-leading order.
We finally wish to note that the introduction of target fragmentation functions
of the form $f_{i,h/P}(\xi,\zeta,k_T^2,\mu^2)$, giving the probability
density to find, at a scale~$\mu$, a parton of momentum fraction~$\xi$ and
a hadron of momentum fraction~$\zeta$ and transverse momentum smaller
than~$k_T$, is not possible in a universal, process-independent way,
because
the contribution of the hard matrix element at finite transverse momenta
$p_T<k_T$ is process-dependent.

\begin{figure}[htb] \unitlength 1mm
\begin{center}
\dgpicture{159}{129}

\put(15,5){\epsfigdg{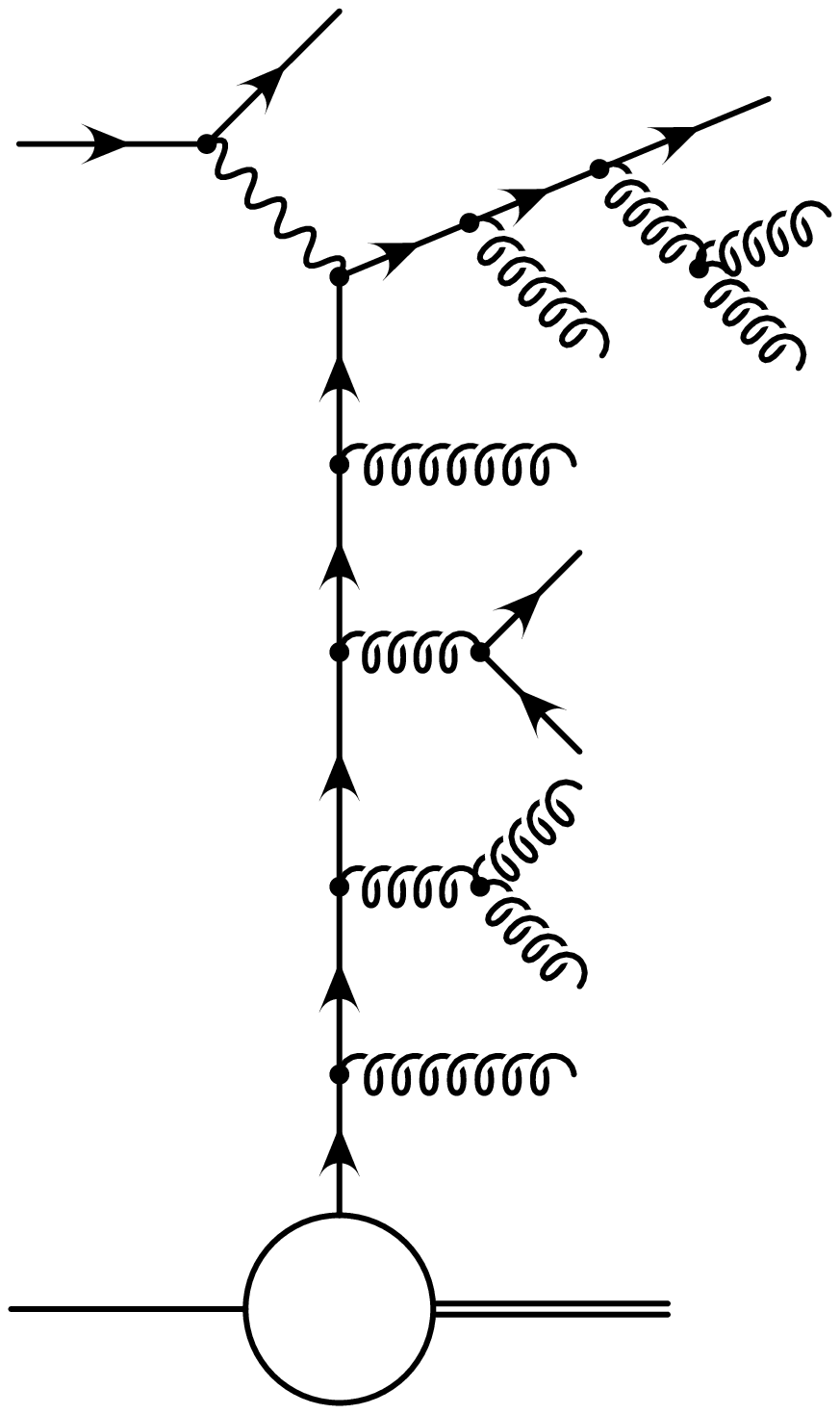}{width=70mm}}

\put(9,11){\LARGE\it P}
\put(12,109){\LARGE\it l}
\put(46,121){\LARGE\it l$\,^\prime$}
\put(41,108){\LARGE $\gamma^*$}

\thinlines

\put(99,5){\vector(0,1){19}}
\put(99,24){\vector(0,-1){19}}
\put(110,14){\begin{minipage}{4.9cm}
                   Non-perturbative target\\ fragmentation region
            \end{minipage}
           }

\put(99,26){\vector(0,1){13}}
\put(99,39){\vector(0,-1){13}}
\put(110,32){\begin{minipage}{4.9cm}
                   Evolution region
            \end{minipage}
           }

\put(99,41){\vector(0,1){34}}
\put(99,75){\vector(0,-1){34}}
\put(110,58){\begin{minipage}{4.9cm}
                   Perturbative target\\ fragmentation region
            \end{minipage}
           }

\put(99,77){\vector(0,1){43}}
\put(99,120){\vector(0,-1){43}}
\put(110,98){\begin{minipage}{4.9cm}
                   Current\\ fragmentation region
            \end{minipage}
           }

\end{picture}
\end{center}
\shiftcaption
\caption[Perturbative and non-Perturbative Fragmentation Regions]
{\labelmm{pnpfr} {\it Perturbative and non-perturbative 
fragmentation regions. The rapidity is increasing from the bottom to 
the top of the figure.
}}   
\end{figure}

The picture for hadron production in deeply inelastic scattering
that suggests itself based on these formal developments is the following
(see also Fig.~\ref{pnpfr}).
Particles produced at large rapidities, i.e.\ in
the {\it current fragmentation region} and at negative, but small rapidities
in the {\it perturbative target fragmentation region}
can be described by means of fragmentation functions. 
Particle production at large negative rapidities, in a region that we
call the {\it non-perturbative target fragmentation region}, is due to
an entirely non-perturbative mechanism, parametrized in terms of the
non-perturbative piece of the
newly introduced target fragmentation functions.
The intermediate region of negative rapidity is called the 
{\it evolution region}. It is described by the perturbative piece 
of the target fragmentation functions, obtained from the inhomogeneous
term in the renormalization group equation.
We note that these regions do overlap; for instance, 
non-perturbative effects will lead to a certain spread in transverse
momentum and hence in rapidity of the target fragments. 
The various regions
are defined in the theoretical formalism by the choice of a particular
factorization scheme.

\dgsb{Organization of the Paper}
The paper is organized as follows.
In the next section, we consider the general formalism for one-particle
inclusive processes. We review some properties of fragmentation functions
and calculate the one-particle inclusive cross section in deeply
inelastic lepton--nucleon scattering in next-to-leading order, 
paying special attention 
to the region of small transverse momenta 
of the produced hadron.
This allows us to isolate the problem 
in the target fragmentation region
and leads naturally to the introduction
of the target fragmentation functions. The QCD corrections for particle 
production in the target fragmentation region are calculated, and it is
demonstrated that a finite cross section can be obtained if the 
renormalization of the phenomenological distribution functions
$f$, $D$ and $M$ is done properly.
In Section~\ref{tffdetails} we describe the concept of
target fragmentation functions in more detail and study some of the properties
of their renormalization group equation.
In Section~\ref{hqff} we review the formalism of heavy-quark fragmentation
functions. We describe how they originate in perturbative QCD 
and solve their renormalization group equation numerically.
The subject of Section~\ref{hqtff} is heavy-quark target 
fragmentation functions. We define a piece that we call ``perturbative'', 
and obtain explicit numerical results from the solution of the
renormalization group equation.
We also briefly study the case of an intrinsic heavy-quark component
of the proton.
Section~\ref{CaseStudy} presents a case study for
heavy-quark production
in the target fragmentation region in deeply inelastic lepton--nucleon
scattering at HERA\footnote{Although HERA is a collider,
and the target fragmentation region is generally not accessible
for measurements because of the obstruction from the beam pipe,
it is instructive to have a comparison to fixed-target experiments
operating at lower energies. We do not include the HERMES experiment
\cite{80} in our study, because its particle
identification capabilities do not seem to allow for the tagging of
hadrons containing heavy quarks.}, E665 and NA47.
We consider perturbative contributions and the case of an intrinsic
heavy-quark component of the proton.
Here the word ``perturbative'' indicates that we include only
those contributions that arise from the radiation of heavy quarks in 
the backward direction represented by the inhomogeneous
term in the evolution equation and by the perturbatively calculable
contributions from the fragmentation of partons scattered in the
backward direction.
A typical experimental result is a differential cross section
$\dd \sigma/\dd p_T \,\dd x_F$; 
we therefore focus our attention on distributions
in the corresponding variables.
The paper closes with a summary, some conclusions and an outlook.
Here some restrictions of the approach and open problems are discussed.
Technical details are relegated to the appendices, where the explicit
analytical results for the cross sections are given.
Moreover, we
discuss
singular functions, convolution formulae 
for distributions, and the Laguerre method
for the solution of integro-differential equations. 


\dgcleardoublepage

\markh{One-Particle Inclusive Processes}
\dgsa{One-Particle Inclusive Processes}
\labelm{opics}
{\it 
In this section we introduce
the formalism of fragmentation functions
and discuss one-particle inclusive processes.
The material is here developed in some detail, because
similar concepts apply to the target fragmentation 
functions, 
and in order to have a concise discussion in that case we already 
dwell on the technical details here.
As an explicit application, we consider in detail the case of the 
next-to-leading-order QCD corrections
in deeply inelastic lepton--nucleon scattering. 
In the current fragmentation region and for large transverse
momenta of the observed particle in the target fragmentation region, 
fragmentation functions and parton densities are sufficient to give
a satisfactory description. For small transverse momenta
in the target fragmentation region
a new collinear singularity appears which cannot be taken care
of by a process-independent redefinition of fragmentation functions and
parton densities. The solution to the problem is the introduction of
a new class of phenomenological distribution functions, the target 
fragmentation functions \cite{77}. It is shown in the explicit example
that all additional singularities can be absorbed
into renormalized distribution functions, leading to a well-defined
cross section in next-to-leading order.
}

\dgsb{Fragmentation Functions}
\labelm{fragmfunc}

The description of one-particle inclusive processes in the framework of
perturbative QCD is based on the factorization 
theorems\footnote{For a review, see Ref.\ \cite{69}.}. 
For large scales, the running coupling constant 
$\alpha_s(\mu_r^2)$ is small,
and thus hard scattering cross sections can be calculated reliably
in perturbative QCD. The fragmentation process, however, involves mass
scales of the order of 
$\Lambda_{\mbox{\scriptsize QCD}}$; therefore the coupling constant
is large and the perturbative picture,
except for special cases involving for example heavy quarks \cite{81},
breaks down. The fragmentation
of partons into hadrons is described phenomenologically by means of 
fragmentation functions
$D_{h/i}(x,\mu^2)$, which are, 
by definition, probability distributions
for a particle~$h$ with a momentum fraction~$x$ 
at a scale~$\mu$ 
within a parton of type $i\in\{q,\overline{q},g\}$, see Fig.~\ref{ffig1}.

\begin{figure}[htb] \unitlength 1mm
\begin{center}
\dgpicture{159}{50}

\put(20,10){\epsfigdg{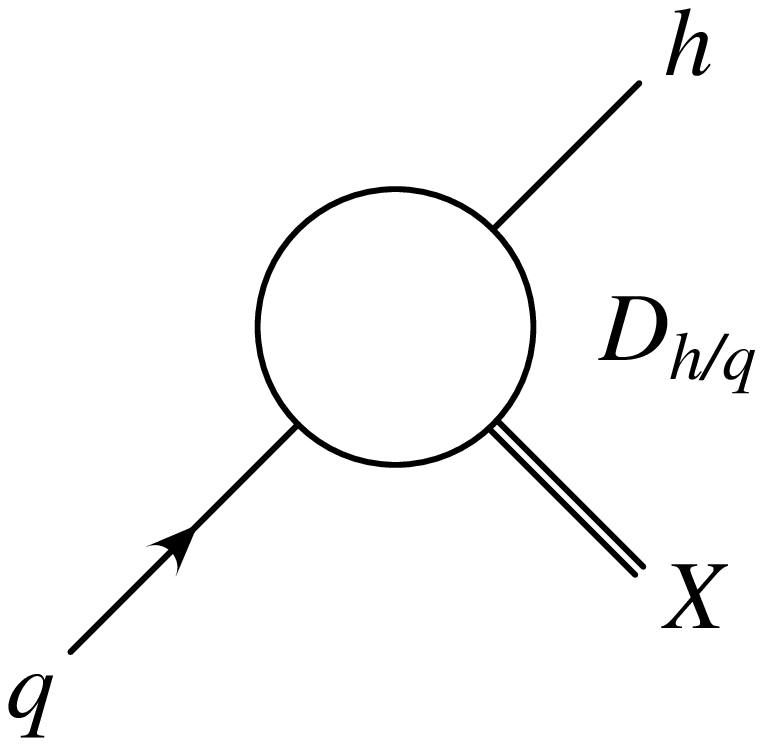}{width=40mm}}
\put(35,0){\lettlab (a)}
\put(90,10){\epsfigdg{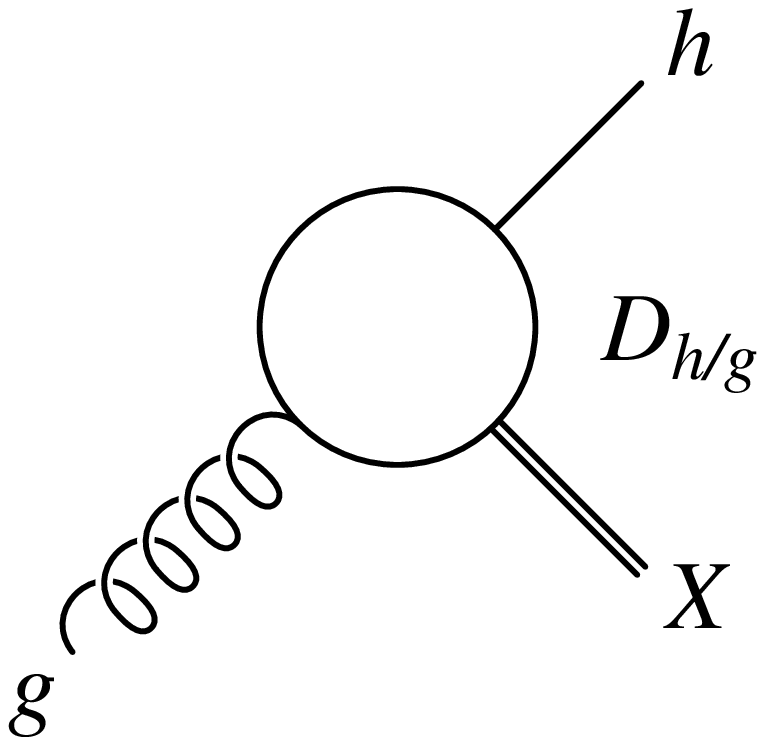}{width=40mm}}
\put(105,0){\lettlab (b)}

\end{picture}
\end{center}
\shiftcaption
\caption[Fragmentation Functions]
{\labelmm{ffig1} {\it Fragmentation functions:
$D_{h/q}(x,\mu^2)$ (a) and 
$D_{h/g}(x,\mu^2)$ (b).
Here $q$ stands for any quark or antiquark.
}}   
\end{figure}

The one-particle-inclusive cross section\footnote{
For a brief survey, see Ref.\ \cite{82}.
},
differential in the 
three-momentum~$\vec{h}$ of the observed hadron~$h$ with four-momentum 
$h=(h^0,\vec{h})$
in the process
\beqm{process}
I \rightarrow h + X,
\eeq
where~$I$ is the initial state,
is given by\footnote{By defining the variable $x=h^0/p_\alpha$, 
this expression can be cast into the more familiar form
$\int(\dd x/x^2)\,\sigma(h^0/x)\,D(x)$, where the angular integration of
$\int\dd p_\alpha\,\delta(p_\alpha^2)$ has been performed by means
of the delta function of solid angles~$\delta^S$.
}
\beqnm{opi}
2h^0\,\frac{\dd \sigma^{I\rightarrow hX}}{\dd \vec{h}}&=&
\sum_{F} \int\dps^{(n_F)}(\underline{p}) 
\,\frac{\dd \sigma^{I\rightarrow F}
(\underline{p})
}{\dps^{(n_F)}(\underline{p})}\nonu
\quad\quad\quad\quad\quad\quad&&\cdot\,
\sum_{\alpha=1}^{n_F} D_{h/F_\alpha}\left(\frac{h^0}{p_\alpha^0}\right)
\,\frac{2}{p_\alpha^0}\,\, \delta^{S}\left(\Omega_\alpha,\Omega_h\right)
\,\frac{1}{(h^0)^{(d-3)}}.
\eeqn
The sum in~$F$ runs over all possible partonic final states
with~$n_F$ partons $F_\alpha\in\{q,\overline{q},g\}$, 
$\alpha=1,\ldots,n_F$, 
$\underline{p}$ denotes the set of the $n_F$ parton momenta~$p_\alpha$,
$d=4-2\epsilon$ is the space-time 
dimension\footnote{The regularization of ultraviolet 
and infrared divergences is
performed in the framework of
dimensional 
regularization (for a review, see Refs.\ \cite{83,84}), 
where all calculations are done
in~$d$
space-time dimensions. Cross sections are finite in the ultraviolet if 
$d<4$ \cite{85}, 
and in the infrared if $d>4$ \cite{86,87,88}. 
The two regions may be connected
by an analytical continuation. Ultraviolet and infrared singularities
can be identified as poles in~$\epsilon$ in the regularized expressions.}, 
and 
$\sigma^{I\rightarrow F}$ is the cross section on the parton level 
for the process
$I\rightarrow F$.
The delta function~$\delta^S$ is defined on the space of test functions
on the $(d-2)$-dimensional sphere of solid angles by
\beq
\varphi(\Omega_0)=\int\dd \Omega\,\delta^S\left(\Omega,\Omega_0\right)
\,\varphi(\Omega).
\eeq

\begin{figure}[htb] \unitlength 1mm
\begin{center}
\dgpicture{159}{52}

\put(50,0){\epsfigdg{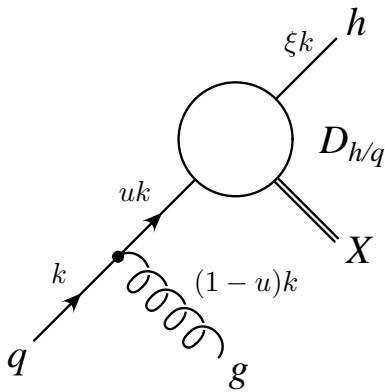}{width=50mm}}

\put(56,14){$k$}
\put(65,25){$uk$}
\put(75,13){$(1-u)k$}
\put(87,45){$\xi k$}

\end{picture}
\end{center}
\shiftcaption
\caption[QCD Correction to a Fragmentation Function]
{\labelmm{ffig2} {\it An example of how a QCD correction 
gives rise to the scale evolution of a fragmentation function.}
}   
\end{figure}


Beyond the leading order of perturbation theory, the parton-level cross section
$\sigma^{I\rightarrow F}$ 
contains collinear 
singularities\footnote{In QCD, analogous to the Bloch--Nordsieck mechanism in 
QED \cite{40,41}, 
soft singularities of the real and virtual corrections
cancel owing to the 
Kinoshita--Lee--Nauenberg theorem \cite{42,43,89}. 
Ultraviolet divergences
are taken care of by means of a redefinition of the wave functions, 
the coupling constants and the masses in the process of renormalization.}
arising from processes where partons are emitted collinear\-ly
from the parent parton
of the hadron; 
for an example, see Fig.~\ref{ffig2}.
The factorization theorems guarantee that the structure of these singularities
is of a universal form, and so they can be consistently absorbed into the
renormalized fragmentation functions $D^r$ by a redefinition \cite{46}
\beqm{dredef}
D_{h/i}(\xi)=\int_\xi^1\frac{\dd u}{u}\,
\left[\delta_{ij}\,\delta(1-u)
+\frac{1}{\epsilon}\frac{\alpha_s(\mu_r^2)}{2\pi}
\frac{\Gamma(1-\epsilon)}{\Gamma(1-2\epsilon)}
\left(\frac{4\pi\mu_r^2}{\mu^2}\right)^\epsilon
P_{j\leftarrow i}(u)\right]
\,D^r_{h/j}\left(\frac{\xi}{u},\mu^2\right).
\eeq
This expression is valid in leading order 
for the \msbar{} factorization
scheme \cite{45,66}. 
Here
$D_{h/i}(x)$ is the bare fragmentation function occurring 
in Eq.~(\ref{opi}), $D^r_{h/j}\left(x,\mu^2\right)$ is the
renormalized (i.e.\ physical, measurable and finite) fragmentation 
function\footnote{We will drop the superscript ``$r$'' in the following. 
If there is a scale argument attached to a distribution function, it is
assumed that it represents a renormalized, physical quantity.},
$\mu_r$~is the renormalization scale, and~$\mu$ is the
factorization scale. The singular terms are those 
proportional to $1/\epsilon$. Because of the process of redefinition, 
which has to be done at a specified scale, 
the renormalized fragmentation function is scale-dependent. 
The scale evolution of the $D_{h/j}\left(x,\mu^2\right)$
is governed by an Altarelli--Parisi type
of evolution equation\footnote{
In higher orders, the distribution functions depend on, in general different,
factorization and renormalization scales (see for example Ref.~\cite{90}).
Here we follow the common practice to identify these scales.
}
\beqm{Drgeq}
\frac{\partial D_{h/i}(x,\mu^2)}{\partial \ln \mu^2}=
\frac{\alpha_s(\mu^2)}{2\pi}\int_x^1\frac{\dd u}{u}\,
\overline{P}_{j\leftarrow i}(u)
\,D_{h/j}\left(\frac{x}{u},\mu^2\right);
\eeq
it can, in leading order, be derived from Eq.~(\ref{dredef})
by taking the derivative with respect to~$\mu^2$ on both sides and
by observing that the bare distribution function is scale-independent.
In the case of the leading-order evolution equation, 
the evolution kernels $\overline{P}_{j\leftarrow i}(u)$ 
for fragmentation functions coincide
with the standard Altarelli--Parisi splitting functions
$P_{j\leftarrow i}(u)$, cf.\ Eq.~(\ref{APsplitting});
this is, however, no longer true beyond the leading order \cite{91}.

The momentum sum rule fulfilled by fragmentation functions is
\beqm{Dmomsum}
\sum_h \int_0^1\dd x \, x\,D_{h/i}(x,\mu^2)=1.
\eeq
This equation expresses
the fact that the momentum of every parton finally ends up
as the momentum of one of the outgoing hadrons.
If the equation is fulfilled for one factorization
scale~$\mu_0$, then the renormalization
group equation ensures that it is true
for all scales~$\mu$.
We will see in Section~\ref{tffmomsr} that a corresponding relation for 
target fragmentation functions holds as well.

\dgsb{Deeply Inelastic Lepton--Nucleon Scattering:\\ 
      The Current Fragmentation Region}
\labelm{dislns}
We now consider the case of one-particle inclusive
cross sections in deeply inelastic lepton--nucleon scattering.
After setting up the formalism, we discuss QCD corrections. 
In this and in the next section we focus our attention on
particle production in the
current fragmentation region. For the produced hadrons
being strictly in the current fragmentation region or having
non-vanishing transverse momentum in the target fragmentation region, 
the matrix elements
can be found in Ref.\ \cite{70}. The set-up of the 
present calculation is such that the limit of small transverse momenta
in the target fragmentation region
can be discussed. Moreover, it will be possible to identify the reason
why the standard formulation fails.

Let us consider the scattering process 
\beq
   l+P \rightarrow 
   l^\prime+h+X,
\eeq
where~$l$ and~$P$ are the incoming charged lepton and nucleon,
respectively, $l^\prime$ is the scattered charged lepton, $h$ is the identified
hadron, and~$X$ denotes anything else in the hadronic final 
state\footnote{
Where possible, we identify the momenta of particles with their
genuine names.}.
The integration over the angles
that determine the relative orientation of the leptonic and hadronic
final states can be performed. 
The remaining lepton phase-space variables are the Bjorken
variables
\beq
\XB=\frac{Q^2}{2\PP q}, \quad y=\frac{\PP q}{\PP l}.
\eeq
Here $q=l-l^\prime$ 
is the momentum transfer and $Q^2=-q^2>0$ is the virtuality of
the exchanged virtual photon. We work in the limit of vanishing lepton mass.
The cross section for the production of an $n$-parton final state, 
differential in~$\XB$, $y$ and in the phase-space variables of the final-state
partons, is
\beqnm{eq12a}
\frac{\mbox{d}\sigma}{\mbox{d}\XB\mbox{d}y\,\dps^{(n)}}&=&
\sum_i\int\frac{\mbox{d}\xi}{\xi}\,P_{i/P}(\xi)\,\,\alpha^2 
\,\frac{1}{2\SH\XB}\,\frac{1}{e^2(2\pi)^{2d}}
\nonu
&&
\quad\,\,\cdot\left(Y^M\,\left(-g^{\mu\nu}\right)
            +Y^L\,
             \frac{4\XB^2}{Q^2}\,\PP^\mu\PP^\nu
       \right) H_{\mu\nu}.
\eeqn
This formula is valid for the case 
of one-photon exchange,
the exchange of weak vector bosons is neglected for simplicity;
$\SH=(\PP+l)^2$ is the square of the total centre-of-mass energy of the
lepton--nucleon scattering process, $\xi$ is the momentum fraction of
the initial parton of the QCD subprocess, and $P_{i/P}(\xi)$ is the
probability distribution function for the parton~$i$ in the nucleon~$P$; 
$P_{i/P}(\xi)$ can be either a parton density $f_{i/P}(\xi)$ or, 
as will be introduced later,
a target fragmentation function $M_{i,h/P}(\xi,\zeta)$, and 
$\alpha=e^2/4\pi$ is the fine structure constant.
The value of~$Q^2$ is given by $S_Hx_By$, and the total
hadronic energy is $W=\sqrt{S_H(1-x_B)y}$, if the nucleon mass is neglected.
A factor of $1/4$ for the average over the
spin degrees of freedom of the incoming particles
is already included.
The last factor, $H_{\mu\nu}$, is the hadron tensor
defined by 
\beq
H_{\mu\nu}=\sum_{\mbox{\scriptsize spins}}
{\overline{{\cal M}_\mu}\,{\cal M}_\nu},
\eeq
where $\epsilon^\mu(\lambda){\cal M}_\mu$ is the matrix element for the process 
\beq
   \gamma^*+\mbox{parton} \rightarrow 
   n\,\,\mbox{partons},
\eeq
with $\epsilon^\mu(\lambda)$ the polarization vector of 
a virtual photon with polarization~$\lambda$. The ratios
\beq
Y^M=\frac{1+(1-y)^2-\epsilon y^2}{2(1-\epsilon)y^2} \quad\mbox{and}\quad
Y^L=\frac{4(1-\epsilon)(1-y)+1+(1-y)^2-\epsilon y^2}{2(1-\epsilon)y^2}
\eeq
specify the $y$-dependence of the contributions from the 
two photon polarizations under consideration.
The projections operating on 
$H_{\mu\nu}$ are the result of the integration
of the lepton tensor (see for example Ref.\ \cite{92}) over the angles
that describe the  orientation of the momentum of the 
outgoing lepton with respect to the momenta of the outgoing hadrons.
The cross section consists of two parts, proportional to 
$Y^M$, the ``metric'' 
contribution, and proportional to 
$Y^L$, the longitudinal 
contribution\footnote{The metric and longitudinal contributions are 
obtained by a contraction of the hadron tensor $H_{\mu\nu}$
with the projection tensors given in Eq.~(\ref{projgl}).}.

Let us consider
the production process in the centre-of-mass
frame of the incoming nucleon and
the incoming virtual photon, so $\vec{\PP}+\vec{q}=\vec{0}$.
The positive $z$-axis is defined by the $q$-direction.
The hadron~$h$ has polar angle~$\vartheta$ 
relative to the virtual photon and energy~$h_0$,
see Fig.~\ref{ctfrpic}.
In this frame, the energy of the incoming nucleon is
\beq
\EE\,=\,\frac{Q}{2}\,\frac{1}{\sqrt{\XB(1-\XB)}}.
\eeq
Two new variables~$v$ and~$z$ can be defined 
by
\addtocounter{footnote}{1}
\footnotetext{\footn{Afoot}Please note the remark in footnote 
\itemr{Bfoot}.
}
\beqm{vdef}
v=\frac{1}{2}(1-\cos\vartheta), \quad z=\frac{\EH}{\EE(1-\XB)},
\eeq
whose range is $v,\,z\in[0,1]$
if the masses of all particles are neglected.
For $v=1$ the angle between the observed hadron and the 
nucleon remnant is zero, and the hadron is 
produced in the 
target remnant direction. The value $v=0$ corresponds to particle 
production in the current direction.
The variables~$v$ and~$z$ are convenient for the discussion of
the soft and collinear regions and for the explicit calculation. 
{}From an experimental point of view, however, 
variables such as~$x_F$ and~$p_T$ are more appropriate. They 
will be defined in Section~\ref{DiscPar}.


We are going to 
calculate in the following 
the differential cross section 
\beq
\frac{\mbox{d}\sigma(l+P \rightarrow 
   l^\prime+h+X
)}{\mbox{d}\XB\,\mbox{d}y\,\mbox{d}z\,\mbox{d}v}.
\eeq
It turns out that QCD corrections to the lowest-order process
for small transverse momenta
require subtractions in the collinear regions $v=0$
and $v=1$
that make this differential
cross section a distribution instead of a function of the variable~$v$.
Anticipating this problem, 
we therefore consider an observable $A(v)$ and
integrate over~$v$:
\beq
\langle A \rangle = \int_0^1 \dd v \, \frac{\dd \sigma}{\dd v} \, A(v),
\eeq
and correspondingly define the expectation value
\beq
{\cal A}\,=\,\frac{\dd \langle A \rangle}{\dd\XB\,\dd y\,\dd z}.
\eeq
Explicitly, it is given by
\beqnm{cala}
{\cal A}&=&\sum_j \int_\XB^1 \frac{\dd u}{u} \sum_F
\int \dps^{(n_F)}(\underline p)\,
\frac{\alpha^2}{2\SH\XB}\,\frac{1}{e^2 (2\pi)^d}
\left[Y^M\,\Pii_M^{\mu\nu}
            +Y^L\,\Pii_L^{\mu\nu}
       \right]\,H_{\mu\nu}
\nonu
&&\quad\quad\quad\quad\quad\quad\quad\cdot
f_{j/P}\left(\frac{\XB}{u}\right)\,
\sum_{\alpha=1}^N\,D_{h/F_\alpha}\left(\frac{\EH}{p_\alpha^0}\right)
\,\frac{\EE}{E_\alpha} \,(1-\XB)\, A(v_\alpha)
,
\eeqn
as can be inferred from Eqs.~(\ref{opi}) and (\ref{eq12a}).
The energy 
$p_\alpha^0$ of the $\alpha^{\mbox{\scriptsize th}}$ parton $F_\alpha$
is assumed to be defined
in the hadronic centre-of-mass frame.
We have defined $v_\alpha=(1-\cos\vartheta_\alpha)/2$, 
where $\vartheta_\alpha$ is the polar
angle of the $\alpha^{\mbox{\scriptsize th}}$ parton in the same frame;
$f_{j/P}$ and $D_{h/F_\alpha}$ are the parton densities
and the fragmentation functions, respectively,
and we have used the following projection operators:
\beqm{projcons}
\labelmmm{projgl}
\Pii_M^{\mu\nu}=\left(-g^{\mu\nu}\right),\quad
\Pii_L^{\mu\nu}=\frac{4\XB^2}{Q^2}\PP^\mu\PP^\nu.
\eeq
{}From Eq.~(\ref{cala}) we infer that
the leading order given by the process of the naive parton model depicted
in Fig.~\ref{lodiagra} is\footnote{``LO'' stands for leading order, 
and ``NLO'' for next-to-leading order, respectively.}
\beqm{eq24}
{\cal A}_{\mbox{\scriptsize LO, $fD$}} = Y^M\,\sum_{i=q,\overline{q}}\,c_i
\,\int_{\XB}^1\frac{\dd u}{u}\,
\int\frac{\dd \rho}{\rho}\,
f_{i/P}
\left(\frac{\XB}{u}\right)
\,D_{h/i}\left(\frac{z}{\rho}\right)\,\delta(1-u)\,\delta(1-\rho)\,A(0),
\eeq
where
\beqnm{cin}
c_i&=&\frac{\alpha^2}{2\SH \XB}\cdot 2\pi \cdot 4(1-\epsilon)\,Q_i^2,
\eeqn
$eQ_i$ being the electric charge of the quark of flavour~$i$.
The trivial integration variable $\rho$ is introduced already here,
because later on the factorization terms will depend on the same
variable.
The integration in $u$ is kept for the same reason. 
To this order, there are no longitudinal contributions. They arise
in the QCD-improved parton model in \porder{\alpha_s}.

\begin{figure}[htb] \unitlength 1mm
\begin{center}
\dgpicture{159}{39}
 
\put(54,0){\epsfigdg{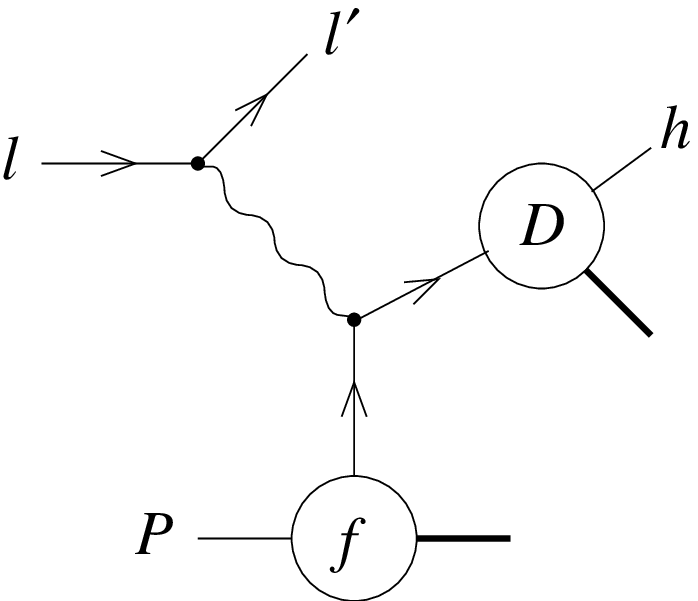}{width=42mm}}

\end{picture}
\end{center}
\shiftcaption
\caption[Leading-Order Contribution in the Current
Fragmentation Region]
{\labelmm{lodiagra} {\it Feynman diagram corresponding
to the leading-order contribution in the
current fragmentation region.
}}   
\end{figure}

\dgsb{QCD Corrections in the Current Fragmentation 
Region}
\labelm{dislnsqcd}

This section gives the details of the calculation of the
\porder{\alpha_s} corrections
to the parton-model process. The virtual one-loop corrections 
to the leading-order QCD subprocess are shown in Fig.~\ref{QCDvirt}, and the
real corrections in Figs.~\ref{QCDcfr1}a to f.

\begin{figure}[htb] \unitlength 1mm
\begin{center}
\dgpicture{159}{30}

\put(25,0){\epsfigdg{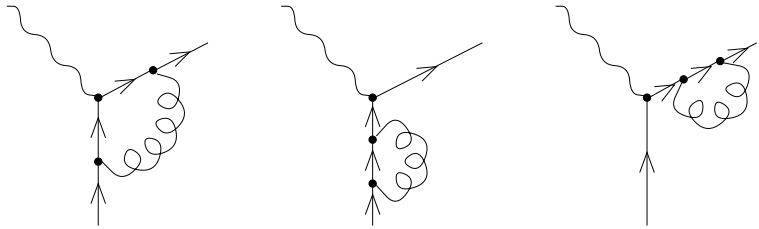}{width=100mm}}

\end{picture}
\end{center}
\shiftcaption
\caption[Virtual QCD Corrections]
{\labelmm{QCDvirt} {\it Feynman diagrams of virtual QCD corrections. 
}}   
\end{figure}

The overall effect of the virtual corrections is to multiply the
leading-order cross section by a constant \cite{45}:
\beqnm{clovirt}
{\cal A}_{\mbox{\scriptsize LO+virt., $fD$}}
&=&Y^M\,\sum_{i=q,\overline{q}}\,c_i\nonu
&&\mathl\cdot
\int_{\XB}^1\frac{\dd u}{u}\,
\int\frac{\dd \rho}{\rho}\,
f_{i/P}
\left(\frac{\XB}{u}\right)
\,D_{h/i}\left(\frac{z}{\rho}\right)\,\delta(1-u)\,\delta(1-\rho)\,A(0)
\nonu
&&\mathl\cdot\Bigg\{
1\,+\,\frac{\alpha_s}{2\pi}
\left(\frac{4\pi\mu_r^2}{Q^2}\right)^\epsilon
\frac{\Gamma(1-\epsilon)}{\Gamma(1-2\epsilon)}C_F
\left(-2\frac{1}{\epsilon^2}-3\frac{1}{\epsilon}-8-\frac{\pi^2}{3}
\right)\Bigg\},
\eeqn
where~$\mu_r$ is the renormalization scale, 
$\alpha_s=\alpha_s(\mu_r^2)$, and~$C_F$ is one of the Casimir invariants
of the colour gauge group $\mbox{SU}(N_C)$, $N_C=3$.
The double and single poles in~$\epsilon$ represent an infrared divergence,
which is cancelled by a contribution from the real corrections similar
to the virtual correction, but of opposite sign.
To this order of perturbation theory, the strong
coupling constant is not renormalized, because 
the loop corrections are the lowest-order
QCD corrections of a QED vertex.

\begin{figure}[htb] \unitlength 1mm
\begin{center}
\dgpicture{159}{113}

\put(5,70){\epsfigdg{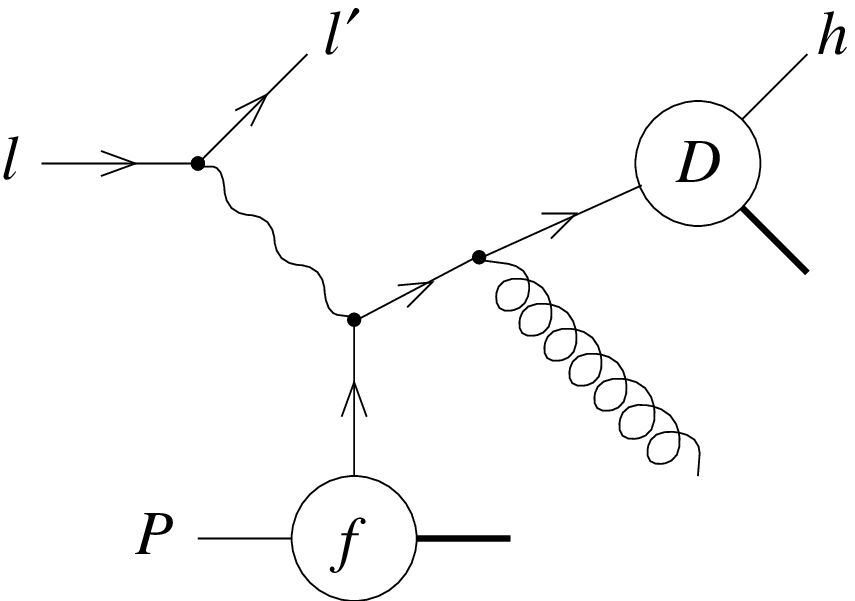}{width=50mm}}
\put(15,60){\lettlab (a)}

\put(60,70){\epsfigdg{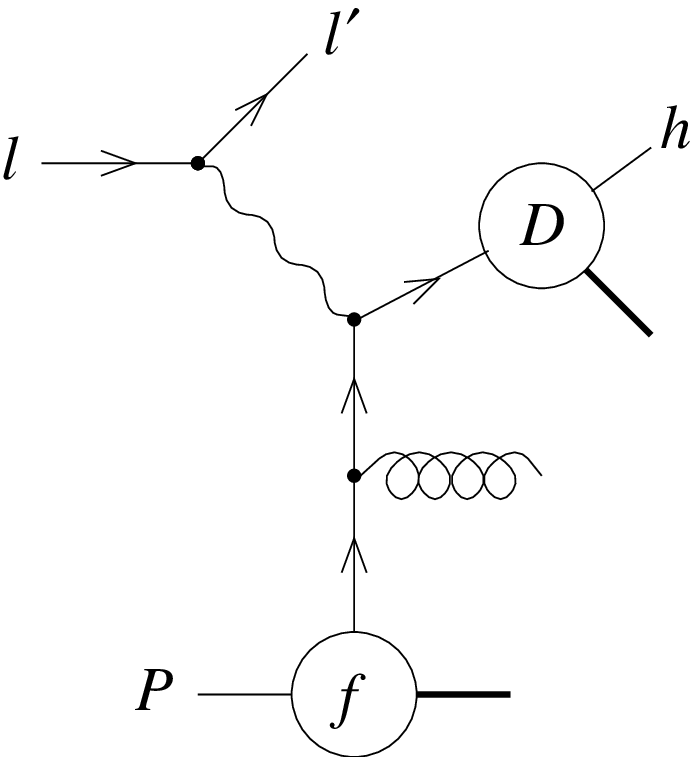}{width=40mm}}
\put(65,60){\lettlab (b)}

\put(105,70){\epsfigdg{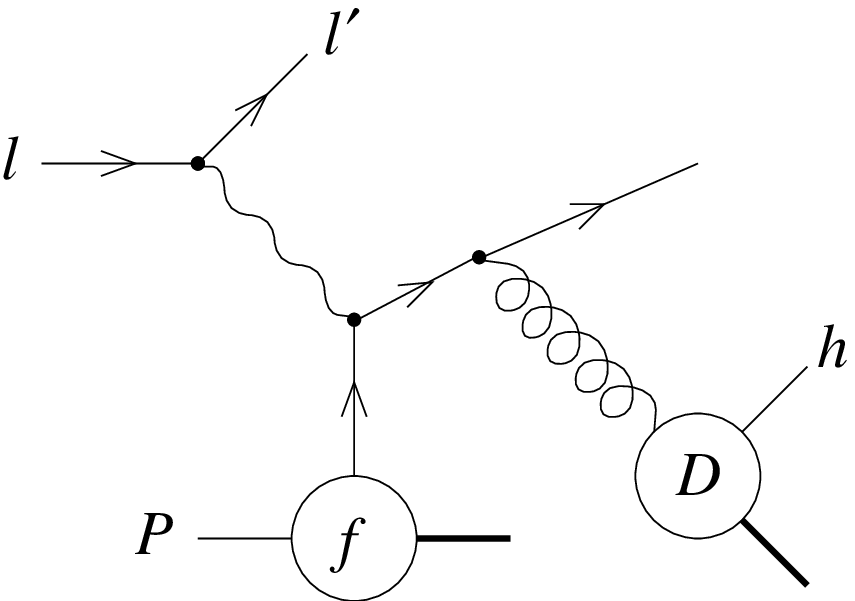}{width=50mm}}
\put(115,60){\lettlab (c)}

\put(5,10){\epsfigdg{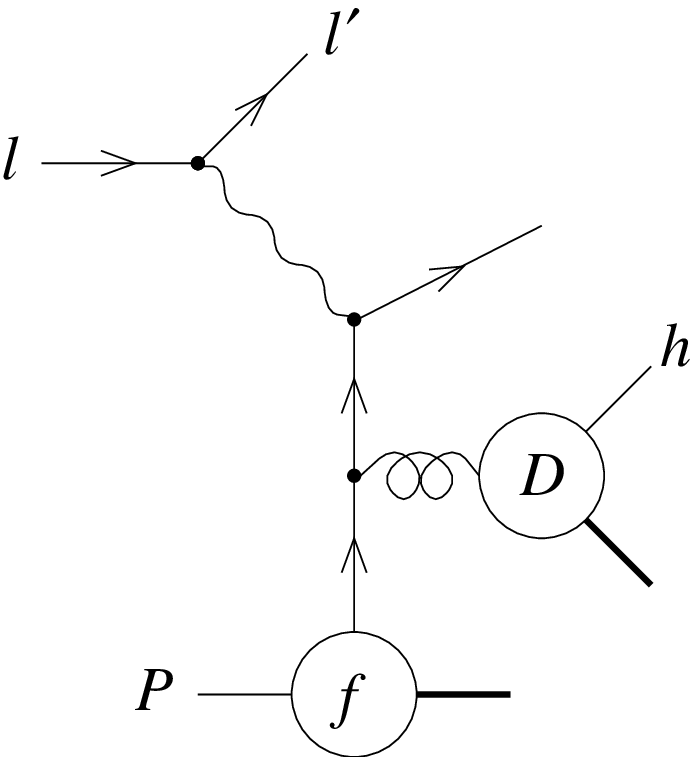}{width=40mm}}
\put(15,0){\lettlab (d)}

\put(55,10){\epsfigdg{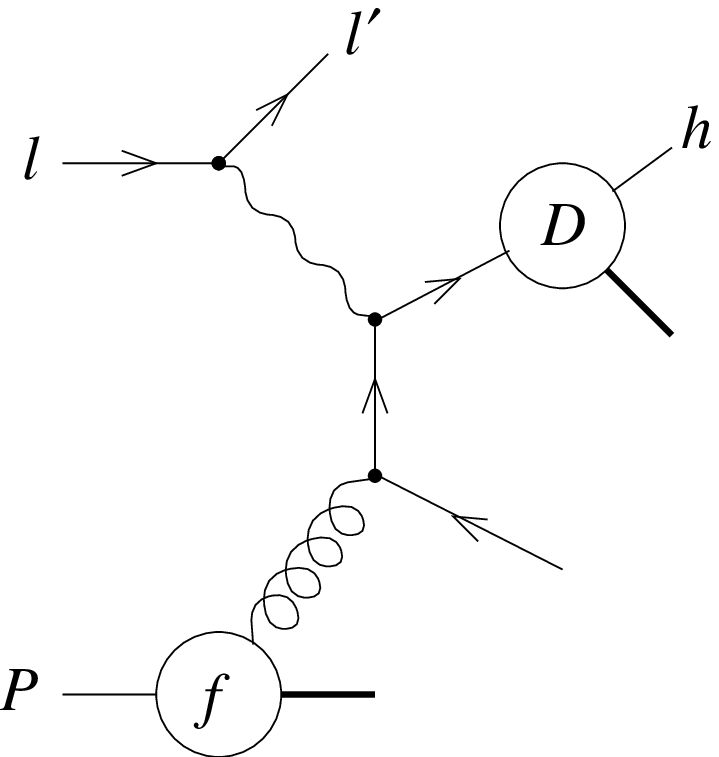}{width=40mm}}
\put(65,0){\lettlab (e)}

\put(105,10){\epsfigdg{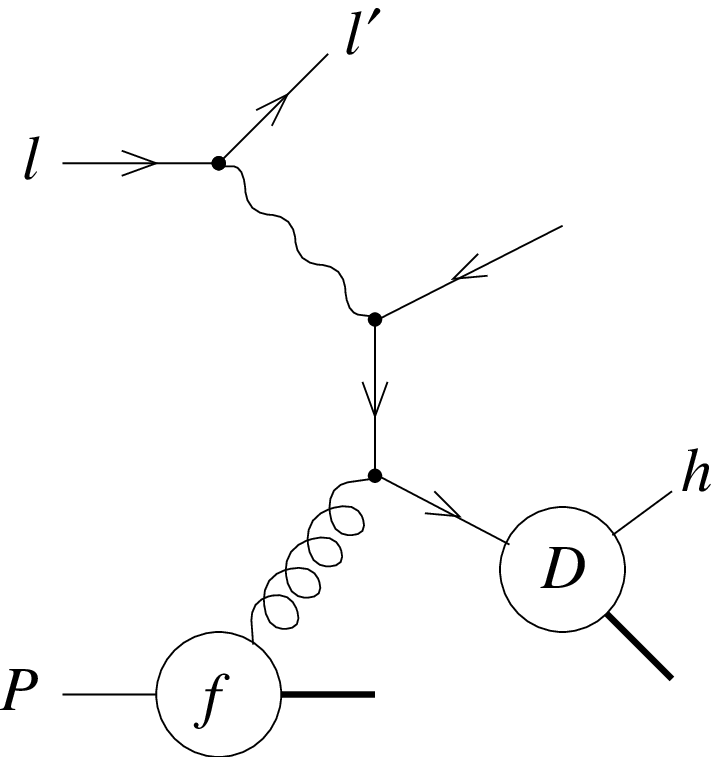}{width=40mm}}
\put(115,0){\lettlab (f)}

\end{picture}
\end{center}
\shiftcaption
\caption[Real QCD Corrections: The Current Fragmentation Region]
{\labelmm{QCDcfr1} {\it Feynman diagrams of real QCD corrections 
for particle production in the current fragmentation region.
}}   
\end{figure}

The metric and longitudinal 
projections of the hadron tensor for the real corrections 
are\footnote{
\footn{footsixteen}
$C_A$ and~$C_F$ are Casimir invariants of the colour gauge group.
For $\mbox{SU}(N_C)$, they are given by $C_A=N_C$ and $C_F=(N_C^2-1)/2N_C$.
For~$N_f$ flavours of quarks, we define $T_R\doteq N_f/2$  and set 
$T_f\doteq T_R/N_f=1/2$.}
(see for example Refs.~\cite{93,94,92,95})
\beqnm{hmnproj}
&&\frac{1}{e^2(2\pi)^{2d}}\Pii_M^{\mu\nu}H_{\mu\nu}(\gamma^*q\rightarrow qg)
\nonu
&&\quad=\,8\pi\, \frac{\alpha_s}{2\pi}\,\mu_r^{2\epsilon}
\,2\pi\,C_F\,Q_q^2\,\cdot 4(1-\epsilon)\,
\left[
  (1-\epsilon)\left(
       \frac{s_{ig}}{s_{qg}}
      +\frac{s_{qg}}{s_{ig}}
              \right)
 +\frac{2Q^2s_{iq}}{s_{ig}s_{qg}}
 +2\epsilon
\right],\nonu
&&{ }\nonu
&&\frac{1}{e^2(2\pi)^{2d}}\Pii_L^{\mu\nu}H_{\mu\nu}(\gamma^*q\rightarrow qg)
\nonu
&&\quad=\,8\pi\, \frac{\alpha_s}{2\pi}\,\mu_r^{2\epsilon}
\,2\pi\,C_F\,Q_q^2\,\cdot 4(1-\epsilon)\,
\left[
4\,\frac{u^2}{Q^2}\,\frac{1}{2}s_{iq}
\right],\nonu
&&{ }\nonu
&&\frac{1}{e^2(2\pi)^{2d}}\Pii_M^{\mu\nu}H_{\mu\nu}(\gamma^*g\rightarrow q\qb)
\nonu
&&\quad=\,8\pi\, \frac{\alpha_s}{2\pi}\,\mu_r^{2\epsilon}
\,2\pi\,T_f\,Q_q^2\,\cdot 4(1-\epsilon)\,
\left[
       \frac{s_{iq}}{s_{i\qb}}
      +\frac{s_{i\qb}}{s_{iq}}
 -\frac{1}{1-\epsilon}\frac{2Q^2s_{q\qb}}{s_{iq}s_{i\qb}}
 -2\frac{\epsilon}{1-\epsilon}
\right],\nonu
&&{ }\nonu
&&\frac{1}{e^2(2\pi)^{2d}}\Pii_L^{\mu\nu}H_{\mu\nu}(\gamma^*g\rightarrow q\qb)
\nonu
&&\quad=\,8\pi\, \frac{\alpha_s}{2\pi}\,\mu_r^{2\epsilon}
\,2\pi\,T_f\,Q_q^2\,\cdot 4(1-\epsilon)\,
\left[
4\,\frac{u^2}{Q^2}\,s_{q\qb}
\right].
\eeqn
The invariants are defined by
$s_{AB}=2p_A p_B$, 
and the variable~$u$ is given by $u=Q^2/(Q^2+\hat{s})$,
where~$\hat{s}$ is the invariant mass squared of the two outgoing
partons.
The momenta~$p_i$, $p_q$, $p_{\qb}$, $p_g$ correspond to
the incident parton (quark or gluon), an outgoing quark, an outgoing
antiquark and an outgoing gluon, respectively. The formulae already contain
the appropriate factors for the average over the colour degrees of freedom
of the incoming partons. An additional factor of $1/(1-\epsilon)$
has been provided for the terms with an incoming gluon,
because gluons have $2(1-\epsilon)$ helicity states in
$(4-2\epsilon)$ space-time dimensions compared with only two in 
the case of quarks.
\noindent
In order to perform the phase-space integrations, suitable parametrizations
of the two-particle phase space $\dps^{(2)}$, depending on 
two independent variables, are needed.
After integration 
over the azimuthal angle relative to the lepton plane one variable
is left over. 
The latter can be chosen such that it is the
one that is actually used in the factorization of the collinear singularities.
Three parametrizations suitable for the consideration of 
various collinear limits
are given explicitly 
in Appendix~\ref{phasespace}.

\begin{figure}[htb] \unitlength 1mm
\begin{center}
\dgpicture{159}{85}

\put(45,0){\epsfigdg{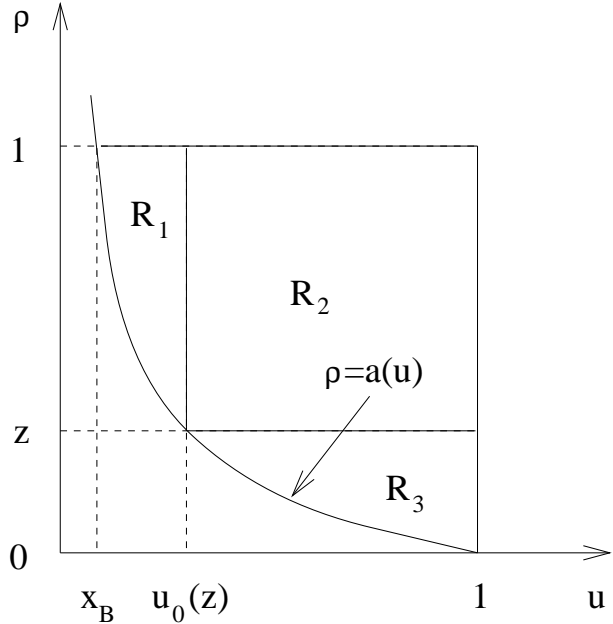}{width=80mm}}

\end{picture}
\end{center}
\shiftcaption
\caption[Phase Space for the Real Corrections]
{\labelmm{PSRC} {\it Phase space for the real corrections (see text).
The two regions corresponding to $\alpha=1$ and 
$\alpha=2$ are marked by~$R_1$ and $R_2$, respectively.
Region~$R_3$ is excluded because the parent parton of the observed
hadron must have an energy larger than that of the hadron itself.
The curve $\rho=a(u)$ corresponds to particle production in the
direction of the target remnant.}
}   
\end{figure}

The singularity structure of the matrix elements for the real corrections
may be discussed
in terms of the
variables~$\rho$ and~$u$ 
of parametrization A from Appendix~\ref{mlp}.
The phase space is shown in Fig.~\ref{PSRC}.
Let~$p_1$ be the momentum of the parent parton of the observed hadron, $p_2$ 
the momentum of the other outgoing parton, and~$p_i$ the momentum 
of the incident parton.
Then~$\rho$ is the energy corresponding to~$p_1$
in the hadronic centre-of-mass frame, 
scaled such that $\rho\in[0,1]$.
The singularities of the matrix elements are located in the phase-space
regions collected in Table~\ref{SPSR}.
In the case of the singularities 
$p_2\,\parallel\,\,p_i$ and $p_1\,\parallel\,\,p_2$, 
the hadron is produced in the current direction;
$p_1\rightarrow 0$ is excluded because the parent parton of the
observed hadron must have a non-vanishing energy, and
if $p_1\,\parallel\,p_i$, then the observed hadron is part of
the target fragments.

\begin{table}[htb]
\begin{center}
\begin{tabular}[h]{|c||c|}
\hline
\rule[-2.5mm]{0mm}{8mm}Singularity & Condition\\ \hline\hline
$p_2\,\parallel\,\,p_i$ & $\rho=1$\\ \hline
$p_1\,\parallel\,\,p_i$ & $\rho=a(u)$\\ \hline
$p_1\,\parallel\,\,p_2$ & $u=1$\\ \hline
$p_1\rightarrow 0$ & $\rho=0$ and $u=1$\\ \hline
$p_2\rightarrow 0$ & $\rho=1$ and $u=1$\\ \hline
\end{tabular}
\end{center}
\shiftcaption
\caption[Singular Phase-Space Regions]
{\labelmm{SPSR} {\it Singular phase-space regions.
The variables are defined in Appendix~\ref{mlp}.
}}
\end{table}

By means of the given phase-space parametrizations, the infrared and collinear
singularities can be factorized and the integrations can
be performed.
The calculation is done by expressing the product of the regulator terms
from the $d$-dimensional phase space,
for instance
of the form of $(1-u)^{-\epsilon}$,
and the singular propagators represented in this case by $1/(1-u)$,
by means of a Laurent series of distributions in~$\epsilon$,
see Appendix~\ref{appdistr}.
In this way the singular terms can be identified, 
and the finite terms are directly given in a form suitable
for the numerical evaluation of arbitrary observables.
The intermediate steps of this straightforward but lengthy calculation
are not given here. The results are presented in two steps.
First we give the divergent contributions and discuss the extent 
to which they can be absorbed into renormalized parton densities
and fragmentation functions.
Then we discuss the structure of the finite contributions including the
terms that compensate scale variations of the factorization scales.
The contributions are denoted by
$B_1^M$, $B_2^M$,
$B_1^L$, $B_2^L$, where
the superscript and subscript in $B_\alpha^X$ stand for
\begin{itemize}
\item[\dgbullet] 
$\alpha$: phase-space region for the integration variable~$u$:\\
$\alpha=1$ (Region $\mbox{R}_1$): 
$u\in\Big[\XB,\XB/(\XB+(1-\XB)z)\Big]$,\\
$\alpha=2$ (Region $\mbox{R}_2$):
$u\in\Big[\XB/(\XB+(1-\XB)z),1\Big]$;
\item[\dgbullet]
$X$: polarization of the exchanged photon:\\
$X=M$: metric contribution,\\
$X=L$: longitudinal contribution. 
\end{itemize}
The contribution of the real corrections to the expectation value~${\cal A}$ 
is given by
\beqm{AsumA}
{\cal A}_{\mbox{\scriptsize real, $fD$}}=
  {\cal A}_{B_1^M}
+ {\cal A}_{B_2^M}
+ {\cal A}_{B_1^L}
+ {\cal A}_{B_2^L}.
\eeq
The singular and finite contributions will be denoted by
${\cal A}_{B_\alpha^X}^s$ and ${\cal A}_{B_\alpha^X}^f$, 
respectively, so that
${\cal A}_{B_\alpha^X}={\cal A}_{B_\alpha^X}^s+{\cal A}_{B_\alpha^X}^f$.
Throughout the calculation the $\overline{\mbox{MS}}$ factorization
scheme is used both for the parton densities and
fragmentation functions. The choice of 
the factorization scheme defines
the finite parts unambiguously. Moreover, one
has to choose three factorization scales, 
one ($\mu_f^2$) for the renormalized parton densities $f^r$, 
another one ($\mu_D^2$) for the fragmentation functions $D^r$, 
and a third one ($\mu_M^2$) for the target
fragmentation functions~$M^r$, which will soon be introduced.
For the singular contributions we obtain
\beqnm{scwo}
{\cal A}_{B^M_1}^s&=&Y^M\,\sum_{i=q,\qb}\,c_i\,\frac{\alpha_s}{2\pi}\nonu
&&\mathl\cdot\,\Bigg\{
\int_{\XB}^{\XB/(\XB+(1-\XB)z)}\,\frac{\dd u}{u}\,
\int_{a(u)}^1\,\frac{\dd \rho}{\rho}\,A(v(\rho,u))\nonu
&&\mathl\cdot\Bigg[
f_{i/P}\left(\frac{\XB}{u}\right)\,
D_{\hh/i}\left(\frac{z}{\rho}\right)
\left(
-\frac{1}{\epsilon}\right)\frac{\Gamma(1-\epsilon)}{\Gamma(1-2\epsilon)}
\left(\frac{4\pi\mu_r^2}{\mu_f^2}\right)^\epsilon\,
P_{q\leftarrow q}(u)\,\delta(1-\rho)
\nonu
&&\mathl\;\;+\,
f_{g/P}\left(\frac{\XB}{u}\right)\,
D_{\hh/i}\left(\frac{z}{\rho}\right)
\left(
-\frac{1}{\epsilon}\right)\frac{\Gamma(1-\epsilon)}{\Gamma(1-2\epsilon)}
\left(\frac{4\pi\mu_r^2}{\mu_f^2}\right)^\epsilon\,
P_{q\leftarrow g}(u)\,\delta(1-\rho)
\Bigg]\nonu
&&\mathl+
\int_{\XB}^{\XB/(\XB+(1-\XB)z)}\,\frac{\dd u}{u}\,
(1-\XB)\,A(1)\nonu
&&\mathl\cdot\Bigg[
f_{i/P}\left(\frac{\XB}{u}\right)\,
D_{\hh/g}\left(\frac{(1-\XB)\,z\,u}{\XB(1-u)}\right)
\left(-\frac{1}{\epsilon}\right)\frac{\Gamma(1-\epsilon)}{\Gamma(1-2\epsilon)}
\left(\frac{4\pi\mu_r^2}{\mu_M^2}\right)^\epsilon\,
\frac{1}{1-u}\,\frac{u}{\XB}\,\hat{P}_{gq\leftarrow q}(u)\nonu
&&\mathl\;\;+
f_{g/P}\left(\frac{\XB}{u}\right)\,
D_{\hh/i}\left(\frac{(1-\XB)\,z\,u}{\XB(1-u)}\right)
\left(-\frac{1}{\epsilon}\right)\frac{\Gamma(1-\epsilon)}{\Gamma(1-2\epsilon)}
\left(\frac{4\pi\mu_r^2}{\mu_M^2}\right)^\epsilon\,
\frac{1}{1-u}\,\frac{u}{\XB}\,\hat{P}_{\qb q\leftarrow g}(u)
\Bigg]
\Bigg\}\nonu
&&\mathl\!\!\!\!+\,\porder{\epsilon},
\nonu
&& {} \nonu
{\cal A}_{B^M_2}^s&=&Y^M\,\sum_{i=q,\qb}\,c_i\,\frac{\alpha_s}{2\pi}\,
\int_{\XB/(\XB+(1-\XB)z)}^1\,\frac{\dd u}{u}\,
\int_{z}^1\,\frac{\dd \rho}{\rho}\,A(v(\rho,u))\nonu
&&\mathl\cdot\Bigg[
f_{i/P}\left(\frac{\XB}{u}\right)\,
D_{\hh/i}\left(\frac{z}{\rho}\right)
\Bigg\{
\frac{\Gamma(1-\epsilon)}{\Gamma(1-2\epsilon)}
\left(\frac{4\pi\mu_r^2}{Q^2}\right)^\epsilon\,
C_F\,\left(2\frac{1}{\epsilon^2}+3\frac{1}{\epsilon}\right)\,
\delta(1-u)\delta(1-\rho)\nonu
&&\mathl\quad\quad\quad
-\frac{1}{\epsilon}\frac{\Gamma(1-\epsilon)}{\Gamma(1-2\epsilon)}
\left[
\left(\frac{4\pi\mu_r^2}{\mu_f^2}\right)^\epsilon\,
P_{q\leftarrow q}(u)\,\delta(1-\rho)
+\left(\frac{4\pi\mu_r^2}{\mu_D^2}\right)^\epsilon\,
P_{q\leftarrow q}(\rho)\,\delta(1-u)
\right]\Bigg\}\nonu
&&\mathl\;+\,
f_{i/P}\left(\frac{\XB}{u}\right)\,
D_{\hh/g}\left(\frac{z}{\rho}\right)
\left(-\frac{1}{\epsilon}\right)\frac{\Gamma(1-\epsilon)}{\Gamma(1-2\epsilon)}
\left(\frac{4\pi\mu_r^2}{\mu_D^2}\right)^\epsilon\,
P_{g\leftarrow q}(\rho)\,\delta(1-u)\nonu
&&\mathl\;+\,
f_{g/P}\left(\frac{\XB}{u}\right)\,
D_{\hh/i}\left(\frac{z}{\rho}\right)
\left(-\frac{1}{\epsilon}\right)\frac{\Gamma(1-\epsilon)}{\Gamma(1-2\epsilon)}
\left(\frac{4\pi\mu_r^2}{\mu_f^2}\right)^\epsilon\,
P_{q\leftarrow g}(u)\,\delta(1-\rho)
\Bigg]\nonu
&&\mathl\!\!\!\!+\,\porder{\epsilon},
\nonu
&& {} \nonu
{\cal A}_{B^L_1}^s&=&0,
\nonu
&& {} \nonu
{\cal A}_{B^L_2}^s&=&0.
\eeqn
Here $A(1)$ is the observable $A$ evaluated for the observed particle
running in the target remnant direction.
One sees immediately that the infrared singularities proportional to
$2/\epsilon^2+3/\epsilon$ cancel in the sum of virtual and real corrections.
Let us for the moment
assume that the observable~$A$ is such that $A(1)=0$, i.e.\  
that it vanishes if the observed hadron is in the target fragmentation
region. Then the only additional singular contributions are those that involve
the Altarelli--Parisi splitting functions $P_{B\leftarrow A}$. 
They can be absorbed into renormalized fragmentation functions $D^r$
(terms $\sim \delta(1-u)$), cf.\ Eq.~(\ref{dredef}), and renormalized
parton densities $f^r$ (terms $\sim \delta(1-\rho)$), given by the expression
\cite{44,45}
\beqm{fredef}
f_{i/P}(\xi)=\int_\xi^1\frac{\dd u}{u}\,
\left[\delta_{ij}\,\delta(1-u)
+\frac{1}{\epsilon}\frac{\alpha_s(\mu_r^2)}{2\pi}
\frac{\Gamma(1-\epsilon)}{\Gamma(1-2\epsilon)}
\left(\frac{4\pi\mu_r^2}{\mu^2}\right)^\epsilon
P_{i\leftarrow j}(u)\right]
\,f^r_{j/P}\left(\frac{\xi}{u},\mu^2\right).
\eeq
The finite terms from the Born contribution, the virtual correction 
and the renormalization of the distribution functions are
\beqnm{ALOVfinite}
{\cal A}^f_{\mbox{\scriptsize LO+virt., $fD$}}
&=&Y^M\,\sum_{i=q,\overline{q}}\,c_i\nonu
&&\mathl\cdot
\int_{\XB}^1\frac{\dd u}{u}\,
\int\frac{\dd \rho}{\rho}\,
f_{i/P}
\left(\frac{\XB}{u},\mu_f^2\right)
\,D_{\hh/i}\left(\frac{z}{\rho},\mu_D^2\right)\,\delta(1-u)\,
\delta(1-\rho)\,A(0)
\nonu
&&\mathl\cdot\Bigg\{
1\,+\,\frac{\alpha_s}{2\pi}\,C_F\,
\left(-8-\frac{\pi^2}{3}
\right)\Bigg\}.
\eeqn

For observables~$A$ with support in the target fragmentation region, 
there are singular terms that cannot be taken into account
by means of a redefinition of~$f$ and~$D$, namely those
proportional to unsubtracted Altarelli--Parisi
splitting functions $\hat{P}_{CB\leftarrow A}$; 
for explicit expressions see Appendix~\ref{APapp}. 
The redefinition of the
distribution 
functions 
$f$ and~$D$
is already dictated by, for example, 
the inclusive cross section in lepton--nucleon
scattering and the one-particle-inclusive cross section in \epem{} scattering.
In order for them to be universal, i.e.\ process-independent functions,
the absorption of additional singular terms is not permitted.
The kinematical configuration in the singular phase-space region
$\rho=a(u)$, see Fig.~\ref{PSRC}, 
is such that the observed
hadron is part of the target remnant, collinear to its parent parton,
with the second outgoing 
parton in the current direction.
The Feynman diagrams with singularities in this region are those from 
Figs.~\ref{QCDcfr1}d and~f. 
It can easily be seen that a term suited to 
absorb these singularities
cannot occur in \porder{\alpha_s^0}
in the standard formulation involving only parton densities and fragmentation
functions.
This problem is tackled in the next section, where it is shown that
all singularities can be absorbed consistently if target fragmentation
functions are introduced.

The finite expectation value of $A$ is given by
\beqnm{asumfin}
{\cal A}^f_{\mbox{\scriptsize total}}
={\cal A}^f_{\mbox{\scriptsize LO+virt., $fD$}}
+{\cal A}^f_{\mbox{\scriptsize real, $fD$}}.
\eeqn
The finite terms ${\cal A}_{B^X_\alpha}^f$ are collected in 
Appendix~\ref{rfee}.
Due to the necessary subtractions of collinear singularities, they
are distributions in the variables~$v$ and~$\rho$. 
In the case of fragmentation functions
$D(\sigma)$ with subtractions 
or delta functions 
at $\sigma=1$, the cross section will be a distribution 
in the variable~$z$, being singular at $z=1$. All convolutions with
regular observables are finite. Care must be taken, however, 
to choose 
the region of integration appropriately (for a discussion see 
Section~\ref{ImpStudy}).
As can be seen from the explicit expressions
in Eq.~(\ref{Arealfinite}), and as is expected, 
the next-to-leading-order contributions provide compensating terms for 
the dependence on the factorization scales $\mu_f$, $\mu_D$ and $\mu_M$, 
the latter being introduced in Section~\ref{QCDTFR}. 
Formally, the factorization-scale dependence of the total cross section
will therefore be of \porder{\alpha_s^2}. In practice, in particular in the
case of heavy-quark production, the scale dependence may still be substantial,
for the reason that the photon--gluon fusion process, arising at 
\porder{\alpha_s}, may give large contributions, and its 
genuine factorization-scale dependence 
of \porder{\alpha_s^2} is not compensated by any term included in 
the present calculation.
Since the leading-order process is of \porder{\alpha_s^0}, there is no 
renormalization-scale dependence in leading order. 
For a numerical study of the scale dependence, 
we refer to Section~\ref{FRDEP}.

\dgsb{Deeply Inelastic Lepton--Nucleon Scattering:\\
The Target Fragmentation Region}
\labelm{dilntfr}
We have shown that in the standard formalism involving parton densities~$f$ and
fragmentation functions~$D$, a collinear singularity, which cannot be absorbed
into~$f$ and~$D$ by a renormalization, is present in the cross section
for the kinematical configuration of the observed hadron collinear with
the target fragments. It turns out that this problem can be resolved
by the introduction of a new class of phenomenological distribution functions, 
the target fragmentation functions.
Briefly, a target fragmentation function
or ``fracture function''
$M_{i,h/P}(\xi,\zeta)$ is a probability density
to find a parton~$i$ with momentum fraction~$\xi$ and a hadron~$h$
with momentum fraction~$\zeta$ in the nucleon~$P$, where the observed
hadron is collinear with the target fragments.
Obviously, the kinematical restrictions on~$\xi$ and~$\zeta$
are $\xi\in[0,1]$, $\zeta\in[0,1]$, and
$\xi+\zeta\leq 1$.
For the purpose of this and the following section, we need
no other properties of target fragmentation functions, and so a more detailed
discussion is postponed to Section~\ref{tffdetails}.
Here and in the next section 
we calculate the cross section in next-to-leading order 
in deeply inelastic lepton--nucleon scattering
for the observed hadron in the target fragmentation region, 
and show that the additional collinear singularity can be absorbed
into the target fragmentation functions by a suitable renormalization, 
thus giving in total a finite result.

\begin{figure}[htb] \unitlength 1mm
\begin{center}
\dgpicture{159}{42}

\put(58,0){\epsfigdg{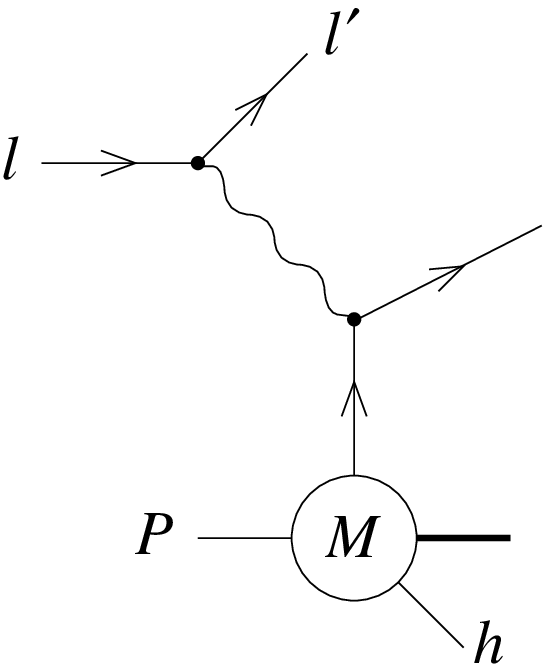}{width=34mm}}
\end{picture}
\end{center}
\shiftcaption
\caption[Leading-Order Contribution in the Target
Fragmentation Region]
{\labelmm{lodiagratarget} {\it Feynman diagram corresponding
to the leading-order contribution in the
target fragmentation region.
}}   
\end{figure}

The lowest-order contribution, given by the graph in 
Fig.~\ref{lodiagratarget}, is
\beqnm{alom}
{\cal A}_{\mbox{\scriptsize LO, $M$}}
&=&Y^M\,\sum_{i=q,\overline{q}}\,c_i
\int_{\XB/(1-(1-\XB)z)}^1\frac{\dd u}{u}
\,M_{i,\hh/P}\left(\frac{\XB}{u},(1-\XB)z\right)
\nonu
&& { } \nonu
&&
\quad\quad\quad\quad\quad
\quad\quad\quad\quad\quad
\cdot\,\delta(1-u)\,(1-\XB)\,A(1).
\eeqn
This expression is similar to Eq.~(\ref{eq24}) for the current fragmentation 
region, with the difference that the product of a parton density and 
a fragmentation function is replaced by a target fragmentation function.
The observable~$A$ is evaluated at $v=1$. The additional factor of
$(1-\XB)$ is explained by 
the fact that the momentum of
the observed hadron and of the other nucleon fragments is given by $(1-\XB)P$, 
after the parton incident in the hard scattering process has been removed
from the nucleon.

\dgsb{QCD Corrections in
the Target Fragmentation
Region}
\labelm{QCDTFR}

As in the case of the current fragmentation region, the next-to-leading-order
QCD corrections consist of virtual and real corrections.
The virtual corrections are given by the same graphs as in Fig.~\ref{QCDvirt}, 
so the sum of the leading order and the virtual corrections is
given by
\beqnm{tlovirt}
{\cal A}_{\mbox{\scriptsize LO+virt., $M$}}
&=&Y^M\,\sum_{i=q,\overline{q}}\,c_i
\int_{\XB/(1-(1-\XB)z)}^1\frac{\dd u}{u}
M_{i,\hh/P}\left(\frac{\XB}{u},(1-\XB)z\right)
\nonu
&&{ } \nonu
&&\mathl\cdot
\,\delta(1-u)\,(1-\XB)\,A(1)
\nonu
&&{ } \nonu
&&\mathl\cdot\Bigg\{
1\,+\,\frac{\alpha_s}{2\pi}
\left(\frac{4\pi\mu_r^2}{Q^2}\right)^\epsilon
\frac{\Gamma(1-\epsilon)}{\Gamma(1-2\epsilon)}C_F
\left(-2\frac{1}{\epsilon^2}-3\frac{1}{\epsilon}-8-\frac{\pi^2}{3}
\right)\Bigg\}.
\eeqn

\begin{figure}[htb] \unitlength 1mm
\begin{center}
\dgpicture{159}{131}

\put(20,80){\epsfigdg{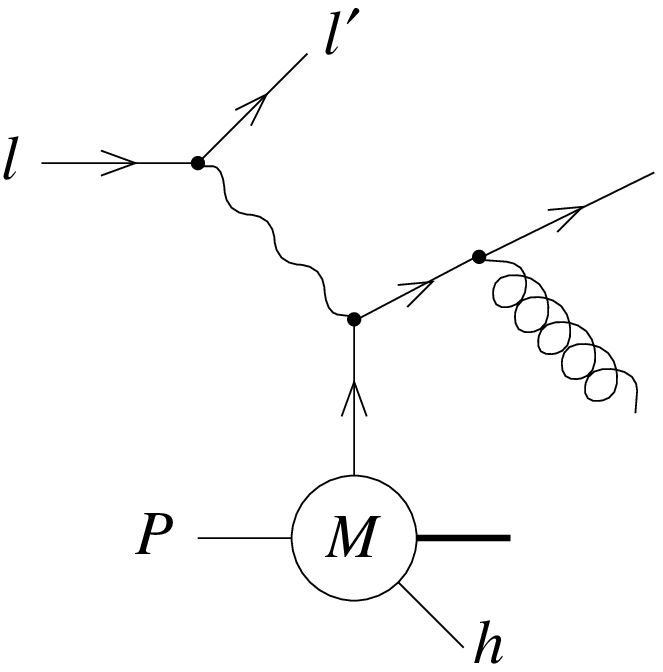}{width=41mm}}
\put(35,70){\lettlab (a)}

\put(90,80){\epsfigdg{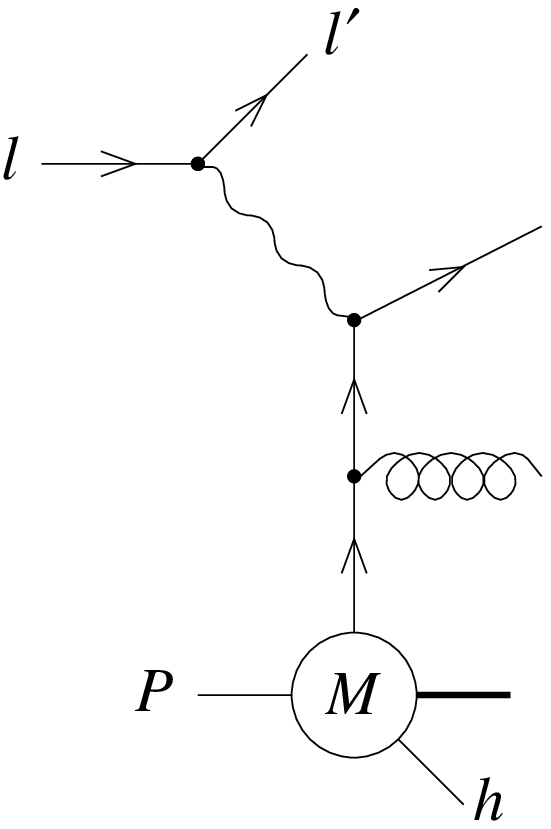}{width=34mm}}
\put(105,70){\lettlab (b)}

\put(30,10){\epsfigdg{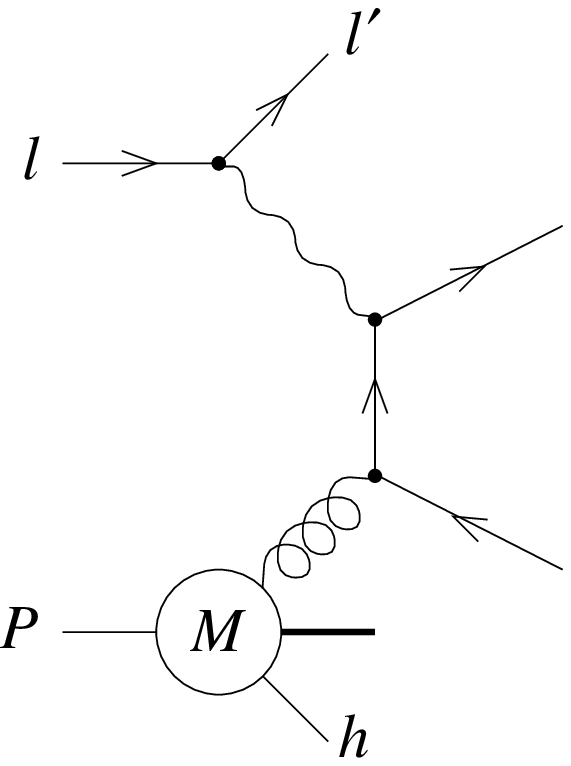}{width=34mm}}
\put(35,0){\lettlab (c)}

\put(100,10){\epsfigdg{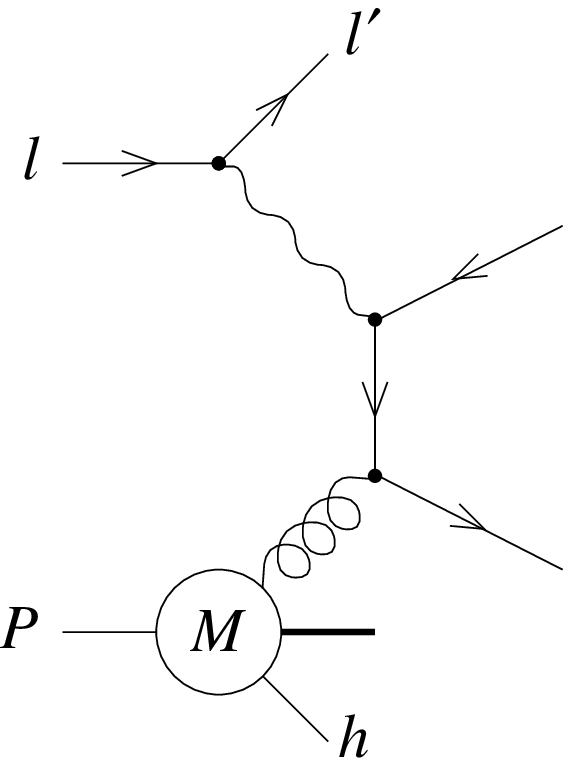}{width=34mm}}
\put(105,0){\lettlab (d)}

\end{picture}
\end{center}
\shiftcaption
\caption[Real QCD Corrections: The Target Fragmentation Region]
{\labelmm{QCDtfr1} {\it Feynman diagrams of real QCD corrections 
for particle production in the target fragmentation region.
}}   
\end{figure}

The graphs for the real corrections are shown in Fig.~\ref{QCDtfr1}.
The contribution of the real corrections to the expectation value~${\cal A}$ 
is given by
\beqm{AsumB}
{\cal A}_{\mbox{\scriptsize real, $M$}}=
  {\cal A}_{C^M}
+ {\cal A}_{C^L}.
\eeq
As before, we state the results for the singular and regular terms
separately, ${\cal A}_{C^X}={\cal A}_{C^X}^s+{\cal A}_{C^X}^f$. 
The singular contributions are
\beqnm{tsar}
{\cal A}_{C^M}^s&=&Y^M\,\sum_{i=q,\qb}\,c_i\,\frac{\alpha_s}{2\pi}\,
\int_{\XB/(1-(1-\XB)z)}^1\,\frac{\dd u}{u}\,A(1)\nonu
&&\mathl\cdot\Bigg[
M_{i,h/P}\left(\frac{\XB}{u},(1-\XB)z\right)
\Bigg\{
\frac{\Gamma(1-\epsilon)}{\Gamma(1-2\epsilon)}
\left(\frac{4\pi\mu_r^2}{Q^2}\right)^\epsilon\,
C_F\,\left(2\frac{1}{\epsilon^2}+3\frac{1}{\epsilon}\right)\,
\delta(1-u)\,(1-\XB)\nonu
&&\quad\quad\quad\mathr\mathr
-\frac{1}{\epsilon}\frac{\Gamma(1-\epsilon)}{\Gamma(1-2\epsilon)}
\left(\frac{4\pi\mu_r^2}{\mu_M^2}\right)^\epsilon\,
P_{q\leftarrow q}(u)\,(1-\XB)
\Bigg\}\nonu
&&\mathl\;+\,
M_{g,h/P}\left(\frac{\XB}{u},(1-\XB)z\right)
\Bigg\{
-\frac{1}{\epsilon}\frac{\Gamma(1-\epsilon)}{\Gamma(1-2\epsilon)}
\left(\frac{4\pi\mu_r^2}{\mu_M^2}\right)^\epsilon\,
P_{q\leftarrow g}(u)\,(1-\XB)\Bigg\}\Bigg]
\nonu
&&\!\!\!\!\!\!\!\!\!+\,\porder{\epsilon},
\nonu
{} \nonu
{\cal A}_{C^L}^s&=&0.
\eeqn
Again, the singularities from the virtual corrections cancel against 
corresponding terms from the real corrections.
The singularities proportional to $-(1/\epsilon)P_{B\leftarrow A}(u)$
have the same structure as those for the case of parton densities.
It can easily be seen that, in order to take into account the additional 
collinear singularities proportional to   
$-(1/\epsilon)\hat{P}_{CB\leftarrow A}(u)$ 
from Section~\ref{dislnsqcd}, the bare target
fragmentation functions $M$ have to be defined in terms of the renormalized 
ones 
$M^r$
as \cite{78}\footnote{The terms involving the splitting functions
$P_{g\leftarrow g}$ and $\hat{P}_{gg\leftarrow g}$
are not needed for the process under consideration.}
\beqnm{Mren}
M_{i,\hh/P}(\xi,\zeta)&&\nonu
&&\mathl\mathl
=\int_{\xi/(1-\zeta)}^1\frac{\dd u}{u}\,
\left[\delta_{ij}\,\delta(1-u)
+\frac{1}{\epsilon}\frac{\alpha_s(\mu_r^2)}{2\pi}
\frac{\Gamma(1-\epsilon)}{\Gamma(1-2\epsilon)}
\left(\frac{4\pi\mu_r^2}{\mu^2}\right)^\epsilon
P_{i\leftarrow j}(u)\right]
\,M^r_{j,\hh/P}\left(\frac{\xi}{u},\zeta,\mu^2\right)\nonu
&&\mathl\mathl
+\int_\xi^{\xi/(\xi+\zeta)}\frac{\dd u}{u}\,
\frac{1}{1-u}
\,\frac{u}{\xi}\,
\frac{1}{\epsilon}\frac{\alpha_s(\mu_r^2)}{2\pi}
\frac{\Gamma(1-\epsilon)}{\Gamma(1-2\epsilon)}
\left(\frac{4\pi\mu_r^2}{\mu^2}\right)^\epsilon\nonu
&&\quad\quad
\cdot\,\hat{P}_{ki\leftarrow j}(u)
\,f_{j/P}\left(\frac{\xi}{u}\right)
\,D_{\hh/k}\left(\frac{\zeta u}{\xi(1-u)}\right).
\eeqn
Of course, as usual, the finite terms 
of the convolution kernels may be chosen in a way in principle
arbitrary, thereby defining a specific factorization scheme. 
The expression just given subtracts the poles in~$\epsilon$
such that no unnatural transcendental numbers
are left over in the finite terms. We define the functions~$M^r$ as given in 
Eq.~(\ref{Mren})
to be the renormalized target fragmentation functions in the
\msbar{} scheme.  
The sum of all terms
\beqnm{asum}
{\cal A}_{\mbox{\scriptsize total}}
={\cal A}_{\mbox{\scriptsize LO+virt., $fD$}}
+{\cal A}_{\mbox{\scriptsize real, $fD$}}
+{\cal A}_{\mbox{\scriptsize LO+virt., $M$}}
+{\cal A}_{\mbox{\scriptsize real, $M$}},
\eeqn
expressed in terms of renormalized quantities, is finite. 
Here the finite contribution 
from the leading order and from the
virtual correction in Eq.~(\ref{tlovirt}) reads
\beqnm{tlovirtfin}
{\cal A}^f_{\mbox{\scriptsize LO+virt., $M$}}
&=&Y^M\,\sum_{i=q,\overline{q}}\,c_i
\int_{\XB/(1-(1-\XB)z)}^1\frac{\dd u}{u}
M_{i,\hh/P}\left(\frac{\XB}{u},(1-\XB)z,\mu_M^2\right)
\nonu
&&{ } \nonu
&&\mathl\quad\quad\quad\cdot
\,\delta(1-u)\,(1-\XB)\,A(1)\,\Bigg\{
1\,+\,\frac{\alpha_s}{2\pi}\,
C_F
\left(
-8-\frac{\pi^2}{3}
\right)\Bigg\},
\eeqn
and the explicit expressions for the finite contributions ${\cal A}_{C^X}^f$
are collected in Appendix~\ref{rfee}.

\dgcleardoublepage

\markh{Target Fragmentation Functions}
\dgsa{Target Fragmentation Functions}
\labelm{tffdetails}
{\it
Target fragmentation functions are joint probability distributions
for an observed particle in the target fragmentation region and a 
parton incident in the hard scattering process.
The expression of the renormalized 
distribution functions in terms of the bare ones
follows from the requirement that the additional collinear singularity
that was calculated in Section~\ref{dislnsqcd} be absorbed. 
{}From this expression a renormalization group equation can be derived.
Scale evolution is driven by two terms, a homogeneous one, reminiscent of
the Altarelli--Parisi equation, and an inhomogeneous one, due
to particle production by the fragmentation of partons emitted in 
the backward direction.
The scale evolution equation can also be motivated by 
an intuitive argument, and
rewritten in a form that allows
an analogy to the Altarelli--Parisi equation to be drawn.
We also briefly discuss momentum sum rules.
It turns out that a momentum sum rule related to the momentum 
fraction of the observed particle is fulfilled, whereas the one
that would naively be expected to hold in relation 
to the momentum fraction of the incoming parton is violated. 
This violation can be traced back to the inhomogeneous term 
in the evolution equation.
}

\dgsb{Definition}
\labelm{tffdef}
Target fragmentation functions $M_{i,h/P}(\xi,\zeta)$ 
\cite{77}\footnote{
See also Ref.~\cite{96}.
},
see Fig.~\ref{TMrgfig},
are probability densities
to find a parton~$i$ with momentum fraction~$\xi$ and a hadron~$h$
with momentum fraction~$\zeta$ in the nucleon~$P$, where the observed
hadron is collinear with the target fragments.
The kinematical restrictions on~$\xi$ and~$\zeta$
are $\xi\in[0,1]$, $\zeta\in[0,1]$, and
$\xi+\zeta\leq 1$.
There is no definition of target fragmentation functions in terms
of the operator product expansion, because the process under 
consideration is not fully inclusive.
In the case of QCD corrections, target fragmentation functions,
very similar to parton densities, have
to be redefined in order to be finite and physical quantities.
As a consequence, they become scale-dependent.
The expression for the bare target fragmentation functions in terms of the
renormalized ones has already been given in Eq.~(\ref{Mren}).

\begin{figure}[htb] \unitlength 1mm
\begin{center}
\dgpicture{159}{42}

\put(50,5){\epsfigdg{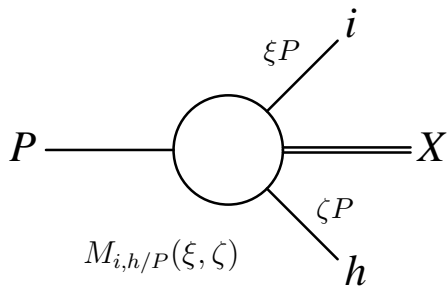}{width=58mm}}

\put(84,35){${\ds\xi P}$}
\put(91,14){${\ds\zeta P}$}
\put(60,8){\large $M_{i,h/P}(\xi,\zeta)$}

\end{picture}
\end{center}
\shiftcaption
\caption[A Target Fragmentation Function]
{\labelmm{TMrgfig} {\it A target fragmentation function:
$P$ is the incoming hadron, $h$ is the observed particle, 
and $i$ is the parton incident in the hard scattering process.
}}   
\end{figure}

\dgsb{Renormalization Group Equation}
\labelm{tffrgen}
The renormalization group 
equation for~$M$, corresponding to Eq.~(\ref{Mren}), is \cite{77}
\beqnm{Mrge}
\frac{\partial M_{i,\hh/P}(\xi,\zeta,\mu^2)}{\partial \ln\mu^2}
&=&\frac{\alpha_s(\mu^2)}{2\pi}\int_{\xi/(1-\zeta)}^1\frac{\dd u}{u}\,
P_{i\leftarrow j}(u)
\,M_{j,\hh/P}\left(\frac{\xi}{u},\zeta,\mu^2\right)\nonu
&&\mathl\mathl\!\!\!\!\!\!\!
+\frac{\alpha_s(\mu^2)}{2\pi}\int_\xi^{\xi/(\xi+\zeta)}\frac{\dd u}{u}\,
\frac{1}{1-u}
\,\frac{u}{\xi}\,
\hat{P}_{ki\leftarrow j}(u)
\,f_{j/P}\left(\frac{\xi}{u},\mu^2\right)
\,D_{\hh/k}\left(\frac{\zeta u}{\xi(1-u)},\mu^2\right),
\eeqn
where the sums over the parton indices 
$j$ and~$k$ are done implicitly.
There are two contributions to the scale evolution: the
homogeneous term is related to the emission of partons in the initial state, 
as indicated in 
Fig.~\ref{Mrgfig}a; the inhomogeneous term comes from a source term 
that originates from the fact that a parton radiated
collinearly from the 
incident parton in the backward direction is the parent parton of the observed 
hadron (see Fig.~\ref{Mrgfig}b).

Intuitively, this scale evolution equation can be derived as follows.
The parton content of the nucleon in the interval $[\xi,\xi+\dd\xi]$
at a scale $\mu^2+\dd\mu^2$ under the assumption that a
hadron is observed in the target fragmentation region 
is given by the contribution at the scale~$\mu^2$ plus the contribution (a)
from the radiation of a parton (Fig.~\ref{Mrgfig}a) 
and (b) from particle production in the backward direction
(Fig.~\ref{Mrgfig}b). Case (a) gives, up to a factor of $\alpha_s/(2\pi)$,  
and suppressing the particle labels and scales,
a contribution
$\int\dd\rho\, M(\rho,\zeta)\int\dd u \,P(u)\,\delta(\xi-\rho u)$,
since $M(\rho,\zeta)\,\dd\rho$ is the probability to find a parton 
in the interval $[\rho,\rho+\dd\rho]$ in the nucleon, 
and $P(u)\,\dd u$ is the probability
of the incident parton splitting
into a parton carrying only a fraction~$u$ of the
parent parton's momentum. Performing the $\rho$-integration by means
of the $\delta$-function, we arrive at 
$\int\dd u / u \,M(\xi/u,\zeta)\,P(u)$. 
The range of integration in~$u$ has to be chosen such that 
the momentum fraction~$\rho$ of the parton before the splitting process
is in the range of~$\xi$ (the parton does not split at all) 
and $(1-\zeta)$ (the parton carries all of the available momentum 
of the nucleon, after removing the momentum fraction~$\zeta$
for the observed hadron). This is the first term 
of Eq.~(\ref{Mrge}).
The second term is obtained from case (b). The contribution is
$\int\dd\rho \,f(\rho)\int\dd u \,\hat{P}(u)
\,\delta(\xi-\rho u)\int \dd z \,D(z)$:
$D(z)\,\dd z$ is the probability for a parton to turn into a hadron with
momentum fraction in the interval $[z,z+\dd z]$, and 
$\hat{P}(u)\,\dd u$ is the probability
for a parton to split into two partons, one of these carrying
a momentum fraction~$u$, under the condition that the second parton is not 
soft (cf.\ Appendix~\ref{APapp}).
By performing the $\rho$-integration by means of the $\delta$-function
and by using the fact that $\dd z=\dd \zeta \,u/[\xi(1-u)]$, 
we arrive at 
$\int\dd u/u \,f(\xi/u)\,\hat{P}(u)\,
\int \dd \zeta \,u/[\xi(1-u)]\, D(\zeta u/[\xi(1-u)])$, which is, after
``dividing'' by $\dd \zeta$ to obtain the distribution differential 
in the momentum fraction of the observed hadron, the second term
of Eq.~(\ref{Mrge}).
The range of integration in~$u$ has to be such that the momentum
fraction of the incident parton~$\rho$ is between
$\xi+\zeta$ (so as to produce a parton with momentum~$\xi P$
and a hadron with momentum~$\zeta P$) and~$1$ (the maximum 
possible momentum fraction). 

\begin{figure}[htb] \unitlength 1mm
\begin{center}
\dgpicture{159}{158}

\put(25,85){\epsfigdg{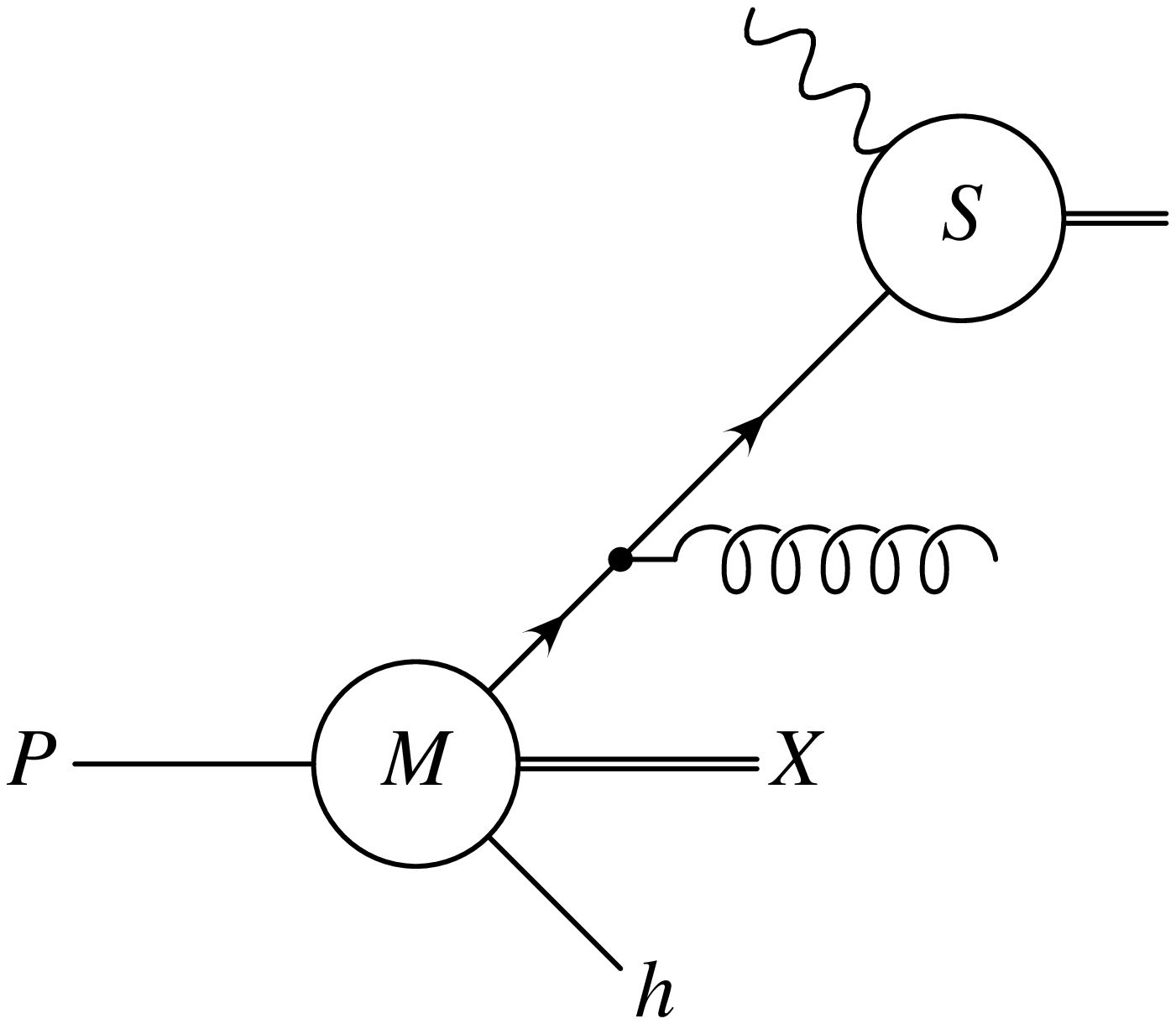}{width=83mm}}
\put(30,85){\lettlab (a)}
\put(25,10){\epsfigdg{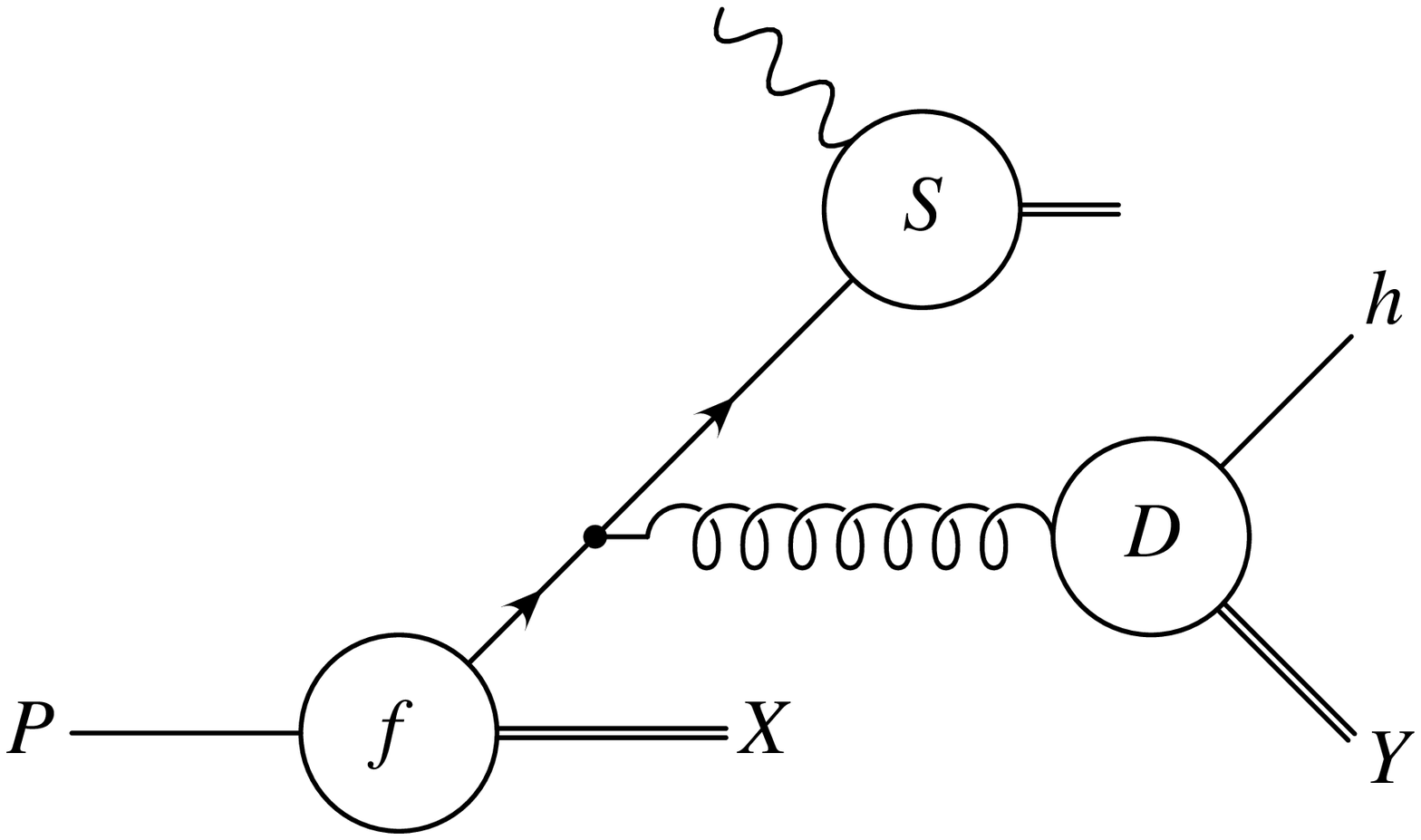}{width=105mm}}
\put(30,0){\lettlab (b)}

\put(57,115){${\ds\frac{\xi}{u}P}$}
\put(68,125){${\ds\xi P}$}
\put(84,124){${\ds\xi\frac{1-u}{u}P}$}
\put(66,94){${\ds\zeta P}$}

\put(58,30){${\ds\frac{\xi}{u}P}$}
\put(69,40){${\ds\xi P}$}
\put(86,39){${\ds\xi\frac{1-u}{u}P}$}
\put(123,38){${\ds\zeta P}$}

\end{picture}
\end{center}
\shiftcaption
\caption[Contributions to the Scale Evolution of Target
Fragmentation Functions]
{\labelmm{Mrgfig} {\it Examples of contributions to the scale evolution
of target fragmentation functions: emission of a parton in the initial state
giving rise to the standard Altarelli--Parisi scale evolution 
$K\otimes M$ (a) 
and the additional source term $\hat{K} \otimes f \otimes D$ 
for particle production in the
backward direction (b).
The hard scattering process is denoted by $S$.
}}   
\end{figure}

The general solution of the inhomogeneous differential equation (\ref{Mrge})
is a sum of a special solution of the inhomogeneous equation and an arbitrary
solution of the homogeneous equation. Let us denote these two terms by
$M^{(P)}$ and $M^{(NP)}$, respectively, such that 
$M=M^{(P)}+M^{(NP)}$. The equations satisfied by these functions are, in 
symbolic form,
\beqm{PNPMeq}
\begin{array}[b]{lclcl}
\partial_{\ln\mu^2}M&=&K\otimes M &+& \hat{K}\otimes f\otimes D,\\
\partial_{\ln\mu^2}M^{(P)}&=&K\otimes M^{(P)} &+& \hat{K}\otimes f\otimes D,\\
\partial_{\ln\mu^2}M^{(NP)}&=&K\otimes M^{(NP)}.&&
\end{array}
\eeq
Here~$K$ and~$\hat{K}$ stand for the evolution kernels.
We may use the boundary condition 
\beqm{bcm}
M^{(P)}(\xi,\zeta,\mu_0^2)=0
\eeq 
for an arbitrary scale~$\mu_0$. This definition then fixes~$M^{(NP)}$ for a 
given~$M$. The superscripts~$(P)$ and~$(NP)$ 
stand for ``perturbative'' and ``non-perturbative'',
respectively. The interpretation we have in mind is that the contribution 
to~$M$ via $M^{(P)}$ comes from the ``perturbative'' production of hadrons,
via the radiation of partons in the backward direction and their subsequent 
fragmentation, corresponding to graphs
such as those of Figs.~\ref{QCDcfr1}d and~f. 
Given the parton densities~$f$ and fragmentation functions
$D$, this contribution can be calculated in perturbation theory.
Loosely speaking, 
$M^{(P)}$ contains all contributions that arise by this mechanism for scales
larger than~$\mu_0$. Everything below this scale is termed 
``non-perturbative'', and has to be determined experimentally
or via explicit phenomenological models. 
It should be clear that there is no strict separation of the perturbative
and non-perturbative contributions, due to the fact that they are defined
by an arbitrary scale 
$\mu_0$\footnote{In the application to the production of heavy quarks,
this scale will be set to a value of the order of the heavy-quark mass.}.

For the solution of the renormalization group equations, it is 
useful to redefine the arguments of the target fragmentation functions
by introducing {\it reduced} target fragmentation functions~$N$ by
\beqm{redtff}
N_{i,h/P}(x,\zeta,\mu^2)\doteq M_{i,h/P}(x(1-\zeta),\zeta,\mu^2).
\eeq
Thus~$x$ is the momentum fraction carried by the parton incident in the 
hard subprocess with respect to the momentum $P-h$. The renormalization
group equation of the functions~$N$ is
\beqnm{Nrge}
\frac{\partial N_{i,\hh/P}(x,\zeta,\mu^2)}{\partial \ln\mu^2}
&=&\frac{\alpha_s(\mu^2)}{2\pi}\int_{x}^1\frac{\dd u}{u}\,
P_{i\leftarrow j}(u)
\,N_{j,\hh/P}\left(\frac{x}{u},\zeta,\mu^2\right)\nonu
&
+&\,\frac{\alpha_s(\mu^2)}{2\pi}
\int_{(1-\zeta)x}^{(1-\zeta)x/\left(x+(1-x)\zeta\right)}\frac{\dd u}{u}\,
\frac{1}{1-u}
\,\frac{u}{x(1-\zeta)}\,
\nonu
&&\mathl\cdot\,
\hat{P}_{ki\leftarrow j}(u)
\,f_{j/P}\left(\frac{x(1-\zeta)}{u},\mu^2\right)
\,D_{\hh/k}\left(\frac{1}{x}\frac{u}{1-u}\frac{\zeta}{1-\zeta},\mu^2\right).
\eeqn
The first term has exactly the same form as the one in the scale evolution 
equation of parton densities~$f$ \cite{97}:
\beqm{frgeq}
\frac{\partial f_{i/P}(x,\mu^2)}{\partial \ln \mu^2}=
\frac{\alpha_s(\mu^2)}{2\pi}\int_x^1\frac{\dd u}{u}\,
P_{i\leftarrow j}(u)
\,f_{j/P}\left(\frac{x}{u},\mu^2\right).
\eeq
Parton densities fulfil the sum rule
\beqm{fsum}
\sum_i \int_0^1 \dd x\,x\,f_{i/P}(x,\mu^2)=1;
\eeq
the $\mu^2$-independence follows from the fact that
\beqm{spfsum}
\sum_i \int_0^1 \dd u\,u\,P_{i\leftarrow j}(u)=0.
\eeq
In the next section it is shown that the corresponding
sum rule for target fragmentation functions is violated.
\dgsb{Momentum Sum Rules}
\labelm{tffmomsr}
In analogy to the sum rule for fragmentation functions
in Eq.~(\ref{Dmomsum}),
target fragmentation functions fulfil a momentum sum rule
\cite{77}
\beqm{tffmsum}
\sum_h\int_0^{1-\xi}\dd \zeta \, \zeta\, M_{i,h/P}\left(\xi,\zeta,\mu^2\right)
=(1-\xi)\,f_{i/P}(\xi,\mu^2),
\eeq
i.e.\ the sum over all possible tagged particles
corresponds to the probability distribution of the parton incident in the
hard subprocess.
Under the assumption that the sum rule 
in Eq.~(\ref{tffmsum}) is valid for some scale~$\mu_0$,
Eq.~(\ref{tffmsum}) can be proved for any~$\mu$ by an application of the 
renormalization group equations (\ref{Drgeq}), 
(\ref{Mrge}) 
and (\ref{frgeq}), and
by means of relations from the ``jet calculus'' of
Refs.\ \cite{98,99}.

It is interesting to note that the momentum sum rule with respect to
the variable~$\xi$ for fixed~$\zeta$ is violated. 
As in the case of parton densities, cf.\  Eq.~(\ref{fsum}), 
the homogeneous term 
does not give a contribution to the evolution of the sum over
parton species~$i$. The inhomogeneous term, 
however, is not zero. Because the convolution kernels $\hat{P}$ are
positive, the inhomogeneous term is positive, and consequently
the integral
\beqm{msumr}
\sum_i \int_0^1 \dd x\,x\,M_{i,h/P}(x,z,\mu^2)
\eeq
is increasing with increasing~$\mu^2$.
This can also be seen by 
considering\footnote{D.~Graudenz, L.~Trentadue, 
G.~Veneziano, unpublished manuscript (1995).} 
the double moments $M(n,m)$, defined by 
\beqm{doubmom}
M(n,m)\doteq\int_0^1\frac{\dd \xi}{\xi}\,\xi^n
            \int_{0}^{1-\xi}\frac{\dd \zeta}{\zeta}\,\zeta^m
            \,M(\xi,\zeta),
\eeq
suppressing the scale argument for the moment.
The Mellin moments of functions~$F$ of only one argument, 
such as~$f$, $D$ and the splitting functions~$P$, are defined by
\beqm{singmom}
F(n)\doteq\int_0^1\frac{\dd \xi}{\xi}\,\xi^n\,F(\xi),
\eeq
and we define the
moments $\hat{P}(n,m)$ of the unsubtracted splitting functions
$\hat{P}$ by
\beqm{hatmom}
\hat{P}(n,m)\doteq\int_0^1\frac{\dd u}{u}\,u^n\,(1-u)^m\,\hat{P}(u).
\eeq
After some formal manipulations, the renormalization group 
equation~(\ref{Mrge}) can be rewritten in terms of the
$M(n,m)$, $f(n)$, $D(n)$, and $P(n,m)$, 
the scale derivative of $M(n,m)$ being a sum of two terms corresponding
to the homogeneous and inhomogeneous terms of the renomalization group
equation, respectively.
The first moment in~$\xi$ (i.e.\ $n=2$) 
of the sum over all parton species~$i$ of the 
homogeneous term, 
being of the same structure as the scale derivative of the sum rule for
parton densities in Mellin space,
is zero. 
The inhomogeneous term, however, turns out to be a 
product of positive definite factors, and thus does not 
generally give a zero result.
As a consequence, the derivative of the first moment with respect to
the factorization scale is non-vanishing.
The interpretation is obvious: the fragmentation of partons
emitted in the backward direction enhances the particle multiplicity
of the target remnant jet.
The momentum sum rule violation causes no problem with respect
to unitarity, as can be seen by considering the regularized, i.e.\ 
$\epsilon$ being non-zero, expression for the cross section.
The factorization terms that are ultimately absorbed into the renormalized
target fragmentation functions, giving rise to the inhomogeneous evolution 
term, are effectively subtracted from the real corrections involving
parton densities and fragmentation functions.

\dgsb{Extended Factorization}
\labelm{exfac}
According to Eqs.~(\ref{cur}) and (\ref{tar}),
the complete one-particle-inclusive cross section is given by
\beqm{opicompl}
\sigma=\sigma^{\mbox{\scriptsize hard}}_{fD}\otimes f\otimes D
+\sigma^{\mbox{\scriptsize hard}}_{M}\otimes M.
\eeq
In the current fragmentation region, only the first term contributes, 
whereas the target fragmentation region receives contributions
from both terms. In the standard approach to particle production
in the current fragmentation region, 
the form of the first term as 
a convolution of a hard scattering cross section 
and distribution functions is guaranteed by the factorization theorems
of perturbative QCD. For particle production in the target fragmentation 
region, the factorization theorems have
to be generalized; we call the form of Eq.~(\ref{opicompl})
the {\it extended factorization conjecture}.
Although there is no proof of this conjecture yet, the general picture,
which is to be expected, is that
the soft infrared divergences cancel in the sum of
virtual and real corrections, and that all remaining collinear singularities
can be absorbed into the renormalized process-independent 
distribution functions~$f^r$, $D^r$ and~$M^r$.
This mechanism works in the explicit next-to-leading-order calculation
to one-loop as 
described in Section~\ref{opics}.
We wish to point out that Eq.~(\ref{opicompl})
provides a method to calculate the {\it total} one-particle
inclusive cross section, without any restrictions 
to specific phase-space regions.

\dgcleardoublepage

\markh{Heavy-Quark Fragmentation Functions}
\dgsa{Heavy-Quark Fragmentation Functions}
\labelm{hqff}
{\it
We now consider the case of heavy quarks in the final state. The heavy-quark
mass$\,$\footnote{
In next-to-leading order, 
some care has to be taken to define the mass of a heavy quark, 
see for example the review in Ref.\ \cite{100}, p.~1433.} 
$m$ is a large scale ($m=4.5\,\GeV$ for bottom quarks
and $1.5\,\GeV$ for 
charm 
quarks),
and it is expected that perturbative
QCD is applicable to determine fragmentation functions
for partons into heavy quarks,
although it fails for other particles in the final
state, such as mesons
built of only light valence quarks. 
Indeed it is possible to calculate the heavy-quark 
fragmentation functions in QCD from first principles. A short review 
is given in Section~\ref{ffpqcd}. 
Heavy-quark fragmentation functions for arbitrary scales can be obtained 
by means of the renormalization group equation. Explicit numerical results
are discussed in Section~\ref{hqffrg}.
}

\dgsb{Fragmentation Functions from Perturbative QCD}
\labelm{ffpqcd}
The heavy-quark mass~$m$ is a scale parameter large enough to justify
the application of perturbative QCD. This is certainly true for the bottom
quark, and probably true for the charm 
quark\footnote{It can be questioned whether $1.5\,\GeV$ is a 
sufficiently large 
scale to have a valid perturbative expansion. 
In case of doubt, it is, in principle, possible to use experimentally
determined fragmentation functions for charmed mesons, 
or fragmentation functions of the type proposed in Ref.~\cite{35}, in the 
applications later on.}.
The fragmentation functions $D_{Q/i}(x,\mu^2)$ for partons
$i$ into a heavy quark~$Q$ depend on two mass scales: the heavy 
quark mass~$m$ and the factorization scale~$\mu$.
For~$\mu$ of the order of~$m$, possible logarithmic
terms $\sim\ln(\mu^2/m^2)$ in the only available scales~$\mu$ and~$m$
must be small, 
and so $D_{Q/i}(x,\mu^2)$ has an expansion in terms of the strong
coupling constant:
\beqm{hqexp}
D_{Q/i}(x,\mu^2)=\sum_{n} \, \left(\frac{\alpha_s(\mu^2)}{2\pi}\right)^n 
D_{Q/i}^{(n)}(x,\mu^2).
\eeq
The coefficients $D_{Q/i}^{(n)}(x,\mu^2)$ may be obtained by 
a comparison of a direct calculation 
of heavy-quark inclusive processes 
with heavy quarks represented by external lines in Feynman diagrams,
with a calculation 
based on the fragmentation function formalism as described in 
Section~\ref{fragmfunc}.

\begin{figure}[htb] \unitlength 1mm
\begin{center}
\dgpicture{159}{88}

\put(30,65){\epsfigdg{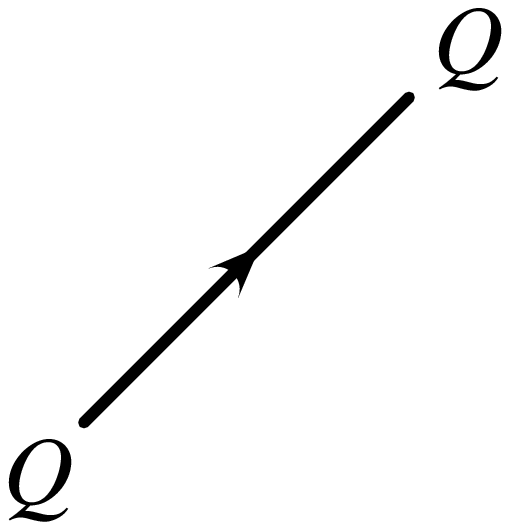}{width=20mm}}
\put(40,55){\lettlab (a)}
\put(85,55){\epsfigdg{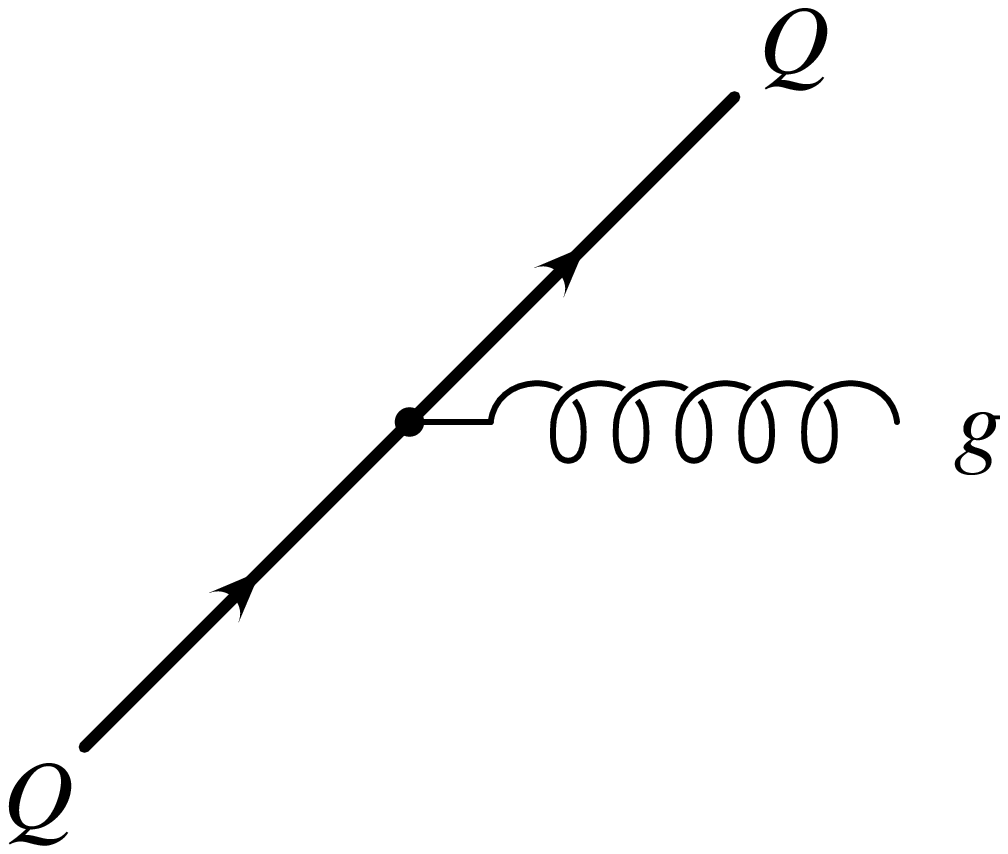}{width=40mm}}
\put(105,55){\lettlab (b)}
\put(20,10){\epsfigdg{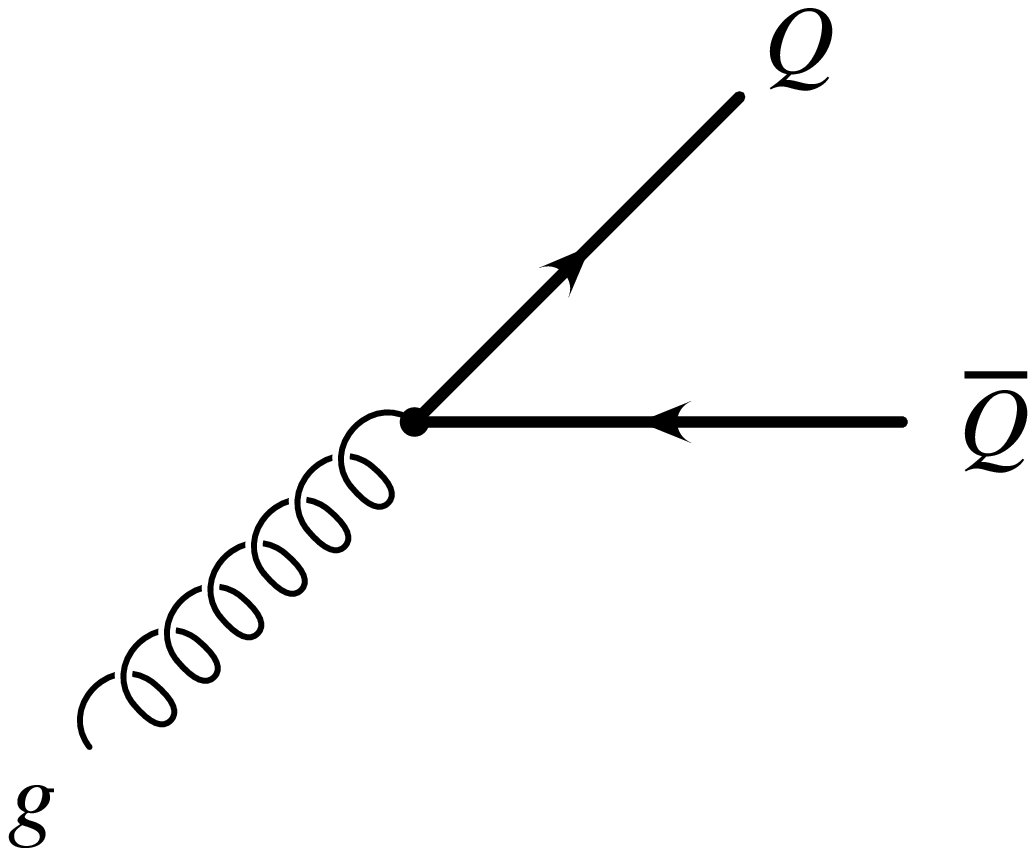}{width=40mm}}
\put(40,0){\lettlab (c)}
\put(75,0){\epsfigdg{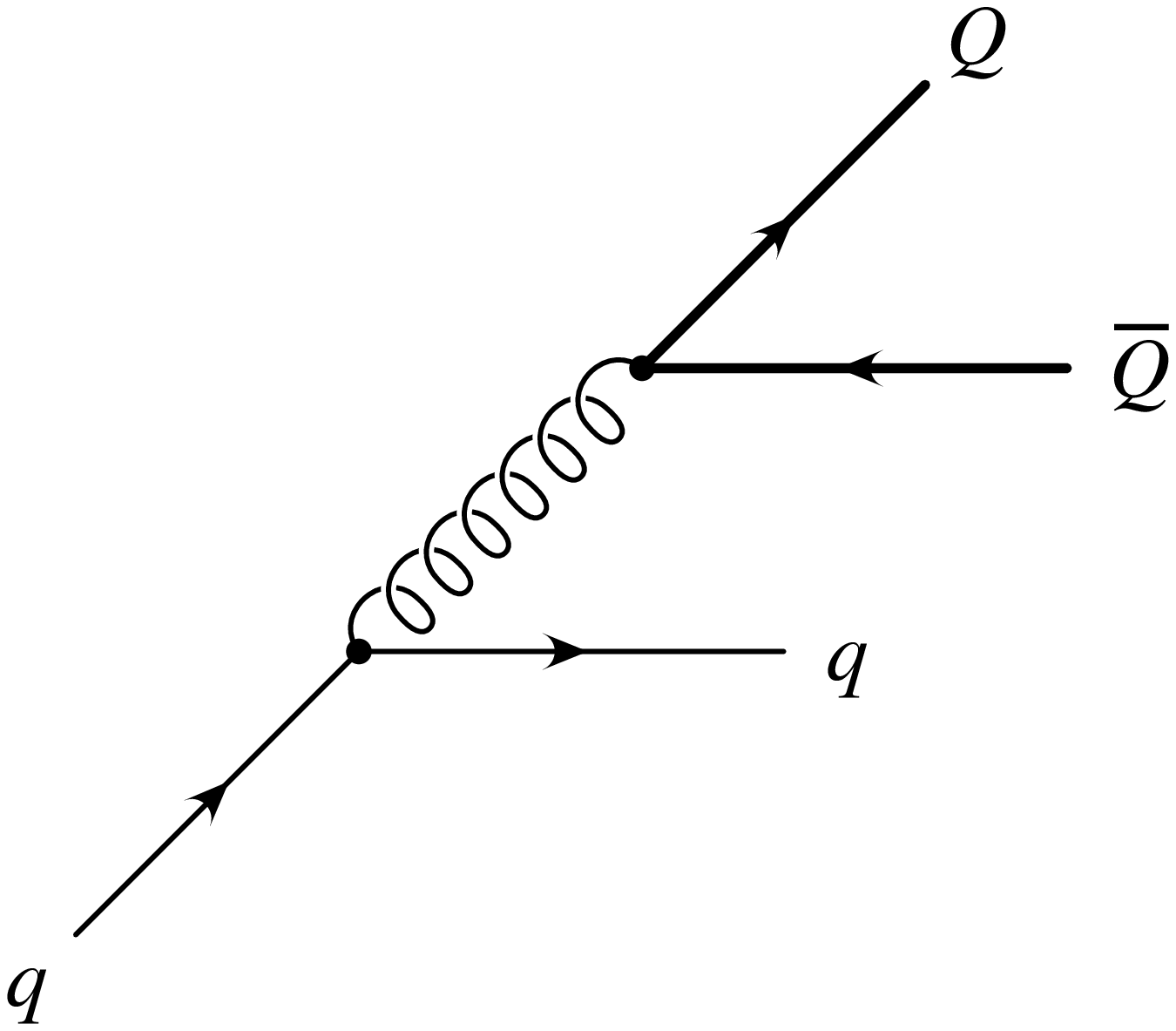}{width=54mm}}
\put(105,0){\lettlab (d)}

\end{picture}
\end{center}
\shiftcaption
\caption[Feynman Diagrams Corresponding
to Heavy-Quark Fragmentation Functions]
{\labelmm{hqvertexfig} {\it Feynman diagrams corresponding
to heavy-quark fragmentation functions:
$D_{Q/Q}(x)$ in \porder{\alpha_s^0} (a),
$D_{Q/Q}(x)$ in \porder{\alpha_s} (b),
$D_{Q/g}(x)$ in \porder{\alpha_s} (c),
$D_{Q/q}(x)$ in \porder{\alpha_s^2} (d).
Here $q$ stands for any light-flavoured quark or antiquark,
or for the heavy antiquark~$\overline{Q}$.
}}   
\end{figure}

The coefficients may be interpreted in terms of the diagrams in 
Fig.~\ref{hqvertexfig}. To \porder{\alpha_s^0}, the only diagram
is the one in Fig.~\ref{hqvertexfig}a describing the direct ``propagation''
of the heavy quark. Obviously, 
\beqnm{DQQcoeff}
D_{Q/Q}^{(0)}(x,\mu^2)&=&\delta(1-x),
\nonu
&& { } \nonu
D_{Q/i}^{(0)}(x,\mu^2)&=&0\quad\mbox{for $i\neq Q$.}
\eeqn
In next-to-leading order, corresponding to the diagrams
in Figs.~\ref{hqvertexfig}b to~d, 
the calculation has been done in Refs.\ \cite{101,37}.
The results are collected in Appendix~\ref{aphqff}.

\dgsb{Solving the Renormalization Group Equation}
\labelm{hqffrg}
In order to obtain $D_{Q/i}(x,\mu^2)$
for scales~$\mu$ very different from the heavy-quark mass~$m$, 
the large logarithms $\sim\ln(\mu^2/m^2)$ must be summed
by means of the renormalization group equation~(\ref{Drgeq}), 
with~$h$ being replaced by~$Q$. In this section we discuss explicit 
numerical results in leading order for the heavy-quark fragmentation functions. 
Numerical results in next-to-leading order have been obtained in 
Refs.\ \cite{102,103}. We restrict ourselves to the leading order, 
because the main focus
of the present work is the issue of target fragmentation functions, where
the evolution kernels are
known in leading order only. On this level a 
next-to-leading-order
input for the heavy-quark fragmentation functions in the inhomogeneous
term of Eq.~(\ref{Mrge}) 
would be inconsistent. Of course, a full next-to-leading-order analysis
would be desirable.

As can be seen, the heavy-quark fragmentation functions $D_{Q/i}(x,\mu^2)$
are singular at $x=1$. In leading order, 
the only singularity is $\delta(1-x)$; in next-to-leading order
additional singularities of the form 
$\subbl{\left(\ln(1-x)/(1-x)\right)}{x}{0}{1}$
come in, cf.\ Eq.~(\ref{eq172a}). 
The splitting functions of the renormalization group equation
(\ref{Drgeq}) contain singular terms, which means that the evolved
contributions will be singular as 
well\footnote{
A direct computation 
by means of Eq.~(\ref{convfor}) shows that actually terms 
$\sim\alpha_s^m\subbl{\left(\ln^{m}(1-x)/(1-x)\right)}{x}{0}{1}$ appear. 
In principle, this means that, for an accurate result,
the terms $\left(\alpha_s \ln(1-x)\right)^k$ should be
resummed for $x\rightarrow 1$ \cite{37}. 
We apply the fragmentation functions later on to get a result for
what we call the ``perturbative piece'' of the target fragmentation function, 
depending on a scale~$\mu_0$. Since we work in leading order, 
due to the lack of
compensating terms in the expansion, the results will
depend considerably on~$\mu_0$. 
Therefore, at this stage of the development of the
formalism, we do not consider it necessary to go into the details of
the subtle problem of $x\rightarrow 1$, whose influence
is probably smaller than the dependence on the
perturbative input scale. For more precise predictions, 
this problem should however be treated along the lines of Ref.\ \cite{37},
where the technical details are explained.}.

\begin{figure}[htb] \unitlength 1mm
\begin{center}
\dgpicture{159}{175}
 
\put(30, 90){\epsfigdg{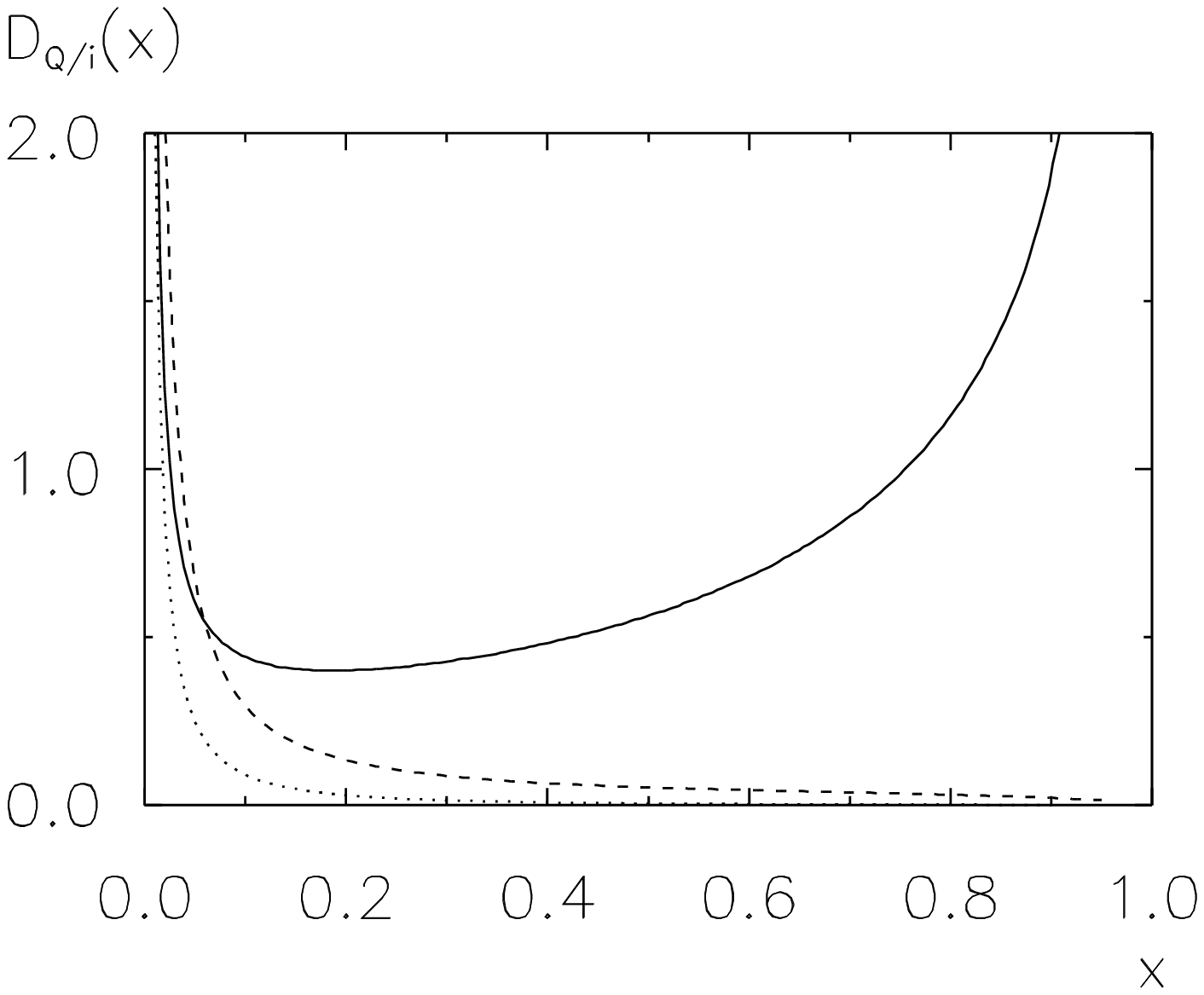}{width=100mm}}
\put(10,105){\lettlab (a)}

\put(30,  0){\epsfigdg{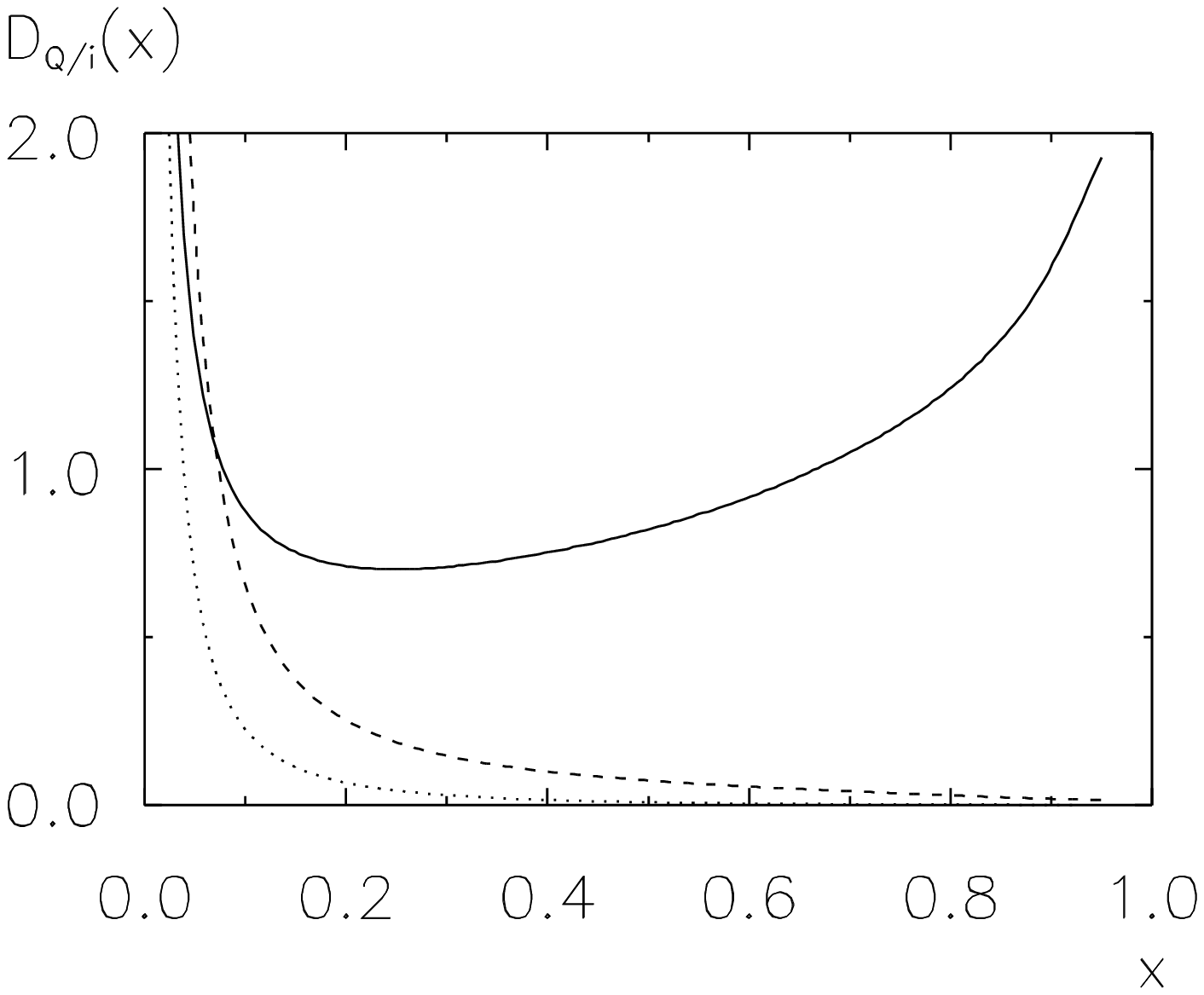}{width=100mm}}
\put(10, 15){\lettlab (b)}

\end{picture}
\end{center}
\shiftcaption
\caption[Heavy-Quark Fragmentation Functions]
{\labelmm{HQFFL} {\it
Bottom (a) and charm (b) quark fragmentation functions
at $\mu=100\,\GeV$.
The input distribution is defined by the distribution in fixed order
at $\mu_0=m$. 
Heavy quark~$Q$ \mbox{(\fullline)}, gluon \mbox{(\dashline)}, 
light quark flavours \mbox{(\dotline)}.
}}   
\end{figure}

\begin{sloppypar}
The leading-order results\footnote{
The numerical solution of the renormalization group equation
for heavy-quark fragmentation functions is described in 
Appendix~\ref{numrgff}.}
for the bottom- and charm-quark fragmentation functions
$D_{Q/i}\left(x,\mu^2\right)$ 
at a scale of $\mu=100\,\GeV$
are
shown in Fig.~\ref{HQFFL}. The fragmentation functions
are set to the perturbative input
$D_{Q/Q}\left(x,\mu_0^2\right)=\delta(1-x)$ from Eq.~(\ref{DQQcoeff}) 
at the factorization
scale $\mu_0=m$. The probability $D_{Q/Q}\left(x,\mu^2\right)$ 
to find the heavy quark within a 
heavy quark is singular,
as $\subbl{(1/(1-x))}{x}{0}{1}$ 
for $x\rightarrow 1$,
up to factors of logarithms $\ln^m(1-x)$.
We have not shown the subtraction term
proportional to $\delta(1-x)$ in the figure.
The contributions of gluons, light quarks and 
heavy antiquarks~$\overline{Q}$ are rather
small at large~$x$, 
because they are of $\porder{\alpha_s}$ and
$\porder{\alpha_s^2}$, respectively. The results at very small~$x$
are not reliable, because 
there is an ambiguity in whether 
the momentum-fraction variable~$x$ refers to the
fraction of momenta or to the fraction of energies.
In next-to-leading order, 
numerical studies indicate that the fragmentation functions go to $-\infty$
for $x\rightarrow 0$ \cite{102,103}.
\end{sloppypar}

\ifnum1=1
\begin{figure}[htb] \unitlength 1mm
\begin{center}
\dgpicture{159}{167}
 
\put(0,93){\epsfigdg{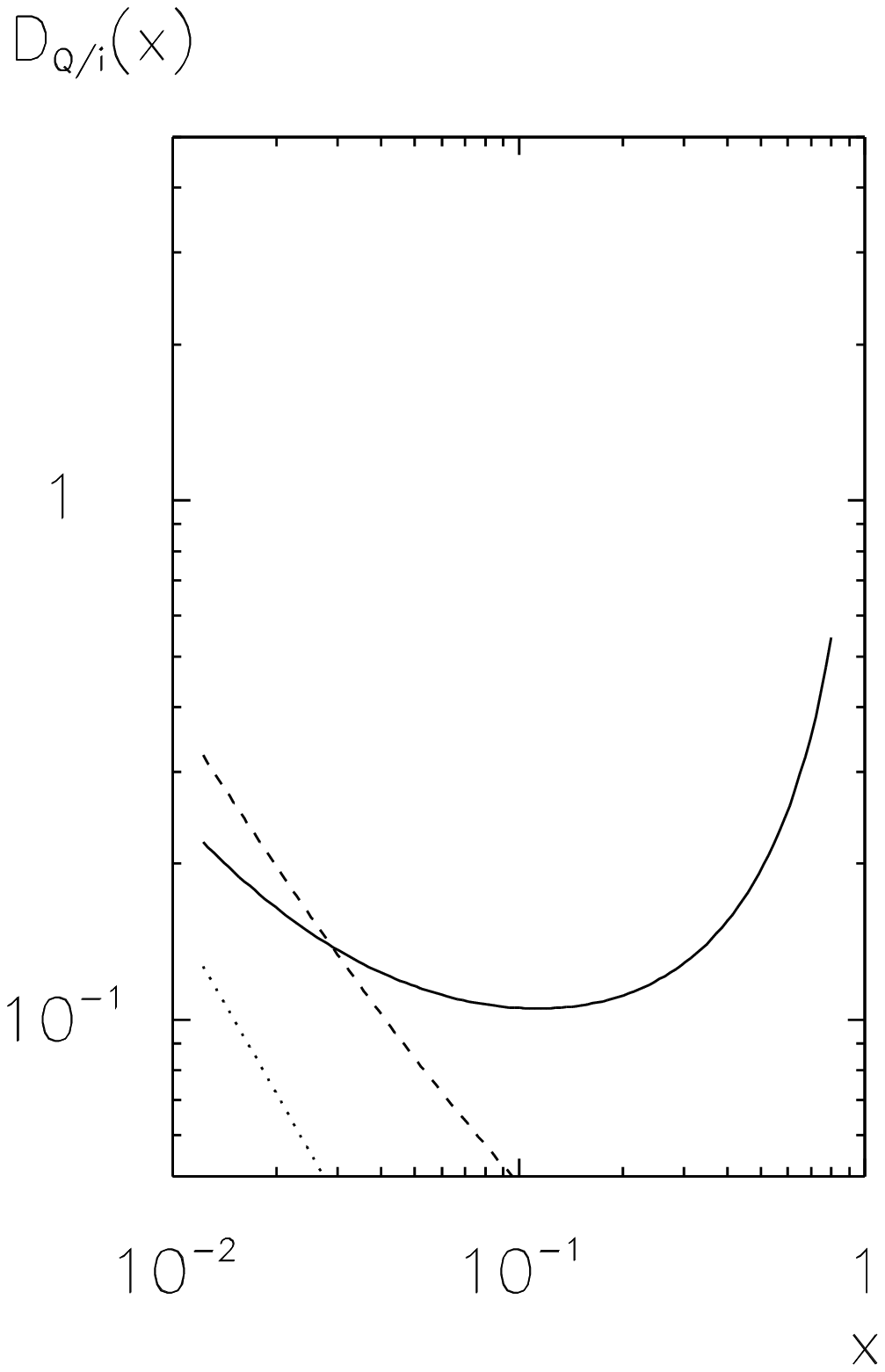}{width=45mm}}
\put(10,85){\lettlab (a)}

\put(55,93){\epsfigdg{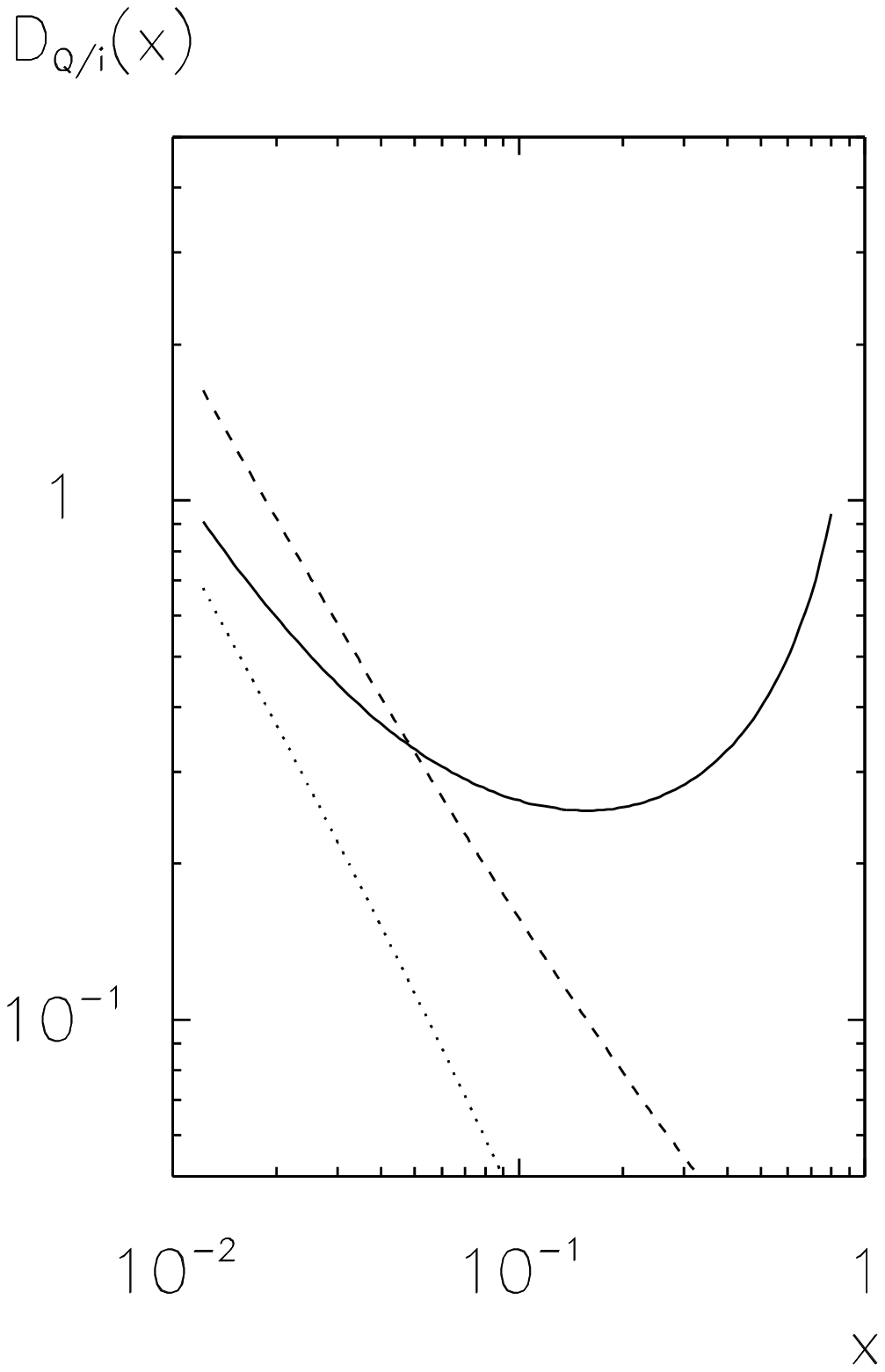}{width=45mm}}
\put(65,85){\lettlab (b)}

\put(110,93){\epsfigdg{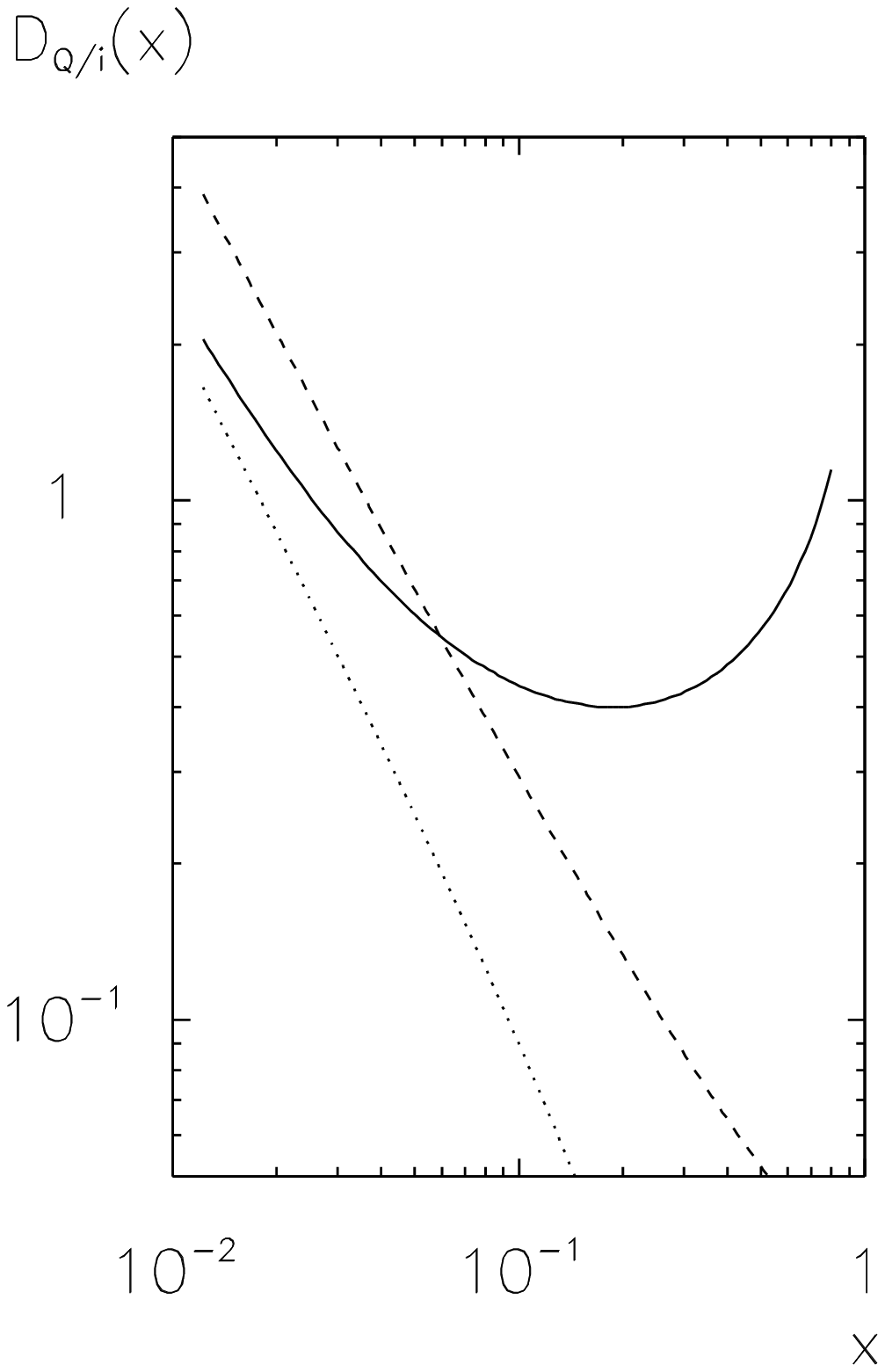}{width=45mm}}
\put(120,85){\lettlab (c)}

\put(0,8){\epsfigdg{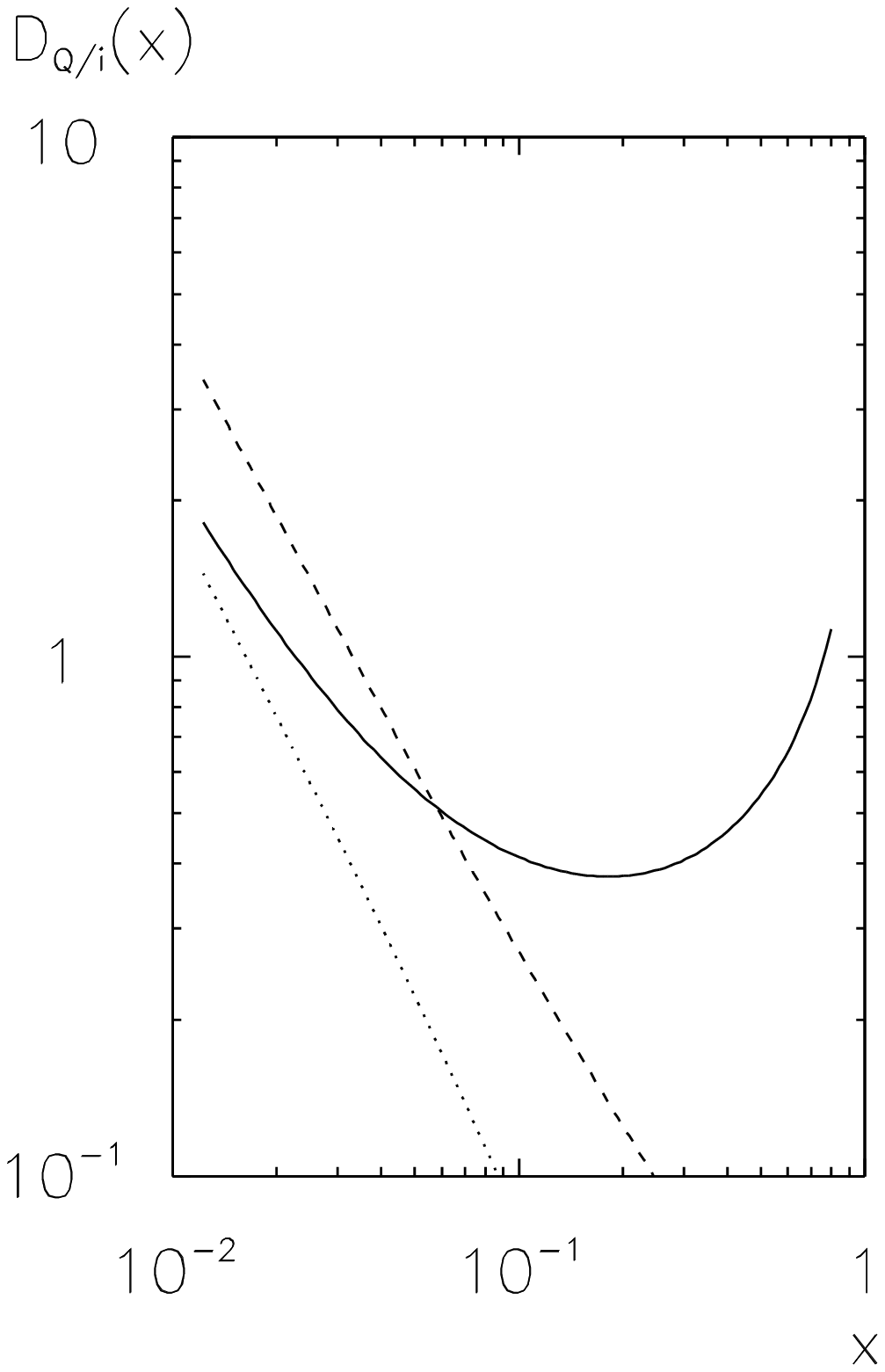}{width=45mm}}
\put(10,0){\lettlab (d)}

\put(55,8){\epsfigdg{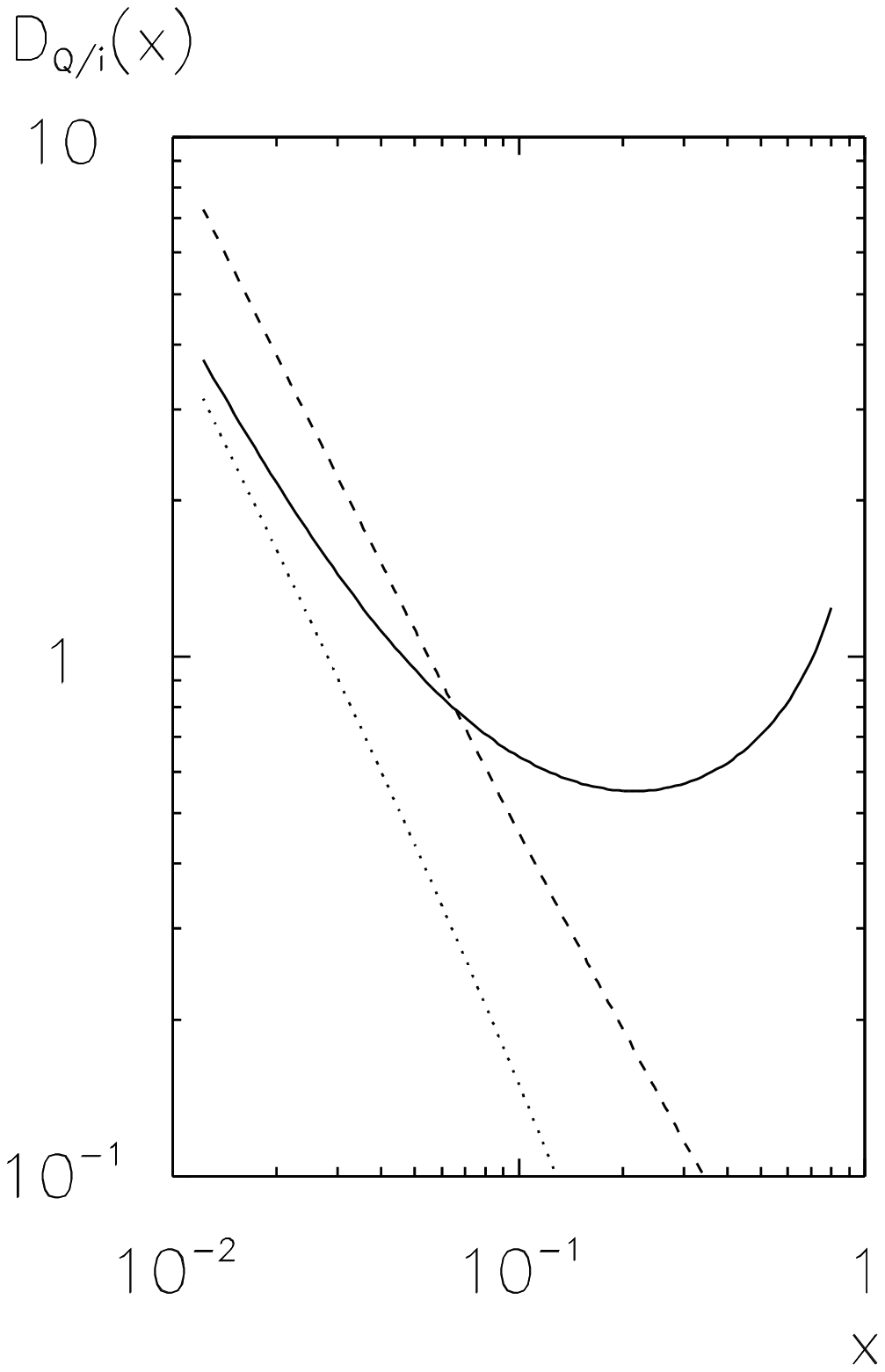}{width=45mm}}
\put(65,0){\lettlab (e)}

\put(110,8){\epsfigdg{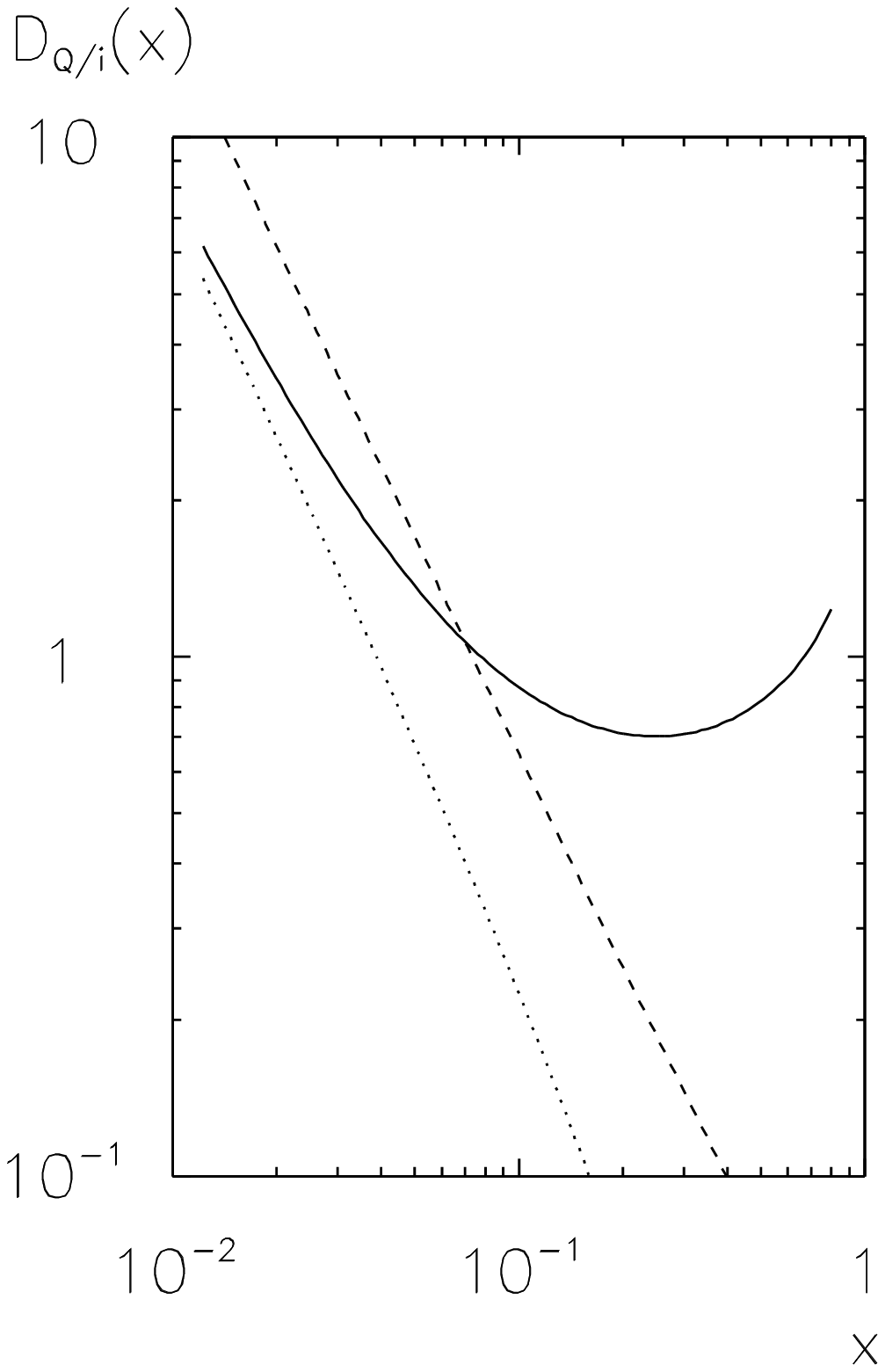}{width=45mm}}
\put(120,0){\lettlab (f)}

\end{picture}
\end{center}
\shiftcaption
\caption[Scale Evolution of Heavy-Quark Fragmentation Functions]
{\labelmm{HQFFA} {\it Scale evolution 
of bottom (a)--(c) and charm (d)--(f) quark fragmentation functions.
The factorization scale is 
$\mu=10\,\GeV$ (a), (d);
$\mu=30\,\GeV$ (b), (e);
$\mu=100\,\GeV$ (c), (f).
The input distribution is defined by the distribution in fixed order
at $\mu_0=m$. 
Heavy quark~$Q$ \mbox{(\fullline)}, gluon \mbox{(\dashline)}, 
light quark flavours \mbox{(\dotline)}.
}}   
\end{figure}
\else
\begin{figure}[htb] \unitlength 1mm
\begin{center}
\dgpicture{159}{153}
 
\put(0,86){\epsfigdg{hpic31.eps}{width=75mm}}
\put(10,80){\lettlab (a)}

\put(83,86){\epsfigdg{hpic32.eps}{width=75mm}}
\put(93,80){\lettlab (b)}

\put(0,6){\epsfigdg{hpic33.eps}{width=75mm}}
\put(10,0){\lettlab (c)}

\put(83,6){\epsfigdg{hpic34.eps}{width=75mm}}
\put(93,0){\lettlab (d)}

\end{picture}
\end{center}
\shiftcaption
\caption[Scale Evolution of Heavy-Quark Fragmentation Functions]
{\labelmm{HQFFA} {\it Scale evolution 
of bottom (a), (b) and charm (c), (d) quark fragmentation functions.
Shown are the ratios 
$D_{Q/i}\left(x,\mu_2^2\right)/D_{Q/i}\left(x,\mu_1^2\right)$
for the factorization scales 
$\mu_2=30\,\GeV$, $\mu_1=10\,\GeV$ (a), (c) and 
$\mu_2=100\,\GeV$, $\mu_1=30\,\GeV$ (a), (c).
The input distribution is defined by the distribution in fixed order
at $\mu_0=m$. 
Heavy quark~$Q$ \mbox{(\fullline)}, gluon \mbox{(\dashline)}, 
light quark flavours \mbox{(\dotline)}.
}}   
\end{figure}
\fi

The scale evolution of heavy-quark fragmentation functions is shown 
in Fig.~\ref{HQFFA} for three different factorization scales. 
Again, the perturbative input is chosen at the scale $\mu_0=m$.
The increase of the distributions at small~$x$ for increasing 
factorization scale can be seen clearly. 

\ifnum1=1
\begin{figure}[htb] \unitlength 1mm
\begin{center}
\dgpicture{159}{167}

\put(0,93){\epsfigdg{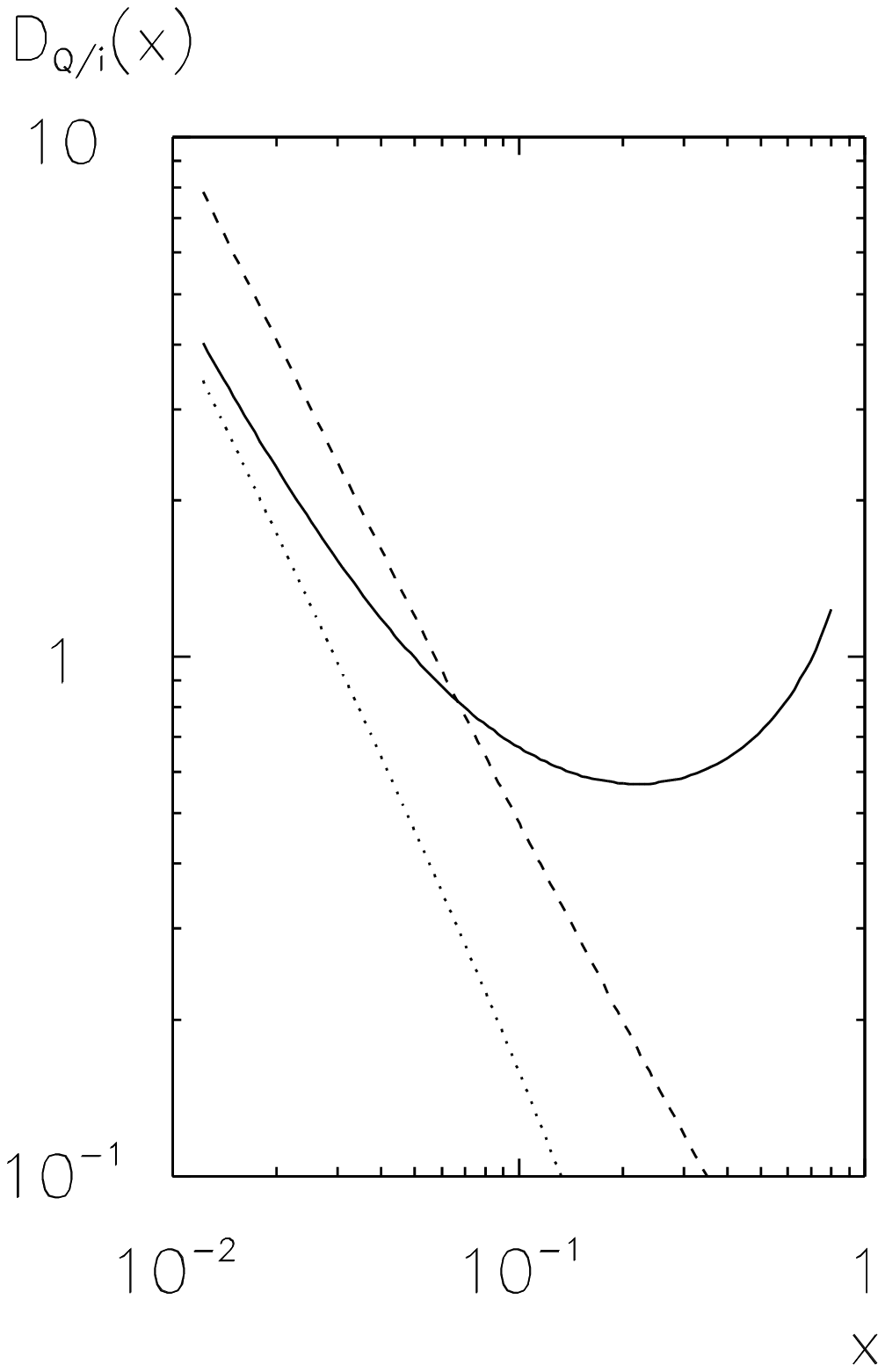}{width=45mm}}
\put(10,85){\lettlab (a)}

\put(55,93){\epsfigdg{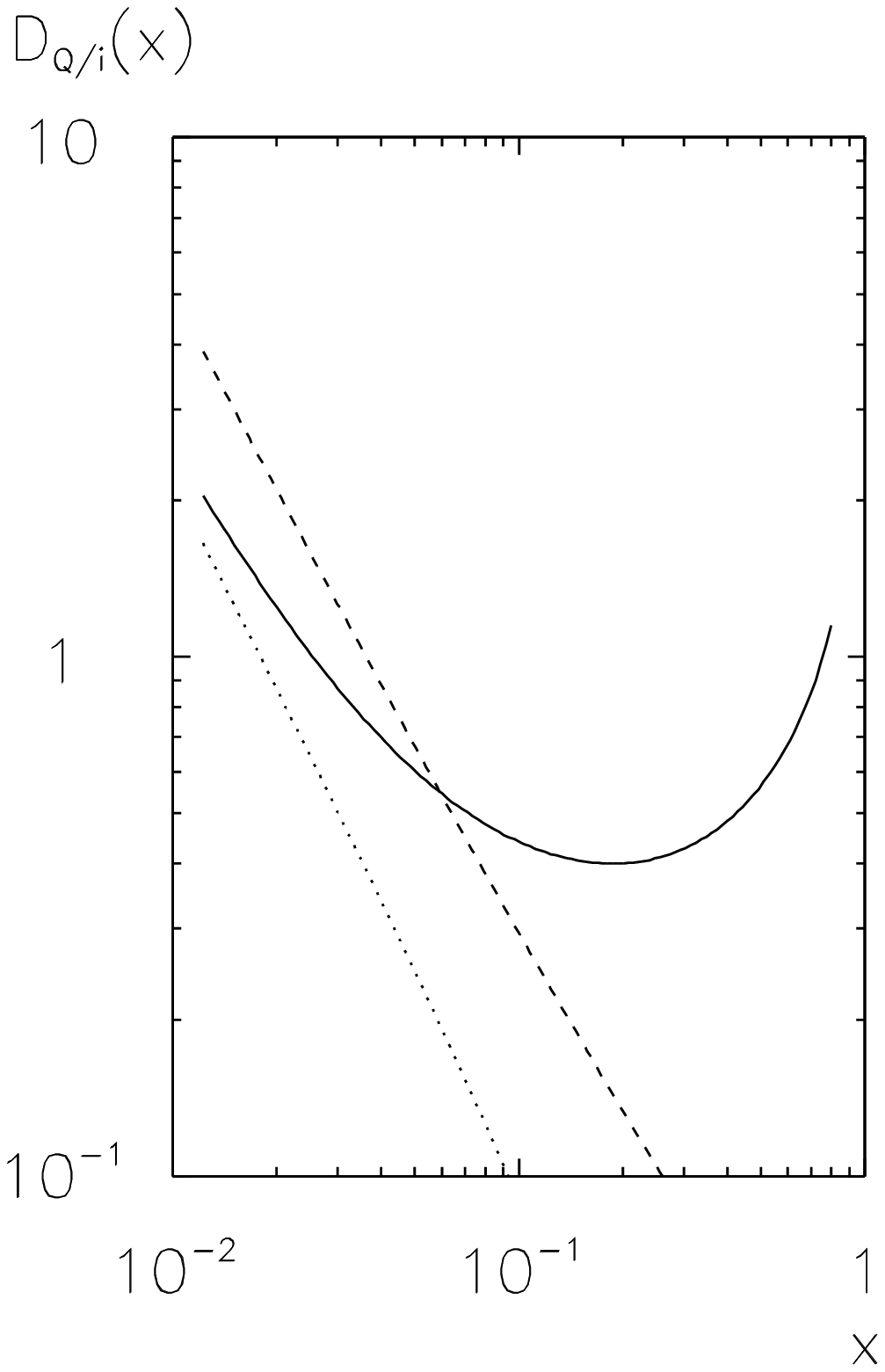}{width=45mm}}
\put(65,85){\lettlab (b)}

\put(110,93){\epsfigdg{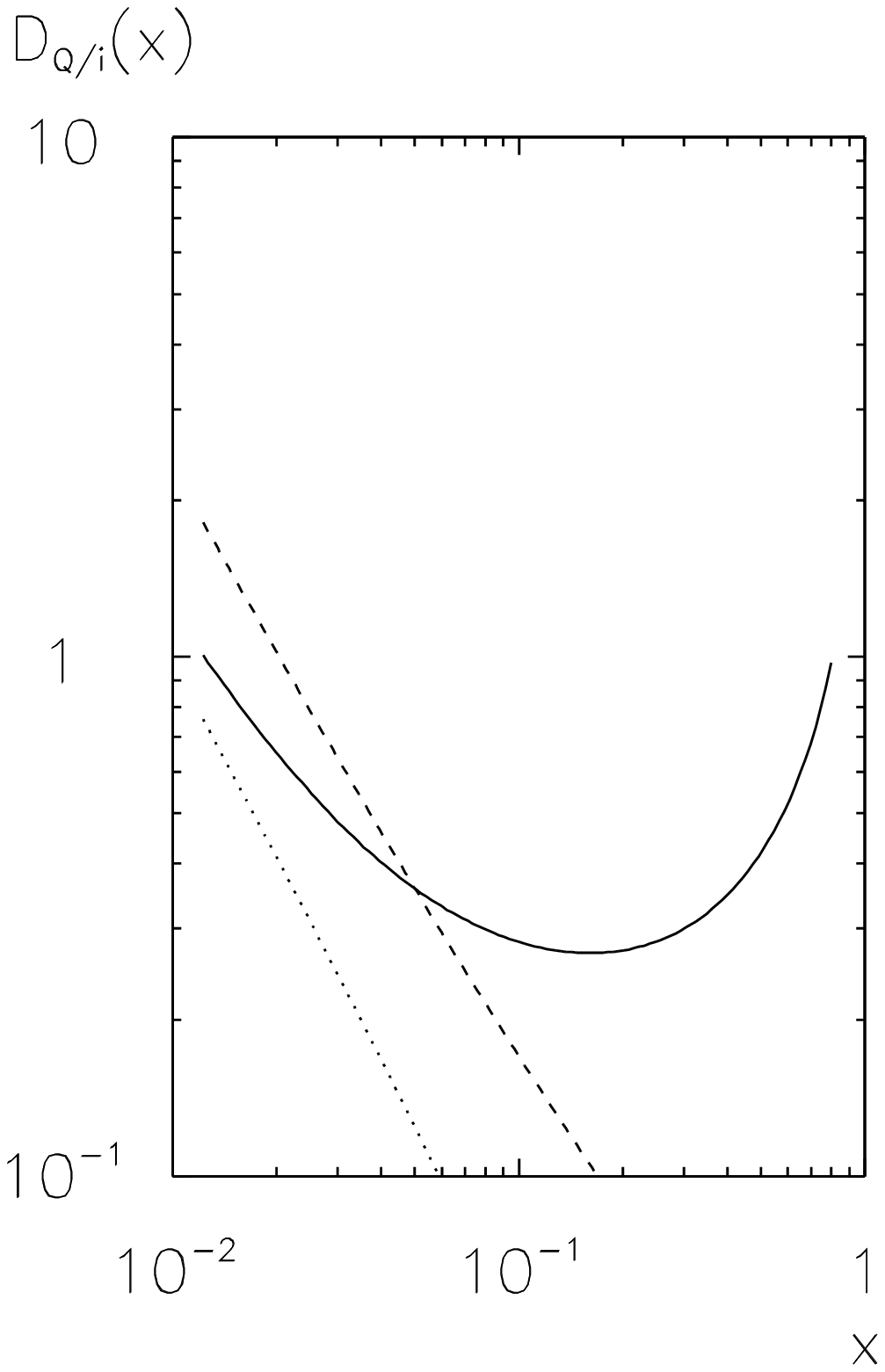}{width=45mm}}
\put(120,85){\lettlab (c)}

\put(0,8){\epsfigdg{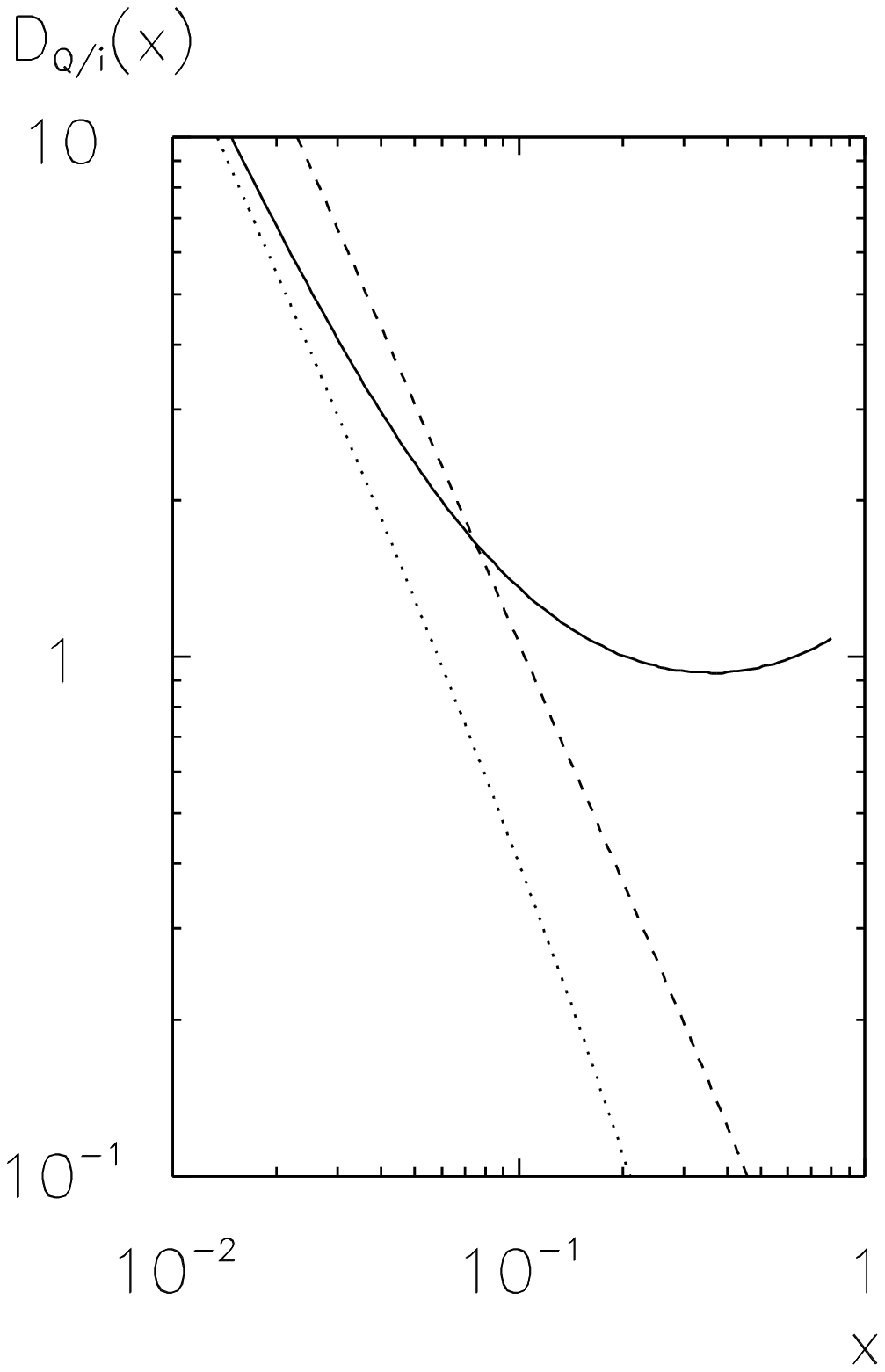}{width=45mm}}
\put(10,0){\lettlab (d)}

\put(55,8){\epsfigdg{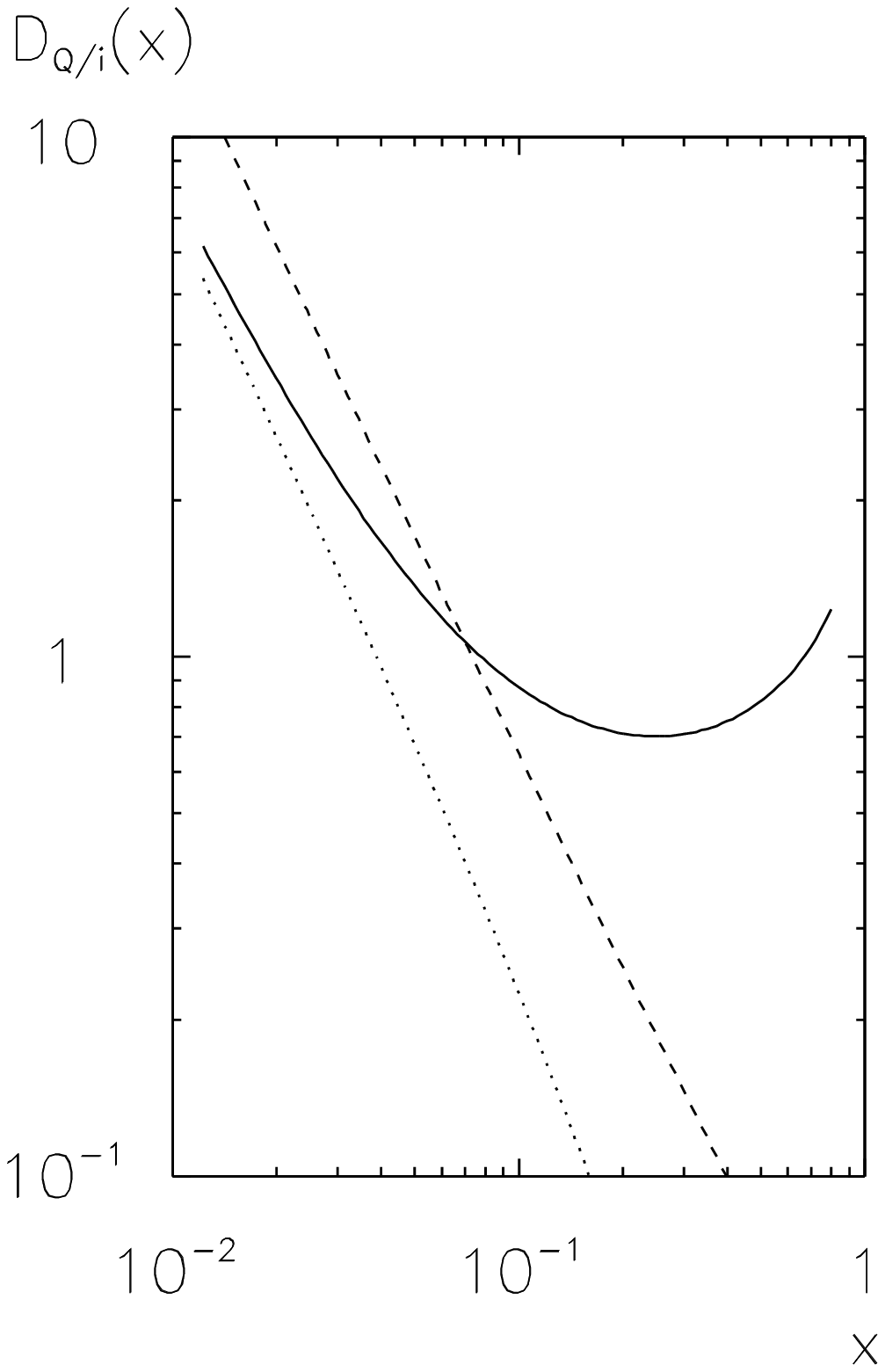}{width=45mm}}
\put(65,0){\lettlab (e)}

\put(110,8){\epsfigdg{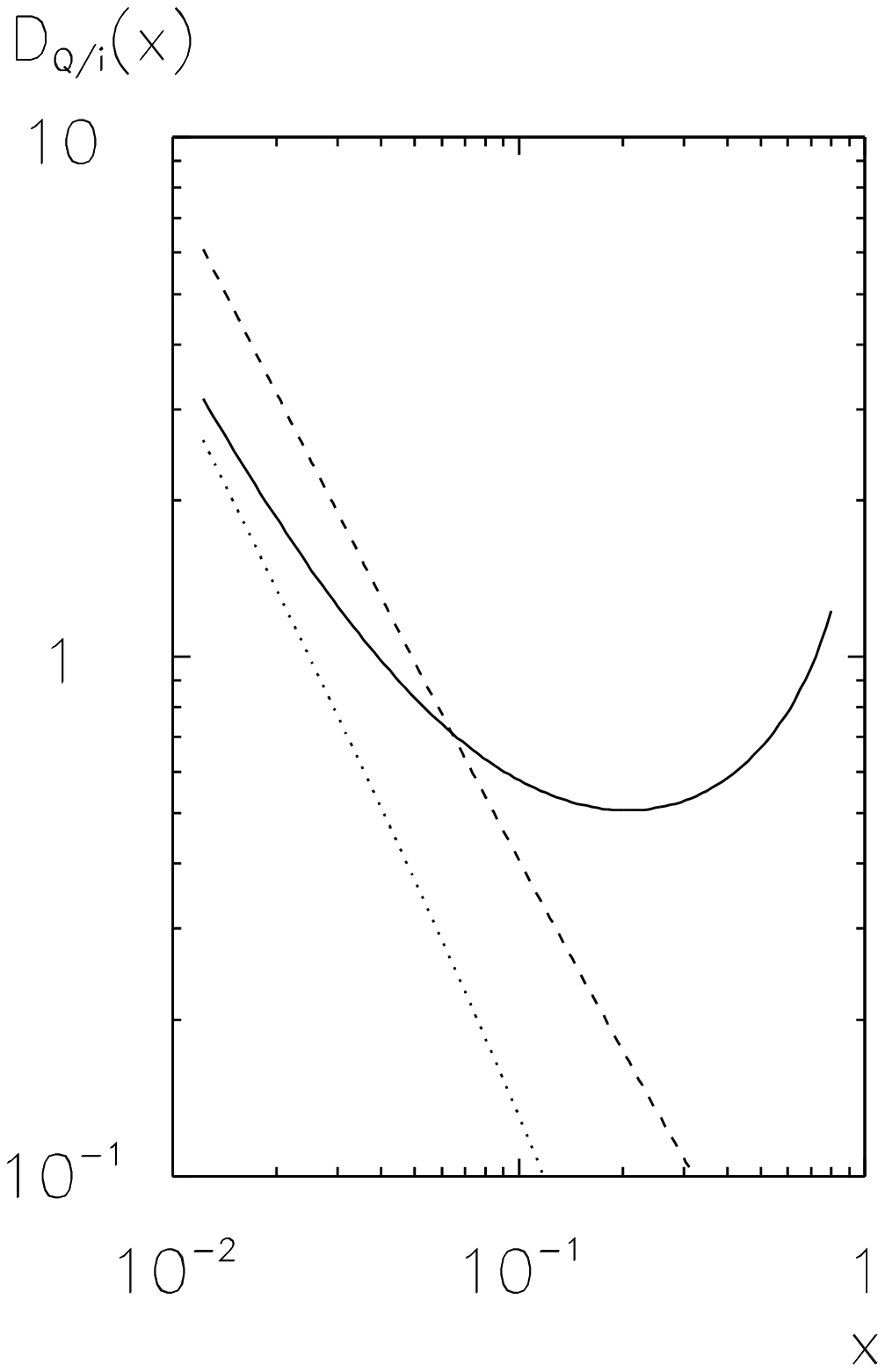}{width=45mm}}
\put(120,0){\lettlab (f)}

\end{picture}
\end{center}
\shiftcaption
\caption[Input-Scale Dependence of the Heavy-Quark Fragmentation Functions]
{\labelmm{HQFFAinp} {\it 
Dependence on the scale $\mu_0$, where the input distribution is
defined by the fixed-order calculation, for
bottom (a)--(c) and charm (d)--(f) 
quark fragmentation functions
at $\mu=100\,\GeV$.
The input scale is 
$\mu_0=m/2$ (a), (d);
$\mu_0=m$ (b), (e);
$\mu_0=2m$ (c), (f).
Heavy quark~$Q$ \mbox{(\fullline)}, gluon \mbox{(\dashline)}, 
light quark flavours \mbox{(\dotline)}.
}}   
\end{figure}
\else
\begin{figure}[htb] \unitlength 1mm
\begin{center}
\dgpicture{159}{153}
 
\put(0,86){\epsfigdg{hpic35.eps}{width=75mm}}
\put(10,80){\lettlab (a)}

\put(83,86){\epsfigdg{hpic36.eps}{width=75mm}}
\put(93,80){\lettlab (b)}

\put(0,6){\epsfigdg{hpic37.eps}{width=75mm}}
\put(10,0){\lettlab (c)}

\put(83,6){\epsfigdg{hpic38.eps}{width=75mm}}
\put(93,0){\lettlab (d)}

\end{picture}
\end{center}
\shiftcaption
\caption[Input-Scale Dependence of Heavy-Quark Fragmentation Functions]
{\labelmm{HQFFAinp} {\it Bottom (a), (b) and charm (c), (d) 
quark fragmentation functions
at $\mu=100\,\GeV$.
Shown are the ratios
$D_{Q/i}^{(2)}\left(x\right)/D_{Q/i}^{(1)}\left(x\right)$
dependending on the scales~$\mu_2$ and $\mu_1$, 
where the input distribution is
defined by the fixed order calculation.
The input scales are 
$\mu_2=m$, $\mu_0=m/2$ (a), (c) and
$\mu_2=2m$, $\mu_0=m$ (b), (d).
Heavy quark \mbox{(\fullline)}, gluon \mbox{(\dashline)}, 
all other quark flavours \mbox{(\dotline)}.
}}   
\end{figure}
\fi

The dependence of heavy-quark fragmentation functions
on the scale $\mu_0$ of the perturbative input
is shown 
in Fig.~\ref{HQFFAinp} for the three different choices 
$\mu_0=m/2$, $\mu_0=m$ and $\mu_0=2m$.
The factorization scale is chosen to be $\mu=100\,\GeV$.
The dependence on the input scale $\mu_0$ is particularly strong 
for small~$x$, and much smaller for large $x>0.1$. The strong dependence
for small~$x$ has its origin in the fact that the contribution
comes 
exclusively 
from the evolution (the input distribution
at~$\mu_0$
is located at $x=1$), and a large evolution span in~$\mu$
leads to a strong enhancement in this region.
At large~$x$, the variation from $\mu_0=m$ to $\mu_0=m/2$ and 
to $\mu_0=2m$
is about $\pm 25\%$. This input-scale dependence is much smaller
in next-to-leading order, owing to a compensating logarithmic term
in the input distribution at the scale~$\mu_0$ \cite{102,103}, 
cf.\ Appendix~\ref{aphqff}.

\dgcleardoublepage

\markh{Heavy-Quark Target Fragmentation Functions}
\dgsa{Heavy-Quark Target Fragmentation Functions}
\labelm{hqtff}
{\it
Having set up the necessary formalism, we now consider target 
fragmentation functions for heavy quarks
for the case of incident protons. Perturbative
heavy-quark target fragmentation functions, corresponding to the
production of heavy quarks via the fragmentation of a parton
emitted in the backward direction, are defined 
in Section~\ref{hqtffpert}. The corresponding renormalization 
group equation is solved numerically in Section~\ref{hqtffren}.
Due to the radiative production
mechanism,
the distribution in the momentum-fraction variable~$z$ of the
observed heavy quark falls off rapidly for $z\rightarrow 1$.
To give a qualitative example for the non-perturbative piece, 
we consider the hypothesis of intrinsic heavy quarks
in Section~\ref{hqtffmic}. Based on a simple
expression for a 
correlated distribution function of the $|uudQ\overline{Q}\rangle$
Fock state, we obtain numerical results for the non-perturbative 
target fragmentation functions via the homogeneous evolution equation.
As is expected, the $z$-distribution is harder for this
particular non-perturbative
piece.
}

\dgsb{Target Fragmentation Functions from Perturbative QCD}
\labelm{hqtffpert}
In Section~\ref{tffdetails} it has been shown that target fragmentation
functions~$M$ may be written as a sum $M=M^{(P)}+M^{(NP)}$ of a
``perturbative'' and a ``non-perturbative'' contribution. The 
perturbative contribution is defined to be zero at an (arbitrarily
chosen) factorization scale~$\mu_0$.
The non-perturbative piece $M^{(NP)}$,
with rescaled arguments as in Eq.~(\ref{redtff}), 
evolves according to the homogeneous
Altarelli--Parisi equation. The perturbative piece $M^{(P)}$ 
may be calculated,
under the assumption that the fragmentation functions
and parton densities are known,
by solving the inhomogeneous renormalization group equation
(\ref{Mrge}). 
The contributions from $M^{(P)}$ are therefore related to 
the production of heavy quarks by means of the mechanism
shown in Fig.~\ref{Mrgfig}b. A complete set of the lowest-order
contributions for heavy-quark production is shown in 
Fig.~\ref{hqvertexfig}, where the ``incoming parton'' has to be attached
to a parton density. From these graphs it is clear that a correlation
of the flavour of an observed heavy quark in the target fragmentation
region and of the type of parton incident in the hard scattering
process is to be expected. If the heavy-quark content of the
nucleon is not too large, then there is a large probability
that a heavy antiquark initiates the hard scattering process, for
the case of an observed heavy quark in the target fragmentation
region, owing to the splitting of the gluons in Figs.~\ref{hqvertexfig}c
and~d
into a $Q\overline{Q}$-pair.

As long as non-perturbative methods to calculate 
target fragmentation functions from 
first principles are not available,
the non-perturbative piece can only be determined by means of a 
measurement, much as is the case for parton densities,
or from phenomenological models.
The perturbative contribution
to the case of heavy quarks in the remnant of a proton 
will be considered in the next section,
and the hypothesis of intrinsic heavy quarks in Section~\ref{hqtffmic}.

\dgsb{Solving the Renormalization Group Equation}
\labelm{hqtffren}
The standard method to solve renormalization group equations of
phenomenological distribution functions is to transform the equation
into the Mellin moment space, where the moments of the 
convoluted expression can be written as the products of the respective
moments. The differential equations for the moments decouple, and can
be solved analytically for complex values of the moment variable.
An inverse Mellin transform allows the transformation 
back into $x$-space.
This method works for the homogeneous case, and for some inhomogeneous 
cases as well, namely in those cases where the expression of the moments
for the inhomogeneous part are known analytically, see for instance
Ref.\ \cite{104}.
In our case, however, the inhomogeneous term of Eq.~(\ref{Mrge})
contains two sets of quantities that are known only numerically, namely the
parton densities and the fragmentation functions. The complicated arguments
of those functions and the unusual limits of the integral 
prevent an expression of the single moments 
in~$x$ for fixed~$z$ of this term in a simple way.
Therefore the 
renormalization group equation (\ref{Mrge}) is solved numerically 
in the form
of Eq.~(\ref{Nrge}) by a discretization 
in the relevant variables\footnote{
The numerical solution of the renormalization group equation
for heavy-quark target fragmentation functions is described in 
Appendix~\ref{numrgtff}.}.

\ifnum1=1
\begin{figure}[htb] \unitlength 1mm
\begin{center}
\dgpicture{159}{83}

\put(0,8){\epsfigdg{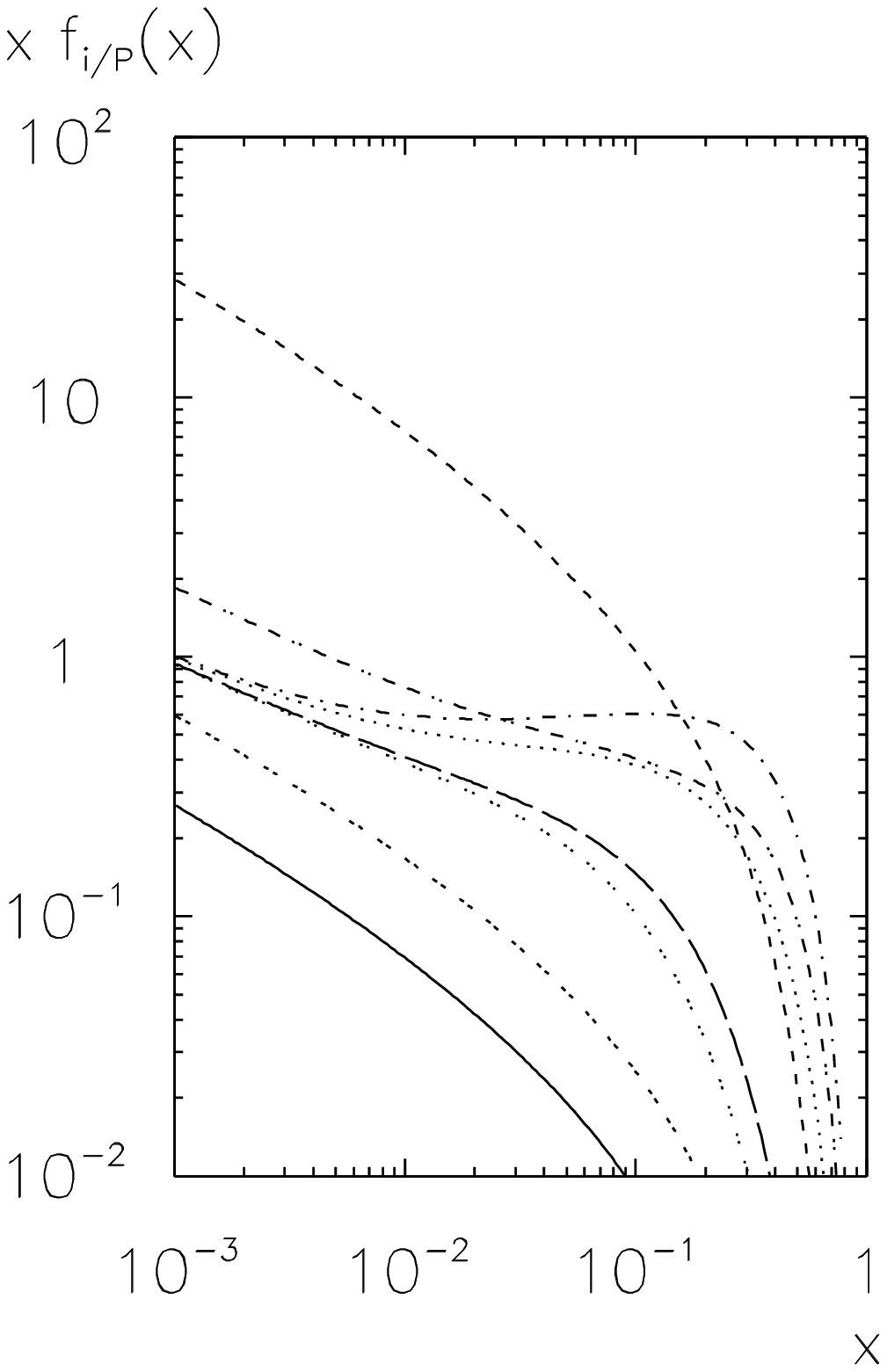}{width=45mm}}
\put(10,0){\lettlab (a)}

\put(55,8){\epsfigdg{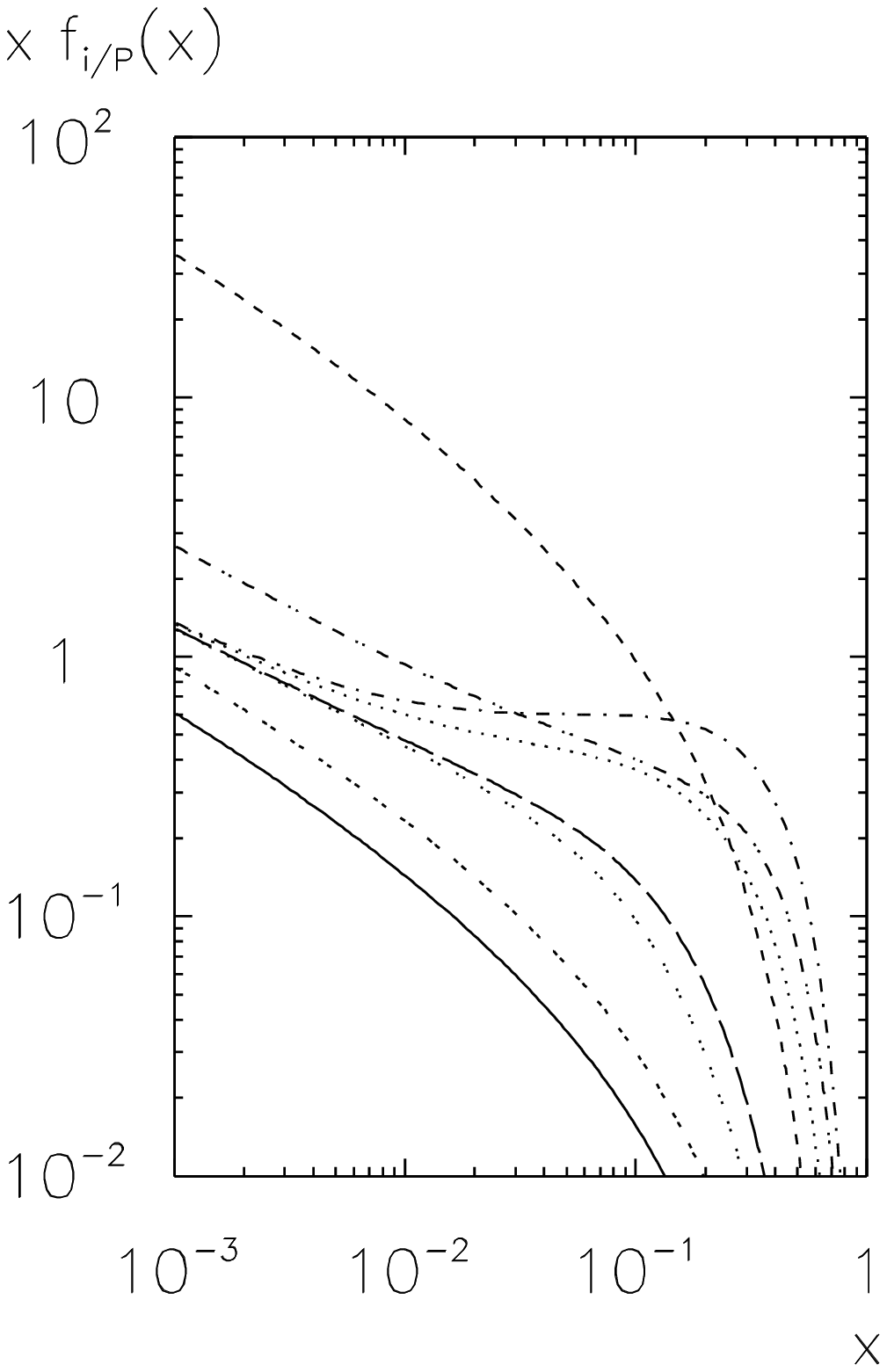}{width=45mm}}
\put(65,0){\lettlab (b)}

\put(110,8){\epsfigdg{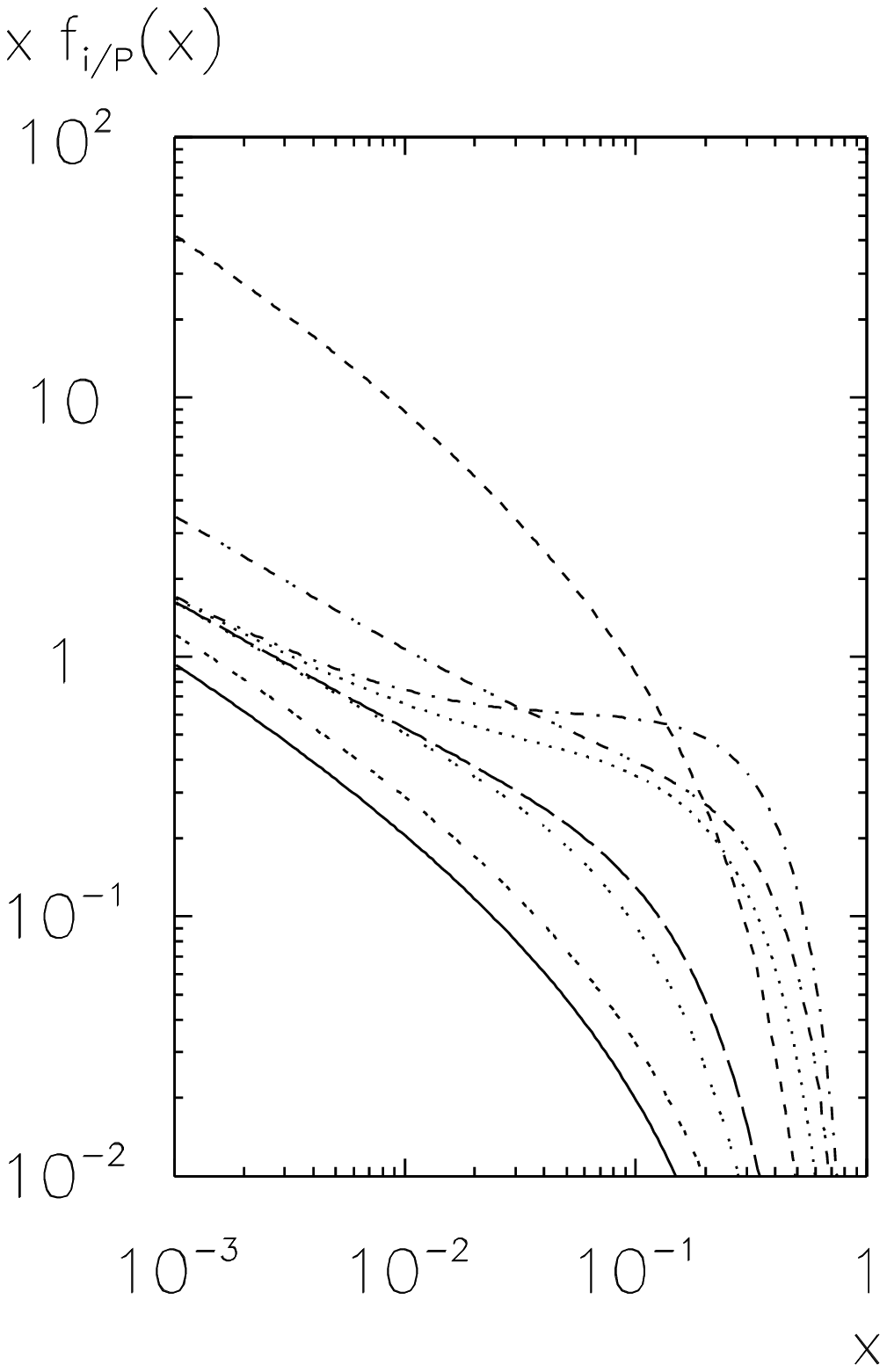}{width=45mm}}
\put(120,0){\lettlab (c)}

\end{picture}
\end{center}
\shiftcaption
\caption[Scale Evolution of the GRV Parton Densities]
{\labelmm{HQTFFP} {\it Scale evolution of the leading-order 
GRV parton densities $x\,f_{i/P}(x,\mu^2)$.
The factorization scale is 
$\mu=10\,\GeV$ (a);
$\mu=30\,\GeV$ (b);
$\mu=100\,\GeV$ (c).
The flavours~$i$ are given by 
$b$, $\overline{b}$ \mbox{(\fullline)}; 
$c$, $\overline{c}$ \mbox{(\dashdashline)}; 
$g$ \mbox{(\dashline)};
$d$ \mbox{(\dotline)};
$\overline{d}$ \mbox{(\longdashline)};
$u$ \mbox{(\dashdotline)};
$\overline{u}$ \mbox{(\dotdotline)};
$F_2$ \mbox{(\dashdashdotdotline)}.
}}   
\end{figure}
\fi

\begin{figure}[htb] \unitlength 1mm
\begin{center}
\dgpicture{159}{167}

\put(0,93){\epsfigdg{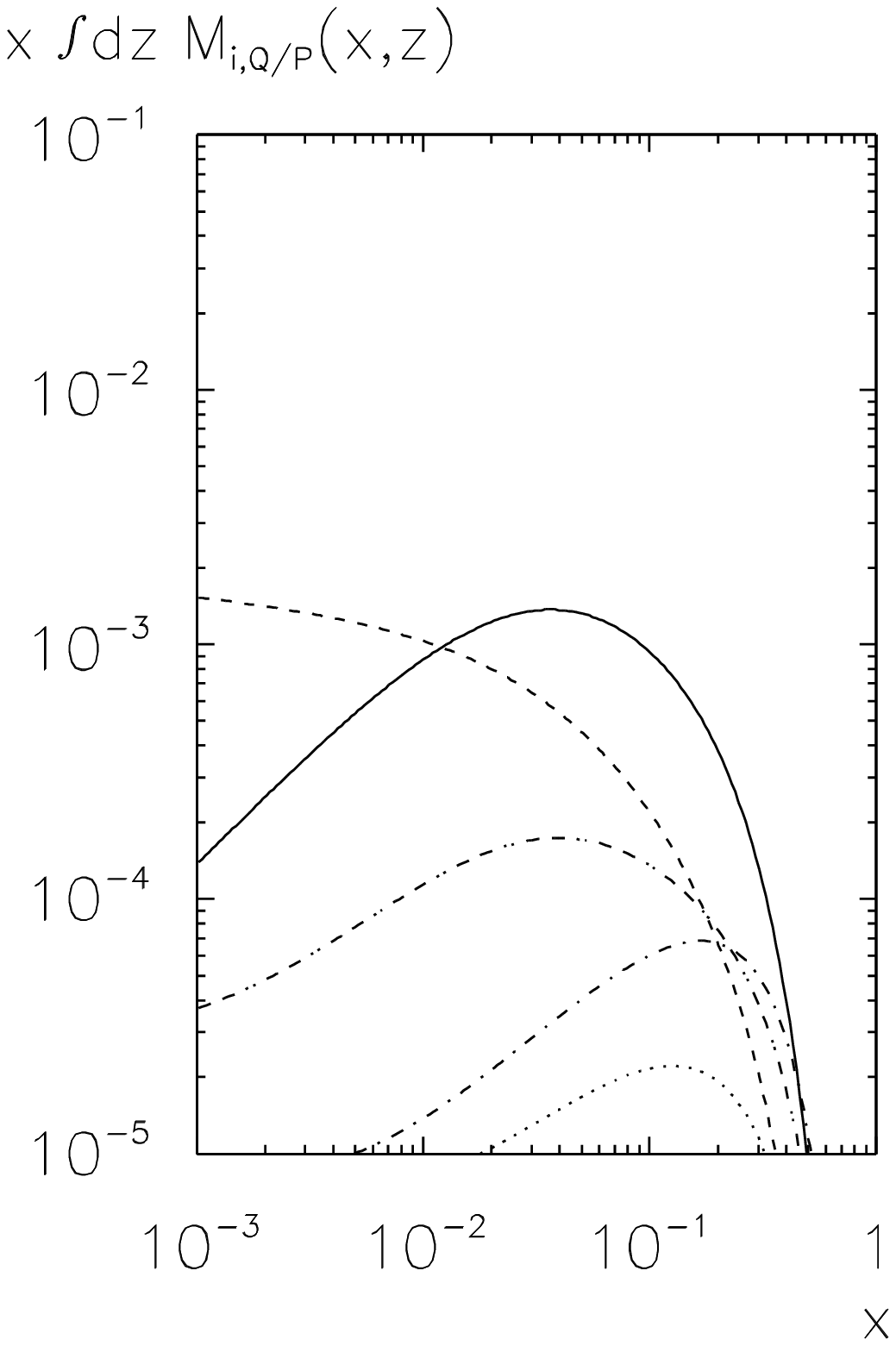}{width=45mm}}
\put(10,85){\lettlab (a)}

\put(55,93){\epsfigdg{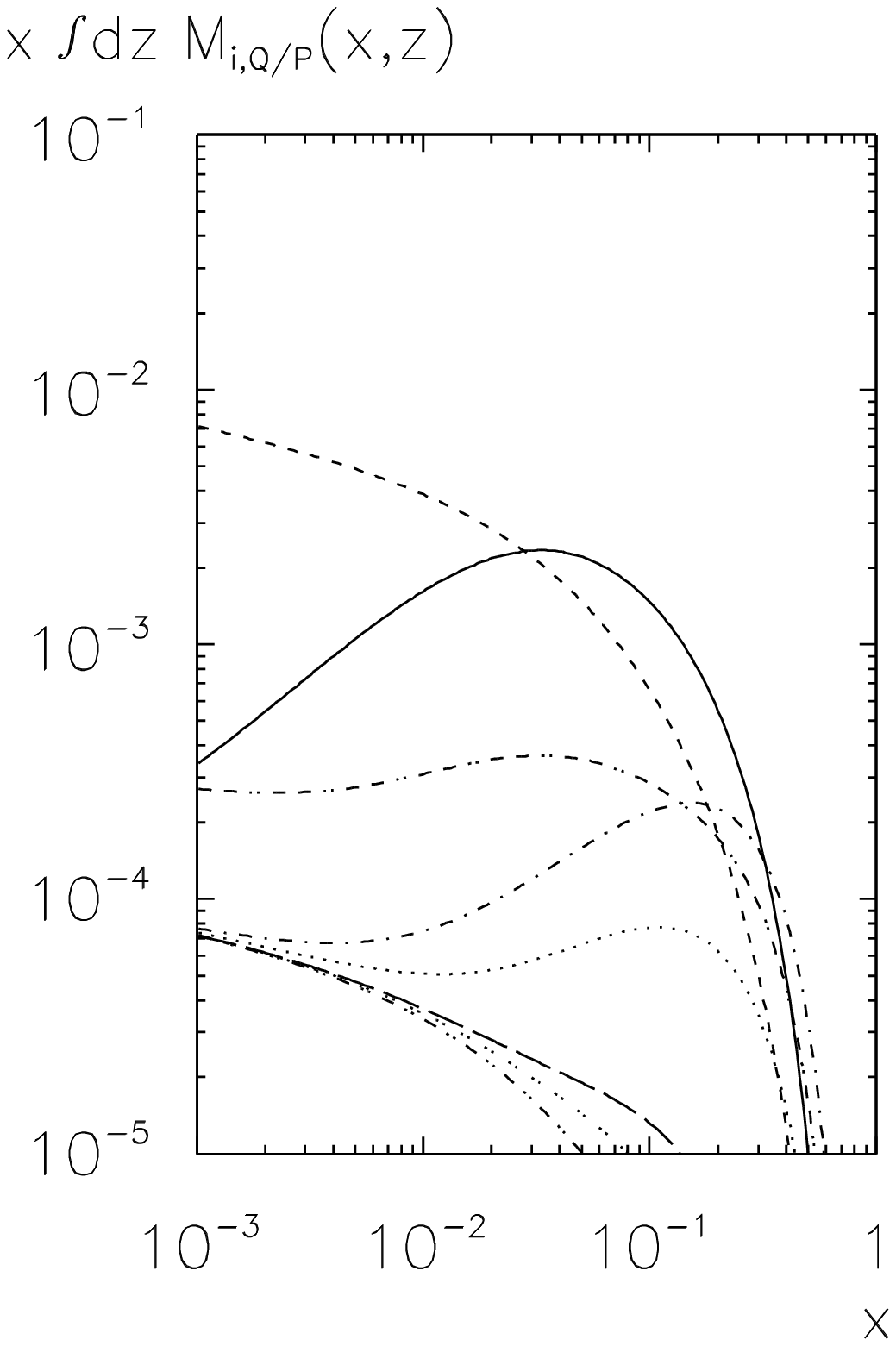}{width=45mm}}
\put(65,85){\lettlab (b)}

\put(110,93){\epsfigdg{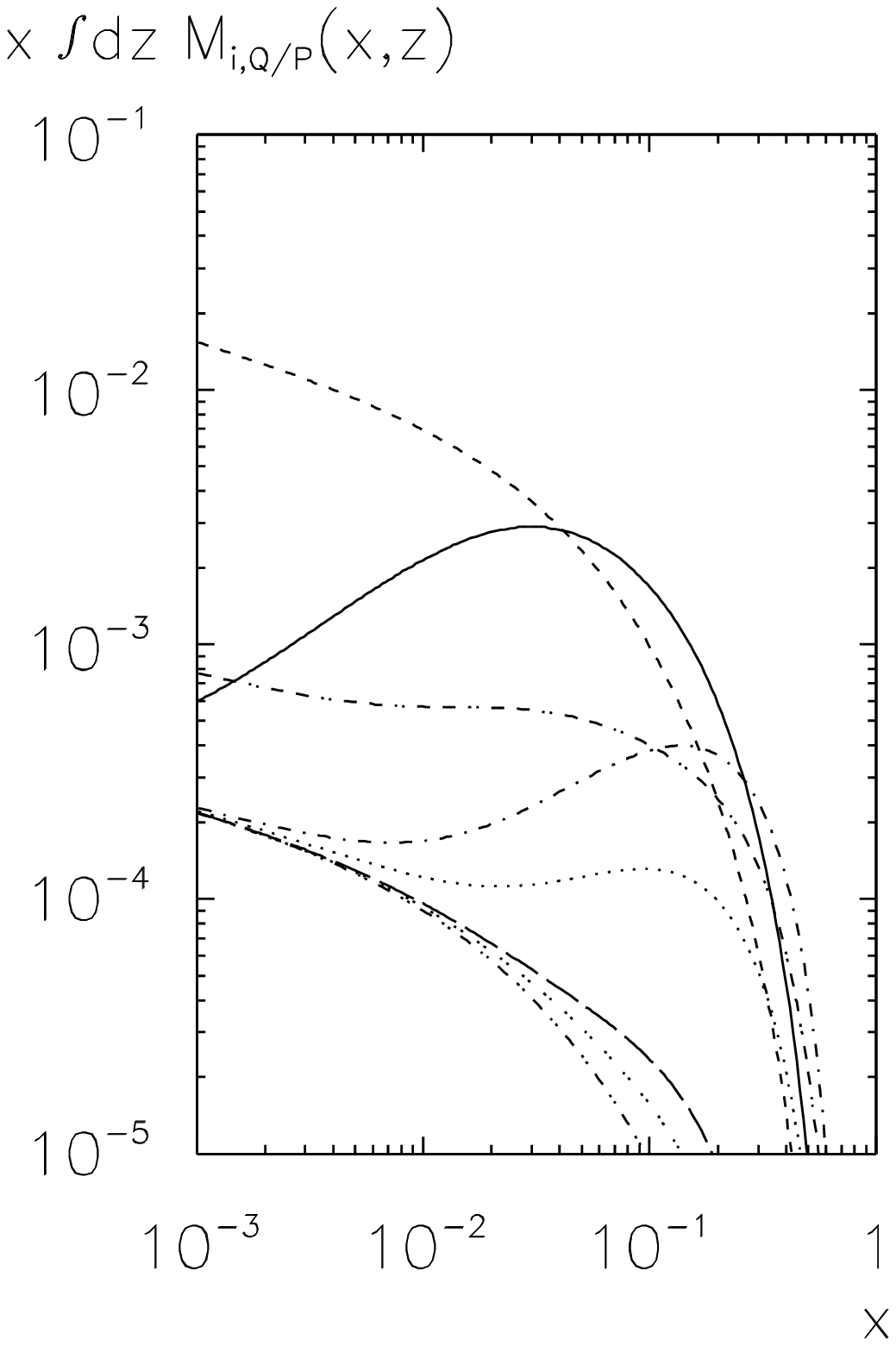}{width=45mm}}
\put(120,85){\lettlab (c)}

\put(0,8){\epsfigdg{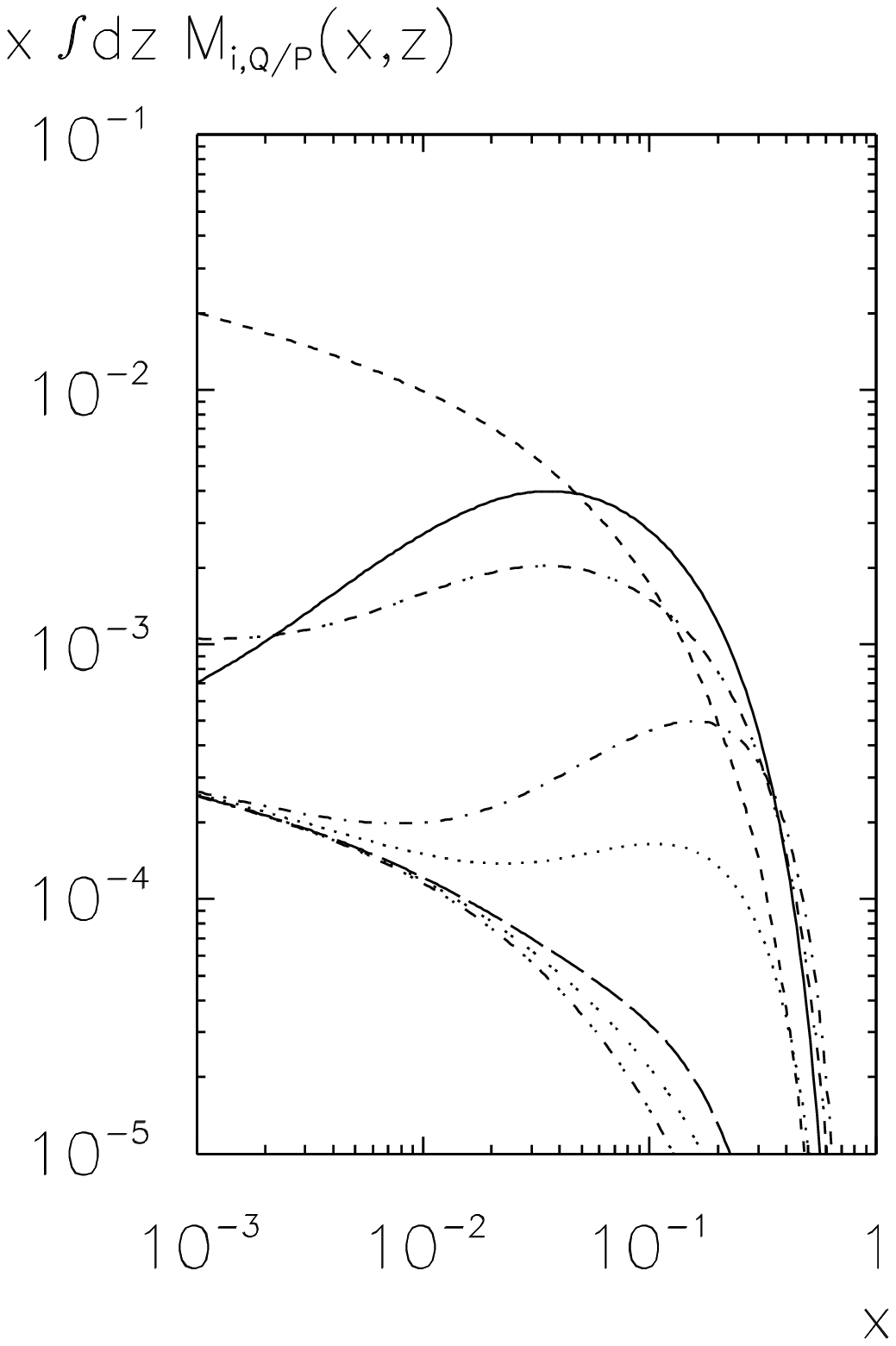}{width=45mm}}
\put(10,0){\lettlab (d)}

\put(55,8){\epsfigdg{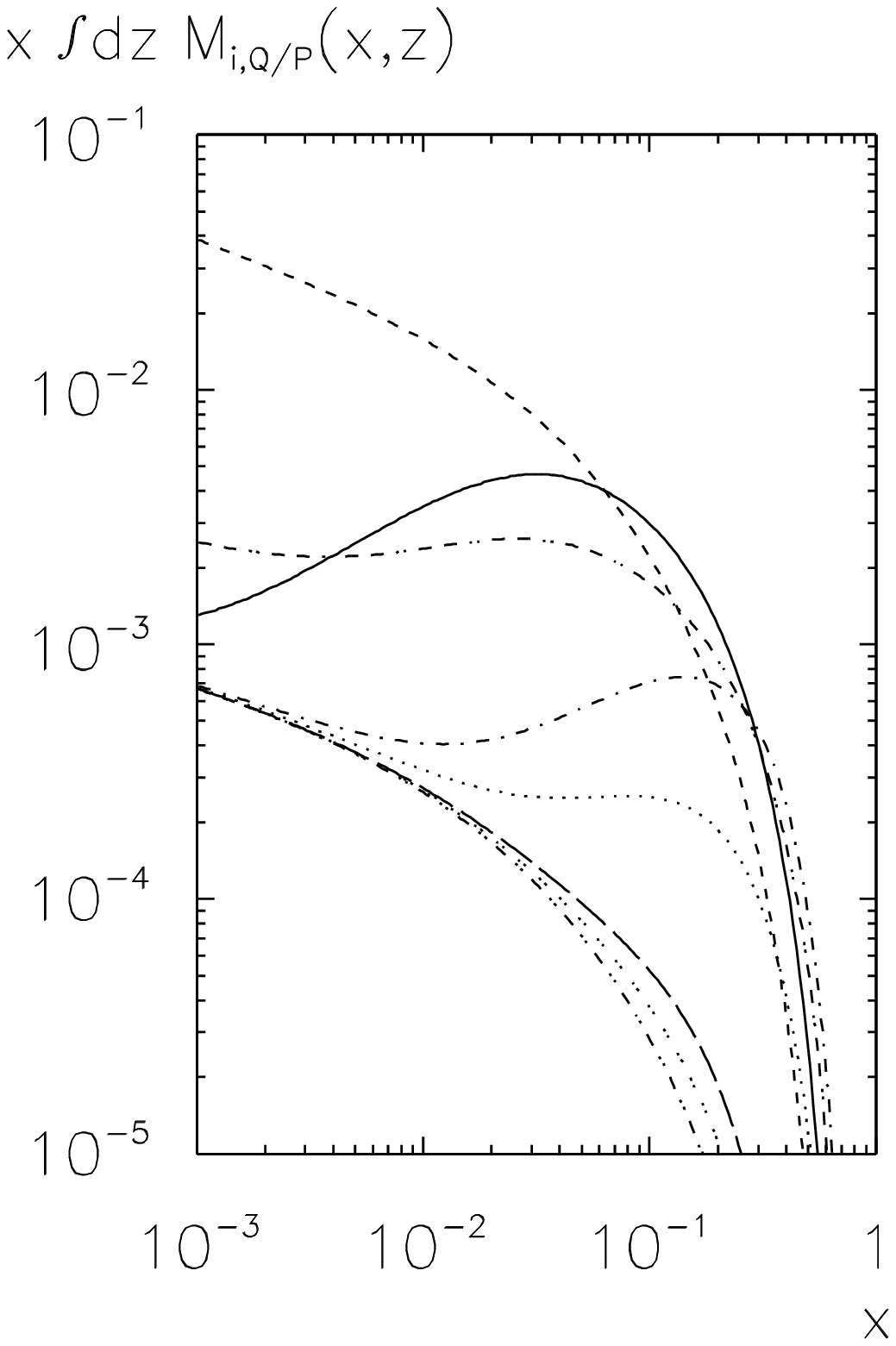}{width=45mm}}
\put(65,0){\lettlab (e)}

\put(110,8){\epsfigdg{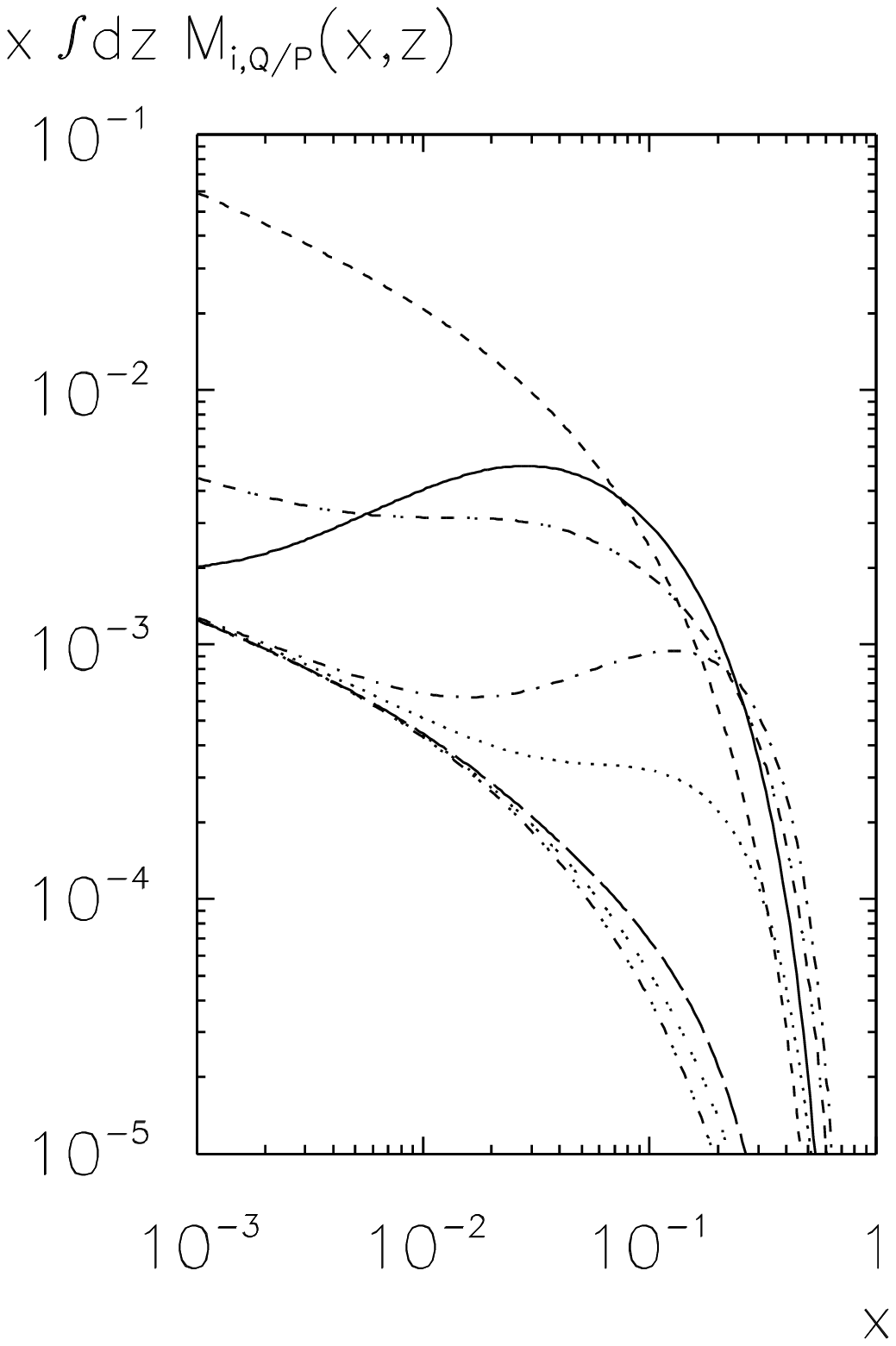}{width=45mm}}
\put(120,0){\lettlab (f)}

\end{picture}
\end{center}
\shiftcaption
\caption[Scale Evolution of Perturbative 
Heavy-Quark Target Fragmentation Functions]
{\labelmm{HQTFFA} {\it Scale evolution of perturbative bottom (a)--(c) and
charm (d)--(f) quark 
target fragmentation functions $x\,\int\dd z\,M^{(P)}_{i,Q/P}(x,z,\mu^2)$, 
where
$z$ is integrated from $0.1$ to $1-x$.
The factorization scale is 
$\mu=10\,\GeV$ (a), (d);
$\mu=30\,\GeV$ (b), (e);
$\mu=100\,\GeV$ (c), (f).
The input distribution is zero 
at $\mu_0=m$. 
The flavours~$i$ are given by 
$\overline{Q}$ \mbox{(\fullline)}, 
$g$ \mbox{(\dashline)}, 
$d$ \mbox{(\dotline)},
$\overline{d}$ \mbox{(\longdashline)},
$u$ \mbox{(\dashdotline)},
$\overline{u}$ \mbox{(\dotdotline)},
$Q$ \mbox{(\dashdotdotline)},
$F_2^{M}$ \mbox{(\dashdashdotdotline)}.
}}   
\end{figure}

The parton densities which we have used 
in the inhomogeneous term
of the renormalization group equation
are given by the 
leading-order 
parametrizations\footnote{The heavy-quark contribution is zero 
in the parametrization from Ref.\ 
\cite{105}. We adopt the suggestion from the
authors of this parametrization
to use the heavy-quark distributions from Ref.\ \cite{106} and the
light quark and gluon distributions from Ref.\ \cite{105}.
\newcounter{grvlabel}
\setcounter{grvlabel}{\value{footnote}}
} 
by Gl\"uck, Reya and Vogt
(GRV)
\cite{106,105}, which we show in Fig.~\ref{HQTFFP}
for comparison with the following distributions. The heavy-quark 
fragmentation functions are those described in Section~\ref{hqff}, 
where the perturbative input is used at
the scale of the heavy-quark mass~$m$.
The scale evolution of perturbative heavy-quark target fragmentation functions
is shown in Fig.~\ref{HQTFFA},
where we have set the input distribution 
for the perturbative heavy-quark target fragmentation functions
to zero
at $\mu_0=m$. 
The plotted quantity is
\beqm{Mzint}
x\int_{z_0}^{1-x}\dd z \,M^{(P)}_{i,Q/P}(x,z,\mu^2),
\eeq
for $z_0=0.1$.
The momentum fraction of the observed heavy quark is therefore between
$0.1$ and $1-x$.
We define $F_2^M$ by
\beqm{F2M}
F_2^M(x,z,\mu^2)\doteq x\sum_{i=q,\overline{q}}Q_i^2\,
M_{i,Q/P}(x,z,\mu^2),
\eeq
in analogy to the leading-order expression
for the structure function
\beqm{F2}
F_2(x,\mu^2)=x\sum_{i=q,\overline{q}}Q_i^2\,f_{i/P}(x,\mu^2)
\eeq
in terms of parton densities.
The perturbative heavy-quark target fragmentation functions 
have the following main features:
\begin{itemize}
\item[\dgbullet]
At fairly large~$x$, i.e.\ $0.05\lesssim x\lesssim 0.3$, 
the distributions are dominated
by $M^{(P)}_{\overline{Q},Q/P}$,
whereas
at small~$x$, i.e.\ $x\lesssim 0.01$, the distributions are dominated
by $M^{(P)}_{g,Q/P}$. 
At small~$x$, the quark distributions converge to a common
limit distribution.
The latter property is comparable to 
the case of parton densities.
The large values of 
$M^{(P)}_{\overline{Q},Q/P}$ can be explained
by the splitting of a gluon into a $Q\overline{Q}$ pair,
see Fig.~\ref{hqvertexfig}c, and the
dominance of $M^{(P)}_{g,Q/P}$ at small~$x$ by 
the $1/u$-term in the $q\rightarrow g$
splitting function.
\item[\dgbullet]
The distributions 
$M^{(P)}_{q,Q/P}$ and $M^{(P)}_{\overline{q},Q/P}$ for light quarks
and $M^{(P)}_{Q,Q/P}$ are
small. Among the light quark distributions, the~$u$- and $d$-distributions
are the largest. This is due to the fact that a~$u$- or $d$-quark
may radiate a gluon in the backward direction 
before the hard scattering process, 
this gluon subsequently splitting into a $Q\overline{Q}$-pair, 
where~$Q$ is observed in the target fragmentation region.
\item[\dgbullet]
The perturbative 
charm-quark target fragmentation functions are much larger than 
the bottom-quark target fragmentation functions. This is because
the charm-quark fragmentation functions are larger than the bottom-quark
ones, and moreover the evolution span is wider
in the charm-quark case, because of the choice $\mu_0=m$.
\end{itemize}

\begin{figure}[htb] \unitlength 1mm
\begin{center}
\dgpicture{159}{167}

\put(0,93){\epsfigdg{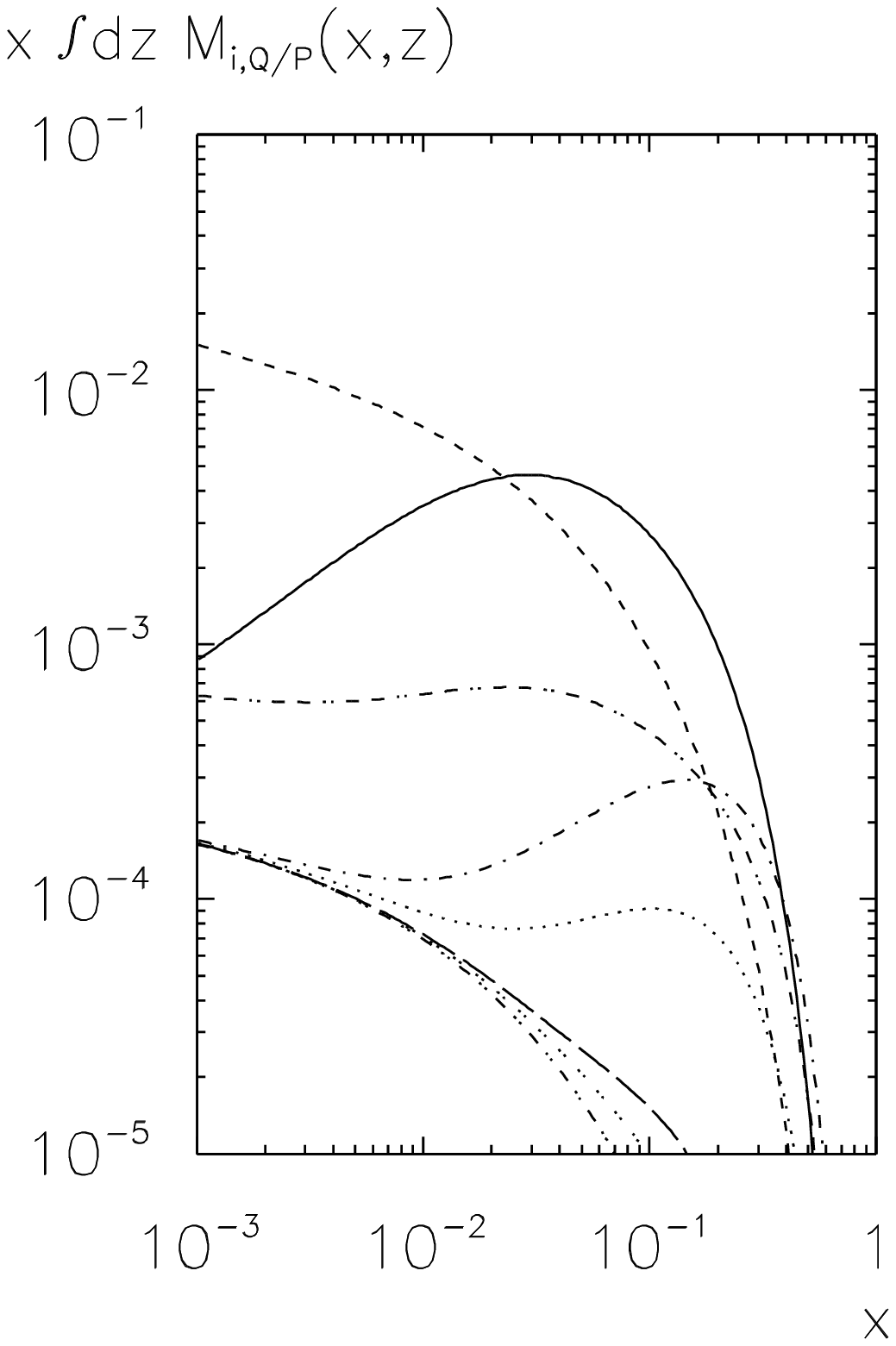}{width=45mm}}
\put(10,85){\lettlab (a)}

\put(55,93){\epsfigdg{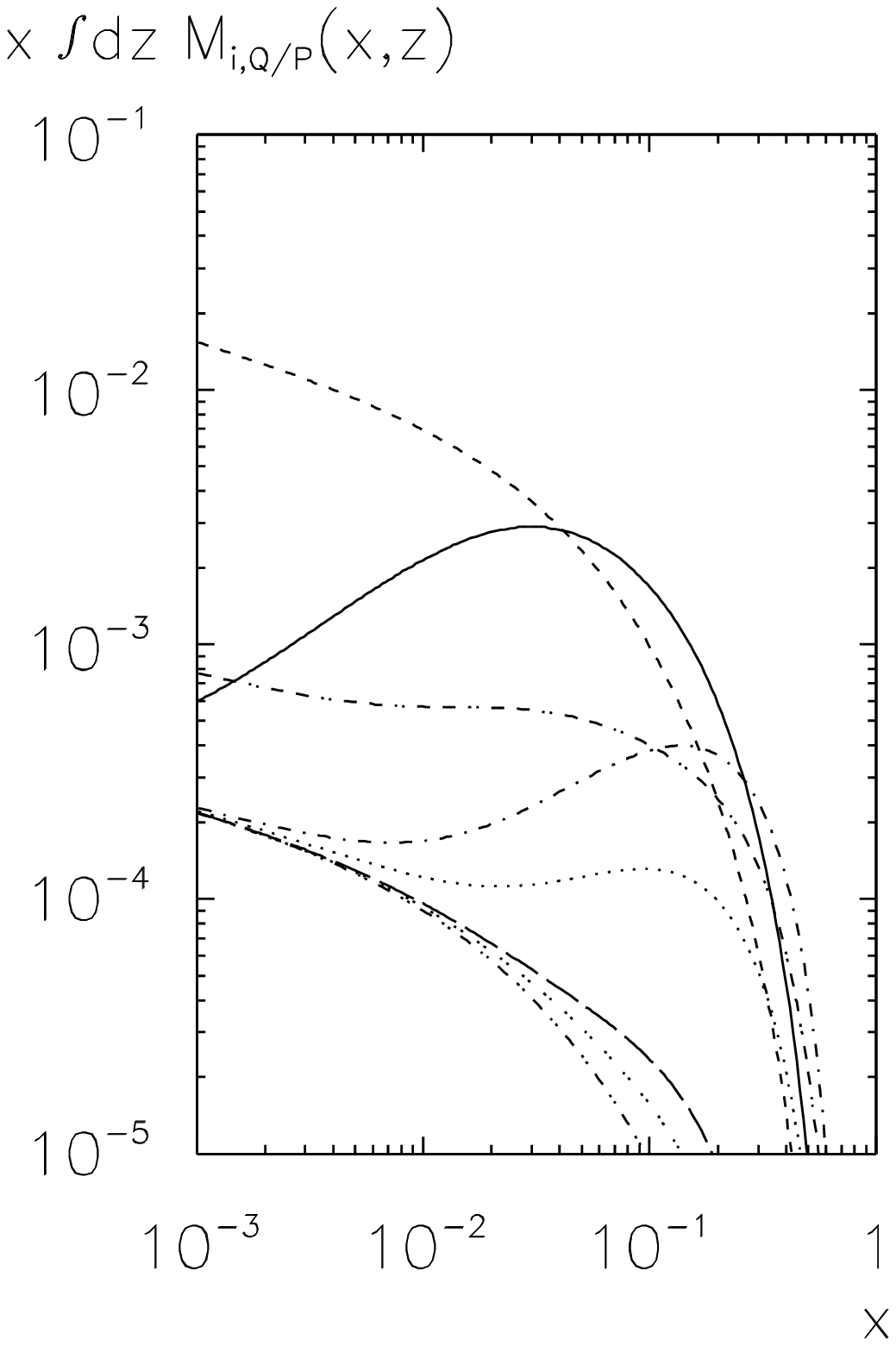}{width=45mm}}
\put(65,85){\lettlab (b)}

\put(110,93){\epsfigdg{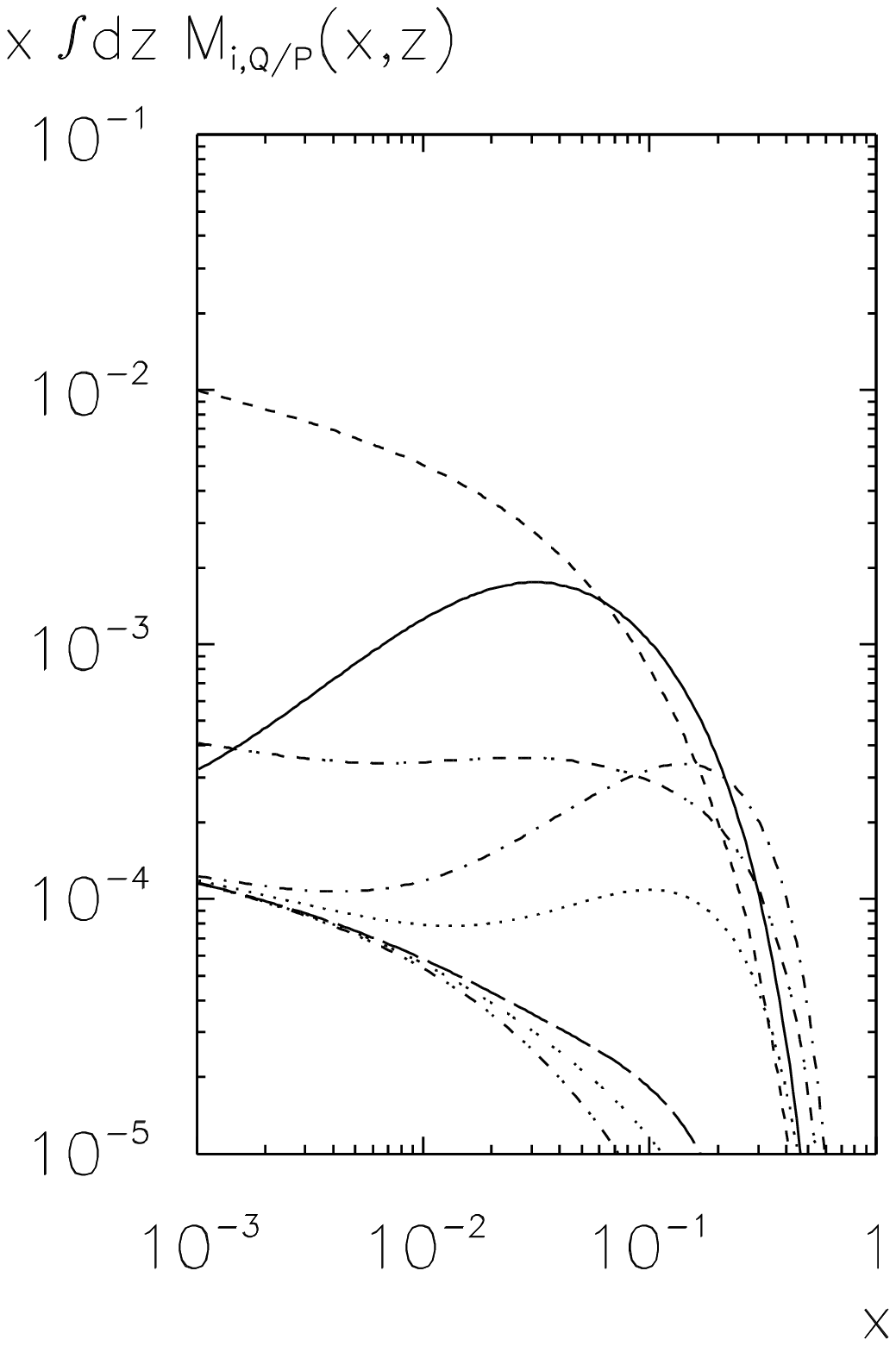}{width=45mm}}
\put(120,85){\lettlab (c)}

\put(0,8){\epsfigdg{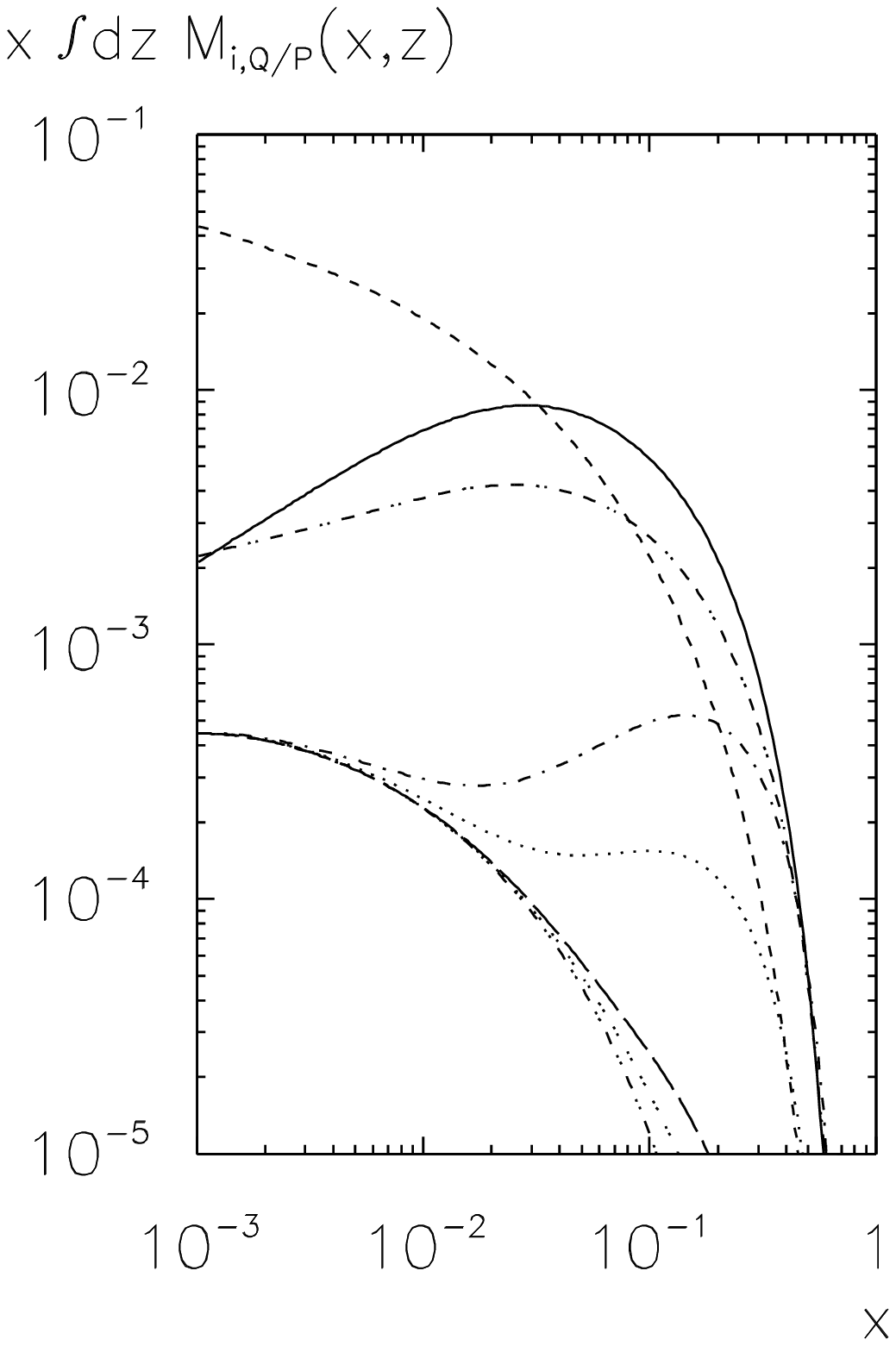}{width=45mm}}
\put(10,0){\lettlab (d)}

\put(55,8){\epsfigdg{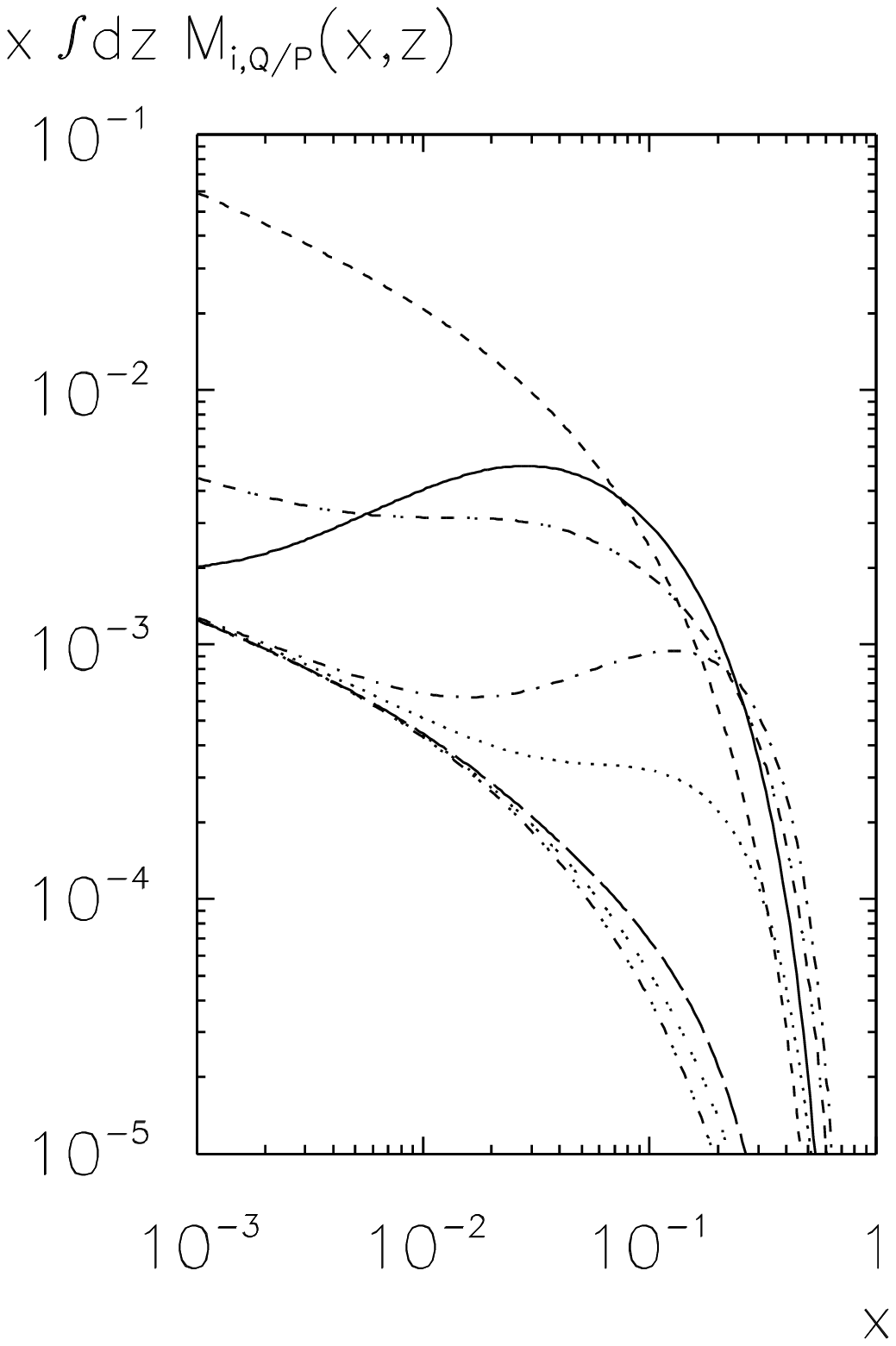}{width=45mm}}
\put(65,0){\lettlab (e)}

\put(110,8){\epsfigdg{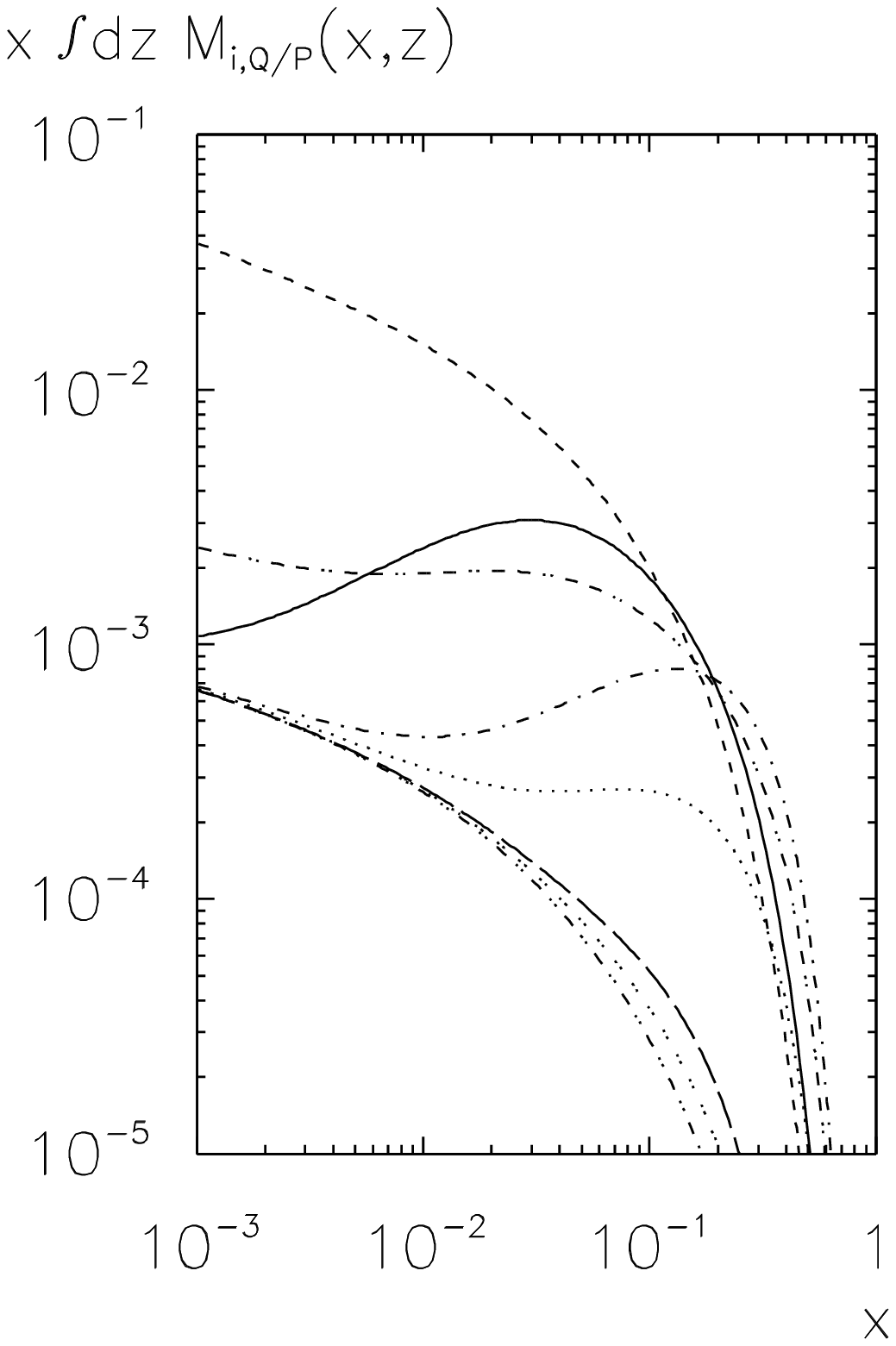}{width=45mm}}
\put(120,0){\lettlab (f)}

\end{picture}
\end{center}
\shiftcaption
\caption[Definition-Scale Dependence of Perturbative 
Heavy-Quark Target Fragmentation Functions]
{\labelmm{IHQTFFA} {\it 
The dependence on the scale~$\mu_0$, where $M^{(P)}$ is set to zero,
for 
perturbative bottom (a)--(c) and charm (d)--(f) quark 
target fragmentation functions $x\,\int\dd z\,M^{(P)}_{i,Q/P}(x,z,\mu^2)$ at 
$\mu=100\,\GeV$, where
$z$ is integrated from $0.1$ to $1-x$.
The definition scale is
$\mu_0=m/2$ (a), (d);
$\mu_0=m$ (b), (e);
$\mu_0=2m$ (c), (f).
The flavours~$i$ are given by 
$\overline{Q}$ \mbox{(\fullline)}, 
$g$ \mbox{(\dashline)}, 
$d$ \mbox{(\dotline)},
$\overline{d}$ \mbox{(\longdashline)},
$u$ \mbox{(\dashdotline)},
$\overline{u}$ \mbox{(\dotdotline)},
$Q$ \mbox{(\dashdotdotline)},
$F_2^{M}$ \mbox{(\dashdashdotdotline)}.
}}   
\end{figure}

In Fig.~\ref{IHQTFFA} 
we show the dependence on the mass scale~$\mu_0$, where $M^{(P)}$
is set to 
zero, i.e.\ the dependence on
the boundary condition for the evolution equation.
This variation introduces an inherent uncertainty
in our predictions. It is clearly unphysical, and would be compensated
by a corresponding variation, proportional to
$\ln \mu_0^2$, of the non-perturbative piece $M^{(NP)}$.

\begin{figure}[htb] \unitlength 1mm
\begin{center}
\dgpicture{159}{167}

\put(0,93){\epsfigdg{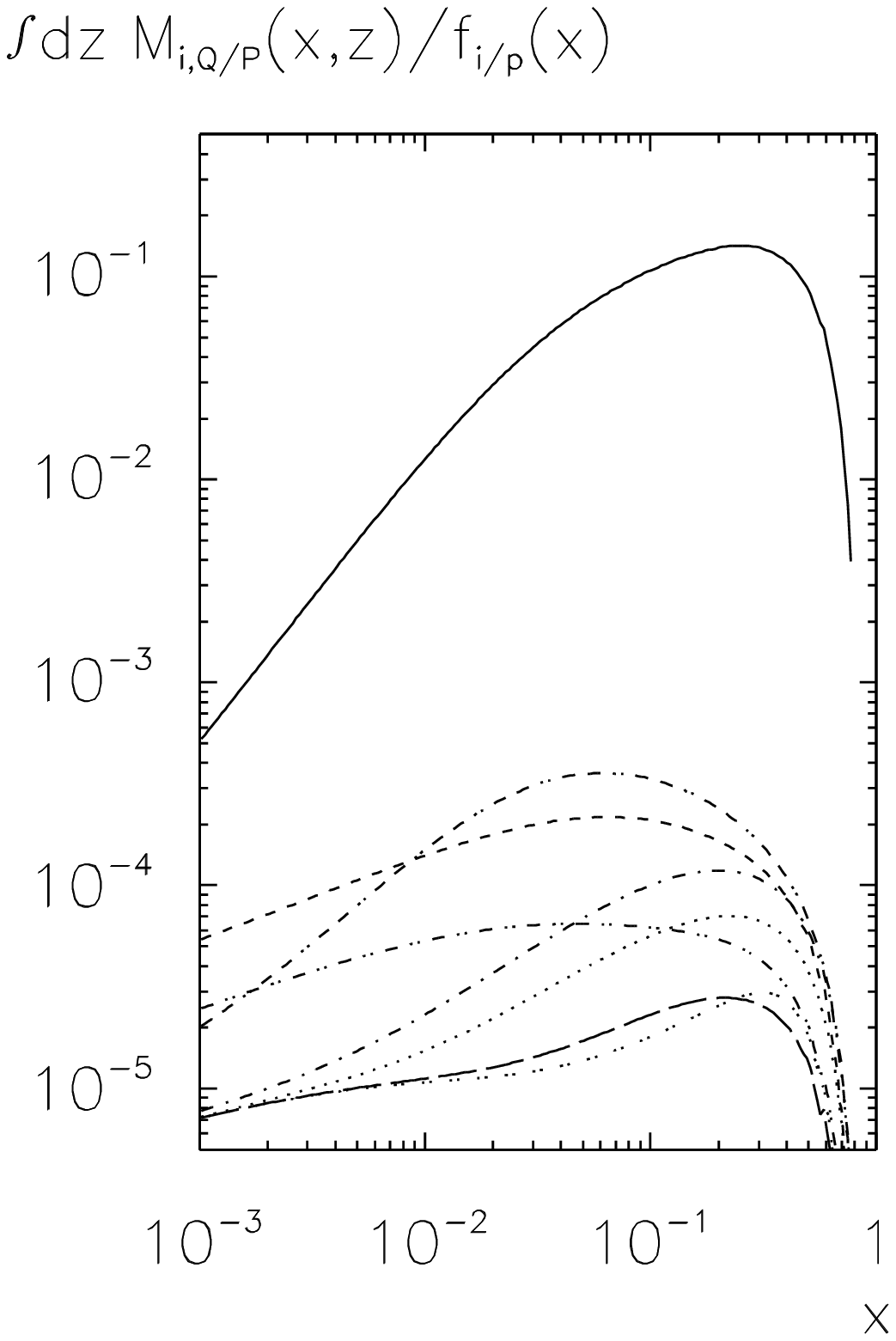}{width=45mm}}
\put(10,85){\lettlab (a)}

\put(55,93){\epsfigdg{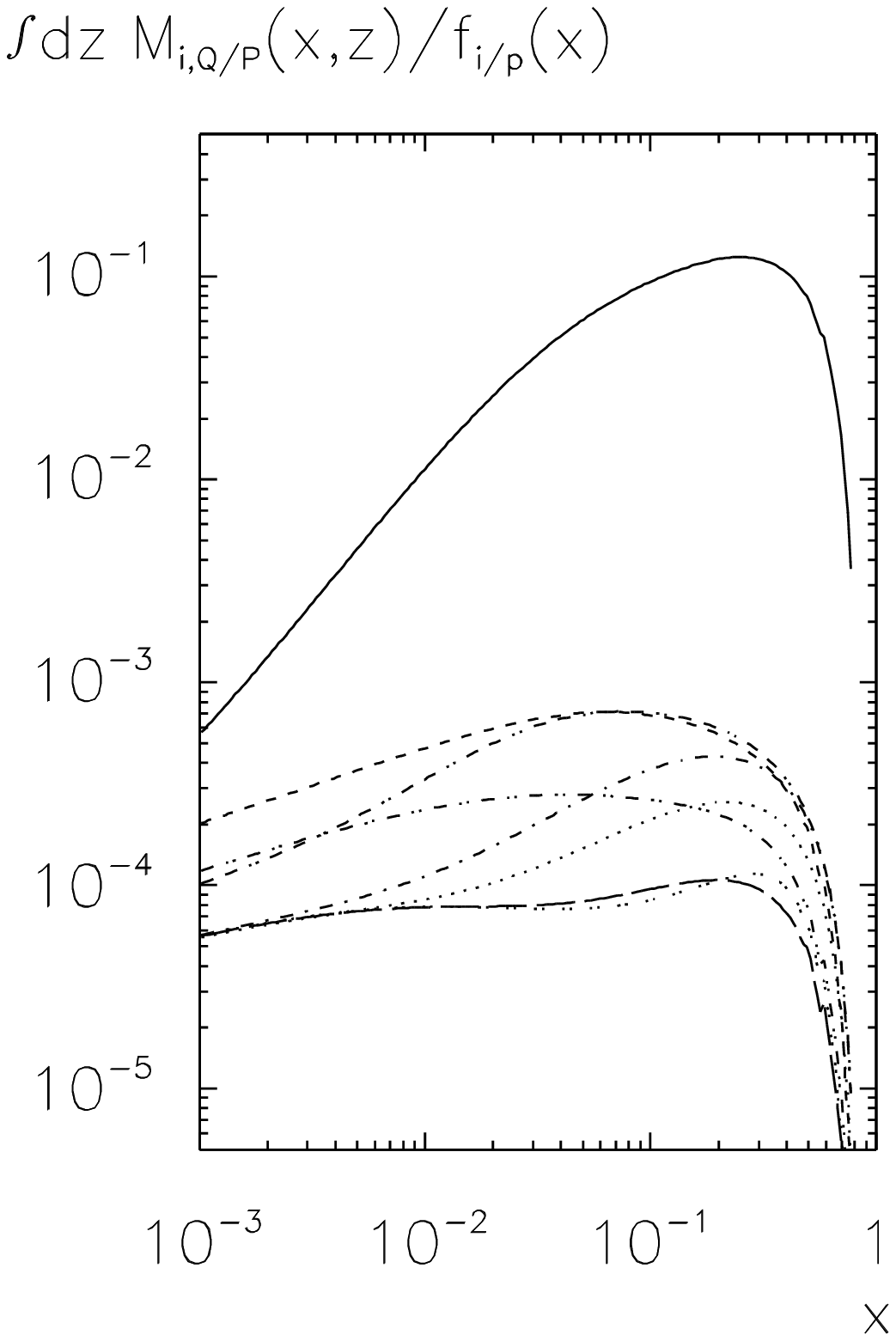}{width=45mm}}
\put(65,85){\lettlab (b)}

\put(110,93){\epsfigdg{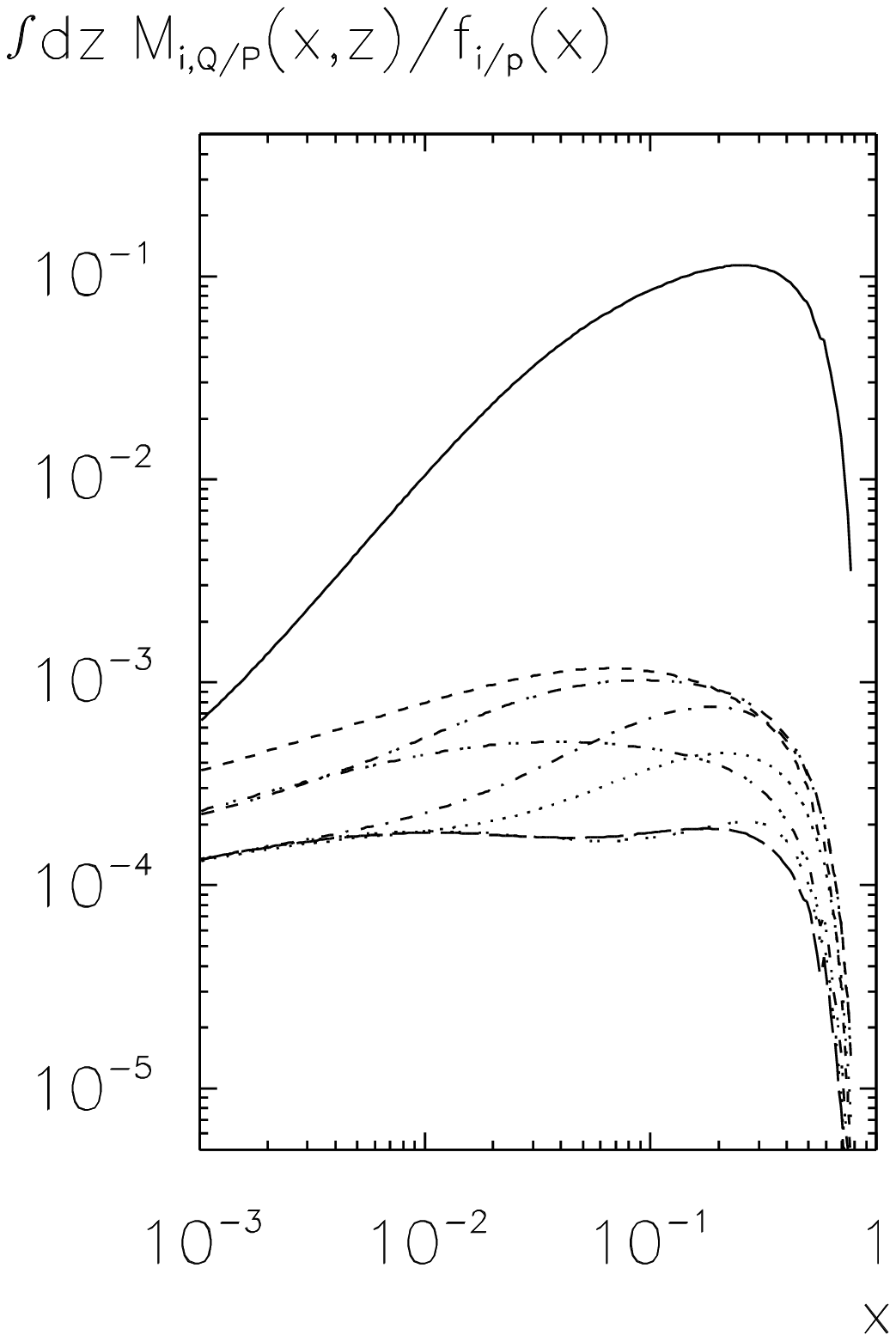}{width=45mm}}
\put(120,85){\lettlab (c)}

\put(0,8){\epsfigdg{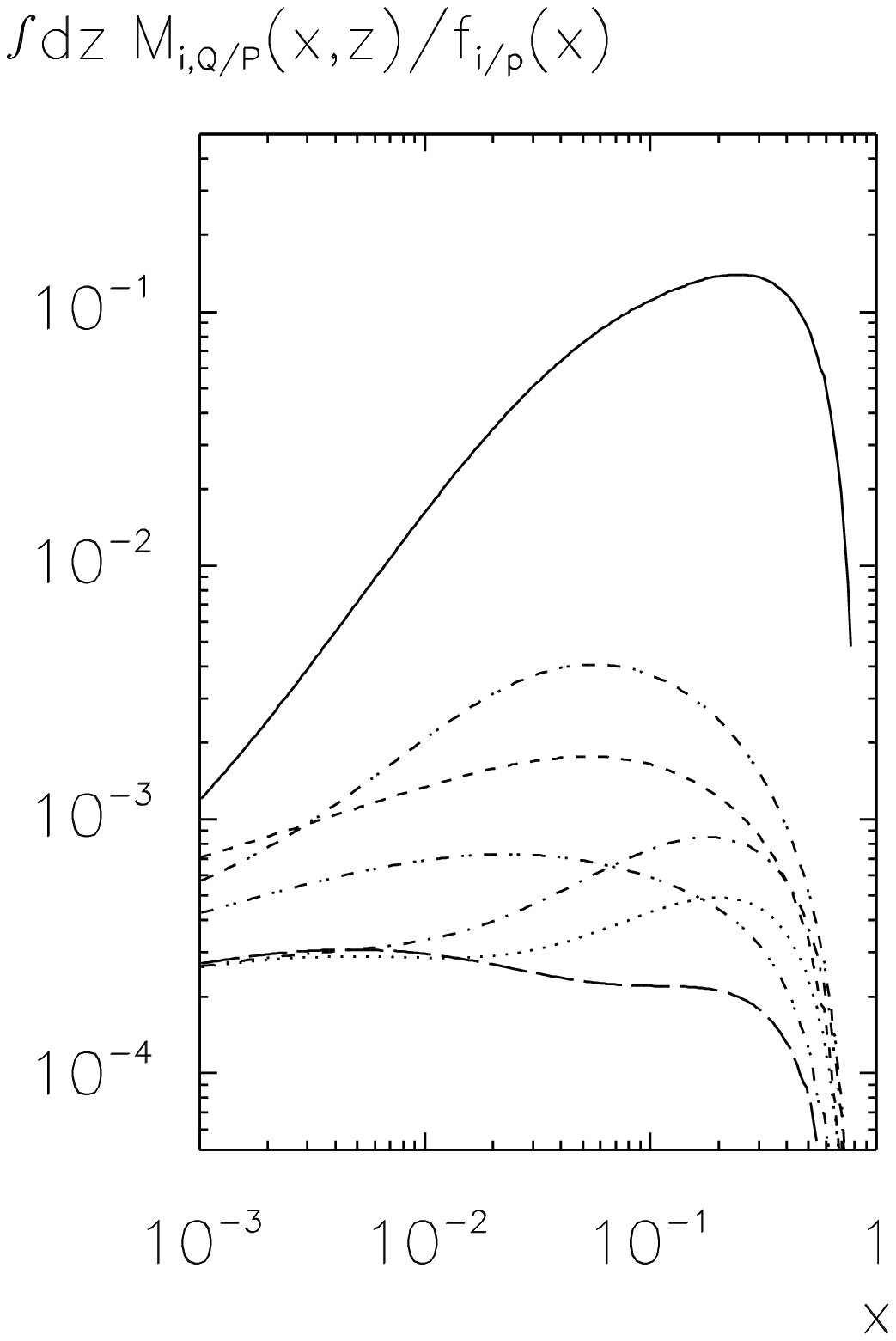}{width=45mm}}
\put(10,0){\lettlab (d)}

\put(55,8){\epsfigdg{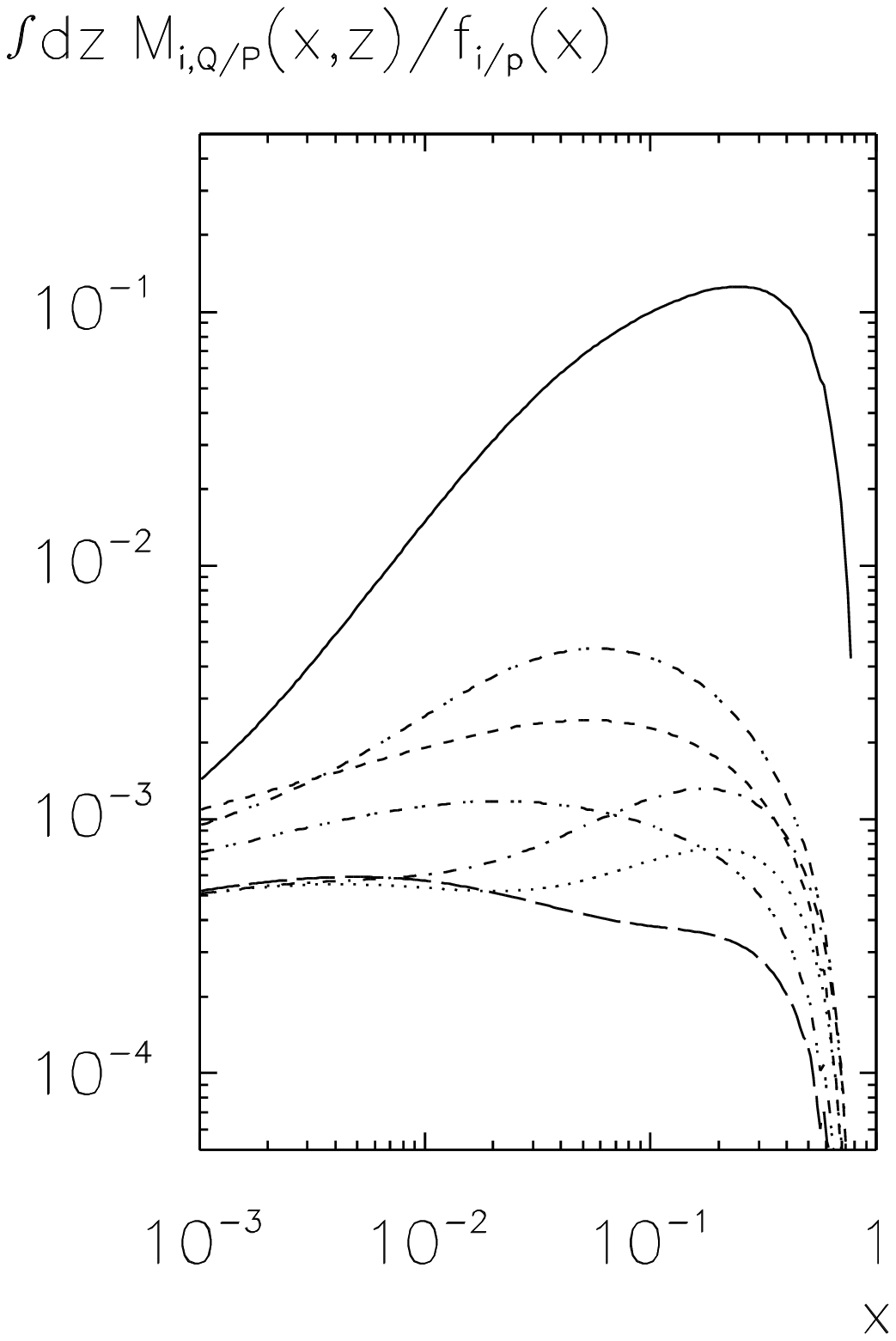}{width=45mm}}
\put(65,0){\lettlab (e)}

\put(110,8){\epsfigdg{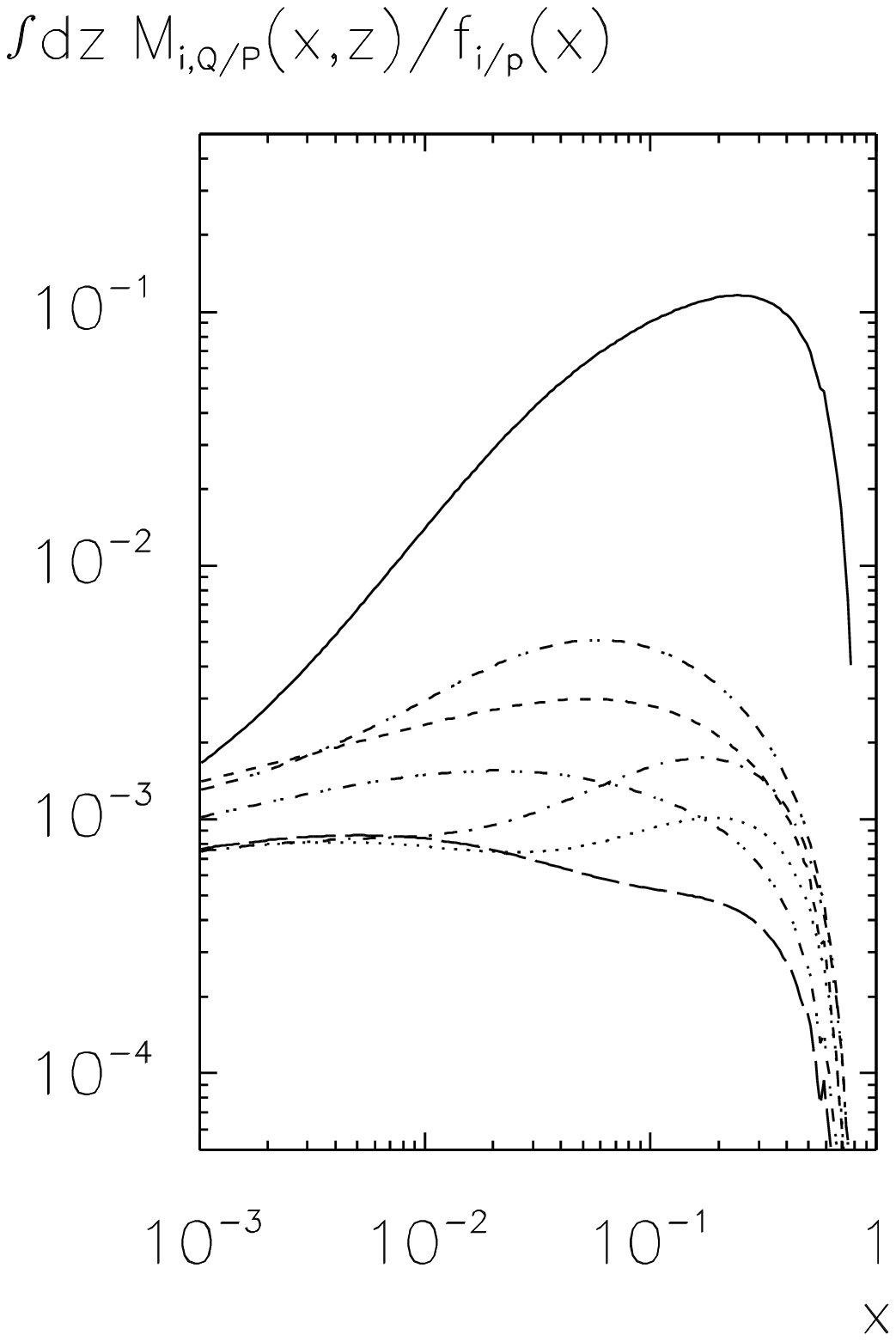}{width=45mm}}
\put(120,0){\lettlab (f)}

\end{picture}
\end{center}
\shiftcaption
\caption[Ratio of Perturbative 
Heavy-Quark Target Fragmentation Functions
and Parton Densities]
{\labelmm{RHQTFFA} {\it Ratio of perturbative bottom (a)--(c)
and charm (d)--(f) quark 
target fragmentation functions $\int\dd z\,M^{(P)}_{i,Q/P}(x,z,\mu^2)$, where
$z$ is integrated from $0.1$ to $1-x$, and parton densities
$f_{i/P}(x,\mu^2)$.
The factorization scale is 
$\mu=10\,\GeV$ (a), (d);
$\mu=30\,\GeV$ (b), (e);
$\mu=100\,\GeV$ (c), (f).
The input distribution for $M^{(P)}$ is zero 
at $\mu_0=m$. 
The flavours~$i$ are given by 
$\overline{Q}$ \mbox{(\fullline)}, $g$ \mbox{(\dashline)}, 
$d$ \mbox{(\dotline)},
$\overline{d}$ \mbox{(\longdashline)},
$u$ \mbox{(\dashdotline)},
$\overline{u}$ \mbox{(\dotdotline)},
$Q$ \mbox{(\dashdotdotline)}, 
$F_2^{M}/F_2$ \mbox{(\dashdashdotdotline)}.
}}   
\end{figure}

As mentioned in the introduction,
the observation of particles in the target fragmentation region
may be exploited to gain information on the hard scattering process itself.
This possibility is illustrated in Fig.~\ref{RHQTFFA}, showing the ratio
of perturbative heavy-quark target fragmentation functions and parton 
densities. Tagging a heavy quark~$Q$ in the backward direction 
clearly enhances those processes that have a heavy antiquark~$\overline{Q}$ 
initiating the hard process. 
The scale dependence of the ratio is large for the
light quark flavours and for the gluon, and nearly absent for the
heavy quark, the reason being that the heavy-quark contributions are 
produced by similar radiation 
mechanisms in both cases (i.e.\ for the parton densities and the
perturbative heavy-quark target fragmentation functions), whereas the
light flavours and the gluon are already non-vanishing at some small scale 
in the case of the parton densities.
Of course, Fig.~\ref{RHQTFFA} can only be taken
as an indication of this fact, since the non-perturbative
contribution is not included, and since there is also the production 
process described by the graphs in Fig.~\ref{QCDcfr1} that contributes
in the collinear limit.

\begin{figure}[htb] \unitlength 1mm
\begin{center}
\dgpicture{159}{167}

\put(0,93){\epsfigdg{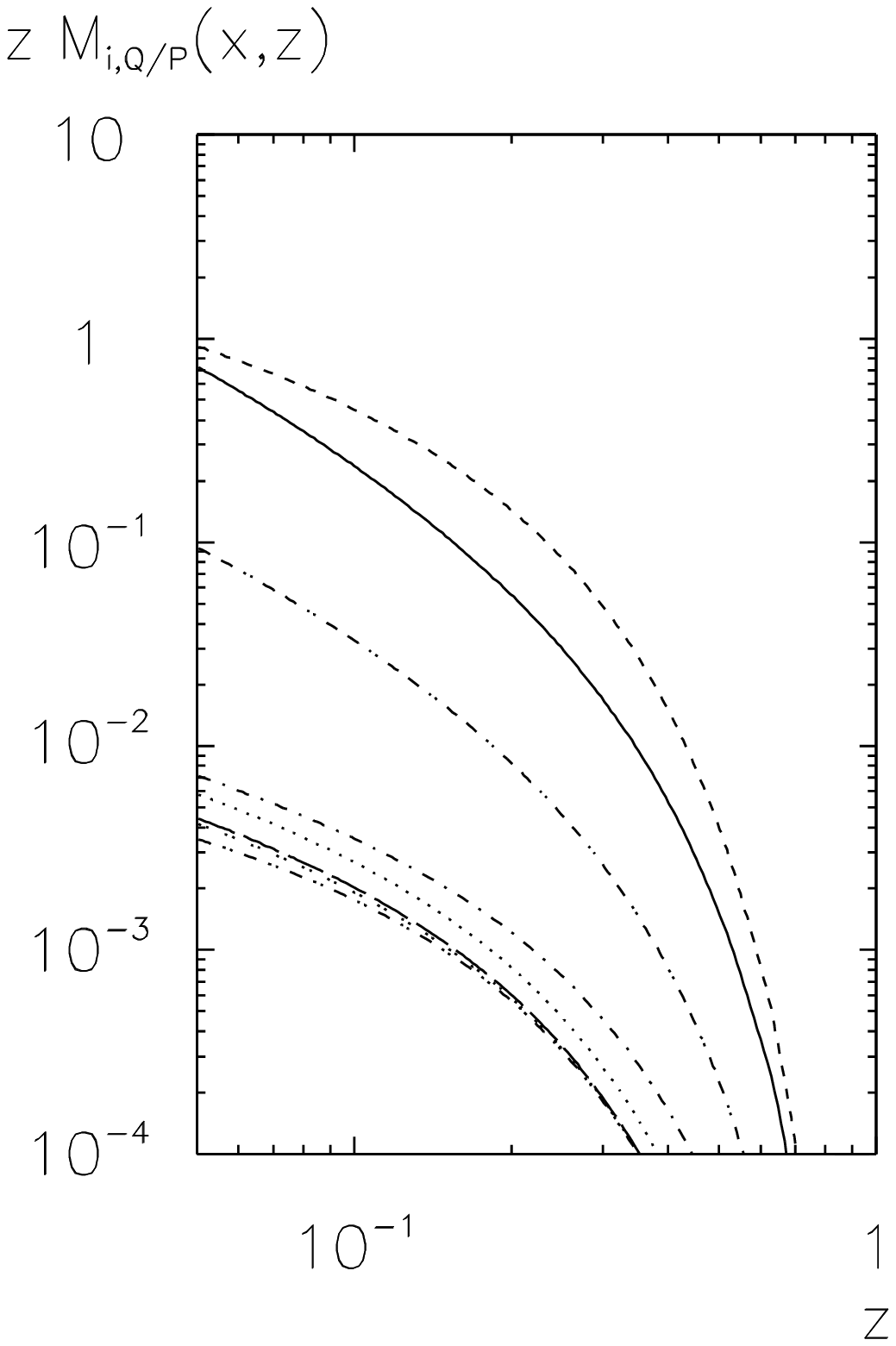}{width=45mm}}
\put(10,85){\lettlab (a)}

\put(55,93){\epsfigdg{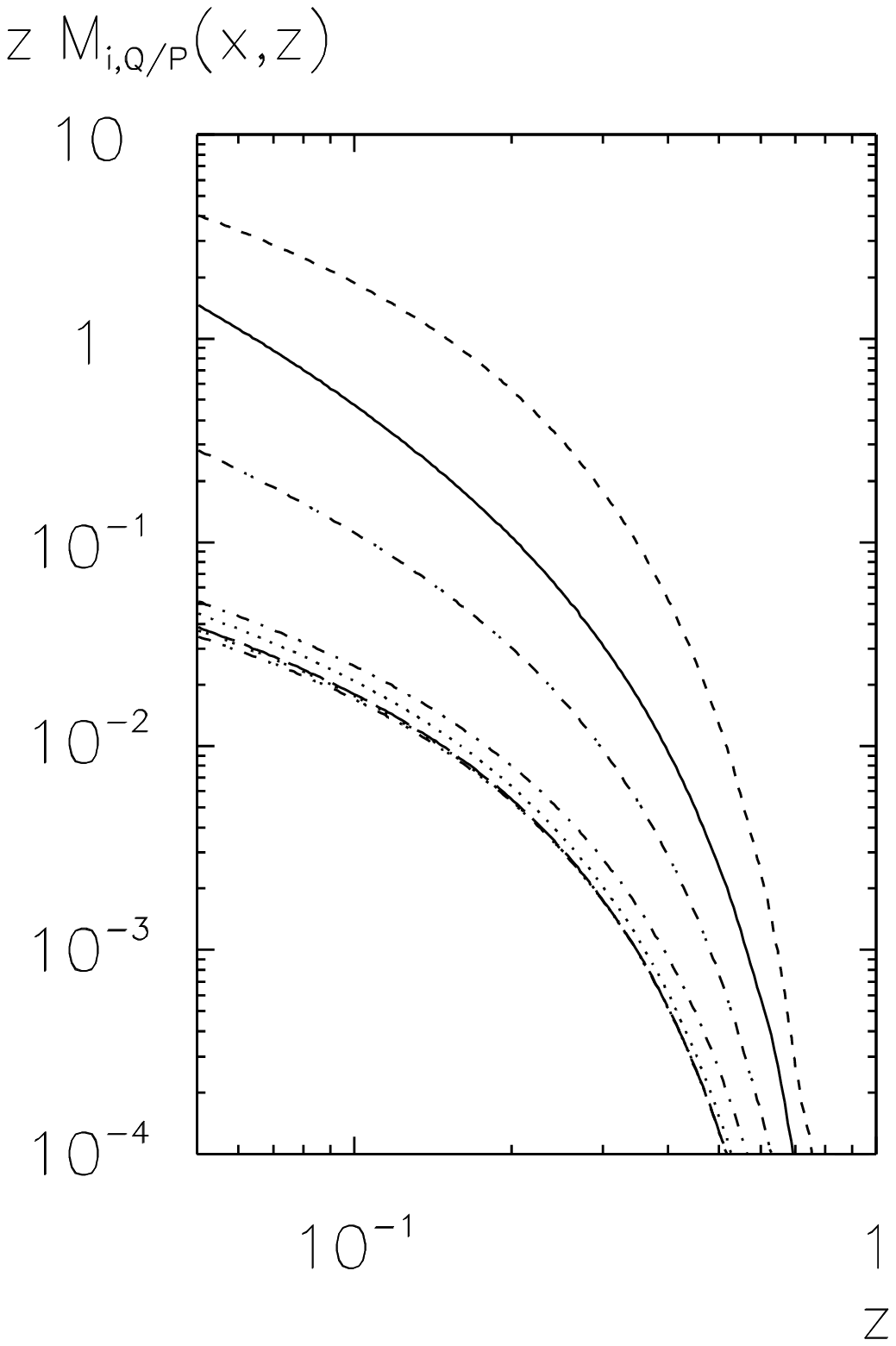}{width=45mm}}
\put(65,85){\lettlab (b)}

\put(110,93){\epsfigdg{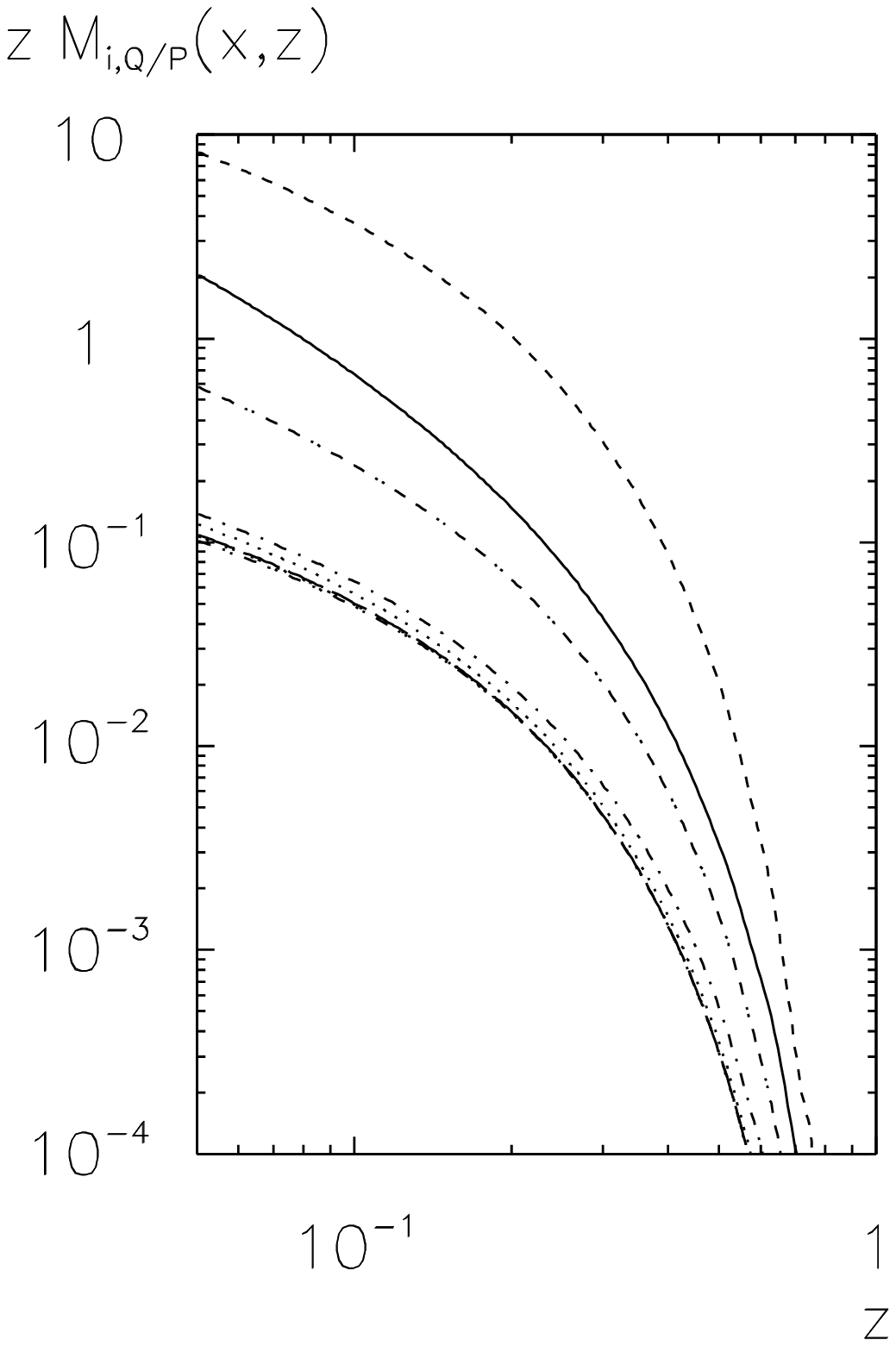}{width=45mm}}
\put(120,85){\lettlab (c)}

\put(0,8){\epsfigdg{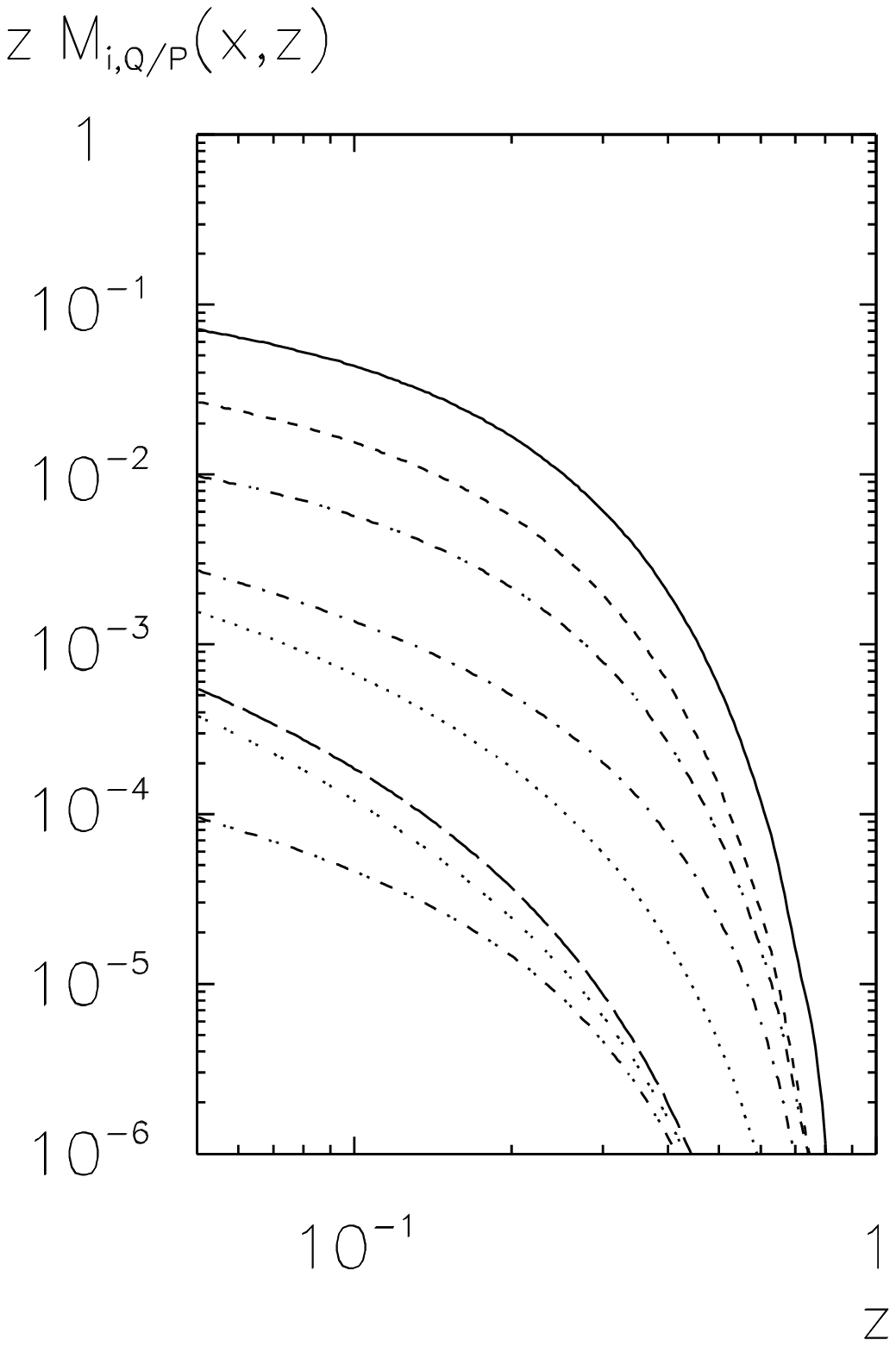}{width=45mm}}
\put(10,0){\lettlab (d)}

\put(55,8){\epsfigdg{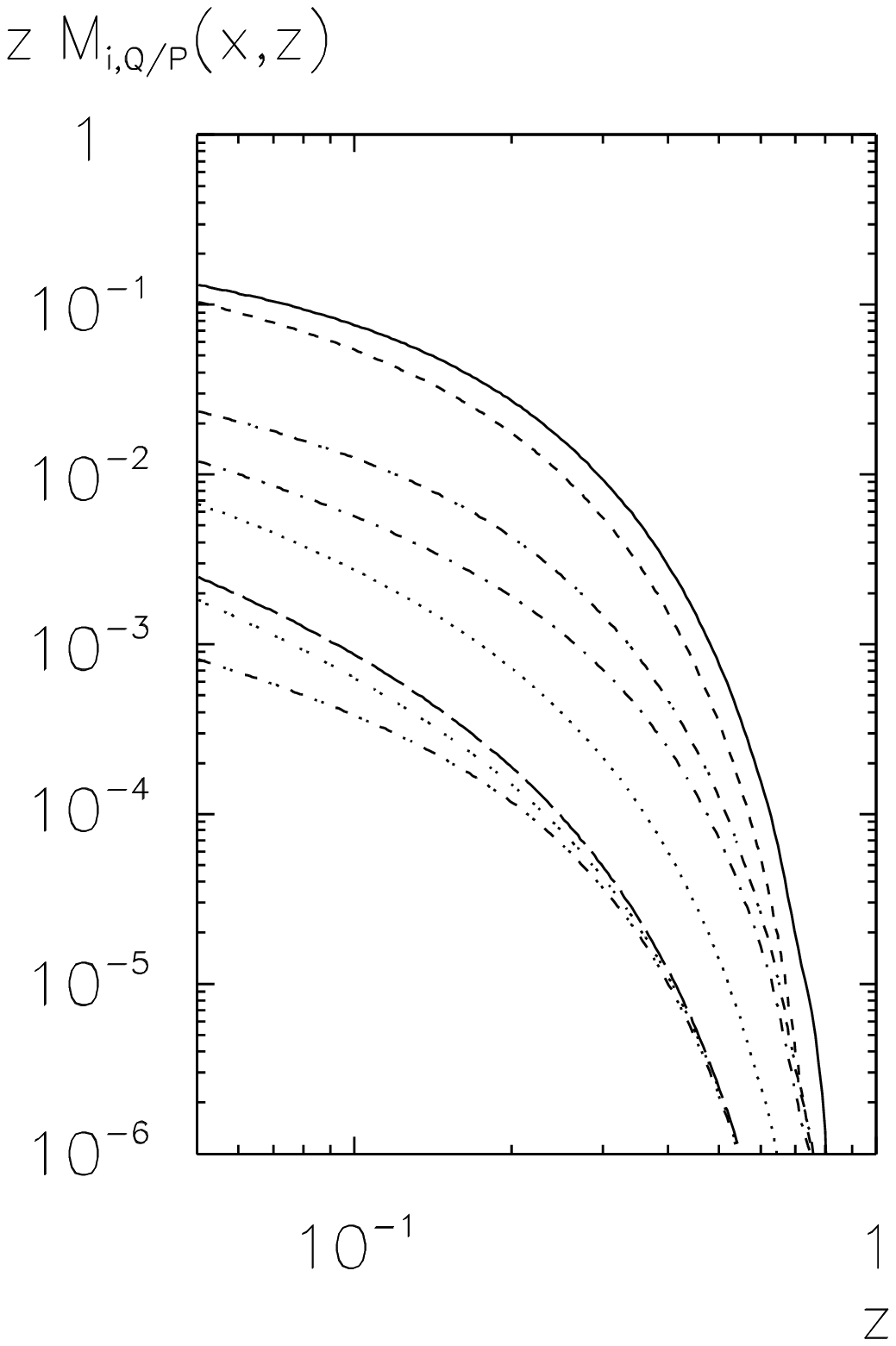}{width=45mm}}
\put(65,0){\lettlab (e)}

\put(110,8){\epsfigdg{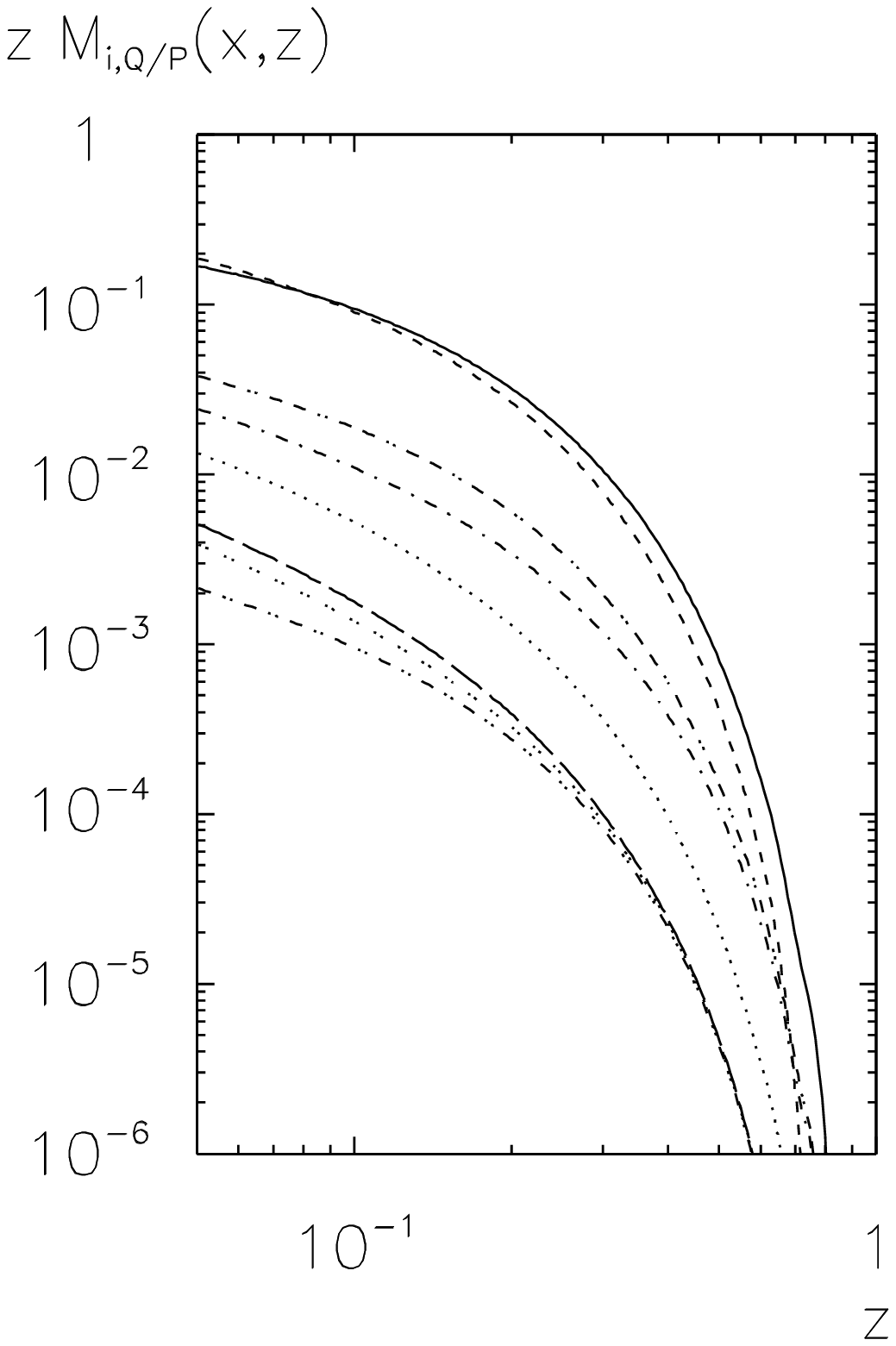}{width=45mm}}
\put(120,0){\lettlab (f)}

\end{picture}
\end{center}
\shiftcaption
\caption[Momentum-Fraction Dependence of Perturbative 
Bottom-Quark Target Fragmentation Functions]
{\labelmm{HQTFFZb} {\it Dependence of perturbative bottom-quark
target fragmentation functions $z\,M^{(P)}_{i,Q/P}(x,z,\mu^2)$
on the momentum fraction~$z$ of the observed
heavy quark. The momentum fraction of the parton incident in the
hard subprocess is $x=0.005$ (a)--(c) and $x=0.05$ (d)--(e), respectively.
The factorization scale is
$\mu=10\,\GeV$ (a), (d);
$\mu=30\,\GeV$ (b), (e);
$\mu=100\,\GeV$ (c), (f).
The input distribution for $M^{(P)}$ is zero 
at $\mu_0=m$. 
The flavours~$i$ are given by 
$\overline{b}$ \mbox{(\fullline)}, 
$g$ \mbox{(\dashline)}, 
$d$ \mbox{(\dotline)},
$\overline{d}$ \mbox{(\longdashline)},
$u$ \mbox{(\dashdotline)},
$\overline{u}$ \mbox{(\dotdotline)},
$b$ \mbox{(\dashdotdotline)},
$F_2^{M}$ \mbox{(\dashdashdotdotline)}.
}}   
\end{figure}

\begin{figure}[htb] \unitlength 1mm
\begin{center}
\dgpicture{159}{167}

\put(0,93){\epsfigdg{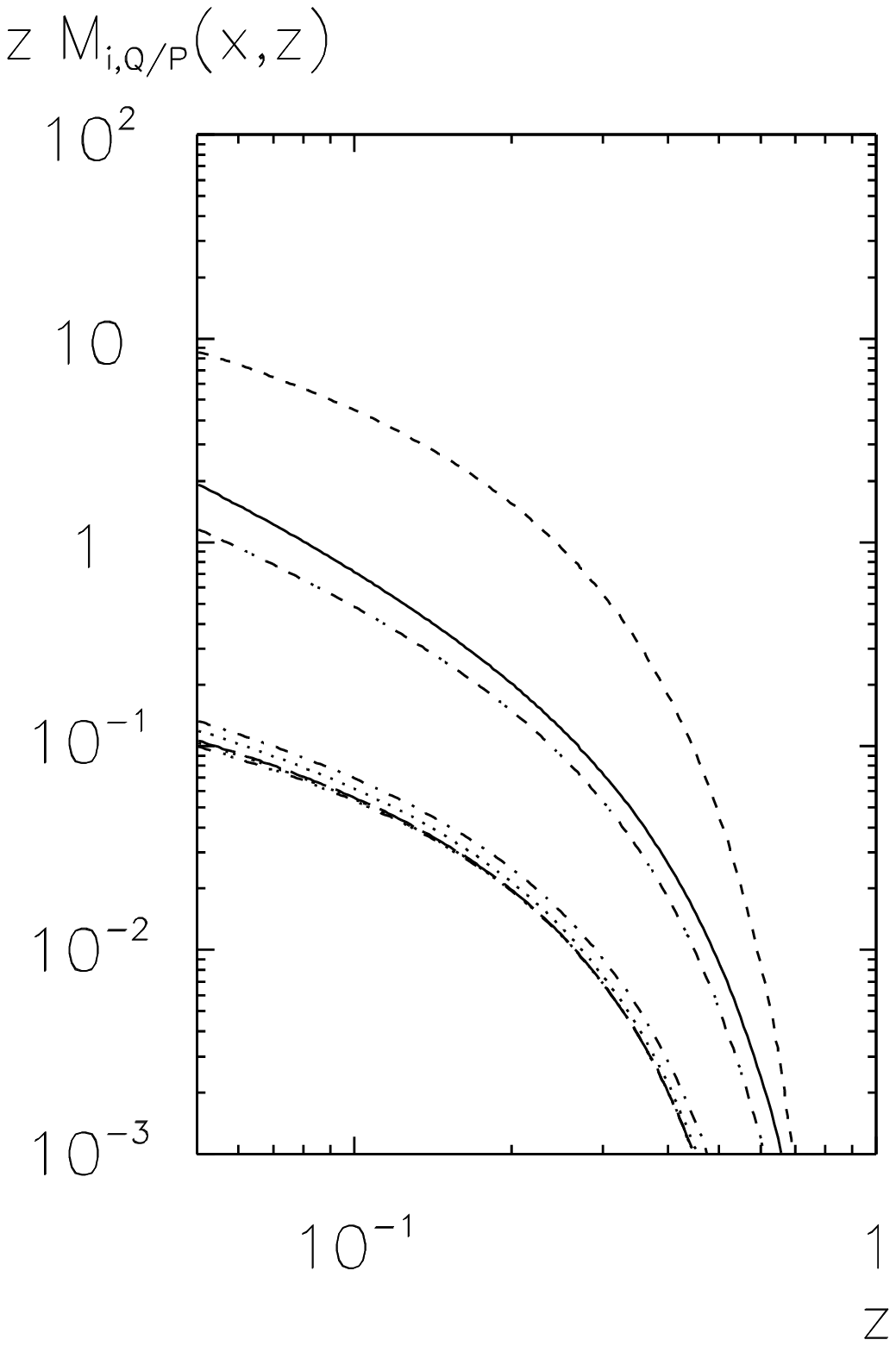}{width=45mm}}
\put(10,85){\lettlab (a)}

\put(55,93){\epsfigdg{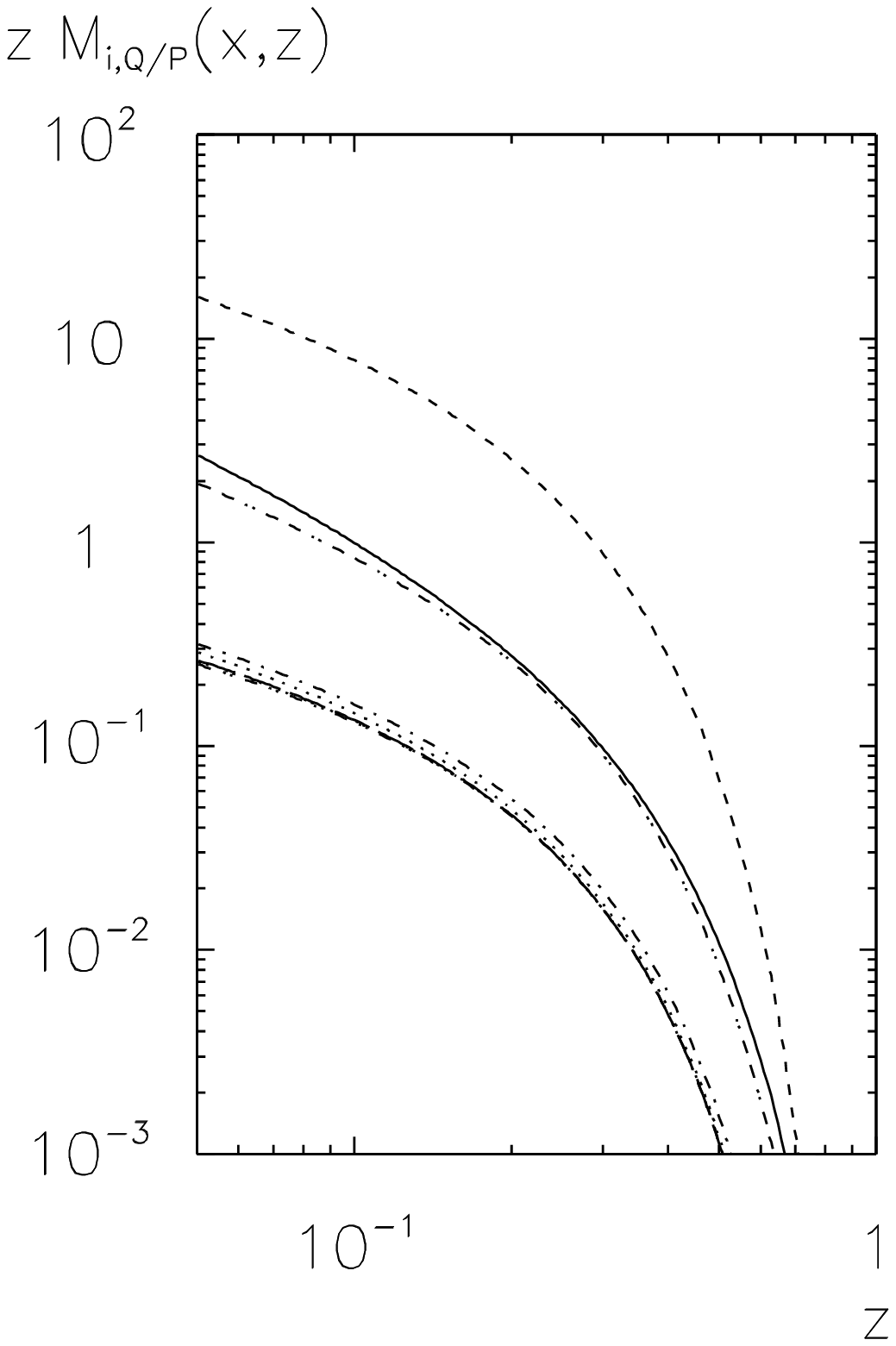}{width=45mm}}
\put(65,85){\lettlab (b)}

\put(110,93){\epsfigdg{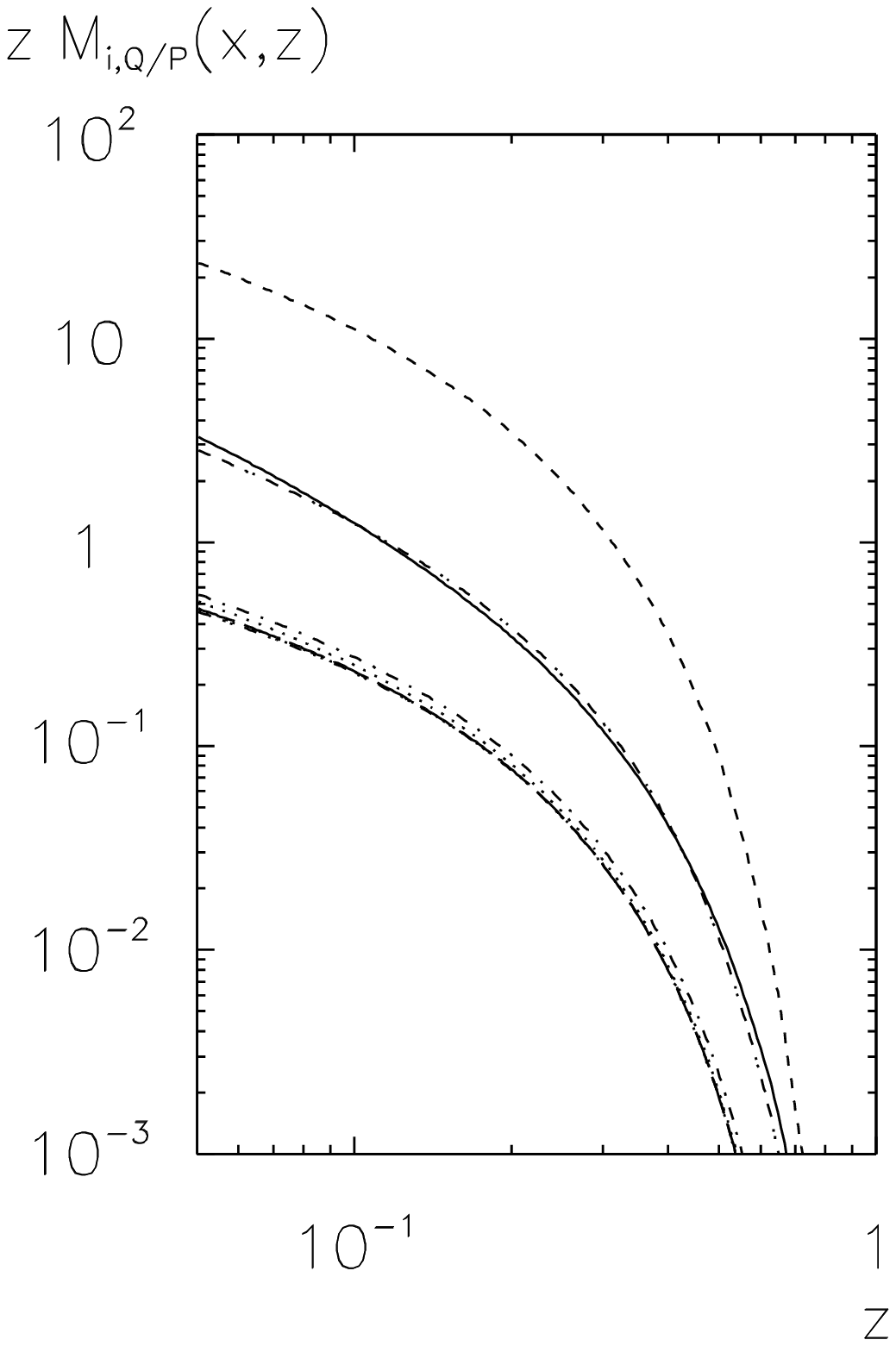}{width=45mm}}
\put(120,85){\lettlab (c)}

\put(0,8){\epsfigdg{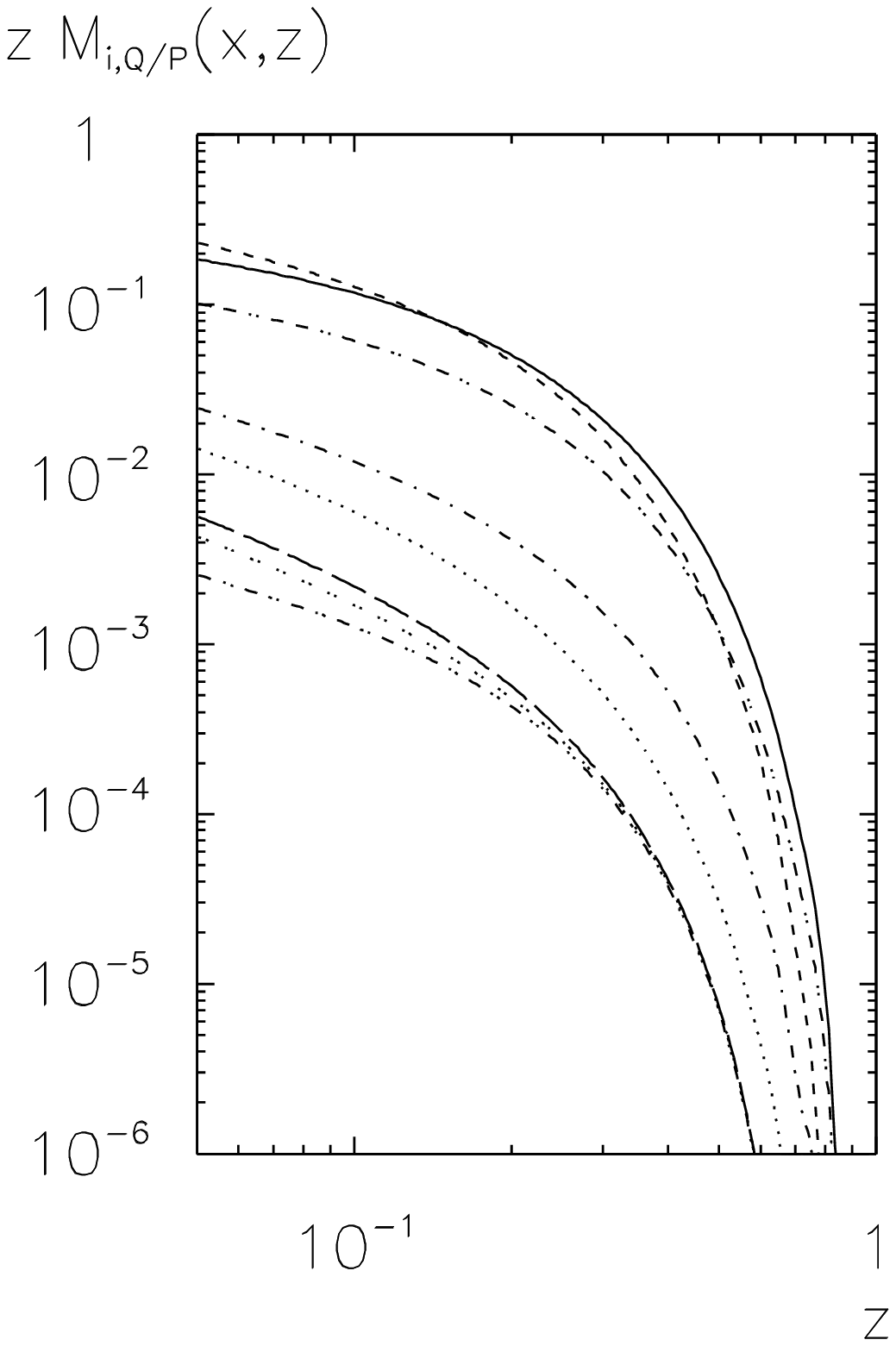}{width=45mm}}
\put(10,0){\lettlab (d)}

\put(55,8){\epsfigdg{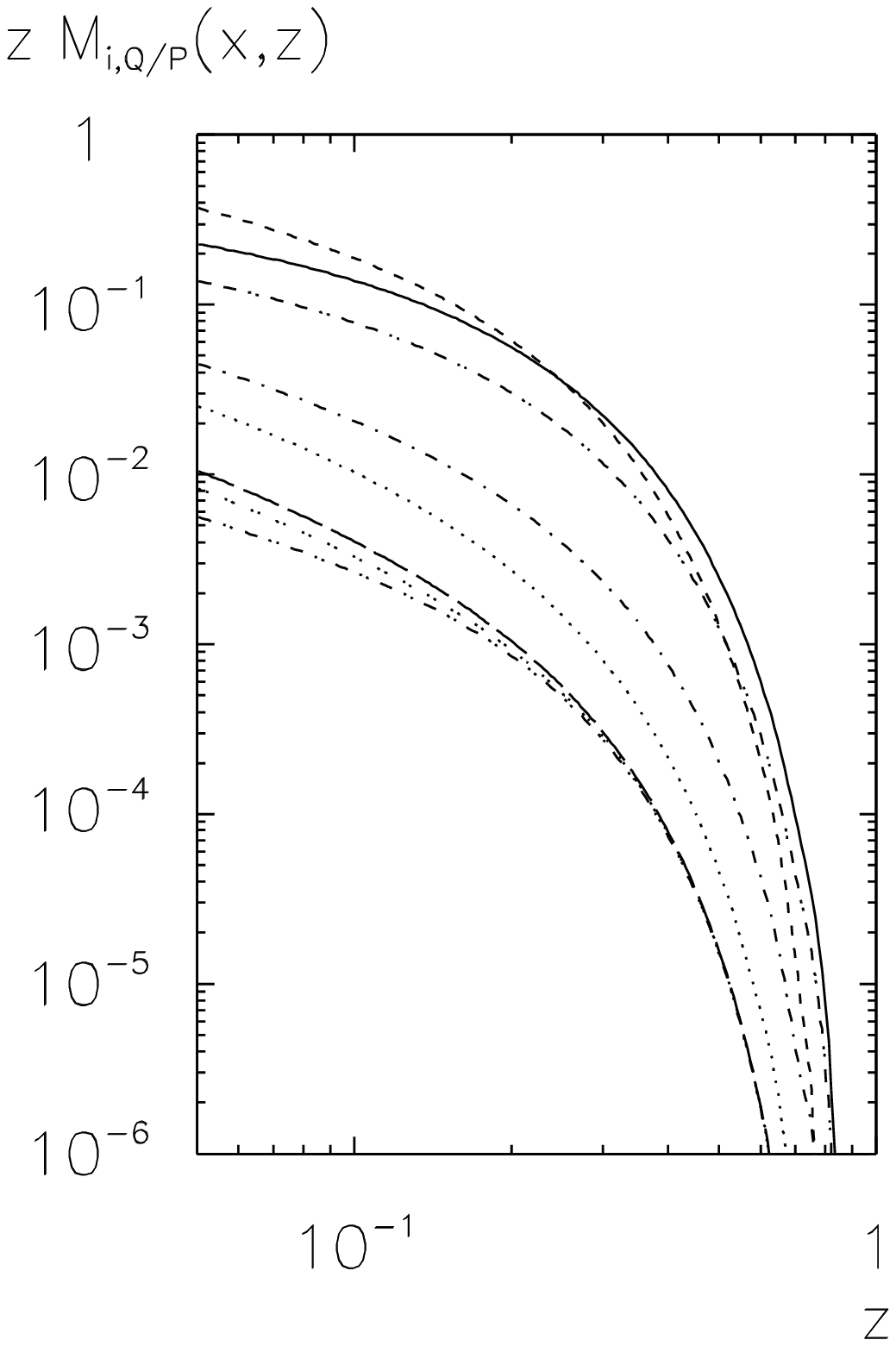}{width=45mm}}
\put(65,0){\lettlab (e)}

\put(110,8){\epsfigdg{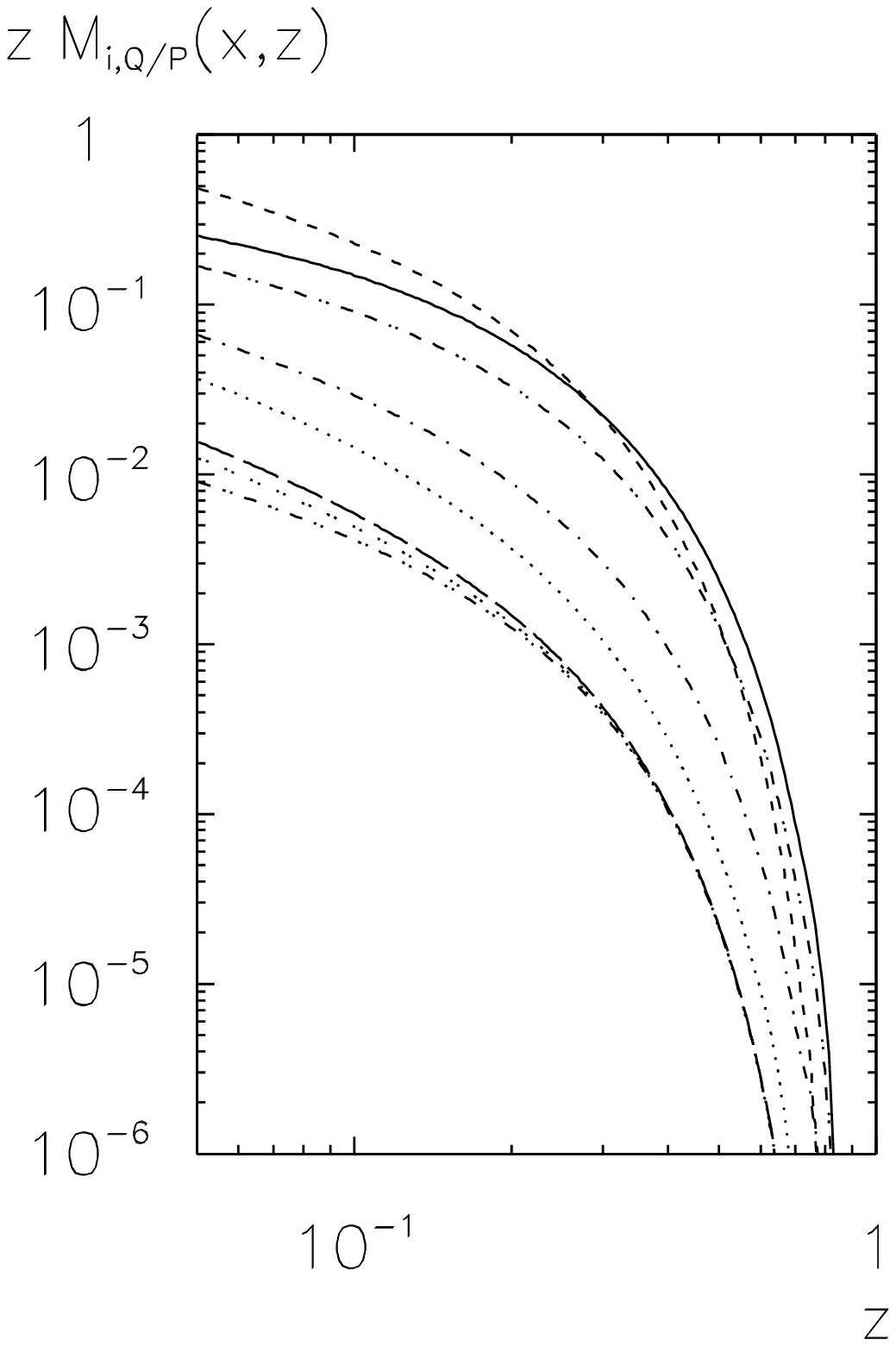}{width=45mm}}
\put(120,0){\lettlab (f)}

\end{picture}
\end{center}
\shiftcaption
\caption[Momentum-Fraction Dependence of Perturbative 
Charm-Quark Target Fragmentation Functions]
{\labelmm{HQTFFZc} {\it Dependence of perturbative charm-quark
target fragmentation functions $z\,M^{(P)}_{i,Q/P}(x,z,\mu^2)$
on the momentum fraction~$z$ of the observed
heavy quark. The momentum fraction of the parton incident in the
hard subprocess is $x=0.005$ (a)--(c) and $x=0.05$ (d)--(e), respectively.
The factorization scale is
$\mu=10\,\GeV$ (a), (d);
$\mu=30\,\GeV$ (b), (e);
$\mu=100\,\GeV$ (c), (f).
The input distribution for $M^{(P)}$ is zero 
at $\mu_0=m$. 
The flavours~$i$ are given by 
$\overline{c}$ \mbox{(\fullline)}, 
$g$ \mbox{(\dashline)}, 
$d$ \mbox{(\dotline)},
$\overline{d}$ \mbox{(\longdashline)},
$u$ \mbox{(\dashdotline)},
$\overline{u}$ \mbox{(\dotdotline)},
$c$ \mbox{(\dashdotdotline)},
$F_2^{M}$ \mbox{(\dashdashdotdotline)}.
}}   
\end{figure}

The dependence of the perturbative target fragmentation functions
on the momentum fraction~$z$ carried by the observed heavy quark is
shown in Figs.~\ref{HQTFFZb} and~\ref{HQTFFZc}.
The dependence is very steep, and the production at small~$z$ 
is clearly favoured,
because 
of the radiative production mechanism.
It is interesting to note that the distributions tend to become harder
for increasing factorization scales.

\begin{figure}[htb] \unitlength 1mm
\begin{center}
\dgpicture{159}{167}

\put(20,93){\epsfigdg{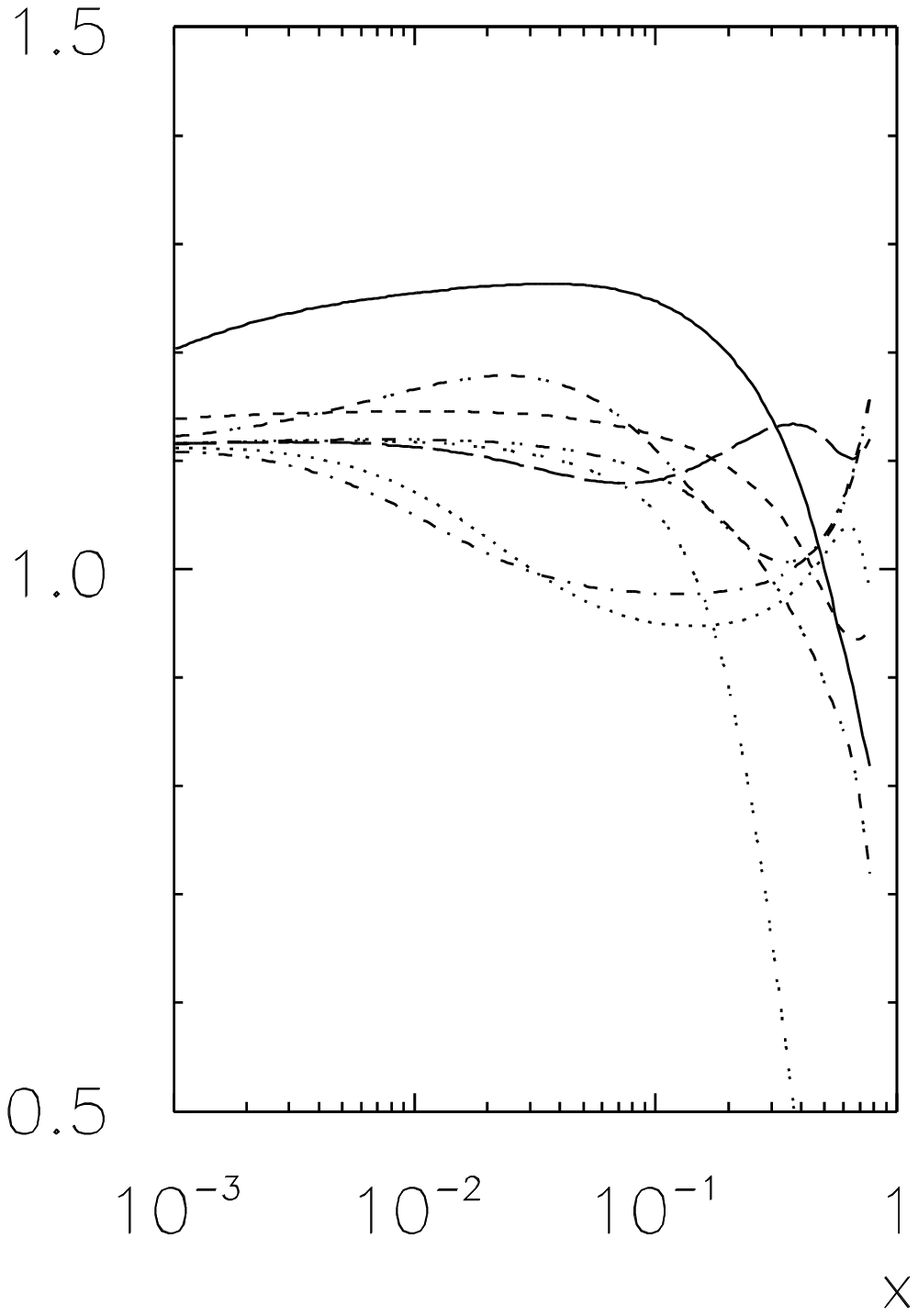}{width=45mm}}
\put(30,85){\lettlab (a)}

\put(85,93){\epsfigdg{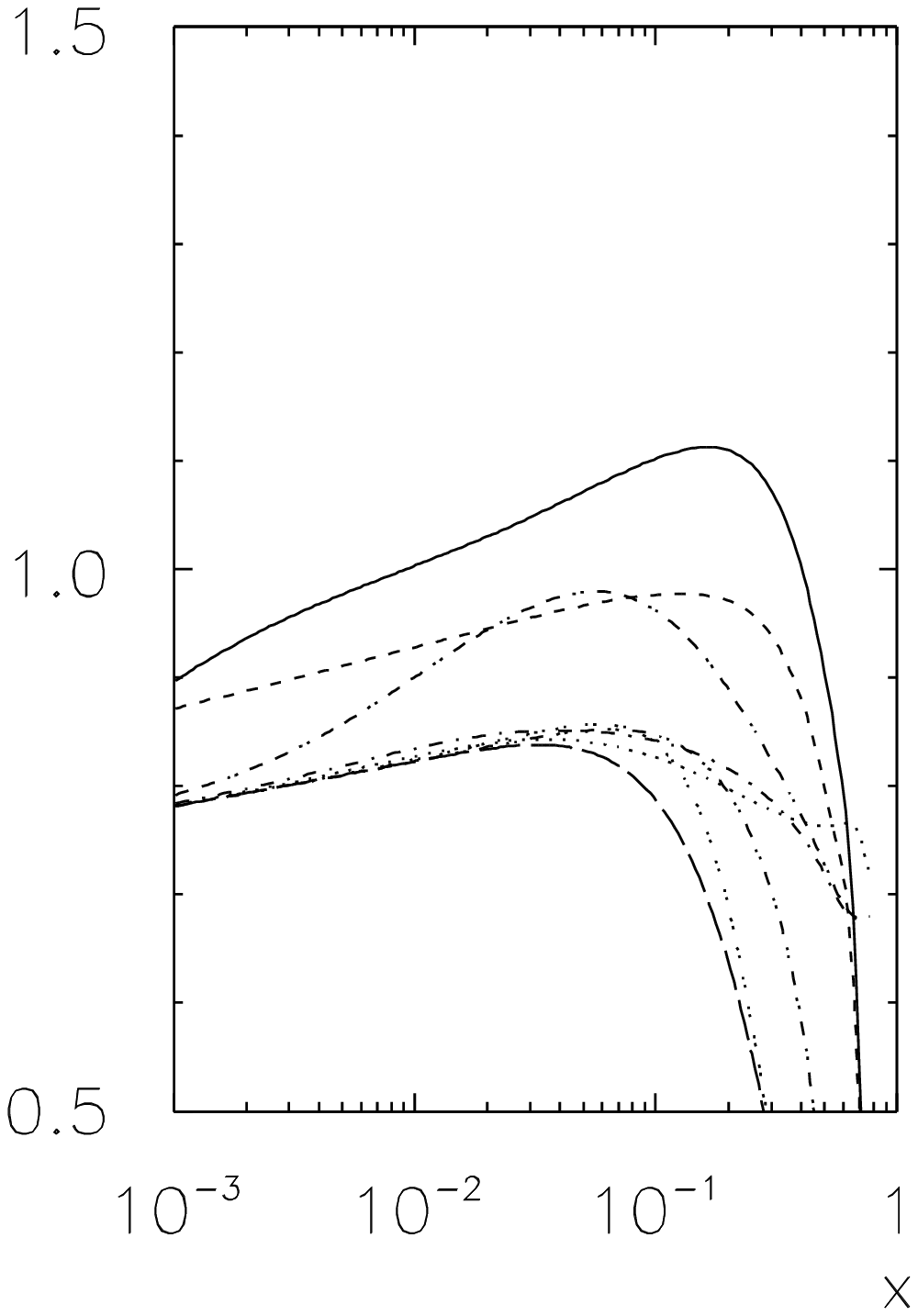}{width=45mm}}
\put(95,85){\lettlab (b)}


\put(20,8){\epsfigdg{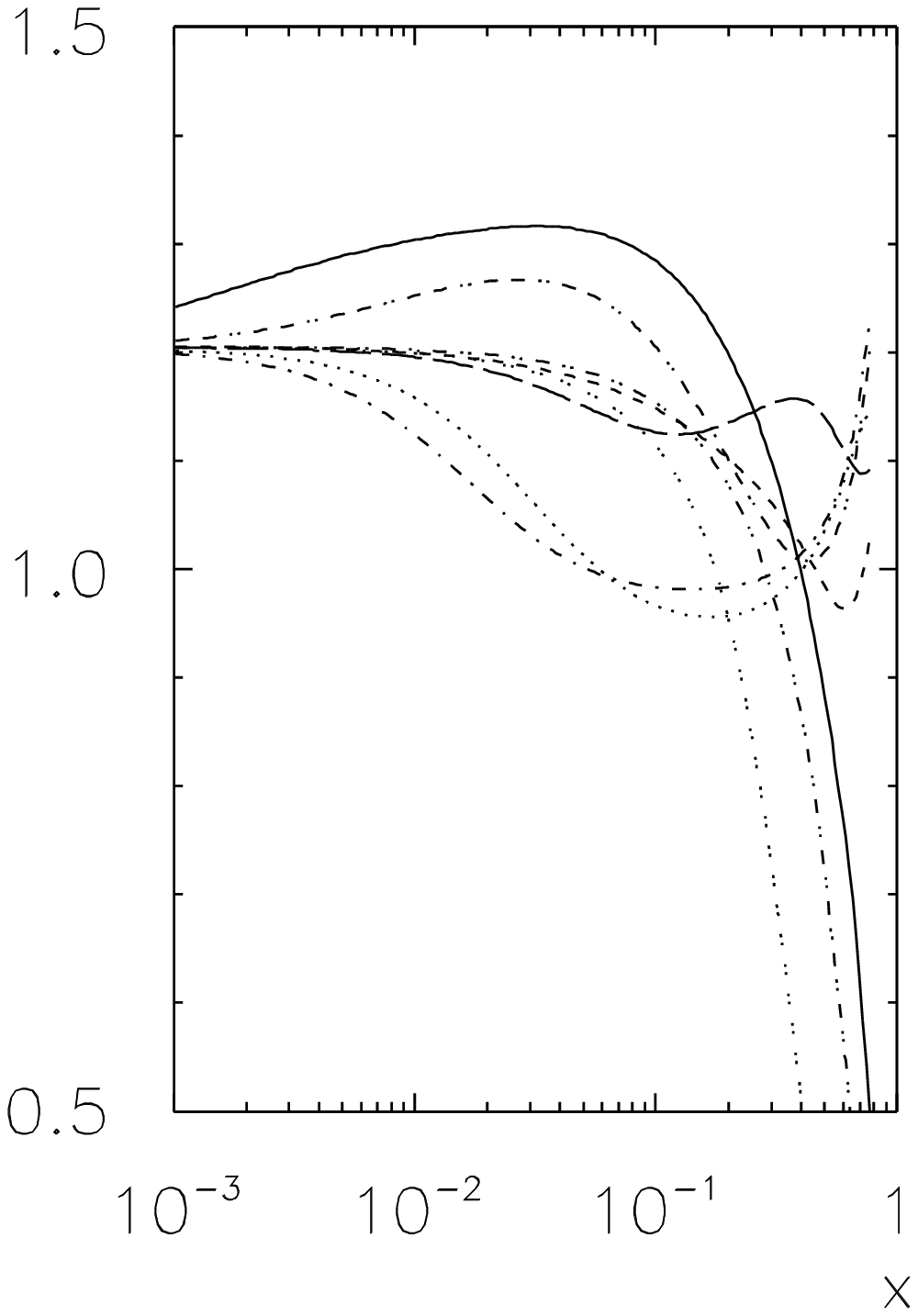}{width=45mm}}
\put(30,0){\lettlab (c)}

\put(85,8){\epsfigdg{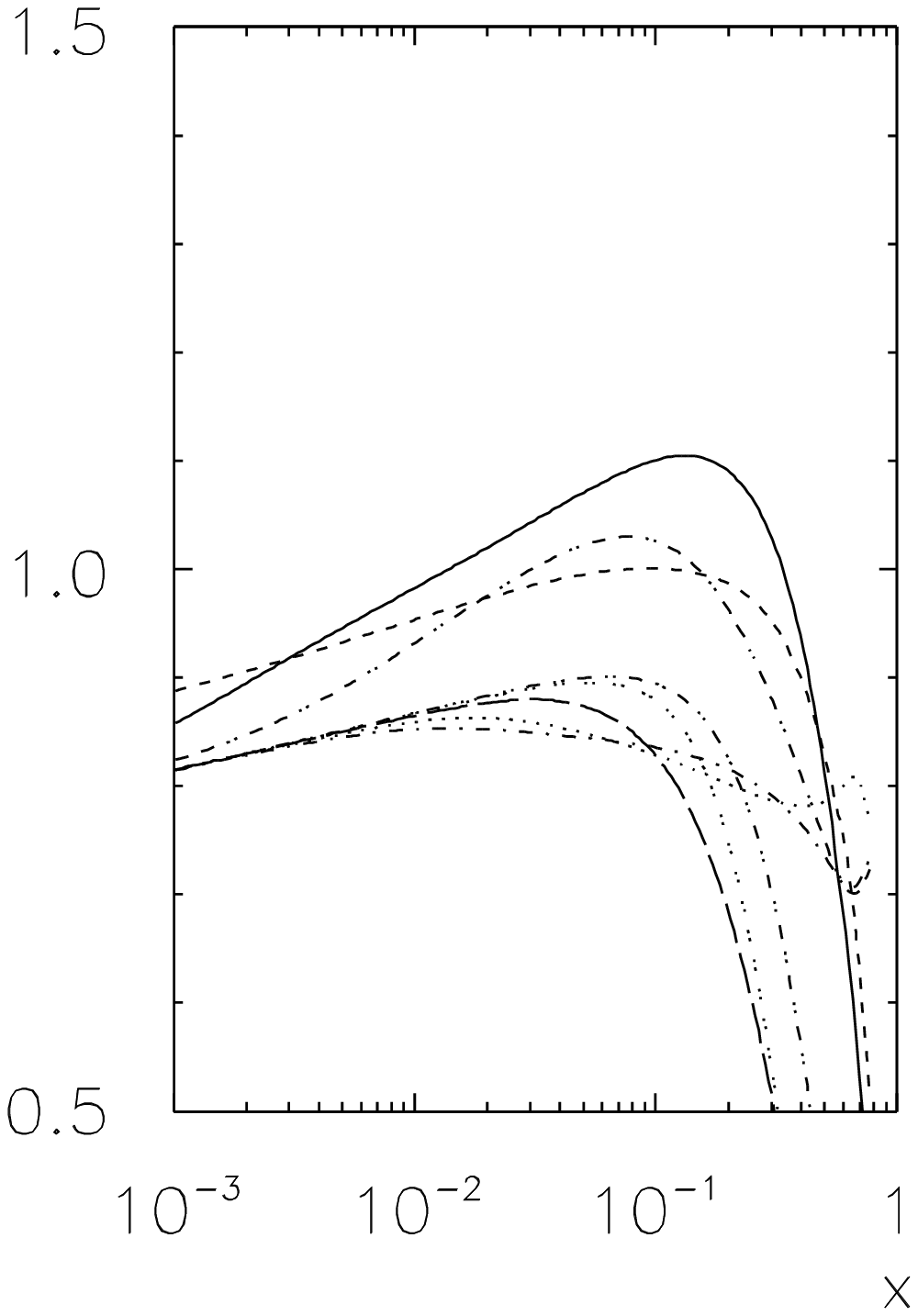}{width=45mm}}
\put(95,0){\lettlab (d)}


\end{picture}
\end{center}
\shiftcaption
\caption[Dependence of Perturbative
Heavy-Quark Target Fragmentation Functions
on the Input Parton Density]
{\labelmm{HQTFFPD} {\it 
Ratios of 
target fragmentation functions $\int\dd z\,M^{(P)}_{i,Q/P}(x,z,\mu^2)$
for different input parton densities: 
CTEQ, leading order (CTEQ~3L) \cite{107} (a), (c)
and MRS, next-to-leading order (MRS~(G)) \cite{108} (b), (d), 
divided by GRV, leading order \cite{106,105}, 
for the
bottom (a), (b) and
charm (c), (d) quark at $\mu=100\,\GeV$.
The input distribution $M^{(P)}$ is zero 
at $\mu_0=m$, and
$z$~is integrated from $0.1$ to $1-x$.
The flavours~$i$ are given by 
$\overline{Q}$ \mbox{(\fullline)}, 
$g$ \mbox{(\dashline)}, 
$d$ \mbox{(\dotline)},
$\overline{d}$ \mbox{(\longdashline)},
$u$ \mbox{(\dashdotline)},
$\overline{u}$ \mbox{(\dotdotline)},
$Q$ \mbox{(\dashdotdotline)},
$F_2^{M}$ \mbox{(\dashdashdotdotline)}.
}}   
\end{figure}

Finally, we show the dependence on the input parton density
in Fig.~\ref{HQTFFPD}. Except for large differences at very large~$x$,
$x\gtrsim 0.2$, 
where the values of $M^{(P)}$ are small, the
uncertainty is of the order of $20-40\%$.

\clearpage

\dgsb{Intrinsic Heavy Quarks}
\labelm{hqtffmic}

The hypothesis of intrinsic heavy quarks in the proton asserts that 
there is a non-vanishing non-perturbative component
$| u u d Q \overline{Q}\rangle$ in a Fock space expansion of the proton
state vector \cite{38,39}. 
For charm quarks, this hypothesis is not in contradiction 
with experimental facts \cite{109}. As a consequence of the large mass
of the heavy quark with respect to the up and down quarks, and 
since
all quarks must have comparable velocities to stay together
in a coherent bound state, 
the fraction of proton momentum carried by the intrinsic heavy
quarks should be comparably
large, eventually 
giving rise to a substantial contribution to $f_{Q/P}(x)$ at large~$x$.
So far, analyses of an intrinsic heavy quark content, 
based on experimental data, are available
for charmed mesons in the final state
in the current fragmentation region:
see for example Ref.\ \cite{109} and
references therein. If the charm quark in these mesons 
comes from the intrinsic component, then it was the quark that initiated 
the hard scattering process, as shown in Fig.~\ref{ihqpic}a.
If the intrinsic component is not negligible,
then there is another region in phase space where it is reasonable to look
for intrinsic heavy quarks, namely the target fragmentation region
\cite{110}. 
In the parton model, 
the scattering process at large $Q^2$ breaks up the proton, and any of the
valence quarks, the heavy antiquark~$\overline{Q}$, or sea partons
may take part in the hard scattering
process. The intrinsic heavy quark~$Q$ is liberated, it fragments and is seen 
in the target fragmentation region, as indicated in Fig.~\ref{ihqpic}b. 
In the case of charm-quark production, it is 
expected to
recombine preferably with valence or sea quarks from the proton
remnant into $D$~mesons and $\Lambda_c$~baryons; the latter,
owing to a leading-particle effect,
occur more frequently than $\overline{\Lambda_c}$~baryons
because of the abundance of~$u$ and~$d$ valence quarks with respect 
to~$\overline{u}$ and~$\overline{d}$ sea quarks.

\begin{figure}[htb] \unitlength 1mm
\begin{center}
\dgpicture{159}{174}

\put(15,100){\epsfigdg{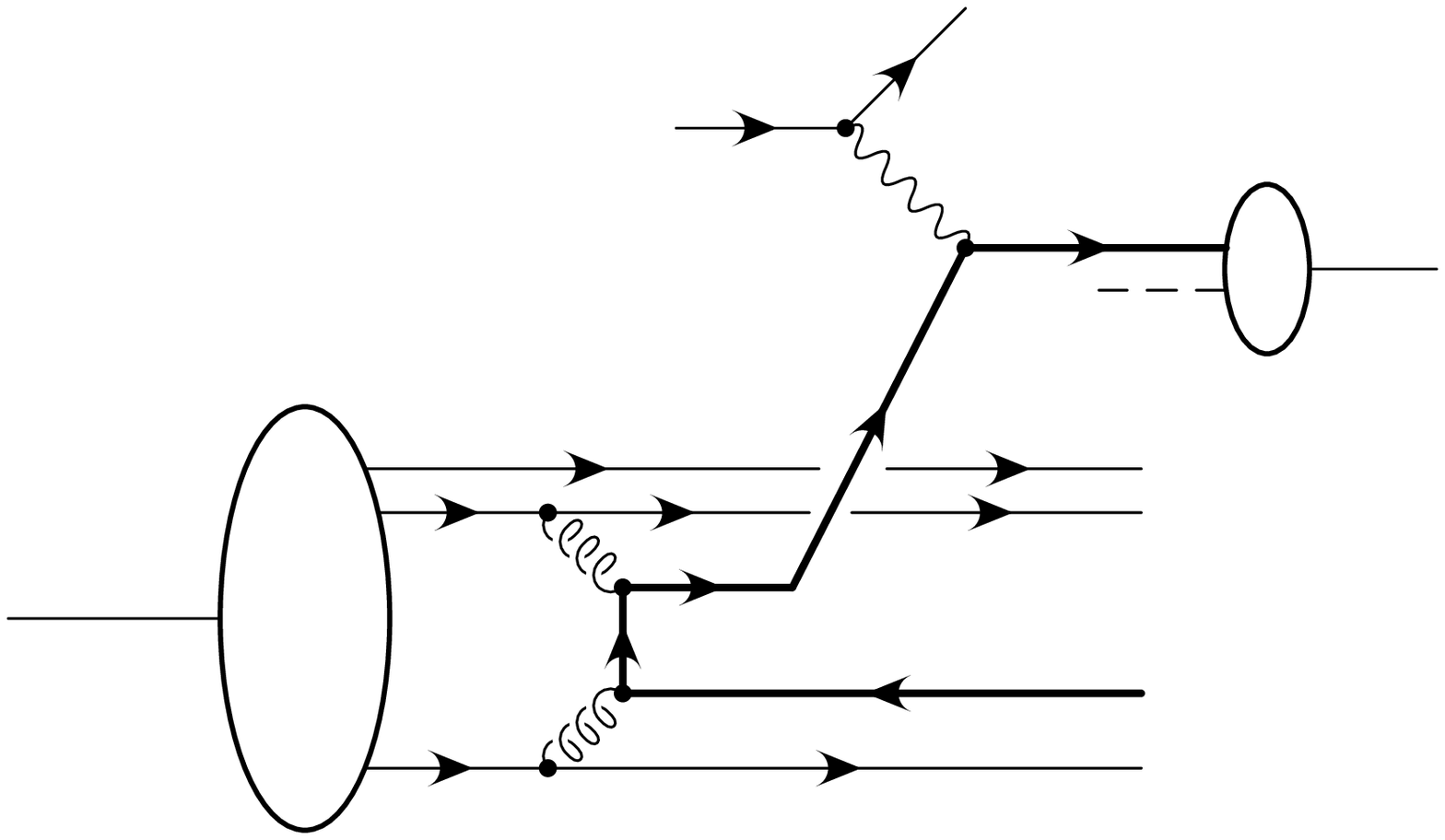}{width=120mm}}
\put(20,90){\lettlab (a)}

\put(15,10){\epsfigdg{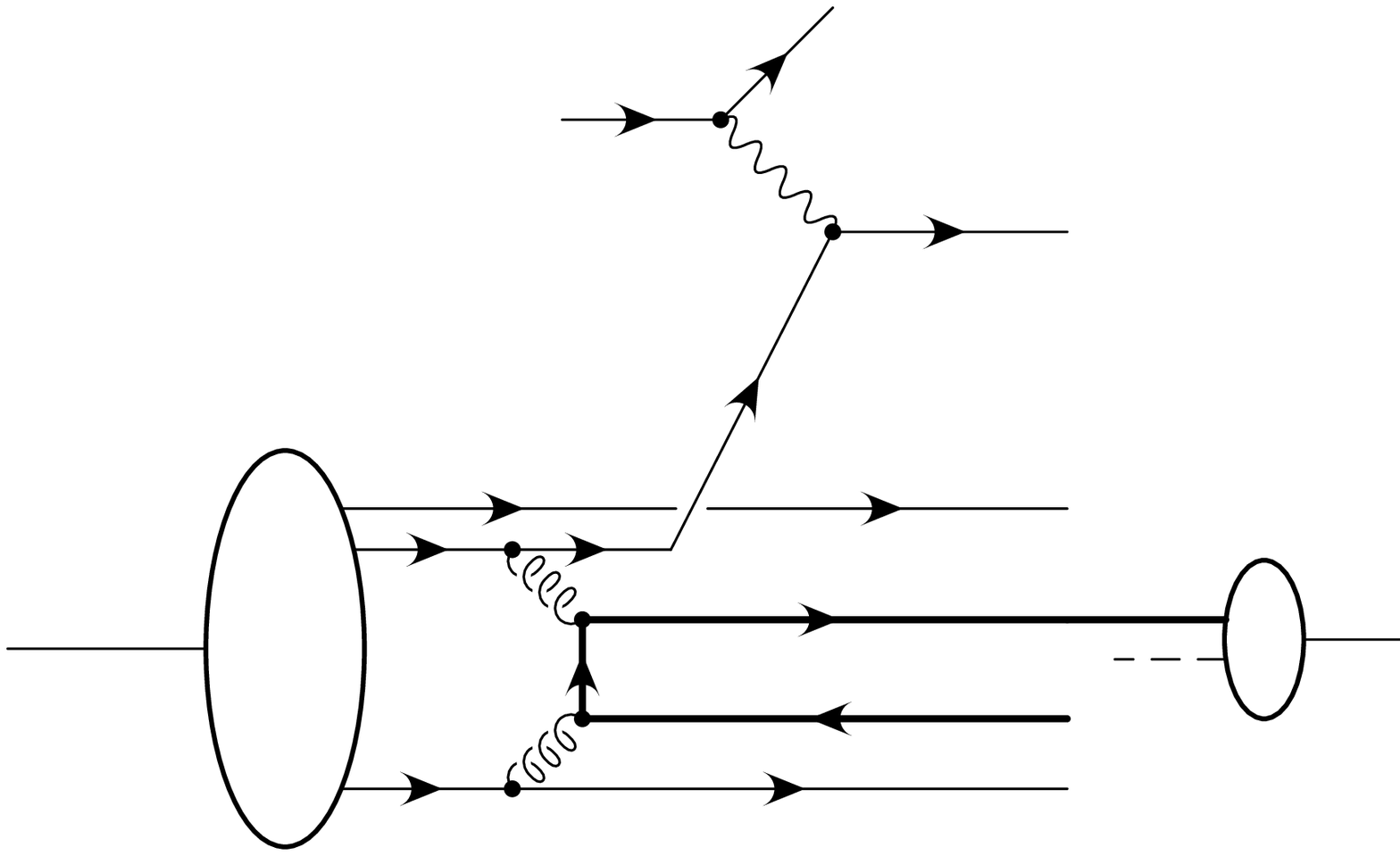}{width=125mm}}
\put(20,0){\lettlab (b)}

\put(10,116){\LARGE\it P}
\put(68,157){\LARGE\it l}
\put(98,168){\LARGE\it l$\,^\prime$}
\put(109,152){\LARGE\it Q}
\put(95,155){\LARGE $\gamma^*$}

\put(10,26){\LARGE\it P}
\put(62,74){\LARGE\it l}
\put(91,84){\LARGE\it l$\,^\prime$}
\put(116,33){\LARGE\it Q}
\put(88,72){\LARGE $\gamma^*$}

\end{picture}
\end{center}
\shiftcaption
\caption[Intrinsic Heavy-Quark Production]
{\labelmm{ihqpic} {\it Intrinsic heavy-quark production
in the current (a) and target (b) fragmentation regions.
The solid line stands for the heavy quark that subsequently
fragments into a hadron.
}}   
\end{figure}

To study this process semi-quantitatively, we employ a simple model
distribution as an example of a non-perturbative 
contribution to a target fragmentation function.
Following Ref.~\cite{38}, we write the correlated distribution function
for the momentum distribution of three light valence quarks $uud$ and
a pair $Q\overline{Q}$ of heavy quarks with momentum fractions
$x_1$, $x_2$, $x_3$ and $x_Q$, $x_{\overline Q}$, respectively, as
\beqm{momdistric}
\frac{\dd P}{\dd x_1 \dd x_2 \dd x_3 \dd x_Q \dd x_{\overline{Q}}}
=36\,\beta\,\frac{x_Q^2 x_{\overline{Q}}^2}{(x_Q+x_{\overline{Q}})^2}
\,\delta\left(1-x_1-x_2-x_3-x_Q-x_{\overline{Q}}\right),
\eeq
where the momentum distribution is symmetric in the momentum fraction 
variables $x_1$, $x_2$ and $x_3$ of the three light quarks.
The normalization factor $\beta$ is not predicted by the model. Instead, 
it has to be obtained from experiment. The value $\beta=1$ 
is a reasonable
value 
for intrinsic charm quarks, close to what is obtained 
from a fit to experimental data \cite{109}.
The distributions should scale as $1/m_Q^2$ with the heavy-quark 
mass. Thus, $\beta=0.1$ is a reasonable assumption for the
bottom quark. 

In Refs.\ \cite{38,39}, the expression in Eq.~(\ref{momdistric})
is integrated over the variables~$x_1$, $x_2$, $x_3$ and 
$x_{\overline{Q}}$ in order to obtain
the single particle distribution for intrinsic heavy quarks in the proton:
\beqm{spdhq}
f^{\IHQ}_{Q/P}(x)=36\,\beta\,\frac{x^2}{2}
\left[
\frac{1}{3}(1+10x+x^2)(1-x)+2x(1+x)\ln x
\right].
\eeq
A normalization factor of $\beta=1$
actually corresponds to an intrinsic heavy-quark content of 
$\int_0^1\dd x\,f^{\IHQ}_{Q/P}(x)=1\%$.
The resulting distributions, evolved in the factorization scale~$\mu$
with the input from Eq.~(\ref{spdhq}) at $\mu_0=m$, are shown in 
Fig.~\ref{HQTPDEN}. Compared with standard parton densities,
cf.\ Fig.~\ref{HQTFFP},
the heavy-quark content is fairly large and peaked at large values of~$x$
of about~$0.3$.

Here, instead, we integrate over only three of the variables
$x_1$, $x_2$, $x_3$, $x_Q$, $x_{\overline{Q}}$, leaving
a pair $(x,z)$ of variables in the distribution 
function. The result
is interpreted as a correlated distribution function 
for a heavy quark~$Q$ and another quark (either a valence quark or the 
heavy antiquark~$\overline{Q}$), i.e.\ a target fragmentation 
function in leading 
order. Here $x=x_1$, $x_2$, $x_3$, $x_{\overline{Q}}$ 
is the momentum fraction of the
quark initiating the hard process, and $z=x_Q$ is the momentum fraction 
of the heavy quark detected in the proton remnant.
For simplicity, we neglect all mass effects\footnote{
A complete next-to-leading-order analysis
of $f^{\sIHQ}$, including mass effects, has been done in 
Ref.~\cite{109}. A~comparable analysis of $M^{\sIHQ}$ is beyond the scope
of the present paper.}, and assume that
the hard scattering process, possibly involving neither of 
the heavy quarks, breaks the coherence of the $Q\overline{Q}$ 
pair and allows them to propagate into the final state without 
recombination. For a detailed discussion, see Ref.\ \cite{110}.
The numerical results 
obtained in this section therefore cannot be more than a crude estimate
for a simple model. They give, however, a good idea of the qualitative 
features 
for the case of intrinsic heavy quarks in the proton.

Performing the integrations, the final result for the 
target fragmentation functions
is 
\beqnm{icfrac}
M^\IHQ_{d,Q/P}(x,z)&=&36\,\beta\,z^2\,\left[
\frac{1}{2}(1-x)^2+2z(1-x)-\frac{5}{2}z^2
+(2z(1-x)+z^2)\ln\frac{z}{1-x}
\right],\nonu
M^\IHQ_{u,Q/P}(x,z)&=&2\,M_{d,Q/P}(x,z),\nonu
M^\IHQ_{\overline{Q},Q/P}(x,z)&=&36\,\beta\,z^2\,
\frac{1}{2}\frac{x^2}{(x+z)^2}(1-x-z)^2, 
\eeqn
these expressions being valid for a factorization scale of the
order of the heavy-quark mass\footnote{S.~Brodsky, private communication.
The input distribution scales with $m$ 
as $1/m^2$ for factorization scales $\mu$ larger
than $m$, whilst 
for $\mu<m$, the distribution is suppressed by a resolution factor
and scales as $1/m^4$.}.
We therefore use them as input distributions 
for a factorization scale of $\mu_0=m$ 
for the
homogeneous part of the evolution equation (\ref{Mrge}),
corresponding to the ``non-perturbative'' distributions.
We wish to note that, in principle, the non-perturbative 
target fragmentation functions should have a dependence,
proportional to $\ln \mu_0^2$, on the unphysical scale $\mu_0$, 
in order to cancel a corresponding
term from the evolution equation of the perturbative 
target fragmentation functions. We assume that the exact
cancellation takes place at $\mu_0=m$.

The $x$-distributions of target fragmentation functions, 
for~$z$ integrated from $0.1$ to $1-x$, are shown for
a set of factorization scales in Fig.~\ref{HQTFFSIC}.
They are of the same order of magnitude as the 
perturbative heavy-quark target fragmentation functions,
cf.\ Fig.~\ref{HQTFFA}.
The $z$-distributions 
are shown in Fig.~\ref{HQTFFZIC}.
It is obvious that the $z$-distribution is harder than in the case of 
the perturbative functions, cf.\ Figs.~\ref{HQTFFZb} and~\ref{HQTFFZc}.
 
\begin{figure}[htb] \unitlength 1mm
\begin{center}
\dgpicture{159}{167}

\put(0,93){\epsfigdg{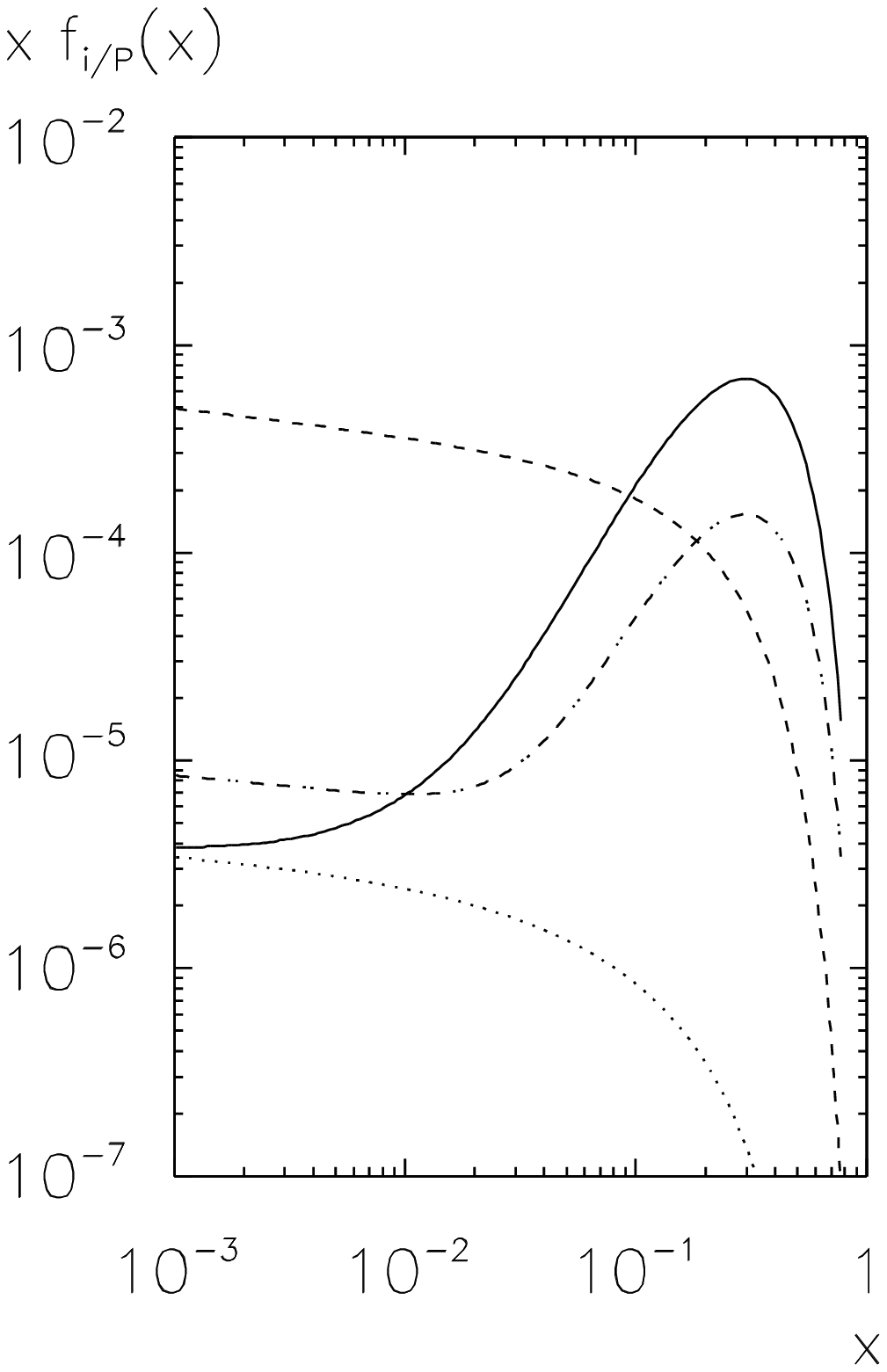}{width=45mm}}
\put(10,85){\lettlab (a)}

\put(55,93){\epsfigdg{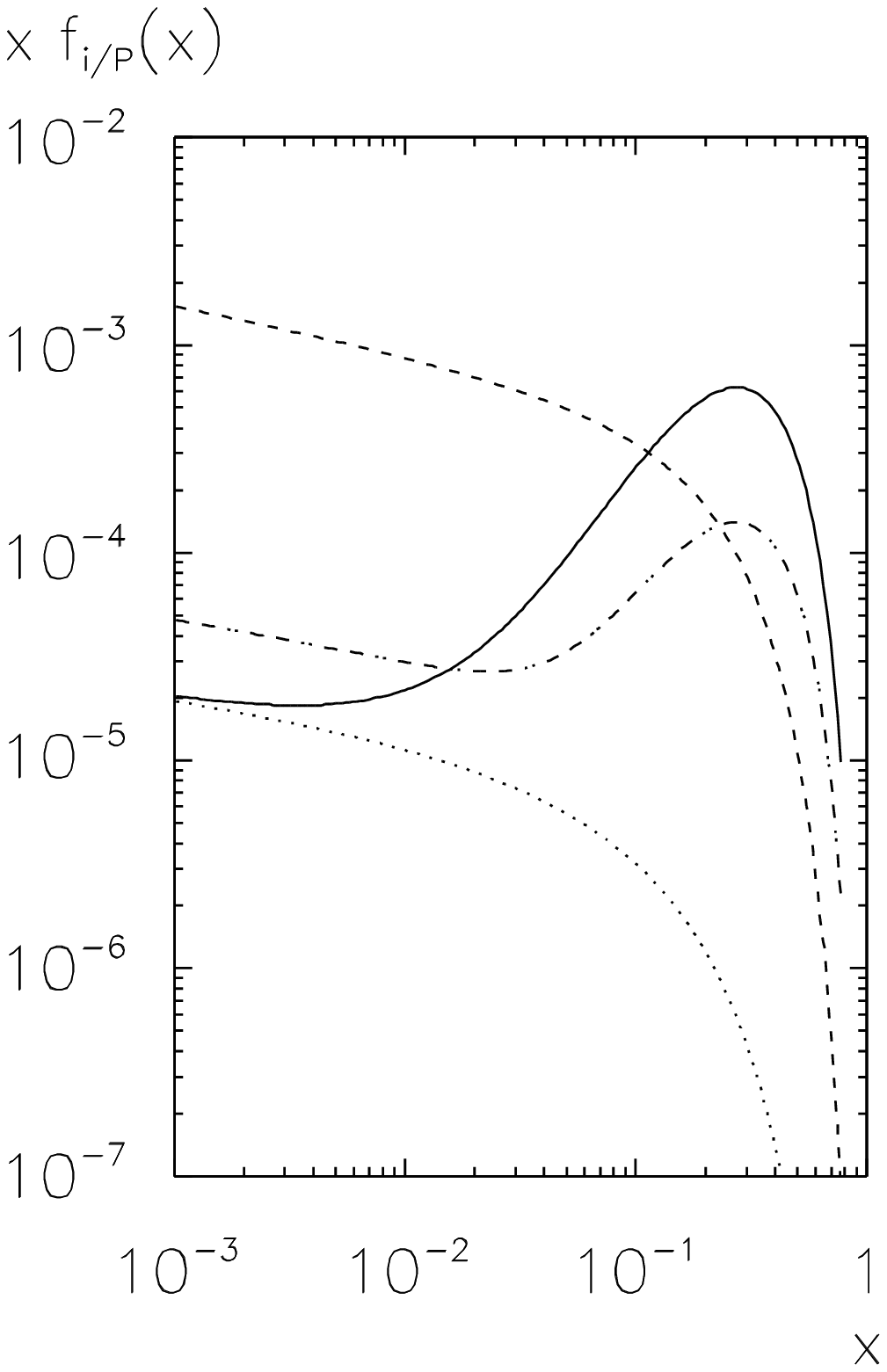}{width=45mm}}
\put(65,85){\lettlab (b)}

\put(110,93){\epsfigdg{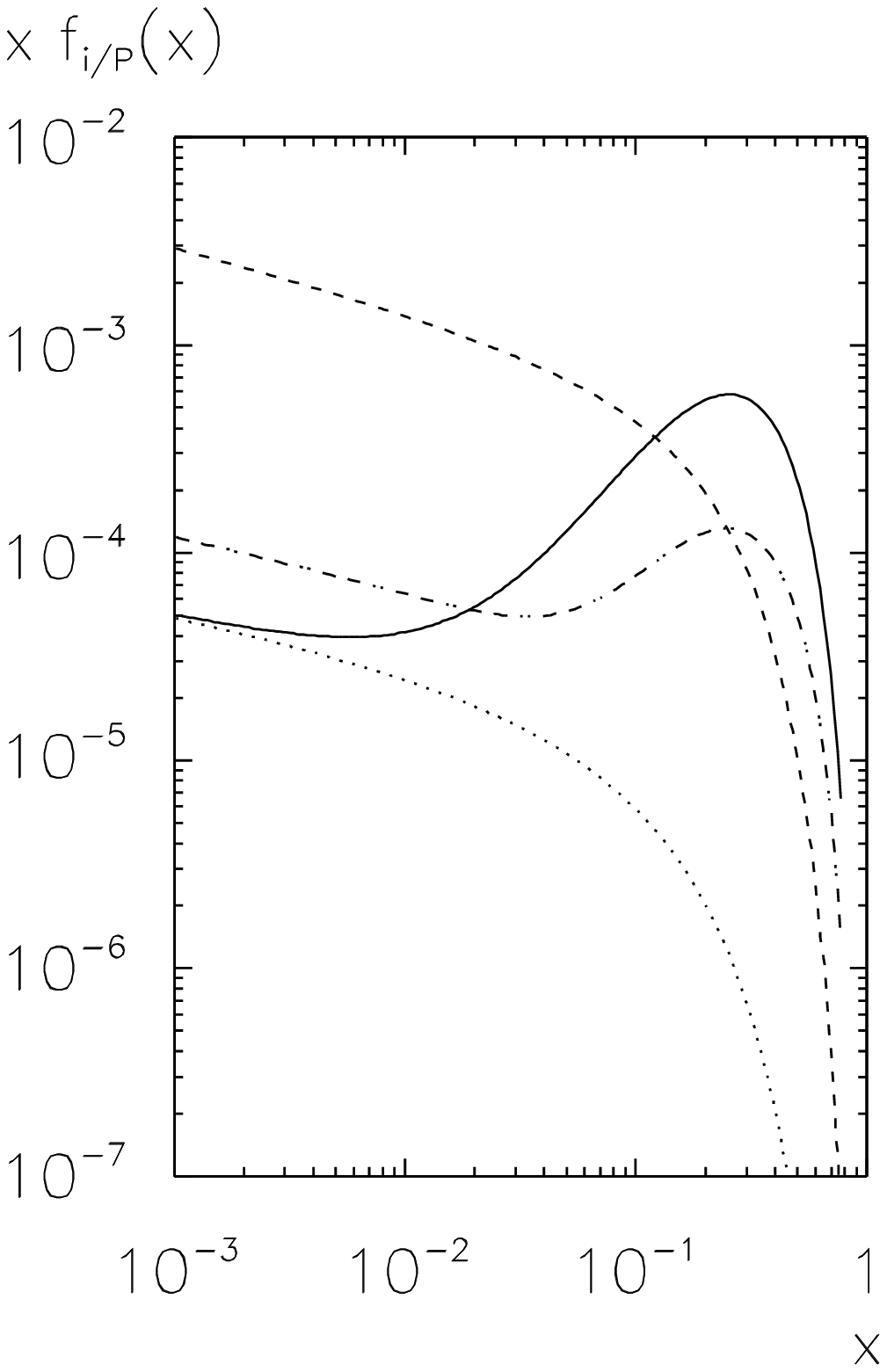}{width=45mm}}
\put(120,85){\lettlab (c)}

\put(0,8){\epsfigdg{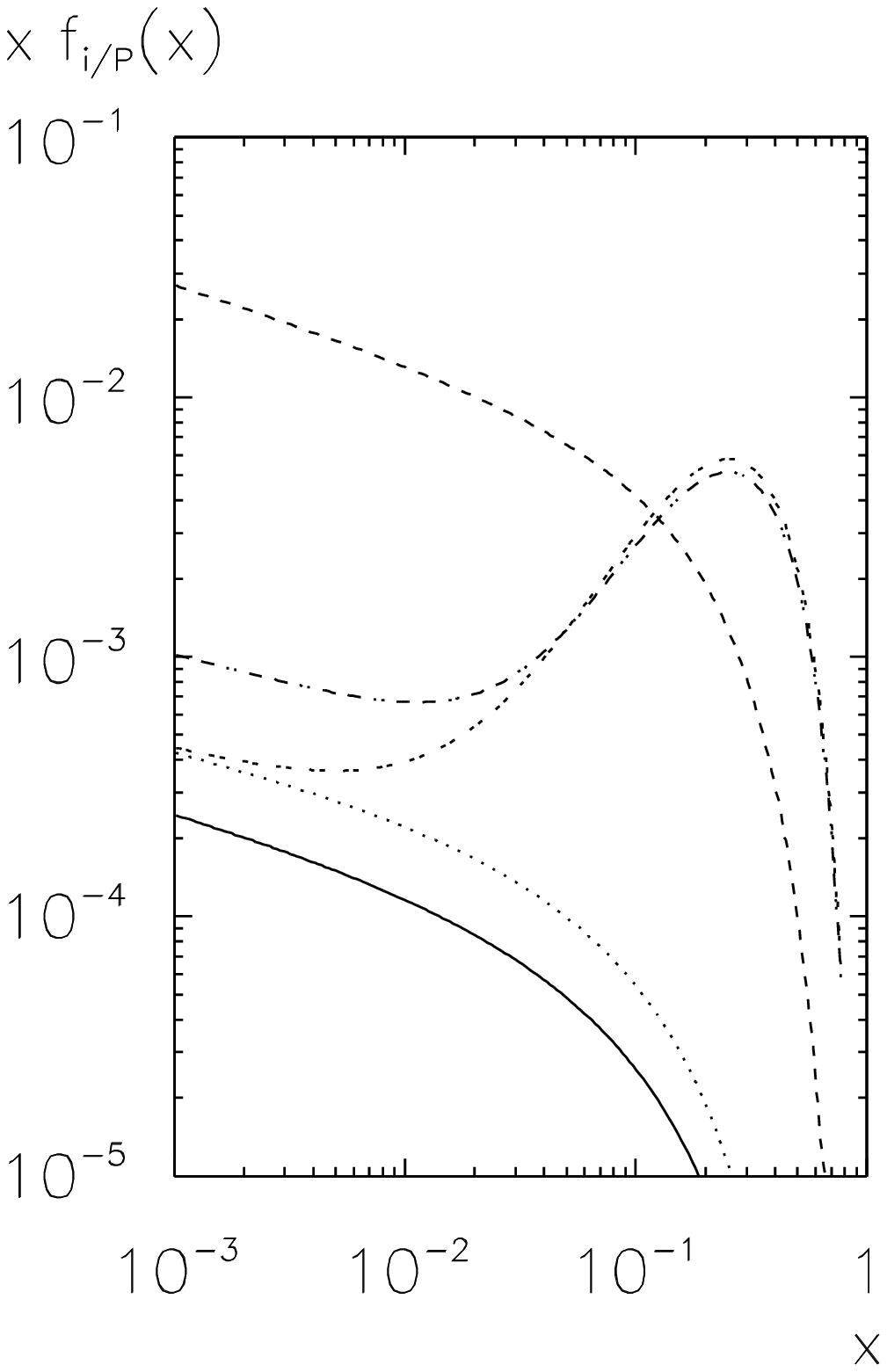}{width=45mm}}
\put(10,0){\lettlab (d)}

\put(55,8){\epsfigdg{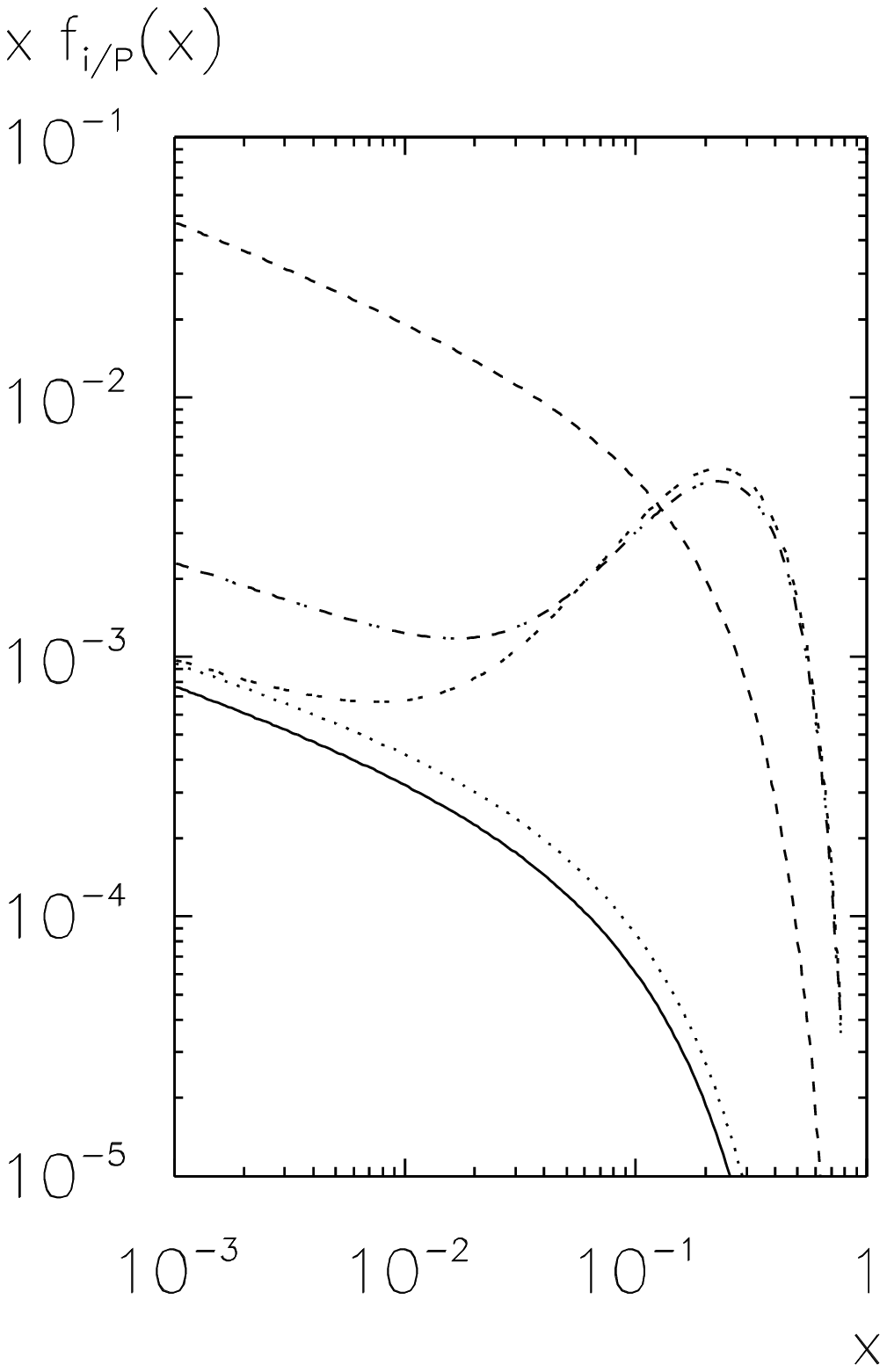}{width=45mm}}
\put(65,0){\lettlab (e)}

\put(110,8){\epsfigdg{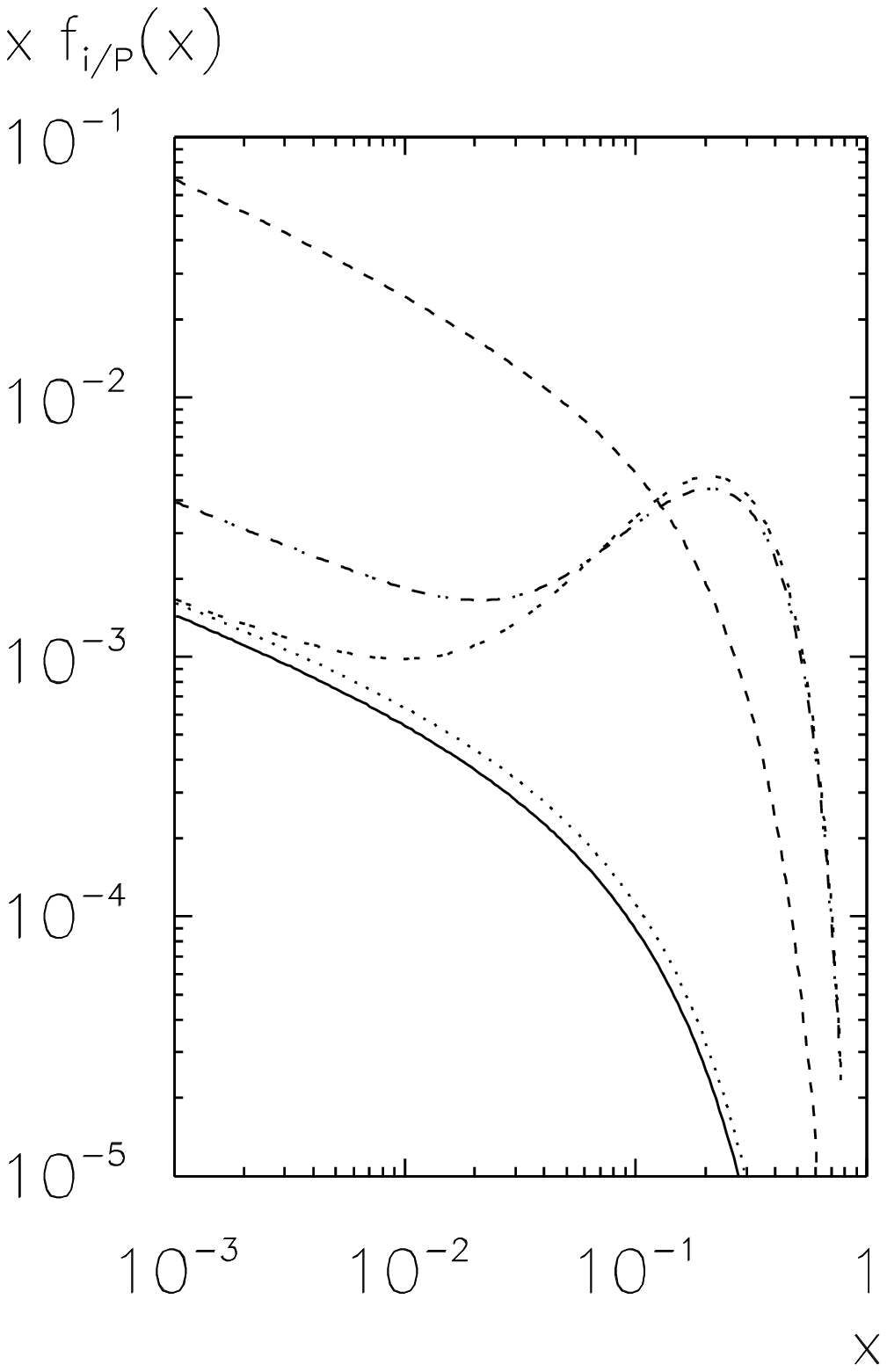}{width=45mm}}
\put(120,0){\lettlab (f)}

\end{picture}
\end{center}
\shiftcaption
\caption[Scale Evolution of Parton Densities for Intrinsic Heavy Quarks]
{\labelmm{HQTPDEN} 
{\it Scale evolution of the parton densities $x\,f^\IHQ_{i/P}(x,\mu^2)$ 
based on intrinsic
bottom (a)--(c) and charm (d)--(f) quark distributions.
The factorization scale is 
$\mu=10\,\GeV$ (a), (d);
$\mu=30\,\GeV$ (b), (e);
$\mu=100\,\GeV$ (c), (f).
The flavours~$i$ are given by 
$b$, $\overline{b}$ \mbox{(\fullline)}; 
$c$, $\overline{c}$ \mbox{(\dashdashline)}; 
$g$ \mbox{(\dashline)};
light quarks \mbox{(\dotline)};
$F_2$ \mbox{(\dashdashdotdotline)}.
}}   
\end{figure}

\begin{figure}[htb] \unitlength 1mm
\begin{center}
\dgpicture{159}{167}

\put(0,93){\epsfigdg{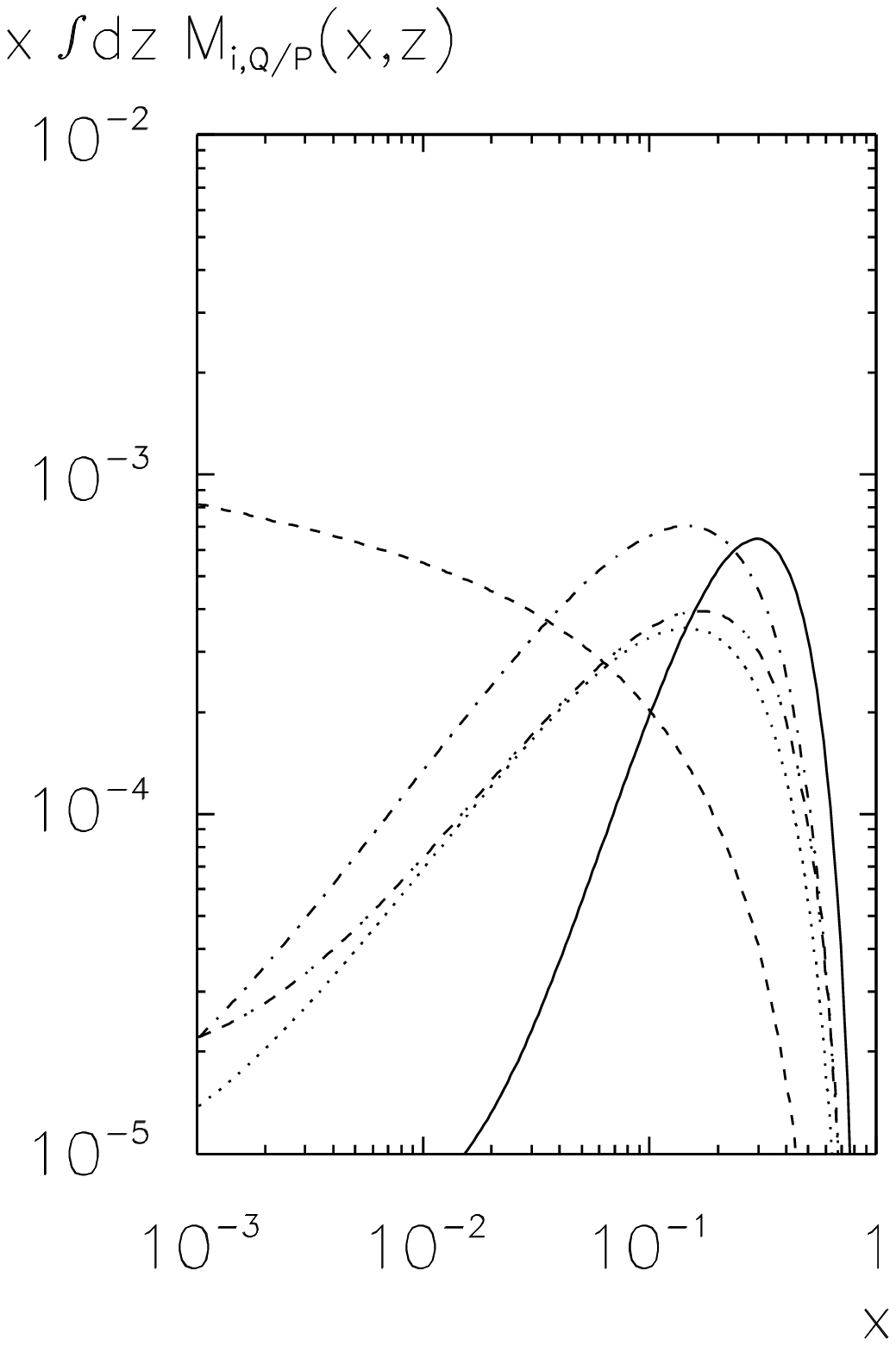}{width=45mm}}
\put(10,85){\lettlab (a)}

\put(55,93){\epsfigdg{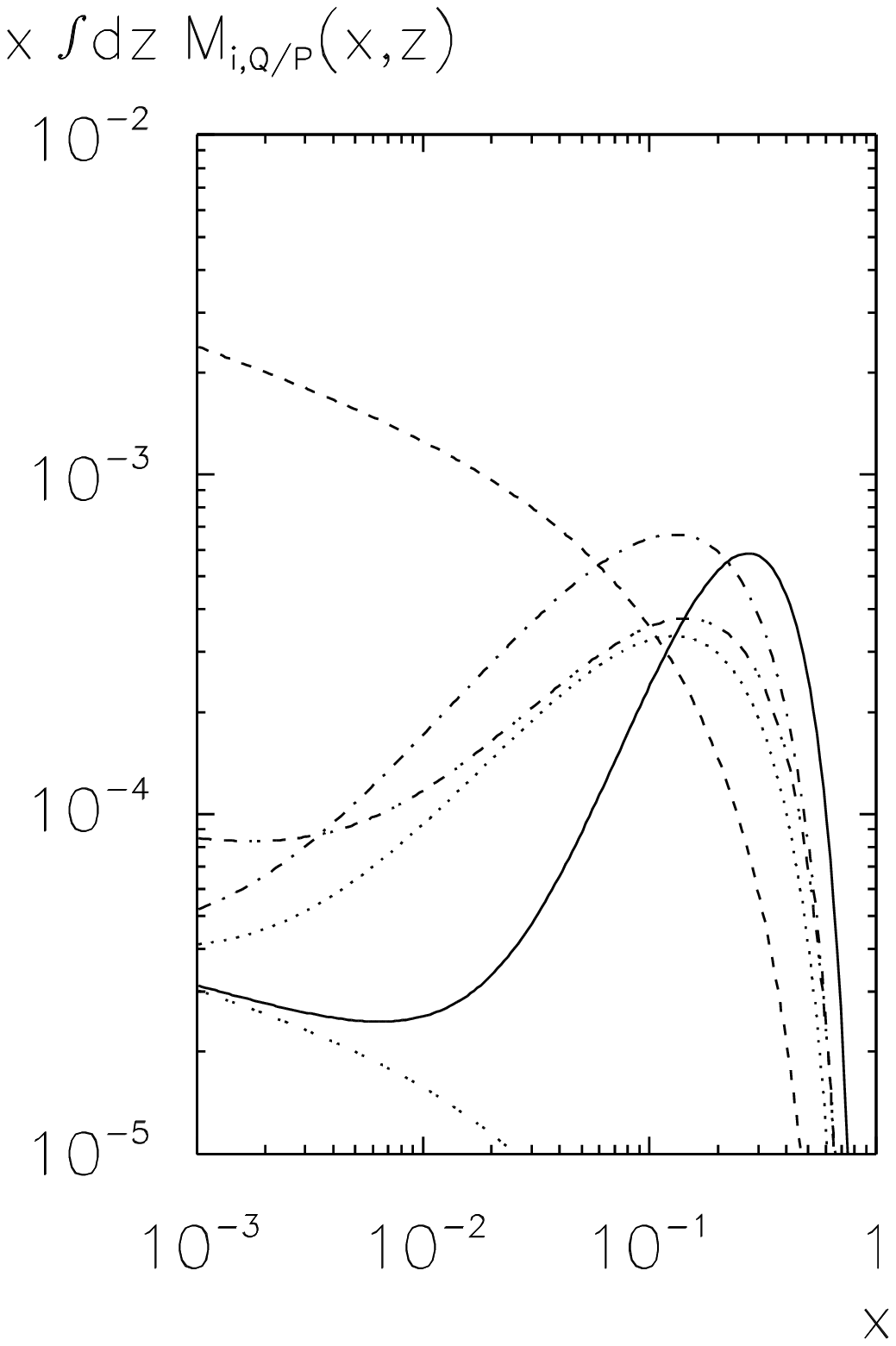}{width=45mm}}
\put(65,85){\lettlab (b)}

\put(110,93){\epsfigdg{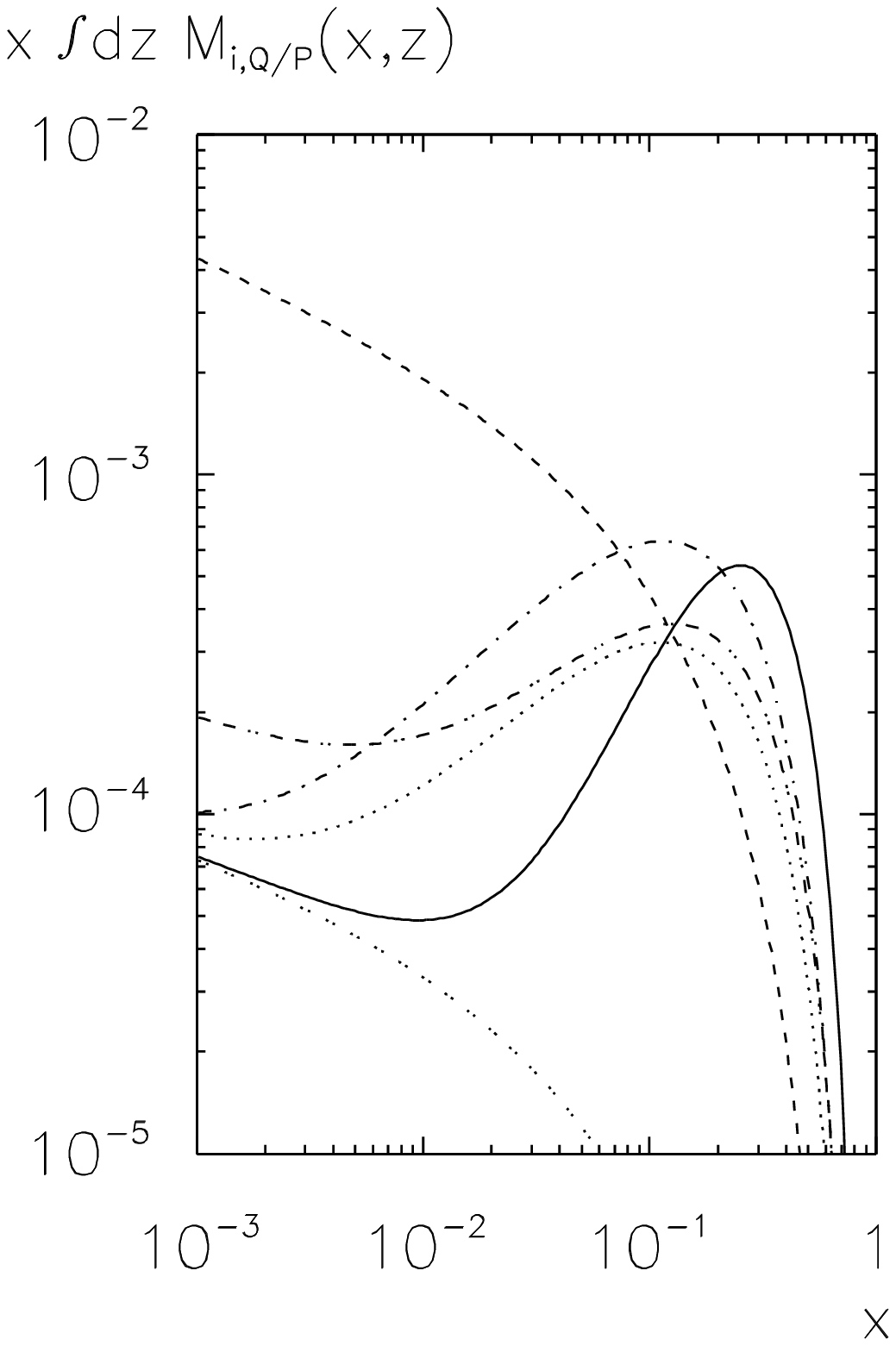}{width=45mm}}
\put(120,85){\lettlab (c)}

\put(0,8){\epsfigdg{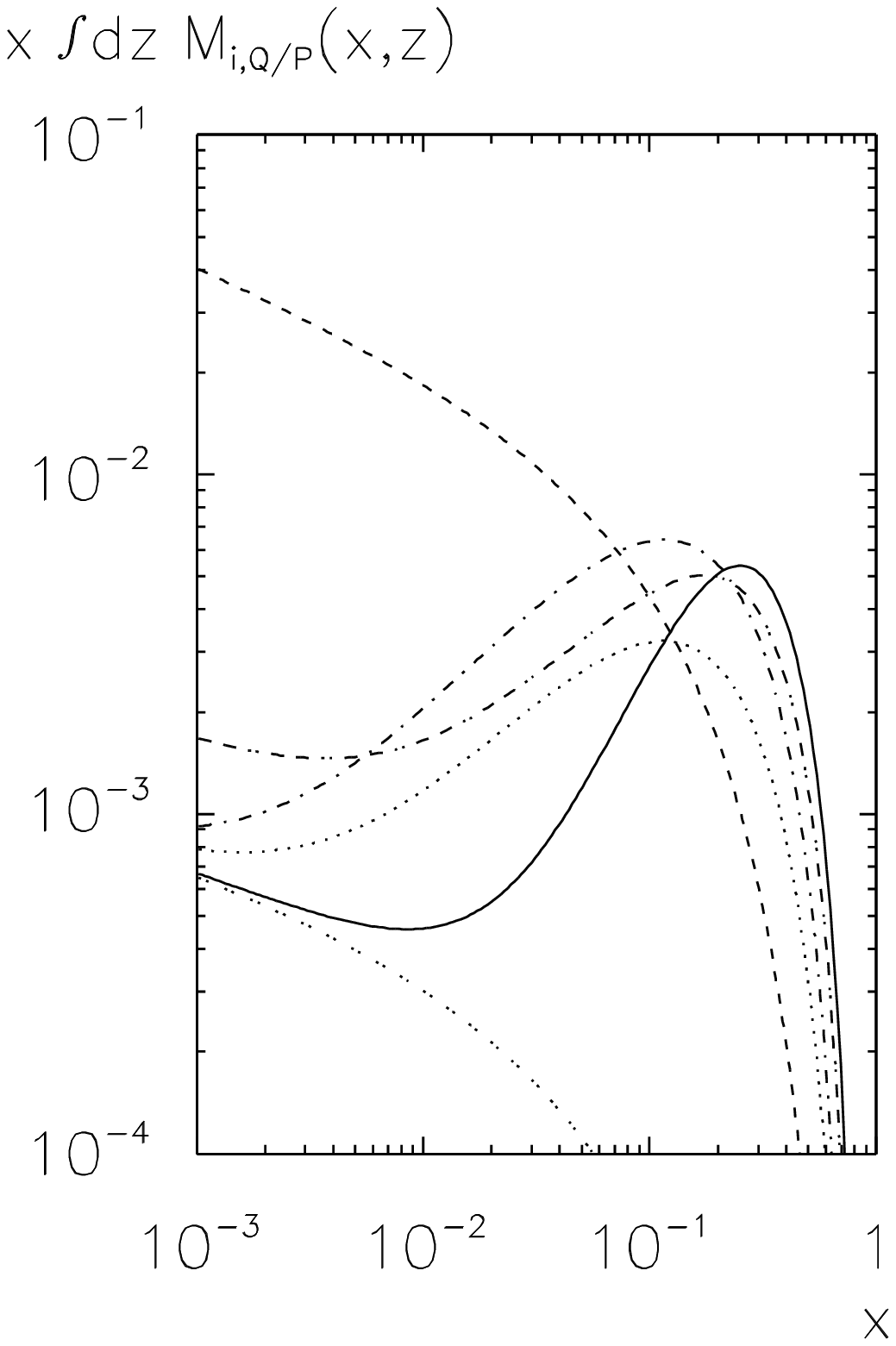}{width=45mm}}
\put(10,0){\lettlab (d)}

\put(55,8){\epsfigdg{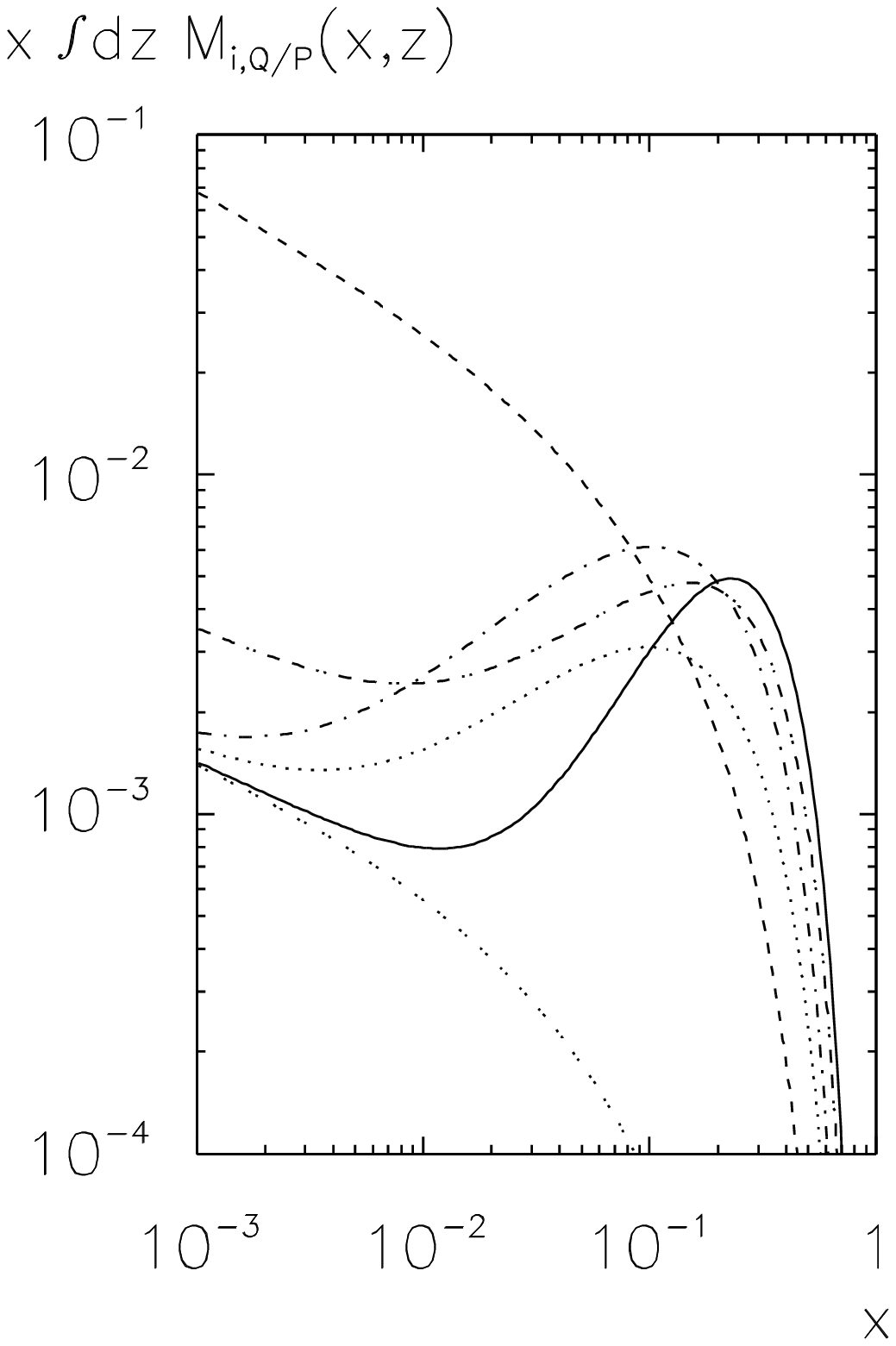}{width=45mm}}
\put(65,0){\lettlab (e)}

\put(110,8){\epsfigdg{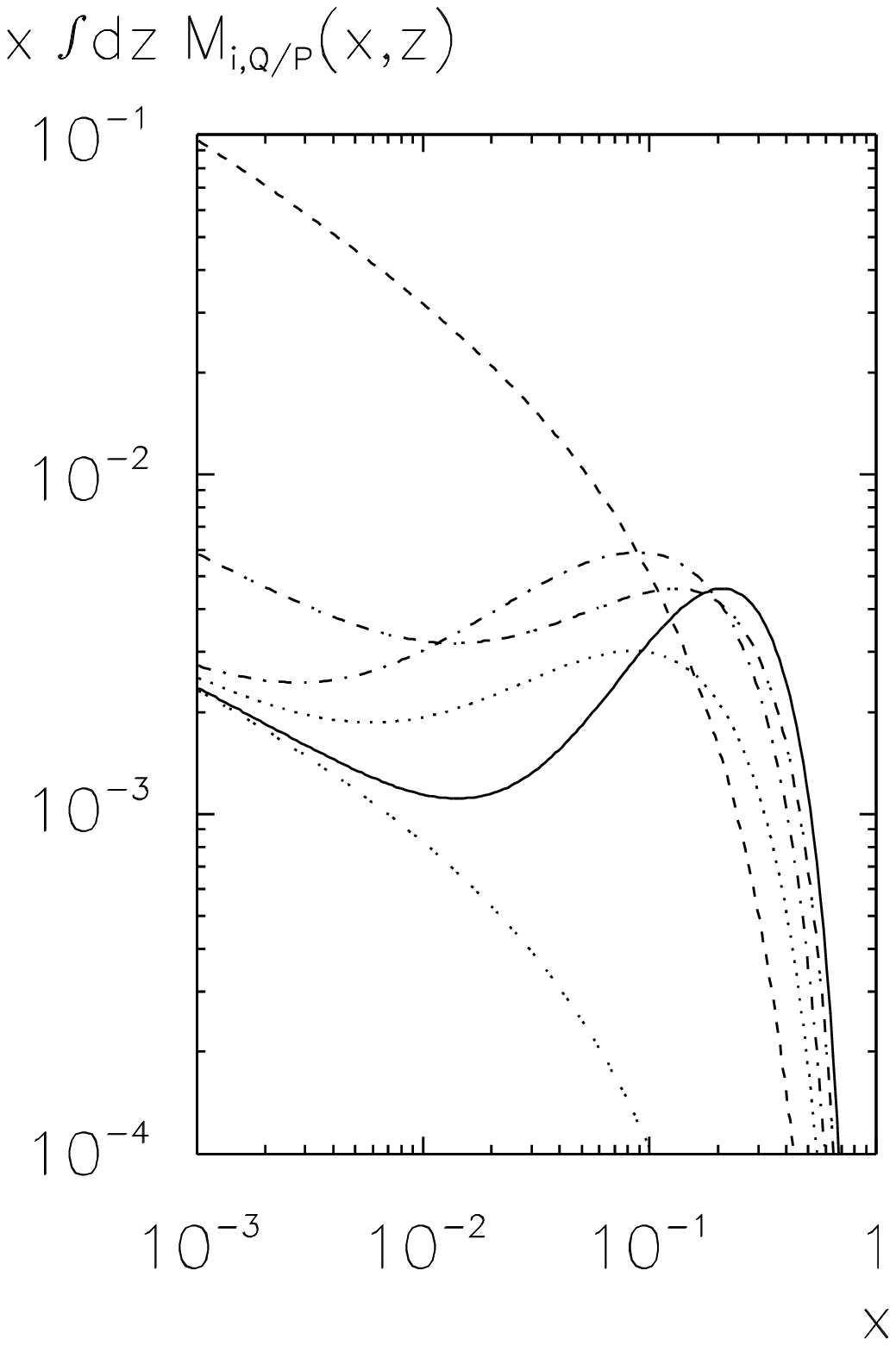}{width=45mm}}
\put(120,0){\lettlab (f)}

\end{picture}
\end{center}
\shiftcaption
\caption[Scale Evolution of
Target Fragmentation Functions
for Intrinsic Heavy Quarks]
{\labelmm{HQTFFSIC} {\it Scale evolution of 
target fragmentation functions $x\,\int\dd z\,M^\IHQ_{i,Q/P}(x,z,\mu^2)$
for intrinsic bottom (a)--(c) and charm (d)--(f) quarks, where
$z$ is integrated from $0.1$ to $1-x$.
The factorization scale is 
$\mu=10\,\GeV$ (a), (d);
$\mu=30\,\GeV$ (b), (e);
$\mu=100\,\GeV$ (c), (f).
The input distribution is defined 
at $\mu_0=m$. 
The flavours~$i$ are given by 
$\overline{Q}$ \mbox{(\fullline)};
$g$ \mbox{(\dashline)};
$d$ \mbox{(\dotline)};
$u$ \mbox{(\dashdotline)};
$\overline{u}$, $\overline{d}$ \mbox{(\dotdotline)};
$F_2^{M}$ \mbox{(\dashdashdotdotline)}.
}}   
\end{figure}

\begin{figure}[htb] \unitlength 1mm
\begin{center}
\dgpicture{159}{167}

\put(0,93){\epsfigdg{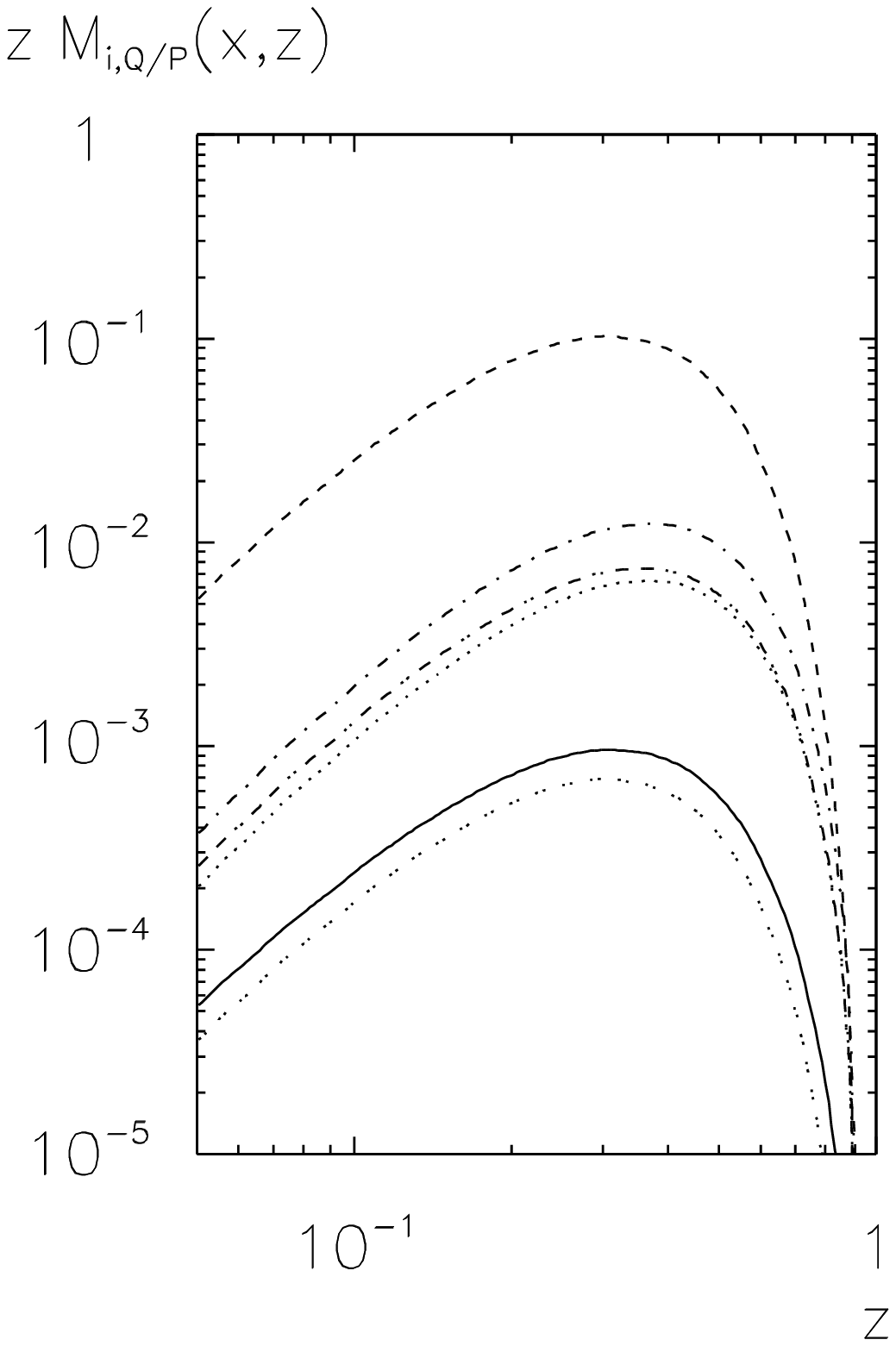}{width=45mm}}
\put(10,85){\lettlab (a)}

\put(55,93){\epsfigdg{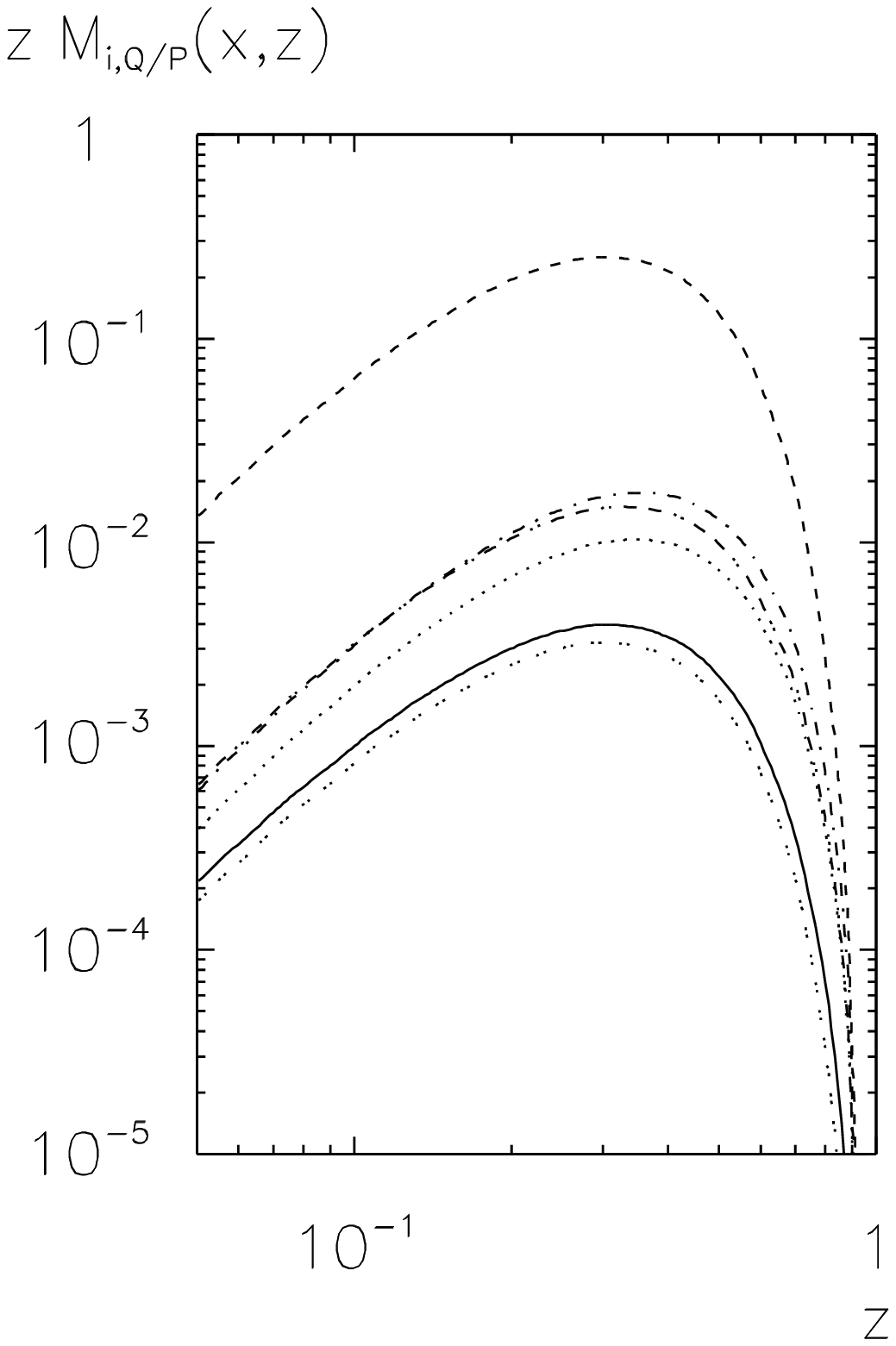}{width=45mm}}
\put(65,85){\lettlab (b)}

\put(110,93){\epsfigdg{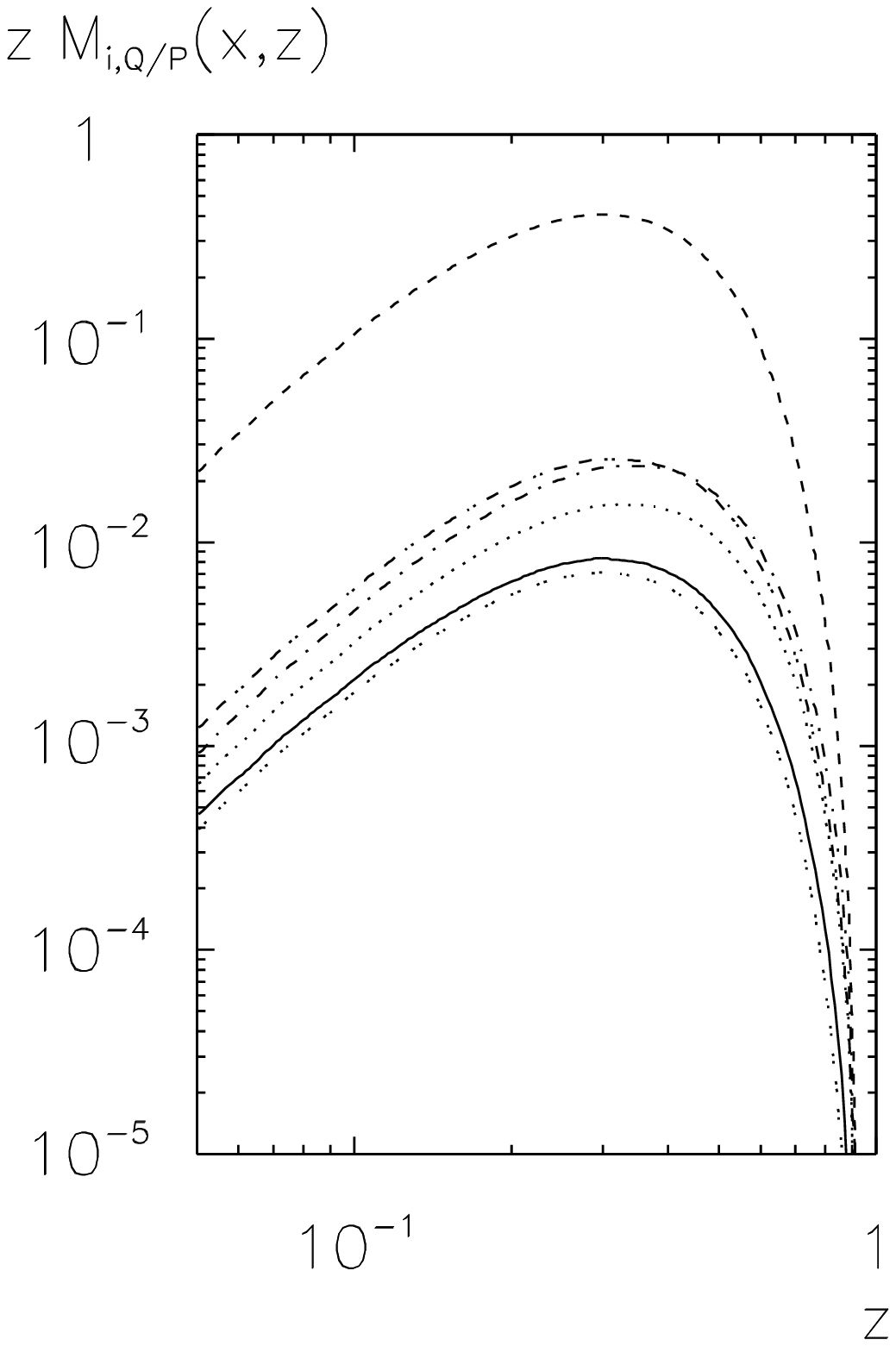}{width=45mm}}
\put(120,85){\lettlab (c)}

\put(0,8){\epsfigdg{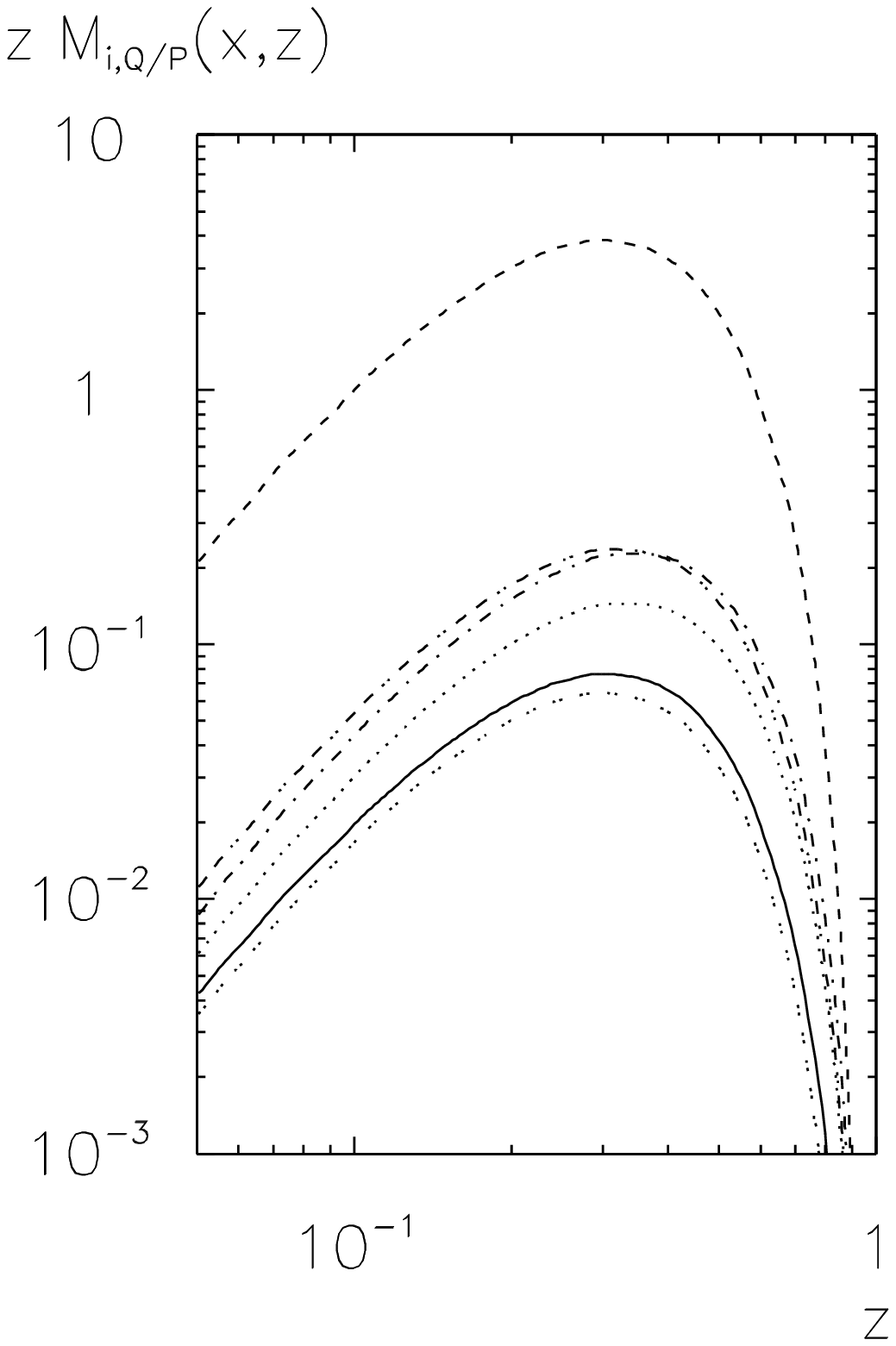}{width=45mm}}
\put(10,0){\lettlab (d)}

\put(55,8){\epsfigdg{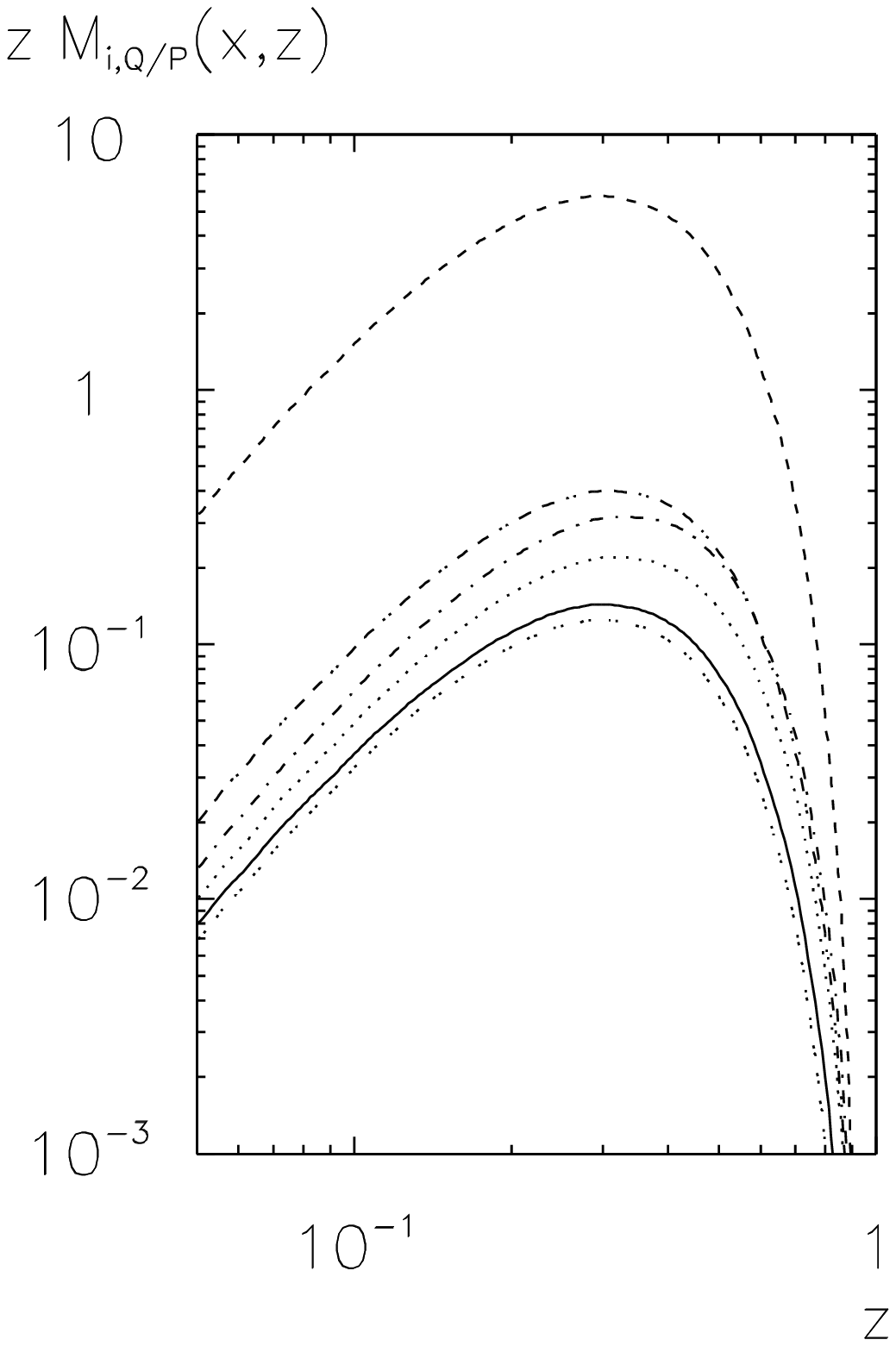}{width=45mm}}
\put(65,0){\lettlab (e)}

\put(110,8){\epsfigdg{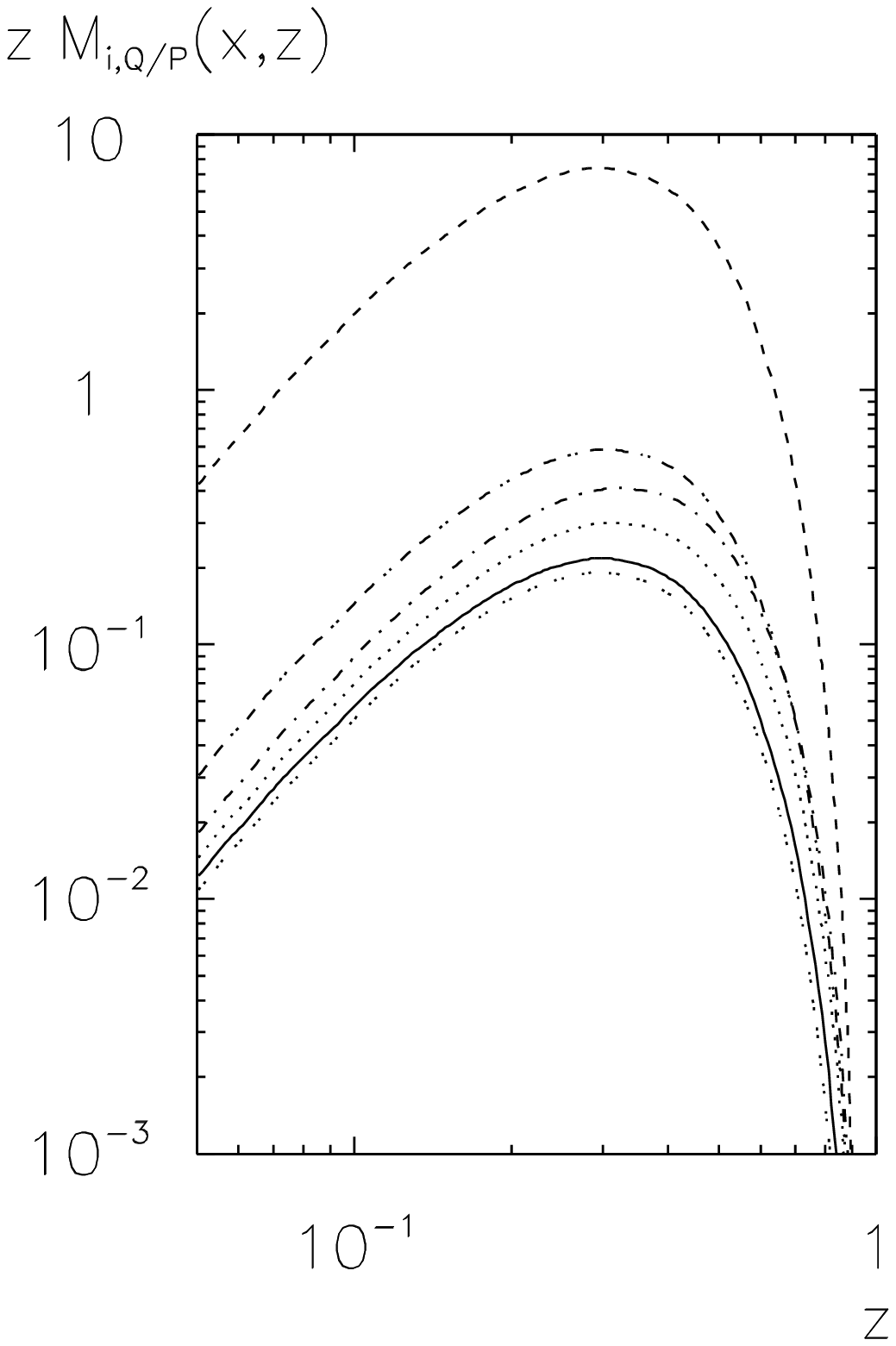}{width=45mm}}
\put(120,0){\lettlab (f)}

\end{picture}
\end{center}
\shiftcaption
\caption[Momentum-Fraction Dependence of
Target Fragmentation Functions for Intrinsic Heavy Quarks]
{\labelmm{HQTFFZIC} {\it Dependence of
target fragmentation functions $z\,M^\IHQ_{i,Q/P}(x,z,\mu^2)$
for intrinsic bottom (a)--(c) and charm (d)--(f) quarks
on the momentum fraction of the observed
heavy quark. The momentum fraction of the parton incident in the
hard subprocess is $x=0.005$.
The factorization scale is
$\mu=10\,\GeV$ (a), (d);
$\mu=30\,\GeV$ (b), (e);
$\mu=100\,\GeV$ (c), (f).
The input distribution for~$M$ is defined
at $\mu_0=m$. 
The flavours~$i$ are given by
$\overline{Q}$ \mbox{(\fullline)};
$g$ \mbox{(\dashline)};
$d$ \mbox{(\dotline)};
$u$ \mbox{(\dashdotline)};
$\overline{u}$, $\overline{d}$ \mbox{(\dotdotline)};
$F_2^{M}$ \mbox{(\dashdashdotdotline)}.
}}   
\end{figure}

\dgcleardoublepage
\markh{A Phenomenological Case Study}
\dgsa{A Phenomenological Case Study}
\labelm{CaseStudy}
{\it 
This section contains the results of a numerical case study for heavy-quark 
production in deeply inelastic lepton--nucleon scattering at
HERA, E665 and NA47. The goal of this study is twofold: (a) to show that
for an actual experimental situation reasonable 
theoretical results may be obtained 
for heavy-quark production in the target fragmentation region, and (b)
to investigate
how the mechanism that cancels singularities works in principle.
Section~\ref{HQPCase} gives a short introduction to the fragmentation
function picture of heavy-quark production as applied in this study.
Section~\ref{ImpStudy} briefly describes how the matrix elements are 
implemented. The parameters for the numerical evaluation
are given in Section~\ref{DiscPar}.
Section~\ref{DiscStudy} serves to give an overview of
cross sections in various phase-space regions.
Transverse momentum distributions and the 
subtraction process in the collinear phase-space regions
are investigated in Section~\ref{DiscSub}.
A comparison of distributions in~$x_F$ in leading and next-to-leading 
order, including the case of intrinsic heavy quarks,
is done in Section~\ref{XFSTUDY}.
The scale dependence of the cross sections is investigated in 
Section~\ref{FRDEP}.
Our results indicate 
that, in deeply inelastic scattering, taking into 
account the fragmentation of the heavy quarks into observable
mesons, the cross section is too small to allow for an experimental
determination of the heavy-quark target fragmentation functions
for the luminosities of the experiments under consideration.
A similar analysis based on the photoproduction process, although not 
treated in the present paper, is expected to be a more realistic scenario.
}

\dgsb{Heavy-Quark Production}
\labelm{HQPCase}
Having obtained the fragmentation and target fragmentation functions 
of heavy quarks, we can now treat heavy-quark production 
in the fragmentation function picture. 
The cross section is written as 
\beqm{hqxsect}
\sigma=
\sum_{i,j}\sigma^{\mbox{\scriptsize hard}}_{fD,ij} \otimes f_{i/P} 
\otimes D_{Q/j}
+\sum_{i}\sigma^{\mbox{\scriptsize hard}}_{M,i} \otimes M_{i,Q/P},
\eeq
cf.\ Eq.~(\ref{star}).
The hard scattering matrix elements are those
from Section~\ref{opics}, where all quarks are treated as massless.
Depending on the available phase space, 
we assume $N_f=4$ or 
$N_f=5$ active flavours in the
photon--gluon fusion process. 
The treatment of
the heavy quarks as massless flavours in the hard scattering process
is justified in two cases.
It may be done as an 
approximation for the case of large transverse momenta,
where the heavy-quark propagators are far off-shell, 
see for example Ref.~\cite{111}.
It is also justified as a technical tool for the extraction of
mass singularities, 
as in the fragmentation function approach
to heavy-quark production,
as developed in Refs.\ \cite{101,37},
where massless partons fragment into heavy quarks.
The region of small transverse momenta
gives rise to terms $\sim\log m^2/Q^2$ (the mass singularities
for $m\rightarrow 0$) for the integrated massive matrix
element, corresponding to the $1/\epsilon$ collinear singularities
in the massless approach. Absorbing the $1/\epsilon$ singularities into
the renormalized target fragmentation functions corresponds to a resummation
of the terms logarithmic in $m^2/Q^2$.
In principle, the universality of the coefficient
of the mass singularities renders the use of ``massless'' heavy
quarks in the hard scattering matrix element a valid procedure\footnote{
It is certainly desirable to extend the formalism in such a way that 
the heavy quarks can be treated as massive, even if the
limit $m\rightarrow 0$ in the matrix element is done only for 
the extraction of the mass singularities in the \msbar{} scheme.
This would affect the renormalization
group equation and the calculation of the hard scattering matrix elements.
The present approach has the advantage that we can treat the
fragmentation of light quarks and gluons into heavy quarks.}.


We will compare our results to the leading-order 
process
$\gamma^*g\rightarrow Q\overline{Q}$, shown in Fig.~\ref{qqbfig},
with massive quarks in the matrix element.
Since the heavy-quark mass~$m$ acts as a regulator of the collinear and
soft singularities, it is possible to integrate the cross section 
over the full phase space without encountering divergent terms.
Although this approach, as already discussed in the introduction, is 
probably not sufficient to fully describe the target fragmentation region, 
the comparison will be instructive.
Since the $\gamma^*g\rightarrow Q\overline{Q}$ matrix element will be studied
only in leading order, the results of the comparison should,
however, be interpreted with great care.

The required projections of the hadron tensor
in $d=4$ space-time dimensions are
(see for example Ref.\ \cite{27}):
\beqnm{hmnprojQ}
&&\frac{1}{e^2(2\pi)^{8}}\Pii_M^{\mu\nu}H_{\mu\nu}(\gamma^*g\rightarrow Q\Qb)
\nonu
&&\quad=\,8\pi\, \frac{\alpha_s}{2\pi}
\,2\pi\,T_f\,Q_Q^2\,\cdot 4\cdot\,
\Bigg[
       \frac{s_{iQ}+2m^2}{s_{i\Qb}}
      +\frac{s_{i\Qb}+2m^2}{s_{iQ}}
 -\frac{2(Q^2-2m^2)s_{Q\Qb}}{s_{iQ}s_{i\Qb}}
\nonu
&&
\quad\quad\quad\quad 
\quad\quad\quad\quad 
\quad\quad\quad
 +2m^2\frac{s_{i\Qb}-s_{Q\Qb}-4m^2}{s_{iQ}^2}
 +2m^2\frac{s_{iQ}-s_{Q\Qb}-4m^2}{s_{i\Qb}^2}
\Bigg],\nonu
&&{ }\nonu
&&\frac{1}{e^2(2\pi)^{8}}\Pii_L^{\mu\nu}H_{\mu\nu}(\gamma^*g\rightarrow Q\Qb)
\nonu
&&\quad=\,8\pi\, \frac{\alpha_s}{2\pi}
\,2\pi\,T_f\,Q_Q^2\,\cdot 4\cdot\,
\left[
4\,\frac{u^2}{Q^2}\,\left(s_{Q\Qb}-m^2\left[
       \frac{s_{iQ}}{s_{i\Qb}}
      +\frac{s_{i\Qb}}{s_{iQ}}
\right]\right)
\right];
\eeqn
for the conventions see Sections~\ref{dislns} and~\ref{dislnsqcd}.
A suitable phase-space parametrization is listed in 
Appendix~\ref{milp}.

\begin{figure}[htb] \unitlength 1mm
\begin{center}
\dgpicture{159}{65}

\put(20,4){\epsfigdg{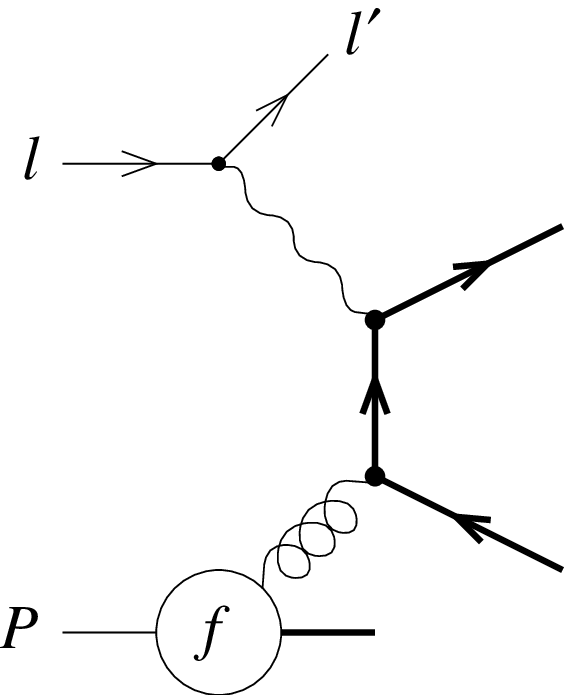}{width=50mm}}
\put(90,4){\epsfigdg{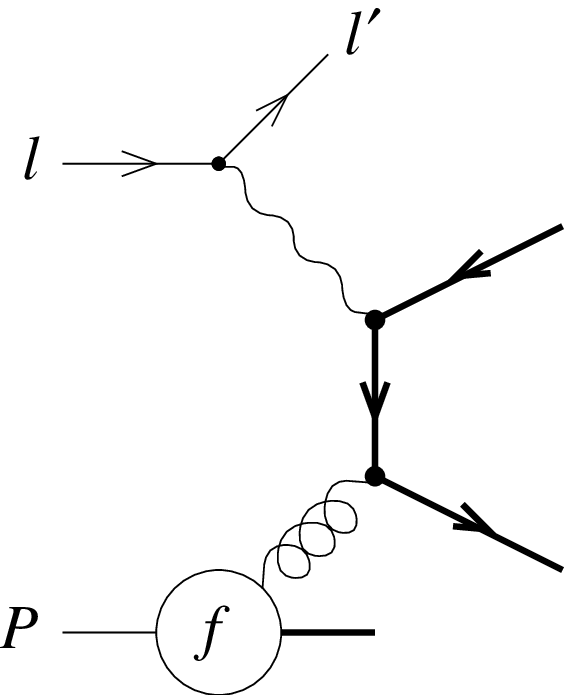}{width=50mm}}

\put(73,45){{$\mbox{\Large\it Q}$}}
\put(73,13){{$\overline{\mbox{\Large\it Q}}$}}
\put(143,45){{$\overline{\mbox{\Large\it Q}}$}}
\put(143,13){{$\mbox{\Large\it Q}$}}

\end{picture}
\end{center}
\shiftcaption
\caption[The Process $\gamma^*g\rightarrow Q\overline{Q}$]
{\labelmm{qqbfig} {\it Feynman diagrams corresponding to the 
process $\gamma^*g\rightarrow Q\overline{Q}$ in leading order.
}}   
\end{figure}

\dgsb{Implementation of the Cross Section Formula}
\labelm{ImpStudy}
The cross section is implemented in a FORTRAN program that 
allows us to determine arbitrary distributions in the variables 
describing the produced heavy quark and the scattered lepton by means
of a Monte Carlo integration\footnote{
The implementation has been cross-checked by a comparison to PROJET
\cite{112} for non-vanishing $p_T$ and by an explicit analytical
integration for the phase-space regions of vanishing $p_T$, for the special
case of $D(x)=\delta(1-x)$. Moreover, the invariance with respect to shifts
of finite contributions between regular and singular terms $D^r$ and $D^s$
of the fragmentation functions, according to the decomposition
of Eq.~(\ref{eq131}), has been verified.
}.
Depending on the particular process at hand (i.e.\ Born terms or 
next-to-leading-order contributions in the current and target fragmentation 
regions), the program generates the corresponding kinematical variables
and calls a user-defined function that has to supply the value of the 
observable to be averaged. In a similar way, the matrix element weights 
of arbitrary observables may be summed up in histogram bins.
The adaptive Monte Carlo integration is based on the
multidimensional integration routine VEGAS \cite{113,114}.

In the case of next-to-leading-order contributions, the user-defined
subroutine is typically called several times for every event, with 
different momenta of the external particles, due to the necessary 
subtractions, as is described in Appendix~\ref{rfef}.
In the case of heavy-quark production, where the heavy-quark fragmentation
functions themselves are singular functions, the final expression for
the cross section is a triple convolution (in the variables~$u$, $\rho$ and
$z$), where all integrals are possibly related to subtractions.
The cross sections derived in this paper, integrated over
physically well motivated regions to be defined later,
have to be finite and positive 
if they are to be physical.
In the current and target fragmentation regions at small transverse
momenta, cancellations of 
tree-level terms and finite ``redefinition'' terms will occur. It is therefore
to be expected that the numerical results will only be meaningful if
the corresponding integration region is sufficiently large. This phenomenon 
will be studied in more detail numerically in Section~\ref{DiscSub}.

\dgsb{Parameters and Variables}
\labelm{DiscPar}

We define the kinematical parameters for a case study
of heavy-quark production at 
HERA\footnote{Please note that a measurement of the heavy-quark content 
of the target fragments is, at present, realistic only
in fixed-target experiments, although, for instance, 
special-purpose detector components may permit the tagging
of for example charm quarks in the target fragmentation region
of the HERA experiments. In any case, it is interesting
to have a comparison to an actual experimental situation 
at energies larger than those achievable in present fixed-target 
experiments.} 
(charm and bottom quarks), E665
(charm quarks) and NA47 (charm quarks) as follows:.
\begin{itemize}
\item[\dgbullet]
HERA: $E_{CM}=296\,\GeV$, $10^{-3}<x_B<1$, 
$W>20\,\GeV$. 
For bottom-quark production, we use $20\,\GeV<Q<100\,\GeV$,
and for charm-quark production  $5\,\GeV<Q<10\,\GeV$.
We moreover require that the observed heavy quark carries 
at least $10\%$ of the proton's momentum, and assume that $5$~quark flavours 
may be produced in the photon--gluon fusion process.
The integrated luminosity can be assumed to be
$\int{\cal L}\, \dd t=250\,\mbox{pb}^{-1}$.
\item[\dgbullet]
E665: $E_{CM}=31\,\GeV$, $0.05<y<0.95$, $2\,\GeV<Q<5\,\GeV$, 
$W>13\,\GeV$. We require that the observed heavy quark carries 
at least $20\%$ of the proton's momentum, and assume that $4$~quark flavours
may be produced in the photon--gluon fusion process.
\item[\dgbullet]
NA47 (SMC, HMC): $E_{CM}=17.3\,\GeV$, $3\times 10^{-3}<x_B<0.7$,
$2\,\GeV<Q<7.7\,\GeV$, $W>8\,\GeV$. 
As in the case of E665, we require that the observed heavy quark carries 
at least $20\%$ of the proton's momentum, and assume that $4$~quark flavours
may be produced in the photon--gluon fusion process.
The integrated luminosity for SMC is approximately
$\int{\cal L}\, \dd t=300\,\mbox{pb}^{-1}$ per year. It is expected
to be larger by a factor of~$5$ 
for the HMC experiment \cite{11,12}.
\end{itemize}
The parton-density parametrizations were taken from 
the packages PAKPDF \cite{115} and PDFLIB \cite{116}.
We use the leading-order GRV distribution
\cite{105}, with the heavy-quark content 
modified$^{\mbox{\scriptsize\arabic{grvlabel}}}$
according to
the distribution from Ref.\ \cite{106}.
The running strong coupling constant $\alpha_s(\mu_r^2)$
with flavour thresholds at the single quark masses\footnote{
See Ref.~\cite{117} and also Ref.~\cite{118}.}
is evaluated using the standard one-loop formula
with $\Lambda_{\mbox{\scriptsize QCD}}=200\,\mbox{MeV}$.
The factorization scales~$\mu_f$, $\mu_D$ and~$\mu_M$ and the 
renormalization scale~$\mu_r$ are set to~$Q$, unless otherwise 
stated\footnote{For our purposes, namely to give an overview 
of cross sections, $Q$ is sufficient as a scale.
It is of course possible to apply specific principles
in order to improve the theoretical prediction, such as
the principle of {\it Fastest Apparent Convergence} (FAC) \cite{119,120},
the {\it Principle of Minimum Sensitivity} (PMS) 
\cite{121,122,123,124},
and the method of {\it Automatic Scale Fixing} (BLM) \cite{125}.
As can be seen from Figs.~\ref{HQTSCALE1} and~\ref{HQTSCALE2}, 
the scale dependence is drastically reduced in next-to-leading order, 
so that the issue of scale choices can be postponed to a later study.}.
For the heavy-quark fragmentation functions, we use the leading order
expressions, with the perturbative input defined at $\mu_{D,0}=m$. 
The perturbative heavy-quark target fragmentation functions are
defined to be zero at $\mu_{M,0}=m$.

We will study distributions in the angular variable~$v$,
in the momentum $P_Q$ of the heavy quark in the hadronic centre-of-mass frame 
given by
\beqm{Qmome}
P_Q=\frac{Q}{2}\,\sqrt{\frac{1-x_B}{x_B}}\,z,
\eeq
in the
transverse momentum~$p_T$ in the hadronic centre-of-mass frame
given by
\beqm{ptdef}
p_T=\frac{Q}{2}\,\sqrt{\frac{1-x_B}{x_B}}\,z_T,
\eeq
where $z_T\in[0,1]$ is defined by
\beqm{ztdef}
z_T\doteq2\,z\,\sqrt{v(1-v)},
\eeq
and in the Feynman variable~$x_F$ \cite{126} given by 
\beqm{xfdef}
x_F\doteq\frac{h_L}{W/2}=z(1-2v),
\eeq
where~$h_L$ is the longitudinal momentum of the heavy quark in 
the hadronic centre-of-mass frame.
The variable~$x_F$
can be used to disentangle the current 
($x_F\geq 0$)  and
target ($x_F<0$) fragmentation regions. 
We will study distributions in~$p_T$
separately in these regions, and
denote the corresponding variables by
$p_T^c$ and $p_T^t$, respectively.
We also consider the variable 
$z_{\zshortparallel}$, as defined in Ref.\ \cite{127} by
\beqm{xpar}
z_{\zshortparallel}\doteq\frac{h^{\mbox{\scriptsize Breit}}_L}{Q/2}
=z\left(1-\frac{v}{x_B}\right),
\eeq
where $h^{\mbox{\scriptsize Breit}}_L$ is the longitudinal momentum in 
the Breit frame\footnote{The Breit frame is defined as the reference
frame where the energy of the exchanged virtual photon is
zero and where the incident nucleon and the virtual photon
are back-to-back. Again, we define the $z$-axis to be in the current
direction.}.
It is easy to see that 
\beqm{zbound}
-\frac{1-x_B}{x_B}\leq z_{\zshortparallel}\leq 1.
\eeq

\dgsb{Cross Sections}
\labelm{DiscStudy}

As a first step, we give an overview of the partial cross sections
for various phase-space regions in Table~\ref{abstab}.
The total cross sections $\sigma_{\mbox{\scriptsize tot}}$, 
corresponding to arbitrary hadronic
final states in deeply inelastic scattering, 
are calculated 
in leading \{1\}\footnote{The numbers in braces \{\ \}
refer to the rows in the tables.}
and next-to-leading \{2\} order
by means of PROJET \cite{112}.
The \porder{\alpha_s} contributions are small, of the order of
$10$--$20\%$. The leading-order cross section for heavy-quark production
$\sigma_{\mbox{\scriptsize $fD$}}$ \{3\} 
is a fraction of the order of $1$--$10\%$ of
$\sigma_{\mbox{\scriptsize tot}}$, depending on the 
experiment.
The leading-order contribution $\sigma_{\mbox{\scriptsize $M$}}$ \{8\}
is much smaller than $\sigma_{\mbox{\scriptsize $fD$}}$ \{3\} in leading order.
The contributions
\porder{\alpha_s}
are given separately
for the current and target 
fragmentation regions. 
In the current fragmentation region, the size of the QCD corrections \{4\}
varies considerably and may be as large as about $40\%$ of the leading-order
term \{3\}, 
because the photon--gluon fusion process contributes significantly.
In the target fragmentation region, the contribution
from  
$\sigma_{\mbox{\scriptsize $fD$}}$ \{5\} is roughly of the same order of
magnitude as the leading-order terms in $\sigma_{\mbox{\scriptsize $M$}}$
\{8\}. The QCD corrections to $\sigma_{\mbox{\scriptsize $M$}}$ \{9\}
are negative and about --$5$ to --$10\%$.

Summing up leading and next-to-leading orders, we arrive at the following 
results. The overall next-to-leading-order correction
is in the range of --$5$ to +$40\%$ \{10, 11\}, where the
current fragmentation region dominates \{12, 13\}.
A comparison of \{12\} with \{14\} and of \{13\} with \{15\} shows
that a considerable fraction of heavy quarks is produced
with a~$p_T$ smaller than $p_{T,\mbox{\scriptsize min}}$, where 
$p_{T,\mbox{\scriptsize min}}\doteq 9\,\GeV$
and $3\,\GeV$ for bottom- and charm-quark production, respectively.

The results from the fragmentation function approach can be compared
with those from the $\gamma^* g\rightarrow Q\overline{Q}$ matrix element, 
shown in rows \{16--21\} in the table.
Comparing \{6\} with \{18\} and \{7\} with \{19\},
one sees that the cross sections for
$p_T>p_{T,\mbox{\scriptsize min}}$ agree well.
We will see later in Section~\ref{DiscSub} that there are considerable 
differences for small~$p_T$.
In the target fragmentation region, comparing \{15\} and \{21\},
the excess in the fragmentation function approach can be attributed
to the evolution of the perturbative target fragmentation functions.

The contribution from intrinsic heavy quarks
in the current fragmentation region \{22\} is small, less
than $1\%$ of the non-intrinsic contribution
\{10\}, except for NA47, where it reaches about $8\%$.
The situation is very different in the target fragmentation region
\{25\}, where the intrinsic heavy-quark component is comparable to the
non-intrinsic heavy-quark component \{13\}, or even dominant.

The flavour decomposition of the 
cross sections is shown in Table~\ref{bbstab}.
The heavy-quark- and gluon-initiated contributions of the total
cross section are shown in 
rows \{1--3\}.
Tagging a heavy quark in the current \{4,~5\} and target \{6,~7\}
fragmentation regions considerably 
increases the fraction of processes initiated
by the heavy quark and antiquark, respectively, 
the absolute cross sections, however, being much smaller.
The fractions in leading order are fairly large \{4--7\}, and they are
reduced in next-to-leading order  \{8,~9, 11,~12\}, due to the
inclusion of gluon-initiated processes \{10,~13\}.
This clearly shows that the tagging of particles in the target fragments
permits to bias the flavour content of the hard scattering process.

Differential distributions in $x_B$ and in the total hadronic final state
energy $W$ are shown in Figs.~\ref{HQTXB} and~\ref{HQTW}, respectively.
For charm-quark production at HERA, 
the dominant contributions come from the
region of small $x_B$.
Figure~\ref{HQTW} shows that the values of the hadronic final state
energy $W$
in the fragmentation function approach, compared with 
the $\gamma^* g\rightarrow Q\overline{Q}$ matrix element,
extend to slightly smaller values.

 \newcommand{\prc}[1]{$#1\%$}
\newcommand{\pr}[1]{$#1$}
\newcommand{\sru}{\rule[-2.0mm]{0mm}{6.3mm}}
\newcommand{\opb}{} 
\newcommand{\clb}{} 
\newcommand{\rb}[1]{\raisebox{1.7ex}[-1.7ex]{#1}} 
\newcommand{\none}[1]{}

\begin{table}[htb]
\begin{center}
\begin{tabular}[h]{|r|c|c|c|c|c|c|c|}
\cline{5-8}
\multicolumn{4}{c|}{}
&\rule[-2.5mm]{0mm}{8mm} Bottom & \multicolumn{3}{|c|}{Charm}\\
\cline{1-8}
\multicolumn{4}{|c|}{\rule[-2.5mm]{0mm}{8mm}Process} 
& HERA & HERA & E665 & NA47 \\ 
\hline
\hline
\sru \opb 1\clb &LO & 
\multicolumn{2}{|c|}{\none{$\sigma_{\mbox{\scriptsize tot}}$}} &
  864  & 23861 & 27434 & 21620 \\
\cline{1-2}\cline{5-8}

\sru \opb 2\clb &NLO & 
\multicolumn{2}{|c|}{\rb{$\sigma_{\mbox{\scriptsize tot}}$}} &
  809  & 21451 & 22743 & 18707 \\
\hline\hline

\sru \opb 3\clb &\porder{\alpha_s^0} & 
\multicolumn{2}{|c|}{$\sigma_{\mbox{\scriptsize $fD$}}$} &
  10.2  & 2307 & 1007 & 433 \\ 
\hline\hline

\sru \opb 4\clb &\none{\porder{\alpha_s}} & 
\none{$\sigma_{\mbox{\scriptsize $fD$}}$} 
& $x_F\geq 0$ &
 --1.29  & 1001 & 344  & 20.8 \\
\cline{1-1}\cline{4-8}

\sru \opb 5\clb &\none{\porder{\alpha_s}} & 
\none{$\sigma_{\mbox{\scriptsize $fD$}}$} 
& $x_F<0$&
 0.144  & 14.4   &  20.2  &  19.2 \\
\cline{1-1}\cline{4-8}

\sru \opb 6\clb &
\rb{\porder{\alpha_s}} & 
\rb{$\sigma_{\mbox{\scriptsize $fD$}}$} 
& $x_F\geq 0$, $p_T>p_{T,\mbox{\scriptsize min}}$ &
 2.46      & 776      & 53.0      & 5.52     \\ 
\cline{1-1}\cline{4-8}

\sru \opb 7\clb &\none{\porder{\alpha_s}} & 
\none{$\sigma_{\mbox{\scriptsize $fD$}}$} 
& $x_F< 0$,  $p_T>p_{T,\mbox{\scriptsize min}} $ &
 0.131      & 11.8      & 4.36      & 0.623    \\ 
\hline\hline

\sru \opb 8\clb &\porder{\alpha_s^0} & 
\multicolumn{2}{|c|}{\none{$\sigma_{\mbox{\scriptsize $M$}}$}} &
 0.456  & 40.1   &  11.6  &  11.7   \\
\cline{1-2}\cline{5-8}

\sru \opb 9\clb &\porder{\alpha_s} & 
\multicolumn{2}{|c|}{\rb{$\sigma_{\mbox{\scriptsize $M$}}$}} &
 --0.0297 & --1.99   & --1.01  & --0.922  \\ \hline\hline

\sru \opb 10\clb &LO & 
\multicolumn{2}{|c|}{\none{$\sigma$}} &
 10.6  & 2347 & 1018 & 444 \\ 
\cline{1-2}\cline{5-8}

\sru \opb 11\clb &NLO & 
\multicolumn{2}{|c|}{\rb{$\sigma$}} &
 9.48 & 3360 & 1381 & 483 \\ 
\hline\hline

\sru \opb 12\clb &\none{NLO} & \none{$\sigma$} & $x_F\geq 0$ &
 8.91 & 3308 & 1351 & 453 \\
\cline{1-1}\cline{4-8}

\sru \opb 13\clb &\none{NLO} & \none{$\sigma$} & $x_F<0$ &
 0.570  &  52.5  & 30.7   & 29.9  \\
\cline{1-1}\cline{4-8}

\sru \opb 14\clb &\rb{NLO} 
& 
\rb{$\sigma$} & $x_F\geq 0$, $p_T\leq p_{T,\mbox{\scriptsize min}}$ &
 6.45 & 2532 & 1298 & 447 \\
\cline{1-1}\cline{4-8}

\sru \opb 15\clb &\none{NLO} 
& 
\none{$\sigma$} & $x_F< 0$, $p_T\leq p_{T,\mbox{\scriptsize min}}$ &
 0.439 & 40.7 & 26.3 & 29.2 \\
\hline\hline

\sru \opb 16\clb &\none{\porder{\alpha_s}} & 
\none{$\sigma_{Q\overline{Q}}$} & $x_F\geq 0$ &
 8.13  & 3427 & 930 & 260 \\
\cline{1-1}\cline{4-8}

\sru \opb 17\clb &\none{\porder{\alpha_s}} & 
\none{$\sigma_{Q\overline{Q}}$} & $x_F<0$&
 0.233  & 30.5  & 18.2  & 10.1  \\
\cline{1-1}\cline{4-8}

\sru \opb 18\clb &\none{\porder{\alpha_s}} & 
\none{$\sigma_{Q\overline{Q}}$} 
& $x_F\geq 0$, $p_T>p_{T,\mbox{\scriptsize min}}$ &
 2.40  & 939 & 50.8 & 4.57 \\
\cline{1-1}\cline{4-8}

\sru \opb 19\clb  &\rb{\porder{\alpha_s}} & 
\rb{$\sigma_{Q\overline{Q}}$} 
& $x_F<0$, $p_T>p_{T,\mbox{\scriptsize min}}$ &
 0.101  & 14.1  & 3.96   & 0.487  \\
\cline{1-1}\cline{4-8}

\sru \opb 20\clb &\none{\porder{\alpha_s}} & 
\none{$\sigma_{Q\overline{Q}}$} 
& $x_F\geq 0$, $p_T\leq p_{T,\mbox{\scriptsize min}}$ &
5.73 & 2488 & 879 & 255 \\
\cline{1-1}\cline{4-8}

\sru \opb 21\clb  &\none{\porder{\alpha_s}} & 
\none{$\sigma_{Q\overline{Q}}$} 
& $x_F<0$, $p_T\leq p_{T,\mbox{\scriptsize min}}$ &
0.132 & 16.4 & 14.2 & 9.61 \\
\hline\hline

\sru \opb 22\clb &\porder{\alpha_s^0} 
& 
\multicolumn{2}{|c|}
{$\sigma_{\mbox{\scriptsize $fD$}}^\IHQ$} &
 0.599$\times 10^{-1}$   & 9.17 & 8.61 & 34.3 \\ 
\hline\hline

\sru \opb 23\clb &\none{\porder{\alpha_s}} 
& 
\none{$\sigma_{\mbox{\scriptsize $fD$}}^\IHQ$} 
& $x_F\geq 0$ &
 0.232$\times 10^{-3}$  & 2.65 & 4.01 & -0.430 \\
\cline{1-1}\cline{4-8}

\sru \opb 24\clb &\rb{\porder{\alpha_s}} 
& 
\rb{$\sigma_{\mbox{\scriptsize $fD$}}^\IHQ$} & $x_F<0$&
 0.451$\times 10^{-3}$  & 0.212  & 0.381 & 0.613  \\ 
\hline\hline

\sru \opb 25\clb &\porder{\alpha_s^0} 
& 
\multicolumn{2}{|c|}{\none{$\sigma_{\mbox{\scriptsize $M$}}^\IHQ$}} &
 0.499  & 60.9  & 72.1   & 124   \\
\cline{1-2}\cline{5-8}

\sru \opb 26\clb &\porder{\alpha_s} 
& 
\multicolumn{2}{|c|}{\rb{$\sigma_{\mbox{\scriptsize $M$}}^\IHQ$}} &
 --0.170$\times 10^{-1}$ & --2.11  & --1.61 & --10.8 \\ \hline

\end{tabular}
\end{center}
\shiftcaption
\caption[Cross Sections]
{
\labelmm{abstab} 
{\it
Cross sections in [pb] for the various experiments and subprocesses.
The 
\porder{\alpha_s} contributions
do not include the corresponding \porder{\alpha_s^0} term. 
Terms denoted by NLO are the sums of \porder{\alpha_s^0}
and \porder{\alpha_s} contributions for the corresponding phase-space
region.
The contributions labelled by
$\sigma_{\mbox{\scriptsize $fD$}}$
and
$\sigma_{\mbox{\scriptsize $M$}}$
correspond to the terms in 
Eqs.~(\ref{cur}) and~(\ref{tar}), 
whereas~$\sigma$ stands for the sum of these two terms.
``IHQ'' stands for intrinsic heavy-quark contributions.
The cross section for massive heavy quarks in the
$\gamma^* g\rightarrow Q\overline{Q}$ matrix element
is denoted by $\sigma_{Q\overline{Q}}$, and 
the total cross section, including all hadronic final states, 
by $\sigma_{\mbox{\scriptsize tot}}$.
}}
\end{table}

\begin{table}[htb]
\begin{center}
\begin{tabular}[h]{|r|c|c|c|c|c|c|c|c|}
\cline{6-9}
\multicolumn{5}{c|}{}
& \multicolumn{1}{|c|}{\rule[-2.5mm]{0mm}{8mm} Bottom} & 
\multicolumn{3}{|c|}{Charm}\\
\hline
\multicolumn{4}{|c|}{\rule[-2.5mm]{0mm}{8mm}Process} 
& $i$ & 
\multicolumn{1}{|c|}{HERA} & 
\multicolumn{1}{|c|}{HERA} & 
\multicolumn{1}{|c|}{E665} & 
\multicolumn{1}{|c|}{NA47} \\ 
\hline\hline

\sru \opb 1\clb &LO & 
\multicolumn{2}{|l|}{\none{$\sigma_{\mbox{\scriptsize tot}}$}} & 
$Q$, $\overline{Q}$ &
 \pr{0.8} & \pr{9.4} & \pr{3.7} & \pr{2.0} \\
 
\cline{1-2}\cline{5-9}

\sru \opb 2\clb & & 
\multicolumn{2}{|c|}{$\sigma_{\mbox{\scriptsize tot}}$} & 
$Q$, $\overline{Q}$ &
 \pr{0.8} & \pr{10.3} & \pr{4.2} & \pr{2.2} \\ 
\cline{1-1}\cline{5-9}

\sru \opb 3\clb &\rb{NLO} & 
\multicolumn{2}{|l|}{\none{$\sigma_{\mbox{\scriptsize tot}}$}} & 
$g$ &
 \pr{-5.4} & \pr{-9.7} & \pr{14.9} & \pr{10.0} \\ \hline
\hline

\sru \opb 4\clb &\none{\porder{\alpha_s^0}} & 
\multicolumn{2}{|c|}{\none{$\sigma_{\mbox{\scriptsize $fD$}}$}} &
$Q$ &
 \pr{69.2} & \pr{94.7} & \pr{98.2} & \pr{96.7} \\
\cline{1-1}\cline{5-9}

\sru \opb 5\clb &\none{\porder{\alpha_s^0}} & 
\multicolumn{2}{|c|}{\rb{$\sigma_{\mbox{\scriptsize $fD$}}$}} &
$\overline{Q}$ &
 \pr{0.2} & \pr{0.5} & \pr{0.0} & \pr{0.0} \\
\cline{1-1}\cline{3-4}\cline{5-9}


\sru \opb 6\clb &\rb{\porder{\alpha_s^0}} & 
\multicolumn{2}{|c|}{\none{$\sigma_{\mbox{\scriptsize $M$}}$}} &
$Q$ &
 \pr{0.3}  & \pr{2.3} & \pr{0.0} &  \pr{0.0} \\ 
\cline{1-1}\cline{5-9}

\sru \opb 7\clb & \none{\porder{\alpha_s^0}} & 
\multicolumn{2}{|c|}{\rb{$\sigma_{\mbox{\scriptsize $M$}}$}} &
$\overline{Q}$ &
 \pr{59.1}  & \pr{80.5} & \pr{98.2} &  \pr{96.5} \\ 
\hline\hline


\sru \opb 8\clb &\none{NLO} & \none{$\sigma$} & \none{$x_F\geq 0$} &
$Q$ &
 \pr{79.9} & \pr{66.5} & \pr{74.7} & \pr{95.1} \\
\cline{1-1}\cline{5-9}

\sru \opb 9\clb &\none{NLO} & \none{$\sigma$} & $x_F\geq 0$ &
$\overline{Q}$ &
 \pr{0.0} & \pr{0.0} & \pr{0.0} & \pr{-0.1} \\
\cline{1-1}\cline{5-9}

\sru \opb 10\clb &\none{NLO} & \none{$\sigma$} & \none{$x_F\geq 0$} &
$g$ &
 \pr{13.3} & \pr{32.7} & \pr{26.3} & \pr{7.3} \\
\cline{1-1}\cline{4-9}

\sru \opb 11\clb &\rb{NLO} & \rb{$\,\,\,\,\,\,\sigma\,\,\,\,\,\,$} & 
\none{$x_F<0$} &
$Q$ &
 \pr{0.5} & \pr{2.0} & \pr{0.4} & \pr{5.0} \\
\cline{1-1}\cline{5-9}

\sru \opb 12\clb &\none{NLO} & \none{$\sigma$} & $x_F<0$ &
$\overline{Q}$ &
 \pr{46.2} & \pr{60.0} & \pr{34.3} & \pr{35.0} \\
\cline{1-1}\cline{5-9}

\sru \opb 13\clb &\none{NLO} & \none{$\sigma$} & \none{$x_F<0$} &
$g$ &
 \pr{10.7} & \pr{21.7} & \pr{60.5} & \pr{58.0} \\ \hline

\end{tabular}
\end{center}
\shiftcaption
\caption[Flavour Decomposition]
{
\labelmm{bbstab} 
{\it
The same as Table~\ref{abstab}, but now shown according to 
the partial cross sections in per cent, for various partons $i$ incident in the
hard scattering process.
Partons $i$ not included in the table add up to \prc{100}.
Entries below \prc{0.05} are marked as \pr{0.0}.
}}
\end{table}

\begin{figure}[htb] \unitlength 1mm
\begin{center}
\dgpicture{159}{185}

\put( 15,100){\epsfigdg{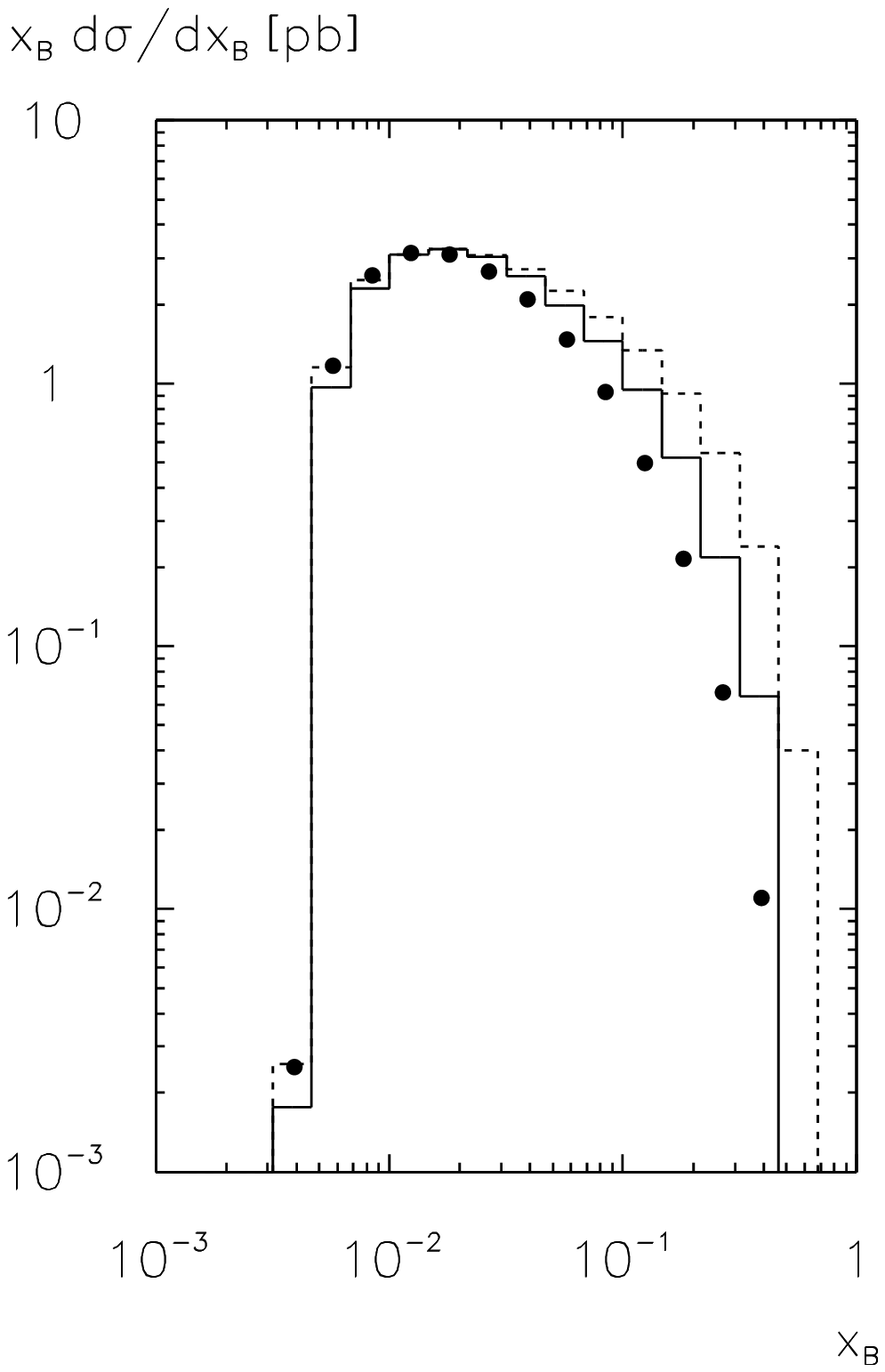}{width=55mm}}
\put( 25,95){\lettlab (a)}

\put( 90,100){\epsfigdg{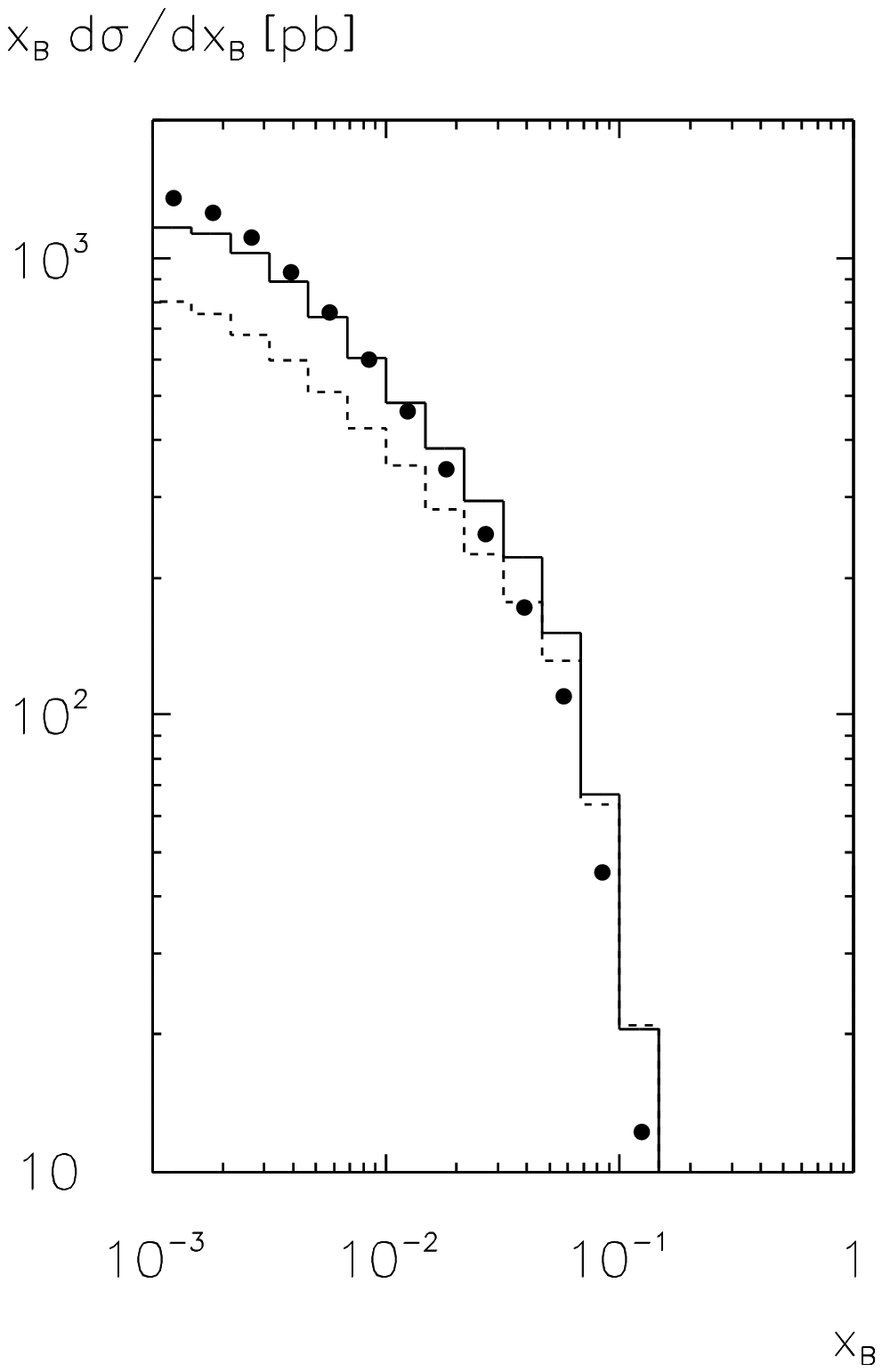}{width=55mm}}
\put(100,95){\lettlab (b)}

\put( 15,5){\epsfigdg{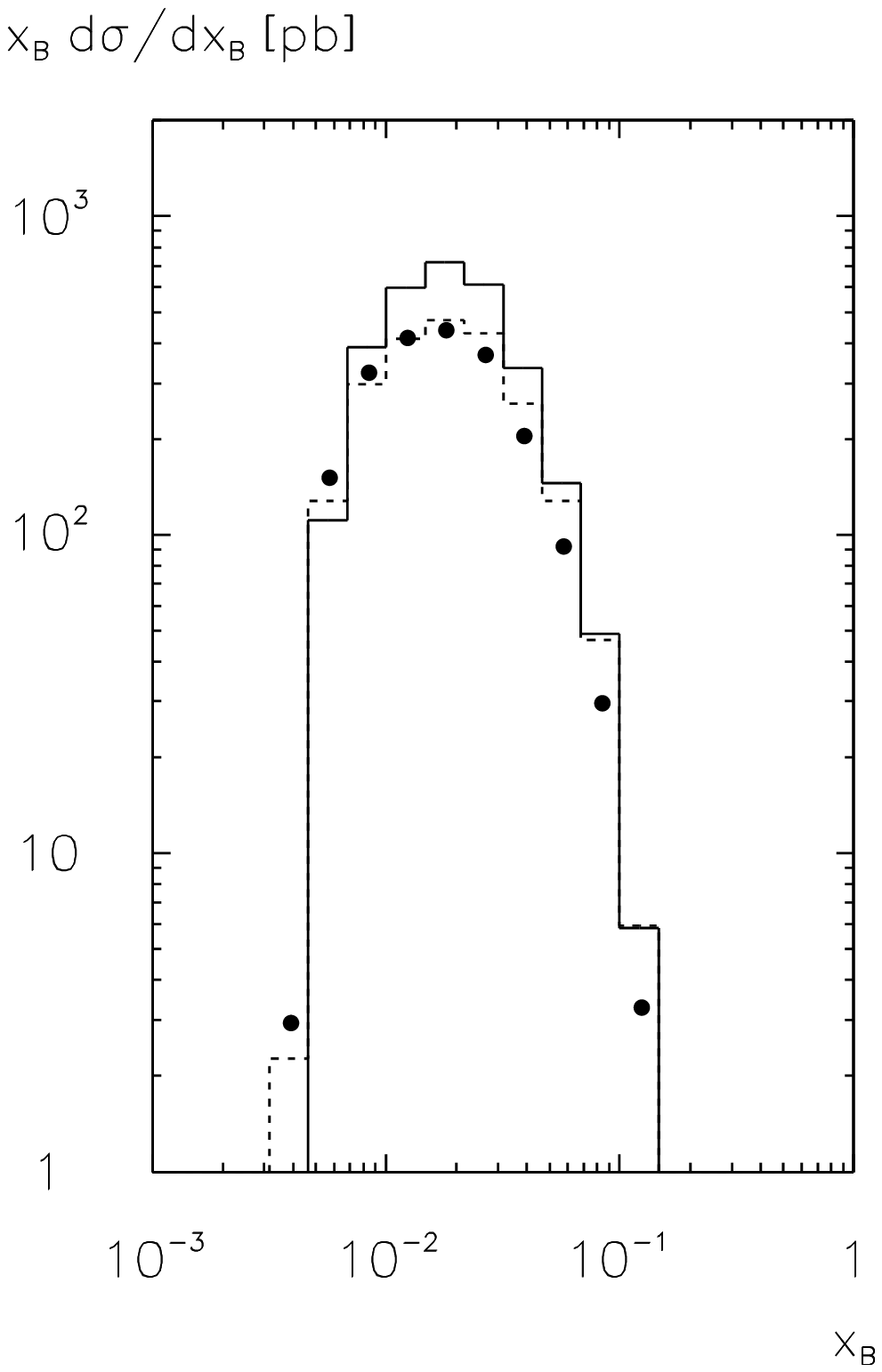}{width=55mm}}
\put( 25,0){\lettlab (c)}

\put( 90,5){\epsfigdg{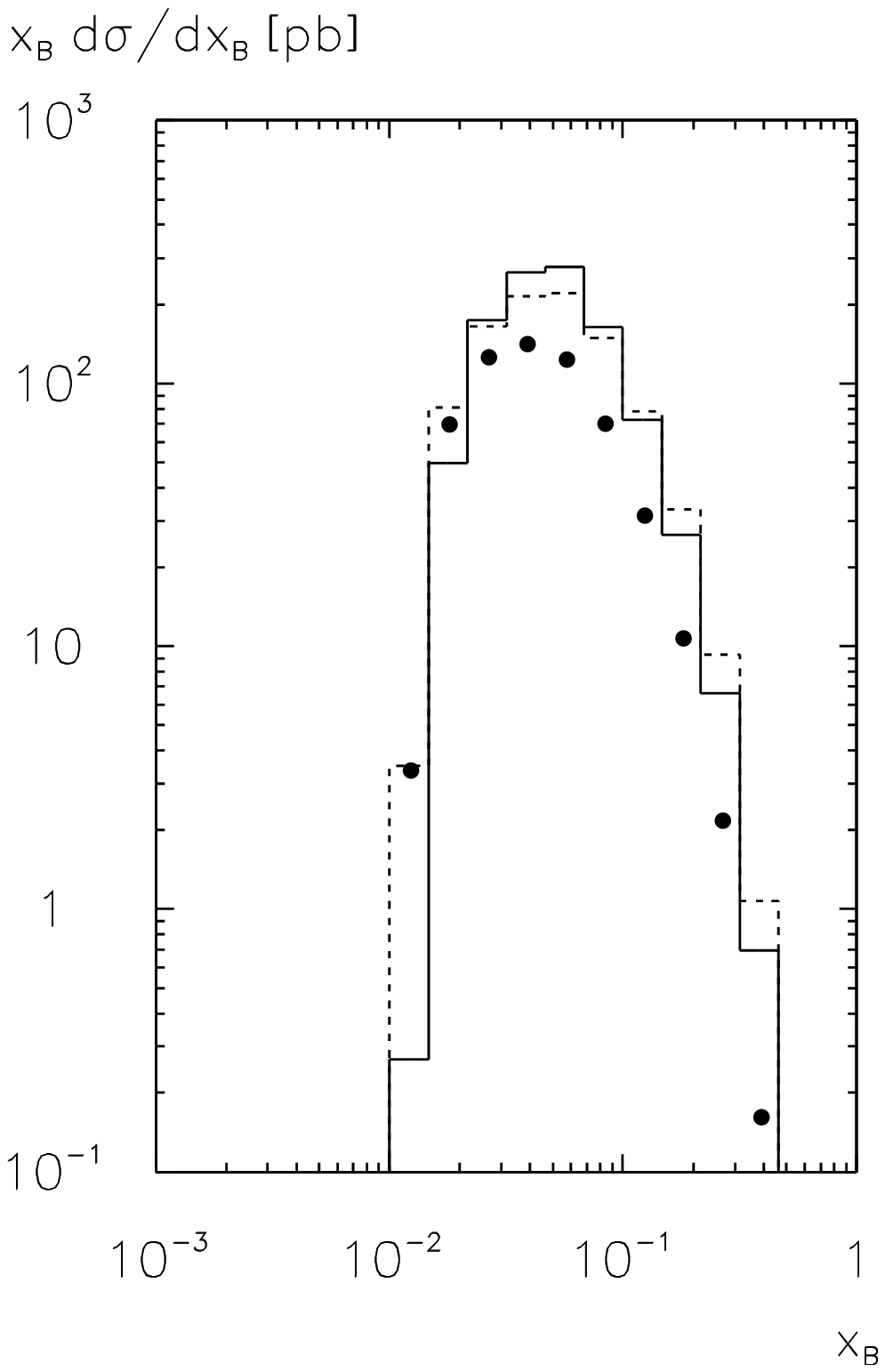}{width=55mm}}
\put(100,0){\lettlab (d)}

\end{picture}
\end{center}
\shiftcaption
\caption[Distributions in~$x_B$ 
for Heavy-Quark Production]
{\labelmm{HQTXB} {\it 
Distributions in~$x_B$ 
for bottom (a) 
and charm (b) quark production at HERA
and for charm-quark production at E665 (c) and NA47 (d),
in leading order \mbox{(\dashline)} 
and in next-to-leading order \mbox{(\fullline)}. 
Also shown is the distribution from the matrix element
$\gamma^*g\rightarrow Q\overline{Q}$ \mbox{(\fullcircle)}.
}}   
\end{figure}

\begin{figure}[htb] \unitlength 1mm
\begin{center}
\dgpicture{159}{185}

\put( 15,100){\epsfigdg{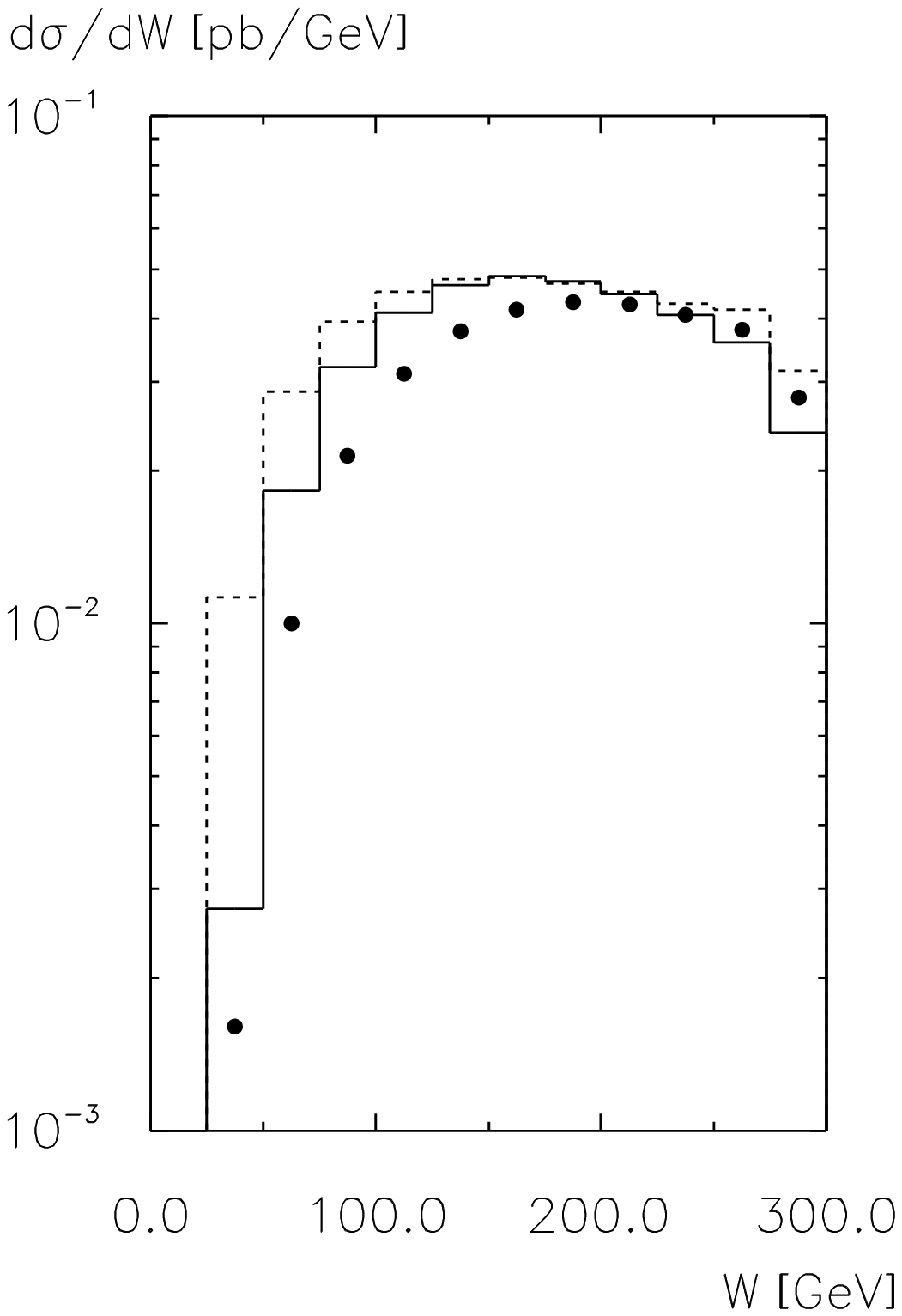}{width=55mm}}
\put( 25,95){\lettlab (a)}

\put( 90,100){\epsfigdg{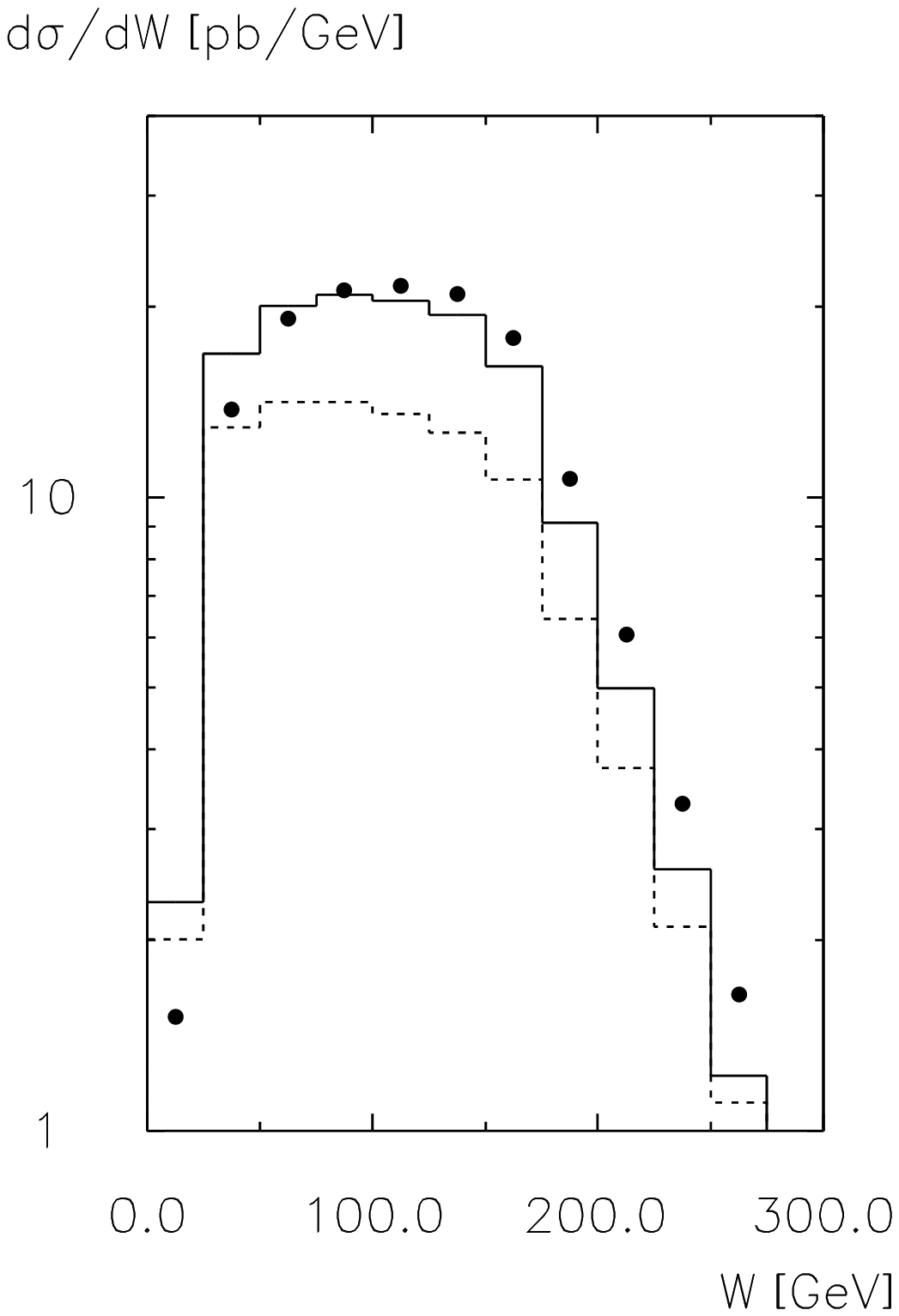}{width=55mm}}
\put(100,95){\lettlab (b)}

\put( 15,5){\epsfigdg{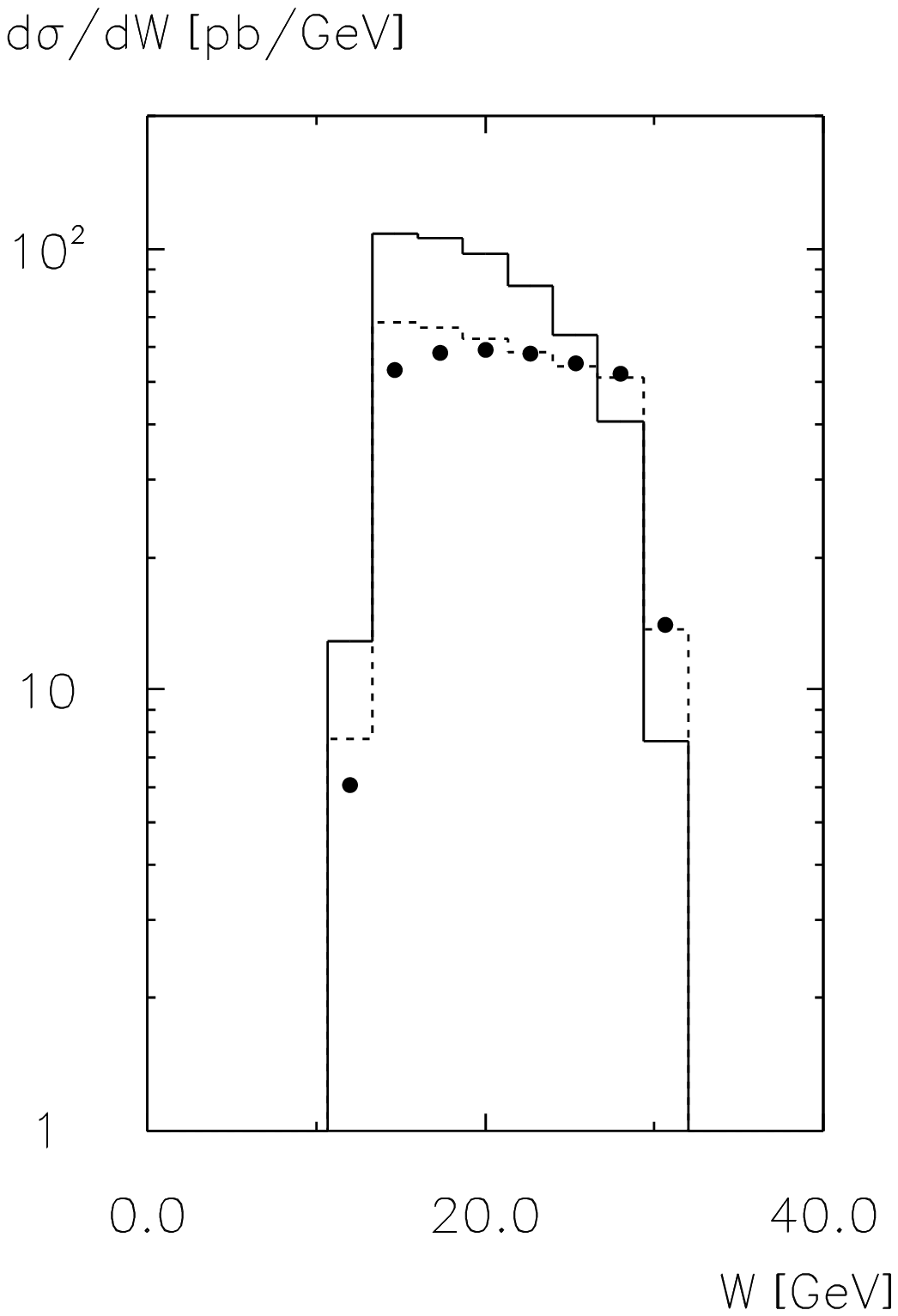}{width=55mm}}
\put( 25,0){\lettlab (c)}

\put( 90,5){\epsfigdg{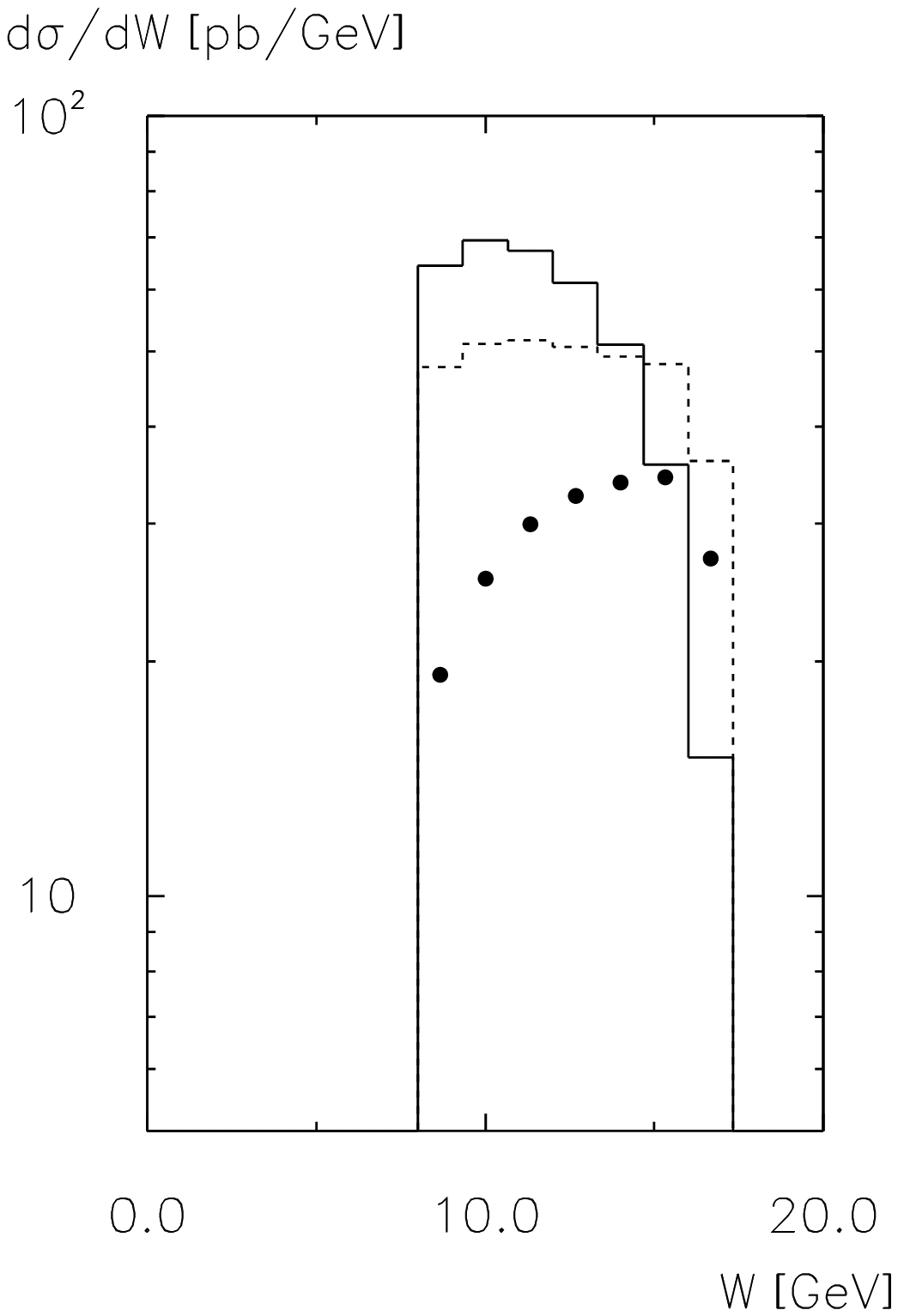}{width=55mm}}
\put(100,0){\lettlab (d)}

\end{picture}
\end{center}
\shiftcaption
\caption[Distributions in~$W$ 
for Heavy-Quark Production]
{\labelmm{HQTW} {\it 
Distributions in~$W$ 
for bottom (a) 
and charm (b) quark production at HERA
and for charm-quark production at E665 (c) and NA47 (d),
in leading order \mbox{(\dashline)} 
and in next-to-leading order \mbox{(\fullline)}. 
Also shown is the distribution from the matrix element
$\gamma^*g\rightarrow Q\overline{Q}$ \mbox{(\fullcircle)}.
}}   
\end{figure}

\clearpage

\dgsb{Distributions in $p_T$ and Collinear Subtractions}
\labelm{DiscSub}
Considering the transverse momentum distributions
in  Figs.~\ref{HQTPT1} and~\ref{HQTPT2}, 
we demonstrate that the mechanism to cancel
collinear singularities in the target fragmentation region actually
works in practice. 
The cross section, differential in~$p_T$, is of the form
$g(p_T)\,(1/p_T)_+$, where
$g(p_T)$ is an integrable function and $(1/p_T)_+$
is the function $1/p_T$ with a subtraction at $p_T=0$.
As already mentioned in Section~\ref{ImpStudy}, the expected mechanism at work 
for finite cross sections is that large contributions from collinear 
singularities at small~$p_T$ cancel against subtraction terms 
of opposite sign
at $p_T=0$. For too small a bin size at $p_T=0$, this mechanism does not work, 
since the positive contributions will not be large enough. This is
demonstrated in Figs.~\ref{HQTPT1}a,~c and \ref{HQTPT2}a,~c for 
the case of the current fragmentation
region. The differential cross section is defined by the cross section
corresponding to a certain bin divided by the bin size, which makes
sense, despite the subtractions, for the bin containing $p_T=0$ as well.
The cross section is strongly rising for 
$p_T^c\rightarrow 0$. 
For a small bin size, Figs.~\ref{HQTPT1}a and~\ref{HQTPT2}a,
the entry in the first bin, 
containing $p_T^c=0$, is negative, representing the
subtraction in the current fragmentation region.
For a sufficiently large bin size, Figs.~\ref{HQTPT1}c and~\ref{HQTPT2}c, 
the entry in the first bin is positive, showing that the perturbative result 
is well-defined in this case. 
This result is expected 
from the factorization theorems of perturbative QCD, applied to the 
absorption of singularities in the case of fragmentation 
functions\footnote{Please note that $p_T^c=0$ stands for the
parton configuration where the observed heavy quark is produced in the
current direction, with a second parton collinear to the
observed heavy quark, possibly emitted from the heavy quark's parent parton.}.

An important result of this section,
illustrated in Figs.~\ref{HQTPT1}b,~d and~\ref{HQTPT2}b,~d, is that a 
similar result can be obtained for the target fragmentation region.
Again, the cross section is strongly rising for 
$p_T^t\rightarrow 0$, Figs.~\ref{HQTPT1}b and~\ref{HQTPT2}b. 
For a small bin size, the entry in the first bin, 
containing $p_T^t=0$, is negative, again representing a
subtraction, in this case in the target fragmentation region.
For a sufficiently large bin size, 
Figs.~\ref{HQTPT1}d and~\ref{HQTPT2}d, the entry in the first bin is 
positive. 
This behaviour is close to the one of jet cross sections, 
which are meaningless unless the ``jet cut'', being effectively
an external mass scale, is large enough.

In Figs.~\ref{HQTPT1} and~\ref{HQTPT2}
we have also included the distributions from the 
corresponding 
$\gamma^*g\rightarrow Q\overline{Q}$
matrix element. We have a fairly good agreement for $p_T\gtrsim2m$.
For small~$p_T$, the distributions for small bin sizes
look very different,
and the results from the fragmentation function approach 
overshoot those from the
$\gamma^*g\rightarrow Q\overline{Q}$
matrix element considerably.
For a larger bin size, 
the results in the charm quark cases are compatible,
owing to the subtractions at $p_T=0$.
The differences remain however for bottom-quark production
in the target fragmentation region
at small transverse momenta.

It is also instructive to have a look at the 
momentum and angular distributions in the
variables $P_Q$ (Fig.~\ref{HQTMOM}) and~$v$ (Fig.~\ref{HQTV1}), respectively.
For $p_T> p_{T,\mbox{\scriptsize min}}$, 
the momentum distributions for the fragmentation function
approach and for the $\gamma^*g\rightarrow Q\overline{Q}$ matrix element
agree well, as is expected.
They are very different for $p_T\le p_{T,\mbox{\scriptsize min}}$:
the fragmentation function approach yields larger cross
sections for small $P_Q$, but, for
E665 and NA47, it falls off more rapidly at large $P_Q$.
The distribution in~$v$,
being directly related to the distribution in 
pseudorapidity 
\beqm{pseudrap}
\eta=\frac{1}{2}\,\ln{\displaystyle\frac{1-v}{v}},
\eeq
is everywhere finite,
albeit strongly peaked
in the current fragmentation region ($v\rightarrow 0$)
and, for HERA, in the target 
fragmentation region ($v\rightarrow 1$) as well.

\begin{figure}[htb] \unitlength 1mm
\begin{center}
\dgpicture{159}{185}

\put( 15,100){\epsfigdg{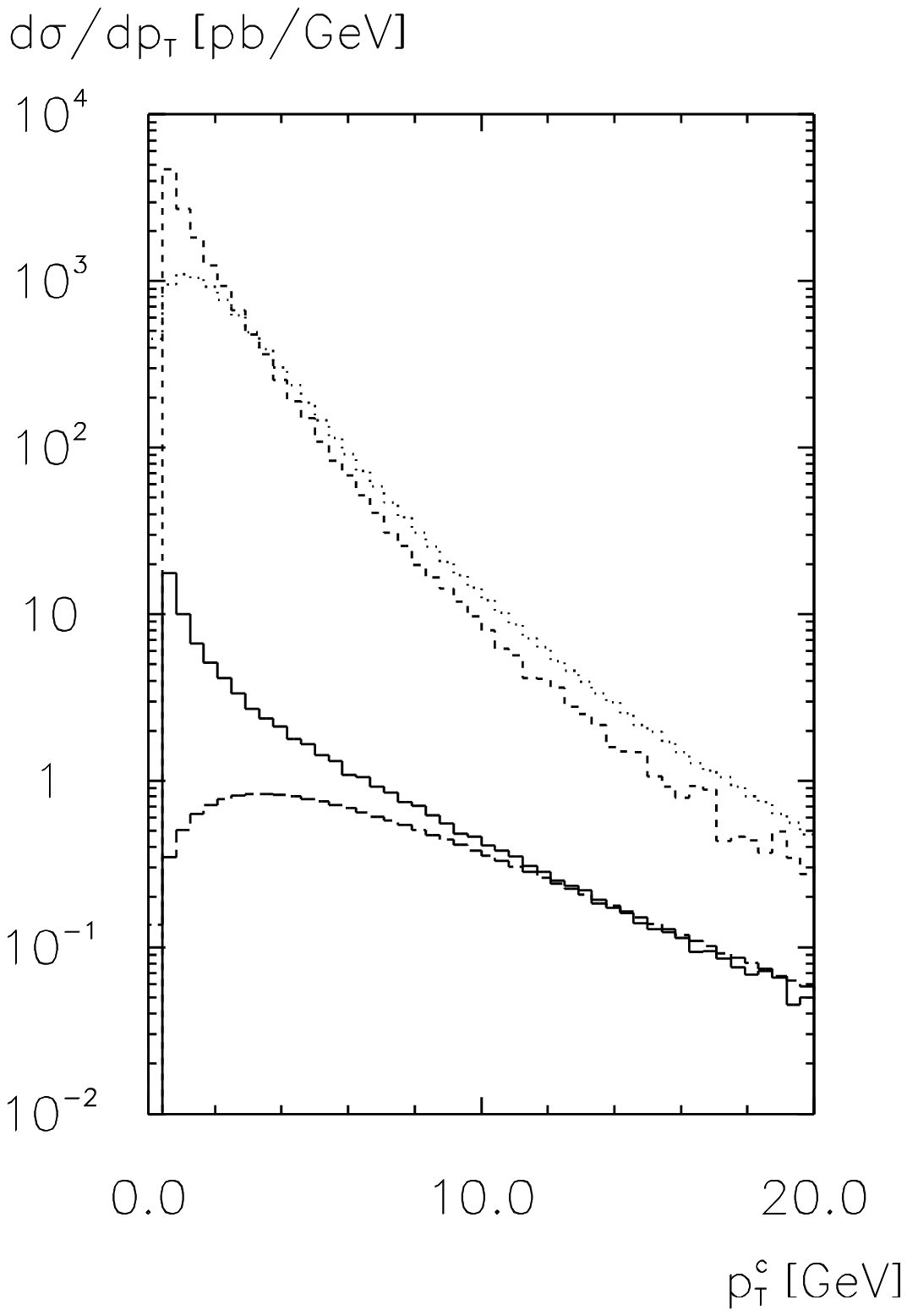}{width=55mm}}
\put( 25,95){\lettlab (a)}

\put( 90,100){\epsfigdg{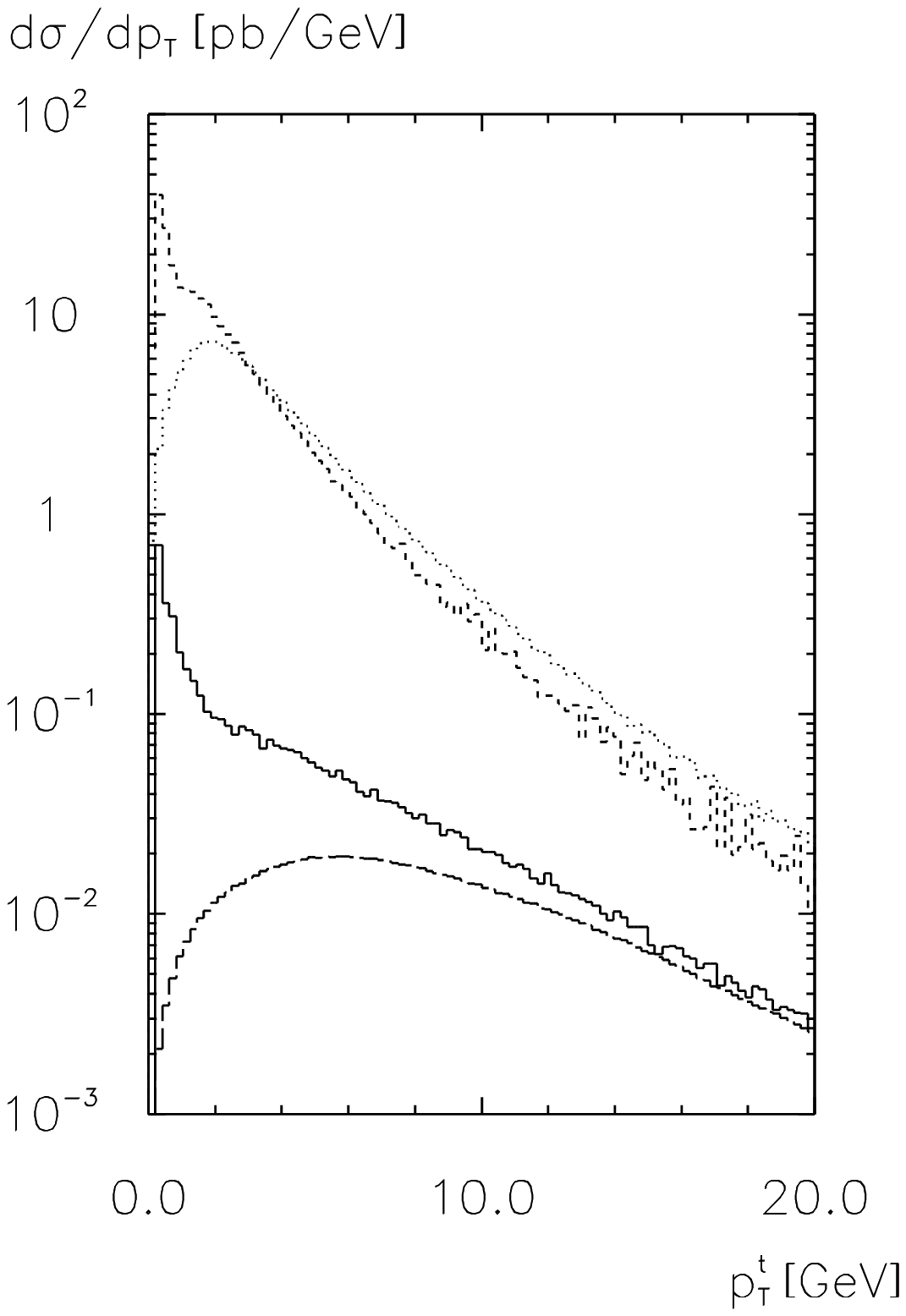}{width=55mm}}
\put(100,95){\lettlab (b)}

\put( 15,5){\epsfigdg{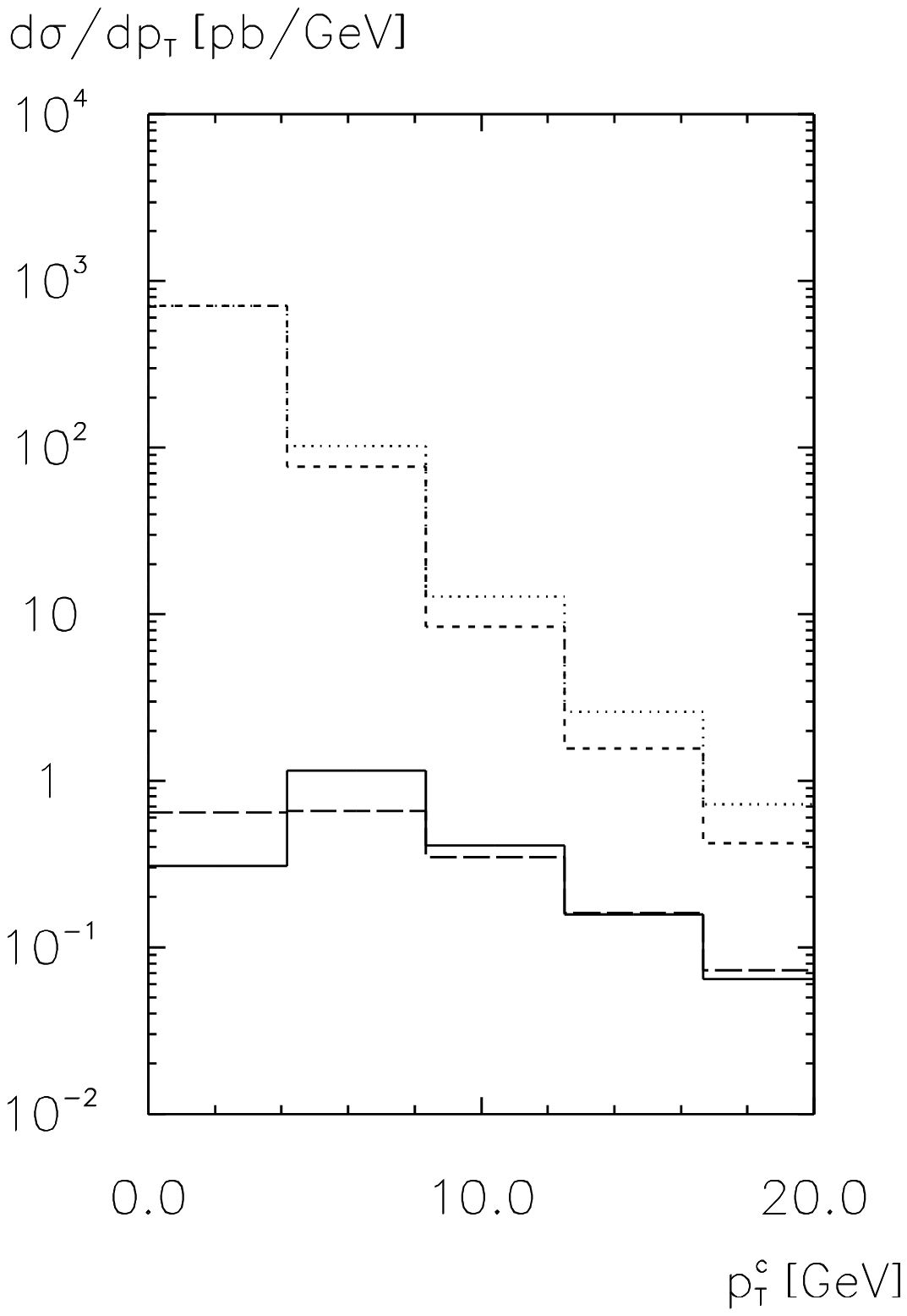}{width=55mm}}
\put( 25,0){\lettlab (c)}

\put( 90,5){\epsfigdg{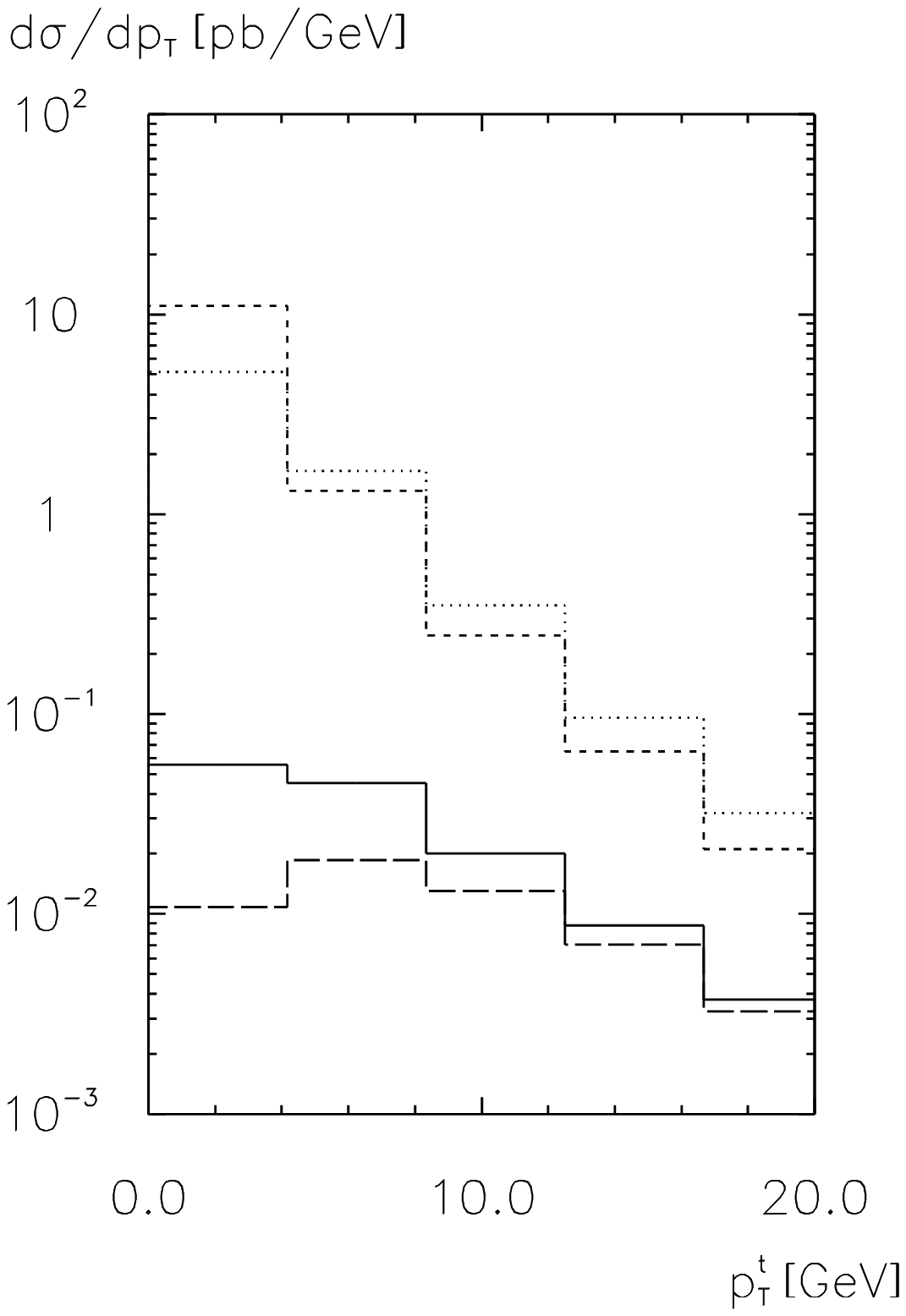}{width=55mm}}
\put(100,0){\lettlab (d)}

\end{picture}
\end{center}
\shiftcaption
\caption[Transverse-Momentum Distributions
for Heavy-Quark Production at HERA]
{\labelmm{HQTPT1} {\it 
Distributions in the transverse momentum~$p_T$ in the current (a), (c)
and target (b), (d) fragmentation regions for bottom \mbox{(\fullline)}
and charm \mbox{(\dashline)} quark production at HERA
for two different bin widths.
Also shown are the distributions from the matrix element
$\gamma^*g\rightarrow Q\overline{Q}$:
bottom \mbox{(\longdashline)}
and charm \mbox{(\dotline)}.
}}   
\end{figure}

\begin{figure}[htb] \unitlength 1mm
\begin{center}
\dgpicture{159}{185}

\put( 15,100){\epsfigdg{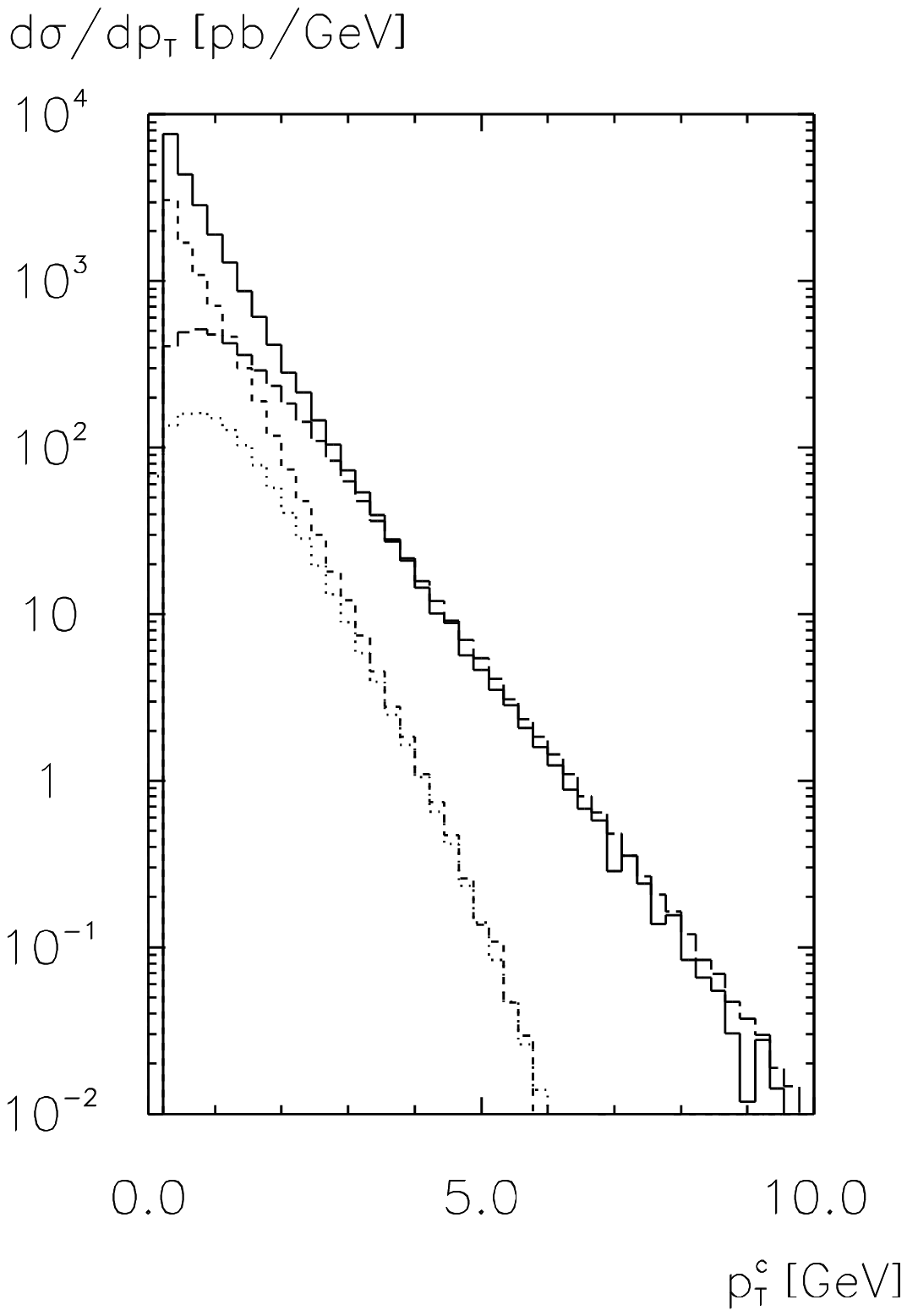}{width=55mm}}
\put( 25,95){\lettlab (a)}

\put( 90,100){\epsfigdg{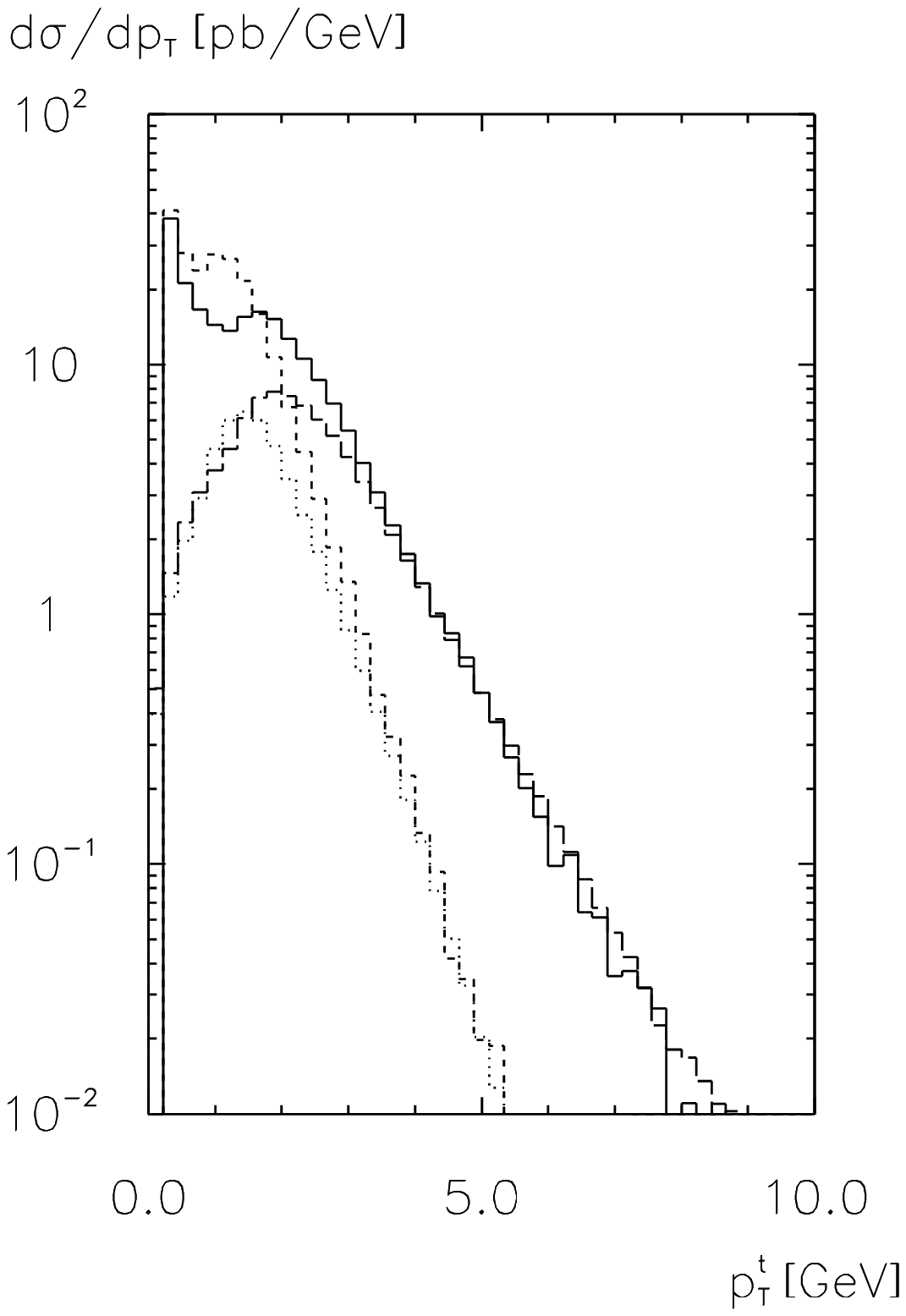}{width=55mm}}
\put(100,95){\lettlab (b)}

\put( 15,5){\epsfigdg{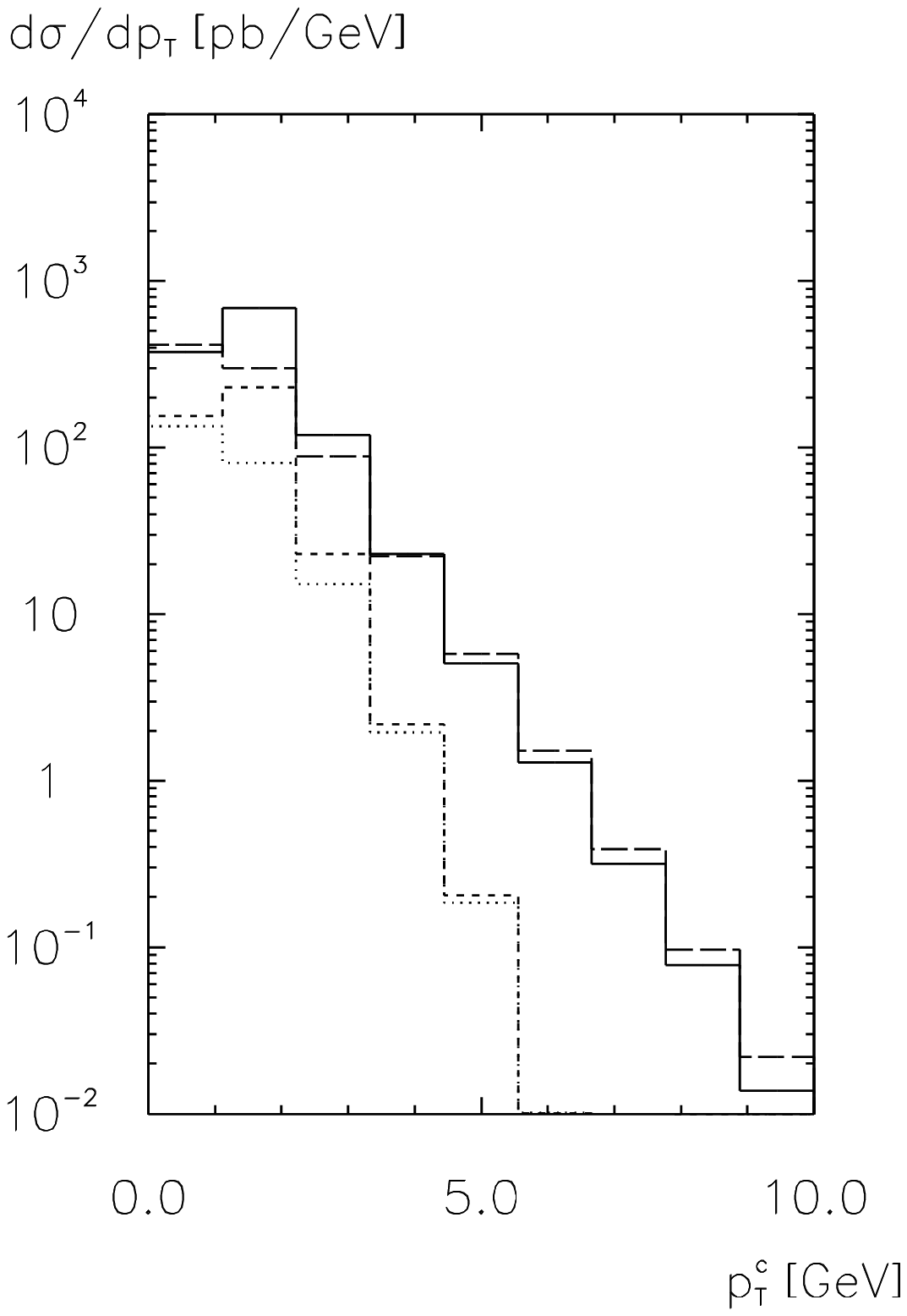}{width=55mm}}
\put( 25,0){\lettlab (c)}

\put( 90,5){\epsfigdg{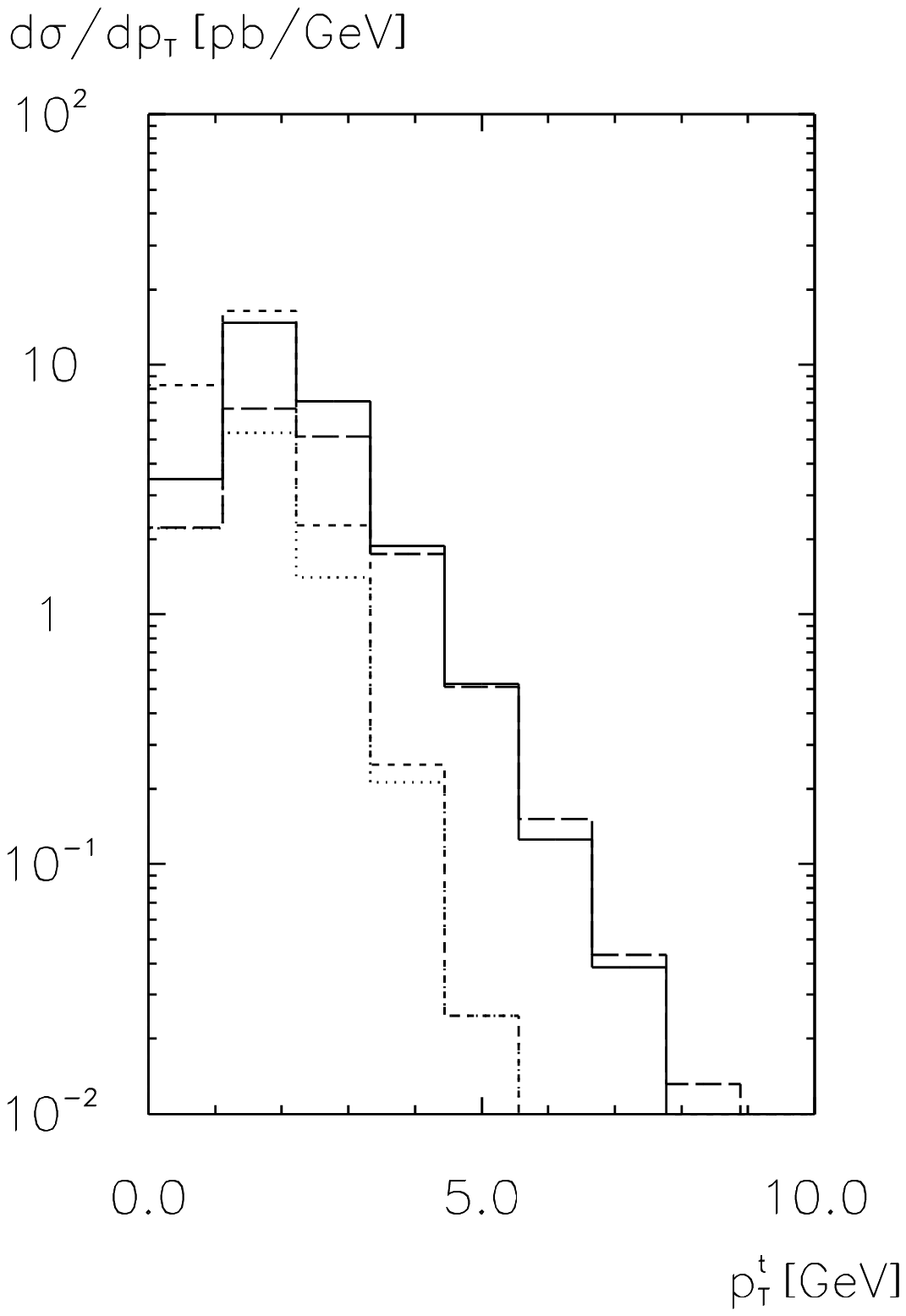}{width=55mm}}
\put(100,0){\lettlab (d)}

\end{picture}
\end{center}
\shiftcaption
\caption[Transverse-Momentum Distributions
for Charm-Quark Production at E665 and NA47]
{\labelmm{HQTPT2} {\it 
Distributions in the transverse momentum~$p_T$
in the current (a), (c) and target 
(b), (d) fragmentation
regions
for charm-quark production at E665 \mbox{(\fullline)}
and NA47 \mbox{(\dashline)}
for two different bin widths. 
Also shown are the distributions from the matrix element
$\gamma^*g\rightarrow c\overline{c}$:
E665 \mbox{(\longdashline)},
NA47 \mbox{(\dotline)}.
}}   
\end{figure}


\begin{figure}[htb] \unitlength 1mm
\begin{center}
\dgpicture{159}{185}

\put( 15,100){\epsfigdg{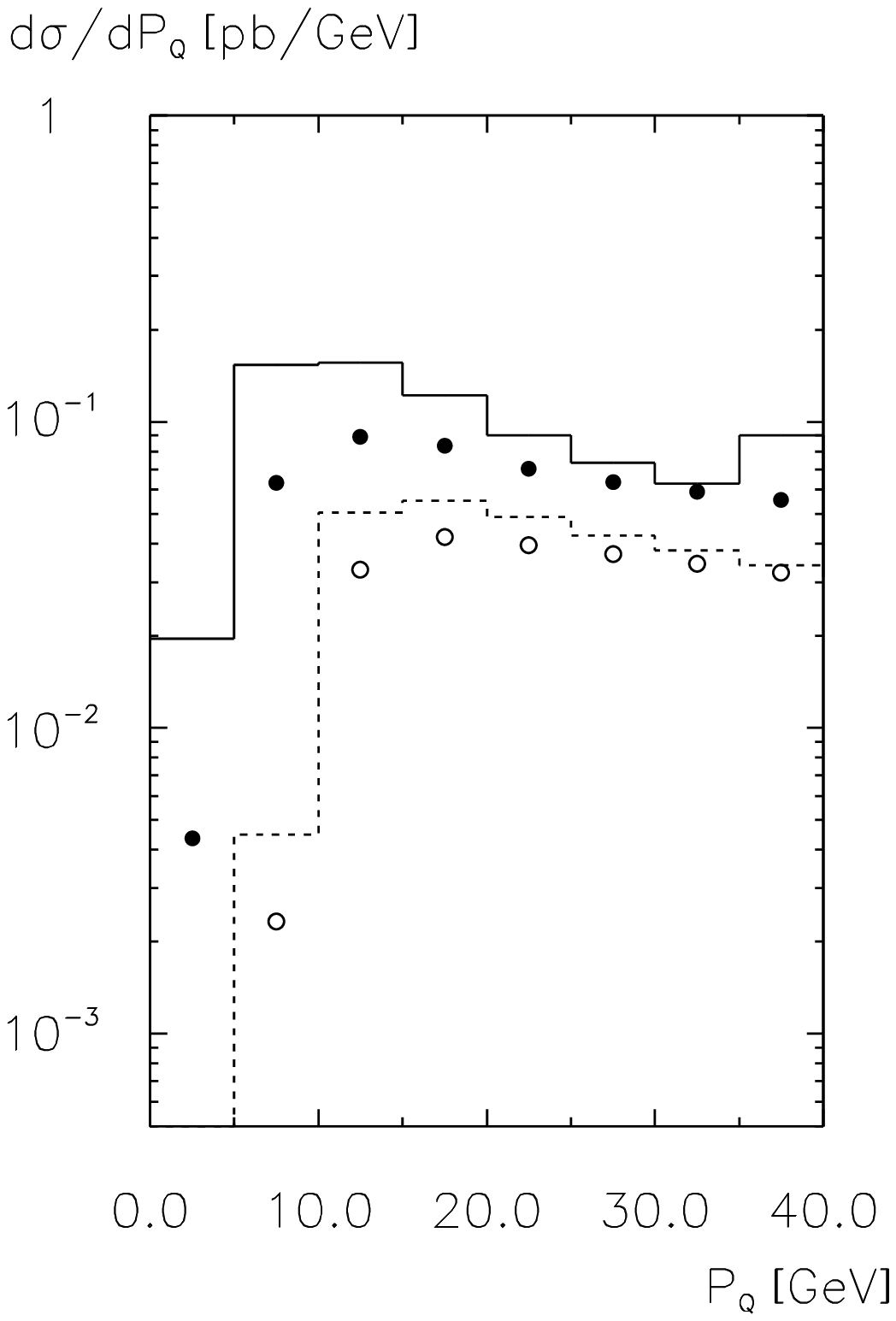}{width=55mm}}
\put( 25,95){\lettlab (a)}

\put( 90,100){\epsfigdg{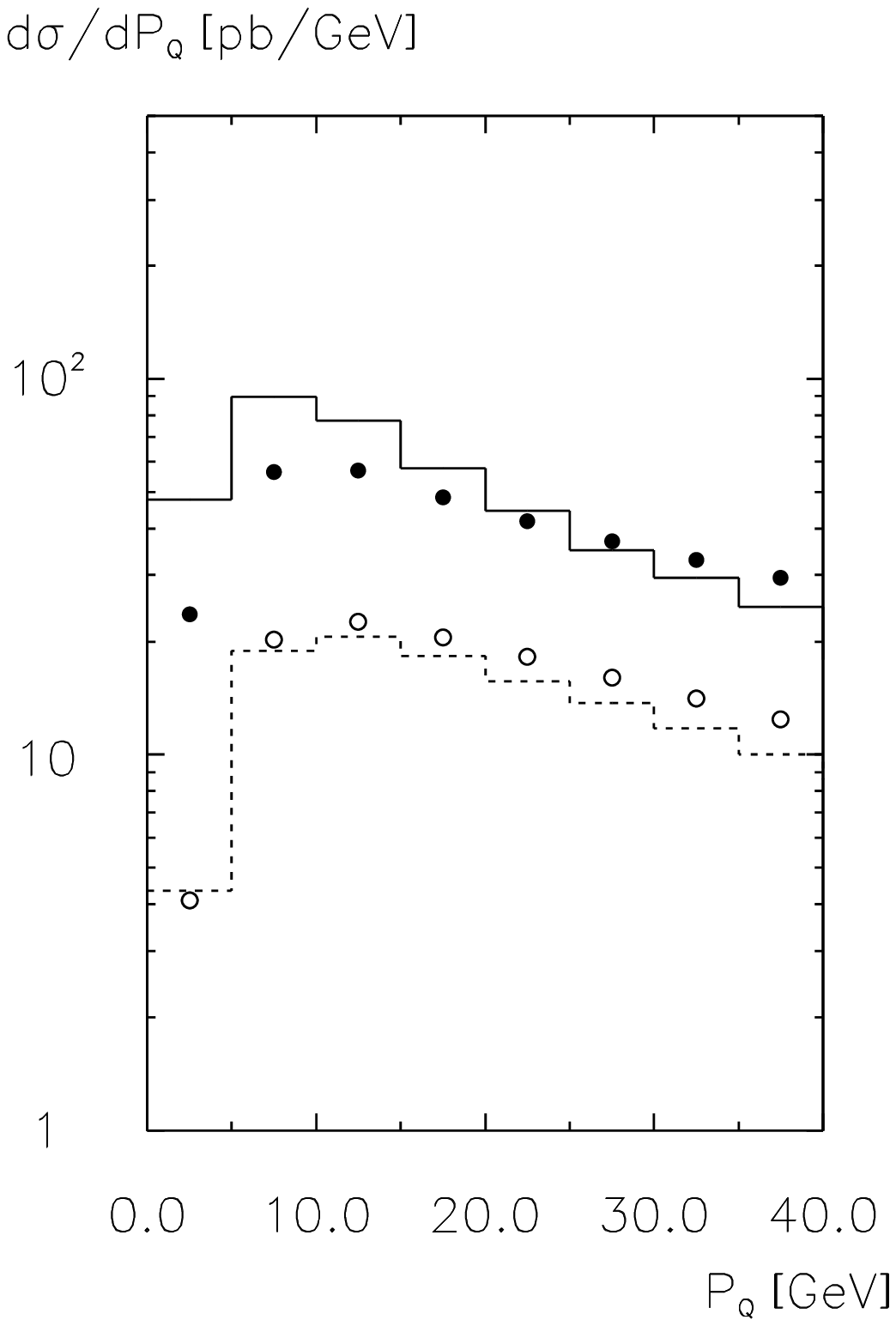}{width=55mm}}
\put(100,95){\lettlab (b)}

\put( 15,5){\epsfigdg{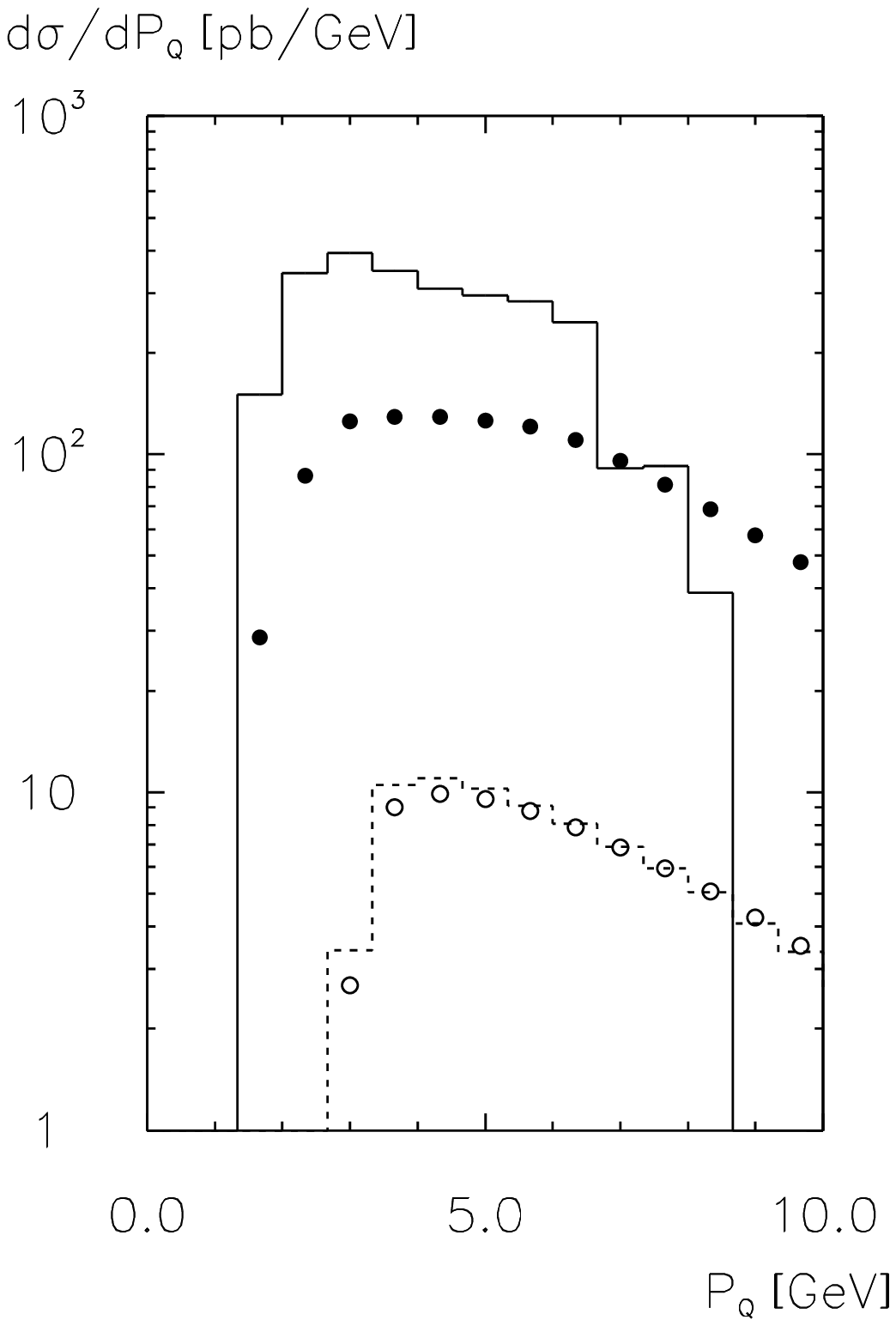}{width=55mm}}
\put( 25,0){\lettlab (c)}

\put( 90,5){\epsfigdg{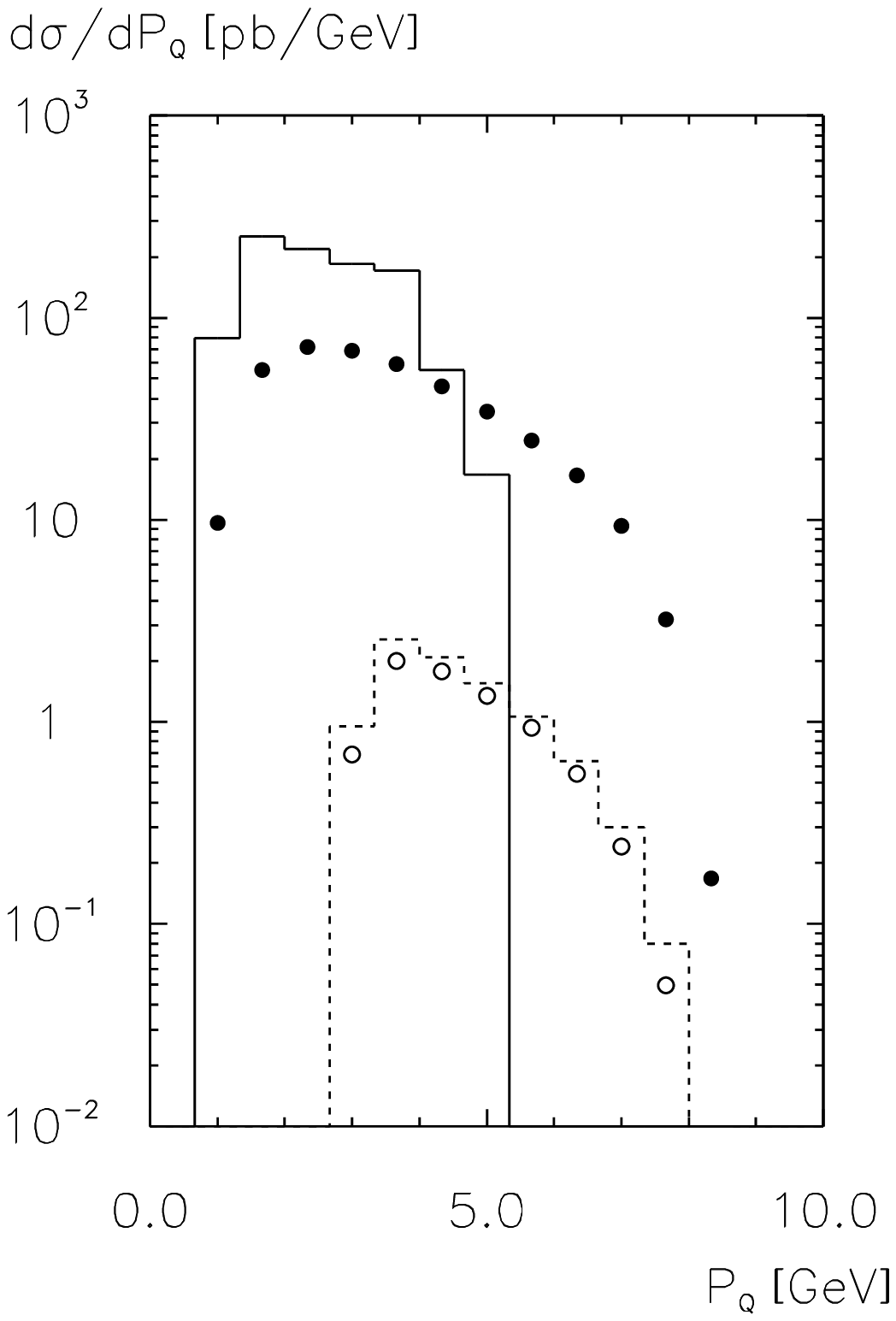}{width=55mm}}
\put(100,0){\lettlab (d)}

\end{picture}
\end{center}
\shiftcaption
\caption[Momentum Distributions
for Heavy-Quark Production]
{\labelmm{HQTMOM} {\it 
Distributions in the momentum~$P_Q$
of the heavy quark
for bottom (a) 
and charm (b) quark production at HERA
and for charm-quark production at E665 (c) and NA47 (d)
up to \porder{\alpha_s},
for $p_T\leq p_{T,\mbox{\scriptsize min}}$ \mbox{(\fullline)} 
and $p_T> p_{T,\mbox{\scriptsize min}}$ \mbox{(\dashline)}. 
Also shown is the distribution from the matrix element
$\gamma^*g\rightarrow Q\overline{Q}$ 
for $p_T\leq p_{T,\mbox{\scriptsize min}}$ \mbox{(\fullcircle)}
and $p_T> p_{T,\mbox{\scriptsize min}}$ \mbox{(\opencircle)}.
The definition of 
$p_{T,\mbox{\scriptsize min}}$ is given in Section~\ref{DiscStudy}.
}}   
\end{figure}

\begin{figure}[htb] \unitlength 1mm
\begin{center}
\dgpicture{159}{90}



\put( 15,5){\epsfigdg{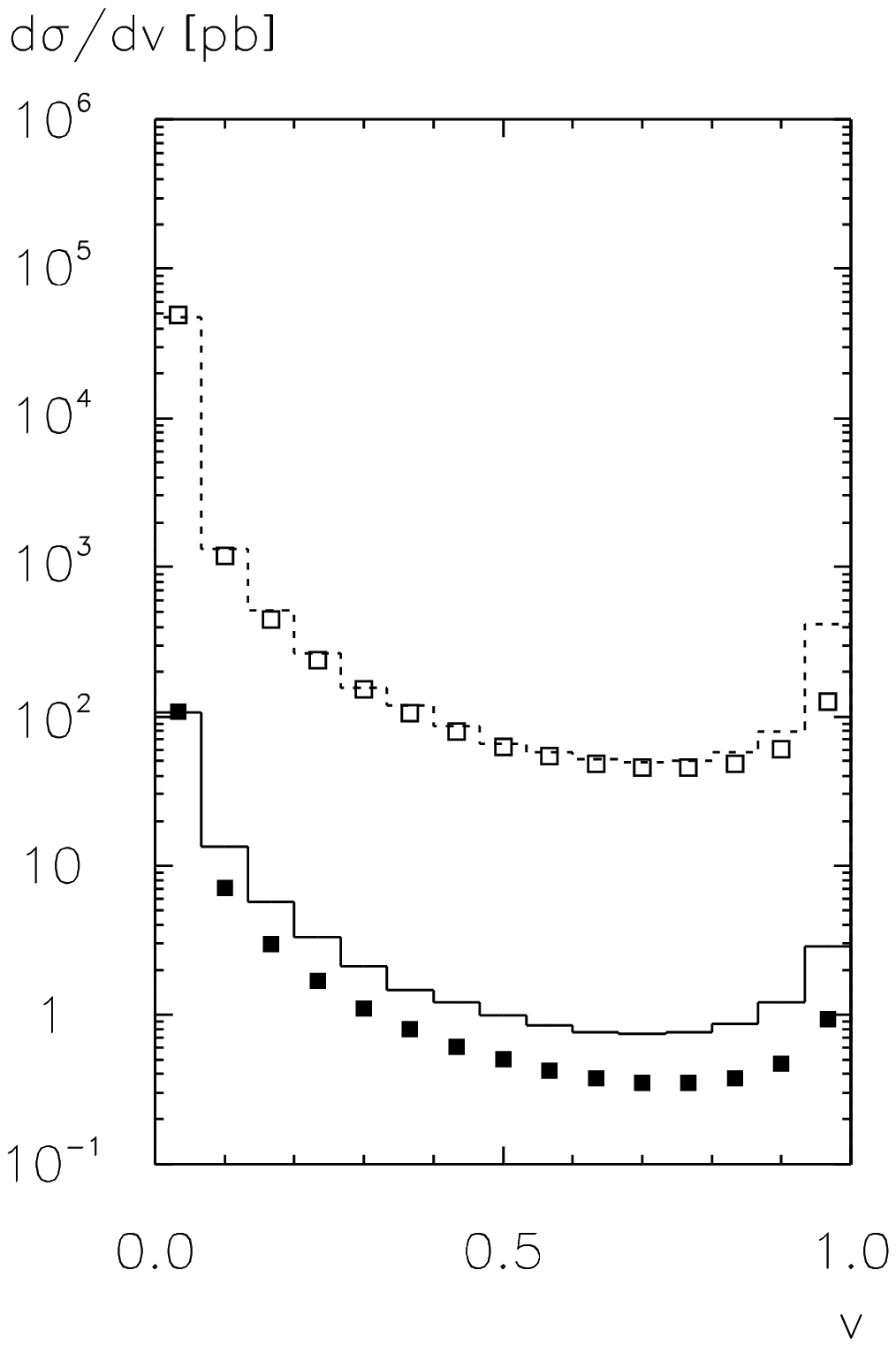}{width=55mm}}
\put( 25,0){\lettlab (a)}

\put( 90,5){\epsfigdg{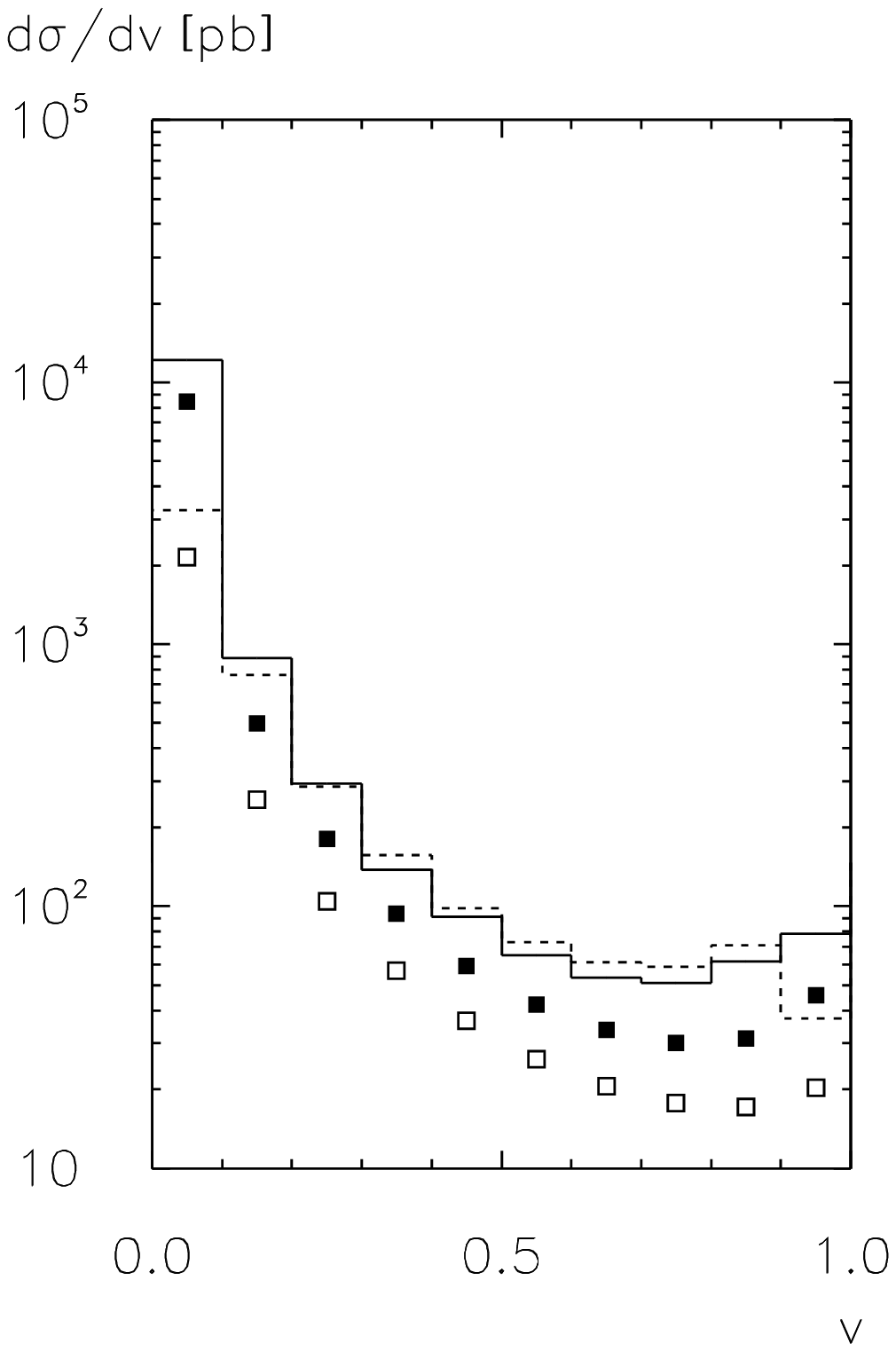}{width=55mm}}
\put(100,0){\lettlab (b)}

\end{picture}
\end{center}
\shiftcaption
\caption[Angular Distributions
for Heavy-Quark Production]
{\labelmm{HQTV1} {\it 
Distributions in the angular variable~$v$
of the heavy quark
for bottom \mbox{(\fullline)} and charm
\mbox{(\dashline)} quark production at HERA (a)
and for charm-quark production at E665 \mbox{(\fullline)}
and NA47 \mbox{(\dashline)} (b)
in next-to-leading order.
The distributions from the \porder{\alpha_s} matrix element
$\gamma^*g\rightarrow Q\overline{Q}$ are:
bottom \mbox{(\fullsquare)} and charm \mbox{(\opensquare)}
quark production at HERA (a) 
and charm-quark
production at E665 \mbox{(\fullsquare)} and NA47 \mbox{(\opensquare)}~(b).
}}   
\end{figure}

\clearpage

\dgsb{Distributions in $x_F$ and $z_{\zshortparallel}$}
\labelm{XFSTUDY}

Differential distributions 
in leading and next-to-leading order 
for the variable~$x_F$ 
are shown in 
Fig.~\ref{HQTXF1}. In the leading-order case, 
the distributions may have a dip at small values of $|x_F|$, due to the
lower cut-off on the energy fraction of the observed particle.
This dip disappears in next-to-leading order, where it is filled 
by events where the heavy quark is produced at large transverse momentum 
with only a small longitudinal momentum component.
The shape of the distribution changes drastically for $x_F>0$ when
going from leading to next-to-leading order, in the cases where the QCD
corrections are large. This effect has its origin in the fact that 
heavy quarks are produced copiously via photon--gluon fusion, which in 
our notation is next-to-leading order. Since the event topology is 
completely different from that in leading order, it is expected that
the distributions look different\footnote{
We note that for E665 and NA47 the entry in the bin containing 
$x_F=1$ is negative. This is due to the collinear subtraction, 
and could in principle be controlled by adjusting the bin size
accordingly. Since we are mainly interested in the target fragmentation
region, this problem shall not concern us here.
}.
Also shown in Fig.~\ref{HQTXF1} is the distribution for the 
$\gamma^* g\rightarrow Q\overline{Q}$ matrix element.
Qualitatively, it is closer to the result from the fragmentation function
approach in next-to-leading order than to the one in leading order.
Quantitatively, mainly in the current fragmentation region, 
there are however large differences.
A comparison for the regions of small and large transverse momenta
is shown in Fig.~\ref{HQTXF2}. We note that the shapes
of the distributions in the target fragmentation region are comparable
for $x_F\lesssim 0.2$.
For $p_T>p_{T,\mbox{\scriptsize min}}$, the results from the fragmentation
function approach and from the 
$\gamma^*g\rightarrow Q\overline{Q}$
matrix element agree very well.

The distributions for the variable $z_{\zshortparallel}$ are shown
in Figs.~\ref{HQTZC} and~\ref{HQTZT} for $z_{\zshortparallel}\geq 0$ and
$z_{\zshortparallel}< 0$, respectively, and split according to 
$p_T\leq p_{T,\mbox{\scriptsize min}}$ and $p_T>p_{T,\mbox{\scriptsize min}}$.
The pattern follows the one already encountered in the case of the 
$x_F$-distributions. For $p_T>p_{T,\mbox{\scriptsize min}}$,
there is reasonable
agreement between the fragmentation function approach and the
$\gamma^* g\rightarrow Q\overline{Q}$ matrix element.
For $p_T\leq p_{T,\mbox{\scriptsize min}}$, in the 
region $z_{\zshortparallel}<0$,
the shapes corresponding to the two approaches are comparable, although
the fragmentation function approach leads to considerably 
larger cross sections.
For $p_T\leq p_{T,\mbox{\scriptsize min}}$ and $z_{\zshortparallel}\geq 0$ 
neither the shapes nor the absolute sizes compare well.  
We wish to remark that the dip in the $z_{\zshortparallel}$-distribution for 
$p_T\leq p_{T,\mbox{\scriptsize min}}$ for 
bottom-quark production at HERA comes from the requirement that
the observed heavy quark 
has to carry a minimum momentum fraction.

We now turn to the case of intrinsic heavy quarks.
In Section~\ref{hqtffmic} we have introduced a model for the non-perturbative
piece of the target fragmentation functions based on the 
hypothesis of intrinsic heavy quarks in the proton. 
We set the parton densities to 
$f=f^{\mbox{\scriptsize (GRV)}}+f^\IHQ$, where
$f=f^{\mbox{\scriptsize (GRV)}}$ is the parton density parametrization
by Gl\"uck, Reya and Vogt\footnote{In principle, the sum violates
various sum rules, taken into account as constraints in the GRV fit.
Since the intrinsic heavy-quark content is small, this violation may be
safely neglected here.}, 
and the target fragmentation functions
to $M=M^{(P)}+M^{\IHQ}$.
In principle, the parton densities $f^\IHQ$ contribute also to 
the perturbative heavy-quark target fragmentation functions $M^{(P)}$.
We neglect this contribution here.

Figure~\ref{HQTXF1I} shows a comparison of the 
$x_F$-distribution of the produced heavy quark in next-to-leading
order
without
and with 
intrinsic heavy quarks. As is expected from the hardness of the corresponding
heavy-quark target fragmentation functions in the momentum-fraction variable
of the observed heavy quarks, the distributions including intrinsic
heavy quarks extend to larger negative 
values of~$x_F$ than those without intrinsic
heavy quarks. For $x_F<0.2$ they are, in the cases studied here, 
always dominant.
These results should, however, be interpreted carefully. First of all, 
the distribution that we used as input in the scale evolution is only a
rough approximation, where mass effects have been neglected.
Moreover, intrinsic heavy quarks are a non-perturbative phenomenon, 
and it is not clear how they fragment if the proton is 
hit by a high-$Q^2$ probe
when neither the heavy quark nor the heavy antiquark
participate in the hard scattering process. 
It is possible that the $Q\overline{Q}$-pair,
being a quantum fluctuation of the $|uud\rangle$ state,
simply recombines, in the cases where a light quark or a gluon initiates the
hard scattering process. The results shown here are based on the naive
assumption that every intrinsic heavy quark in the proton Fock state
$|uudQ\overline{Q}\rangle$ is seen in the final state. It is beyond the scope
of the present study to consider this complicated non-perturbative
problem. It would, however, be very interesting to analyse 
this aspect of experimental
data, in order to see whether there is any excess with
respect to the perturbative contribution.

\begin{figure}[htb] \unitlength 1mm
\begin{center}
\dgpicture{159}{185}

\put( 15,100){\epsfigdg{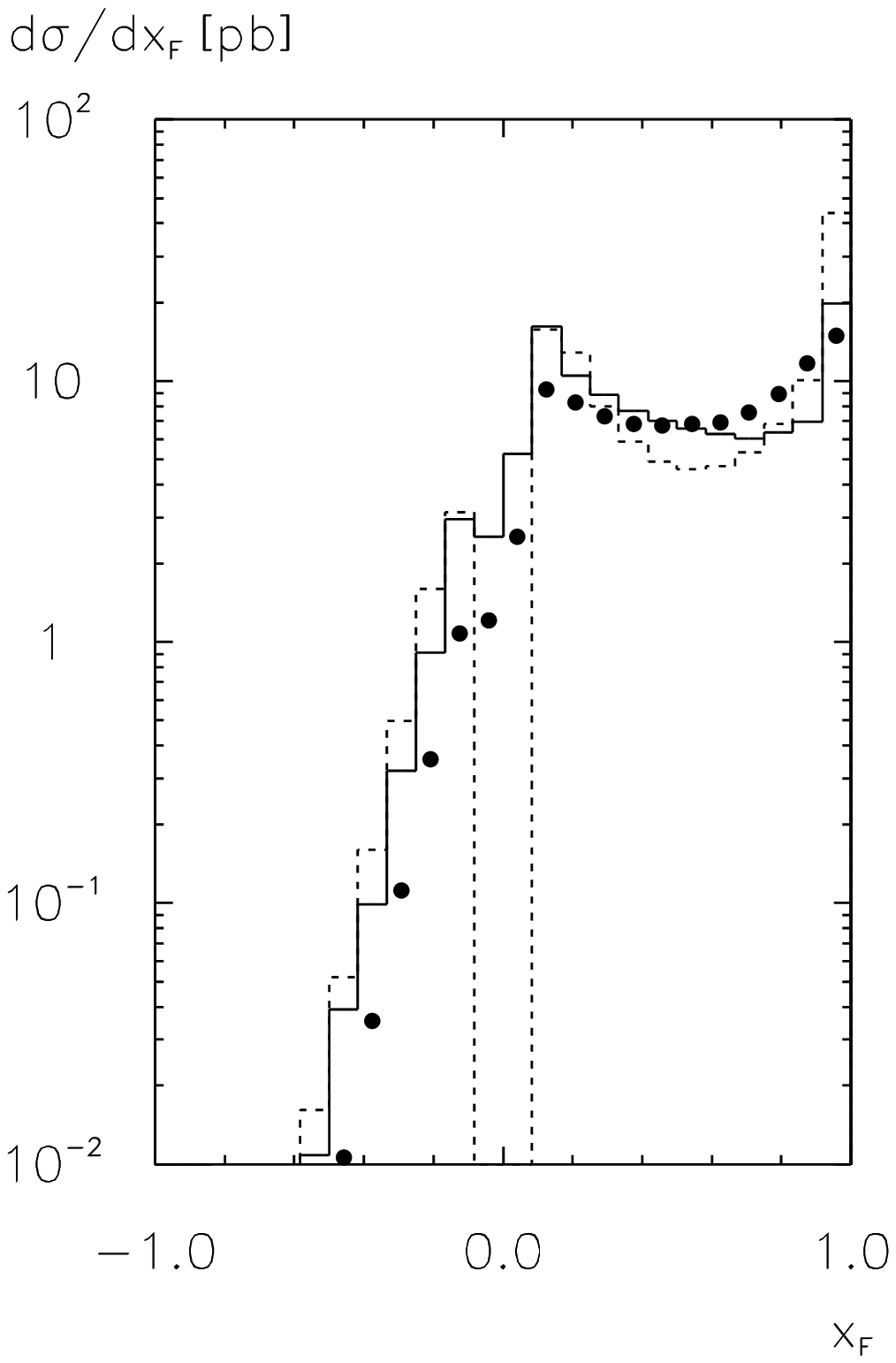}{width=55mm}}
\put( 25,95){\lettlab (a)}

\put( 90,100){\epsfigdg{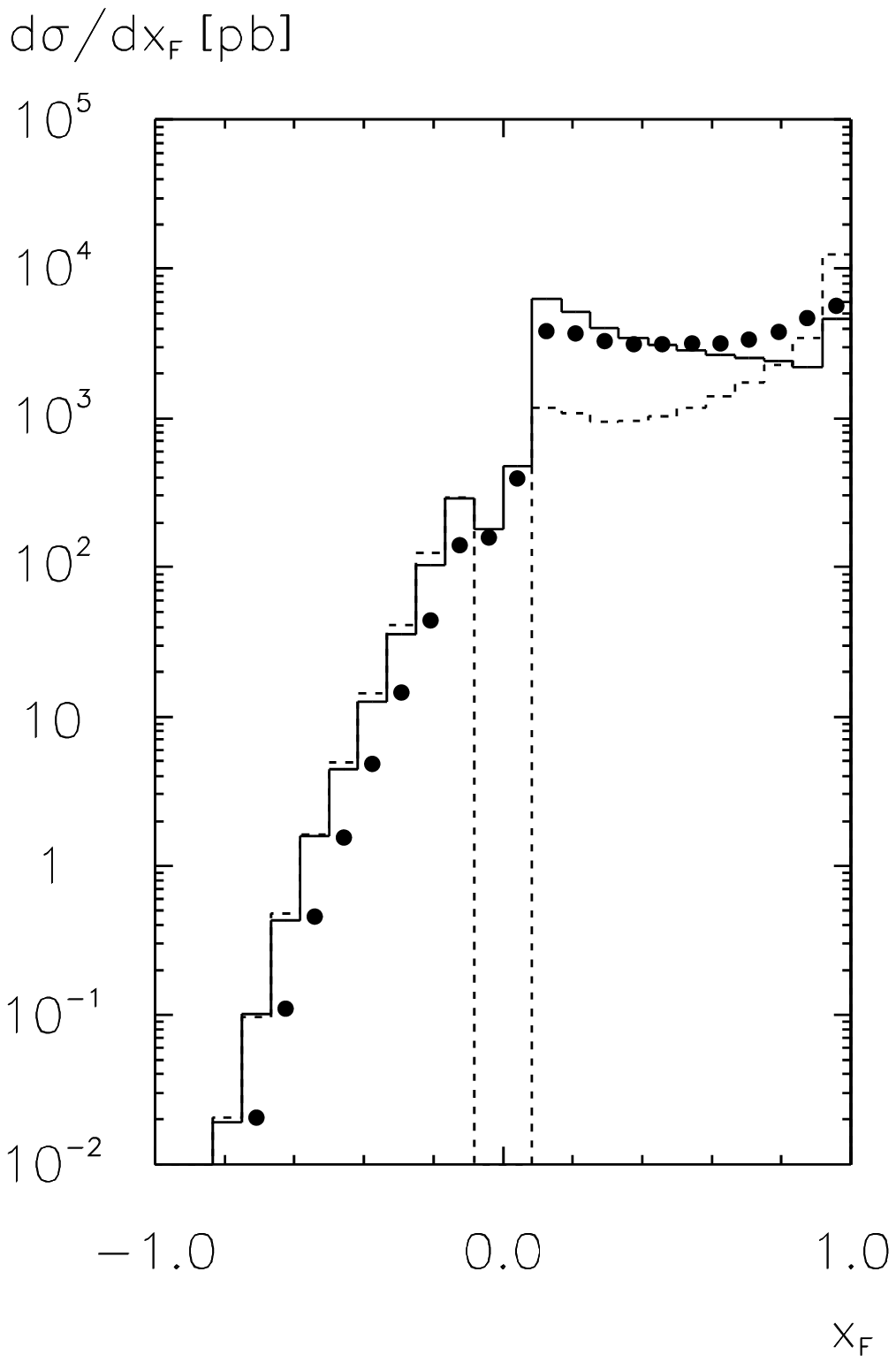}{width=55mm}}
\put(100,95){\lettlab (b)}

\put( 15,5){\epsfigdg{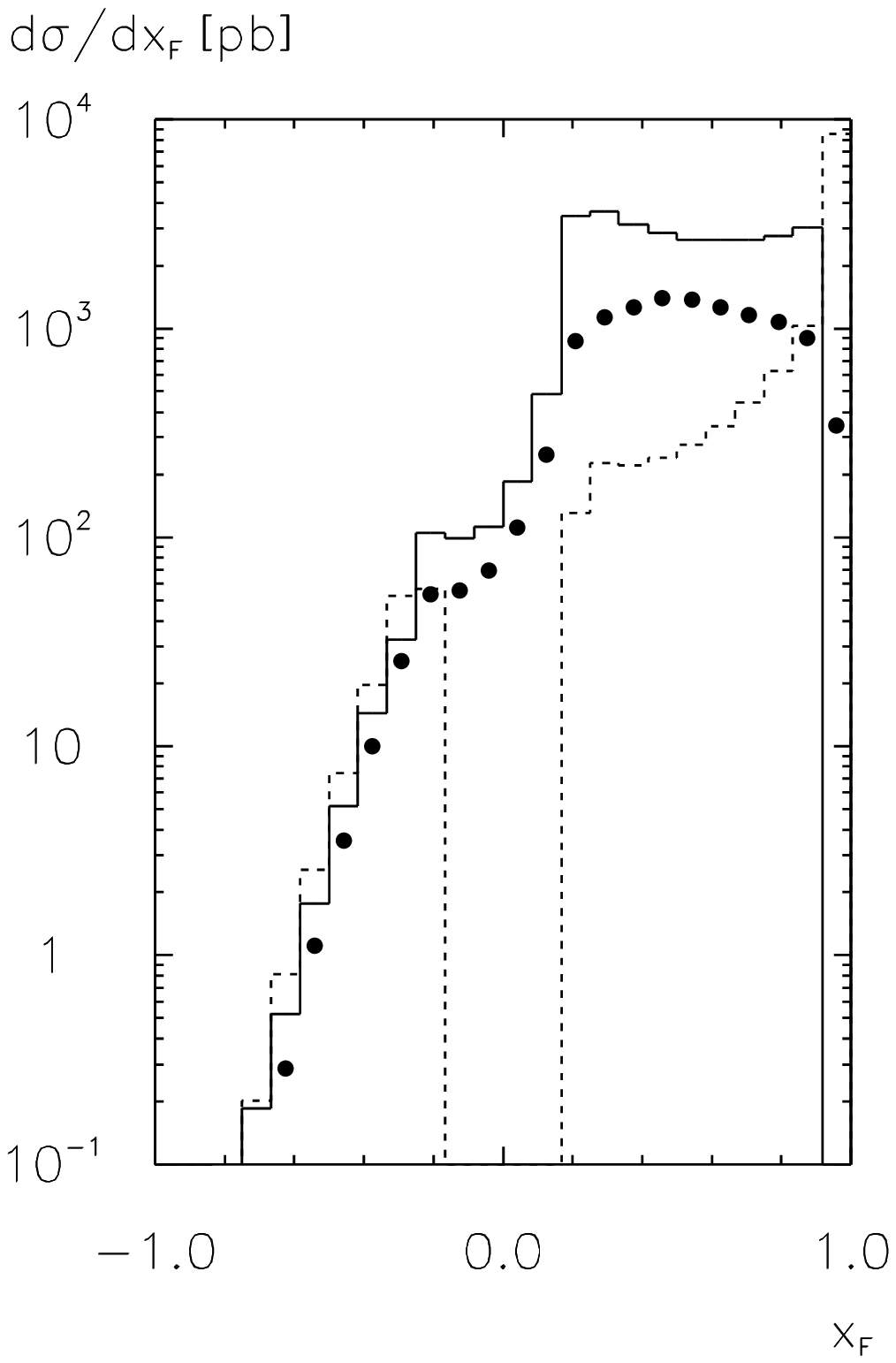}{width=55mm}}
\put( 25,0){\lettlab (c)}

\put( 90,5){\epsfigdg{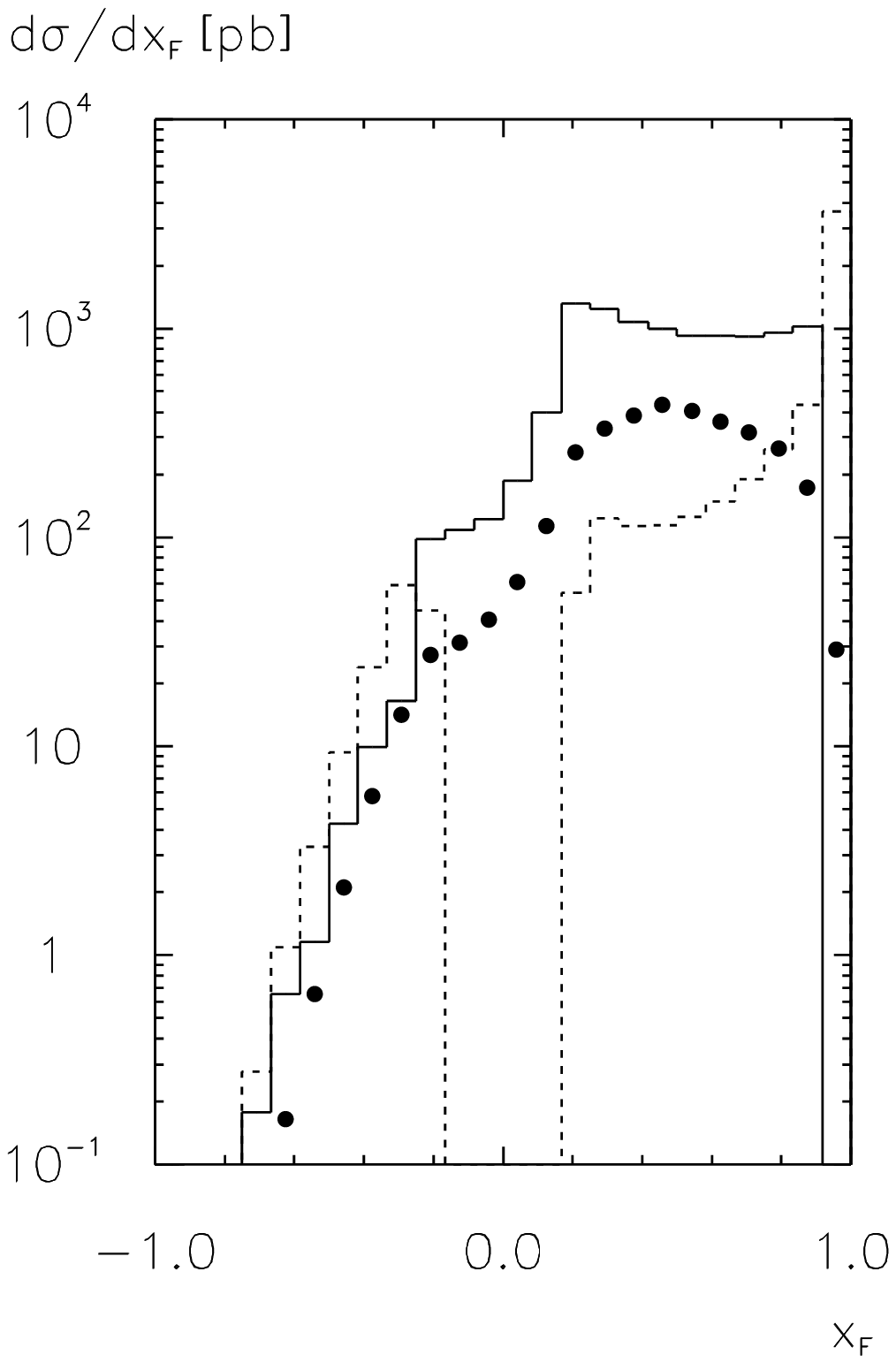}{width=55mm}}
\put(100,0){\lettlab (d)}

\end{picture}
\end{center}
\shiftcaption
\caption[Distributions in~$x_F$ 
for Heavy-Quark Production; LO vs.\ NLO]
{\labelmm{HQTXF1} {\it 
Distributions in~$x_F$ 
for bottom (a) 
and charm (b) quark production at HERA
and for charm-quark production at E665 (c) and NA47 (d),
in leading order \mbox{(\dashline)} 
and in next-to-leading order \mbox{(\fullline)}. 
Also shown is the distribution from the matrix element
$\gamma^*g\rightarrow Q\overline{Q}$ \mbox{(\fullcircle)}.
}}   
\end{figure}

\begin{figure}[htb] \unitlength 1mm
\begin{center}
\dgpicture{159}{185}

\put( 15,100){\epsfigdg{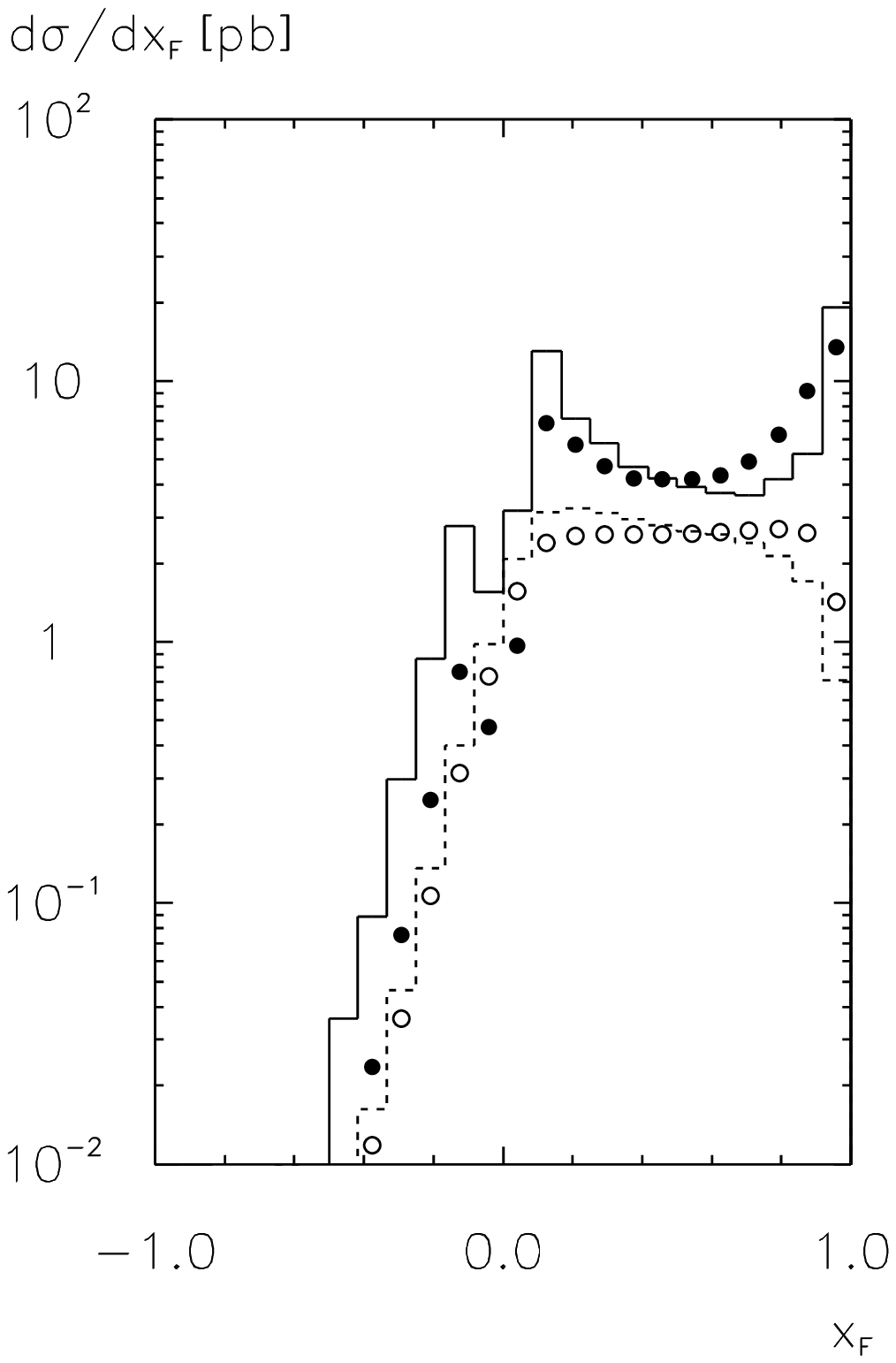}{width=55mm}}
\put( 25,95){\lettlab (a)}

\put( 90,100){\epsfigdg{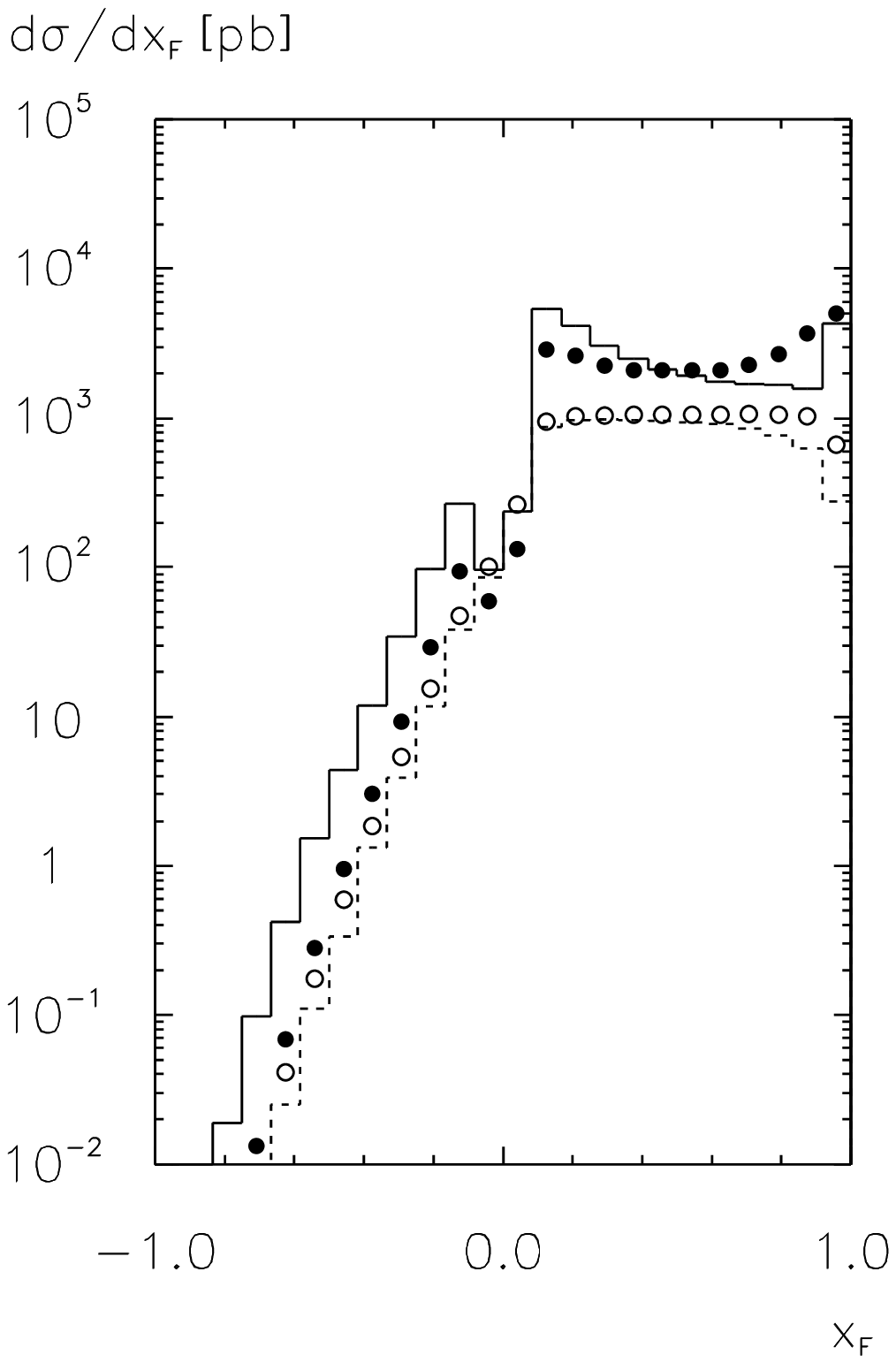}{width=55mm}}
\put(100,95){\lettlab (b)}

\put( 15,5){\epsfigdg{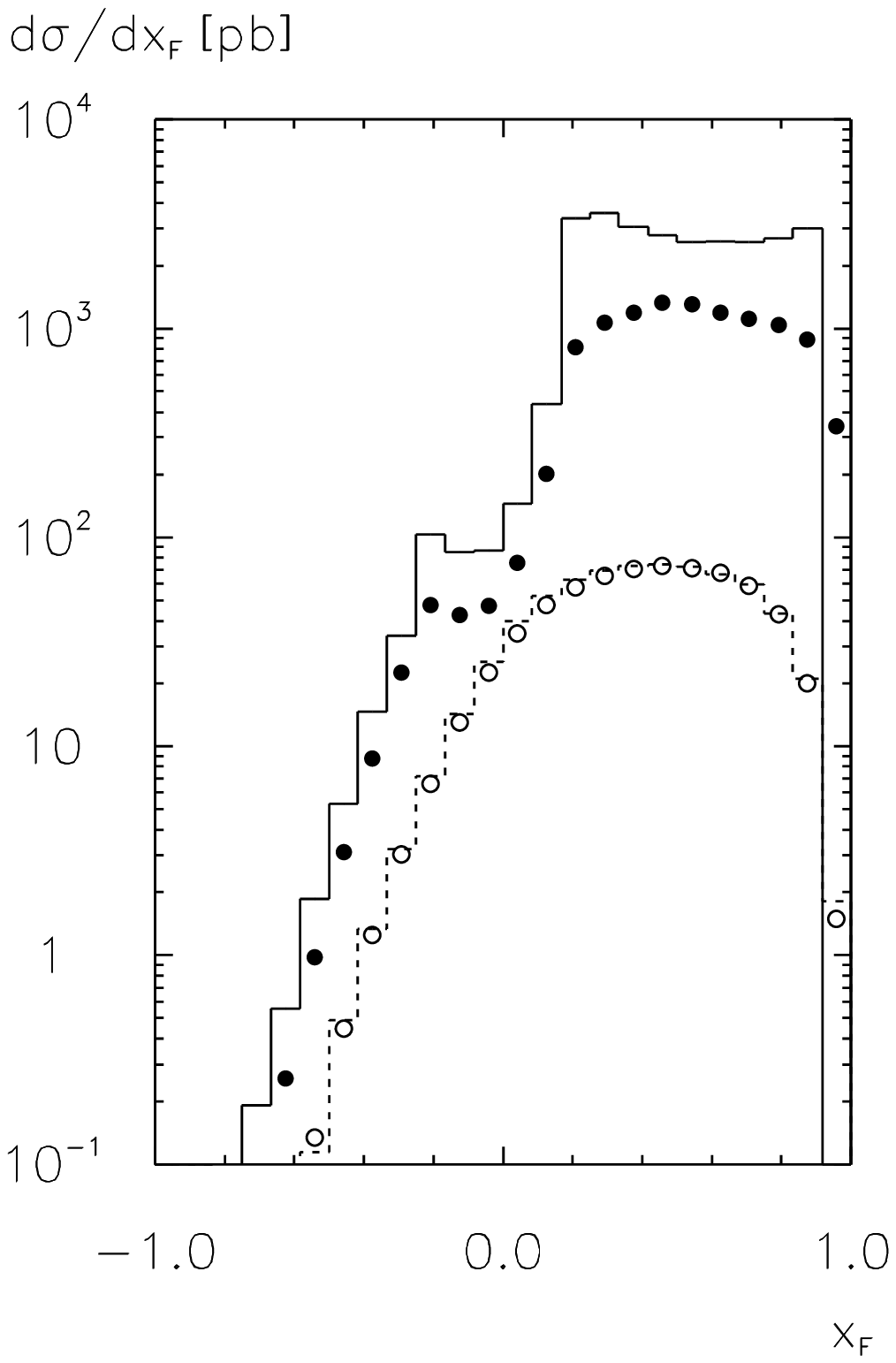}{width=55mm}}
\put( 25,0){\lettlab (c)}

\put( 90,5){\epsfigdg{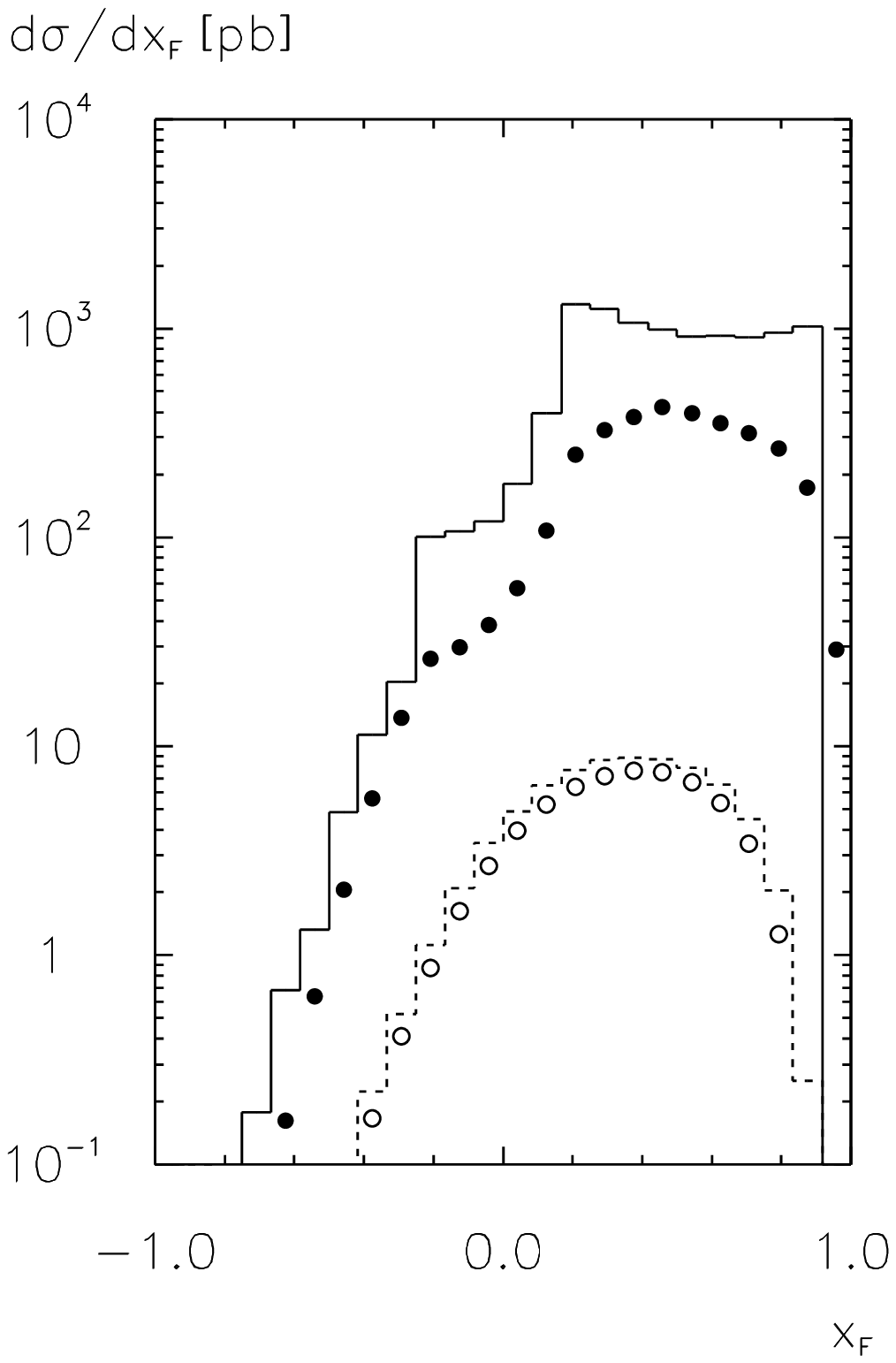}{width=55mm}}
\put(100,0){\lettlab (d)}

\end{picture}
\end{center}
\shiftcaption
\caption[Distributions in~$x_F$ 
for Heavy-Quark Production; Small $p_T$ vs.\ Large~$p_T$]
{\labelmm{HQTXF2} {\it 
Distributions in~$x_F$ 
for bottom (a) 
and charm (b) quark production at HERA
and for charm-quark production at E665 (c) and NA47 (d)
up to \porder{\alpha_s},
for $p_T\leq p_{T,\mbox{\scriptsize min}}$ \mbox{(\fullline)} 
and $p_T> p_{T,\mbox{\scriptsize min}}$ \mbox{(\dashline)}. 
Also shown is the distribution from the matrix element
$\gamma^*g\rightarrow Q\overline{Q}$ 
for $p_T\leq p_{T,\mbox{\scriptsize min}}$ \mbox{(\fullcircle)}
and $p_T> p_{T,\mbox{\scriptsize min}}$ \mbox{(\opencircle)}.
The definition of 
$p_{T,\mbox{\scriptsize min}}$ is given in Section~\ref{DiscStudy}.
}}   
\end{figure}

\begin{figure}[htb] \unitlength 1mm
\begin{center}
\dgpicture{159}{185}

\put( 15,100){\epsfigdg{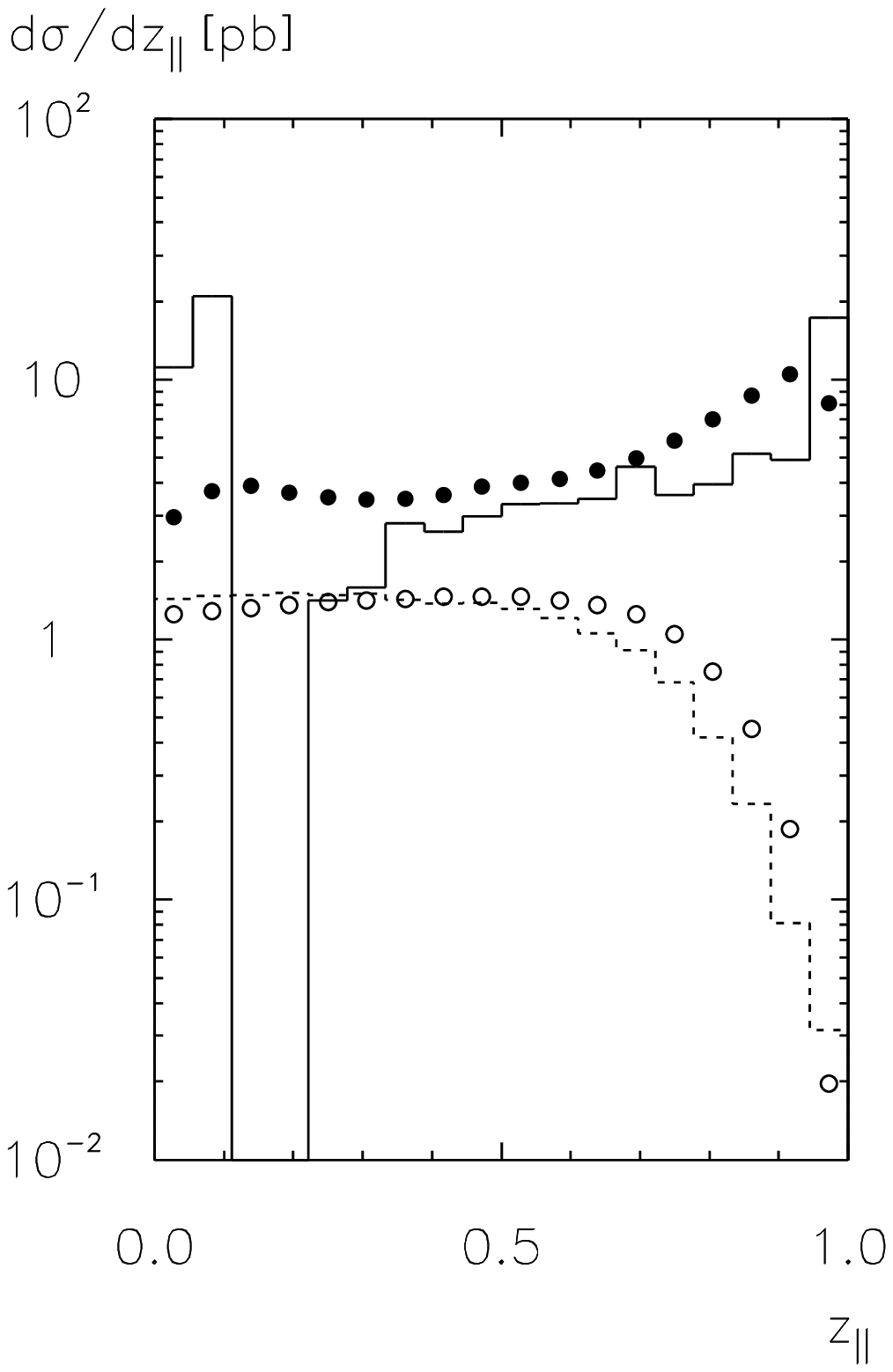}{width=55mm}}
\put( 25,95){\lettlab (a)}

\put( 90,100){\epsfigdg{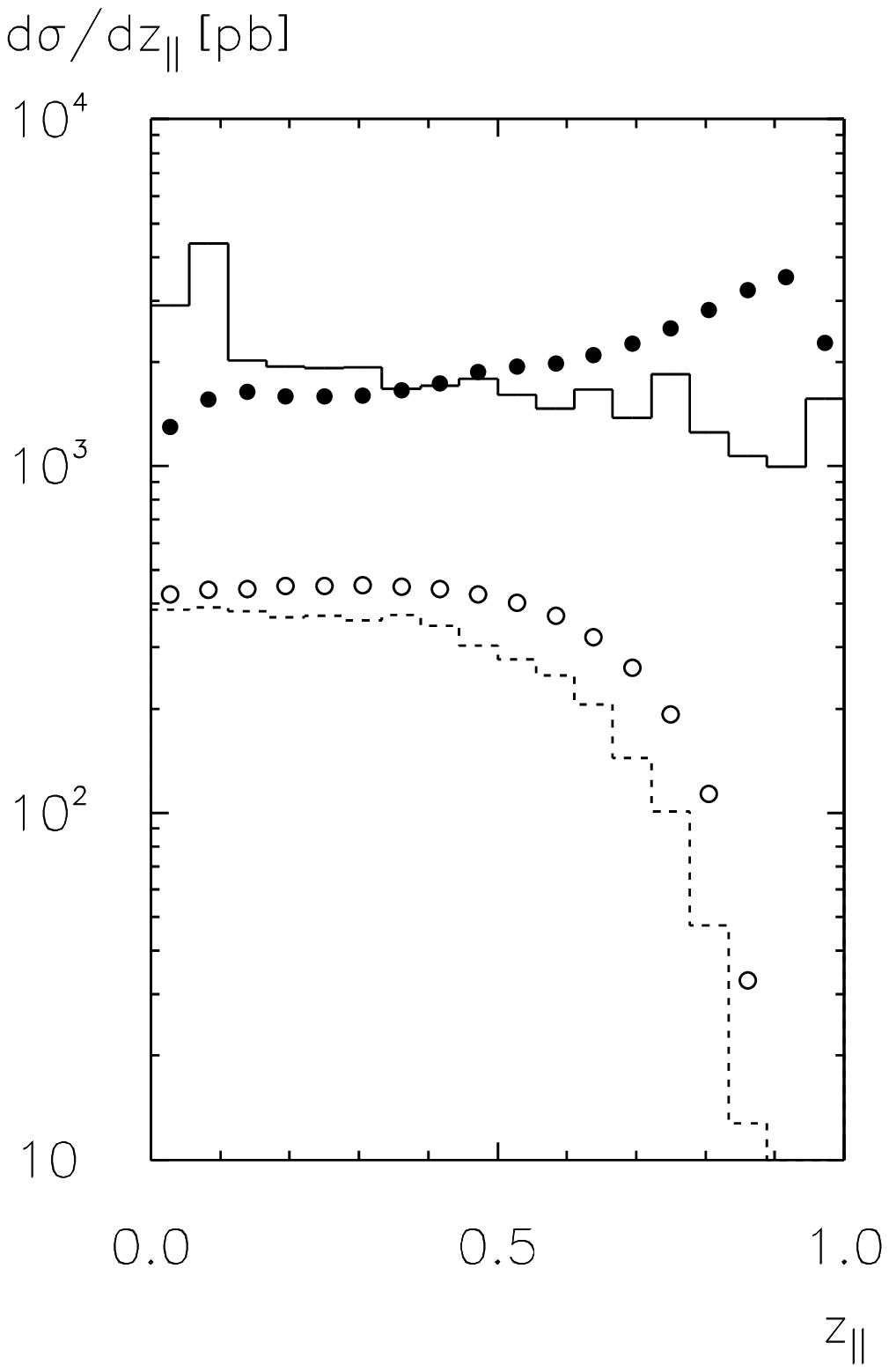}{width=55mm}}
\put(100,95){\lettlab (b)}

\put( 15,5){\epsfigdg{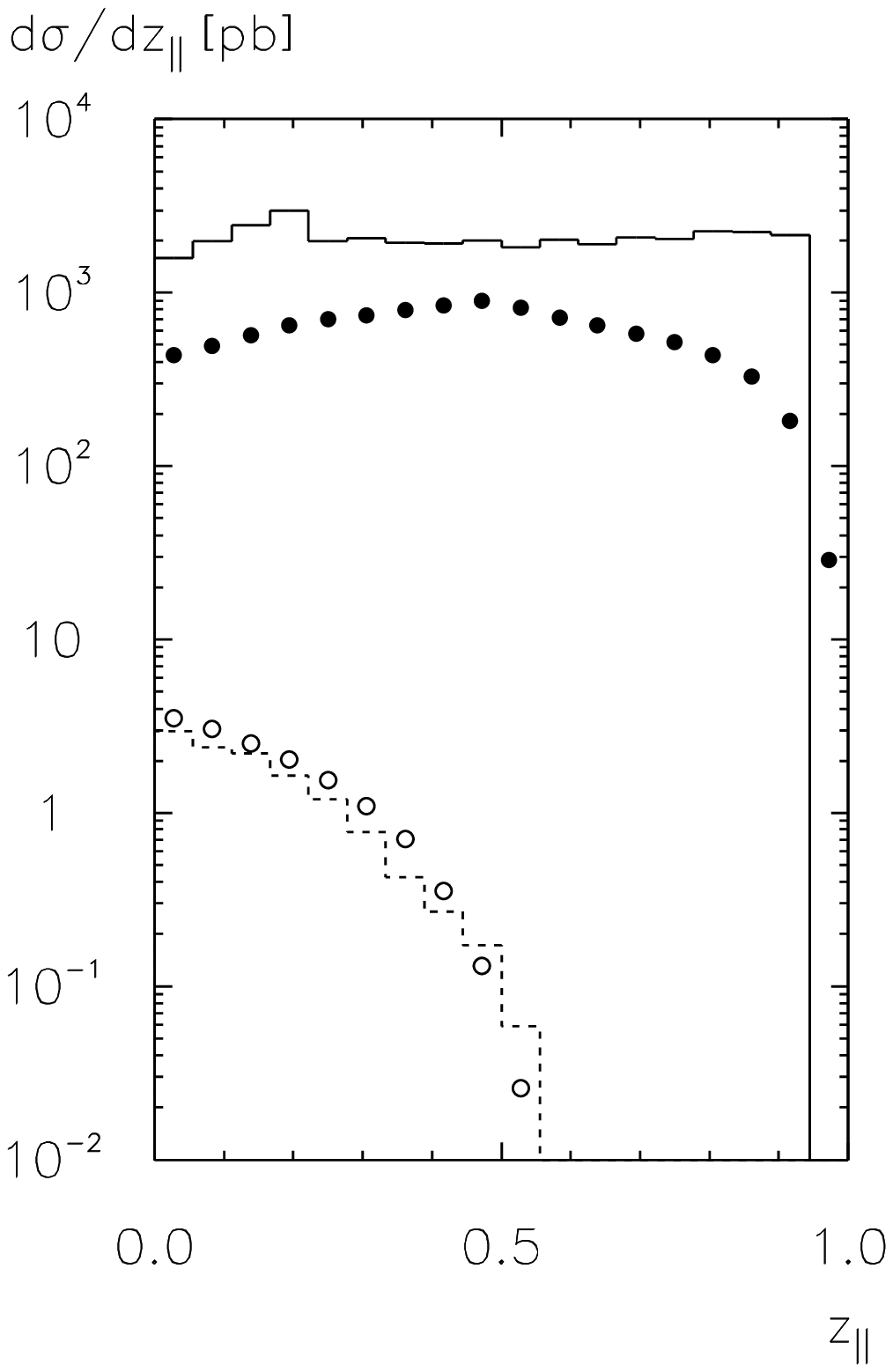}{width=55mm}}
\put( 25,0){\lettlab (c)}

\put( 90,5){\epsfigdg{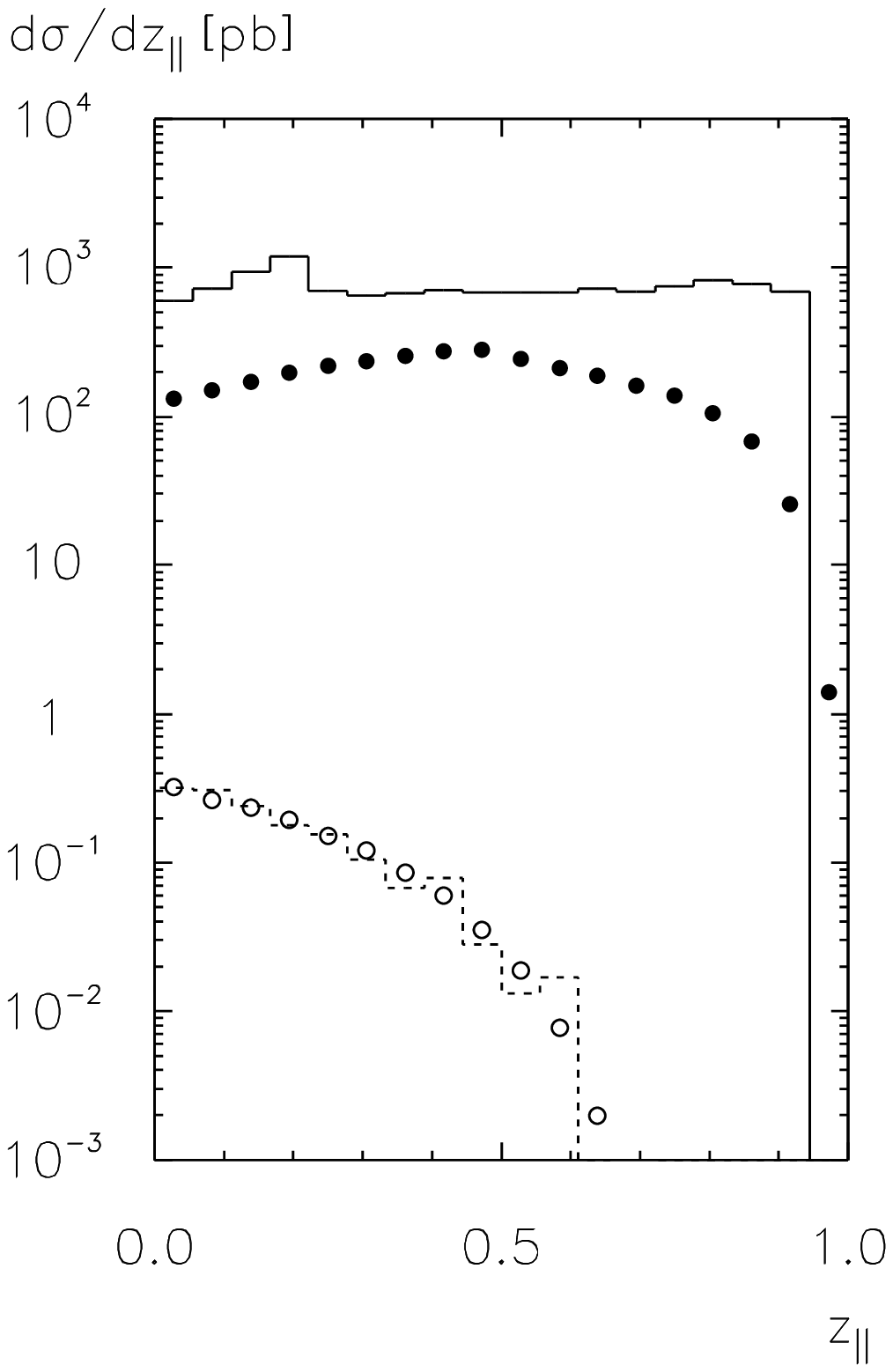}{width=55mm}}
\put(100,0){\lettlab (d)}

\end{picture}
\end{center}
\shiftcaption
\caption[Distributions in $z_{\zshortparallel}$ 
for Heavy-Quark Production: $z_{\zshortparallel}\geq 0$]
{\labelmm{HQTZC} {\it 
Distributions in~$z_{\zshortparallel}$ 
for bottom (a) 
and charm (b) quark production at HERA
and for charm-quark production at E665 (c) and NA47 (d)
for $z_{\zshortparallel}\geq 0$
up to \porder{\alpha_s},
for $p_T\leq p_{T,\mbox{\scriptsize min}}$ \mbox{(\fullline)} 
and $p_T> p_{T,\mbox{\scriptsize min}}$ \mbox{(\dashline)}. 
Also shown is the distribution from the matrix element
$\gamma^*g\rightarrow Q\overline{Q}$ 
for $p_T\leq p_{T,\mbox{\scriptsize min}}$ \mbox{(\fullcircle)}
and $p_T> p_{T,\mbox{\scriptsize min}}$ \mbox{(\opencircle)}.
The definition of
$p_{T,\mbox{\scriptsize min}}$ is given in Section~\ref{DiscStudy}.
}}   
\end{figure}

\begin{figure}[htb] \unitlength 1mm
\begin{center}
\dgpicture{159}{185}

\put( 15,100){\epsfigdg{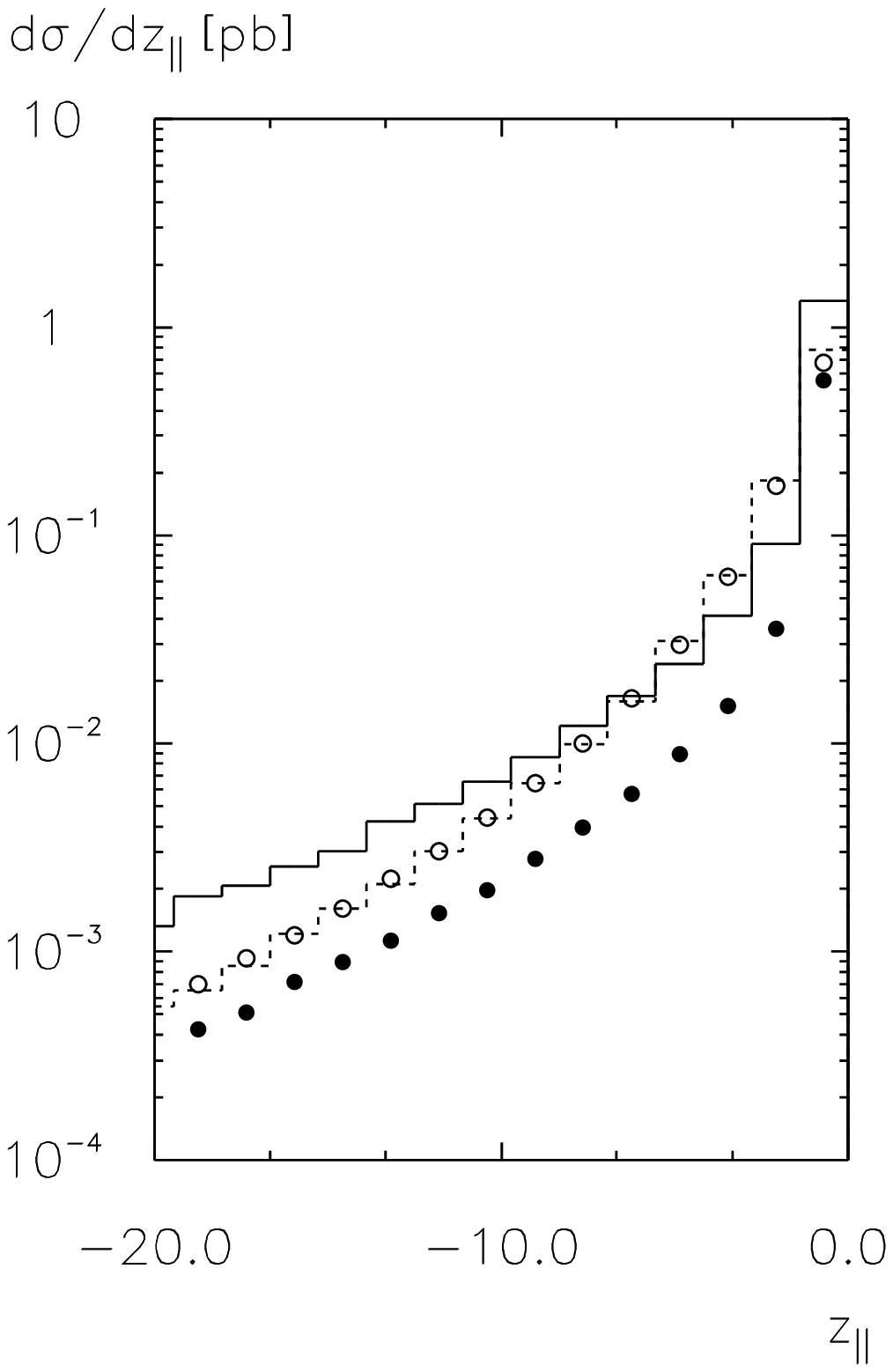}{width=55mm}}
\put( 25,95){\lettlab (a)}

\put( 90,100){\epsfigdg{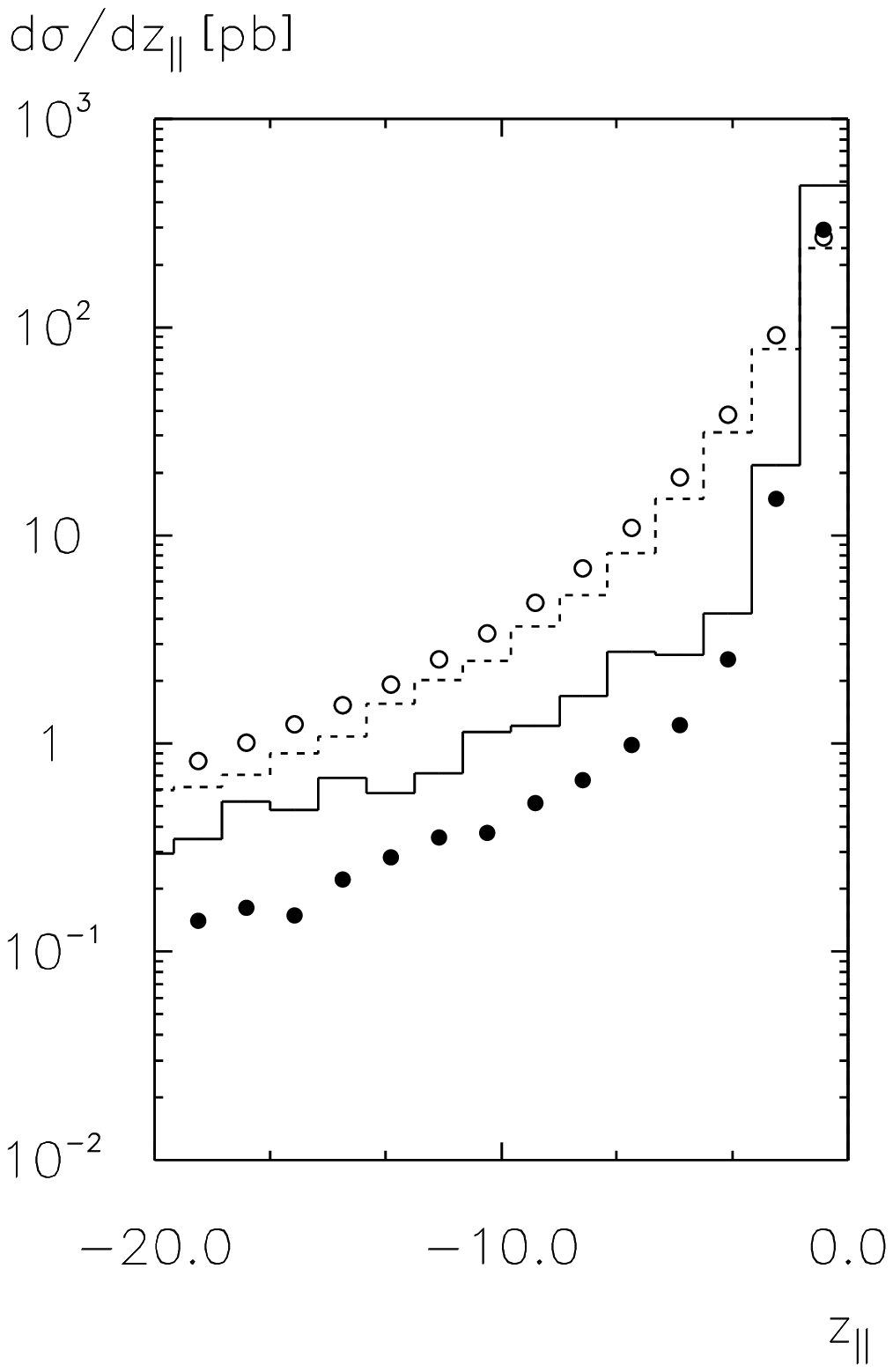}{width=55mm}}
\put(100,95){\lettlab (b)}

\put( 15,5){\epsfigdg{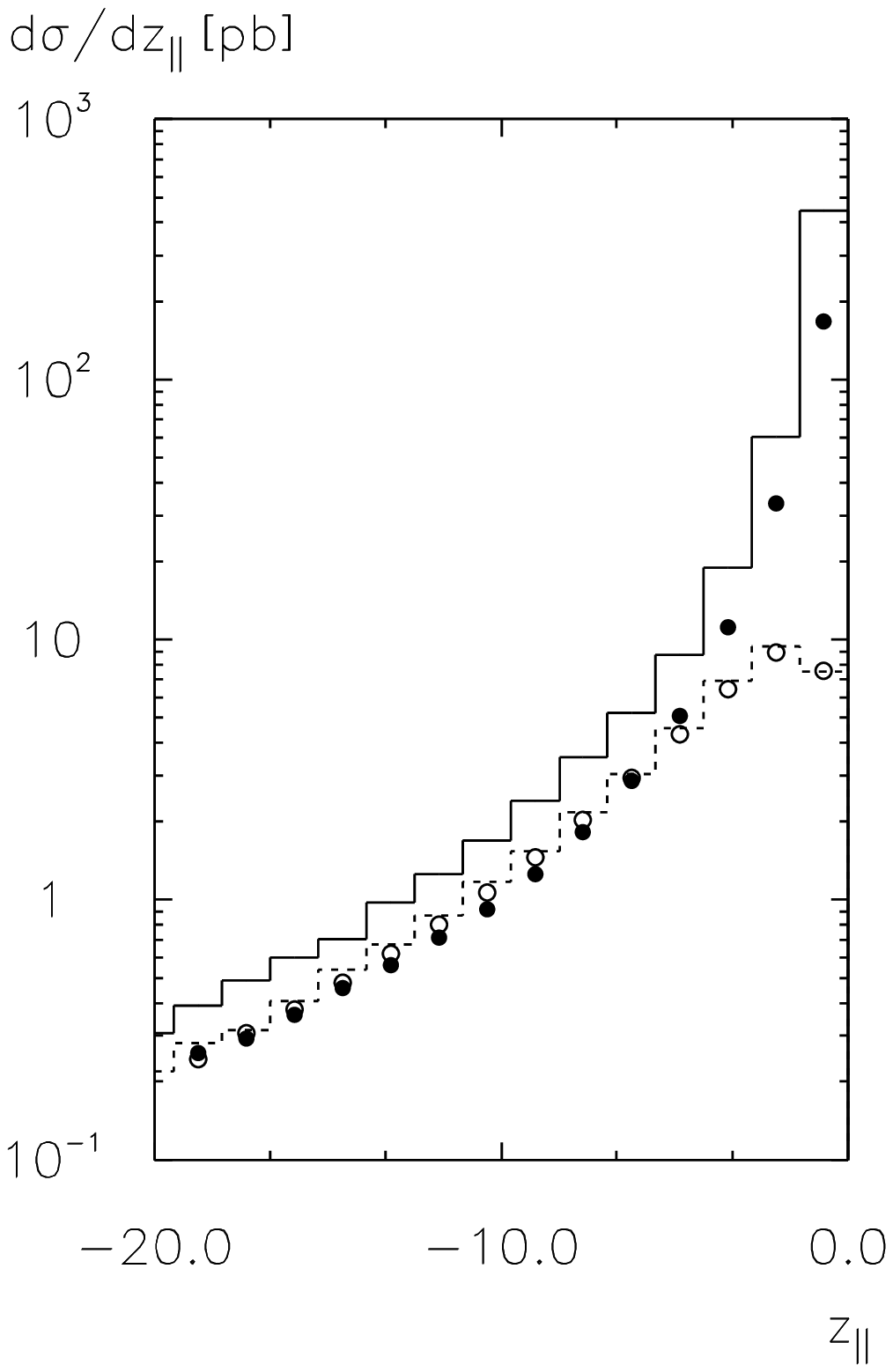}{width=55mm}}
\put( 25,0){\lettlab (c)}

\put( 90,5){\epsfigdg{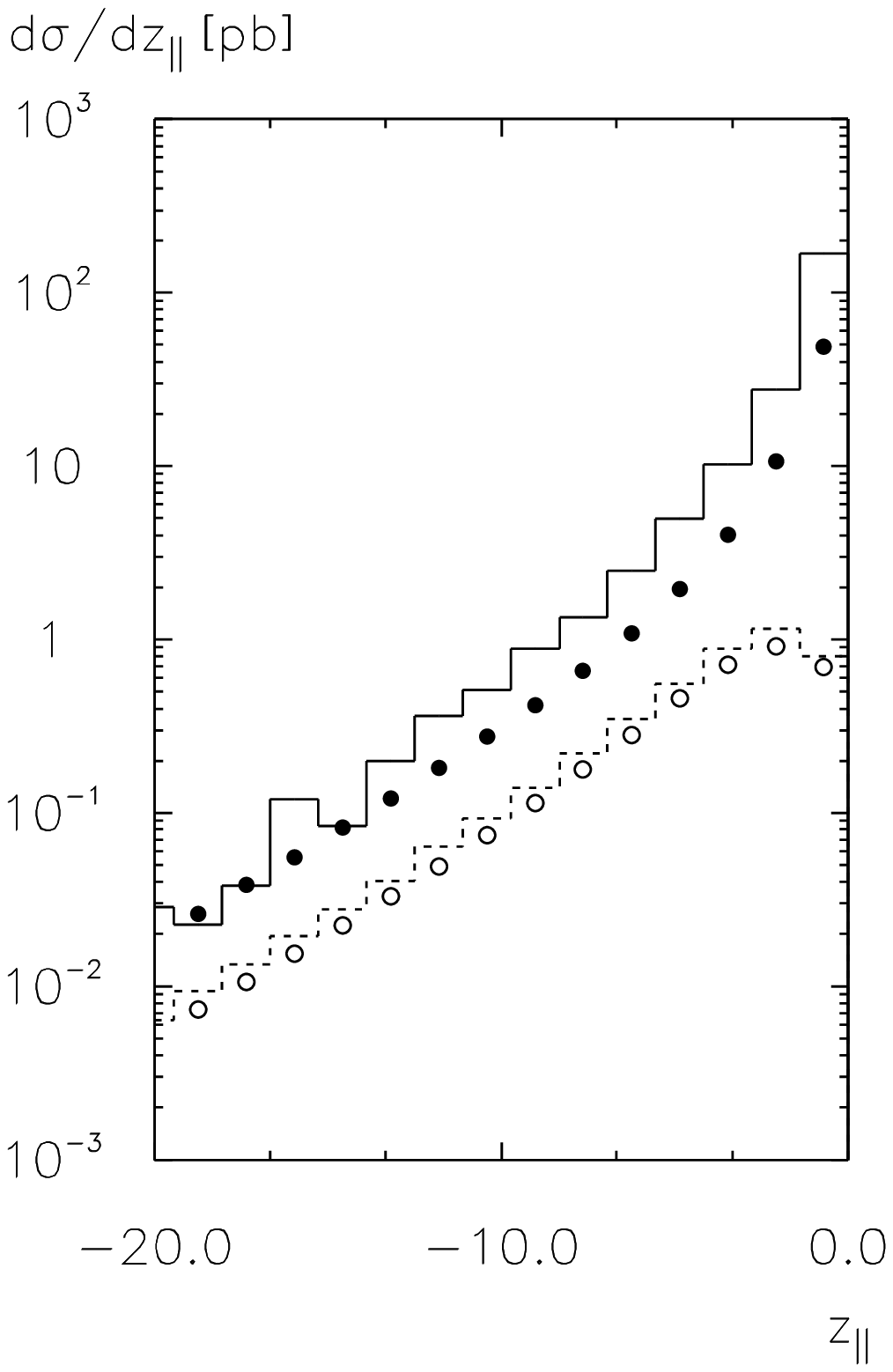}{width=55mm}}
\put(100,0){\lettlab (d)}

\end{picture}
\end{center}
\shiftcaption
\caption[Distributions in~$z_{\zshortparallel}$ 
for Heavy-Quark Production: $z_{\zshortparallel}< 0$]
{\labelmm{HQTZT} {\it 
Distributions in~$z_{\zshortparallel}$ 
for bottom (a) 
and charm (b) quark production at HERA
and for charm-quark production at E665 (c) and NA47 (d)
for $z_{\zshortparallel}< 0$
up to \porder{\alpha_s},
for $p_T\leq p_{T,\mbox{\scriptsize min}}$ \mbox{(\fullline)} 
and $p_T> p_{T,\mbox{\scriptsize min}}$ \mbox{(\dashline)}. 
Also shown is the distribution from the matrix element
$\gamma^*g\rightarrow Q\overline{Q}$ 
for $p_T\leq p_{T,\mbox{\scriptsize min}}$ \mbox{(\fullcircle)}
and $p_T> p_{T,\mbox{\scriptsize min}}$ \mbox{(\opencircle)}.
The definition of
$p_{T,\mbox{\scriptsize min}}$ is given in Section~\ref{DiscStudy}.
}}   
\end{figure}

\begin{figure}[htb] \unitlength 1mm
\begin{center}
\dgpicture{159}{185}
 
\put( 15,100){\epsfigdg{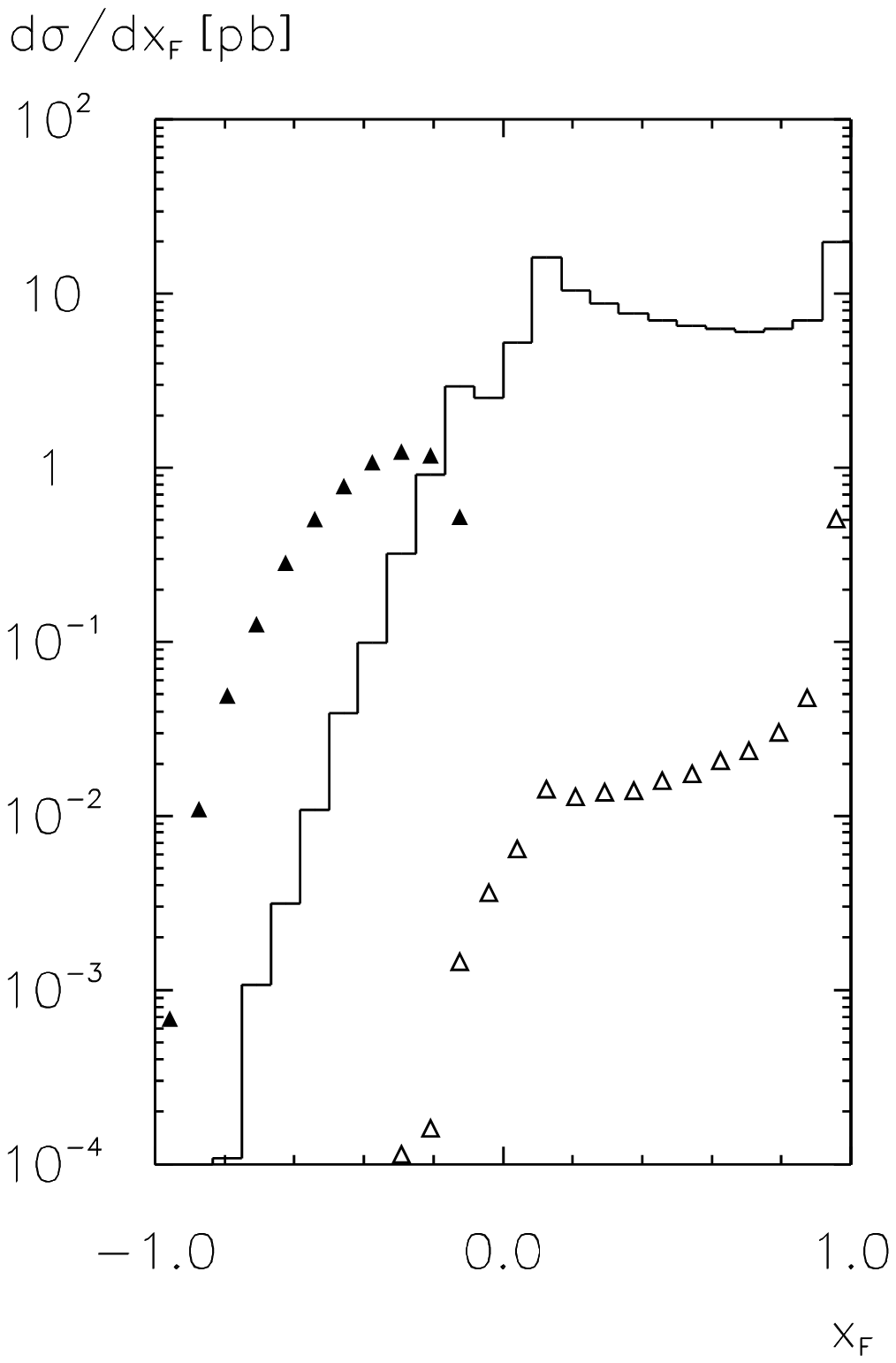}{width=55mm}}
\put( 25,95){\lettlab (a)}
  
\put( 90,100){\epsfigdg{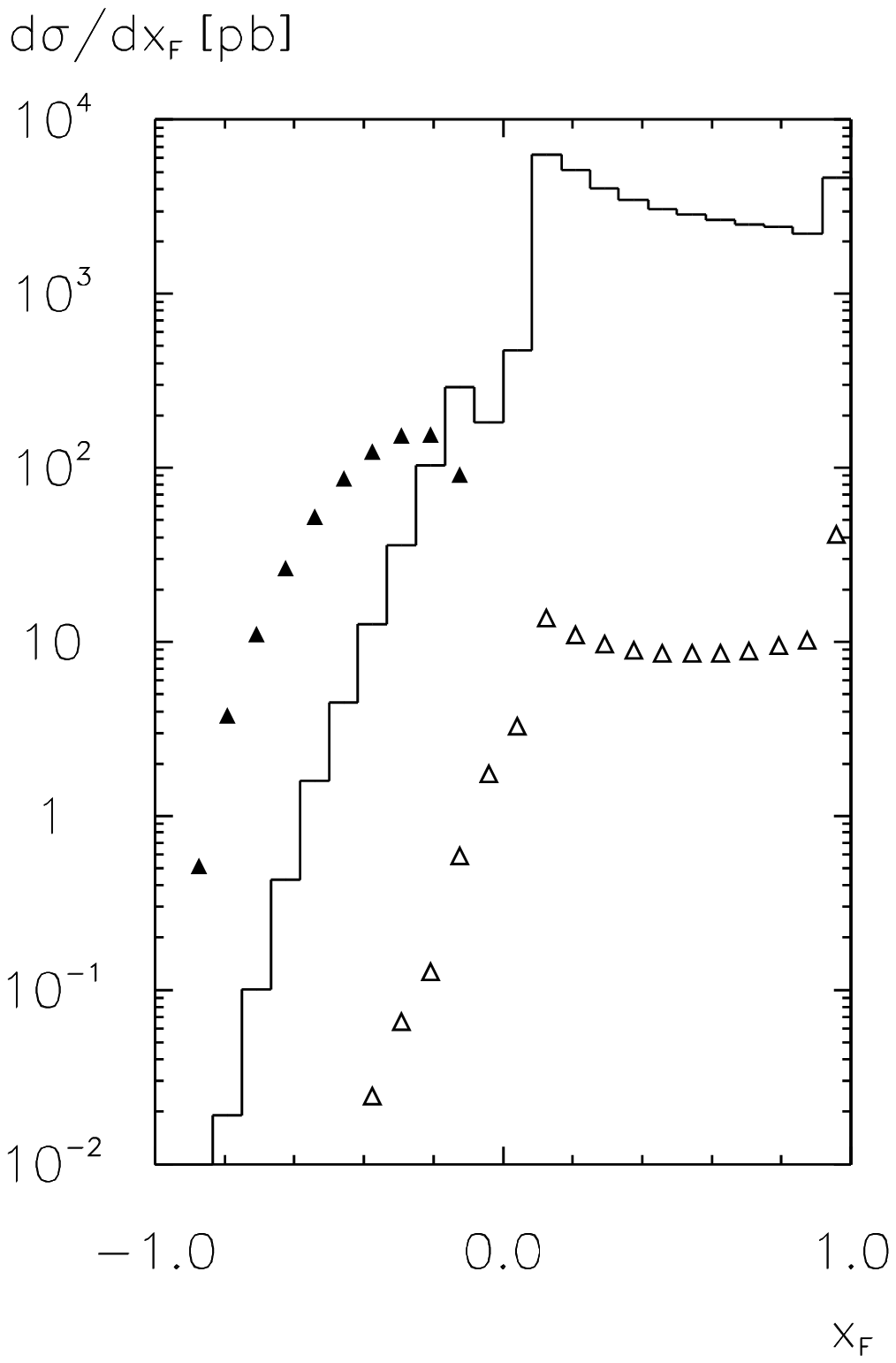}{width=55mm}}
\put(100,95){\lettlab (b)}
   
\put( 15,5){\epsfigdg{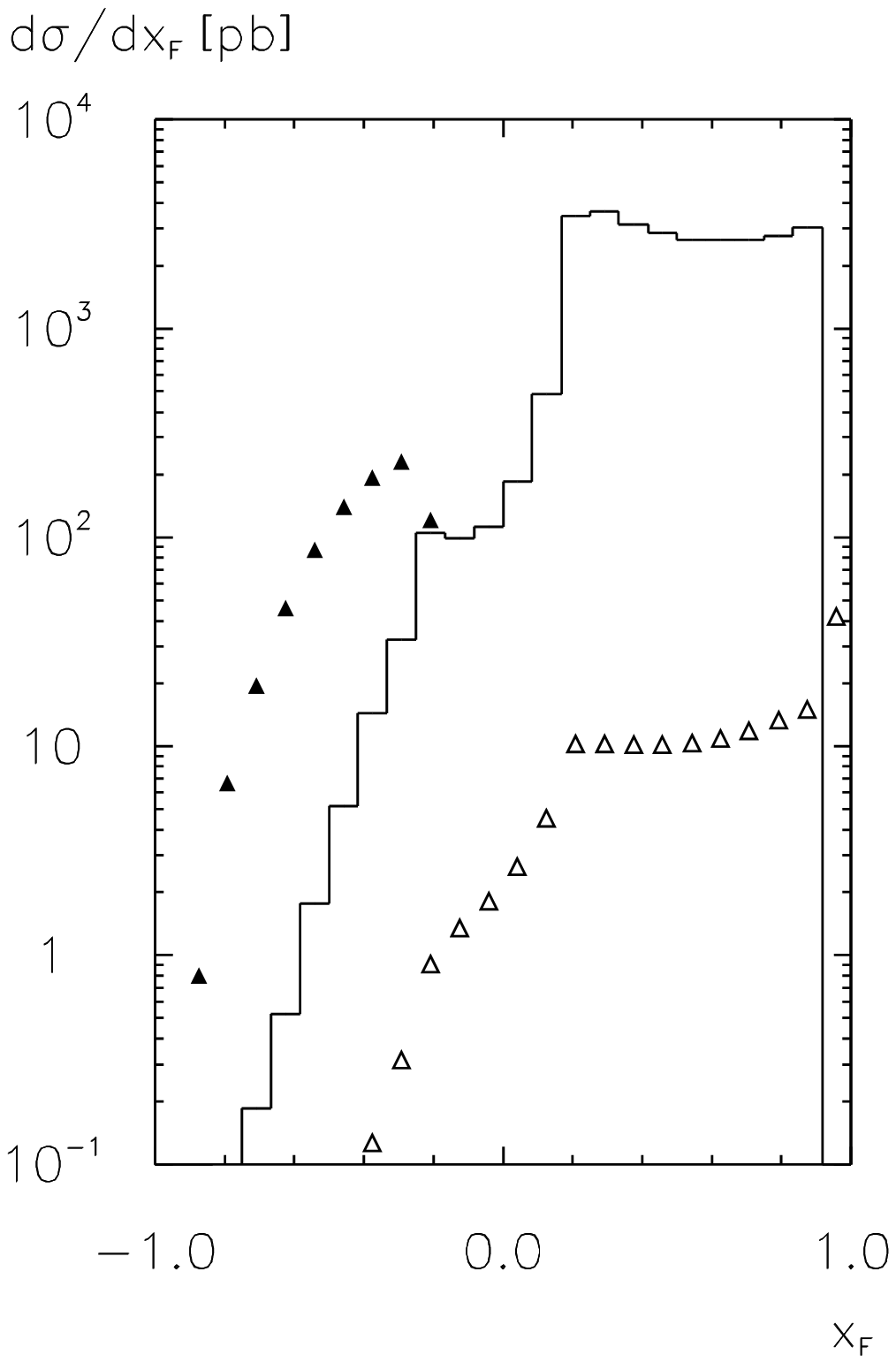}{width=55mm}}
\put( 25,0){\lettlab (c)}
    
\put( 90,5){\epsfigdg{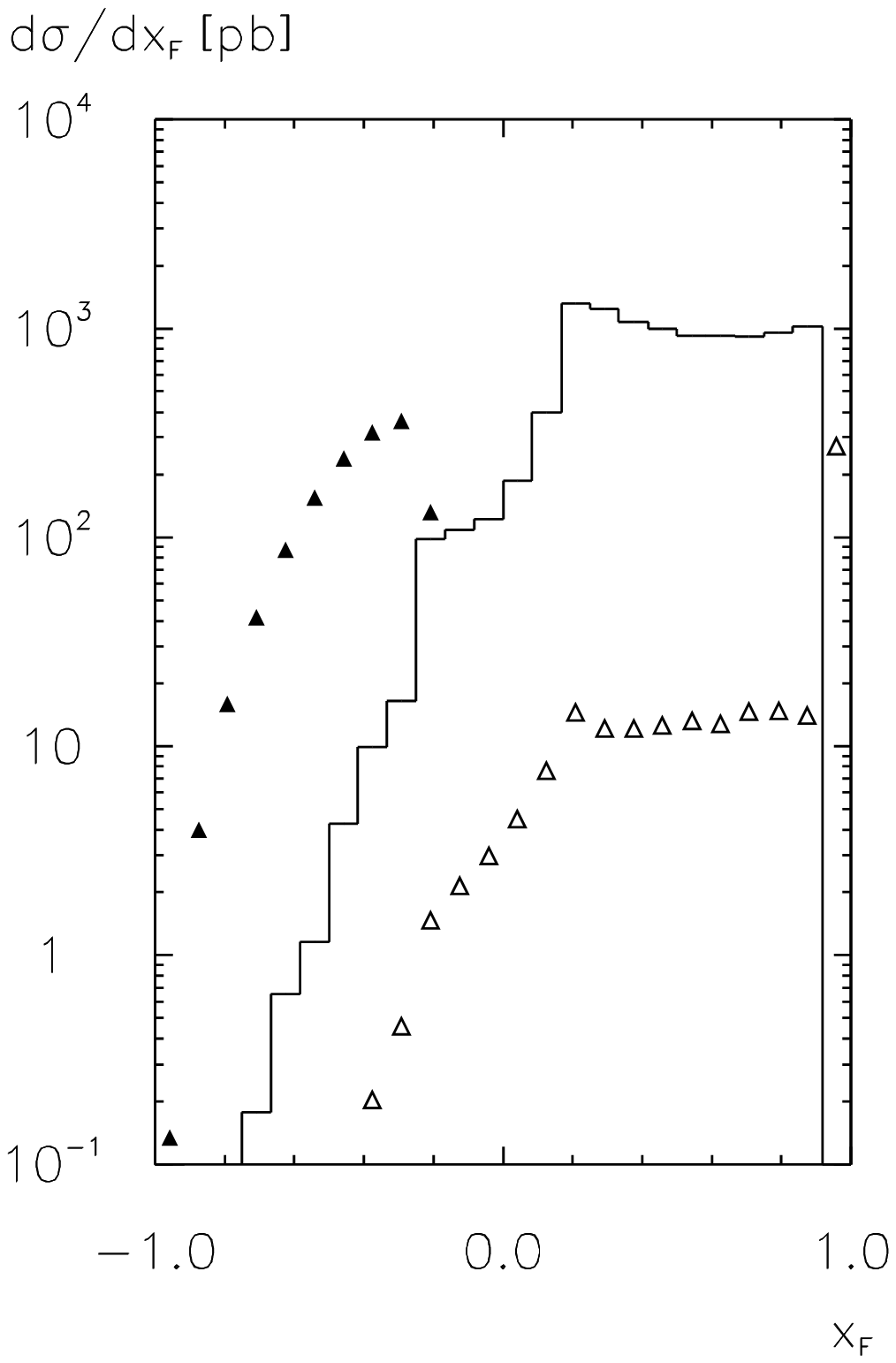}{width=55mm}}
\put(100,0){\lettlab (d)}
     
\end{picture}
\end{center}
\shiftcaption
\caption[Distributions in~$x_F$ 
for Intrinsic Heavy-Quark Production]
{\labelmm{HQTXF1I} {\it 
Distributions in~$x_F$ 
for bottom (a) 
and charm (b) quark production at HERA
and
for charm-quark production at E665 (c) 
and NA47 (d)
in next-to-leading order.
Contributions without \mbox{(\fullline)} 
and with
intrinsic heavy quarks
from $\sigma_{\mbox{\scriptsize $M$}}^\IHQ$ 
\mbox{(\fulltriangle)}
and $\sigma_{\mbox{\scriptsize $fD$}}^\IHQ$ 
\mbox{(\opentriangle)}.       
The three histograms have to be added up to give the total 
differential production cross section.
}}   
\end{figure}

\clearpage

\dgsb{Factorization- and Renormalization-Scale Dependence}
\labelm{FRDEP}

In this section we study the factorization and renormalization-scale dependence
of the cross sections. In general, leading-order cross sections strongly
depend on the factorization and renormalization scales, because the 
distribution functions and coupling constants are themselves
scale-dependent, 
whereas the lowest-order 
parton-level scattering cross sections,
excluding the coupling constants, are scale-independent.
This situation is improved in higher orders, where compensating terms arise
in the mass-factorized and renormalized parton-level scattering cross sections, 
giving an overall reduction of the sensitivity to scale variations.

In our case, the leading-order cross section is of \porder{\alpha_s^0}, 
and thus independent of the strong coupling constant, and therefore 
renormalization-scale independent. To compensate for the 
renormalization-scale dependence 
of the next-to-leading order, contributions of the
next-to-next-to-leading order would have to be incorporated, which we have
not
done. 

Regarding the factorization scales, the situation looks better. We have three
factorization scales, namely~$\mu_f$, $\mu_D$ and~$\mu_M$, 
for parton densities, 
fragmentation functions and target fragmentation functions, respectively.
A look at the explicit expressions for the cross sections in 
Appendix~\ref{rfee} shows that compensating terms in the form
of a product of 
the Born cross section, a splitting function and a logarithm of 
the factorization scale
are present; a variation in the leading-order 
cross section from a scale variation of the distribution functions
is thus
compensated to first order in the coupling constant. The remaining scale
variation is due to higher-order variations of the distribution
functions, i.e.\ their resummation, and due to the genuine scale variation 
of the next-to-leading-order
cross section, excluding the compensating terms.

We now study the various scale dependences for the production of bottom and
charm quarks at HERA, E665 and NA47 
numerically, by varying the 
factorization scales
$\mu_f$, $\mu_D$, $\mu_M$ and the renormalization scale~$\mu_r$
as~$\rho\,Q$, where~$\rho$ is a parameter between $0.5$ and 
$2$\footnote{(This is a footnote, not an exponent.) 
The curves do not always cross at~$\rho=1$
due to small shifts from the interpolation and smoothing of the Monte Carlo
results.}.
In order to stay away from very small scales, we moreover require
the scales to be larger than $1.5\,\GeV$.
The choice of~$p_T$ as a factorization scale is not possible in the case under
study, as we are interested in particular
in the limit $p_T\rightarrow 0$.
We discuss the case of bottom-quark production at HERA in 
detail, cf.~Figs.~\ref{HQTSCALE1}a and~b,
by studying the scale dependence separately in the current and 
target fragmentation regions. 

In the current fragmentation region, 
the dependence on~$\mu_f$ and~$\mu_D$ is large in leading order, and is
reduced substantially in next-to-leading order. Please note that the
renormalization-scale dependence is fairly small, although it 
is not compensated from a higher order term. This is due to the
fact that the 
$\mu_r$-dependent matrix element itself is
of \porder{\alpha_s}, and the
cross section is dominated by the renormalization-scale
independent \porder{\alpha_s^0}-term. 

In the target fragmentation region, the leading-order term depends
only on~$\mu_f$, and varies by about a factor of two in the range
of~$\rho$ under consideration. This situation is considerably improved 
in next-to-leading order, where the cross-section variation due to 
a variation of~$\mu_M$
is only about $\pm 5\%$. The dependence on~$\mu_f$, $\mu_D$ and~$\mu_r$ in the
target fragmentation region is not compensated in next-to-leading
order, because the leading order does not depend on any of these scales. Again,
the dependence on these scales is small, because the next-to-leading-order
correction is small, cf.\ Table~\ref{abstab}. We have also studied the case when
either~$\sigma_M$ or $\sigma_{fD}$ in \porder{\alpha_s} is not included.
We then obtain a large variation when varying~$\mu_M$. This shows that
both terms,
corresponding to the homogeneous and inhomogeneous evolution
contributions,
are equally important to partially cancel the $\mu_M$-dependence.

The scale dependence for charm-quark production at HERA, E665 and NA47 is
shown in Figs.~\ref{HQTSCALE1}c,~d and Fig.~\ref{HQTSCALE2}. We do not repeat 
the detailed 
discussion just given for bottom-quark production, and simply note
that the results follow a similar pattern.
We want to mention, however, that the scale compensation in the case 
of charm-quark production apparently
does not work for the scale~$\mu_D$ in the current fragmentation region. 
This is however due to the fact that the scale variation
in leading order is particularly small.
We have checked that the variation in next-to-leading order is much smaller
if only the compensating terms, i.e.\ those consisting of a product of
the leading-order cross section, 
a splitting function and a logarithm containing~$\mu_D$, are included.

\ifnum0=1
\begin{table}[hb]
\begin{center}

\begin{tabular}{|c|c|c|c|c|c|}\cline{3-6}
\multicolumn{2}{c|}{} & \rule[-4.0mm]{0mm}{10mm} 
\makebox[1.7cm]{$\mu_f$} & 
\makebox[1.7cm]{$\mu_D$} & 
\makebox[1.7cm]{$\mu_M$} & 
\makebox[1.7cm]{$\mu_r$} \\\cline{3-6}\hline

\rule[-2.5mm]{0mm}{8mm}
& \mbox{\dashline} 
& $\rho\,Q$ & $Q$ & -- & -- \\\cline{2-6}
\rule[-2.5mm]{0mm}{8mm}
\raisebox{1.7ex}[-1.7ex]{current LO}
& \mbox{\dotline} 
& $Q$ & $\rho\,Q$ & -- & -- \\
\hline

\rule[-2.5mm]{0mm}{8mm}
& \mbox{\dashdotline} 
& $Q$ & $Q$ & -- & $\rho\,Q$ \\\cline{2-6}
\rule[-2.5mm]{0mm}{8mm}
current NLO
& \mbox{\dotdotline} 
& $\rho\,Q$ & $Q$ & -- & $Q$ \\\cline{2-6}
\rule[-2.5mm]{0mm}{8mm}
& \mbox{\dashdotdotline} 
& $Q$ & $\rho\,Q$ & -- & $Q$ \\\cline{2-6}
\hline
\rule[-2.5mm]{0mm}{8mm}
target LO
& \mbox{\dashline} 
& -- & -- & $\rho\,Q$ & -- \\\hline
\rule[-2.5mm]{0mm}{8mm}
& \mbox{\dotline} 
& $Q$ & $Q$ & $Q$ & $\rho\,Q$ \\\cline{2-6}
\rule[-2.5mm]{0mm}{8mm}
& \mbox{\longdashline} 
& $\rho\,Q$ & $Q$ & $Q$ & $Q$ \\\cline{2-6}
\rule[-2.5mm]{0mm}{8mm}
\raisebox{1.7ex}[-1.7ex]{target NLO}
& \mbox{\dashdotline} 
& $Q$ & $\rho\,Q$ & $Q$ & $Q$ \\\cline{2-6}
\rule[-2.5mm]{0mm}{8mm}
& \mbox{\dotdotline} 
& $Q$ & $Q$ & $\rho\,Q$ & $Q$ \\\cline{2-6}
\hline
\end{tabular}

\end{center}
\shiftcaption
\caption[Scale Dependences.]
{\labelmm{TABCURTAR} {\it Scale dependences. A ``--'' indicates
that the cross section does not depend on the
corresponding scale. 
}}
\end{table}
\fi

\begin{figure}[htb] \unitlength 1mm
\begin{center}
\dgpicture{159}{185}

\put( 15,100){\epsfigdg{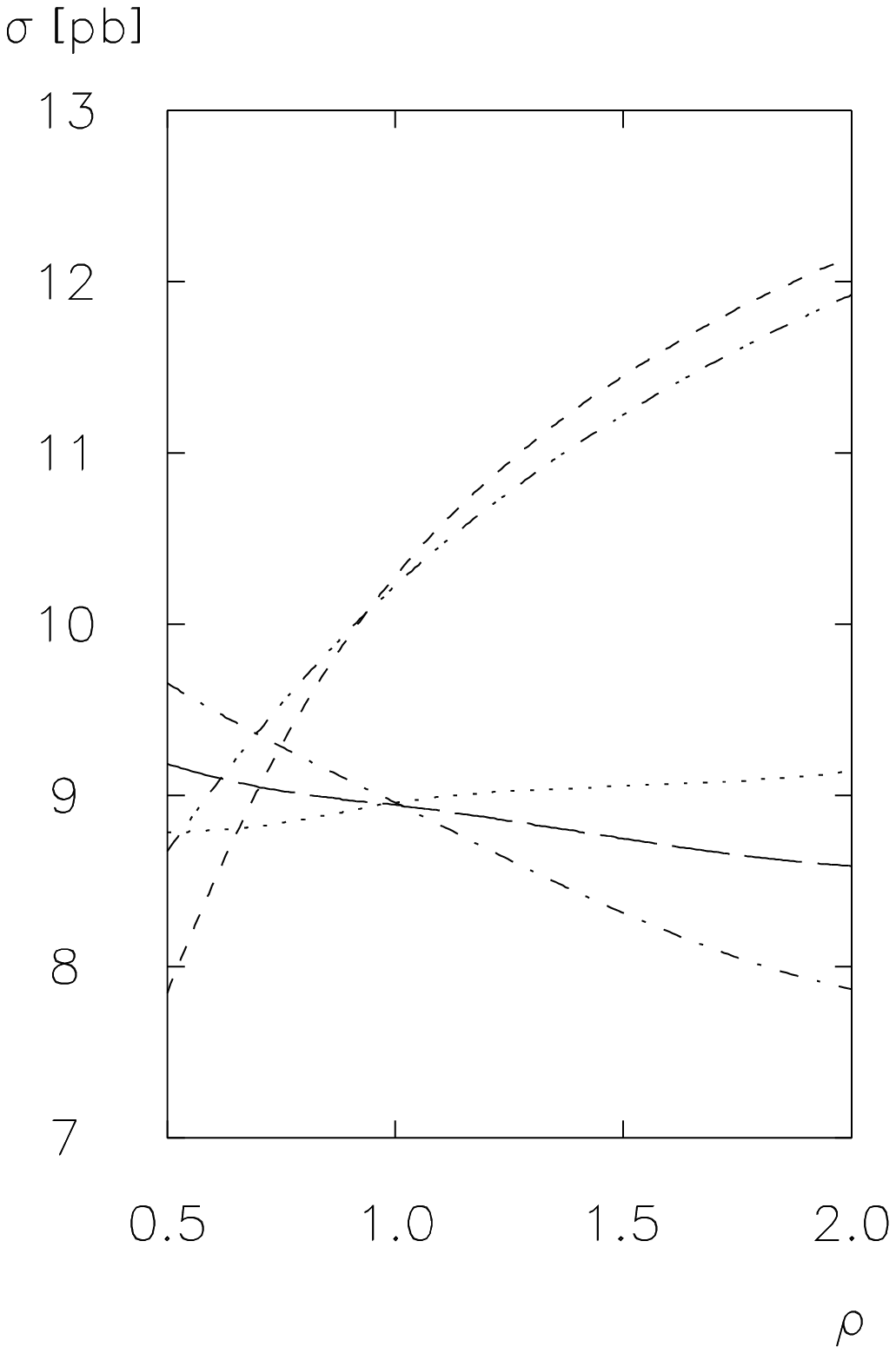}{width=55mm}}
\put( 25,95){\lettlab (a)}

\put( 90,100){\epsfigdg{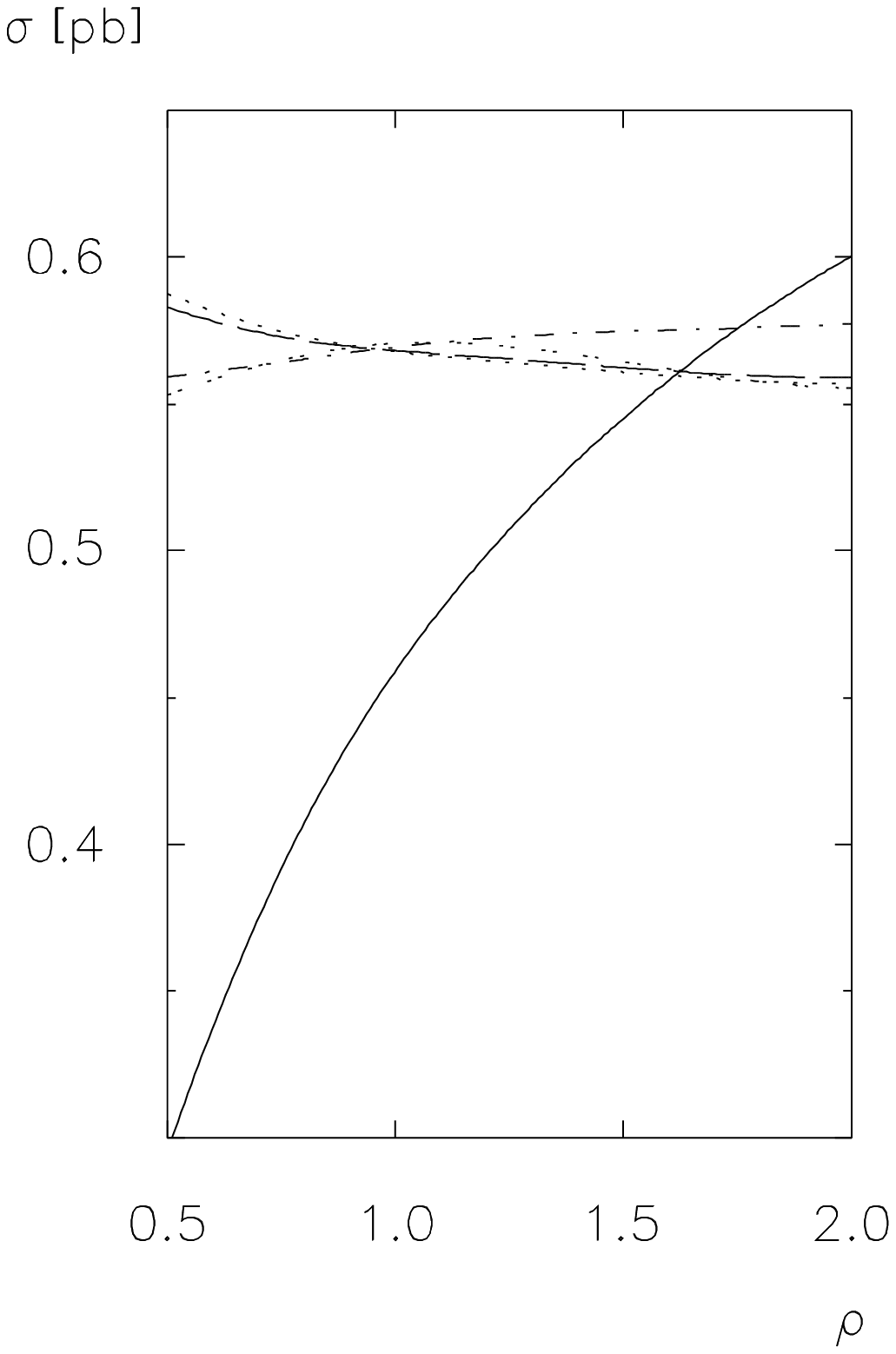}{width=55mm}}
\put(100,95){\lettlab (b)}

\put( 15,5){\epsfigdg{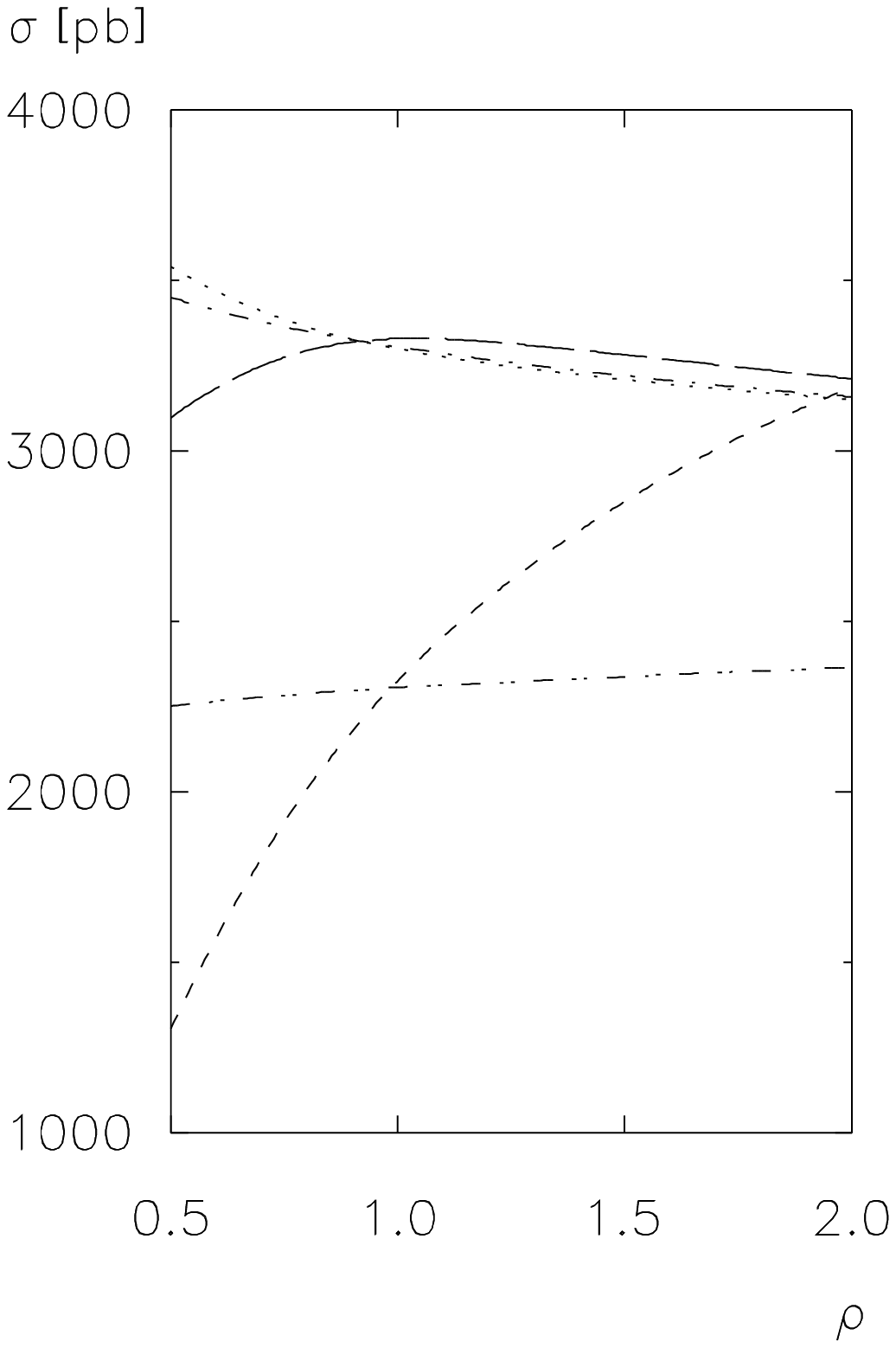}{width=55mm}}
\put( 25,0){\lettlab (c)}

\put( 90,5){\epsfigdg{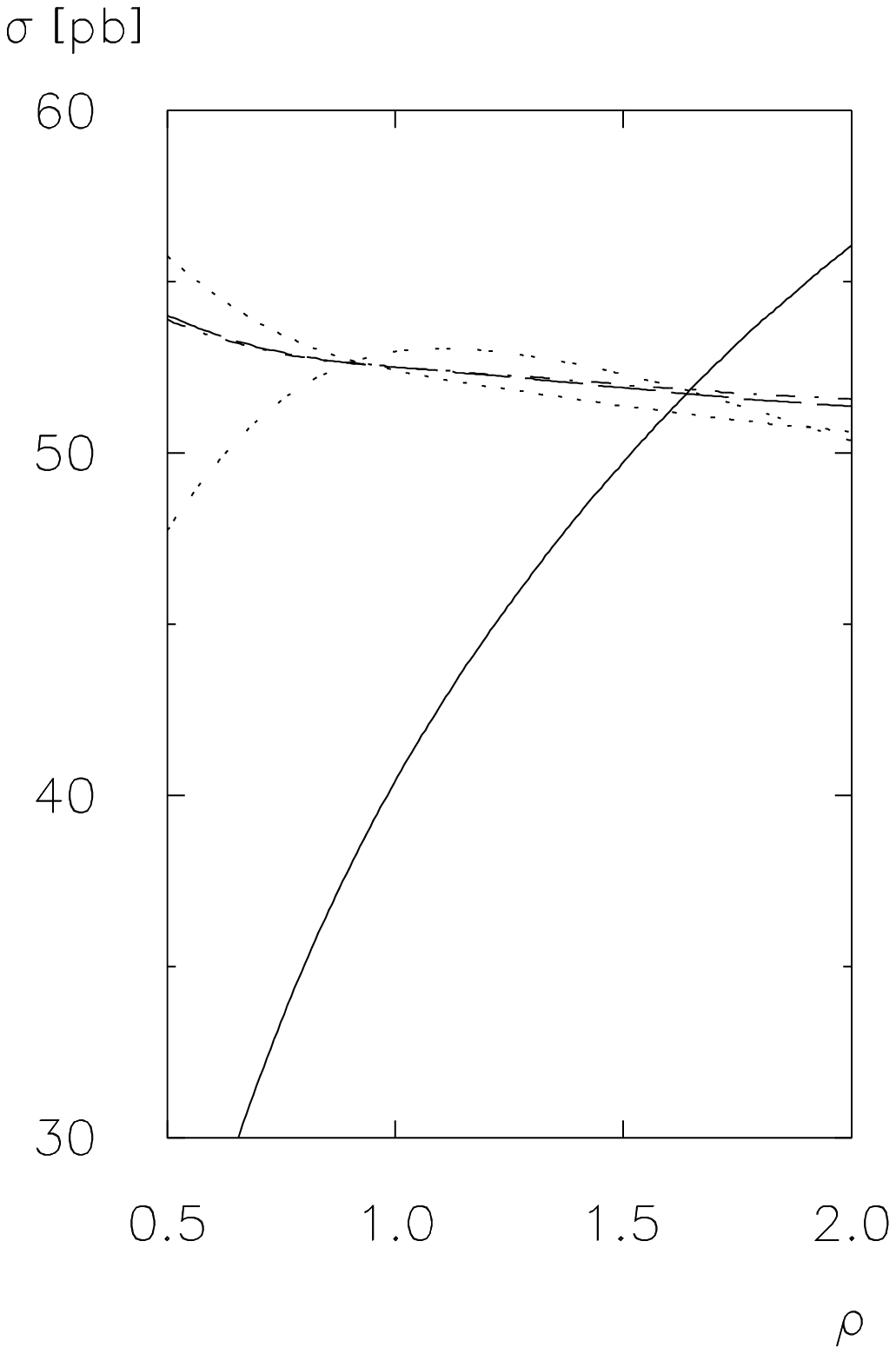}{width=55mm}}
\put(100,0){\lettlab (d)}

\end{picture}
\end{center}
\shiftcaption
\caption[Scale Dependence for
Heavy-Quark Production at HERA]
{\labelmm{HQTSCALE1} {\it 
Scale dependence for bottom (a), (b) and charm (c), (d)
quark production at HERA
in the current (a), (c) and target (b), (d) fragmentation regions.
The particular scale set to $\rho\,Q$ is given by
$\mu_f$ \mbox{(\dashline)}, $\mu_D$ \mbox{(\dashdotdotline)},
$\mu_M$ \mbox{(\fullline)} in leading order and
$\mu_r$ \mbox{(\dotline)}, $\mu_f$ \mbox{(\longdashline)},
$\mu_D$ \mbox{(\dashdotline)}, $\mu_M$ \mbox{(\dotdotline)}
in next-to-leading order; 
the other scales are fixed to be equal to~$Q$.
}}   
\end{figure}

\begin{figure}[htb] \unitlength 1mm
\begin{center}
\dgpicture{159}{185}

\put( 15,100){\epsfigdg{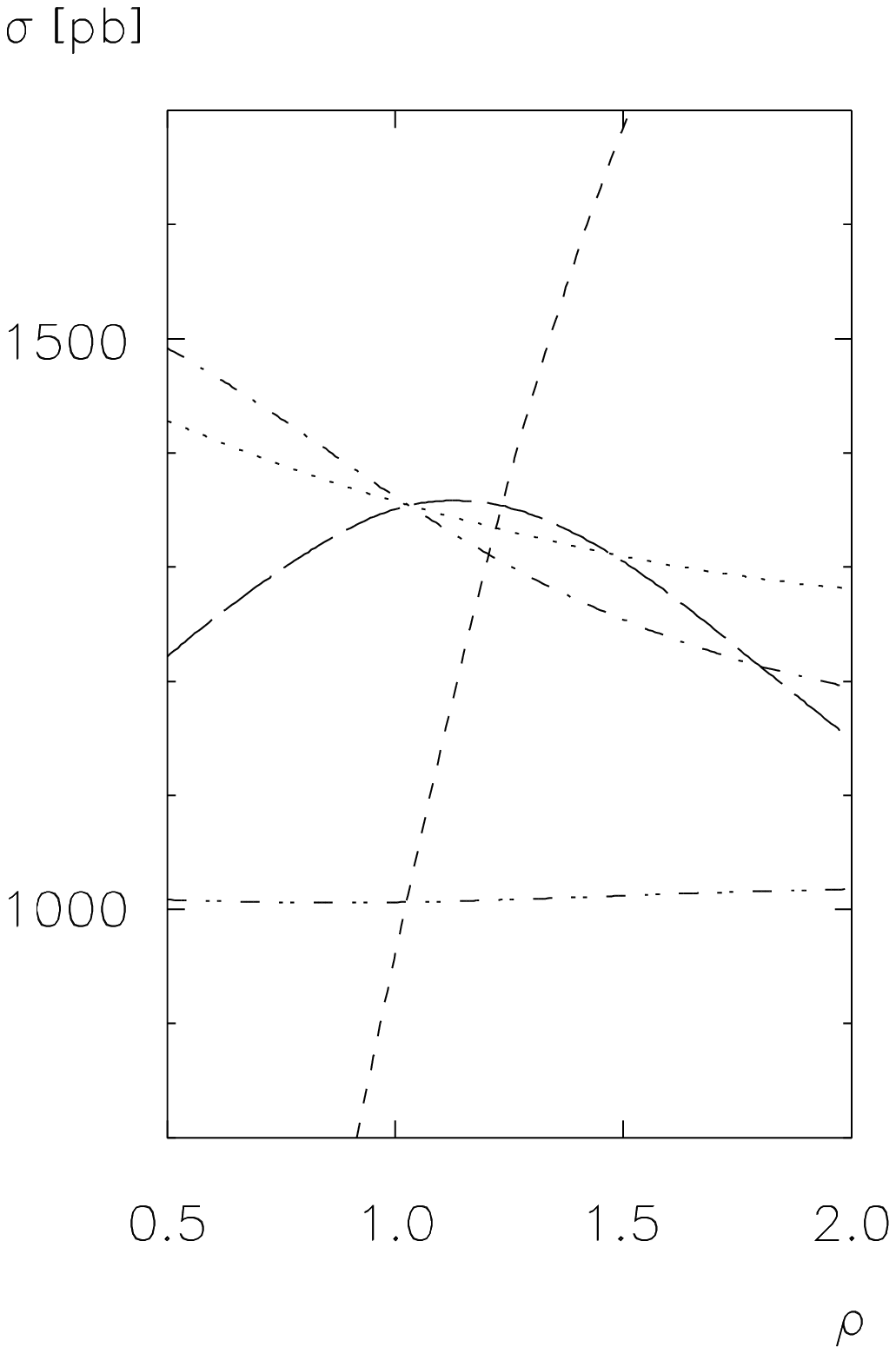}{width=55mm}}
\put( 25,95){\lettlab (a)}

\put( 90,100){\epsfigdg{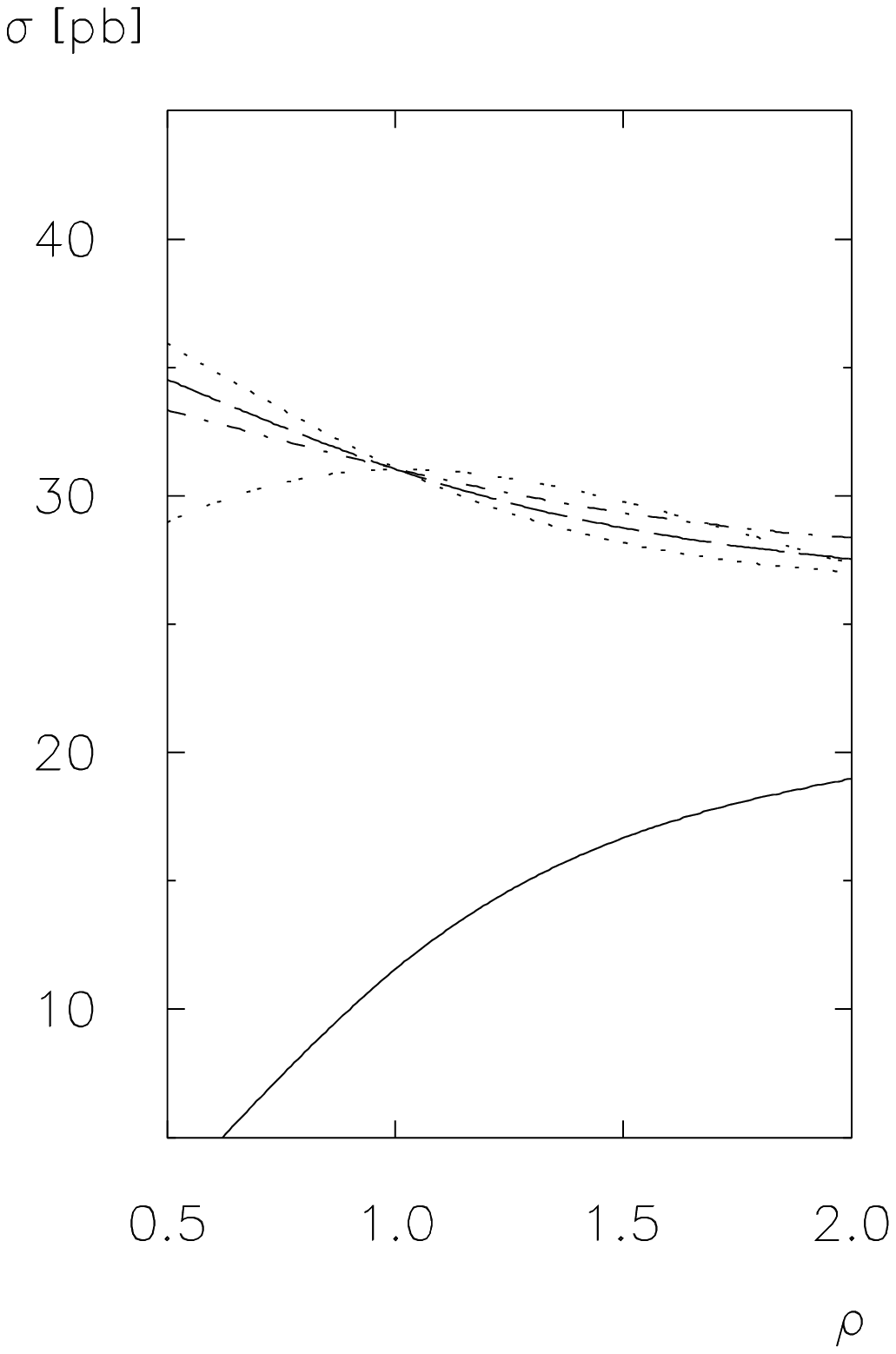}{width=55mm}}
\put(100,95){\lettlab (b)}

\put( 15,5){\epsfigdg{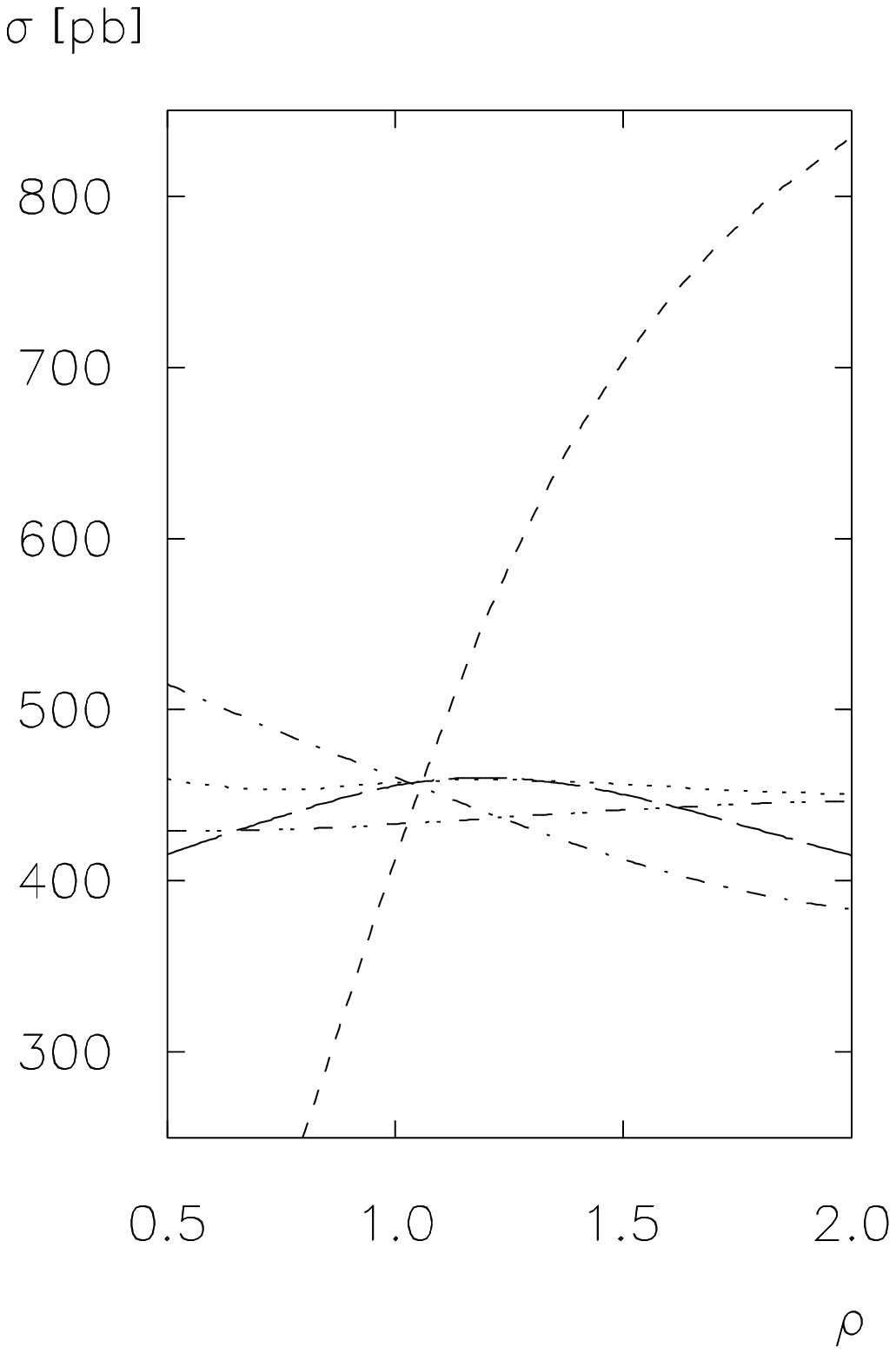}{width=55mm}}
\put( 25,0){\lettlab (c)}

\put( 90,5){\epsfigdg{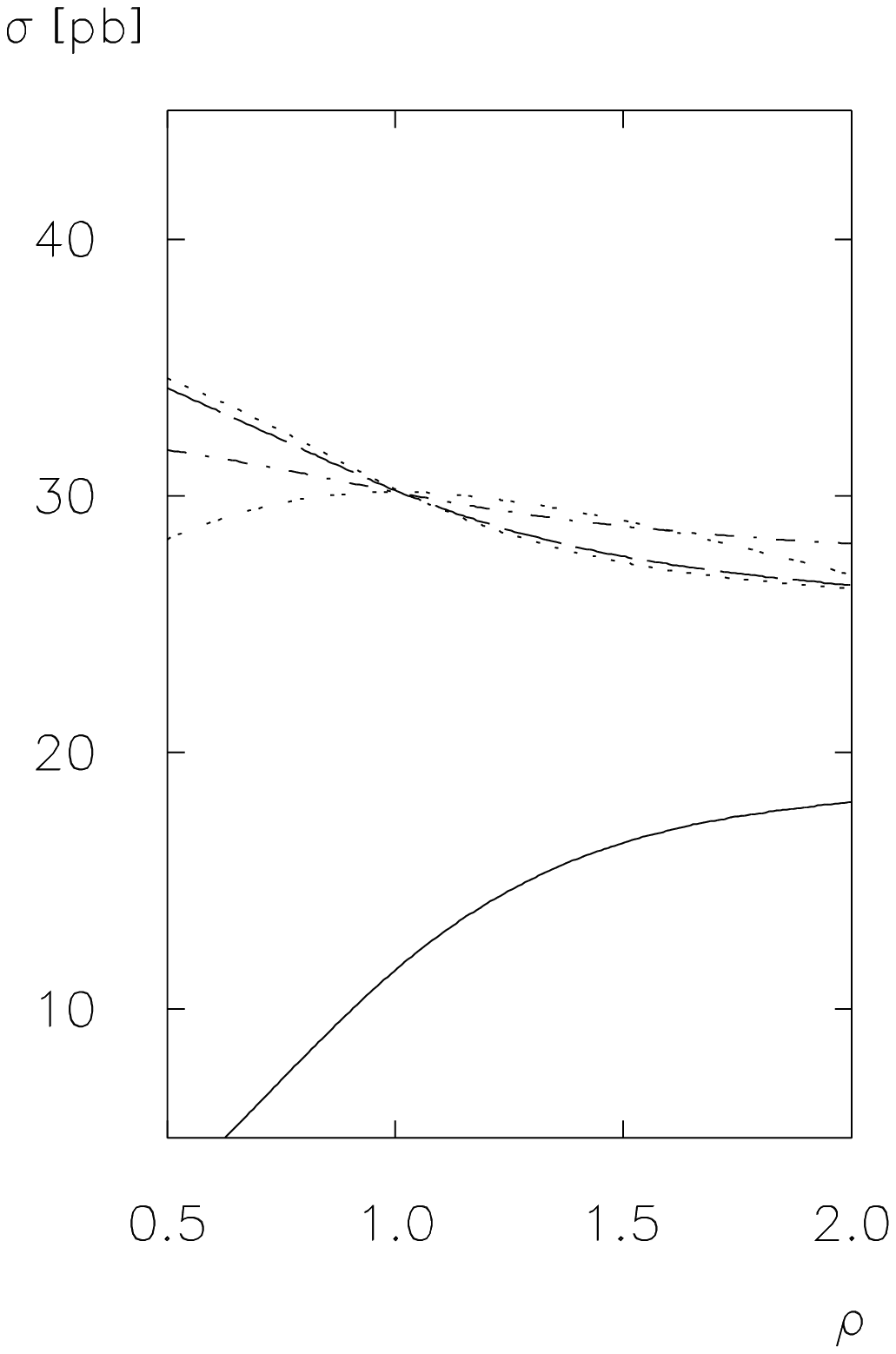}{width=55mm}}
\put(100,0){\lettlab (d)}

\end{picture}
\end{center}
\shiftcaption
\caption[Scale Dependence for
Charm-Quark Production at E665 and NA47]
{\labelmm{HQTSCALE2} {\it 
Scale dependence for charm-quark production at E665 (a), (b)
and NA47 (c), (d)
in the current (a), (c) and target (b), (d) fragmentation regions.
The particular scale set to $\rho\,Q$ is given by
$\mu_f$ \mbox{(\dashline)}, $\mu_D$ \mbox{(\dashdotdotline)},
$\mu_M$ \mbox{(\fullline)} in leading order and
$\mu_r$ \mbox{(\dotline)}, $\mu_f$ \mbox{(\longdashline)},
$\mu_D$ \mbox{(\dashdotline)}, $\mu_M$ \mbox{(\dotdotline)}
in next-to-leading order; 
the other scales are fixed to be equal to~$Q$.
}}   
\end{figure}


\dgcleardoublepage

\markh{Summary, Open Problems and Conclusions}
\dgsa{Summary, Open Problems and Conclusions}
\labelm{SCOO}

\dgsb{Summary}
\labelm{SC}
We have presented a formalism to describe the production of particles
in the target fragmentation region of deeply inelastic 
lepton--nucleon scattering
in the framework of perturbative QCD.
An explicit one-loop calculation shows the necessity to introduce new
phenomenological distribution functions, the target fragmentation functions or
``fracture functions'' \cite{77}. Their renormalization group equation
is inhomogeneous owing to a source term that describes the ``perturbative
production'' of particles in the backward direction. It is shown 
that, to one-loop, all 
collinear singularities
can be
consistently absorbed into renormalized distribution functions.
This is also true of a new singularity, which does
not appear in the
case when the observed particle is produced strictly
in the current fragmentation region
or, at finite~$p_T$, in the target fragmentation region.
It is conjectured that this mechanism works to all orders in perturbation 
theory.
The finite cross section is a convolution of a mass-factorized parton-level
scattering cross section with parton densities, fragmentation functions
and target fragmentation functions.

The formalism has been applied in the case of the production of heavy quarks.
The renormalization group equations 
for heavy-quark fragmentation functions, perturbative 
heavy-quark target fragmentation
functions and target fragmentation functions
for an intrinsic heavy-quark content of the proton
have been solved numerically. In a case study
of deeply inelastic lepton--nucleon scattering,
we have investigated in detail the production of bottom and charm quarks at 
HERA, E665 and NA47. Cross sections have been presented for various phase-space
regions of the produced heavy quark.
Restricting the phase space to the target fragmentation region, it has
been shown that the perturbative piece from the evolution of the
target fragmentation functions contributes significantly.

Using the example of $p_T$-distributions, we have demonstrated how the
subtraction process in the singular phase-space regions works in 
practice.
{}From a theoretical point of view, the
calculated cross sections are finite; all divergences are being absorbed into
renormalized parton densities, fragmentation functions and target fragmentation
functions. The finite differential cross section in~$p_T$ is, 
however, a distribution
(in the mathematical sense), with a singularity at $p_T=0$.
Therefore it has 
to be integrated over a certain region around $p_T=0$ in order to 
give rise to a well-defined numerical prediction.
For realistic parameters, reasonable numerical results for 
heavy-quark production cross sections in the target fragmentation region
can be obtained. It should be stressed that the results discussed here
only include the perturbative part of the target fragmentation functions,
obtained by the corresponding inhomogeneous renormalization group equation 
with a specific boundary condition at a small input scale, 
and a non-perturbative part based on a model assumption
assuming intrinsic heavy quarks in the proton. The true non-perturbative
piece of these
functions, which are in principle
process-independent,
must be obtained from 
experiment. 

The shape of the $x_F$-distributions in the current fragmentation
region is modified considerably if next-to-leading-order
contributions are included, whereas the shape in the target 
fragmentation region is stable. In particular in the case of
charm-quark production,
the next-to-leading-order corrections
for $x_F\geq 0$ are large.

We have also studied $x_F$-distributions for a model based on the
hypothesis of intrinsic heavy quarks in the proton. According to 
Refs.\ 
\cite{38,39}, the heavy quarks~$Q$, $\overline{Q}$ from a Fock space
component $|uudQ\overline{Q}\rangle$ of the proton 
should carry a large momentum fraction.
If the proton is hit by a large-$Q^2$ probe, these heavy quarks may 
fragment in the target fragmentation region, even if they themselves
do not participate in the hard scattering subprocess.
As expected, the 
$x_F$-distribution in the target fragmentation region is much more pronounced
and extends to larger negative values of~$x_F$ if the intrinsic component is
included, compared with the case where the heavy quark is produced via
the hard matrix element or by evolution of the perturbative 
target fragmentation function.

The factorization-scale dependence follows the expected pattern, namely
that compensating terms lead to a smaller
overall scale dependence in next-to-leading order. 
The dependence on the factorization scales~$\mu_f$ and~$\mu_D$ in the current
fragmentation region and on~$\mu_M$ in the target fragmentation region is
reduced in next-to-leading order, as expected, with the exception
of the $\mu_D$-dependence in the cases of E665 and NA47, due to
a very small $\mu_D$-dependence in leading order.
Since in leading order the process under 
consideration does not depend on the 
strong coupling constant, the renormalization-scale dependence 
arises first in next-to-leading order, and would be
compensated only by contributions in next-to-next-to-leading order.

We add a short remark on the possibility to observe the heavy quarks in 
the target fragmentation region of
an actual experiment. In Ref.\ \cite{11} it is assumed that the acceptance
for a $D^0$~meson is about $0.4$. Assuming that about half of all 
produced charm quarks fragment into a~$D^0$, and using the branching ratio
of about $4\%$ of the decay $D^0\rightarrow K^-\pi^+$, we arrive
at an overall probability of $0.8\%$ 
that a produced charm quark will be detected.
Assuming the same value in the target fragmentation region\footnote{
In the target fragmentation region, leading particle effects are expected, 
and the fragmentation of, for example, a charm quark into a $D^0$~meson
will take place with a different probability from that in the
current fragmentation region. See also Ref.\ \cite{128}.
}
as well, the total number of
reconstructed charm-quark events in the target fragmentation region
of the HMC experiment, based on an
integrated luminosity of 1500~pb$^{-1}$, will be $360$. This is probably
not enough to attempt a fit of target fragmentation functions, but it shows
that a similar study in photoproduction\footnote{
Up to now, experimental studies in photoproduction
are available in the current fragmentation region only, 
see for instance Refs.\ \cite{129,130,131,132,133}.
We note that, for the resolved contribution
in the current fragmentation region, 
the target fragmentation functions 
for real photons have to be introduced.
}, where the expected
charm-quark production cross section 
would be expected to be much higher,
may well be feasible.

In order to study the mechanism for particle production
in the target fragmentation region, as discussed in this
paper, the
production of mesons 
not containing heavy 
valence quarks,
such as $\pi^+$, $\pi^-$ and $\pi^0$, 
could also be considered; there,
the cross sections are expected to be larger. 
In this case, the perturbative
piece of the target fragmentation function can be obtained by using 
experimentally determined pion fragmentation functions, 
as recently given, for instance, in Ref.~\cite{134}.

\dgsb{Restrictions of the Approach}
\labelm{RA}
For the case of regular fragmentation functions, the finite cross section
is a distribution in the angular variable~$v$. If the fragmentation functions
themselves
are singular, as 
in the case of heavy-quark fragmentation functions, 
then there is an additional subtraction related to the
energy-fraction variable~$z$. In any case, in next-to-leading order, 
there are subtractions at small 
transverse momenta~$p_T$, which means that a meaningful prediction can be made
only for the integral of the 
differential cross section $\dd \sigma/\dd p_T$ over 
$p_T$ from~$0$ to some 
$p_{T,\mbox{{\scriptsize max}}}$, where 
$p_{T,\mbox{{\scriptsize max}}}$ is not allowed to be
small. This certainly limits the range of applicability of the 
presented results.
This particular restriction is, however, already present in the 
standard formulation for the current fragmentation region.

The approach is formulated in terms of fragmentation functions and therefore
incorporates leading twist effects only. 
Similarly, it cannot be expected that 
non-perturbative 
leading-particle effects can be described.

The case study for heavy-quark production in deeply inelastic scattering
had to be restricted to the perturbative contributions
and to a model for intrinsic heavy quarks in the proton, because the
non-perturbative piece $M^{(NP)}$ of the heavy-quark
target fragmentation functions
is not yet known. A limitation of the numerical investigation 
is that we have used the target fragmentation functions
in the leading-logarithmic approximation only. To have a consistent 
renormalization-group-improved
next-to-leading-order prediction, the target fragmentation functions
should be evolved with next-to-leading-logarithmic accuracy.

\dgsb{Open Problems}
The question of extended factorization of the form conjectured in 
Eq.~(\ref{opicompl}) is a central issue related to the problem 
of universality of target fragmentation functions. In Ref.\ \cite{135}
it is shown in the framework of an explicit toy model
for the case of {\it diffractive} hadron--hadron scattering
that factorization breaks down when there are {\it two}
strongly interacting particles in the initial state, due to the exchange
of soft gluons in the final state. This cannot happen for the
case of deeply inelastic lepton--nucleon 
scattering. The process considered in 
Ref.\ \cite{135} is diffractive, so the target is essentially untouched.
The toy model makes no prediction of what happens if the target 
nucleon fragments, as is required, for example, 
in the case of heavy-quark
or meson production in the target fragmentation region.

The formalism developed in this paper may 
be applied
to other processes
as well,
such as the 
photoproduction\footnote{In diffractive photoproduction, the situation is
different due to the hadronic component of the incident real photon,
and the factorization assumption might not apply.
}
of hadrons or heavy quarks
in lepton--nucleon scattering, or one-particle-inclusive
Drell--Yan-like processes, i.e.\ $pp\rightarrow \mu^+\mu^-hX$.
Lepton--nucleon scattering is very interesting because it 
permits 
the direct measurement 
of the non-perturbative part $M^{(NP)}$ of the target fragmentation
functions.
Photoproduction is particularly attractive because of the
large cross sections, but it introduces a large uncertainty
due to the poorly constrained gluon content 
of the photon structure function.

We have considered the problem of the production of heavy quarks in the
target fragmentation region, but we have not touched the problem of 
how these
quarks eventually fragment into hadrons. 
The fragmentation into mesons 
containing a heavy quark 
may be treated by adding a
convolution of the fragmentation function of a heavy quark into
a meson 
\cite{35} 
with the cross section for heavy-quark 
production as determined in this paper.
However, it is to be expected that, in particular in the 
target fragmentation region, leading particle effects may play a 
r\^{o}le\footnote{
See for example  Ref.\ \cite{136}.
}, 
although they should be of higher twist. The experimental study of 
the production of mesons containing heavy quarks in this region of 
phase space may shed some light on this issue.

As has been stressed in Section~\ref{RA}, the cross section, being a 
distribution in 
the phase-space variables of the observed heavy quark, 
is only well-defined if integrated over a certain 
phase-space region. This problem is related to a similar and familiar
phenomenon in jet physics. There, only sufficiently inclusive 
infrared-safe quantities are physically meaningful, and the 
limit of small jet cuts leads to meaningless results. It is interesting, 
however, that in the present case of one-particle-inclusive 
processes the subtractions have to be done in terms of in principle
{\it measurable} quantities such as 
the transverse momentum of the observed hadron, 
contrary to the case of jet physics, where the subtractions
are done in terms of unobservable parton momenta.
This problem deserves further study.

Two formulations of the theory of one-particle-inclusive
processes go back to A.H.~Mueller; one based on an analysis
of Regge poles \cite{137,138}\footnote{For 
a review, see Ref.\ \cite{139}.
}, 
the other one related to the formalism of cut vertices 
\cite{140,141,142,143}. The approach mentioned first is based on the
crossing of the observed particle into the initial state with a subsequent
application of the optical theorem and Regge phenomenology. The second
approach can be formulated within perturbative QCD. There should be 
a possibility to relate the concept of target fragmentation functions
to expectation values of operators corresponding to cut vertices.

For a complete next-to-leading-logarithmic calculation, the scale evolution 
equation (\ref{Mrge}) 
should be known in next-to-leading order. This would
require the calculation of the splitting functions related to the 
inhomogeneous term on the two-loop level, 
in analogy to the corresponding calculation 
for parton densities and fragmentation functions \cite{144,91}. 
As long as the scales 
$\mu_e$, where target fragmentation functions are extracted from 
experimental data, and~$\mu_p$, where they are used for predictions, 
are not too different, i.e.\ as long as $\alpha_s\ln(\mu_e^2/\mu_p^2)$
is small, the leading-order evolution equation should be 
sufficient.

\dgsb{Outlook}
A measurement of the non-perturbative part $M^{(NP)}$ of the target 
fragmentation functions would be a new interesting test of QCD,
since 
QCD predicts the scale evolution according to Eq.~(\ref{Mrge}).
Fixed-target experiments allow in principle a detector with 
a solid-angle coverage of $4\pi$ because of the strong forward boost
and therefore permit measurements in the target fragmentation region; 
for a recent proposal, see Ref.~\cite{11,12}. Because of their 
clean experimental signature, 
the tagging of charmed and strange mesons is particularly attractive.

Target fragmentation functions may also be applied in 
hard diffractive processes, e.g.\ 
in diffractive deeply inelastic electron--proton scattering, 
where the proton, either essentially untouched or excited, is the
observed particle. In the Ingelman--Schlein picture
\cite{145,146},
the exchanged object with vacuum quantum numbers
initiates the hard scattering process, and fragments.
{}From experimental determinations of the structure functions of the exchanged
object, it is possible to obtain explicit parametrizations for
diffractive target fragmentation functions, 
cf.\ Ref.\ \cite{147}. A~similar concept
has been developed in Ref.\ \cite{148}.

As mentioned in the introduction,
an interesting possibility is that the tagging of certain particles
in the target fragmentation region may reveal additional information
about the hard scattering process. The tagging of, for example, 
a proton in the backward 
direction is expected to enhance the event sample  of 
gluon-initiated events. 
Similar conclusions are possible for other 
tagged particles, see Ref.~\cite{77}. 
The possibility to constrain the hard scattering process may be 
helpful to measure the part of the spin of the proton 
that is carried by gluons
and strange quarks
\cite{19}. For this application, however, the present study
has
to be generalized to the case of strange quarks in the polarized case.
Since the strange-quark mass is too small to permit the calculation
of perturbative 
heavy-quark fragmentation functions, one would have to go back 
to measured fragmentation functions of strange hadrons
to obtain the perturbative target fragmentation functions.

\dgsb{Conclusions}
It is possible to describe particle production in the current and
target fragmentation regions in a unified way by an extension
of the standard QCD formalism involving parton densities and fragmentation
functions, as developed in this paper. 
{}From the theoretical side, the extended factorization 
conjecture remains to be proved, and the extension of the formalism
to a true next-to-leading-logarithmic framework is to be done.
Experimentally, in particular in the case of heavy-quark production, 
a rich phenomenology is waiting to be explored. It will be interesting to 
confront the theoretical ideas
presented in this paper with actual experimental
data.

\vspace{0.5cm}

\dgcleardoublepage
\newcommand{\lbk}{\vspace{2mm}}
\markh{}
\dgsm{Acknowledgements}
\addcontentsline{toc}{section}{Acknowledgements}

\medskip

\noindent
I wish to thank
\begin{itemize}
\item[\dgbulleta] 
L.~Trentadue and G.~Veneziano, who invented the beautiful concept
of target fragmentation functions, for numerous discussions,
\item[\dgbulleta] 
H.~Joos, G.~Kramer and P.M.~Zerwas for helpful conversations,
\item[\dgbulleta] 
G.~Altarelli for a discussion on Ref.\ \cite{70},
\item[\dgbulleta]
W.~Buchm\"uller for a discussion about diffractive scattering,
\item[\dgbulleta]
H.~He{\ss}ling for some remarks related to the experimental aspects
of the backward direction at HERA,
\item[\dgbulleta] 
E. Laenen for a discussion about heavy-quark production,
\item[\dgbulleta] 
O.~Nachtmann for drawing my attention to 
Refs.\ \cite{76,73,74},
\item[\dgbulleta] 
P.~Nason for a discussion about heavy-quark fragmentation functions,
\item[\dgbulleta] 
D.~Soper for a discussion about factorization in QCD,
\item[\dgbulleta] 
A.~Vogt for some remarks regarding the heavy-quark content of
the proton,
\item[\dgbulleta] 
the Theory Divisions of the Argonne National Laboratory,
the Lawrence Berkeley National Laboratory and the Stanford
Linear Accelerator Center for the warm hospitality extended
to me during visits in the autumn of 1995, 
and
E.L.~Berger, S.J.~Brodsky, M.~Derrick, D.~Geesaman, I.~Hinchliffe, 
R.~Vogt
and 
A.~White
for valuable discussions about target fragmentation functions,
heavy-quark production and diffractive physics,
\item[\dgbulleta] 
E.~Gianolio, L.~L\"{o}nnblad and R.~Sommer
for support related to UNIX, work stations and computing,
\item[\dgbulleta] 
the CERN computer centre, 
in particular E.~McIntosh, B.~Panzer--Steindel and
H.~Renshall, for technical support.
\end{itemize}
I am grateful to 
S.J.~Brodsky, G.~Kramer, H.~Spiesberger, L.~Trentadue
and G.~Veneziano
for comments on the manuscript.
The numerical calculations for this paper were done on the
CERNSP and on the CERN CS2 cluster
of the GPMIMD--2 ESPRIT project financed by the European Union.
This work was supported in part by a Habilitan\-den\-sti\-pen\-di\-um from the
Deutsche For\-schungs\-ge\-mein\-schaft.

\dgcleardoublepage
\begin{appendix}

\markh{Distributions}
\dgsa{Distributions}
\labelm{distribu}
In this section, the distributions used in the text are defined and some
of their 
properties are exhibited. Attention is paid in particular to convolution
formulae of distributions involving the ``$+$''~prescriptions and to 
expressions involving a variable transformation 
useful for the solution
of the renormalization group equations.

\dgsb{Definition of Singular Functions}
\labelm{appdistr}
Apart from the familiar $\delta$-function\footnote{All relations discussed
in this section may also be derived in a more rigorous mathematical framework,
cf.\ Ref.\ \cite{149}.} defined by
\beq
\delta_c[\varphi]\doteq\int\dd x\, \delta(x-c)\,\varphi(x)\doteq\varphi(c),
\eeq
singular functions with ``$+$''~prescriptions appear frequently. 
They arise in Laurent expansions of the form
\beqm{lexpa}
x^{-1-\epsilon}=\frac{\Gamma^2(1-\epsilon)}{\Gamma(1-2\epsilon)}
\,\left[
-\frac{1}{\epsilon}\,\delta(x)+\subbl{\left(\frac{1}{x}\right)}{x}{0}{1}
+\epsilon\left(-\subbl{\left(\frac{\ln x}{x}\right)}{x}{0}{1}
-\frac{\pi^2}{6}\,\delta(x)
\right)
\right]
+\porder{\epsilon^2}
\eeq
for $x\in[0,1]$,
and they are defined by a subtraction
via \cite{149}
\beq
\subabc{D}{a}{b}{c}[\varphi]
\doteq\int_a^b\dd x \,D(x)\,\left(\varphi(x)-\varphi(c)\right).
\eeq
This subtraction is sufficient as long as the singularity of~$D$ at~$c$
is not worse than $\sim 1/x^{1+\rho}$, $\rho<1$. For $\rho\geq 1$, 
higher orders in the Taylor expansion of~$\varphi$ have to be subtracted:
\beqm{jetsubtr}
\subabc{D}{a}{b}{c,n}[\varphi]
\doteq\int_a^b\dd x \,D(x)\,\left(\varphi(x)-\varphi^{[n]}(c,x)\right), 
\eeq
where 
\beqm{jetdef}
\varphi^{[n]}(c,x)=\sum_{k=0}^n \frac{1}{k!}\,\partial^k\varphi(c)
(x-c)^k
\eeq
is the $n$-jet of~$\varphi$ at~$c$. For the rest of this section, 
we need only the
case $n=0$; however, in Appendix~\ref{lagrecon} we will briefly discuss 
a situation with $n>1$.
For subtractions at the boundaries of the integration region we use the 
short-hand notation
\beqnm{shnd}
\suba{D}{a}{b}[\varphi]
&\doteq&\int_a^b\dd x \,D(x)\,\left(\varphi(x)-\varphi(a)\right),
\nonu
\subb{D}{a}{b}[\varphi]
&\doteq&\int_a^b\dd x \,D(x)\,\left(\varphi(x)-\varphi(b)\right).
\eeqn
If a function depends on several variables or is given in an explicit form, 
we include the variable relevant to the subtraction in the
subscript as in $\subbl{(1/x)}{x}{0}{1}$.

The singular functions~$D_+$ can be rewritten for a changed domain of
their definition. To this end, we define the characteristic function
$\chi_A$ for a set~$A$ by
\beq
\chi_A(x)\doteq
\left\{
\begin{array}{cl}
1 ,& \mbox{if}\quad x\in A\\
0, & \mbox{else}
\end{array}
\right. .
\eeq
Then, for $d\in[a,b]$,
\beqnm{Dshift}
\suba{D}{a}{b}&=&\suba{D}{a}{d}\,+\,\chi_{[d,b]}\,D\,-\,
\int_d^b\dd x\,D(x)\,\delta_a,
\nonu
\subb{D}{a}{b}&=&\subb{D}{d}{b}\,+\,\chi_{[a,d]}\,D\,-\,
\int_a^d\dd x\,D(x)\,\delta_b.
\eeqn
This leads for instance to the expressions
\beqm{xtrf}
\begin{array}[b]{lcl}
{\ds \subbl{\left(\frac{1}{1-x}\right)}{x}{0}{1}}
&=&{\ds\subbl{\left(\frac{1}{1-x}\right)}{x}{d}{1}
+\chi_{[0,d]}(x)\frac{1}{1-x}+\delta(1-x)\ln(1-d),}\\
{ }&&\\
{\ds \subbl{\left(\frac{\ln(1-x)}{1-x}\right)}{x}{0}{1}}
&=&{\ds\subbl{\left(\frac{\ln(1-x)}{1-x}\right)}{x}{d}{1}
+\chi_{[0,d]}(x)\frac{\ln(1-x)}{1-x}}
\\&{ }&
\\
&&{\ds \!\!\!+\,\delta(1-x)\frac{1}{2}\ln^2(1-d)}.
\end{array}
\eeq
Similarly, a regular function~$f$ can be written as a ``$+$''~distribution
as
\beqnm{regdis}
f(x)\,\chi_{[a,b]}(x)&=&\suba{f}{a}{b}(x)+\delta(x-a)\,\int_a^b\dd v \, f(v),
\nonu
f(x)\,\chi_{[a,b]}(x)&=&\subb{f}{a}{b}(x)+\delta(b-x)\,\int_a^b\dd v \, f(v).
\eeqn

It turns out to be useful to perform a variable transformation
$y=\ln(1/x)$ in order to express the scale evolution equation
for fragmentation functions
in terms of convolutions of Laguerre polynomials. 
This transformation has also advantages for the numerical evaluation
of convolutions.
More precisely, 
define the function 
$y\mapsto\tilde{f}(y)$
 for a function
$x\mapsto f(x)$ by
\beqm{st3}
\tilde{f}(y)\doteq xf(x), \quad y=\ln\frac{1}{x}, 
\quad x=\mbox{e}^{-y}.
\eeq
If $x\in[0,1]$, then $y\in[0,\infty]$.
We have to give a meaningful definition of the quantity~$\tilde{D}$ in the
case where~$D$ is a distribution. In order to do so, we 
consider the convolution of the distributions~$D$ and~$\tilde{D}$ with
the test functions~$\varphi$ and~$\tilde{\varphi}$, respectively,
where~$\tilde{\varphi}$ is defined by Eq.~(\ref{st3}). The definition 
of the convolution ``$\otimes$'' is given in Appendix~\ref{convdistr},
Eqs.~(\ref{conv1}) and~(\ref{conv2}).
$\left(D\otimes \varphi\right)(x)$ and 
$\left(\tilde{D}\otimes \tilde{\varphi}\right)(y)$
are regular functions of~$x$ and~$y$,
$\left(D\otimes \varphi\right)^\sim$
is well defined by Eq.~(\ref{st3}), and we may require that 
\beqm{st4}
\left(D\otimes \varphi\right)^\sim=\tilde{D}\otimes \tilde{\varphi}
\eeq
for any test function~$\varphi$.
An explicit calculation then shows that
\beqnm{tf1}
\tilde{\delta}_1&=&\delta_0,
\nonu&&\nonu
\left(\subb{D}{\xi}{1}\right)^\sim&=&\suba{\tilde{D}}{0}
{\ln(1/\xi)}.
\eeqn

\dgsb{Convolution of Distributions}
\labelm{convdistr}
The convolution of two functions~$f$, $g$ defined on $[0,1]$ is given
by
\beqm{st2}
\labelmmm{conv1}
\left(f\otimes g\right)(x)\doteq\int_x^1\frac{\dd u}{u}\,f(u)\,
g\left(\frac{x}{u}\right).
\eeq
This definition is also applicable in the case when~$f$ is a distribution
on $[0,1]$, 
by writing
\beqnm{dco}
\left(f\otimes g\right)(x)&=&\int_0^1\frac{\dd u}{u}\,f(u)\,
\chi_{[x,1]}(u)\,g\left(\frac{x}{u}\right)\nonu
&=&f\left[u\mapsto\frac{1}{u}\,\chi_{[x,1]}(u)\,
g\left(\frac{x}{u}\right)\right],
\eeqn
and by interpreting the latter expression in terms of an application of a 
distribution to a test function.
This simple prescription fails in the case of two distributions
$f$ and~$g$. It is, however, possible to give a well-defined
meaning to $f\otimes g$ by applying it to a test function~$\varphi$:
\beqnm{st1}
\left(f\otimes g\right)[\varphi]
&=&
\int_0^1\dd x\,\int_x^1\frac{\dd u}{u}\,f(u)\,
g\left(\frac{x}{u}\right)\,\varphi(x)
\nonu
&=&\int_0^1\dd u\,f(u)\,
\int_0^1\dd w\,g(w)\,\varphi(uw).
\eeqn
Now $f\otimes g$ is rewritten in terms of a twofold application of 
distributions to a test function, which is a well-defined procedure.
{}From Eq.~(\ref{st1}), it follows that $f\otimes g=g\otimes f$.

Convolutions for functions $\tilde{f}$, $\tilde{g}$ 
(cf.\ Eq.~(\ref{st3})) can be 
defined by\footnote{We use the same symbol as in Eq.~(\ref{st2}).
The meaning of ``$\otimes$'' will be clear from the context.}
\beqm{conv2}
\left(\tilde{f}\otimes\tilde{g}\right)(y)=
\int_0^y\dd z \,\tilde{f}(z)\,\tilde{g}(y-z).
\eeq
Again, this expression is well defined for a distribution~$\tilde{f}$ by
\beqnm{deftcon}
\left(\tilde{f}\otimes \tilde{g}\right)(y)&=&\int_0^\infty\dd z\,
\tilde{f}(z)\,
\chi_{[0,y]}(z)\,\tilde{g}(y-z)\nonu
&=&\tilde{f}\left[z\mapsto\chi_{[0,y]}(z)\,
\tilde{g}(y-z)\right].
\eeqn
If~$\tilde{f}$ and~$\tilde{g}$ are distributions, then
\beqnm{deffg}
\left(\tilde{f}\otimes \tilde{g}\right)[\varphi]
&=&\int_0^\infty\dd y\,\int_0^y\dd z\,\tilde{f}(z)\,
\tilde{g}(y-z)\,\varphi(y)
\nonu
&=&\int_0^\infty\dd z\,\tilde{f}(z)\,
\int_0^\infty\dd v\,\tilde{g}(v)\,
\varphi(z+v).
\eeqn
 
We wish to derive explicit expressions for the cases where
$\tilde{f}$, $\tilde{g}$ are of the forms~$\delta_0$ or
$\suba{\tilde{D}}{0}{M}$. The case of $\delta$-functions is straightforward,
and one easily obtains
$\delta_0\otimes\tilde{h}=\tilde{h}$
for any regular or singular function~$\tilde{h}$. 
The case of distributions $\suba{\tilde{D}}{0}{M}$
is slightly more complicated.
By defining
\beqnm{firstmoment}
\tilde{D}^{M1}(y)&\doteq&\int_y^M\dd z\,\tilde{D}(z),
\nonu
\left(\tilde{D}\ltimes\tilde{g}\right)(y)&\doteq&
\left(\suba{\tilde{D}}{0}{y}\otimes\tilde{g}\right)(y)
=\int_0^y\dd z \tilde{D}(z)\,\left[
\tilde{g}(y-z)-\tilde{g}(y)
\right],
\eeqn
we obtain for a regular function~$\tilde{g}$ and for $y<M$
\beqnm{dpreg}
\left(\suba{\tilde{D}}{0}{M}\otimes\tilde{g}\right)(y)
&=&\int_0^M\dd z\,\suba{\tilde{D}}{0}{M}(z)\,\chi_{[0,y]}(z)\,\tilde{g}(y-z)
\nonu
&=&\int_0^M\dd z\,\tilde{D}(z)\,\left[
\chi_{[0,y]}(z)\,\tilde{g}(y-z)
-\chi_{[0,y]}(0)\,\tilde{g}(y)
\right]
\nonu
&=&\int_0^y\dd z\,\tilde{D}(z)\,\left[
\tilde{g}(y-z)
-\tilde{g}(y)
\right]
-\int_y^M\dd z\,\tilde{D}(z)\,\tilde{g}(y)
\nonu
&=&\left(\tilde{D}\ltimes\tilde{g}\right)(y)-\tilde{D}^{M1}(y)\,
\tilde{g}(y).
\eeqn
Note that $\tilde{D}^{M1}(M-z)=\porder{z}$.
In order to state the result for 
$\suba{\tilde{E}}{0}{M}\otimes\suba{\tilde{F}}{0}{M}$, 
we introduce the following notation:
\beqnm{EFDef}
\left(\tilde{E}\wedge\tilde{F}\right)(y)
&\doteq&\int_0^y\dd z\,
\left(\tilde{E}(z)\tilde{F}(y-z)-\tilde{E}(y)\tilde{F}(y-z)
-\tilde{E}(z)\tilde{F}(y)\right),
\nonu
c_{\tilde{E}\tilde{F}}^M&\doteq&\int_0^M\dd z\,\tilde{E}(z)\int_{M-z}^M\dd w
\,\tilde{F}(w)
\nonu
&=&\int_0^M\dd z\,\tilde{E}(z)\,\tilde{F}^{M1}(M-z)
=\left(\tilde{E}\otimes\tilde{F}^{M1}\right)(M).
\eeqn
The quantity
$c_{\tilde{E}\tilde{F}}^M$ is symmetric in $\tilde{E}$ and~$\tilde{F}$.
A lengthy calculation then yields, after performing suitable subtractions,
\beqnm{cfbas}
\suba{\tilde{E}}{0}{M}\otimes\suba{\tilde{F}}{0}{M}
&=&\suba{\left(\tilde{E}\wedge\tilde{F}-\tilde{E}\tilde{F}^{M1}
-\tilde{E}^{M1}\tilde{F}\right)}{0}{M}-c_{\tilde{E}\tilde{F}}^M\delta_0.
\eeqn
The final result for the convolution of two arbitrary linear
combinations
\beqm{frc}
\begin{array}[b]{lclclcl}
\tilde{K}&=&\tilde{K}^\delta\delta_0&+&\suba{\tilde{K}^s}{0}{M}&+&\tilde{K}^r,
\\
\tilde{L}&=&\tilde{L}^\delta\delta_0&+&\suba{\tilde{L}^s}{0}{M}&+&\tilde{L}^r
\end{array}
\eeq
is
\beqnm{convfor}
\tilde{K}\otimes\tilde{L} &=& \quad\left(\tilde{K}^\delta\tilde{L}^\delta
-c_{\tilde{K}^s\tilde{L}^s}^M\right)\delta_0
\nonu
&&+\suba{\left(\tilde{K}^\delta\tilde{L}^s+\tilde{K}^s\tilde{L}^\delta
+\tilde{K}^s\wedge\tilde{L}^s-\tilde{K}^s\tilde{L}^{sM1}
-\tilde{K}^{sM1}\tilde{L}^s\right)}{0}{M}
\nonu
&&
+\left(\tilde{K}^\delta-\tilde{K}^{sM1}\right)\tilde{L}^r
+\left(\tilde{L}^\delta-\tilde{L}^{sM1}\right)\tilde{K}^r
+\tilde{K}^r\otimes\tilde{L}^r
\nonu
&&
+\,\tilde{K}^s\ltimes\tilde{L}^r
+\tilde{L}^s\ltimes\tilde{K}^r.
\eeqn
This is again a linear combination of a term proportional to~$\delta_0$, 
a term with a ``$+$''~prescription of the form $\suba{D}{0}{M}$,
and a regular term. The decomposition in Eq.~(\ref{frc}) is therefore stable
under 
convolutions. Consequently, heavy-quark fragmentation functions
can be parametrized in this way, stable under scale evolution by means
of the renormalization group equation.

\dgcleardoublepage
\markh{Phase-Space Parametrizations}
\dgsa{Phase-Space Parametrizations}
\labelm{phasespace}

The $d$-dimensional phase-space element $\dps^{(N)}(\underline{p},
\underline{m})$
for~$N$ outgoing particles with momenta $\underline{p}=(p_1,\ldots,p_N)$
and masses $\underline{m}=(m_1,\ldots,m_N)$
is defined by
\beq
\dps^{(N)}(\underline{p},\underline{m})=(2\pi)^d\,
\delta\left(\sum_{\alpha=1}^N p_\alpha - w \right)\,\prod_{\alpha=1}^N
\frac{\dd p_\alpha\,\delta(p_\alpha^2-m_\alpha^2)}{(2\pi)^{d-1}},
\eeq
where~$w$ is the sum of all outgoing momenta.
We use the short-hand notation 
$\dps^{(N)}(\underline{p})$ if all masses are zero.

\dgsb{Massless Partons}
\labelm{mlp}
The one-particle phase space for a massless parton is
\beqm{opps}
\int\dps^{(1)}(p_1)=2\pi\,\frac{\XB}{Q^2}\,\delta(\xi-\XB)
=2\pi\,\frac{1}{Q^2}\,\delta(1-u),
\eeq
where $u=\XB/\xi$, the variables being defined
as in Section~\ref{dislns}. The energy of~$p_1$ is given by
$E_1=P_0(1-\XB)$.
\begin{sloppypar}
Now we give three different parametrizations of the
phase space $\dps^{(2)} (p_1,p_2)$ of two massless
partons with momenta~$p_1$ and
$p_2$. In the following, energies and angles
are defined in the hadronic centre-of-mass frame, 
cf.\ the remarks in Section~\ref{dislns}.
The azimuthal angle has been integrated out.
\end{sloppypar}
\begin{itemize}

\item[\dgbullet] {\it Parametrization A (cf.\ Fig.~\ref{PSRC}):}\\
The integration variable~$\rho$
is the energy~$E_1$
of the parton with momentum~$p_1$, scaled by a multiple of the proton 
momentum:
\beq
\rho\,\doteq\,\frac{E_1}{\EE(1-\XB)}.
\eeq
This parametrization is used for those contributions where a 
collinear singularity has to be absorbed into a fragmentation function~$D$;
its explicit form is
\beqnm{exps1}
\int\dps^{(2)}(p_1,p_2)\,&=&\,\int\,\frac{1}{8\pi}\,
\frac{(4\pi)^\epsilon}{\Gamma(1-\epsilon)}\,
\left(Q^2\right)^{-\epsilon}\,
\frac{u(1-\XB)}{u-\XB}\,
\nonu
&&\mathl\mathl\cdot\,(1-\XB)^{-2\epsilon}\,u^{-\epsilon}\,
(1-u)^{-\epsilon}\,(u-\XB)^{2\epsilon}\,
(\rho-a(u))^{-\epsilon}\,(1-\rho)^{-\epsilon}\,\dd \rho.
\eeqn
Here 
\beq
a(u)\doteq\frac{\XB}{1-\XB}\frac{1-u}{u}.
\eeq
The inverse function of $u\mapsto a(u)$ is denoted by 
$\rho\mapsto u_0(\rho)$ and is given by
\beqm{u0r}
u_0(\rho)=\frac{\XB}{\XB+(1-\XB)\rho}.
\eeq
The energies and invariants are given by
\beqnm{exps2}
E_1&=&\EE(1-\XB)\rho,\nonu
E_2&=&\EE(1-\XB)(1-\rho+a(u)),\nonu
s_{12}&=&Q^2\frac{1-u}{u},\nonu
s_{i1}&=&Q^2\frac{1-\XB}{u-\XB}(\rho-a(u)),\nonu
s_{i2}&=&Q^2\frac{1-\XB}{u-\XB}(1-\rho).
\eeqn
The angular variable $v_1=(1-\cos\vartheta_1)/2$ is
\beq
v_1=v(\rho,u)=\frac{\XB(1-u)}{u-\XB} \frac{1-\rho}{\rho}.
\eeq
The range of integration is restricted to
$\rho\in[a(u),1]$;
however, in order to ensure that the energy of the parent parton 
is larger than that of the observed hadron itself,
the additional condition $\rho \geq z$ must be satisfied, 
for the case that the parent parton's momentum is~$p_1$.
\addtocounter{footnote}{1}
\footnotetext{\footn{Bfoot}Please note that the remark in footnote 
\itemr{Afoot}
does not apply in this case.
}

\item[\dgbullet] {\it Parametrization B:}\\
The integration variable~$w$
is related to an angular variable in the centre-of-mass system of the
virtual photon and the incoming parton. 
Its relation to~$\rho$ is
\beq
w\doteq\frac{1-\rho}{1-a(u)}.
\eeq
This parametrization is used for the contributions that involve a
target fragmentation function;
it is given by 
\beqnm{exps3}
\int\dps^{(2)}(p_1,p_2)\,=\,\int\,\frac{1}{8\pi}\,
\frac{(4\pi)^\epsilon}{\Gamma(1-\epsilon)}\,
\left(Q^2\right)^{-\epsilon}\,
(1-u)^{-\epsilon}\,u^\epsilon\,
(w(1-w))^{-\epsilon}
\dd w.
\eeqn
The energies and invariants are
\beqnm{exps4}
E_1&=&\EE(1-\XB)(1-(1-a(u))w),\nonu
E_2&=&\EE(1-\XB)(a(u)+(1-a(u))w),\nonu
s_{12}&=&Q^2\frac{1-u}{u},\nonu
s_{i1}&=&Q^2\frac{1}{u}(1-w),\nonu
s_{i2}&=&Q^2\frac{1}{u}w.
\eeqn
The angular variable~$v_1$ is
\beq
v_1=v(w,u)=\frac{a(u)w}{1-(1-a(u))w}.
\eeq
The range of integration is restricted to
$w\in[0,1]$.

\item[\dgbullet] {\it Parametrization C:}\\
This parametrization is convenient for the contributions where 
the observed hadron originates from a parton that is collinear 
to the incoming parton.
A variable~$u^\prime$ is introduced by
\beq
u^\prime=1-\frac{1-\XB}{\XB}\,\rho\,u.
\eeq
The parametrization is given by 
\beqn
\int\dps^{(2)}(p_1,p_2)\,&=&\,\int\,\frac{1}{8\pi}\,
\frac{(4\pi)^\epsilon}{\Gamma(1-\epsilon)}\,
\left(Q^2\right)^{-\epsilon}\,
\frac{\XB}{u-\XB}
\,(u-u^\prime)^{-\epsilon}\,
(1-u)^{-\epsilon}\,
\nonu
&&\!\!\!\!\!\!\!\!\!\!\!\!\cdot\left(
 1-\frac{\XB}{1-\XB}\frac{1-u^\prime}{u}
\right)^{-\epsilon}
(u-\XB)^{2\epsilon}\,
\XB^{-\epsilon}\,(1-\XB)^{-\epsilon}\,\dd u^\prime.
\eeqn
The energies and invariants are
\beqn
E_1&=&\EE\,\XB \frac{1-u^\prime}{u},\nonu
E_2&=&\EE(1-\XB)\left(
 1-\frac{\XB}{1-\XB}\frac{u-u^\prime}{u}
\right),\nonu
s_{12}&=&Q^2\frac{1-u}{u},\nonu
s_{i1}&=&Q^2\frac{\XB}{u-\XB}\frac{u-u^\prime}{u},\nonu
s_{i2}&=&Q^2\frac{1-\XB}{u-\XB}\left(
 1-\frac{\XB}{1-\XB}\frac{1-u^\prime}{u}
\right).
\eeqn
The angular variable~$v_1$ is
\beq
v_1=v(u^\prime,u)=\frac{1-u}{u-\XB} \frac{(1-\XB)u-\XB(1-u^\prime)}{1-u^\prime}.
\eeq
The range of integration is restricted to
$u^\prime\in[1-u(1-\XB)/\XB,u]$.
It has to be further restricted by
\beq
u^\prime\,\leq\,1-\frac{1-\XB}{\XB}\,z\,u
\eeq
in order to avoid that the outgoing hadron
has an energy larger than its parent parton's.
\end{itemize}

\dgsb{Massive Partons}
\labelm{milp}
The two-particle phase space for massive partons of mass~$m$ 
in $d=4$ space-time dimensions can 
be written as
\beqm{twopm}
\int\dps^{(2)}(p_1,p_2,m,m)\,=\,\int\,\frac{1}{8\pi}\,
\sqrt{1-\frac{4m^2}{w^2}}\,\dd r.
\eeq
Here $w=p_1+p_2$, and $r=(1-\cos\chi)/2$, where~$\chi$ is a polar
angle of~$p_1$ in the centre-of-mass frame
of~$p_1$ and~$p_2$.
The azimuthal angle is already integrated out.
The range of~$r$ is $[0,1]$.
We briefly summarize some expressions for the energies
in the centre-of-mass frame
of~$p_1$ and~$p_2$
in the case of $w=\xi P+q$:
\beqnm{enw}
E_w&=&Q\sqrt{\frac{\xi-x_B}{x_B}},
\nonu
E_P&=&\frac{Q}{2}\,\frac{1}{\sqrt{x_B(\xi-x_B)}},
\nonu
E_{1}&=&E_w/2,
\nonu
E_{2}&=&E_w/2.
\eeqn
The expressions for the invariants $s_{AB}$ follow easily.
We have the following phase-space restrictions on~$Q^2$ and~$\xi$:
\beqnm{prqx}
Q^2&\geq&4m^2\,\frac{x_B}{1-x_B},
\nonu
\xi&\geq&x_B\left(1+\frac{4m^2}{Q^2}\right).
\eeqn

\dgcleardoublepage
\markh{Finite Contributions of the Real Corrections}
\dgsa{Finite Contributions of the Real Corrections}
\labelm{realfinite}

\dgsb{Explicit Expressions}
\labelm{rfee}
This appendix contains the explicit results of the finite contributions
from the real 
cor\-rect\-ions\footnote{The calculations have been done with the help of the
algebraic manipulation programs ``Maple'' \cite{150}, 
``Mathematica'' \cite{151} and ``REDUCE'' \cite{152}, and by 
using the package ``Tracer'' \cite{153} for trace calculations
of $\gamma$-matrices.}.
We have dropped terms that vanish for $\epsilon\rightarrow 0$.
The results are given by
\beqnm{Arealfinite}
{\cal A}_{B^M_1}^f&=&Y^M\,\sum_{i=q,\qb}\,c_i\,\frac{\alpha_s}{2\pi}\,
\int_{\XB}^{\XB/(\XB+(1-\XB)z)}\,\frac{\dd u}{u}\,
\int_{a(u)}^1\,\frac{\dd \rho}{\rho}\,A(v(\rho,u))\nonu
&&\mathl\cdot\Bigg[
f^r_{i/P}\left(\frac{\XB}{u},\mu_f^2\right)\,
D^r_{\hh/i}\left(\frac{z}{\rho},\mu_D^2\right)\,
\left\{
-\ln\frac{\mu_f^2}{Q^2}\,P_{q\leftarrow q}(u)\,\delta(1-\rho)
+C_F\,\Phi_{1qq}^M
\right\}\nonu
&&\mathl\;\;+
\,f^r_{i/P}\left(\frac{\XB}{u},\mu_f^2\right)\,
D^r_{\hh/g}\left(\frac{z}{\rho},\mu_D^2\right)\,
\left\{
-\ln\frac{\mu_M^2}{Q^2}\,\hat{P}_{gq\leftarrow q}(u)\,\delta(\rho-a(u))
+C_F\,\Phi_{1qg}^M
\right\}\nonu
&&\mathl\;\;+
\,f^r_{g/P}\left(\frac{\XB}{u},\mu_f^2\right)\,
D^r_{\hh/i}\left(\frac{z}{\rho},\mu_D^2\right)\nonu
&&\mathl\;\;
\cdot\left\{
-\ln\frac{\mu_M^2}{Q^2}\,\hat{P}_{\qb q\leftarrow g}(u)\,\delta(\rho-a(u))
-\ln\frac{\mu_f^2}{Q^2}\,P_{q\leftarrow g}(u)\,\delta(1-\rho)
+T_f\,\Phi_{1gq}^M
\right\}\Bigg],
\nonu
{ }\nonu
{\cal A}^f_{B^M_2}&=&Y^M\,\sum_{i=q,\qb}\,c_i\,\frac{\alpha_s}{2\pi}\,
\int_{\XB/(\XB+(1-\XB)z)}^1\,\frac{\dd u}{u}\,
\int_{z}^1\,\frac{\dd \rho}{\rho}\,A(v(\rho,u))\nonu
&&\mathl\cdot\Bigg[
f^r_{i/P}\left(\frac{\XB}{u},\mu_f^2\right)\,
D^r_{\hh/i}\left(\frac{z}{\rho},\mu_D^2\right)\nonu
&&\mathl\;\;\cdot\left\{
-\ln\frac{\mu_f^2}{Q^2}\,P_{q\leftarrow q}(u)\,\delta(1-\rho)
-\ln\frac{\mu_D^2}{Q^2}\,P_{q\leftarrow q}(\rho)\,\delta(1-u)
+C_F\,\Phi_{2qq}^M
\right\}\nonu
&&\mathl\;+
\,f^r_{i/P}\left(\frac{\XB}{u},\mu_f^2\right)\,
D^r_{\hh/g}\left(\frac{z}{\rho},\mu_D^2\right)\,
\left\{
-\ln\frac{\mu_D^2}{Q^2}\,P_{g\leftarrow q}(\rho)\,\delta(1-u)
+C_F\,\Phi_{2qg}^M
\right\}\nonu
&&\mathl\;+
\,f^r_{g/P}\left(\frac{\XB}{u},\mu_f^2\right)\,
D^r_{\hh/i}\left(\frac{z}{\rho},\mu_D^2\right)\,
\left\{
-\ln\frac{\mu_f^2}{Q^2}\,P_{q\leftarrow g}(u)\,\delta(1-\rho)
+T_f\,\Phi_{2gq}^M
\right\}
\Bigg],
\nonu
{ }\nonu
{\cal A}^f_{C^M}&=&Y^M\,\sum_{i=q,\qb}\,c_i\,\frac{\alpha_s}{2\pi}\,
\int_{\XB/(1-(1-\XB)z)}^1\,\frac{\dd u}{u}\,A(1)\nonu
&&\mathl\cdot\Bigg[
M^r_{i,h/P}\left(\frac{\XB}{u},(1-\XB)z,\mu_M^2\right)\,
\bigg\{
-\ln\frac{\mu_M^2}{Q^2}\,P_{q\leftarrow q}(u)\,(1-\XB)
+C_F\,\Phi_{q}^M
\bigg\}\nonu
&&\mathl\;+
\,M^r_{g,h/P}\left(\frac{\XB}{u},(1-\XB)z,\mu_M^2\right)\,
\bigg\{
-\ln\frac{\mu_M^2}{Q^2}\,P_{q\leftarrow g}(u)\,(1-\XB)
+T_f\,\Phi_{g}^M
\bigg\}
\Bigg],
\nonu
{ }\nonu
{\cal A}^f_{B^L_1}&=&Y^L\,\sum_{i=q,\qb}\,c_i\,\frac{\alpha_s}{2\pi}\,
\int_{\XB}^{\XB/(\XB+(1-\XB)z)}\,\frac{\dd u}{u}\,
\int_{a(u)}^1\,\frac{\dd \rho}{\rho}\,A(v(\rho,u))\nonu
&&\mathl\cdot\Bigg[
f^r_{i/P}\left(\frac{\XB}{u},\mu_f^2\right)\,
D^r_{\hh/i}\left(\frac{z}{\rho},\mu_D^2\right)\,
C_F\,\Phi_{1qq}^L\nonu
&&\mathl+
\,f^r_{i/P}\left(\frac{\XB}{u},\mu_f^2\right)\,
D^r_{\hh/g}\left(\frac{z}{\rho},\mu_D^2\right)\,
C_F\,\Phi_{1qg}^L\nonu
&&\mathl+
\,f^r_{g/P}\left(\frac{\XB}{u},\mu_f^2\right)\,
D^r_{\hh/i}\left(\frac{z}{\rho},\mu_D^2\right)\,
T_f\,\Phi_{1gq}^L
\Bigg],
\nonu
{ }\nonu
{\cal A}^f_{B^L_2}&=&Y^L\,\sum_{i=q,\qb}\,c_i\,\frac{\alpha_s}{2\pi}\,
\int_{\XB/(\XB+(1-\XB)z)}^1\,\frac{\dd u}{u}\,
\int_{z}^1\,\frac{\dd \rho}{\rho}\,A(v(\rho,u))\nonu
&&\mathl\cdot\Bigg[
f^r_{i/P}\left(\frac{\XB}{u},\mu_f^2\right)\,
D^r_{\hh/i}\left(\frac{z}{\rho},\mu_D^2\right)\,
C_F\,\Phi_{2qq}^L\nonu
&&\mathl+
\,f^r_{i/P}\left(\frac{\XB}{u},\mu_f^2\right)\,
D^r_{\hh/g}\left(\frac{z}{\rho},\mu_D^2\right)\,
C_F\,\Phi_{2qg}^L\nonu
&&\mathl+
\,f^r_{g/P}\left(\frac{\XB}{u},\mu_f^2\right)\,
D^r_{\hh/i}\left(\frac{z}{\rho},\mu_D^2\right)\,
T_f\,\Phi_{2gq}^L
\Bigg],
\nonu
{ }\nonu
{\cal A}^f_{C^L}&=&Y^L\,\sum_{i=q,\qb}\,c_i\,\frac{\alpha_s}{2\pi}\,
\int_{\XB/(1-(1-\XB)z)}^1\,\frac{\dd u}{u}\,A(1)\nonu
&&\mathl\cdot\Bigg[
M^r_{i,h/P}\left(\frac{\XB}{u},(1-\XB)z,\mu_M^2\right)\,
C_F\,\Phi_{q}^L\nonu
&&\mathl+
\,M^r_{g,h/P}\left(\frac{\XB}{u},(1-\XB)z,\mu_M^2\right)\,
T_f\,\Phi_{g}^L
\Bigg],
\eeqn

\noindent
where the functions~$\Phi$ are given by

\beqn
\Phi^M_{1qq}&=&
\delta(1-\rho)\,\left[
2\,\frac{\ln(1-u)}{1-u}
+1-u-(1+u)\ln(1-u)-\frac{1+u^2}{1-u}\ln\frac{u-\XB}{1-\XB}
                \right]
\nonu
&&\mathl\!\!\!\!\!
+\,2\left(\frac{1}{1-\rho}\right)_{+\rho[0,\underline{1}]}\,
\frac{1}{1-u}
-\frac{1}{1-u}\,(1+\rho)
-\left(\frac{1}{1-\rho}\right)_{+\rho[0,\underline{1}]}\,(1+u)
\nonu
&&\mathl\!\!\!\!\!
+\,(1-\rho)\frac{\XB}{u-\XB}
\left(1+\frac{u(1-\XB)}{u-\XB}\right)
-2\frac{u\XB}{u-\XB}
+2,
\nonu
{ }\nonu
\Phi^M_{1qg}&=&\delta(\rho-a(u))\,
\left[1-u+\frac{1+u^2}{1-u}\ln\frac{1-u}{u}
\right]
+\left(\frac{1}{\rho-a(u)}\right)_{+\rho[\underline{a(u)},1]}
\frac{1+u^2}{1-u}
\nonu
&&\mathl\!\!\!\!\!
+\,\frac{(1-\XB)^2}{(u-\XB)^2}\frac{u^2}{1-u}\rho
-\frac{\XB(1-\XB)}{(u-\XB)^2}u
-2\frac{1-\XB}{u-\XB}\frac{u^2}{1-u},
\nonu
{ }\nonu
\Phi^M_{1gq}&=&
\delta(\rho-a(u))\,
\Bigg[2u(1-u)+(1-2u+2u^2)\ln\frac{1-u}{u}
\Bigg]
\nonu
&&\mathl\!\!\!\!\!
+\,\,\delta(1-\rho)\,
\Bigg[2u(1-u)+(1-2u+2u^2)\ln\frac{(1-u)(1-\XB)}{u-\XB}
\Bigg]
\nonu
&&\mathl\!\!\!\!\!
+\left(\frac{1}{\rho-a(u)}\right)_{+\rho[\underline{a(u)},1]}
(1-2u+2u^2)
\nonu
&&\mathl\!\!\!\!\!
+\left(\frac{1}{1-\rho}\right)_{+\rho[0,\underline{1}]}
(1-2u+2u^2)
-2\frac{1-\XB}{u-\XB}u,
\nonu
{ }\nonu
\Phi^M_{2qq}&=&\delta(1-u)\,\delta(1-\rho)\,\frac{\pi^2}{3}\nonu
&&\mathl\!\!\!\!\!+
\,\delta(1-\rho)\,\left[
2\left(\frac{\ln(1-u)}{1-u}\right)_{+u[0,\underline{1}]}
+1-u-(1+u)\ln(1-u)-\frac{1+u^2}{1-u}\ln\frac{u-\XB}{1-\XB}
                \right]
\nonu
&&\mathl\!\!\!\!\!+
\,\delta(1-u)\,\left[
2\left(\frac{\ln(1-\rho)}{1-\rho}\right)_{+\rho[0,\underline{1}]}
+1-\rho-(1+\rho)\ln(1-\rho)+\frac{1+\rho^2}{1-\rho}\ln\rho
                \right]
\nonu
&&\mathl\!\!\!\!\!
+\,2\left(\frac{1}{1-\rho}\right)_{+\rho[0,\underline{1}]}\,
\left(\frac{1}{1-u}\right)_{+u[0,\underline{1}]}
-\left(\frac{1}{1-u}\right)_{+u[0,\underline{1}]}\,(1+\rho)
-\left(\frac{1}{1-\rho}\right)_{+\rho[0,\underline{1}]}\,(1+u)
\nonu
&&\mathl\!\!\!\!\!
+\,(1-\rho)\frac{\XB}{u-\XB}
\left(1+\frac{u(1-\XB)}{u-\XB}\right)
-2\frac{u\XB}{u-\XB}
+2,
\nonu
{ }\nonu
\Phi^M_{2qg}&=&\delta(1-u)\,
\Bigg[\rho+(\ln \rho+\ln(1-\rho))\left(\rho+\frac{2}{\rho}-2\right)
\Bigg]
\nonu
&&\mathl\!\!\!\!\!
+\left(\frac{1}{1-u}\right)_{+u[0,\underline{1}]}
\left(\rho+\frac{2}{\rho}-2\right)
\nonu
&&\mathl\!\!\!\!\!
+2-2\frac{\XB u}{u-\XB}
+\frac{\XB}{u-\XB}\rho
-\frac{\XB(1-\XB)u}{(u-\XB)^2}(1-\rho)
+\frac{1}{\rho-a(u)}\frac{1+u^2}{1-u}
-2\frac{1}{\rho}\frac{1}{1-u},
\nonu
{ }\nonu
\Phi^M_{2gq}&=&
\delta(1-\rho)\,
\Bigg[2u(1-u)+(1-2u+2u^2)\ln\frac{(1-u)(1-\XB)}{u-\XB}
\Bigg]
\nonu
&&\mathl\!\!\!\!\!
+\,\frac{1}{\rho-a(u)}
(1-2u+2u^2)
+\left(\frac{1}{1-\rho}\right)_{+\rho[0,\underline{1}]}
(1-2u+2u^2)
-2\frac{1-\XB}{u-\XB}u,
\nonu
{ }\nonu
\Phi^M_{q}&=&(1-\XB)\Bigg[
\frac{7}{2}\,\delta(1-u)\,
-\frac{3}{2}\,\left(\frac{1}{1-u}\right)_{+u[0,\underline{1}]}
+2\,\left(\frac{\ln(1-u)}{1-u}\right)_{+u[0,\underline{1}]}
\nonu
&&\mathl\!\!\!\!\!
\quad\quad
\quad\quad
+\,3-u-(1+u)\ln(1-u)-\frac{1+u^2}{1-u}\ln u
\Bigg],
\nonu
{ }\nonu
\Phi^M_{g}&=&(1-\XB)
\left(\ln\frac{1-u}{u}-1\right)(1-2u+2u^2),
\nonu
{ }\nonu
\Phi^L_{1qq}&=&
2\frac{(1-\XB)^2}{(u-\XB)^2}u^3\rho
-2\frac{\XB(1-\XB)}{(u-\XB)^2}u^2(1-u),
\nonu
{ }\nonu
\Phi^L_{1qg}&=&
2\frac{(1-\XB)^2}{(u-\XB)^2}u^3(1-\rho),
\nonu
{ }\nonu
\Phi^L_{1gq}&=&
4\frac{1-\XB}{u-\XB}u^2(1-u),
\nonu
{ }\nonu
\Phi^L_{2qq}&=&\Phi^L_{1qq},
\nonu
{ }\nonu
\Phi^L_{2qg}&=&\Phi^L_{1qg},
\nonu
{ }\nonu
\Phi^L_{2gq}&=&\Phi^L_{1gq},
\nonu
{ }\nonu
\Phi^L_{q}&=&(1-\XB)\,u,
\nonu
{ }\nonu
\Phi^L_{g}&=&(1-\XB)\,2u(1-u).
\eeqn

\dgsb{The Case of Singular Fragmentation Functions}
\labelm{rfef}
The expressions given in the previous section can be applied directly
in a numerical evaluation in the case where the fragmentation functions
are regular functions. In the application that we have in mind, the 
production of heavy quarks, the fragmentation functions $D_{Q/i}(z/\rho)$
are singular for $\rho\rightarrow z$. A direct application of the given
formulae by means of the convolution formula in Eq.~(\ref{convfor}) fails
for the contributions from the separate phase-space regions~$R_1$ and~$R_2$, 
due to a logarithmic divergence at $\rho=z$ after having performed the
$u$-integration. It turns out that these dangerous terms cancel out,
as they should,
if the contributions from the phase space regions~$R_1$ and~$R_2$
(see Fig.~\ref{PSRC}) are added.
In order to see this explicitly, the distributions
$\delta(\rho-a(u))$ and $\subbl{1/(\rho-a(u))}{\rho}{a(u)}{1}$
have to be expressed in terms of distributions in the variable~$u$.
The corresponding relations are given by
\beqm{rhou}
\begin{array}[b]{ccl}
\delta(\rho-a(u))&=&{\ds\frac{1-\XB}{\XB}}(u_0(\rho))^2\,\,\delta(u-u_0(\rho)),
\\
{ }
\\
{\ds\subal{\left(\frac{1}{\rho-a(u)}\right)}{\rho}{a(u)}{1}}
&=& 
{\ds\frac{1-\XB}{\XB}}(u_0(\rho))^2 
\Bigg[
{\ds\subal{\left(\frac{1}{u-u_0(\rho)}\right)}{u}{u_0(\rho)}{1}
+\frac{1}{u_0(\rho)}}
\\
\rule{0mm}{1.5mm}
\\
&&
\quad\quad\quad\quad\quad
\quad\quad-\delta(u-u_0(\rho))\,
{\ds\ln\left(\frac{1-\rho}{\rho}u_0(\rho)\right)}
\Bigg].
\end{array}
\eeq
The expressions for the cross sections are integrated in~$\rho$ from 
$z$ to~$1$ and in~$u$ from $u_0(\rho)$ to~$1$. The subtractions in the
variable~$u$ are performed first, leaving a regular expression depending on 
the variable~$\rho$. Then the subtractions in~$\rho$ are performed. 
In cases where a ``$+$''~prescription in $z/\rho$ from the 
heavy-quark fragmentation functions is multiplied by a 
``$+$''~prescription in~$\rho$ from the matrix element, the convolution 
formula from Eq.~(\ref{cfbas}) has to be applied.
The resulting expression is singular in the variable~$z$. An integration
over~$z$ involving an observable therefore contains an additional subtraction.
The explicit expressions which are needed for this procedure are given in the
following.
We consider the integrals
\beqm{II}
I(d,s)\doteq\int_{z_0}^1\frac{\dd z}{z}\int_z^1\frac{\dd \sigma}{\sigma}
\,d(\sigma)\,s\left(\frac{z}{\sigma}\right)\,C\left(\frac{z}{\sigma},z\right),
\eeq
where $\sigma\mapsto d(\sigma)$ and $\rho\mapsto s(\rho)$ 
are distributions with
possible singularities at $\sigma=1$ and $\rho=1$, and~$C$ is a 
test function standing for the product of the remaining regular terms
of the matrix elements, the parton densities and the observable
under consideration.
In order to achieve a compact notation, we introduce the following 
short-hand notation:
\beqm{sh1}
M\doteq\ln{\displaystyle \frac{1}{z_0}},\quad
\zeta\doteq \ln{\displaystyle \frac{1}{z}},\quad
\mu\doteq{\displaystyle\frac{1}{\zeta}}
\,\ln{\displaystyle \frac{1}{\sigma}},\quad
\lambda\doteq{\displaystyle\frac{\zeta}{M}},\quad
\rho\doteq {\displaystyle\frac{z}{\sigma}}.
\eeq
We obtain
\beqnm{dsC}
d(\sigma)&=&
\delta(1-\sigma),
\quad
s(\rho)\,\,\,\,=\,\,\,\,
\delta(1-\rho):
\nonu
I(d,s)&=&\int_0^1\dd \lambda\int_0^1\dd \mu\,\,
C(1,1);
\nonu&&\nonu
d(\sigma)&=&
\delta(1-\sigma),
\quad
s(\rho)\,\,\,\,=\,\,\,\,
S(\rho):
\nonu
I(d,s)&=&\int_0^1\dd \lambda\int_0^1\dd \mu\,\,
M\,S(z)\,C(z,z);
\nonu&&\nonu
d(\sigma)&=&
\delta(1-\sigma),
\quad
s(\rho)\,\,\,\,=\,\,\,\,
\subb{S}{z_0}{1}(\rho):
\nonu
I(d,s)&=&\int_0^1\dd \lambda\int_0^1\dd \mu\,\,
M\,S(z)\,\Big[C(z,z)-zC(1,1)\Big];
\nonu&&\nonu
d(\sigma)&=&
D(\sigma),
\quad
s(\rho)\,\,\,\,=\,\,\,\,
\delta(1-\rho):
\nonu
I(d,s)&=&\int_0^1\dd \lambda\int_0^1\dd \mu\,\,
M\,D(z)\,C(1,z);
\nonu&&\nonu
d(\sigma)&=&
D(\sigma),
\quad
s(\rho)\,\,\,\,=\,\,\,\,
S(\rho):
\nonu
I(d,s)&=&\int_0^1\dd \lambda\int_0^1\dd \mu\,\,
M\zeta\,D(\sigma)\,S(\rho)\,C(\rho,z);
\nonu&&\nonu
d(\sigma)&=&
D(\sigma),
\quad
s(\rho)\,\,\,\,=\,\,\,\,
\subb{S}{z_0}{1}(\rho):
\nonu
I(d,s)&=&\int_0^1\dd \lambda\int_0^1\dd \mu\,\,
\Big\{
M\zeta\,D(\sigma)\,S(\rho)\,\Big[C(\rho,z)-C(1,z)\Big]\nonu
&&\quad\quad\quad\quad\quad\quad
 + M\zeta\,\Big[D(\sigma)-\rho D(z)\Big]\,S(\rho)\,C(1,z)\nonu
&&\quad\quad\quad\quad\quad\quad
 - M D(z) S^i_{z_0}(z)\,C(1,z)
\Big\};
\nonu&&\nonu
d(\sigma)&=&
\subb{D}{z_0}{1}(\sigma),
\quad
s(\rho)\,\,\,\,=\,\,\,\,
\delta(1-\rho):
\nonu
I(d,s)&=&\int_0^1\dd \lambda\int_0^1\dd \mu\,\,
M\,D(z)\,\Big[C(1,z)-zC(1,1)\Big];
\nonu&&\nonu
d(\sigma)&=&
\subb{D}{z_0}{1}(\sigma),
\quad
s(\rho)\,\,\,\,=\,\,\,\,
S(\rho):
\nonu
I(d,s)&=&\int_0^1\dd \lambda\int_0^1\dd \mu\,\,
\Big\{
M\zeta\,D(\sigma)\,\Big[S(\rho)\,C(\rho,z)-\sigma S(z)\,C(z,z)\Big]
\nonu
&&\quad\quad\quad\quad\quad\quad
-M\,D^i_{z_0}(z)\,S(z)\,C(z,z)
\Big\};
\nonu&&\nonu
d(\sigma)&=&
\subb{D}{z_0}{1}(\sigma),
\quad
s(\rho)\,\,\,\,=\,\,\,\,
\subb{S}{z_0}{1}(\rho):
\nonu
I(d,s)&=&\int_0^1\dd \lambda\int_0^1\dd \mu\,\,
\Big\{
\Big[M\zeta\,\Big(D(\sigma)\,S(\rho)-\rho D(z)\,S(\rho)
-\sigma D(\sigma)\,S(z)\Big)
\nonu
&&\quad\quad\quad\quad\quad
-M\left(D(z)S^i_{z_0}(z)+D^i_{z_0}(z)\,S(z)\right)\Big]
\,\cdot\,\Big[C(1,z)-zC(1,1)\Big]
\nonu
&&\quad\quad\quad\quad\quad
-M\,z D(z)\,S^i_{z_0}\left(\frac{z_0}{z}\right)\,C(1,1)
\nonu
&&\quad\quad\quad\quad\quad
+M\zeta\,D(\sigma)\,\Big[
S(\rho)\Big(C(\rho,z)-C(1,z)\Big)
-\sigma S(z)\Big(C(z,z)-C(1,z)\Big)
\Big]
\nonu
&&\quad\quad\quad\quad\quad
-M\,D^i_{z_0}(z)\,S(z)\,\Big[C(z,z)-C(1,z)\Big]
\Big\}.
\eeqn
Here we have defined
\beqm{SIZ}
F^i_{x_0}(x)\doteq\int_{x_0}^x\dd\xi\, F(\xi),
\eeq
cf.\ Eq.~(\ref{xmoment}).
Similarly, the integral of the $u$-integration may be written as
\beqm{JJ}
J(e)\doteq\int_{u_0(\rho)}^1\frac{\dd u}{u}
\,e(u)\,F(u),
\eeq
where $u\mapsto e(u)$
is a distribution with possible singularities at 
$u=u_0(\rho)$ and $u=1$. Using the short-hand notation
\beqnm{sh2}
T(\rho)\doteq\ln\frac{1}{u_0(\rho)},
\quad \nu\doteq\frac{1}{T(\rho)}\ln\frac{1}{u},
\eeqn
we obtain:
\beqnm{eF}
e(u)&=&
\delta(1-u):
\nonu
J(e)&=&\int_0^1\dd \nu\,
F(1);
\nonu&&\nonu
e(u)&=&
E(u):
\nonu
J(e)&=&\int_0^1\dd \nu\,
T(\rho)\,E(u)\,F(u);
\nonu&&\nonu
e(u)&=&
\subb{E}{u_0(\rho)}{1}(u):
\nonu
J(e)&=&\int_0^1\dd \nu\,
T(\rho)\,E(u)\Big[F(u)-uF(1)\Big];
\nonu&&\nonu
e(u)&=&
\suba{E}{u_0(\rho)}{1}(u):
\nonu
J(e)&=&\int_0^1\dd \nu\,
T(\rho)\,E(u)\Big[F(u)-\frac{u}{u_0(\rho)}F(u_0(\rho))\Big].
\eeqn
The expressions given here 
are particularly useful for a numerical evaluation.

\dgcleardoublepage
\markh{Splitting Functions}
\dgsa{Splitting Functions}
\labelm{SplitFunc}

\dgsb{Altarelli--Parisi Splitting Functions}
\labelm{APapp}


The Altarelli--Parisi splitting functions $P_{B\leftarrow A}(u)$ are 
the splitting probabilities for a parton~$A$ into a parton~$B$, 
where parton~$B$ carries a fraction~$u$ of parton $A$'s momentum.
These splitting functions are singular at $u=1$ because 
a soft singularity has to be subtracted.
The unsubtracted splitting functions $\hat{P}_{CB\leftarrow A}(u)$
are the splitting probabilities for a parton~$A$ into partons
$B$ and~$C$, where, again, $B$ carries a fraction~$u$ of parton $A$'s momentum.
They are only applied in cases where parton~$C$ is not soft, 
so there are no subtractions at $u=1$. In the case
considered in the present paper, 
a heavy-quark 
fragmentation function is attached to parton~$C$, and therefore~$C$
is not allowed to be soft, because the observed heavy-quark has
to have a non-vanishing energy.

\begin{figure}[htb] \unitlength 1mm
\begin{center}
\dgpicture{159}{83}
 
\put(30, 45){\epsfigdg{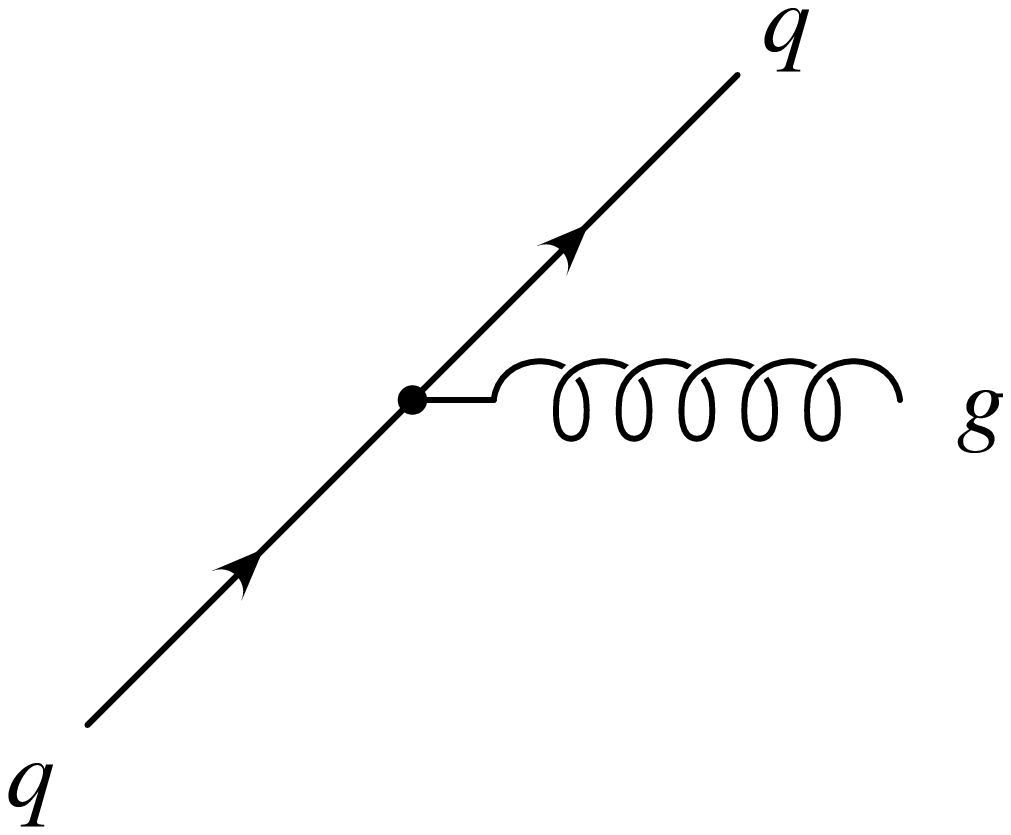}{width=40mm}}
\put(80, 45){\epsfigdg{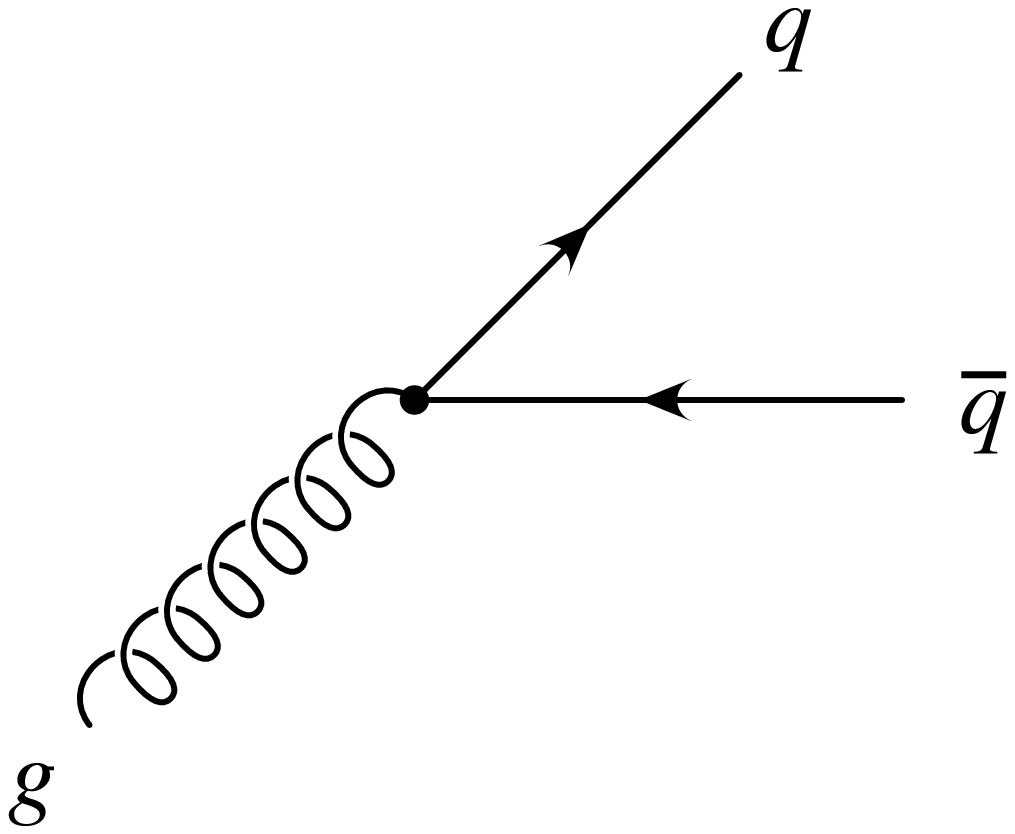}{width=45mm}}
\put(30,  0){\epsfigdg{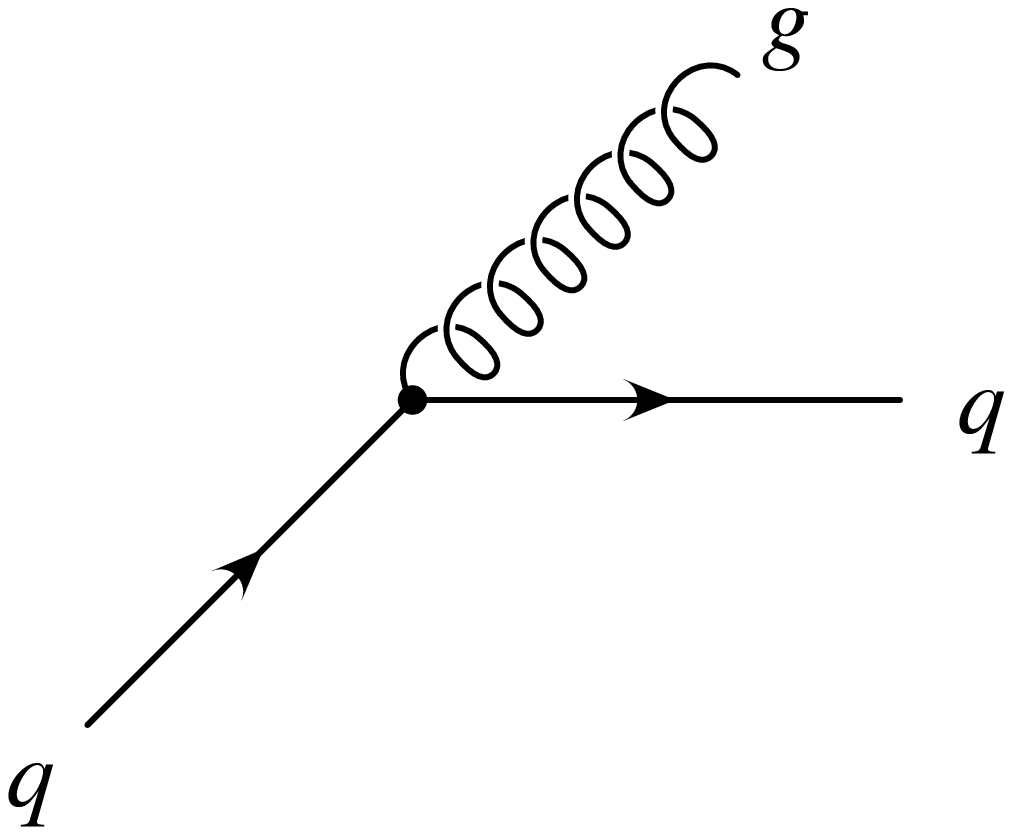}{width=40mm}}
\put(80,  0){\epsfigdg{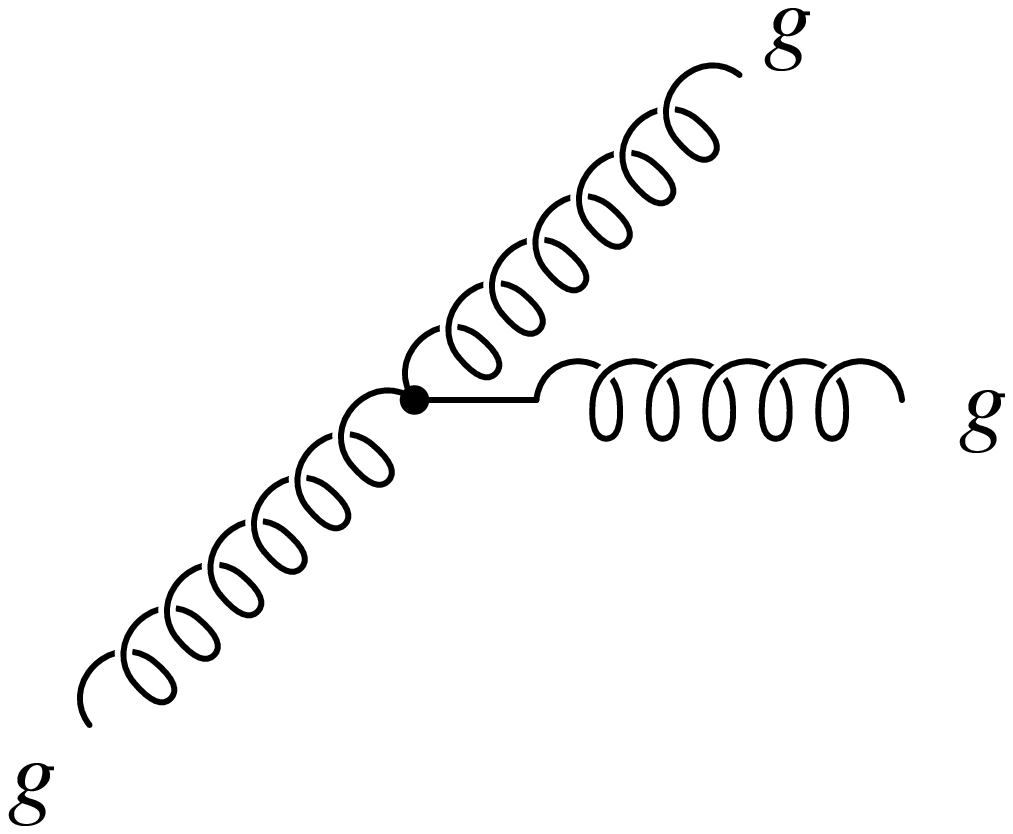}{width=45mm}}
\put(55,45){(a)}
\put(105,45){(b)}
\put(55, 0){(c)}
\put(105, 0){(d)}
\end{picture}
\end{center}
\shiftcaption
\caption[Vertices Corresponding to 
Altarelli--Parisi Splitting Functions]
{\labelmm{psplitfig} {\it Vertices corresponding to 
Altarelli--Parisi splitting functions: $P_{q\leftarrow q}(u)$ (a),
$P_{q\leftarrow g}(u)$ (b), $P_{g\leftarrow q}(u)$ (c),
$P_{g\leftarrow g}(u)$ (d).}}   
\end{figure}

Explicitly, the 
subtracted and unsubtracted Altarelli--Parisi splitting functions 
corresponding to the vertices in Fig.~\ref{psplitfig} are
given by$^{\mbox{\scriptsize\itemr{footsixteen}}}$ \cite{45,99}:
\beqnm{APsplitting}
P_{q\leftarrow q}(u)&=&C_F\left[2\subbl{\left(\frac{1}{1-u}\right)}{u}{0}{1}
+\frac{3}{2}\,\delta(1-u)-1-u\right],\nonu
{}\nonu
P_{q\leftarrow g}(u)&=&T_f\left[1-2u+2u^2
\right],\nonu
{}\nonu
P_{g\leftarrow q}(u)&=&C_F\left[2\frac{1}{u}-2+u
\right],\nonu
{}\nonu
P_{g\leftarrow g}(u)&=& 2C_A\left[
           \subbl{\left(\frac{1}{1-u}\right)}{u}{0}{1} +\frac{1}{u}+u(1-u)-2
                               \right]
\nonu
{}\nonu
&&
                               +\left(
           \frac{11}{6}C_A-\frac{2}{3}T_R
                                \right)\delta(1-u),\nonu
{}\nonu
\hat{P}_{gq\leftarrow q}(u)&=&C_F\left[2\frac{1}{1-u}-1-u
\right],\nonu
{}\nonu
\hat{P}_{\qb q\leftarrow g}(u)&=&T_f\left[1-2u+2u^2
\right],\nonu
{}\nonu
\hat{P}_{qg\leftarrow q}(u)&=&C_F\left[2\frac{1}{u}-2+u
\right],\nonu
{}\nonu
\hat{P}_{gg\leftarrow g}(u)&=& 2C_A\left[
           \frac{1}{1-u}+\frac{1}{u}+u(1-u)-2
                               \right].
\eeqn


The explicit expressions for a range of integration in~$u$ from 
$\alpha$ to~$1$ are 
\beqnm{APsplitt}
P_{q\leftarrow q}(u)&=&C_F
\left[2\subbl{\left(\frac{1}{1-u}\right)}{u}{\alpha}{1}
+\left(\frac{3}{2}+2\ln(1-\alpha)\right)\,\delta(1-u)-1-u\right],\nonu
{}\nonu
P_{g\leftarrow g}(u)&=& 2C_A\left[
\subbl{\left(\frac{1}{1-u}\right)}{u}{\alpha}{1} +\frac{1}{u}+u(1-u)-2
                               \right]
\nonu
{}\nonu
&&
                               +\left(
           \frac{11}{6}C_A-\frac{2}{3}T_R+2C_A \ln(1-\alpha)
                                \right)\delta(1-u).
\eeqn



\dgsb{Heavy-Quark Fragmentation Functions}
\labelm{aphqff}

Heavy-quark fragmentation functions can be considered to be 
scale-dependent splitting functions for partons into heavy quarks,
cf.\ Fig.~\ref{hqvertexfig}.
They have been calculated in
next-to-leading order
in Refs.\ \cite{101,37} and are given in the \msbar{} scheme by
\beqm{eq172a}
\begin{array}[b]{lcl}
D_{Q/Q}(x,\mu^2)&=&\delta(1-x)+C_f\,{\ds\frac{\alpha_s}{2\pi}}\,
\subbl{\left[
{\ds\frac{1+x^2}{1-x}}\left(\ln{\ds\frac{\mu^2}{m^2}}-2\ln(1-x)-1\right)
\right]}{x}{0}{1}
+\,\porder{\alpha_s^2},
\\
&&{}\\
D_{Q/q}(x,\mu^2)&=&\porder{\alpha_s^2},
\\
&&{}\\
D_{Q/g}(x,\mu^2)&=&T_f\,{\ds\frac{\alpha_s}{2\pi}}
\left(
x^2+(1-x)^2
\right)\ln{\ds\frac{\mu^2}{m^2}}+\,\porder{\alpha_s^2}.
\end{array}
\eeq
Here~$m$ is the mass of the heavy quark, and~$q$ stands for any
light-flavoured quark or antiquark
or for the heavy antiquark~$\overline{Q}$.
The origin of the terms logarithmic in the scale~$\mu$ can be understood
by considering the renormalization group equation 
(\ref{Drgeq}) in the case of the leading-order input distribution
from Eq.~(\ref{DQQcoeff}).
These expressions are valid as long as $\alpha_s\ln(\mu^2/m^2)$ is 
a small quantity. The heavy-quark fragmentation functions for arbitrary
scales can be obtained by means of the renormalization group equation,
see Section~\ref{hqffrg}.

\dgcleardoublepage
\markh{Numerical Solution of \ldots}
\dgsa{Numerical Solution of the Renormalization Group Equations}
\labelm{numrg}
Here we describe the numerical solution of the renormalization group equations,
based on the methods of Laguerre polynomials and direct discretization.
\dgsb{Fragmentation Functions}
\labelm{numrgff}
For the heavy-quark fragmentation functions, we 
solve the renormalization group
equation by two different methods.
In the first method, the heavy-quark fragmentation function is written in the
form\footnote{For the conventions related to distributions
see Appendix~\ref{distribu}.}
\beqm{eq131}
\label{Dexpansion}
D_{Q/i}(x,\mu^2)=D^\delta(\mu^2)\,\delta(1-x) + \subb{D^s}{\xiM}{1}(x,\mu^2)
+ \chi_{[0,\xiM]}(x)\,D^r(x,\mu^2).
\eeq
The functions $D^s(x,\mu^2)$ and $D^r(x,\mu^2)$ represent the
singular and regular part of the functional dependence and
are transformed into 
functions  $\tilde{D}^s(y,\mu^2)$ and $\tilde{D}^r(y,\mu^2)$ by means
of the variable transformation $y=\ln(1/x)$, cf.\ Eqs.~(\ref{st3})
and~(\ref{st4}).
For fixed $\mu$, the functional dependence in~$y$ of
$\tilde{D}^s(y,\mu^2)$ and $\tilde{D}^r(y,\mu^2)$ is parametrized
by means of $N_{\mbox{\scriptsize spline}}=50$ equidistant support points.
For the evolution, 
the functions are reconstructed by means of a third-order spline 
interpolation \cite{154}. The evolution equation is solved
in $N_{\mbox{\scriptsize int}}=500$ equidistant 
steps\footnote{
The CPU time needed for a complete parametrization
of a fragmentation function is about 8$\,$h on a ``thin'' SP2 node or
16$\,$h on one 100$\,$MHz HyperSparc processor.}
in the variable
$\ln\mu$ from the starting scale~$\mu_0$ up to $\mu=200\,\GeV$, 
where the convolutions $\overline{P}\otimes D$
from Eq.~(\ref{Drgeq})
are performed by a numerical integration using the convolution 
formula in Eq.~(\ref{convfor}). The 
$N_{\mbox{\scriptsize spline}}$
values of the parametrizations
are stored for 
$N_{\mbox{\scriptsize scale}}=25$
values of the scale parameter $\ln\mu$.
For a fast numerical evaluation in the applications, the
values of the functions $\tilde{D}^s$ and~$\tilde{D}^r$ 
are obtained by a quadratic
interpolation in both variables~$y$ and $\ln\mu$.
The lower boundary of the integration region of the singular term
$D^s$ in Eq.~(\ref{Dexpansion})
is chosen to be~$\xiM$, being of the order of $10^{-3}$. The region 
of~$x$ below a certain~$\xiM$ is not interesting in the applications
because the heavy quark as a massive particle is not allowed to be soft.
The range of integration $[\xi,1]$ in the convolution of 
the mass-factorized parton-level cross section and the fragmentation 
function is in general different from $[\xiM,1]$.
In the case of distributions, one may use
Eq.~(\ref{Dshift}) to rewrite everything in terms of the parametrized 
quantities.
The integration
region $[\xiM,\xi]$ must therefore be subtracted. For this reason, 
$\tilde{D}^{sM1}$, defined in Eq.~(\ref{firstmoment}),
is parametrized as well. The 
$\tilde{D}^{sM1}$ can be related to 
the integral
\beq
\label{xmoment}
D^{s\xiM1}(x)=\int_{\xiM}^x\dd u\,D^s(u)
\eeq
by $D^{s\xiM1}(x)=\tilde{D}^{s\ln({1/\xiM})1}(\ln({1/x}))$.

Alternatively, the heavy-quark fragmentation functions are
expanded in terms of Laguerre polynomials, cf.\ Appendix~\ref{lagintegro}.
The Laguerre coefficients are evolved, and the functional dependence
is reconstructed. Because of the small number of Laguerre coefficients,
of the order of $N_{\mbox{\scriptsize Laguerre}}=20$,
and the explicit convolution equations (\ref{lagconv}) 
and~(\ref{dmulag}), the method
can be implemented very efficiently.
It was employed so as to have a cross 
check and to make sure that the discretization in~$x$ and~$\mu$
used by the other method did not lead to intolerable errors.
The method is, however, not very reliable to parametrize functions at 
values of~$x$ very close to~$1$, because of strong oscillations of the sum 
of approximating polynomials. For the parametrizations employed for the
numerical study, we have therefore relied on the 
method of direct discretization.

\dgsb{Target Fragmentation Functions}
\labelm{numrgtff}
The renormalization group equations for the heavy-quark
target fragmentation functions are solved by a discretization 
in the relevant variables
$y=\ln(1/x)$, $\ln\mu$
and $\ln(1/\zeta)$, for $N_{\mbox{\scriptsize spline}}=50$,
$N_{\mbox{\scriptsize scale}}=25$ and $N_\zeta=25$ support 
points, respectively. Again, for the evolution, the 
interpolation in~$y$ is done by means of a third-order
spline interpolation routine.
The evolution in $\ln\mu$ is done in $N_{\mbox{\scriptsize int}}=500$
steps\footnote{The CPU time needed for a complete parametrization
of a target fragmentation function
is about 250$\,$h on a ``thin'' SP2 node
or 480$\,$h on one 100$\,$MHz HyperSparc processor.}.
It has been checked that the obtained result was stable 
with respect to changes in $N_{\mbox{\scriptsize spline}}$
and $N_{\mbox{\scriptsize int}}$.
The variable $\zeta$ is a fixed parameter in Eq.~(\ref{Nrge}).
The final parametrization of the $N_{\mbox{\scriptsize parton}}=13$
distributions is given in terms of the 
values of the function~$M$ on the 
$N_{\mbox{\scriptsize spline}}\cdot N_{\mbox{\scriptsize scale}}
\cdot N_\zeta$ support points. For efficiency reasons, the final 
interpolations are done by a quadratic interpolation in the 
variables~$y$, $\ln \mu$ and $\ln(1/\zeta)$.

\dgcleardoublepage
\markh{The Laguerre Method for the Solution of \ldots}
\dgsa{The Laguerre Method for the Solution\\
of Integro-Differential Equations}
\labelm{lagintegro}
In this section we describe the Laguerre method for the solution of
homogeneous 
integro-differential equations. We use this method for
cross checks of the direct method based on the discretization
in the scaling and scale variables~$x$ and~$\mu$, respectively, 
for the evolution of the heavy-quark
fragmentation functions. Because these particular fragmentation functions are 
distributions, the explicit convolution formula in Eq.~(\ref{convfor})
is quite complicated, and thus 
the accuracy of the discretized direct method
has to be checked. We first give some basic properties of
Laguerre polynomials, then describe the method, and finally discuss
the prescription on how the singular functions can be reconstructed from 
the Laguerre coefficients.

\dgsb{Laguerre Polynomials}
\labelm{lagpols}
In this section we define Laguerre
polynomials and state some of their properties. 
Moreover, we give explicit expressions
for the Laguerre coefficients of various regular and singular functions.
For more details and for the omitted proofs, see 
Refs.\ \cite{155,156,157}.

The Laguerre polynomials $L_n(y)$ are orthogonal polynomials on $[0,\infty]$
with respect to the measure $\dd y\,\exp(-y)$. They satisfy
\beqm{lortho}
\int_0^\infty \dd y\,\mbox{e}^{-y}\,L_m(y)\,L_n(y)\,=\,\delta_{mn}.
\eeq
The indices $n,m$ run through the integers $\sN$ including~$0$.
An explicit expression is given by 
\beqn
L_n(y)&=&\mbox{e}^y\,\frac{1}{n\mbox{!}}\,\partial_y^n\,
\left(
\mbox{e}^{-y} \, y^n
\right).
\eeqn
Here $\partial_y^n$ is the $n^{\mbox{th}}$ derivative with respect to~$y$.
In order to evaluate convolutions of series in Laguerre polynomials
of the type of 
Eq.~(\ref{conv2}), the following identity
is needed \cite{157}:
\beqm{lagconv}
\int_0^y\dd t\,L_m(t)\,L_n(y-t)=L_{m+n}(y)-L_{m+n+1}(y).
\eeq
Expressions for Laguerre polynomials $L_n(y)$ multiplied
by powers of~$y$ are calculated in the following way.
We define the coefficients $s_{nk}^{(\alpha)}$ for
$\alpha>0$ by\footnote{The $s_{nk}^{(\alpha)}$ are
implicitly assumed to be zero for values of the indices outside
the explicit domain of definition.}
\beqm{eqc246}
 y^\alpha\,L_n(y)\,=\,\sum_{k=0}^{n+\alpha}\,s_{nk}^{(\alpha)}\,L_k(y).
\eeq
The $s_{nk}^{(\alpha)}$ can be expressed in terms
of a function $p_r^{(\alpha)}(n)$ by
$s_{nk}^{(\alpha)}=p_{k-n}^{(\alpha)}(n)$. The
$p_r^{(\alpha)}(n)$ satisfy the recursion relation
\beqm{prec}
\begin{array}[b]{lcl}
p_0^{(0)}(n)\,&=&\,1,\\
\\
p_r^{(\alpha)}(n)\,&=&\,0, \quad \mbox{if} \quad |r|>\alpha,\\
\\
p_r^{(\alpha+1)}(n)\,&=&\,
-(n+r)\,p_{r-1}^{(\alpha)}(n)\,
+\,\left(2(n+r)+1\right)\,p_{r}^{(\alpha)}(n)\,
\\
&&-(n+r+1)\,p_{r+1}^{(\alpha)}(n).
\end{array}
\eeq

The Laguerre polynomials, being orthogonal polynomials on $[0,\infty]$,
cf.\ Eq.~(\ref{lortho}), can be used as a basis for a series 
expansion
\beqm{lagseries}
F(y)\,=\,\sum_{n=0}^\infty\,F_n\,L_n(y)
\eeq
of sufficiently regular\footnote{
For the precise properties required, 
see Ref.\ \cite{155}; basically, $F(y)$ must be integrable 
in a neighbourhood of zero and of at most polynomial growth for
$y\rightarrow\infty$.
} 
functions~$F$.
The Laguerre coefficients $F_n$ are given by
\beqm{lagcoeff}
F_n\,=\,\int_0^\infty\,\dd y\,\mbox{e}^{-y}\,L_n(y)\,F(y).
\eeq
A generating function for the Laguerre coefficients of a function 
$g$ can be obtained from the Mellin moments\footnote{For the 
definition of the ``$\sim$''~operation, see Appendix~\ref{appdistr}.}
\beqm{mellin}
\hat{g}(s)=\int_0^1\dd x\,x^s\,g(x)\nonu
=\int_0^\infty \dd y \, \mbox{e}^{-ys}\,\tilde{g}(y)
\eeq
of this function by the formula \cite{158} 
\beqm{mellin1}
\frac{1}{1-z}\,\hat{g}\left(\frac{1}{1-z}\right)=\sum_{n=0}^\infty
\tilde{g}_n\,z^n,
\eeq
where
\beqm{gmellin}
\tilde{g}(y)\,=\,\sum_{n=0}^\infty\,\tilde{g}_n\,L_n(y).
\eeq

We need the Laguerre coefficients of the following functions:
\beqn
F^{(a)}(y)&=&\left(x\mapsto x^\beta\right)^\sim(y)\,:\nonu
F^{(a)}_n&=&\frac{(\beta+1)^n}{(\beta+2)^{n+1}},\nonu
{}\nonu
F^{(b)}(y)&=&\left(x\mapsto x^\alpha\ln(1-x)\right)^\sim(y)\,:\nonu
F^{(b)}_n&=&\sum_{k=0}^n{n\choose k}(-1)^k(-1)\frac{1}{(\alpha+2)^k}
\nonu
&&\cdot\,\left\{\sum_{i=1}^{k+1}(\alpha+2)^{i-2}\zeta_i(\alpha+2)
-\sum_{i=2}^{k+1}(\alpha+2)^{i-2}\zeta(i)
\right\}.
\eeqn
The Riemann $\zeta$-function and the function $\zeta_k(r)$ are defined by
\beq
\zeta(s)=\sum_{k=1}^\infty\frac{1}{k^s},\quad 
\zeta_k(r)=\sum_{i=1}^r\frac{1}{i^k}.
\eeq

As for functions, Laguerre coefficients may be defined for distributions 
as well, by interpreting 
Eq.~(\ref{lagcoeff}) as the application of a distribution
to the test function $\exp(-y)\,L_n(y)$. This is trivial 
for the $\delta$-function:
\beqn
T^{(a)}(y)&=&\left(x\mapsto\delta(1-x)\right)^\sim(y):
\nonu
T^{(a)}_n&=&1.
\eeqn
In the case when $F$ is given by a ``$+$''~prescription, i.e.\ $F=D_+$, 
Eq.~(\ref{lagcoeff}) is defined only if the function~$D$ is integrable
on $[\rho,\infty]$ for some $\rho>0$. This condition is fulfilled 
for the following functions, owing to the fact that the ``$\sim$''~operation
multiplies functions $f(x)$ by~$x$: 
\beqn
T^{(b)}(y)&=&\left(x\mapsto
\subbl{\left(\frac{1}{1-x}\right)}{x}{0}{1}\right)^\sim(y):
\nonu
T^{(b)}_n&=&\sum_{l=1}^n{n \choose l}\,(-1)^l\,\zeta(l+1)-\delta_{n0},
\nonu{}\nonu
T^{(c)}(y)&=&\left(x\mapsto
\subbl{\left(\frac{\ln(1-x)}{1-x}\right)}{x}{0}{1}\right)^\sim(y):
\nonu
T^{(c)}_n&=&\sum_{r=0}^n{n \choose r}\,(-1)^r\,
\cdot\left\{
\begin{array}{l}
\zeta(r+2)-\zeta_{r+1}^{[1]}+r+1-\sum_{s=2}^{r+1}\zeta(s), 
\mbox{\ if\ } r>0\\
1,\mbox{\ if\ } r=0
\end{array}
\right.\!\!\!\!.
\eeqn
Here $\delta_{ij}$ is the Kronecker symbol and
$\zeta_k^{[1]}$, being defined by
\beqnm{zetaa}
\zeta_k^{[1]}=-\frac{1}{(k-1)!}\int_0^1\dd t\,
\frac{\ln^{k-1}{\displaystyle\frac{1}{t}}}{1-t}\,\ln(1-t)+\zeta(k+1),
\eeqn
can be shown to be\footnote{
$\zeta_k^{[1]}$ as given in Ref.\ \cite{158} contains a misprint.}
\beqnm{zetab}
\zeta_k^{[1]}=\left(1+\frac{1}{2}k\right)\,\zeta(k+1)
-\frac{1}{k-1}\sum_{i=1}^{k-2}i\,\zeta(i+1)\,\zeta(k-i).
\eeqn
The integral in Eq.~(\ref{zetaa}) can be evaluated explicitly \cite{159,160} 
by using the expression
\beqm{eulint}
\partial_x^k\,\partial_y^l\,B(x,y)=\int_0^1\dd t\,\ln^k t\,
\ln^l (1-t)\,t^{x-1}\,(1-t)^{y-1}
\eeq
and by the recursive evaluation in~$k$ of the expression
$\partial_x^{k-1}\Big|_{x=1}\,\partial_y\Big|_{y=0}B(x,y)$ 
involving the Euler beta function.

\dgsb{The Evolution Equation}
\labelm{labevolve}
The Laguerre method \cite{158,161} allows for the quick numerical 
solution of equations of the type
\beqm{ideq}
\partial_t F(x,t)=\int_x^1 \frac{\dd u}{u} \, K(u,t) 
\,F\left(\frac{x}{u},t\right),
\eeq
where~$K$ is the convolution kernel and~$F$ the function or distribution 
to be evolved.
As a first step, the equation is rewritten in terms of the functions
(or distributions)
$\tilde{K}$ and~$\tilde{F}$ via the transformation in Eqs.~(\ref{st3}) and
(\ref{st4}), yielding
\beqm{tildeideq}
\partial_t \tilde{F}(y,t)=\int_0^y \dd z \,\tilde{K}(z,t) 
\,\tilde{F}\left(y-z,t\right).
\eeq
$\tilde{K}$ and~$\tilde{F}$ are then expanded in Laguerre series
as in Eq.~(\ref{lagseries})
with coefficients
$\tilde{K}_m(t)$ and $\tilde{F}_n(t)$.
In the case when~$F$ is a singular function, the corresponding expansion
has to be understood as a formal series, with coefficients defined 
by Eq.~(\ref{lagcoeff}).
The Laguerre coefficients of $\partial_t \tilde{F}(y,t)$ can be 
obtained by means of Eq.~(\ref{lagconv}), and they read
\beqm{dmulag}
\partial_t \tilde{F}_n(t)
=\sum_{k=0}^n\,K_k(t)\,
\left[F_{n-k}(t)-F_{n-k-1}(t)\right]
=\sum_{k=0}^n\,F_k(t)\,
\left[K_{n-k}(t)-K_{n-k-1}(t)\right],
\eeq
where $K_{-1}(t)\doteq 0$ and $F_{-1}(t)\doteq 0$.
In practice, 
the series for $\tilde{K}(z,t)$ and $\tilde{F}(z,t)$
are truncated after 
a finite number of~$N$ terms. As can be seen 
from Eq.~(\ref{dmulag}), this allows for the calculation of 
$\partial_t \tilde{F}_n(t)$ for $n\leq N$. 
It can be shown that
in the case of the Altarelli--Parisi scale evolution \cite{97},
Eq.~(\ref{dmulag}) may be solved explicitly \cite{158,161}, 
i.e.\ the evolution kernel $E_{kl}(t,t_0)$ of 
\beqm{Ecomplete}
\tilde{F}_k(t)=\sum_l\, E_{kl}(t,t_0)\, \tilde{F}_l(t_0)
\eeq 
can be written down in closed form.
Since we are interested in a cross check at all intermediate scales, 
this explicit solution would not be an advantage, so 
the integro-differential
equation is solved by a discretization in~$t$, iterating
the expression in Eq.~(\ref{dmulag}) for very small $\Delta t$.

\dgsb{Reconstruction of the Functional Dependence}
\labelm{lagrecon}
The Laguerre method has been applied in the case of parton distribution 
functions \cite{162,163,164}. These functions are regular, and therefore
they can be obtained directly from the Laguerre coefficients by means
of the Laguerre series in Eq.~(\ref{lagseries}).
In the case of scale-evolved singular functions $F(x,t)$, the situation is 
different. As has already been stressed, in this case the series expansion
has to be considered to be a formal one. For example, if
$F(x)=\subbl{\left(1/(1-x)\right)}{x}{0}{1}$, 
then inserting the corresponding
$\tilde{F}_n$ into Eq.~(\ref{lagseries}) would lead to a rapidly oscillating
function, since a function with a singular behaviour $\sim(1/y)$
cannot be expanded in terms of polynomials. Instead, it is possible
to extract the singular term as a multiplicative factor and
express the remainder in terms of the Laguerre coefficients.

Consider the general case of a function~$G$ defined on $[0,\infty]$, 
which is singular at the origin like $\sim 1/y^{\beta+\rho}$, 
$\rho<1$. Define the function~$g$ by $g(y)=y^\alpha G(y)$, where
$\alpha>\beta$. The Laguerre coefficients of~$g$ are given by
\beqnm{msubtr}
g_n
&=&\sum_{k=0}^{n+\alpha} s_{nk}^{(\alpha)}
\,{\left(\subabc{G}{0}{\infty}{0,\beta}\right)}_n,
\eeqn
cf.\ Eqs.~(\ref{jetsubtr}) and~(\ref{eqc246}).
The function~$G$ is then given by
\beqm{fff}
G(y)=\frac{1}{y^\alpha}\,\sum_{n=0}^\infty\,g_n\,L_n(y).
\eeq
It should be noted that for truncated series the Laguerre coefficients
of $\subabc{G}{0}{\infty}{0,\beta}$ have to be known up to the order of
$N+\alpha$ if~$g$ is to be determined up to order~$N$.

In our particular case, the singularities of the fragmentation functions
$D_{Q/i}(x)$ are of the form $\subbl{\left(\ln^m(1-x)/(1-x)\right)}{x}{0}{1}$.
For $\tilde{D}_{Q/i}(y)$, this singular behaviour translates into
$\subal{\left(\ln^m y/y\right)}{y}{0}{\infty}$, 
and so $\alpha=1$ can be used. A possible term $\sim\delta(1-x)$ translates
into a term $\sim\delta(y)$, and drops out after the multiplication by
$y^\alpha$ for $\alpha>0$ in Eq.~(\ref{msubtr}).

\end{appendix}


\dgcleardoublepage 

\newcommand{\scs}{\rm}
\newcommand{\bibitema}[1]{\bibitem[#1]{#1}}
\newcommand{\bibbeginlong}{\begin{thebibliography}{XXX\,1988\,\,}}
\newcommand{\bibbeginshort}{\begin{thebibliography}{XXX}}

\newcommand{\bibitemb}[4]{\begin{sloppy}
{\bibitem[#1]{#1} {\bf #2:} {\it #3,} {#4.}}
\end{sloppy}}

\newcommand{\nbibitemb}[4]{}

\markh{}
\bibbeginshort 

\addcontentsline{toc}{section}{References}



%

   \nbibitemb{Bjo67} 
      {J.D.~Bjorken}
      {Inequality for Backward Electron-- and Muon--Nucleon Scattering
       at High Momentum Transfer}
      {Phys.\ Rev.\ \underline{163} (1967) 1767}

   \nbibitemb{Bjo69} 
      {J.D.~Bjorken}
      {Asymptotic Sum Rules at Infinite Momentum}
      {Phys.\ Rev.\ \underline{179} (1969) 1547}



   \nbibitemb{BL89} 
      {S.J.~Brodsky, G.P.~Lepage}
      {Exclusive Processes in Quantum Chromodynamics}
      {in: Perturbative Quantum Chromodynamics, ed.\ A.H.~Mueller, 
       World Scientific (Singapore, 1989)}



   \nbibitemb{Bro71} 
      {S.J.~Brodsky}
      {Heavy-Quark Physics in Quantum Chromodynamics}
      {published in: Proceedings of the Lake Louise Winter Inst.\ (1991) 1}

   \nbibitemb{Bro89} 
      {S.J.~Brodsky}
      {Tests of Quantum Chromodynamics in
       Exclusive and Inclusive Electroproduction}
      {preprint SLAC-PUB-5013 (June 1989)}

   \nbibitemb{Bro93} 
      {S.J.~Brodsky}
      {The Challenges of Exclusive Processes in QCD}
      {preprint SLAC-PUB-6356 (September 1993)}

   \nbibitemb{CGRM84} 
      {P.~Chiappetta, Y.~Gabellini, T.~Grandon, J.L.~Meunier}
      {Soft Gluons in Semiinclusive Deep Inelastic Experiments}
      {Nucl.\ Phys.\ \underline{B239} (1984) 410}


 



   \nbibitemb{FG72} 
      {H.~Fritzsch, M.~Gell-Mann}
      {Current Algebra: Quarks and What Else?}
      {in: Proceedings of the XVI International Conference on High 
       Energy Physics, Chicago 1972, Vol. 2}

   \nbibitemb{FGL73} 
      {H.~Fritzsch, M.~Gell-Mann, H.~Leutwyler}
      {Advantages of the Color Octet Gluon Picture}
      {Phys.\ Lett.\ \underline{47B} (1973) 365}



   \nbibitemb{Gel64} 
      {M.~Gell-Mann}
      {A Schematic Model of Baryons and Mesons}
      {Phys.\ Lett.\ \underline{8} (1964) 214}










   \nbibitemb{HN65} 
       {M.Y.~Han, Y.~Nambu}
       {Three-Triplet Model with Double $\mbox{SU}(3)$ Symmetry}
       {Phys.\ Rev.\ \underline{139} (1965) B1006}









    \nbibitemb{Lon92} 
      {L.~L\"{o}nnblad}
      {Ariadne Version 4: A Program for Simulation of QCD Cascades
       Implementing the Color Dipole Model}
      {Comput.\ Phys.\ Commun.\ \underline{71} (1992) 15}

   \nbibitemb{MOS95} 
      {R.~Meng, F.I.~Olness, D.E.~Soper}
      {Semi-Inclusive Deeply Inelastic Scattering at Small~$q_T$}
      {preprint CTEQ--409 (November 1995)}


   \nbibitemb{MRS78} 
      {A.~Mendez, A.~Raychaudhuri, V.J.~Stenger}
      {QCD Effects in Semiinclusive Neutrino Processes}
      {Nucl.\ Phys.\ \underline{B148} (1979) 499}




   \nbibitemb{Pan68} 
      {W.K.H.~Panofsky}
      {Low~$q^2$ Electrodynamics, Elastic and Inelastic Electron
       (and Muon) Scattering}
      {in: Proceedings of the 
       $14^{\mbox{th}}$ 
       International Conference
       on High-Energy Physics, Vienna (1968)}


   \nbibitemb{Tay75} 
      {R.E.~Taylor}
      {Deep Inelastic Electron and Muon Scattering}
      {in: Proceedings of the EPS International Conference,
       Palermo (1975)}

   \nbibitemb{ZK92}
      {P.M.~Zerwas, H.A.~Kastrup (eds.)}
      {QCD -- 20 Years Later}
      {World Scientific (Singapore, 1993)}
    
   \nbibitemb{Zwe64}
      {G.~Zweig}
      {An $\mbox{SU}_3$ Model for Strong Interaction Symmetry and
       its Breaking}
      {reports CERN 8182/TH.401, 8419/TH.412 (1964)}

%
\bibitemb{1} 
      {V.~Ammosov et al.}
      {Properties of $K^0$ and $\Lambda$ Inclusive Production
       in Charged-Current Antineutrino--Nucleon Interactions}
      {Nucl.\ Phys.\ \underline{B162} (1980) 205}

\bibitemb{2} 
      {D.~Allasia et al.}
      {Production of Neutral Strange Particles in
       $\overline{\nu}_\mu d_2$ and $\nu_\mu d_2$
       Charged Current Interactions}
      {Nucl.\ Phys.\ \underline{B224} (1983) 1}

\bibitemb{3} 
      {G.T.~Jones et al.}
      {Polarization of $\Lambda$~Hyperons Produced Inclusively
       in~$\nu p$ and $\overline{\nu} p$ Charged Current Interactions}
      {Z.~Phys.\ \underline{C28} (1985) 23}

\bibitemb{4}
      {S.~Willocq et al.}
      {Neutral Strange Particle Production
       in Antineutrino--Neon Charged Current Interactions}
      {Z.~Phys.\ \underline{C53} (1992) 207}

\bibitemb{5} 
      {J.A.~Appel}
      {Hadroproduction of Charm Particles}
      {Annu.\ Rev.\ Part.\ Sci.\ \underline{42} (1992) 367}

\bibitemb{6} 
      {M.~Aguilar-Benitez et al.}
      {Inclusive Properties of $D$~Mesons Produced
       in $360\,\GeV$ $\pi^- p$ Interactions}
      {Phys.\ Lett.\ \underline{161B} (1985) 400}

\bibitemb{7} 
      {M.~Aguilar-Benitez et al.}
      {Charm Hadron Properties 
       in $360\,\GeV/c$ $\pi^- p$-Interactions}
      {Z.~Phys.\ \underline{C31} (1986) 491}

\bibitemb{8} 
      {M.~Aguilar-Benitez et al.}
      {Charm Hadron Properties 
       in $400\,\GeV/c$ $pp$ Interactions}
      {Z.~Phys.\ \underline{C40} (1988) 321}

\bibitemb{9} 
      {S.~Barlag et al.}
      {Production Properties of $D^0$, $D^+$, $D^{\ast+}$ and
       $D^+_s$ in $230\,\GeV/c$ $\pi^-$ and $K^-$--Cu Interactions}
      {Z.~Phys.\ \underline{C49} (1991) 555}

\bibitemb{10} 
      {G.A.~Alves et al.}
      {Feynman-$x$ and Transverse Momentum Dependence of
       $D^\pm$ and $D^0$, $\overline{D^0}$ Production 
       in $250\,\GeV$ $\pi^-$--Nucleon Interactions}
      {Phys.\ Rev.\ Lett.\ \underline{69} (1992) 3147}

\bibitemb{11} 
      {K.G.~Mallot et al.}
      {Letter of Intent: 
       Semi-Inclusive Muon Scattering from a Polarized Target} 
      {CERN/SPSLC 95--27 (March 1995)}

\bibitemb{12} 
      {G.~Baum et al.}
      {A Proposal for a Common Muon and Proton Apparatus
       for Structure and Spectroscopy}
      {CERN/SPSLC 96--14 (March 1996)}

\bibitemb{13} 
      {A.~Bartl, H.~Fraas, W.~Majerotto} 
      {Quark and Diquark Fragmentation into Mesons and Baryons}
      {Phys.\ Rev.\ \underline{D26} (1982) 1061}

\bibitemb{14} 
      {B.~Andersson, G.~Gustafson, G.~Ingelman, T.~Sj\"{o}strand}
      {Parton Fragmentation and String Dynamics}
      {Phys.\ Rep. \underline{97} (1983) 31}

\bibitemb{15}
      {B.~Webber}
      {A QCD Model for Jet Fragmentation Including Soft Gluon Interference}
      {Nucl.\ Phys.\ \underline{B238} (1984) 492}

\bibitemb{16} 
      {B.~Andersson, G.~Gustafson, L.~L\"{o}nnblad, U.~Pettersson}
      {Coherence Effects in Deep Inelastic Scattering}
      {Z.~Phys.\ \underline{C43} (1989) 625}

\bibitemb{17}
      {G.~Ingelman}
      {LEPTO Version 6.1: The Lund Monte Carlo for Deep Inelastic 
       Lepton--Nucleon Scattering}
      {in: Proceedings of the HERA Workshop 1991, 
       eds. W.~Buchm\"uller, G.~Ingelman (Hamburg 1991)}

\bibitemb{18} 
      {M.~Seymour}
      {HERWIG: A Monte Carlo Event Generator for Simulating 
       Hadron Emission Reactions With Interfering Gluons.
       Version 5.1}
      {Comput.\ Phys.\ Commun.\ \underline{67} (1992) 465}

\bibitemb{19} 
      {J.~Ellis, D.~Kharzeev, A.~Kotzinian}
      {The Proton Spin Puzzle and~$\Lambda$ Polarization in 
       Deep-Inelastic Scattering}
      {preprint CERN--TH/95--135 (June 1995)}

\bibitemb{20} 
      {S.~Frixione, M.L.~Mangano, P.~Nason, G.~Ridolfi}
      {On the Determination of the Gluon Density of the Proton
       from Heavy-Flavour Production at HERA}
      {preprint CERN--TH.6864/93 (April 1993)}

\bibitemb{21} 
      {S.~Frixione, M.L.~Mangano, P.~Nason, G.~Ridolfi}
      {Charm and Bottom Production: Theoretical Results
       versus Experimental Data}
      {preprint CERN--TH.7292/94 (June 1994)}

\bibitemb{22} 
      {S.~Frixione, M.L.~Mangano, P.~Nason, G.~Ridolfi}
      {Total Cross Sections for Heavy Flavour Production
       at HERA}
      {preprint CERN--TH.7527/94 (December 1994)}

\bibitemb{23} 
      {M.~Cacciari, M.~Greco}
      {Charm Photoproduction via Fragmentation}
      {preprint DESY 95--103 (May 1995)}

\bibitemb{24} 
      {S.~Frixione, P.~Nason, G.~Ridolfi}
      {Differential Distributions for Heavy Flavour Production
       at HERA}
      {preprint CERN--TH/95--143 (May 1995)}

\bibitemb{25} 
       {B.A.~Kniehl, M.~Kr\"amer, G.~Kramer, M.~Spira}
       {Cross Sections for Charm Production in $ep$ Collisions:
        Massive versus Massless Scheme}
       {preprint DESY 95--098 (May 1995)}

\bibitemb{26} 
      {E.~Laenen, S.~Riemersma, J.~Smith, W.L.~van~Neerven}
      {\porder{\alpha_s} Corrections to Heavy-Flavour Inclusive
       Distributions in Electroproduction}
      {Nucl.\ Phys.\ \underline{B392} (1993) 229}

\bibitemb{27} 
      {E.~Laenen, S.~Riemersma, J.~Smith, W.L.~van~Neerven}
      {Complete \porder{\alpha_s} Corrections to Heavy-Flavour
       Structure Functions in Electroproduction}
      {Nucl.\ Phys.\ \underline{B392} (1993) 162}

\bibitemb{28} 
      {M.A.G.~Aivazis, F.I.~Olness, Wu-Ki Tung}
      {Leptoproduction of Heavy Quarks~I: General Formalism
       and Kinematics of Charged Current and Neutral Current Production
       Processes}
      {Phys.\ Rev.\ \underline{D50} (1994) 3085}

\bibitemb{29} 
      {M.A.G.~Aivazis, J.C.~Collins, F.I.~Olness, Wu-Ki Tung}
      {Leptoproduction of Heavy Quarks II: 
       A Unified QCD Formulation of Charged and Neutral
       Current Processes from Fixed-Target to Collider Energies}
      {Phys.\ Rev.\ \underline{D50} (1994) 3102}

\bibitemb{30}
       {M.~Gl\"{u}ck, E.~Reya, M.~Stratmann}
       {Heavy Quarks at High Energy Colliders}
       {Nucl.\ Phys.\ \underline{B422} (1994) 37}

\bibitemb{31} 
      {P.~Agrawal, F.I.~Olness, S.T.~Riemersma, Wu-Ki Tung}
      {Leptoproduction of Heavy Quarks}
      {Southern Methodist Univ.\ preprint SMU--HEP/95--04 (May 1995)}

\bibitemb{32} 
      {B.W.~Harris, J.~Smith}
      {Heavy-Quark Correlations in Deep-Inelastic
       Electroproduction}
      {State Univ.\ of New York at Stony Brook 
       preprint ITP--SB--95--08 (March 1995)}

\bibitemb{33} 
       {G.~Kramer, B.~Lampe, H.~Spiesberger}
       {Scheme and Scale Dependence of Charm
        Production in Neutrino Scattering}
       {preprint DESY 95--201 (November 1995)}

\bibitemb{34} 
      {S.J.~Brodsky, P.~Hoyer, W.K.~Tang}
      {Systematics of Heavy Quark Production at HERA}
      {Phys.\ Rev.\ \underline{D52} (1995) 6285}

\bibitemb{35} 
      {C.~Peterson, D.~Schlatter, I.~Schmitt, P.M.~Zerwas}
      {Scaling Violations in Inclusive \epem{} Annihilation Spectra}
      {Phys.\ Rev.\ \underline{D27} (1983) 105}

\bibitemb{36} 
      {R.~Odorico}
      {Hadronic Production of Charm Flavour Excitation}
      {Nucl.\ Phys.\ \underline{B209} (1982) 77}

\bibitemb{37} 
      {B.~Mele, P.~Nason}
      {The Fragmentation Function for Heavy Quarks in QCD}
      {Nucl.\ Phys.\ \underline{B361} (1991) 626}

\bibitemb{38} 
      {S.J.~Brodsky, P.~Hoyer, C.~Peterson, N.~Sakai}
      {The Intrinsic Charm of the Proton}
      {Phys.\ Lett.\ \underline{93B} (1980) 451}

\bibitemb{39} 
      {S.J.~Brodsky, C.~Peterson, N.~Sakai}
      {Intrinsic Heavy-Quark States}
      {Phys.\ Rev.\ \underline{D23} (1981) 2745}

\bibitemb{40} 
      {F.~Bloch, A.~Nordsieck}
      {Note on the Radiation Field of the Electron}
      {Phys.\ Rev.\ \underline{52} (1937) 54}

\bibitemb{41}
      {D.R.~Yennie, S.C.~Frautschi, H.~Suura}
      {The Infrared Divergence Phenomena and High-Energy Processes}
      {Ann.\ Phys.\ \underline{13} (1961) 379}

\bibitemb{42} 
      {T.~Kinoshita}
      {Mass Singularities of Feynman Amplitudes}
      {J.~Math.\ Phys.~\underline{3} (1962) 650}

\bibitemb{43} 
      {T.D.~Lee, M.~Nauenberg}
      {Degenerate Systems and Mass Singularities}
      {Phys.\ Rev.\ \underline{133} (1964) 1549}

\bibitemb{44} 
      {G.~Altarelli, R.K.~Ellis, G.~Martinelli}
      {Leptoproduction and Drell--Yan Processes
       Beyond the Leading Approximation in Chromodynamics}
      {Nucl.\ Phys.\ \underline{B143} (1978) 521,
      Erratum Nucl.\ Phys.\ \underline{B146} (1978) 544}

\bibitemb{45} 
      {G.~Altarelli, R.K.~Ellis, G.~Martinelli}
      {Large Perturbative Corrections to the Drell--Yan Process in QCD}
      {Nucl.\ Phys.\ \underline{B157} (1979) 461}

\bibitemb{46} 
      {R.~Baier, K.~Fey}
      {Finite Corrections to Quark Fragmentation Functions in 
       Perturbative QCD} 
      {Z.~Phys.\ \underline{C2} (1979) 339}

\bibitemb{47} 
      {H.D.~Politzer}
      {Reliable Perturbative Results for Strong Interactions?}
      {Phys.\ Rev.\ Lett.\ \underline{30} (1973) 1346}

\bibitemb{48}
      {D.J.~Gross, F.~Wilczek}
      {Ultraviolet Behaviour of Nonabelian Gauge Theories}
      {Phys.\ Rev.\ Lett.\ \underline{30} (1973) 1343}

\bibitemb{49} 
      {D.J.~Gross, F.~Wilczek}
      {Asymptotically Free Gauge Theories I}
      {Phys.\ Rev.\ \underline{D8} (1973) 3633}

\bibitemb{50} 
      {D.J.~Gross, F.~Wilczek}
      {Asymptotically Free Gauge Theories II}
      {Phys.\ Rev.\ \underline{D9} (1974) 980}

\bibitemb{51}
      {K.G.~Wilson}
      {On Products of Quantum Field Operators at Short Distances}
      {Cornell report (1964)}

\bibitemb{52}
      {K.G.~Wilson}
      {Non-Lagrangian Models of Current Algebra}
      {Phys.\ Rev.\ \underline{179} (1969) 1499}

\bibitemb{53} 
      {R.A.~Brandt, G.~Preparata}
      {Operator Product Expansions near the Light Cone}
      {Nucl.\ Phys.\ \underline{B27} (1971) 541}

\bibitemb{54}
      {K.G.~Wilson, W.~Zimmermann}
      {Operator Product Expansions and Composite Field Operators
       in the General Framework of Quantum Field Theory}
      {Commun.\ Math.\ Phys.\ \underline{24} (1972) 87}

\bibitemb{55}
      {W.~Zimmermann}
      {Normal Products and the Short Distance Expansion
       in the Perturbation Theory of Renormalizable
       Interactions}
      {Ann.\ Phys.\ \underline{77} (1973) 570}

\bibitemb{56} 
      {N.~Christ, B.~Hasslacher, A.H.~Mueller}
      {Light-Cone Behaviour of Perturbation Theory}
      {Phys.\ Rev.\ \underline{D6} (1972) 3543}

\bibitemb{57} 
      {A.~Buras}
      {Asymptotic Freedom in Deep Inelastic Processes in the
       Leading Order and Beyond}
      {Rev.\ Mod.\ Phys.\ \underline{52} (1980) 199}

\bibitemb{58}
      {E.~Reya}
      {Perturbative Quantum Chromodynamics}
      {Phys.\ Rep.\ \underline{69} (1981) 195}

\bibitemb{59} 
      {G.~Altarelli}
      {Partons in Quantum Chromodynamics}
      {Phys.\ Rep.\ \underline{81} (1982)~1}

\bibitemb{60} 
      {R.P.~Feynman}
      {Photon Hadron Interactions}
      {W.A.~Benjamin, Inc. (Reading, Mass., 1972)}

\bibitemb{61} 
      {H.D.~Politzer}
      {QCD off the Light Cone and the Demise of the Transverse
       Momentum Cut-Off}
      {Phys.\ Lett.\ \underline{70B} (1977) 430}

\bibitemb{62} 
      {D.~Amati, R.~Petronzio, G.~Veneziano}
      {Relating Hard QCD Processes Through Universality
       of Mass Singularities}
      {Nucl.\ Phys.\ \underline{B140} (1978) 54} 

\bibitemb{63} 
      {D.~Amati, R.~Petronzio, G.~Veneziano}
      {Relating Hard QCD Processes Through Universality
       of Mass Singularities (II)}
      {Nucl.\ Phys.\ \underline{B146} (1978) 29}

\bibitemb{64} 
      {R.K.~Ellis, H.~Georgi, M.~Machacek, H.D.~Politzer, G.G.~Ross}
      {Factorization and the Parton Model in QCD}
      {Phys.\ Lett.\  \underline{78B} (1978) 281}

\bibitemb{65} 
      {R.K.~Ellis, H.~Georgi, M.~Machacek, H.D.~Politzer, G.G.~Ross}
      {Perturbation Theory and the Parton Model in QCD}
      {Nucl.\ Phys.\ \underline{B152} (1979) 285}

\bibitemb{66} 
      {W.~Furmanski, R.~Petronzio}
      {Lepton--Hadron Processes Beyond Leading Order in
       Quantum Chromodynamics}
      {Z.~Phys.\ \underline{C11} (1982) 293}

\bibitemb{67} 
      {J.C.~Collins, D.E.~Soper}
      {The Theorems of Perturbative QCD}
      {Annu.\ Rev.\ Nucl.\ Part.\ Sci.\ \underline{37} (1987) 383}

\bibitemb{68} 
      {J.C.~Collins, D.E.~Soper, G.~Sterman}
      {Soft Gluons and Factorization}
      {Nucl.\ Phys.\ \underline{B308} (1988) 833}

\bibitemb{69} 
      {J.C.~Collins, D.E.~Soper, G.~Sterman}
      {Factorization of Hard Processes in QCD}
      {in: Perturbative Quantum Chromodynamics, ed.\ A.H.~Mueller,
       World Scientific (Singapore, 1989)}   

\bibitemb{70} 
      {G.~Altarelli, R.K.~Ellis, G.~Martinelli, S.Y.~Pi}
      {Processes Involving Fragmentation Functions Beyond the
       Leading Order in QCD}
      {Nucl.\ Phys.\ \underline{B160} (1979) 301}

\bibitemb{71} 
      {N.~Sakai}
      {Factorization Breaking in Deep Inelastic Neutrino
       Hadron Production in Perturbative QCD}
      {Phys.\ Lett.\ \underline{85B} (1979) 67}

\bibitemb{72} 
      {P.~Chiappetta, J.P.~Guillet}
      {QCD Effects in Semiinclusive Deep Inelastic Scattering
       from a Polarized Target}
      {Z.~Phys.\ \underline{C32} (1986) 209}

\bibitemb{73} 
      {J.D.~Stack}
      {Generalized Scaling Laws for the Electroproduction
       of Hadrons}
      {Phys.\ Rev.\ Lett.\ \underline{28} (1972) 57}

\bibitemb{74} 
      {O.~Nachtmann}
      {Some Consequences of Light-Cone Dominance for Inclusive
       Electroproduction} 
      {Phys.~Rev.\ \underline{D6} (1972) 1718}

\bibitemb{75} 
      {A.V.~Kiselev, V.A.~Petrov}
      {Hadron Production in Hard Processes}
      {Sov.\ J.~Part.\ \&~Nucl.\ \underline{19} (1988) 21}

\bibitemb{76} 
      {J.~Ellis}
      {Light-Cone Analysis of Inclusive Electromagnetic Processes}
      {Phys.~Lett.\ \underline{35B} (1971) 537}

\bibitemb{77} 
      {L.~Trentadue, G.~Veneziano}
      {Fracture Functions: an Improved Description of Inclusive
       Hard Processes in QCD}
      {Phys.\ Lett.\ \underline{B323} (1994) 201}

\bibitemb{78}
       {D.~Graudenz}
       {One-Particle Inclusive Processes in Deeply Inelastic
        Lepton--Nucleon Scattering}
       {Nucl.\ Phys.\ \underline{B432} (1994) 351}

\bibitemb{79} 
      {D.~de Florian, C.A.~Garcia Canal, R.~Sassot}
      {Factorization in Semi-Inclusive Polarized Deep
       Inelastic Scattering}
      {preprint hep-ph/9510262 (October 1995)}

\bibitemb{80} 
       {HERMES Collaboration}
       {Technical Design Report}
       {DESY report (July 1993)}

\bibitemb{81} 
      {J.P.~Ma}
      {Calculating Fragmentation Functions from Definitions}
      {Phys.\ Lett.\ \underline{B332} (1994) 398}

\bibitemb{82} 
      {E.L.~Berger}
      {Semi-Inclusive Inelastic Electron Scattering from Nuclei}
      {in: Proceedings of the NPAS Workshop 1987, Stanford}

\bibitemb{83} 
      {G.~Leibbrandt}
      {Introduction to the Technique of Dimensional Regularization}
      {Rev.\ Mod.\ Phys.\ \underline{47} (1975) 849}

\bibitemb{84} 
      {S.~Narison}
      {Techniques of Dimensional Regularization
       and the Two-Point Function of QCD and QED}
      {Phys.\ Rep.\ \underline{84} (1982) 263}

\bibitemb{85} 
      {G.~'t Hooft, M.~Veltman}
      {Regularization and Renormalization of Gauge Fields}
      {Nucl.\ Phys.\ \underline{B44} (1972) 189}

\bibitemb{86} 
      {R.~Gastmans, R.~Meuldermans}
      {Dimensional Regularization of the Infrared Problem}
      {Nucl.\ Phys.\ \underline{B63} (1973) 277}

\bibitemb{87} 
      {W.J.~Marciano, A.~Sirlin}
      {Dimensional Regularization of Infrared Divergences}
      {Nucl.\ Phys.\ \underline{B88} (1975) 86}

\bibitemb{88} 
      {W.J.~Marciano}
      {Dimensional Regularization and Mass Singularities}
      {Phys.\ Rev.\ \underline{D12} (1975) 3861}

\bibitemb{89} 
      {G.~Sterman}
      {Kinoshita's Theorem in Yang--Mills Theories}
      {Phys.\ Rev.\ \underline{D14} (1976) 2123}

\bibitemb{90}
      {J.~Ch\'{y}la}
      {Parton Distributions Beyond the Leading Order}
      {Phys.\ Rev.\ \underline{D48} (1993) 4385}

\bibitemb{91} 
      {G.~Curci, W.~Furmanski, R.~Petronzio}
      {Evolution of Parton Densities Beyond Leading Order}
      {Nucl.\ Phys.\ \underline{B175} (1980) 27}

\bibitemb{92}
      {Ch.~Rumpf, G.~Kramer, J.~Willrodt}
      {Jet Cross Sections in Leptoproduction from QCD}
      {Z.~Phys.\ \underline{C7} (1981) 337}

\bibitemb{93} 
      {A.~Mendez}
      {QCD Predictions for Semi-Inclusive and Inclusive Leptoproduction}
      {Nucl.\ Phys.\ \underline{B145} (1978) 199}

\bibitemb{94} 
      {R.D.~Peccei, R.~R\"uckl}
      {Energy Flow and Energy Correlations
       in Deep Inelastic Scattering}
      {Nucl.\ Phys.\ \underline{B162} (1980) 125}

\bibitemb{95} 
      {J.G.~K\"orner, E.~Mirkes, G.A.~Schuler}
      {QCD Jets at HERA. I:\ \porder{\alpha_s} Radiative Corrections
       to Electroweak Cross Sections and Jet Rates}
      {Int.\ J.~Mod.\ Phys.\ \underline{A4} (1989) 1781}

\bibitemb{96} 
      {L.~Trentadue}
      {Fracture Functions}
      {in: Proceedings of the ``QCD 94'' Conference, 
       Montpellier (July 1994)}

\bibitemb{97} 
      {G.~Altarelli, G.~Parisi}
      {Asymptotic Freedom in Parton Language}
      {Nucl.\ Phys.\ \underline{B126} (1977) 298}

\bibitemb{98} 
      {K.~Konishi, A.~Ukawa, G.~Veneziano}
      {A Simple Algorithm for QCD Jets}
      {Phys.\ Lett.\ \underline{78B} (1978) 243}

\bibitemb{99} 
      {K.~Konishi, A.~Ukawa, G.~Veneziano}
      {Jet Calculus: A Simple Algorithm for Resolving QCD Jets}
      {Nucl.\ Phys.\ \underline{B157} (1979) 45}

\bibitemb{100} 
      {The Particle Data Group}
      {Review of Particle Properties}
      {Phys.\ Rev.\ \underline{50} (1994) 1173}

\bibitemb{101} 
      {B.~Mele, P.~Nason}
      {Next-to-Leading QCD Calculation of the Heavy Quark
       Fragmentation Function}
      {Phys.\ Lett.\ \underline{B245} (1990) 635}

\bibitemb{102} 
      {P.~Nason}
      {Heavy Quark Production} 
      {in: Heavy Flavours, 
      eds.\ A.~Buras, M.~Lindner, 
      World Scientific (Singapore, 1993)}

\bibitemb{103} 
      {M.~Cacciari, M.~Greco}
      {Large-$p_\perp$ Hadroproduction of Heavy Quarks}
      {Nucl.\ Phys.\ \underline{B421} (1994) 530}

\bibitemb{104}
       {M.~Gl\"{u}ck, E.~Reya, A.~Vogt}
       {Parton Structure of the Photon Beyond the Leading Order}
       {Phys.\ Rev.\ \underline{D45} (1992) 3986}

\bibitemb{105}
       {M.~Gl\"{u}ck, E.~Reya, A.~Vogt}
       {Dynamical Parton Distributions of the Proton
        and Small~$x$ Physics}
       {Z.~Phys.\ \underline{C67} (1995) 433}

\bibitemb{106}
       {M.~Gl\"{u}ck, E.~Reya, A.~Vogt}
       {Parton Distributions for High Energy Collisions}
       {Z.~Phys.\ \underline{C53} (1992) 127}

\bibitemb{107} 
      {CTEQ Collaboration}
      {Global QCD Analysis and the CTEQ Parton Distributions}
      {Phys.\ Rev.\ \underline{D51} (1995) 4763}

\bibitemb{108} 
      {A.~Martin, R.~Roberts, J.~Stirling}
      {Pinning Down the Glue in the Proton}
      {Phys.\ Lett.\ \underline{B354} (1995) 155}

\bibitemb{109} 
      {B.W.~Harris, J.~Smith, R.~Vogt}
      {Reanalysis of the EMC Charm Production Data with Extrinsic and
       Intrinsic Charm at NLO}
      {State Univ.\ of New York at Stony Brook
       preprint ITP--SB--95--15 (October 1995)}

\bibitemb{110} 
      {S.J.~Brodsky, P.~Hoyer, A.H.~Mueller, W.K.~Tang}
      {New QCD Production Mechanism for Hard Processes at Large~$x$}
      {Nucl.\ Phys.\ \underline{B369} (1992) 519}

\bibitemb{111} 
      {B.~Lampe}
      {One Loop QCD Corrections to the Inclusive
       Production Cross-Section for Heavy Quarks
       in $p\overline{p}$ Collisions at High Transverse Energies}
      {Fortschr.\ Phys.\ \underline{40} (1992) 329}

\bibitemb{112}
       {D.~Graudenz}
       {PROJET: Jet Cross Sections in Deeply Inelastic
        Scattering, Version 4.1}
       {Comput.\ Phys.\ Commun.\ \underline{92} (1995) 65}

\bibitemb{113} 
      {G.P.~Lepage}
      {A New Algorithm for Adaptive Multidimensional Integration}
      {J.~Comput.\ Phys.\ \underline{27} (1978) 192}

\bibitemb{114} 
      {G.P.~Lepage}
      {VEGAS --- An Adaptive Multi-Dimensional Integration Program}
      {Cornell preprint CLNS-80/447 (March 1980)}

\bibitemb{115}
      {K.~Charchu{\l}a}{The Package PAKPDF Version 1.1 of
       Parametrizations of Parton Distribution Functions
       in the Proton} 
      {Comput.\ Phys.\ Commun.\ \underline{69} (1992) 360}

\bibitemb{116} 
      {H.~Plothow-Besch}
      {PDFLIB: a Library of All Available Parton Density Functions
       of the Nucleon, the Pion and the Photon
       and the Corresponding $\alpha_s$ Calculations}
      {Comput.\ Phys.\ Commun.\ \underline{75} (1993) 396}

\bibitemb{117} 
      {W.J.~Marciano}
      {Flavour Thresholds and $\Lambda$ in the Modified Minimal-Subtraction
       Scheme}
      {Phys.\ Rev.\ \underline{D29} (1984) 580}

\bibitemb{118} 
      {W.~Bernreuther}
      {Decoupling of Heavy Quarks in Quantum Chromodynamics}
      {Ann.\ Phys.\ \underline{151} (1983) 127}

\bibitemb{119}
       {G.~Grunberg}
       {Renormalization Group Improved Perturbative QCD}
       {Phys.\ Lett.\ \underline{95B} (1980) 70, 
        Erratum Phys.\ Lett.\ \underline{110B} (1982) 501}

\bibitemb{120}
       {G.~Grunberg}
       {Renormalization Scheme Independent QCD and QED:
        The Method of Effective Charges}
       {Phys.\ Rev.\ \underline{D29} (1984) 2315}

\bibitemb{121} 
      {P.M.~Stevenson}
      {Resolution of the Renormalization Scheme Ambiguity
       in Perturbative QCD}
      {Phys.\ Lett.\ \underline{100B} (1981) 61}

\bibitemb{122} 
      {P.M.~Stevenson}
      {Optimized Perturbation Theory}
      {Phys.\ Rev.\ \underline{D23} (1981) 2916}

\bibitemb{123} 
      {P.M.~Stevenson}
      {Sense and Nonsense in the Renormalization-Scheme-Dependence
       Problem}
      {Nucl.\ Phys.\ \underline{B203} (1982) 472}

\bibitemb{124} 
      {P.M.~Stevenson}
      {Optimization and the Ultimate Convergence of QCD Perturbation Theory}
      {Nucl.\ Phys.\ \underline{B231} (1984) 65}

\bibitemb{125} 
      {S.J.~Brodsky, G.P.~Lepage, P.B.~Mackenzie}
      {On the Elimination of Scale Ambiguities in Perturbative
       Quantum Chromodynamics}
      {Phys.\ Rev.\ \underline{D28} (1983) 228}

\bibitemb{126} 
      {R.P.~Feynman}
      {Very High-Energy Collisions of Hadrons}
      {Phys.\ Rev.\ Lett.\ \underline{23} (1969) 1415}

\bibitemb{127} 
      {K.H.~Streng, T.F.~Walsh, P.M.~Zerwas}
      {Quark and Gluon Jets in the Breit Frame of
       Lepton--Nucleon Scattering}
      {Z.~Phys.\ \underline{C2} (1979) 237}

\bibitemb{128} 
      {S.J.~Brodsky, J.F.~Gunion, D.E.~Soper}
      {Physics of Heavy-Quark Production in Quantum Chromodynamics}
      {Phys.\ Rev.\ \underline{D36} (1987) 2710}

\bibitemb{129} 
      {J.C.~Anjos et al.}
      {Charm Photoproduction}
      {Phys.\ Rev.\ Lett.\ \underline{62} (1989) 513}

\bibitemb{130} 
      {J.C.~Anjos et al.}
      {Photon--Gluon-Fusion Analysis of
       Charm-Photoproduction}
      {Phys.\ Rev.\ Lett.\ \underline{65} (1990) 2503}

\bibitemb{131} 
      {M.P.~Alvarez et al.}
      {Branching Ratios and Properties of $D$-Meson
       Decays}
      {Z.~Phys.\ \underline{C50} (1991) 11}

\bibitemb{132} 
      {M.P.~Alvarez et al.}
      {Study of Charm Photoproduction Mechanisms}
      {Z.~Phys.\ \underline{C60} (1993) 53}

\bibitemb{133}
      {ZEUS Collaboration}
      {Study of $D^*(2010)^\pm$ Production 
       in $ep$ collisions at HERA}
      {Phys.\ Lett.\ \underline{B349} (1995) 225}

\bibitemb{134} 
      {J.~Binnewies, B.A.~Kniehl, G.~Kramer}
      {Next-to-Leading Order Fragmentation Functions
       for Pions and Kaons}
      {Z.~Phys.\ \underline{C65} (1995) 471}

\bibitemb{135} 
      {A.~Berera, D.E.~Soper}
      {Diffractive Jet Production in a Simple Model with Applications to 
       DESY HERA}
      {Phys.\ Rev.\ \underline{D50} (1994) 4328}

\bibitemb{136} 
      {S.J.~Brodsky, R.~Vogt}
      {QCD and Intrinsic Heavy Quark Predictions for Leading
       Charm and Beauty Hadroproduction}
      {preprint SLAC-PUB-6468 (April 1994)}

\bibitemb{137} 
      {A.H.~Mueller}
      {$O(2,1)$ Analysis of Single-Particle Spectra at High Energy}
      {Phys.\ Rev.\ \underline{D2} (1970) 2963}

\bibitemb{138} 
      {R.N.~Cahn, E.W.~Colglazier}
      {Unified Mueller Picture of Deep-Inelastic
       Inclusive Leptoproduction}
      {Phys.\ Rev.\ \underline{D8} (1973) 3019}

\bibitemb{139} 
      {D.~Horn, F.~Zachariasen}
      {Hadron Physics at Very High Energies} 
      {W.~Benjamin, Inc.\ 
      (Reading, Mass., 1973), 
      Chapter 6}

\bibitemb{140} 
      {A.H.~Mueller}
      {Cut Vertices and their Renormalization:
       A Generalization of the Wilson Expansion}
      {Phys.\ Rev.\ \underline{D10} (1978) 3705}

\bibitemb{141} 
      {S.~Gupta, A.H.~Mueller}
      {High-Energy Predictions in Quantum Chromodynamics}
      {Phys.\ Rev.\ \underline{D20} (1979) 118}

\bibitemb{142} 
      {A.H.~Mueller}
      {Perturbative QCD at High Energies}
      {Phys.\ Rep.\ \underline{73} (1981) 237}

\bibitemb{143} 
      {L.~Baulieu, E.G.~Floratos, C.~Kounnas}
      {Parton Model Interpretation of the Cut Vertex Formalism}
      {Nucl.\ Phys.\ \underline{B166} (1980) 321}

\bibitemb{144} 
      {E.G.~Floratos, D.A.~Ross, C.T.~Sachrajda}
      {Higher Order Effects in Asymptotically Free Gauge Theories.
       2.~Flavor Singlet Wilson Operators and Coefficient Functions}
      {Nucl.\ Phys.\ \underline{B152} (1979) 493}

\bibitemb{145}
      {G.~Ingelman and P.E.~Schlein}
      {Jet Structure in High Mass Diffractive Scattering}
      {Phys.\ Lett.\ \underline{152B} (1985) 256}

\bibitemb{146}
      {E.L.~Berger, J.C.~Collins, D.E.~Soper and G.~Sterman}
      {Diffractive Hard Scattering}
      {Nucl.\ Phys.\ \underline{B286} (1987) 704}

\bibitemb{147} 
      {D.~Graudenz, G.~Veneziano}
      {Estimating Diffractive Higgs Boson Production at LHC
       from HERA Data}
      {Phys.\ Lett.\ \underline{B365} (1996) 302}

\bibitemb{148} 
      {A.~Berera, D.E.~Soper}
      {Behaviour of Diffractive Parton Distribution Functions}
      {Pennsylvania State Univ. preprint PSU/TH/163 
       (September 1995)}

\bibitemb{149} 
      {I.M.~Gel'fand, G.E.~Shilov}
      {Generalized Functions, Vol.\ I}
      {Academic Press (New York, London, 1964)}

\bibitemb{150} 
      {B.W.~Char et al.}
      {Maple V Reference Manual}
      {Springer Verlag (Berlin, 1991)}

\bibitemb{151} 
      {S.~Wolfram}
      {Mathematica}
      {Addison Wesley (Reading, Mass., 1991)}

\bibitemb{152} 
      {A.C.~Hearn}
      {REDUCE User's Manual, Version 3.4.1}
      {ZIB Publication M~2.032.04 (July 1992)}

\bibitemb{153} 
      {M.~Jamin, M.E.~Lautenbacher}
      {Tracer Version 1.1: A Mathematica Package for $\gamma$-Algebra
       in Arbitrary Dimensions}
      {Comput.\ Phys.\ Commun.\ \underline{74} (1993) 265}

\bibitemb{154} 
      {W.H.~Press, B.P.~Flannery, S.A.~Teukolsky, W.T.~Vetterling}
      {Numerical Recipes in C} 
      {Cambridge University Press (Cambridge 1988)}

\bibitemb{155} 
      {N.N.~Lebedev}
      {Special Functions \& Their Applications}
      {Dover Publications, Inc. (New York, 1972)}

\bibitemb{156} 
       {H.~Hochstadt}
       {The Functions of Mathematical Physics}
       {Dover Publications, Inc. (New York, 1986)}

\bibitemb{157} 
      {M.~Abramowitz, I.A.~Stegun (eds.)}
      {Pocketbook of Mathematical Functions}
      {Verlag Harri Deutsch (Thun, Frankfurt/M., 1984)}

\bibitemb{158} 
      {W.~Furmanski, R.~Petronzio}
      {A Method of Analyzing the Scaling Violation
       of Inclusive Spectra in Hard Processes}
      {Nucl.\ Phys.\ \underline{B195} (1982) 237}

\bibitemb{159} 
      {H.~Bateman}
      {Higher Transcendental Functions I--III}
      {McGraw-Hill (1953)}

\bibitemb{160} 
      {I.~Gradstein, I.~Ryshik} 
      {Tables of Series, Products, and Integrals, Vols. I and II}
      {Verlag Harri Deutsch (Thun, Frankfurt/M., 1981)}

\bibitemb{161}
      {G.P.~Ramsey}
      {The Laguerre Method for Solving Integro-Differential Equations}
      {J.~Comput.\ Phys.\ \underline{60} (1985) 97}

\bibitemb{162} 
       {S.~Kumano, J.T.~Londergan}
       {A FORTRAN Program for the Numerical Solution of the
        Altarelli--Parisi Equations by the Laguerre Method}
       {Comput.\ Phys.\ Commun.\ \underline{69} (1992) 373}

\bibitemb{163} 
       {R.~Kobayashi, M.~Konuma, S.~Kumano}
       {FORTRAN program for a numerical solution of the nonsinglet
        Altarelli--Parisi equation}
       {Saga University preprint SAGA-HE-63-94 (August 1994)}

\bibitemb{164} 
       {R.~Kobayashi, M.~Konuma, S.~Kumano, M.~Miyama}
       {Numerical Solution of Altarelli--Parisi Equations}
       {Saga University preprint SAGA-HE-74-94 (December 1994)}
 
\end{thebibliography}

\dgcleardoublepage

\pagestyle{empty}
{
\addcontentsline{toc}{section}{List of Figures}
\listoffigures
\addcontentsline{toc}{section}{List of Tables}
\listoftables
}

\immediate\closeout\itemfile

\end{document}